\documentclass[logos,parttoc,mainlanguage=english,morelanguage=french]{orsay-thesis}

\pdfoutput=1

\usepackage[latin1]{inputenc}
\usepackage[T1]{fontenc}
 \usepackage{ae}
 \usepackage{aecompl}
\usepackage{subeqnarray}
\usepackage{theorem}
\usepackage{appendix} 
\usepackage{graphics}
\usepackage{graphicx}
\usepackage{enumerate}
\usepackage{verbatim}
\usepackage{bigcenter}
\usepackage{color}
\usepackage{pstricks}
\usepackage{bezier}
\usepackage[hang,footnotesize,bf]{caption}
\usepackage{hyperref}
\usepackage{cite}
\hypersetup{ 
pdfstartview=FitV,
colorlinks=true, 
linkcolor=blue, 
citecolor=red, 
filecolor=black, 
urlcolor=magenta,
pdfauthor={Julien Baglio},
pdfsubject={Thèse de doctorat/PhD thesis},
pdfkeywords={Standard Model, Higgs, Supersymmetry, QCD, theoretical uncertainties},
pdftitle={Phenomenology of the Higgs at the hadron colliders: from the
Standard Model to Supersymmetry.}}

\newcommand{\lsim}{\raisebox{-0.13cm}{~\shortstack{$<$ \\[-0.07cm] $\sim$}}~} 
\newcommand{\gsim}{\raisebox{-0.13cm}{~\shortstack{$>$ \\[-0.07cm] $\sim$}}~}
\newcommand{\tb}{\tan \beta}
\newcommand{\cotan}{{\,\rm cotan\,}}

\newcommand{\shat}{\hat{s}}
\newcommand{\lhc}{\rm lHC}

\newcommand{\be} {\begin{equation}}
\newcommand{\ee} {\end{equation}}
\newcommand{\beq} {\begin{eqnarray}}
\newcommand{\eeq} {\end{eqnarray}}
\newcommand{\bsea} {\begin{subeqnarray}}
\newcommand{\esea} {\end{subeqnarray}}

\newcommand{\D}{\mathcal{D}}

\newcommand{\G}{\mathcal G}

\newcommand{\A}{\mathcal A}

\newcommand{\qed}{\nobreak \ifvmode \relax \else
      \ifdim\lastskip<1.5em \hskip-\lastskip
      \hskip1.5em plus0em minus0.5em \fi \nobreak
      \vrule height0.75em width0.5em depth0.25em\fi}

\makeatletter
\DeclareRobustCommand*{\bfseries}{%
  \not@math@alphabet\bfseries\mathbf
  \fontseries\bfdefault\selectfont
  \boldmath
}
\makeatother

\numberwithin{equation}{section}

\author{Julien \textsc{Baglio}}

\title[french]{Phénoménologie du Higgs auprès des collisionneurs
  hadroniques :

 du Modèle Standard à la Supersymétrie.}
\title[english]{Phenomenology of the Higgs at the hadron colliders: from the
Standard Model to Supersymmetry.}

\keywords[french]{Modèle Standard, Higgs, Supersymétrie,
  Chromodynamique quantique, incertitudes théoriques}
\keywords[english]{Standard Model, Higgs, Supersymmetry, QCD, theoretical uncertainties}

\ordernumber{2011PA112193}

\date{Lundi 10 octobre 2011}

\addcommissionmember[Examinateur]{Pr.}{Guido}{Altarelli}
\addcommissionmember[Directeur de thèse]{Dr.}{Abdelhak}{Djouadi}
\addcommissionmember[Examinateur]{Pr.}{Ulrich}{Ellwanger}
\addcommissionmember[Rapporteur]{Dr.}{Louis}{Fayard}
\addcommissionmember[Rapporteur]{Dr.}{Massimiliano}{Grazzini}
\addcommissionmember[Examinateur]{Pr.}{Margarete}{M\"{u}hlleitner}
\addcommissionmember[Examinateur]{Dr.}{Gavin}{Salam}

\begin{document}

\maketitle%

\pagestyle{empty}
\pagenumbering{roman}


\selectlanguage{french}

\begin{abstract}[french]
Cette thèse, conduite dans le contexte de la recherche du boson de
Higgs, dernière pièce manquante du mécanisme de brisure de la symétrie
électrofaible et qui est une des plus importantes recherches auprès
des collisionneurs hadroniques actuels, traite de la phénoménologie de
ce boson à la fois dans le Modèle Standard (SM) et dans son extension
supersymétrique minimale (MSSM). Après un résumé de ce qui constitue
le Modèle Standard dans une première partie, nous présenterons nos
prédictions pour la section efficace inclusive de production du boson
de Higgs dans ses principaux canaux de production auprès des deux
collisionneurs hadroniques actuels que sont le Tevatron au Fermilab et
le grand collisionneur de hadrons (LHC) au CERN, en commençant par le
cas du Modèle Standard. Le principal résultat présenté est l'étude la
plus exhaustive possible des différentes sources d'incertitudes
théoriques qui pèsent sur le calcul : les incertitudes d'échelles vues
comme une mesure de notre ignorance des termes d'ordre supérieur dans
un calcul perturbatif à un ordre donné, les incertitudes reliées aux
fonctions de distribution de partons dans le proton/l'anti--proton
(PDF) ainsi que les incertitudes reliées à la valeur de la constante
de couplage fort, et enfin les incertitudes provenant de l'utilisation
d'une théorie effective qui simplifie le calcul des ordres supérieurs
dans la section efficace de production. Dans un second temps nous
étudierons les rapports de branchement de la désintégration du boson
de Higgs en donnant ici aussi les incertitudes théoriques qui pèsent
sur le calcul. Nous poursuivrons par la combinaison des sections
efficaces de production avec le calcul portant sur la désintégration
du boson de Higgs, pour un canal spécifique, montrant quelles en sont
les conséquences intéressantes sur l'incertitude théorique
totale. Ceci nous amènera à un résultat significatif de la
thèse qui est la comparaison avec l'expérience et notamment les
résultats des recherches du boson de Higgs au Tevatron. Nous irons
ensuite au-delà du Modèle Standard dans une troisième partie où nous
donnerons quelques ingrédients sur la supersymétrie et sa mise en
application dans le MSSM où nous avons cinq bosons de Higgs, puis nous
aborderons leur production et désintégration en se focalisant sur les
deux canaux de production principaux par fusion de gluon et fusion de
quarks $b$. Nous présenterons les résultats significatifs quant à la
comparaison avec aussi bien le Tevatron que les résultats très récents
d'ATLAS et CMS au LHC qui nous permettront d'analyser l'impact de ces
incertitudes sur l'espace des paramètres du MSSM, sans oublier de
mentionner quelques bruits de fond du signal des bosons de
Higgs. Tout ceci va nous permettre de mettre en avant le
deuxième résultat très important de la thèse, ouvrant une nouvelle
voie de recherche pour le boson de Higgs standard au LHC. La dernière
partie sera consacrée aux perspectives de ce travail et notamment
donnera quelques résultats préliminaires dans le cadre d'une étude
exclusive, d'un intérêt primordial pour les expérimentateurs.
\end{abstract}
\pagebreak\strut\newpage


\selectlanguage{english}

\begin{abstract}[english]
This thesis has been conducted in the context of one of the utmost important searches at current hadron colliders, that is the search for the Higgs boson, the remnant of the electroweak symmetry breaking. We wish to study the phenomenology of the Higgs boson in both the Standard Model (SM) framework and its minimal Supersymmetric extension (MSSM). After a review of the Standard Model in a first part and of the key reasons and ingredients for the supersymmetry in general and the MSSM in particular in a third part, we will present the calculation of the inclusive production cross sections of the Higgs boson in the main channels at the two current hadron colliders that are the Fermilab Tevatron collider and the CERN Large Hadron Collider (LHC), starting by the SM case in the second part and presenting the MSSM results, where we have five Higgs bosons and focusing on the two main production channels that are the gluon gluon fusion and the bottom quarks fusion, in the fourth part. The main output of this calculation is the extensive study of the various theoretical uncertainties that affect the predictions: the scale uncertainties which probe our ignorance of the higher--order terms in a fixed order perturbative calculation, the parton distribution functions (PDF) uncertainties and its related uncertainties from the value of the strong coupling constant, and the uncertainties coming from the use of an effective field theory to simplify the hard calculation. We then move on to the study of the Higgs decay branching ratios which are also affected by diverse uncertainties. We will present the combination of the production cross sections and decay branching fractions in some specific cases which will show interesting consequences on the total theoretical uncertainties. We move on to present the results confronted to experiments and show that the theoretical uncertainties have a significant impact on the inferred limits either in the SM search for the Higgs boson or on the MSSM parameter space, including some assessments about SM backgrounds to the Higgs production and how they are affected by theoretical uncertainties. One significant result will also come out of the MSSM analysis and open a novel strategy search for the Standard Higgs boson at the LHC. We finally present in the last part some preliminary results of this study in the case of exclusive production which is of utmost interest for the experimentalists.  
\end{abstract}

\pagebreak\strut\newpage



\selectlanguage{french}
\section*{Remerciements}
\vfill
Trois années ont passé depuis que j'ai poussé pour la première fois
les portes du Laboratoire de Physique Théorique d'Orsay,
chaleureusement accueilli par son directeur Henk Hilhorst que je
remercie beaucoup. Trois années
d'une activité intense, aussi bien dans mes recherches scientifiques
au LPT et au CERN, dans le groupe de physique théorique, où j'ai passé
quelques mois à partir de la seconde année, que dans mes activités
hors recherche au sein de l'université Paris-Sud 11. J'ai appris
beaucoup et rencontré un certain nombre de personnes dont je vais me
rappeler pour longtemps, si je ne les énumère pas ici qu'elles veuillent
bien me pardonner cela ne signifie pas que je les ai pour autant
oubliées.

Tout ceci n'aurait pu se faire sans les encouragements, les conseils
et les discussions passionnées avec Abdelhak Djouadi, mon directeur de
thèse qui a guidé ainsi mes premiers pas de professionnel dans ma
carrière de physicien théoricien des particules élémentaires. Je l'en
remercie profondément et j'espère qu'il aura apprécié notre
collaboration autant que moi, aussi bien lors de notre travail qu'en
dehors.

Je voudrais aussi remercier Rohini Godbole avec qui j'ai
collaboré sur la passionnante physique du Higgs au Tevatron. Je ne
peux non plus oublier Ana Teixeira pour son soutien constant et les
nombreuses discussions passionnantes aussi bien scientifiques
que personnelles que nous avons eues ensemble. Ma première année en
tant que doctorant lui doit beaucoup.

Je remercie aussi tous les membres de mon jury de thèse et en
particulier mes deux rapporteurs qui m'ont certainement maudit d'avoir
écrit autant, non seulement pour le temps qu'ils auront pris pour
assister à ma soutenance et lire ma thèse, mais aussi pour toutes
leurs judicieuses remarques et questions.

Aussi bien le LPT que le CERN se sont révélés des lieux très
enrichissants pour le début de ma carrière scientifique. Je voudrais
profiter tout d'abord de ces quelques mots pour remercier les équipes
administratives des deux laboratoires pour leur aide au jour le jour,
toujours avec le sourire, et pour toute leur aide dans mes divers
voyages scientifiques. Je remercie aussi tous les chercheurs de ces
deux laboratoires pour toutes les discussions que j'ai eues et qui
m'ont beaucoup appris. Je pense tout particulièrement à Asmâa Abada et
à Grégory Moreau d'un côté, à Géraldine Servant et Christophe Grojean
qui m'a invité à venir au CERN, de l'autre. Je ne peux
bien sur pas oublier les doctorants et jeunes docteurs du groupe de
physique théorique du CERN, Sandeepan Gupta, Pantelis Tziveloglou et
tous les autres, ainsi que Léa Gauthier, doctorante au CEA, que j'ai
rencontrée au CERN : les magnifiques randonnées autour de Genève que
nous avons faites ont été salutaires. Enfin je remercie aussi tous mes
camarades doctorants et jeunes docteurs du SINJE à Orsay pour tous les
merveilleux moments que nous avons passés et toutes les discussions
passionnées et passionnnantes, je ne vous cite pas tous mais le
c\oe{}ur y est. Je pense quand même tout particulièrement à mes
camarades ayant partagé mon bureau et bien plus, Adrien Besse et
Cédric Weiland, mais aussi à Guillaume Toucas, Blaise Goutéraux et
Andreas Goudelis. Jérémie Quevillon qui va prendre ma succession
auprès de mon directeur de thèse n'est pas non plus oublié. Mes amis
de Toulouse eux aussi sont loin d'avoir été oubliés et ont fortement
contribué non seulement à rendre exceptionnel mon stage de Master 2
mais aussi ma première année de thèse, de loin en loin : merci à
Ludovic Arnaud, Gaspard Bousquet, Arnaud Ralko, Clément Touya, Fabien
Trousselet, mais aussi mes deux tuteurs Nicolas Destainville et Manoel
Manghi.

Je ne peux terminer sans exprimer ma profonde gratitude à ma famille
et à mes amis de longue date, qui se reconnaîtront. Anne, Charles,
Elise, Gaetan, Lionel, Mathieu, Matthieu, Patrick, Pierre,
Rayna, Sophie, Yiting et tous ceux que je n'ai pas cités mais
qui sont dans mes pensées, ces mots sont pour vous ! Le mot de la fin
revient à ma fiancée, Camille : sans ton profond amour et ton soutien
constant, ces trois dernières années auraient été bien différentes, et
certainement pas aussi fécondes. Merci pour tout.

\vfill

\selectlanguage{english}

\section*{Acknowledgments}
\vfill
Three years have now passed since my first steps in the Laboratoire de
Physique Th\'eorique at Orsay, where I have been warmly welcomed by
its director Henk Hilhorst that I thank a lot. They have been very
intense, both in the laboratory and at the CERN Theory Group in
Geneva, where I spent some months starting from the second year. I
have learnt much, either within these labs or outside, encountered
many people that I will remember for a long time. If some of you are
not cited in these acknowledgments, please be kind with me: that does
not mean I have forgotten you.

This would have never been possible without the constant
encouragement, advices and fruitful discussions with Dr. Abdelhak
Djouadi, my thesis advisor, who guided my first steps in theoretical
particle physics research. I hope he got as much great time as I had
working with him and more than that.

I also would like to thank Pr. Rohini Godbole whom I worked with from
time to time on Higgs physics at the Tevatron. I cannot also forget
Dr. Ana Teixeira for her constant support and all the great discussions
on various topics we had together. My first year as a PhD candidate
was scientifically exciting thanks to her.

I am very grateful to all the members in the jury for my defence, for
the time they would took and the useful comments. In particular I
would like to thank my two referees who certainly have cursed me for
the length of the thesis.

The LPT environnement as well as the CERN Theory Group have been
proven to be very fruitful environnements for the beginning of my
career. I then would like to thank the administrative staff from both
laboratories for their constant help in day--to--day life and support
when I had to travel for various workshops, conferences or seminars. I
would like to thank all the members of these two groups for the very
passionate discussions we had and where I have learnt a lot. I dedicate
special thanks to Asm\^aa Abada and Gr\'egory Moreau on the one side,
G\'eraldine Servant and also Christophe Grojean, who invited me to come
by, on the other side. I cannot forget the PhD
candidates and post-doctoral researchers from the CERN Theory Group,
Sandeepan Gupta, Pantelis Tziveloglou and all the others, not to
forget L\'ea Gauthier, who is a PhD candidate at the CEA and was at
CERN at that time: the hiking we did in the Jura and Alps around
Geneva were great. I also would like to thank all my SINJE fellows at
the LPT, with whom I had so many great time and passionate
discussions; you are not all cited but I do not forget you. I dedicate
special thanks to my office (and more than office) friends Adrien
Besse and C\'edric Weiland, and also to Blaise Gout\'eraux, Andreas
Goudelis and Guillaume Toucas. The next PhD candidate, J\'er\'emie
Quevillon, who will follow my path, is also thanked for the
discussions we had. I finally cannot forget my friends from Toulouse,
where I did my Master 2 internship and whom I collaborated with during
my first PhD thesis year from time to time: many thanks to Ludovic
Arnaud, Gaspard Bousquet, Arnaud Ralko, Cl\'ement Touya, Fabien
Trousselet, and also to my two internship advisors Nicolas
Destainville and Manoel Manghi.

I now end this aknowledgments by expressing my deep gratitude and love
to my family and long--time friends who will recognize
themselves. Anne, Charles, Elise, Gaetan, Lionel, Mathieu,
Matthieu, Patrick, Pierre, Rayna, Sophie, Yiting and all the
others, these words are for you! The last word is for Camille, my
fiancee: without your deep love and constant support these three years
would have been without doubts completely different and not as
fruitful.

\vfill

\newpage


\strut\newpage
\pagebreak\strut\newpage

\begingroup
\hypersetup{linkcolor=black}
\tableofcontents
\vfill
\pagebreak
\listoffigures
\vfill
\pagebreak
\pagebreak\strut\newpage
\listoftables
\vfill 
\pagebreak
\selectlanguage{french}

\vfill
\section*{Liste des publications}

Cette page donne la liste de tous mes articles concernant le
travail réalisé depuis 3 ans.

\selectlanguage{english}

\noindent
This page lists all the papers that I have written for 3 years in the
context of my PhD work.\smallskip

\paragraph{Articles publiés (published papers) :\newline}

\hspace{3mm}{\it Predictions for Higgs production at the Tevatron and
  the associated uncertainties}, J. B. et A. Djouadi, JHEP {\bf 10}
(2010) 064;

{\it Higgs production at the lHC}, J. B. et A. Djouadi, JHEP {\bf 03}
(2011) 055;

{\it The Tevatron Higgs exclusion limits and theoretical
  uncertainties: A Critical appraisal}, J. B., A. Djouadi, S. Ferrag
et R. M. Godbole, Phys.Lett.{\bf B699} (2011)
368-371; erratum Phys.Lett.{\bf B702} (2011) 105-106;

{\it Revisiting the constraints on the Supersymmetric Higgs sector at
  the Tevatron}, J. B. et A. Djouadi, Phys.Lett.{\bf B699} (2011)
372-376;

{\it The left-right asymmetry of the top quarks in associated
  top--charged Higgs bosons at the LHC as a probe of the parameter
  $\tan\beta$}, J.B {\it et al.}, Phys.Lett.{\bf B705} (2011) 212-216.

\paragraph{Articles non--publiés (unpublished papers) :\newline}

\hspace{3mm}{\it Implications of the ATLAS and CMS searches in the
  channel $pp \to$ Higgs $\to \tau^+\tau^-$ for the MSSM and SM Higgs
  bosons}, J. B. et A. Djouadi, arXiv:1103.6247 [hep-ph] (soumis à
Phys.Lett.{\bf B});

{\it Clarifications on the impact of theoretical uncertainties on the
  Tevatron Higgs exclusion limits}, J. B., A. Djouadi et
R. M. Godbole, arXiv:1107.0281 [hep-ph].

\paragraph{Rapport de collaboration (review collaboration report)
  :\newline}

\hspace{3mm}{\it Handbook of LHC Higgs Cross Sections: 1. Inclusive
  Observables}, LHC Higgs Cross Section Working Group, S. Dittmaier
{\it et al.}, arXiv:1101:0593 [hep-ph].

\paragraph{Comptes--rendus de conférences (proceedings) :\newline}

\hspace{3mm}{\it Higgs production at the Tevatron: Predictions and
  uncertainties}, J. B., ICHEP 2010, Paris (France), PoS {\bf
  ICHEP2010} (2010) 048;

{\it The Supersymmetric Higgs bounds at the Tevatron and the LHC},
J.B., XLVI$^{e}$ Rencontres de Moriond, EW interactions \& unified
theory, La Thuile (Italie), arXiv:1105.1085 [hep-ph].

\vfill
\vfill
\endgroup

\pagebreak\strut\newpage
\pagebreak
\ 
\vspace{-14cm}
{\it \begin{verse}
Cette thèse est dédiée à mon père et à mes deux grand-pères, disparus
bien trop tôt.
\end{verse}}
\pagebreak\strut\newpage
\pagebreak
\includegraphics[scale=0.4]{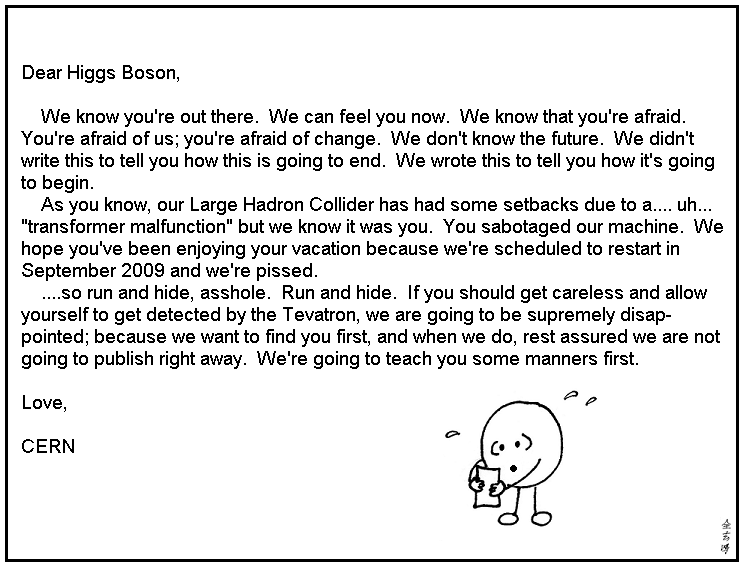}

\vspace{-5mm}
{\tiny (From {\tt http://abstrusegoose.com/118})}
\newline
\vspace{1cm}
{\it \footnotesize \begin{verse}
Et maintenant, apprends les vérités qui me restent à te découvrir,\\
Tu vas entendre de plus claires révélations.\\
Je n'ignore pas l'obscurité de mon sujet ;
\end{verse}}
{\tiny Lucrèce, dans \underline{De rerum natura}, v. 902-943 livre I}
\newline
\vspace{1cm}
{\it \footnotesize \begin{verse}
Les amoureux fervents et les savants austères\\
Aiment également, dans leur mûre saison,\\
Les chats puissants et doux, orgueil de la maison,\\
Qui comme eux sont frileux et comme eux sédentaires.
\end{verse}}
{\tiny Charles Baudelaire, dans \underline{Les Fleurs du Mal}}
\pagebreak\strut\newpage
\pagebreak\strut\newpage
\vfill
\selectlanguage{english}

\pagestyle{fancy}



\pagenumbering{arabic}
\selectlanguage{english}
\section*{Introduction\markboth{Introduction}{Introduction}}
\addcontentsline{toc}{section}{\protect\numberline{}Introduction}

In this thesis, we wish to present some predictions for the Higgs
boson(s) study at the two largest hadron colliders currently in
activity: the Fermilab Tevatron collider and the CERN Large Hadron
Collider (LHC). Our focus will be on the inclusive production cross
sections and the decay branching fractions, first in the Standard
Model which in itself is the topic of part~\ref{part:one} and then in
its minimal supersymmetric extension which is the topic of
part~\ref{part:three}.

The study of the fundamental mechanisms of Nature at the elementary
level has a long story and has known many milestones in the past sixty
years. Physicists have built a theory, nowadays known as the Standard
Model, to describe the elementary particles and their interactions,
that are those of the strong, weak and electromagnetic, the two last
being unified in a single electroweak interaction. It relies on the
elegant concept of gauge symmetry within a quantum field theory
framework and has known many experimental successes: despite decades
of effort to surpass this model it is still the one that describes
accurately nearly all the known phenomena\footnote{We leave aside the
  neutrino mass issue.}. One of its key concepts is the spontaneous
breakdown of electroweak symmetry: indeed in order to give mass to the
weak bosons that mediate the weak interaction, a scalar field is
introduced in the theory whose vacuum breaks the electroweak symmetry
and gives mass to the weak bosons. In fact it also gives masses to the
fermions and one piece of this mechanism remains to be discovered: the
Higgs boson, the ``Holy Grail'' of the Standard Model. Its discovery
is one of the main goal of current high energy colliders.

It is then of utmost importance to give theoretical predictions for
the production cross sections and decay branching fractions of the
Higgs boson at current colliders to serve as a guideline for
experiments. However, the hadronic colliders are known to be very
difficult experimental environments because of the huge hadronic, that
is Quantum ChromoDynamics (QCD), activity. This is also true on a
theoretical side, which means that an accurate description of all
possible sources of theoretical uncertainties is needed: this is
precisely the main output of this thesis. We shall mention that in the
very final stage of this thesis new results have been presented in the
HEP--EPS 2011 conference; our work is to be read in the light of the
results that were available before these newest experimental output
which will be briefly commented in the conclusion.

Part~\ref{part:one} is entirely devoted to a review of the Standard
Model. In section~\ref{section:Intro} we will draw a short history of the
Standard Model and list its main milestones of the past sixty years,
followed by a description of its main concepts. We will go into more
details about the Higgs mechanism, which spontaneously breaks
electroweak symmetry, in section~\ref{section:Higgs}: we will review
some reasons to believe that either the Higgs mechanism itself or
something which looks like the Higgs mechanism is needed, and then how
the Higgs boson emerges from the electroweak symmetry breaking and
what are its couplings to fermions and bosons of the Standard Model.

Part~\ref{part:two} is the core of the Standard Model study of this
thesis. Indeed the Higgs boson remains to be discovered and is one of
the major research programs at current high energy colliders. The old
CERN Large Electron Positron (LEP) collider has put some bounds on the
possible value of the Higgs boson mass, which is above 114.4 GeV in
the Standard Model at 95\%CL. We will review in
section~\ref{section:HiggsBounds} the current experimental and
theoretical bounds on the Higgs mass. We then give our predictions for
the Standard Model Higgs boson inclusive production cross section at
the Tevatron in the two main production channels that are the
gluon--gluon fusion and the Higgs--strahlung processes, giving all the
possible sources of theoretical uncertainties: the scale uncertainty
viewed as an estimation of the unknown higher--order terms in the
perturbative calculation; the parton distribution functions (PDFs)
uncertainties related to the non--perturbative QCD processes within
the proton, and its related strong coupling constant issue; the
uncertainty coming from the use of an effective theory approach to
simplify the hard calculation in the gluon--gluon fusion process. We
will specifically address the issue of the combination of all the
uncertainties in section~\ref{section:SMHiggsTevTotal}. We will then
move on to the same study at the LHC, concentrating on its current run
at a 7 TeV center--of--mass energy that we will name as the lHC for
littler Hadron Collider; we will still give some predictions for the
designed LHC at 14 TeV. We will finish this part~\ref{part:two} by the
Higgs boson decay branching fractions predictions in
section~\ref{section:SMHiggsDecay}, together with a detailed study of
the uncertainties that affect these predictions. It will be followed
by the combination of the production cross sections and decay
branching fractions into a single prediction, first at the Tevatron in
section~\ref{section:SMHiggsFinalTev} and then at the lHC in
section~\ref{section:SMHiggsFinalLHC}. We will then study the impact
of our uncertainties on the Tevatron Higgs searches in
section~\ref{section:SMHiggsTevExclusion} and in particular put into
question the Tevatron exclusion limits that are debated within the
community.

Even if the Standard Model is a nice theory with great experimental
successes, it suffers from some problems, both on the theoretical and
experimental sides. It is known for example that the Higgs boson mass
is not predicted by the Standard Model, and even not protected: higher
order corrections in the perturbative calculation of the Higgs boson
mass have the tendency to drive the mass up to the highest acceptable
scale of the theory which means that we need a highly fine--tuning of
the parameters to cancel such driving. It is known as the naturalness
problem of the Standard Model. They are several ways to solve such a
problem, and one of them is particularly elegant and relies on a new
symmetry between bosons and fermions: supersymmetry. This theoretical
concept, born in the 1970s, has many consequences when applied to the
Standard Model of particle physics and is actively searched at current
high energy colliders. This will be the topic of part~\ref{part:three}
in which we will review some of the reasons that drive the theorists
to go beyond the Standard Model and in particular what makes
supersymmetry interesting in this view in
section~\ref{section:SUSYIntro}, then move on to the description of
the mathematical aspects of supersymmetry in
section~\ref{section:FormalSUSY}. We will finish this
part~\ref{part:three} by a very short review of the minimal
supersymmetric extension of the Standard Model, called the MSSM, in
section~\ref{section:MSSM}. We will in particular focus on the Higgs
sector of the theory and show that the MSSM needs two Higgs doublets
to break the electroweak symmetry breaking and has thus a rich Higgs
sector as five Higgs boson instead of a single one are present in the
spectrum: two neutral $CP$--even, one $CP$--odd and two charged Higgs
bosons.

After this review of supersymmetry and the MSSM we will reproduce in
part~\ref{part:four} the same outlines that have been developed in
part~\ref{part:two} in the Standard Model case. We will first review
the neutral Higgs sector at hadron colliders in
section~\ref{section:MSSMHiggsIntro} and show that we can have a quite
model--independent description for our predictions in the sense that
they will hardly depend on most of the (huge) parameters of the MSSM
but two of them, the mass of the $CP$--odd Higgs boson $A$ and the
ratio $\tb$ between the vacuum expectation values of the two Higgs
doublets. We will then give in section~\ref{section:MSSMHiggsTev} our
theoretical predictions for the neutral Higgs bosons inclusive
production cross section at the Tevatron in the two main production
channels that are the gluon--gluon fusion and the bottom quark
fusions, the bottom quark playing a very important role in the MSSM at
hadron colliders. We will reproduce the same study at the lHC in
section~\ref{section:MSSMHiggsLHC} before giving the implications of
our study on the $[M_A,\tb]$ parameter space in
section~\ref{section:MSSMHiggsExp}. We will first give in this last
section our predictions for the main MSSM decay branching fractions
and in particular the di--tau branching fraction that is of utmost
importance for experimental searches. We we will then compare our
predictions together with their uncertainties to the experimental
results obtained at the Tevatron and at the lHC that has now been
running for more than a year at 7 TeV and given impressive results. We
will see that the theoretical uncertainties have a significant impact on
the Tevatron results, less severe at the lHC. We will finish
section~\ref{section:MSSMHiggsExp} by a very important outcome of our
work: the possibility of using the MSSM neutral Higgs bosons searches
in the di--tau channel for the Standard Model Higgs boson in the
gluon--gluon fusion production channel followed by the di--tau decay
channel in the low Higgs boson mass range 115--140 GeV.

Finally, we will give an outlook and draw some conclusions in
part~\ref{part:five} together with some perspectives for future
work. These rest on the next step on the road of the experiments, that
is an exclusive study of the Higgs bosons production channels. We
shall give some early results in section~\ref{section:Exclusive} on
the Standard Model Higgs boson at the lHC in the $gg\to H\to WW\to
\ell \nu \ell \nu$ search channel together with an exclusive study of
the main Standard Model backgrounds. This is also the current roadmap
of the Higgs bosons theoretical community and this work is done in the
framework of a collaboration on this topic.

\vfill


\part{A brief review of the Standard Model of particle physics}
\label{part:one}

\section{Symmetry principles and the zoology of the Standard Model}

The Standard Model (SM) of particle physics is the current description
of the fundamental constituents of our universe together with the
interactions that occur between them. The SM was born in its current
form in the seventies, after nearly twenty years of many experiments and
theoretical reflexions on how to build a somewhat simple and elegant
model to describe accurately the experimental results on the one hand
and to make powerful predictions in order to have a falsifiable theory
on the other hand. Its frameworks are relativistic quantum field theory
and group theory to classify the different interactions. It also needs
the key concept of spontaneous (electroweak) symmetry breaking in
order to account for the masses of the different fields in the theory,
the (weak) bosons as well as the matter fermions. Other reasons also
push for such a theoretical concept and will be presented in the next
sections.

We will in this section present a short review of the major historical
points in the birth of the SM, and present its theoretical
fundations. The focus on the electroweak symmetry breaking, in
particular its minimal realization through the Brout--Englert--Higgs
mechanism, will be discussed in the next section.

\label{section:Intro}

\subsection{A brief history of the Standard
  Model \label{section:IntroHistory}}

This subsection will sketch the different historical steps that have
lead to the current form of the theory that describes the elementary
particles and their interactions among each other, called the Standard
Model (SM). This model has a very rich history over more than fifty
years of the XX$^{\rm th}$ century, not to mention all the diverse and
fruitful efforts made before to attain this level of description of
the elementary world. We will only select some (of the) outstanding
events, both from the theoretical and experimental sides, to present
the twisted path leading to the current Standard Model of particle
physics.

\paragraph{The birth of modern QED\newline}

The first attempt to decribe electromagnetic phenomena in the
framework of special relativity together with quantum mechanics can be
traced back in the 1920s. In particular Dirac was the first to
describe the quantization of the electromagnetic fields as an ensemble
of harmonic oscillators, and introduced the famous
creation--annihilation operators~\cite{Dirac:1927dy}. In 1932 came
Fermi with a first description of quantum
electrodynamics~\cite{RevModPhys.4.87}, but physicists were blocked by
the infinite results that did arise in the calculations beyond the
first order in perturbation theory.

Years after, the difficulty was solved by Bethe in
1947~\cite{Bethe:1947id} with the
concept of renormalization, that is the true physical quantities are
not the bare parameters of the theory, and thus the infinite that
arise are absorbed in the physical quantities, leaving finite results
in the end. This leads to the modern Quantum ElectroDynamics (QED)
with the key concept of gauge symmetry and renormalization, that was 
formulated by Feynman, Schwinger and
Tomonaga~\cite{Tomonaga:1946zz,Dyson:1949bp,Feynman:1950ir} in the
years 1950s and awarded by a Nobel prize in 1965. This is the first
quantum field theory available and has been the root of all the SM
ideas for the key concepts of gauge symmetry and renormalizability.

\paragraph{P violation and $V-A$ weak theory\newline}

It was long considered in physics that the parity symmetry was
conserved: if we repeated an experiment with the experimental
apparatus mirror reversed, the results would be the same as for the
initial set--up. This assessment is true for any experiment involving
electromagnetism or strong interaction, but that is not the case for
weak interaction.

It was first proposed by Yang and Lee in 1956 that the weak
interaction might indeed not respect
P--symmetry~\cite{Lee:1956qn}. This was observed in 1957 by
Chien-Shiung Wu (``Madam Wu'') in the beta desintegration of cobalt 60
atoms~\cite{Wu:1957my}. Yang and Lee were then awarded the 1957 Nobel
prize for their theoretical developments on this concept.

Up until that period, the weak interaction, that shapes the decay of
unstable nucleii, was described by the Fermi theory in which the
fermions interact through a four--particles vertex. The discovery of
the $P$--violation lead to the construction of an effective $V-A$
theory where the tensor structure of the thory is correct and does
respect the charge and parity violations. This $V-A$ theory was later
on replaced by the electroweak theory, see below.

\paragraph{The quark description\newline}

In the first half of the XX$^{\rm th}$ century the pattern of
elementary particles was simple: the electron (and its antiparticle
the positron, postulated by Dirac in 1931 and discovered in 1932 by
Anderson), the proton and the neutron were the only known elementary
particles at that time. The neutrino, first postulated by Pauli in its
famous letter in 1930 to save the energy--momentum conservation in
beta decay reactions\footnote{The original name was ``neutron'' for
  neutral particle. Chadwick discovered in 1932 what would be the
  neutron, thus Fermi proposed the name ``neutrino'' meaning ``little
  neutral one'' in italian.} was discovered only in 1956.

Experimental particle physicists discovered numerous new particles
(the ``hadrons'')  in the 1950s and 1960s after the discovery of the
pion in 1947, predicted by Yukawa in 1935, thus casting some doubts on
the elementary nature both of the ``older'' particles such as the
neutron and the proton and on the new zoo discovered. Gell--Man and
Zweig proposed in 1964 a model of constituant particles of these
hadrons and mesons that could explain the pattern seen by
experimentalists, using only a limited number of new constituant particles: the
quarks~\cite{GellMann:1964nj,Zweig:1981pd}. They introduce the $SU(3)$
flavor symmetry with the three up, down and strange quarks. One year
later the charm quark was proposed to improve the description of weak
interactions between quarks, and in 1969 deep inelastic scattering
experiments at the Stanford Linear Accelerator Center (SLAC)
discovered point--like objects within the
proton~\cite{Breidenbach:1969kd}, an
experimental proof of the compositeness of the hadrons. It is
interesting to note that the term used for these new point--like
objects was ``parton'', proposed by Feynman, as the community was not
entirely convinced that they were indeed the Gell--Mann's
quarks. Nowadays ``parton'' is still a word used in particle physics
to name the different constituants of the hadrons (the quarks,
antiquarks and gluons, the later being the bosons of the strong
interaction).

The (nearly) final word on the quark model was given in 1974 when the
$J/\Psi$ meson was discovered~\cite{Aubert:1974js,Augustin:1974xw} and
thus proved the existence of the 
charm quark, which was proposed by Glashow, Iliopoulos and Maiani in
the GIM mechanism~\cite{Glashow:1970gm} in 1970 to explain the
universality of weak interaction in 
the quark sector, preventing flavor changing neutral currents. The
heaviest quark, that is the top quark, was finally 
discovered in 1995 at the Fermilab Tevatron
collider~\cite{Abe:1995hr,Abachi:1995iq}.

\paragraph{CP violation and the concept of generation\newline}

To explain both the universality and the $u\longleftrightarrow d$
transitions in weak interactions, Cabibbo introduced in 1963 what is
known as the Cabibbo angle~\cite{Cabibbo:1963yz} and was used to
write in the mass eigenstates basis the weak eigenstate for the
down quark $d$. A year later, Cronin and his collaborators discovered
that not only C and P symmetries are broken by weak interactions, but
also the combined CP symmetry~\cite{Christenson:1964fg}, studing the
$K^0\overline K^0$ oscillations: the probability of oscillating from
$K^0$ state into $\overline K^0$ state is different from that of the
$\overline K^{0} \to K^0$, indicating that $T$ time reversal symmetry
is violated. As the combined CPT is assumed to be conserved, this
means that CP is violated.

As mentioned a few lines above, the GIM mechanism introduced a fourth
quark, the charm quark $c$. It then restores universality in the weak
coupling for the quarks, as we have now two weak eigenstates

\beq
|d'\rangle = \cos \theta_{\rm c} |d\rangle + \sin\theta_{\rm
  c}|s\rangle\nonumber\\
|s'\rangle = -\sin \theta_{\rm c} |d\rangle + \cos\theta_{\rm
  c}|s\rangle
\eeq

coupled to respectively the $u$ quark and the $c$ quark. We thus have
two generations in the quark sector, the first one is the $(u,d)$
doublet and the second one is the $(c,s)$ doublet. However, as
explained in 1973 by Kobayashi and Maskawa extending the work
initiated by Cabibbo, this is not sufficient to explain the CP
violation observed by the 1964 experiment. Only with three generations
could be introduced some CP violating effects through a phase angle,
and thus extending the Cabbibo angle to what is known as the
Cabibbo--Kobayashi--Maskawa (CKM)
matrix~\cite{Kobayashi:1973fv}. Kobayashi and Maskawa were awarded
the 2008 Nobel prize for this result\footnote{Unfortunately the Nobel
  committee failed to recognize the important pionnering work from
  Cabibbo.}.

\paragraph{Yang--Mills theory and spontaneous symmetry
  breaking\newline}

We have seen a few lines above that the Fermi theory describing the
weak interactions had been refined by the $V-A$ picture to take into
account the $P$ violation. Still the $V-A$ theory was known to be an
effective theory as the theory was not renormalizable and did not
allow for calculations beyond the first order in perturbation
theory. The only gauge theory that was available at that time was QED,
an abelian gauge theory, which obviously is not the right description
of weak processes as it describes only light--matter interactions.

The first step toward the solution was set--up in 1954, when
Yang and Mills developed a formulation of non--abelian gauge
theories~\cite{Yang:1954ek}  in order to provide (initially) an
explanation for the strong interaction at the hadron level (that we
call nuclear interaction). Unfortunately the theory was not a success
at first, as the gauge bosons must remain massless to preserve the
symmetry of the theory, thus meaning that the weak interaction should
be long--range; experimentally that is not the case.

The key result to solve this contradiction and then still use the
elegant description of gauge theory is given in 1964 by Brout,
Englert, Higgs, Guralnik, Hagen and Kibble after some important work
on the concept of symmetry breaking from Nambu and Goldstone: the
spontaneously gauge symmetry
breaking~\cite{Higgs:1964ia,Englert:1964et,Guralnik:1964eu,Higgs:1966ev}
described by the Brout--Englert--Higgs mechanism. This will
be presented in the following in details, but we can already remind
the reader that the most important result is that it allows for the
use of a Yang--Mills theory together with a description of massive
gauge bosons for any gauge theory.

\paragraph{Interlude: from nuclear force to strong interaction\newline}

Before arriving to the final electroweak description that constitutes
the heart of the SM, we recall the road leading to the description of
the strong interaction between the quarks.

As stated above, Yang--Mills theory in 1954 was the first attempt to
describe the interaction between the hadrons, that we call nuclear
interaction, in a gauge formulation. After the introduction of the
quark model by Gell--Mann in 1964 (see above) and the discovery of the
quarks in 1969 (see above), it has been proposed that the quarks must
have a new quantum charge, called color, to accomodate for the Pauli
exclusion principle within some baryons~\cite{Greenberg:1964pe}. This
was experimentally observed in the SLAC experiments in 1969 which
discovered point--like objects within the nucleon, as discussed
earlier.

With the help of the discovery of asymptotic
freedom~\cite{Gross:1973id,Politzer:1973fx} in 1973 by Wilczek, Gross
and Politzer (who share the 2004 Nobel prize for this result), that
states that at very high energy quarks are free, and with a $SU(3)$
gauge Yang--Mills theory, Quantum ChromoDynamics (QCD) was firmly
established in the 1970s as being the theory of the strong
interactions, with the gluons as the gauge bosons. Evidence of gluons
was discovered in three jet events at PETRA in
1979~\cite{Berger:1979cj}, giving further credits to QCD.

The nuclear interaction between the hadrons is then a residual force
originating from the strong interaction between quarks (and
gluons). However, as the strong coupling is indeed very strong at
large distance (that is the confinement), preventing from the use of
perturbation theory, an analytical description of the strong
interaction within the hadrons at low energies is still to be
found. This problem is now studied within the framework of lattice
gauge theories which give spectacular results.

\paragraph{The weak neutral currents and the path to electroweak theory\newline}

As stated above it was known that the $V-A$ theory for the weak
interaction was an effective theory, with difficulties calculating
beyond the first order in perturbation theory. With the advent of
Yang--Mills theory and the Brout--Englert--Higgs mechanism, describing
the weak interaction with a gauge theory and in the same time allowing
for massive weak bosons as dictated by the experiments, the weak
interaction being a short distance interaction, it would be possible
to account for a renormalizable description of the weak interaction.

During the 1960s there were many attempts to carry on this roadmap,
trying lots of different gauge groups to account for the QED on the
one hand, the weak interaction on the other hand, as both interactions
play a role for lepton particles such as the electron. The gauge
theory that did emerge was the $SU(2)\times U(1)$ model where the weak
and electromagnetic interactions are unified in a single gauge theory
description\footnote{It is actually not a complete unified theory as the algebra
describing the electroweak interaction is a product of two Lie
algebras. Nevertheless as the decription of the weak and
electromagnetic interactions are intimely connected through the
pattern of the electroweak symmetry breaking, see below, this can be
viewed as at least a partial unification.}, with contributions
notabely from Glashow~\cite{Glashow:1961tr},
Salam~\cite{Salam:1964ry} and Weinberg~\cite{Weinberg:1967tq}. This
model together with the Brout--Englert--Higgs mechanism predicts in
particular that there should be a neutral weak boson $Z^0$ to be
discovered and thus neutral currents.

A very important theoretical discovery was made in 1971 by 't Hooft,
who demonstrated that the electroweak theory is indeed
renormalizable~\cite{'tHooft:1971fh,'tHooft:1971rn}. A year after,
together with Veltman, he also described for the first time the
dimensional regularization scheme for renormalizability
calculation~\cite{'tHooft:1972fi}. That gave
prominant credits to the electroweak model 
as renormalizability allows for finite calculation at any order in perturbation
theory. Physicists started to actively look for neutral currents which
would be a great success for the electroweak theory.

The thrilling experimental discovery of these neutral currents happened
in 1973 at CERN in the Gargamelle detector~\cite{Hasert:1973ff}. This
established the electroweak theory as the correct theory describing
the weak and electromagnetic interactions in an unified gauge theory
view, preserving renormalizability and allowing for massive weak
bosons. Glashow, Weinberg and Salam were awarded the 1979 Nobel prize
for the electroweak model,  't Hooft was awarded the 1999 Nobel prize
for the proof of its renormalizability.

The electroweak theory together with QCD describing strong
interactions is what is called the Standard Model (SM) of particle physics.

\paragraph{$W$ and $Z$ bosons: the experimental success\newline}

The final paragraph of this short history of the SM is
devoted to the direct discovery of the massive $W$ and $Z$ in
1983. Indeed after the great success of the year 1973 where the weak
neutral currents, predicted by the electroweak theory, were discovered
at CERN, the weak bosons remained to be discovered.

The UA1 and UA2 experiment at the SPS proton--antiproton collider at
CERN, conducted by Van der Meer and Rubbia, did the great discovery in
1983~\cite{Arnison:1983mk,Arnison:1983rp}. They were awareded the 1984
Nobel prize, an unexpected fast acknowledgement from the Nobel
committee.\bigskip

We are arrived at the end of our path along the history of the SM. This
subsection does not pretend to cover all its aspects, sketching only
some landmarks. What remains to be done? As put in light above the
Brout--Englert--Higgs mechanism plays a key role in the formulation of
the SM. However the Higgs boson related to this mechanism has yet to
be found, despite nearly thirty years of intense research both at the
CERN and Fermilab colliders. This motivates not only this thesis, but
also the very exciting research program of current high energy
colliders! We are all waiting for decisive answers in the coming
years...

\subsection{Gauge symmetries, quarks and
  leptons \label{section:IntroSymmetries}}

Now that we have seen the major pages of the history of the Standard
Model (SM), we review its content. Most of this subsection is based on
Ref.~\cite{Quigg:1983} and the lecture notes of a master course on particle
physics and gauge theories~\cite{Binetruy:2008, Davier:2008}. As
briefly discussed in the historical part of this section, the key
concept behing the SM is that of symmetries. We will thus give the
symmetry algebra of the SM that shapes the interactions between the
fundamental particles.

\paragraph{The particle content\newline}

Each interaction is described by a quantum field theory based on
a Lie algebra which describes the gauge symmetry of the
interaction. The Lie algebra of the SM is $\left(SU(2)_{L}\times
U(1)_{Y}\right) \times SU(3)_c$~:

\begin{itemize}
\item{$SU(3)_c$ describes the strong interaction among the quarks
    which are the colored fermions of the theory. This is the QCD
    theory.}
\item{$SU(2)_L \times U(1)_Y$ describes the electroweak interaction
    between the quarks and the leptons. $Y$ is the hypercharge of the
    fermions that are charged under $U(1)_Y$. This is the electroweak
    theory.}
\end{itemize}

Each fermion is classified in irreducible representations of each
individual Lie algebra, that is according to its quantum numbers $C,
I, Y$ which are respectively the color, the weak isospin and the
hypercharge. The electric charge is then given by the
Gell--Mann--Nishijima relation $\displaystyle
Q=I^{W}_3+\frac{Y}{2}$. The SM fermion part is organised in three
generations, which are three times a replication of the same
structure: one charged lepton and one neutrino on the one hand, two
quarks on the other hand.

The weak interaction maximally violates the parity symmetry: $SU(2)_L$
Lie algebra acts only on left--handed fermions. Apart from the
neutrinos which are only left--handed in the SM and
massless\footnote{Experimentally it has been proved that the neutrinos
  are indeed massive. The existence of the right--handed neutrinos is
  then still an open question.}, all others fermions are both left and
right--handed. Table~\ref{table:SMfermioncontent} below gives the
fermionic content of the SM \footnote{The value of the $b$ and
   $c$--quark masses will be discussed in more details in
   section~\ref{section:SMHiggsDecayResult}.}.

\begin{table}[!h]
{\small%
\let\lbr\{\def\{{\char'173}%
\let\rbr\}\def\}{\char'175}%
\renewcommand{\arraystretch}{1.35}
\vspace*{2mm}
\begin{center}
\begin{tabular}{|c|c|c|ccc|c|}\hline
Type & Name & Mass~[GeV] & Spin & Charge &
$\left(I^W,I^W_3\right)_{L}$ & $SU(3)_c$ rep. \\ \hline
 & $\nu_e$ & $< 2\times 10^{-6}$ & $1/2$ & $0$ & $(1/2,1/2)$
 & $\mathbf{1}$ \\
 & $e$ & $5.11\times 10^{-4}$ & $1/2$ & $-1$ & $(1/2,-1/2)$
 & $\mathbf{1}$ \\ \cline{2-7}
 & $\nu_\mu$ & $< 1.9\times 10^{-4}$ & $1/2$ & $0$ & $(1/2,1/2)$
 & $\mathbf{1}$ \\
 LEPTONS & $\mu$ & $1.06\times 10^{-1}$ & $1/2$ & $-1$ & $(1/2,-1/2)$
 & $\mathbf{1}$ \\ \cline{2-7}
 & $\nu_\tau$ & $< 1.82\times 10^{-2}$ & $1/2$ & $0$ & $(1/2,1/2)$
 & $\mathbf{1}$ \\
 & $\tau$ & $1.777$ & $1/2$ & $-1$ & $(1/2,-1/2)$
 & $\mathbf{1}$ \\ \hline
 & $u$ & $(1.5\leq m\leq 3.0)\times 10^{-3}~(\overline{\rm MS})$ &
 $1/2$ & $2/3$ & $(1/2,1/2)$
 & $\mathbf{3}$ \\
 & $d$ & $(3.0\leq m\leq 7.0)\times 10^{-3}~(\overline{\rm MS})$ &
 $1/2$ & $-1/3$ & $(1/2,-1/2)$
 & $\mathbf{3}$ \\ \cline{2-7}
 & $c$ & $1.28~(\overline{\rm MS})$ & $1/2$ & $2/3$ & $(1/2,1/2)$
 & $\mathbf{3}$ \\
 QUARKS & $s$ & $9.5\times 10^{-2}~(\overline{\rm MS})$ & $1/2$ &
 $-1/3$ & $(1/2,-1/2)$ & $\mathbf{3}$ \\ \cline{2-7}
 & $t$ & $173.1$ & $1/2$ & $2/3$ & $(1/2,1/2)$
 & $\mathbf{3}$ \\
 & $b$ & $4.16~(\overline{\rm MS})$ & $1/2$ & $-1/3$ & $(1/2,-1/2)$
 & $\mathbf{3}$ \\ \hline

\end{tabular} 
\end{center} 
\caption[The fermionic content of the Standard Model]{The fermionic
  content of the Standard Model. Apart from the neutrinos which are
  left--handed only, all other fields are left and right--handed, the
  latter being $SU(2)_L$ gauge singlets.}
\label{table:SMfermioncontent}
\vspace*{-2mm}
}
\end{table}

In addition to the fermionic content there are the gauge bosons mediating
the interactions and one scalar boson, the Higgs boson that we have
already seen in the brief history of the SM. Before the electroweak
symmetry breaking, we have in the electroweak sector:

\begin{itemize}
\item{The three $W_\mu^{a}$ weak bosons that are the generators of the
    $SU(2)_L$ gauge algebra.}
\item{The neutral $B_\mu$ boson that is the generator of the $U(1)_Y$
    gauge algebra.}
\item{The scalar Higgs doublet $\Phi=(\phi^+,\phi^0)$.}
\end{itemize}

We also have 8 gluons $g$ as generators of the non--broken $SU(3)_c$
color symmetry algebra. After electroweak symmetry breaking the electroweak
bosons states mix and give two charged weak bosons $W^{\pm}$, a neutral
weak boson $Z^0$ and the photon $\gamma$ which remains massless. We
also retain one scalar degree of freedom: the Higgs boson.

\paragraph{The SM lagrangian\newline}

As stated in the historical subsection, in order to retain the gauge
symmetry of the SM all fields must remain massless. More precisely,
explicit mass terms for fermions in the form $m \overline{\psi} \psi$
would not break the $SU(3)_c$ gauge symmetry but would violate
$SU(2)_L$ symmetry, as such a term is indeed a product $m
(\overline{\psi}_L \psi_R + \overline{\psi}_R \psi_L)$. One of the
terms in the product does not change under $SU(2)_L$ transformation,
which thus means that the product is not invariant. Morever the same
argument applies for weak bosons  (sketched in a different way as we
are dealing with bosons). How to account for the mass of the weak
bosons and of the fermions, which manifestely exist according to
experiments?

The solution within the SM is precisely the Brout--Englert--Higgs
mechanism that will be described in the next section: we add to the
theory a scalar doublet which will generate through spontaneous
symmetry breaking the masses of the weak bosons, and also of the
fermions through Yukawa interactions.

We will only write down the lagrangian of the unbroken electroweak
symmetry for one generation, as the two others are an exact
mathematical replica (apart from the Yukawa sector). The broken
lagrangian will be described in the following section when dealing
with weak gauge bosons masses. We use the notation $L=(\nu_e, e)_L$,
$Q=(u,d)_L$ for the left doublets, and we have $e_R$, $u_R$ and $d_R$
for the right--handed electron, up and down quarks. The Higgs doublet
is $\Phi = (\phi^+,\phi_0)$.

We use $Y$ for the generator of the $U(1)_Y$ algebra and the three
generators $T^a$ of the $SU(2)_L$ algebra are taken as the half of the
$2\times 2$ Pauli matrices

\beq
T^a= \frac{1}{2} \tau^a \, ; \quad 
\tau_1= \left( \begin{array}{cc} 0 & 1 \\ 1 & 0 \end{array} \right) \, , \ 
\tau_2= \left( \begin{array}{cc} 0 & -i \\ i & 0 \end{array} \right) \, , \ 
\tau_3= \left( \begin{array}{cc} 1 & 0 \\ 0 & -1 \end{array} \right)
\eeq

with the commutation relations between these generators given by

\beq
[T^a,T^b]=i\epsilon^{abc} T_c~~ {\rm and}~~ [Y, Y]=0 
\eeq

where $\epsilon^{abc}$ is the usual antisymmetric tensor. For the
strong sector we use the eight Gell--Mann $3\times 3$ $U^{a}$ matrices
as the generators of $SU(3)_c$ symmetry associated with the eight
gluons $G^{a}_\mu$, with

\beq
[U^a,U^b]=if^{abc} U_c \ \ \ {\rm with} \ \ \  {\rm Tr}[U^a U^b]=
\frac12 \delta_{ab} 
\eeq

where $f^{abc}$ are the structure constants of the Lie algebra. We
then have the field strengths for the gauge bosons given  by

\beq
G_{\mu \nu}^a &=& \partial_\mu G_\nu^a -\partial_\nu G_\mu^a +g_s \, 
f^{abc} G^b_\mu G^c_\nu \nonumber \\
W_{\mu \nu}^a &=& \partial_\mu W_\nu^a -\partial_\nu W_\mu^a +g \, 
\epsilon^{abc} W^b_\mu W^c_\nu \nonumber \\ 
B_{\mu \nu} &=& \partial_\mu B_\nu -\partial_\nu B_\mu 
\eeq

$g_s$ and $g$ are, respectively, the coupling constants of $SU(3)_c$
and $SU(2)_L$ gauge algebras. The non--abelian structure of the strong
and weak interaction leads to non--trivial interactions among
respectively the gluons and the weak bosons, with triple and quartic
couplings.

In order to have a locally gauge invariant lagrangian we have to write
a covariant derivative for the fermions. The minimal coupling between
the fermionic matter fields and the gauge bosons, with $g_Y$ as the
$U(1)_Y$ coupling constant, is defined as

\beq
D_{\mu} \psi = \left( \partial_\mu -\imath g_s U_a G^a_\mu -\imath g
  T_a W^a_\mu -\imath g_Y \frac{Y_q}{2} B_\mu \right) \psi
\label{eq:mattercoupling}
\eeq

In the covariant derivative~\ref{eq:mattercoupling} it has to be
understood that only the relevant part is to be taken: $g_S$ and $g$
couplings are irrelevant for the case of a right--handed electron for
example.

The SM Lagrangian is then written in the following way:

\beq
{\cal L}_{\rm SM} &=& {\cal L}_{\rm gauge} + {\cal L}_{\rm fermion} +
{\cal L}_{\rm scalar}\nonumber\\
&&\nonumber\\
 &  {\cal L}_{\rm gauge}  = & -\frac{1}{4} G_{\mu \nu}^a G^{\mu \nu}_a 
-\frac{1}{4} W_{\mu \nu}^a W^{\mu \nu}_a -\frac{1}{4}
B_{\mu \nu}B^{\mu \nu} \nonumber\\
 & {\cal L}_{\rm fermion}  =  & \bar{L}\, i D_\mu
\gamma^\mu \, L + \bar{e}_{R} \, i D_\mu \gamma^\mu \, e_{R} \ 
+ \bar{Q}\, i D_\mu \gamma^\mu \, Q + \nonumber\\
&&\bar{u}_{R} \, i D_\mu 
\gamma^\mu \, u_{R} \ + \bar{d}_{R} \, i D_\mu \gamma^\mu \,
d_{R}\nonumber\\
&& \left(-\lambda_e \bar{L}\Phi e_R -\lambda_{ij}\bar L_{i}\Phi q_j
  +~{\rm h.c.}\right)\nonumber\\
& {\cal L}_{\rm scalar} = &
\left(D_{\mu}\Phi\right)^{\dagger}\left(D_\mu\Phi\right) -
V(\Phi),~V(\Phi) = \mu^2 \Phi^{\dagger}\Phi +
\lambda\left(\Phi^{\dagger}\Phi\right)^2
\label{eq:smlagrangian}
\eeq

$V(\Phi)$ is a potential term for the scalar field, and will be
studied in the following section. The Yukawa interactions in ${\cal
  L}_{\rm fermions}$ give rise to the mass of the fermions coupled to
the Higgs field through $\lambda_e$ for the electron (more generally
the charged leptons), and $(\lambda_{ij})_{1\leq i,j\leq 3}$ for the 6
quarks. The diagonalization of the $\lambda_{ij}$ matrix gives the CKM
matrix which is then involved in the couplings between the weak bosons
and the quarks. We also remind the reader that the $\lambda_{ij}\bar
L_{i}\Phi q_j$ is a formal form actually involving both $\bar
L_{i}\Phi q_j$ and $\bar L_i \tilde\Phi q_j$ terms where $\tilde \Phi
= \imath \tau_2 \Phi^*$ in order to give a mass to up--type quarks.

\vfill
\pagebreak

\section{The Brout--Englert--Higgs mechanism}

\label{section:Higgs}

After the presentation of the lagrangian of the SM in the previous
section we now turn our attention to the electroweak symmetry
breaking. The first question to be answered is: why do we need the
electroweak symmetry breaking in the description of the SM? For
example, as already stated, we now through the experiment that weak
bosons should be massive, as the weak interaction is that of
short--range; but explicit mass terms are prohibited if we want to
preserve gauge symmetry principles.

This will be discussed in details in the following subsection. We will
then present the spontaneous electroweak symmetry breaking using one
scalar weak doublet, which is known as the Brout--Englert--Higgs
mechanism~\cite{Higgs:1964ia,Englert:1964et,Guralnik:1964eu,Higgs:1966ev}.
We will deduce the weak bosons masses and the Higgs boson
couplings to fermions and bosons.

\subsection{Why do we need the electroweak symmetry breaking?}

\subsubsection{The unitarity puzzle \label{section:UnitarityWW}}

As presented in the historical part, the effective Fermi theory was
the first attempt to describe the weak interaction in a quantum
framework. The neutron and proton interactions were described in an
effective four--point vertex, which in the end results in a cross
section which grows with the center--of--mass energy $\sqrt s$. This
means that the theory at some point has lost unitarity and cannot make
reliable predictions: a reliable quantum theory should arrange for
having the sum of the probabilities of every possible event to be
equal to one, hence conserving the unitarity.

When taking into account the mass of the $W$ boson we restore
unitarity at the Fermi scale. Nevertheless we still face a potential
disaster at higher energy: if we study for example the scattering of
two $W$ bosons we obtain a growth of the cross section with higher
center--of--mass energy, still destroying unitarity beyond the Fermi
scale. We thus have to cure the theory in order to restore unitarity
in all processes, then preserving its prediction power and the ability
to calculate reliable quantities.

The crucial point that has been developed in 1973 in
Refs.~\cite{LlewellynSmith:1973ey,Cornwall:1973tb} is
that of the following: if we impose the SM (without the Higgs field as
introduced in standard textbooks by requiring spontaneous symmetry
breaking) to remain unitary at the Born level in perturbation theory
we end up precisely with two requirements:

\begin{enumerate}[$\bullet$]
\item{The fermion--bosons interactions should be that of Yang--Mills
    theory.}
\item{The theory needs at least a scalar doublet whose couplings to
    fermions and bosons are precisely that of the Higgs mechanism
    type. The self--couplings are also that of the Higgs type.}
\end{enumerate}

This very important result should be interpretated as proving the
existence of something to restore unitarity in weak processes that is,
either the Higgs doublet and subsequently the Higgs boson itself, or
something  which {\it looks exactly like the Higgs doublet and
  reproducing its properties}. This second proposition is what should
remain in mind when dealing with electroweak symmetry breaking: the
$SU(2)_L\times U(1)_Y$ symmetry has to be broken to restore unitarity,
even if we do not think about the weak bosons masses puzzle. Indeed,
if the gauge symmetry were preserved we would not have included new
symmetry breking scalar degree of freedom, hence not have had the
precise type of fermion--scalar and boson-scalar interactions that are
those of gauge breaking type.

\begin{figure}[!t]
\begin{center}
\includegraphics[scale=0.7]{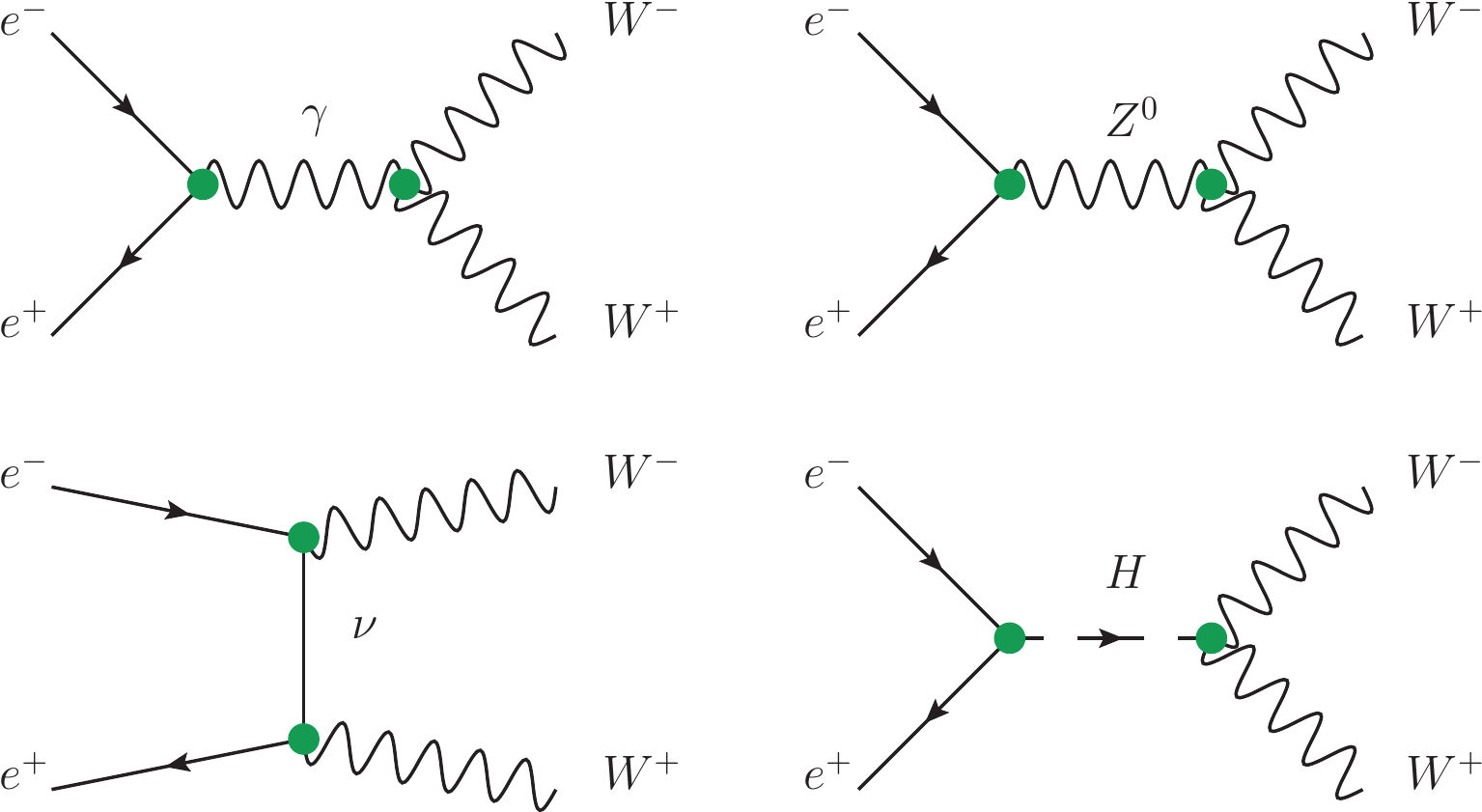} 
\end{center}
\vspace*{-.5cm}
\caption[Feynman diagrams at the Born level for the process $e^+e^-\to
W^+W^-$]{Feynman diagrams at the tree--level for $WW$ production in
  $e^+ e^-$ annihilation.}
\label{fig:unitarity} 
\end{figure}

To demonstrate this intimate connection between the scalar degrees of
freedom and the unitarity requirement, we present the case of the $WW$
production through electron--positron annihilation. The four Feynman
diagrams are depicted in Fig.~\ref{fig:unitarity}.

Let us suppose for a while that the fourth Feynman diagram involving
the Higgs boson exchange does not exist. If we are interested in the
production of longitudinal $W$ bosons, as the electron are massive we
need to cancel the cross section divergence arising because the
electron can have an helicity distinct from that of its chirality. If
this problem might seem to be not so critical, the electron mass
being small, this is certainly not the case if we replace the electron
by the top quark, a process that should be described by the
theory. This divergence is then proportional to $m_e$ and has to be
cancelled by a new contribution also proportional to $m_e$. To avoid
some tedious questions when using particles with spin higher than
$3/2$, this new contribution should be of spin zero.

This two last conditions are precisely fullfilled with the Higgs boson
contribution of the fourth Feynman diagram that we now take into
account and that restores the unitarity of this process!

This very important result, leading to the conclusion that something
should be discovered at hadron colliders at the electroweak scale,
wether it is the SM Higgs boson or its beyond--the--SM collegues, or
anything else which looks like the Higgs boson (e.g. composite Higgs
boson, extra--dimensions mimick, etc.), has also been extended in
Ref.~\cite{Gunion:1990kf} for arbitrary scalar multiplets. We should
then keep in mind the intimate link between unitarity requirement and
the existence of (a) new scalar degree(s) of freedom in the
electroweak theory.

\subsubsection{Masses and gauge
  invariance \label{section:UnitarityMasses}}

So far we have addressed the formal problem of requiring that the
theory remains predictive and thus at least preserve unitarity in high
energy processes. The last subsubsection has shown that this precise
requirement of preserving unitarity leads to the existence of either
the Higgs mechanism (and the Higgs boson), or the existence of
something which looks like the Higgs boson.

We take here the other way around by looking at what the experiment
tells us. We want to use gauge symmetry formalism in order to describe
the fundamental interactions at the high energy scale as it has been
proved to be a very powerful tool in the description of quantum
electromagnetic processes. The problem that we have is that Nature
gives us massive weak bosons, the interaction being of short length
scale; it turns out that explicit mass terms are prohibited in the
lagrangian as to preserve gauge invariance.

If we do not have any mass terms nor take into account the scalar part
in the SM lagrangian~\ref{eq:smlagrangian}, this gauge transformations
preserve the lagrangian:

\beq
\displaystyle L(x) \to L'(x) & = & e^{\imath \alpha_a (x) T^a + \imath \beta
  (x) Y + \imath \gamma_{a}(x)U^{a}} L(x) \nonumber \\
\displaystyle R(x) \to R'(x) & = & e^{\imath \beta (x) Y +\imath
  \gamma_{a}(x)U^{a}} R(x) \nonumber \\
\displaystyle W^{a}_{\mu}(x) \to {W'}^{a}_{\mu}(x) & = &
W^{a}_{\mu}(x) -\frac{1}{g} \partial_{\mu} \alpha^a(x) -
\epsilon_{abc}\alpha^{b}(x) W^{c}_{\mu}(x) \nonumber\\
\displaystyle G^{a}_{\mu}(x) \to {G'}^{a}_{\mu}(x)  & = &
G^{a}_{\mu}(x) -\frac{1}{g_s} \partial_{\mu} \gamma^a(x) -
\epsilon_{abc}\gamma^{b}(x) G^{c}_{\mu}(x) \nonumber\\
B_{\mu}(x) \to B'_{\mu}(x) & = &  B_{\mu}(x) -
\frac{1}{g_Y} \partial_{\mu} \beta (x)
\eeq

It is of course understood that the $SU(3)_c$ gauge tranformation acts
only on the quark and gluons fields.

Let us suppose that we now add explicit mass terms in the
lagrangian~\ref{eq:smlagrangian} (again we exclude at this stage the
scalar part as we are interested in proving that we need it, without
explicit mass terms), taking as an example the hypercharge boson
$B_\mu$ (which is exactly like the case of a massive photon). We write
down its gauge transformation as:

\beq
\frac{1}{2}M_B^2 B_\mu B^\mu \to \frac{1}{2}M_B^2 \left(B_\mu -
  \frac{1}{g_Y} \partial_\mu \beta\right) \left(B^\mu -
  \frac{1}{g_Y} \partial^\mu \beta\right) \neq \frac{1}{2}M_B^2 B_\mu
B^\mu
\eeq  

which does not preserve $U(1)_Y$ gauge invariance. This means that we
cannot account for the mass of the weak bosons while preserving gauge
invariance with explicit mass terms in the lagrangian.

We also have a similar problem when dealing with fermions. Indeed we
obviously observe a massive electron, not to mention the heavy
quarks. We would then include a mass term $-m_f \overline{\psi}_f
\psi_f$ in the lagrangian. This does not break $SU(3)_c$ nor $U(1)_Y$,
but if we rewrite this mass term using chirality degrees of freedom we
have:

\beq
- m_f \overline{\psi}_f \psi_f = -m_f \overline{\psi}_f \left(
  \frac{1}{2} (1-\gamma_5)+\frac{1}{2}(1+ 
\gamma_5) \right) \psi_f= -m_f(\overline{\psi}_{f R}
\psi_{f L}+\overline{\psi}_{f L} \psi_{f R}) 
\eeq

which manifestly violates $SU(2)_L$ gauge symmetry since $\Psi_{f L}$
is a member of a doublet while $\Psi_{f R} $ is a member of a
singlet.

Thus, the incorporation by hand of mass terms for gauge bosons and 
fermions leads to a manifest breakdown of the local $SU(2)_L\times
U(1)_Y$ gauge invariance. We then have at a first look either to give
up the fact that we do observe masses for weak bosons and for
fermions, which of course is unacceptable, or we have to give up the
principle of (exact) gauge symmetry.

The spontaneous electroweak symmetry breaking is a solution which
avoids both of the problematic answers given above: we still preserve
the gauge symmetry of the theory, but the vacuum breaks the
electroweak symmetry and gives rise to massive gauge bosons and
fermions in the spectrum. In the following subsection we will sketch
the Higgs mechanism which is the simplest procedure to do so.

\subsection{The spontaneous electroweak symmetry
  breaking \label{section:HiggsPotential}} 

We have seen that we are bound to add something to the gauge content
of the SM in order to both preserve unitarity and give masses to the
weak bosons, which are known to be massive. The most simple solution
is known as the Brout--Englert--Higgs mechanism~\cite{Higgs:1964ia,
  Englert:1964et,Guralnik:1964eu,Higgs:1966ev} and will be
presented in this subsection. We will see how to obtain the weak
bosons masses, and then we will review the Higgs boson coupling to
fermions and bosons that will be used in the following sections when
dealing with the SM Higgs boson production and decay at the hadron
colliders.

\subsubsection{Weak bosons masses and electroweak
  breaking \label{section:HiggsMasses}}

\paragraph{The Higgs mechanism\newline}

This subsubsection is built upon Refs.~\cite{Gunion:1990} and
\cite{Djouadi:2005gi}.
We start by taking the full SM lagrangian~\ref{eq:smlagrangian}, that
is not only the gauge and fermion parts but also the scalar part. We
thus add a new $SU(2)_L$ scalar doublet to the usual content of the
theory:
\beq
\Phi = \left( \begin{matrix}\phi^+\\ \phi^0\end{matrix}\right)\nonumber
\eeq
with $Y=+1$ hypercharge. $\Phi$ is also a color singlet. We remind the
reader that the scalar part of the SM lagrangian~\ref{eq:smlagrangian}
is
$${\cal L}_{\rm scalar} =
\left(D_{\mu}\Phi\right)^{\dagger}\left(D_\mu\Phi\right) -
V(\Phi),~V(\Phi) = \mu^2 \Phi^{\dagger}\Phi +
\lambda\left(\Phi^{\dagger}\Phi\right)^2$$

\begin{figure}[!t]
\begin{center}
\includegraphics[scale=0.6]{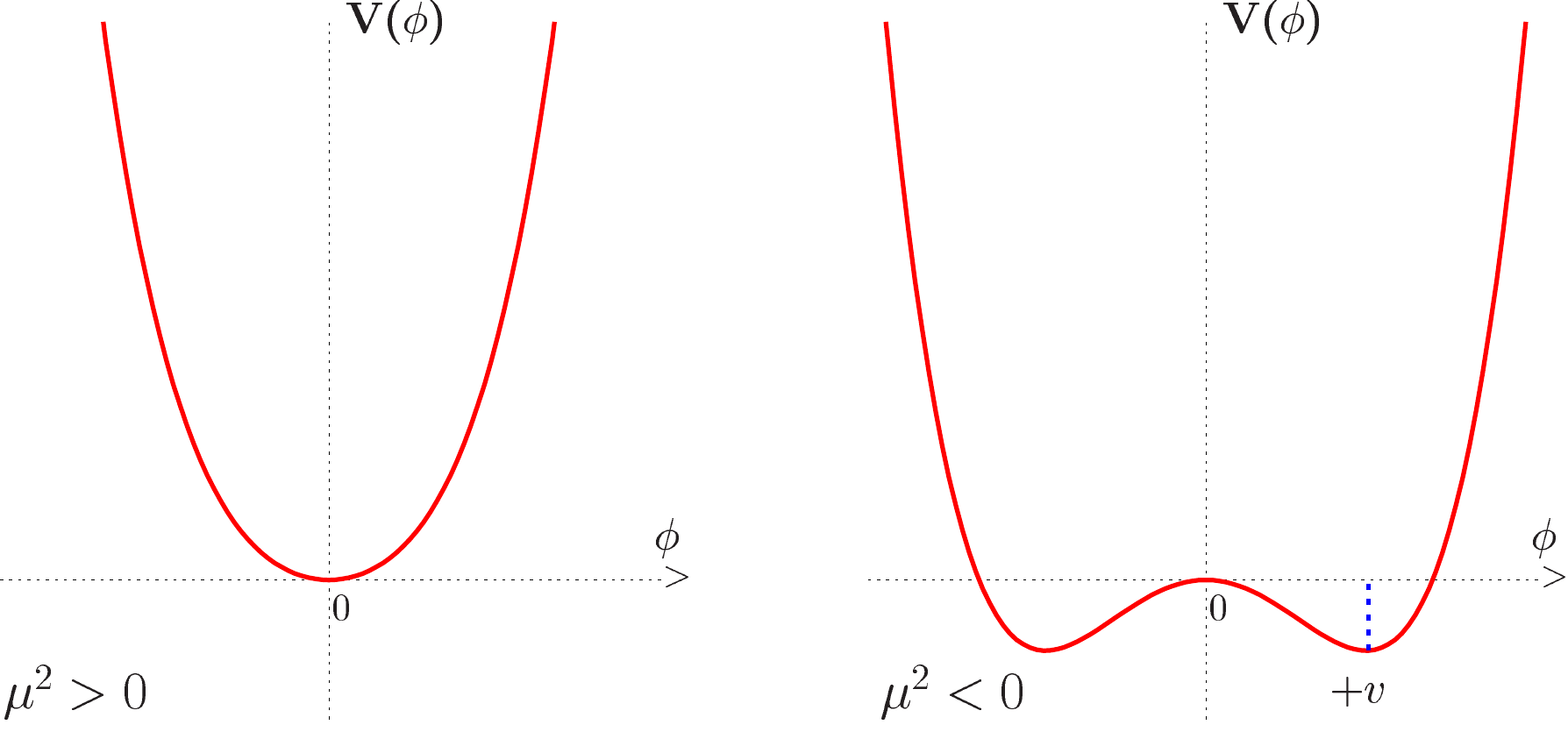} 
\end{center}
\vspace*{-.5cm}
\caption[Higgs potential in the case of a real scalar field, depending
on the sign of the mass term]{Higgs potential in the case of a real
  scalar field, depending on the sign of the mass term. This figure is
taken from Ref.~\cite{Djouadi:2005gi}.}
\label{fig:higgspot} 
\end{figure}

The vacuum expectation value of the Higgs doublet is given by the
minimum of $V(\Phi)$. We first note that $\lambda > 0$ is required to
have a potential bounded from below, thus insuring a stable vacuum. We
then have the two situations depicted in Fig.~\ref{fig:higgspot} for a
one dimensional scalar field.

If $\mu^2>0$ we then have a positive potential everywhere, whose
minimum is $\langle 0 | \Phi | 0 \rangle = 0$: we have not achieved
anything new.

We will then suppose that $\mu^2 < 0$, which is the hypothesis
building the Brout--Englert--Higgs mechanism~\cite{Higgs:1964ia,
  Englert:1964et,Guralnik:1964eu,Higgs:1966ev} that we apply in
the SM. In that case we obtain a minimum

\beq
\phi_0^2 \equiv \langle 0 | |\Phi|^2 | 0 \rangle = - \frac{\mu^2}{2
  \lambda} \equiv \frac{v^2}{2}
\eeq

In the case of the $SU(2)_L$ doublet, the potential is depicted in
Fig.~\ref{fig:higgspot2} below.

\begin{figure}[!t]
\begin{center}
\includegraphics[scale=0.45]{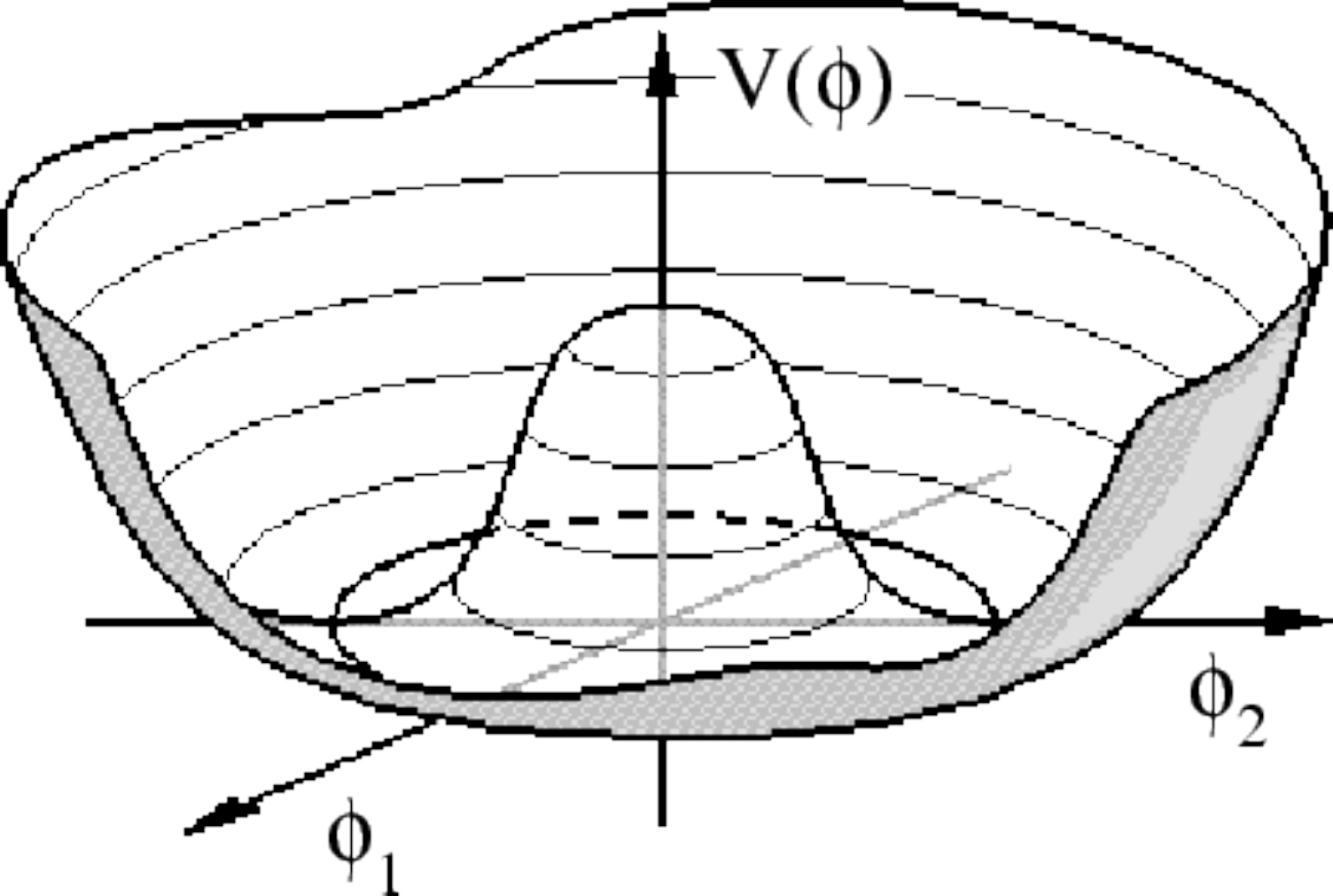} 
\end{center}
\vspace*{-.5cm}
\caption[Higgs potential in the case of the SM]{Higgs potential in the
  SM, also known as the ``mexican hat''.}
\label{fig:higgspot2} 
\end{figure}

The vacuum expectation value (vev) still preserves a $U(1)$ symmetry,
but has broken the $SU(2)$ symmetry: we thus have a spontaneously
breaking of the original SM gauge algebra down to a residual $U(1)$
symmetry which, as seen later, is the electromagnetic $U(1)$ gauge
algebra.

To make a physical interpretation of the lagrangian we will expand
$\Phi$ around its minimum which will give us the physical degree of
freedom. The neutral component of the doublet field $\Phi$ will 
develop a vacuum expectation value 

\beq
\langle \, \Phi \, \rangle_0 \equiv  \langle \, 0 \, | \, \Phi \, | \, 0\, 
\rangle =\left( \begin{array}{c} 0\\ \displaystyle \frac{v}{\sqrt{2}} \end{array} \right) \ \ 
{\rm with} \ \ v= \left( - \frac{\mu^2}{\lambda} \right)^{1/2} 
\eeq

We have chosen this direction so as to preserve $U(1)_{\rm EM}$. We
then write the field $\Phi$ in terms of four fields $\theta_{1,2,3}(x)$ and 
$H(x)$ around the minimum, and at first order we have:

\beq
\Phi(x) = \left( \begin{array}{c} \theta_2 + i \theta_1 \\ 
\frac{1}{\sqrt{2}} ( v + H)  - i \theta_3 \end{array} \right) = 
\displaystyle e^{i \theta_a (x) \tau^a (x)/v} \,  \left( \begin{array}{c} 0 \\ 
\frac{1}{\sqrt{2}} (v + H(x) \, )  \end{array} \right) 
\eeq

The second expression is crucial: as the theory remains gauge
invariant we can always make a $SU(2)_L$ gauge transformation with the
parameter $-\theta^a(x)/v$, which means that we chose a particular
gauge, called the unitary gauge or U--gauge:

\beq
\displaystyle \Phi(x) \to  e^{- i \theta_a (x) \tau^a (x) } \,  \Phi(x) = \frac{1}{\sqrt{2}}
\left( \begin{array}{c} 0 \\  v + H (x)  \end{array} 
\right) 
\eeq

We have to expand the scalar part of the
lagrangian~\ref{eq:smlagrangian} in this gauge, we obtain:

\beq
V(\Phi) & = & \frac{1}{2} \mu^2 (v+H(x))^2 +
\frac{1}{4}\lambda(v+H(x))^4 \nonumber \\
 & = & -\mu^2H(x)^2+\lambda v H(x)^3+\frac{1}{4}\lambda H(x)^4 +~{\rm constant}\nonumber
\eeq
and
\beq
|D_\mu \Phi)|^2= \left| \left( \partial_\mu -\imath g \frac{\tau_a}{2} W_\mu^a 
- \imath g_Y \frac{1}{2} B_\mu \right)\Phi  \right|^2 \hspace*{6cm} \nonumber \\
= \frac{1}{2} \left| \left( \begin{array}{cc} \partial_\mu - \frac{\imath}{2}(
g W_\mu^3 + g_Y B_\mu) & -\frac{\imath g}{2}(W_\mu^1 -\imath W^2_\mu) \\ -
\frac{\imath g}{2} (W_\mu^1 + \imath W^2_\mu) & \partial_\mu + \frac{\imath}{2} (g W_\mu^3  
- g_Y B_\mu) \end{array} \right) \left( \begin{array}{c} 
0 \\ v+H  \end{array} \right) \right|^2 \nonumber \\
= \frac{1}{2} (\partial_\mu H)^2 +
\frac{1}{8}g^2(v+H)^2|W_\mu^1-\imath W_\mu^2|^2
+ \frac{1}{8}(v+H)^2 |g W_\mu^3- g_Y B_\mu|^2 \nonumber
\eeq

We define the physical fields $W^{\pm}_{\mu}$, $Z_\mu$ and $A_\mu$ as:

\beq
W^\pm = \frac{1}{\sqrt{2}} (W^1_\mu \mp \imath W^2_\mu) \  , \ 
Z_\mu = \frac{g W^3_\mu- g_Y B_\mu}{\sqrt{g^2+g_Y^2}} \ , \ 
A_\mu = \frac{g W^3_\mu+ g_Y B_\mu}{\sqrt{g^2+g_Y^2}} \  
\label{eq:WZA-fields}
\eeq

With this new basis we finally obtain the physical interpretation of
the scalar part of the lagrangian:

\beq
{\cal L}_{\rm scalar} & = & \frac{1}{2} \left(\partial_\mu H\right)^2 -
  \frac{1}{2} (-2\mu^2)H^2 + \nonumber\\
 & & \left( \frac{g^2 v^2}{4}\right)W^{+}_\mu W^{- \mu} +
 \frac{1}{2}\left( \frac{(g^2+g_Y^2)v^2}{4}\right)Z_{\mu} Z^{\mu} +
 \nonumber\\
 & & \left(\frac{g^2 v}{2}\right) H W^{+}_\mu W^{- \mu} +
 \left(\frac{(g^2+g_Y^2) v}{4}\right) H Z_{\mu} Z^{\mu} +
 \nonumber \\
 & & \left(\frac{g^2}{4}\right) H^2 W^{+}_\mu W^{- \mu} +
 \left(\frac{(g^2+g_Y^2)}{8}\right) H^2 Z_{\mu} Z^{\mu} +
 \nonumber \\
 & & - \left(\lambda v\right) H^3 - \left(\frac{\lambda}{4}\right) H^4
 +~{\rm constant}
\label{eq:scalarlagrangianbroken}
\eeq

The spontaneous breakdown of $SU(2)_L \times U(1)_Y$ has achieved our
goal of obtaining masses for the weak bosons. They can be read in the
second line of Eq.~\ref{eq:scalarlagrangianbroken} as:

\beq
M_W =\frac{gv}{2} \  , \ M_Z= \frac{v  \sqrt{g^2+g_Y^2}}{2}\ , 
 \ M_A=0 
\label{eq:weakmasses}
\eeq
not to mention the Higgs mass itself $M_H^2 = -2\mu^2$.

The photon $A_\mu$ remains massless as the remaining $U(1)_{\rm EM}$
is unbroken while the three weak bosons $W^{\pm}$ and $Z^0$ are now
massive.  The three degree of freedom $\theta^a_\mu$ of the original
Higgs doublet, which are Goldstone bosons, have been absorbed by the
$W^\pm$ and $Z^0$ bosons to obtain their longitudinal components,
which means that the number of degree of freedom is still conserved
when looking at the lagrangian before and after the electroweak
symmetry breaking. We are left with one scalar degree of freedom: the
Higgs boson.

The remaining lines of Eq.~\ref{eq:scalarlagrangianbroken} give the
couplings of the Higgs boson $H$ with the gauge bosons, and also its
trilinear and quartic self--couplings.\bigskip

If we recall the fermionic part of the SM lagrangian which also
contains Higgs field couplings to fermion, we can also generate the
fermion masses using $\Phi$ and $\tilde \Phi = \imath \tau_2
\Phi^*$. Starting from 
\beq
{\cal L}_{\rm Yukawa}= -\lambda_{e} \, \bar{L} \, \Phi \, e_{R}  
- \lambda_{d} \, \bar{Q} \, \Phi \, d_{R}
- \lambda_u \, \bar{Q} \, \tilde{\Phi} \, u_R    \ + \ { h.\, c.} 
\eeq
where we have left out the CKM issue and only delt with the first
generation, again we expand the expression around the vev:
\beq
{\cal L}_{\rm Yukawa} &=& -\left(\frac{\lambda_e v}{\sqrt{2}}\right)
\bar{e} e - \left(\frac{\lambda_u v}{\sqrt{2}}\right)
\bar{u} u - \left(\frac{\lambda_d v}{\sqrt{2}}\right)
\bar{d} d \nonumber\\
 & & -\left(\frac{\lambda_e}{\sqrt{2}}\right)
\bar{e} H e - \left(\frac{\lambda_u}{\sqrt{2}}\right)
\bar{u} H u - \left(\frac{\lambda_d}{\sqrt{2}}\right)
\bar{d} H d
\label{eq:yukawabroken}
\eeq

We then have the fermion masses:
\beq
m_e= \frac{\lambda_e\, v}{\sqrt{2}} \ \ , \ 
m_u= \frac{\lambda_u\, v}{\sqrt{2} }\ \ , \ 
m_d= \frac{\lambda_d\, v}{\sqrt{2}} 
\eeq
 
and also the Higgs--fermion--fermion couplings $\lambda_f /
\sqrt{2}$. We thus have generated the masses of both the weak vector
bosons $W^\pm, Z$ and the fermions, while preserving the $SU(2)_L
\times U(1)_Y$ gauge symmetry, which is now spontaneously broken or
hidden, together with a preservation of both the color $SU(3)_c$ and
electromagnetic $U(1)_{\rm EM}$ gauge symmetries. This constitutes the
Higgs mechanism applied to the Standard Model. The SM refers actually
to the gauge symmetries together with the Higgs mechanism.

\paragraph{The fundamental parameters\newline}

We list in this small paragraph the main fundamental parameters of the
SM that will be used in the following sections.

\begin{enumerate}[$\bullet$]
\item{{\it Weak coupling and the Fermi constant:} The weak couplings
    in the lagrangian are 
    $g$ and $g_Y$ for the $SU(2)_L\times U(1)_Y$ gauge algebra. This
    can be related to the Fermi constant $G_F=1.6637\times 10^{-5}$
    GeV$^{-2}$, by taking the low energy limit of the electroweak
    theory:
\beq
\frac{g^2}{8 M_W^2} = \frac{G_F}{\sqrt 2}
\eeq 
using the $W$ mass given in Eq.~\ref{eq:weakmasses}, we have the value
of the vev
\beq
v=(G_F\sqrt 2)^{-1/2} \simeq 246~{\rm GeV}
\eeq}
\item{{\it The Weinberg angle:} we can relate the $W$ and $Z$ bosons
    masses through what is called the Weinberg angle $\theta_W$:
\beq
\sin\theta_W \equiv \frac{g_Y}{\sqrt{g^2+g_Y^2}} = 0.23150\nonumber\\
 g_Y = g\tan\theta_W ~,~ M_Z = \frac{M_W}{\cos\theta_W}
\eeq
This leads to
\beq
\rho \equiv \frac{M_W^2}{M_Z^2\cos^2\theta_W} = 1
\eeq
and this value is hardly changed when taking into account radiative
corrections, a statement which can be related to a global $SU(2)$
custodial symmetry in the SM. This imposes constraints on any theory
that goes beyond the SM.}
\item{{\it Strong and electromagnetic constants:} the electromagnetic
    coupling constant is related to the $SU(2)_L\times U(1)_Y$
    coupling constants through $e=g\sin\theta_W$. We have at the scale
    $M_Z$:
\beq
\alpha_s(M_Z^2) & \equiv & \frac{g_s^2}{4\pi} = 0.1172~{\rm
  (LEP2)}\nonumber\\
\alpha(M_Z^2) & \equiv & \frac{e^2}{4\pi} = 1/(128.951)
\eeq
The value of the strong coupling constant will be particularly
important when dealing with Higgs boson production and decay at hadron
colliders; this will be discussed in more details in the following
part~\ref{part:two}.}
\item{{\it $W$ and $Z$ masses:} the current measured values of the $W$
    and $Z$ masses are
\beq
M_W = 80.420~{\rm GeV} ~ , ~ M_Z=91.1876~{\rm GeV}
\eeq}
\end{enumerate}

\subsubsection{SM Higgs boson
  couplings \label{section:HiggsCouplings}}

We finish this presentation of the SM by giving the Higgs couplings to
gauge bosons and fermions in a compact form. This couplings will play
a crucial role in the following parts as they dictate the dynamics of
the SM Higgs boson in production and decays.

The couplings of the Higgs boson $H$ with the gauge bosons as well as its
trilinear and quartic self--couplings can be read in
Eq.~\ref{eq:scalarlagrangianbroken} , while the SM Higgs boson
coupling to fermions can be read in Eq.~\ref{eq:yukawabroken}. These
tree--level couplings are depicted in Fig.~\ref{fig:higgscouplings}
below.

\begin{figure}[!t]
\begin{center}
\includegraphics[scale=0.7]{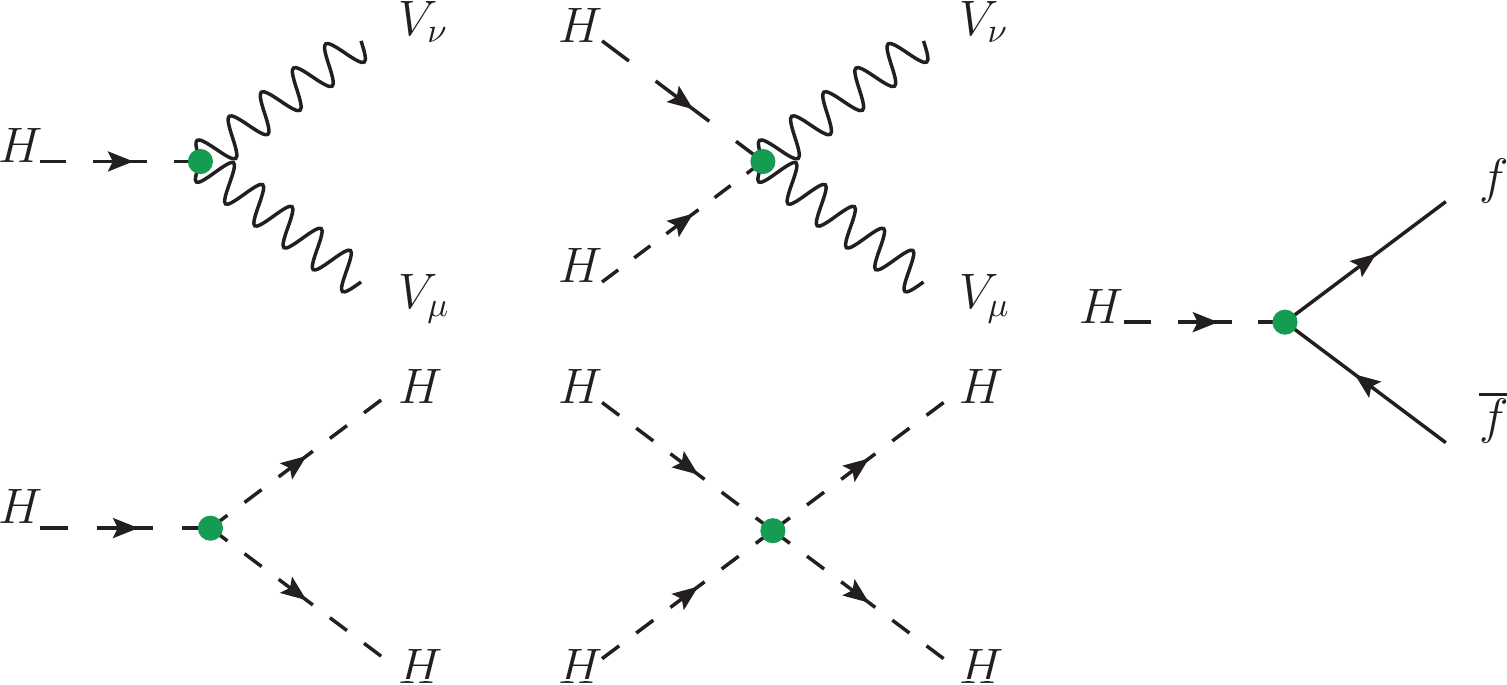} 
\end{center}
\vspace*{-.5cm}
\caption[Tree--level SM Higgs boson couplings to gauge bosons and
fermions]{Tree--level SM Higgs boson couplings to gauge bosons and
  fermions.}
\label{fig:higgscouplings} 
\end{figure}

When taking into account the number of identical particles in the
vertex, this gives
\beq
g_{H V V} = \frac{2 M_V^2}{v} ~(\times \imath g_{\mu \nu}) & ~ , ~ & g_{H
  H V V} = \frac{2 M_V^2}{v^2} ~(\times \imath g_{\mu \nu}) \nonumber\\
g_{H H H} = - \frac{3 M_H^2}{v} ~(\times \imath) & ~ , ~ & g_{H H H H} = -
\frac{3 M_H^2}{v^2} ~(\times \imath) \nonumber\\
g_{H f\bar f} = -\frac{m_f}{v} ~(\times \imath) & &
\label{eq:higgscouplingsSMbosons}
\eeq

\begin{figure}[!h]
\begin{center}
\includegraphics[scale=0.7]{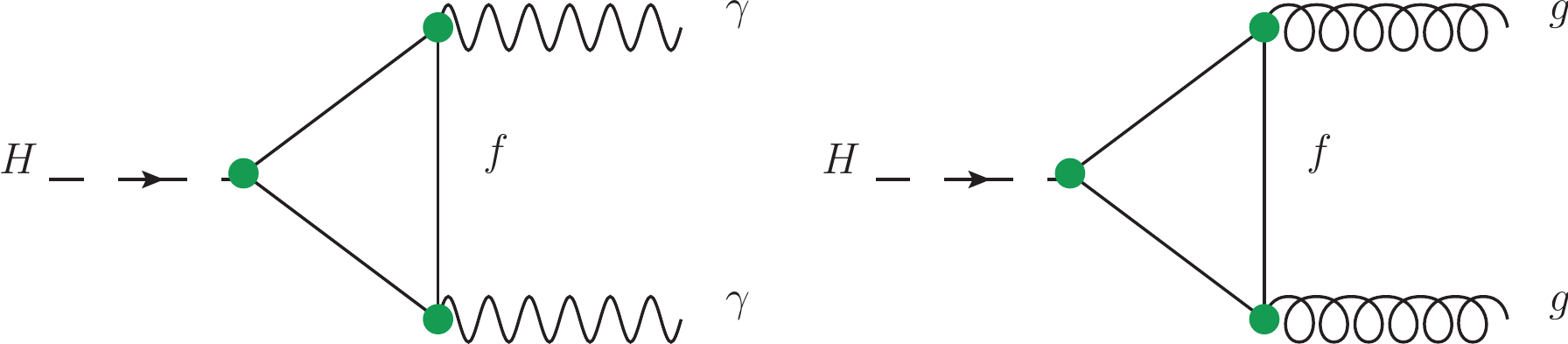} 
\end{center}
\vspace*{-.5cm}
\caption[One--loop SM Higgs boson couplings to the photons and the
gluons]{One--loop SM Higgs boson couplings to the photons and the
  gluons.}
\label{fig:higgscouplings2} 
\end{figure}

There are two additionnal couplings which are of utmost importance for
the study of the Higgs boson, either within or beyond the SM. Indeed,
even if the Higgs is not charged nor colored, it may have a coupling
to the photon and the gluons at the one--loop level, as shown in
Fig.~\ref{fig:higgscouplings2}. This couplings come through a
fermionic triangular loop which carries the absent quantum number of
the Higgs boson. 

The gluon--gluon--Higgs coupling will play a major role in the
production of the Higgs boson at hadron colliders. Now that we have
ended describing the Standard Model and the Higgs mechanism, we are
ready to study the SM Higgs production and decay at the two hadron
colliders in activity.

\pagebreak\strut
\thispagestyle{empty}
\newpage\pagebreak
\vfill


\part{SM Higgs production and decay at hadron colliders}
\label{part:two}

\section{Where can the SM Higgs boson be hiding?}

The section~\ref{section:Higgs} was devoted to the Higgs mechanism
itself and its remnant particle, the Higgs boson. This key particle,
predicted by the Standard Model and at the core of the electroweak
symmetry breaking, has been yet to be discovered for nearly 40
years. Large particle physics experiments have promoted its search as
prominent, such as in the LEP era~\cite{Barate:2003sz} and at the Tevatron
collider. It is also one of the key searches at the newest Large
Hadron Collider (LHC) at CERN.

Before starting to study the Higgs boson production at the two current
hadronic colliders and then its decay, we will review the current
theoretical and experimental bounds on its mass which is not
predicted by the Standard Model.

\label{section:HiggsBounds}

\subsection{Theoretical bounds on the Higgs
  mass \label{section:HiggsboundsTheory}}

The Higgs boson mass is not predicted by the Standard Model. Hence the
only prediction that can be made is on the range of validity for $M_H$
as to be sure that the Standard Model is still a valid and consistent
theory. The theoretical bounds on the Higgs mass are precisely derived
on these requirements: we probe the mass range in which the SM does
not break down and where perturbation theory is still applicable,
before new phenomena occur. These limits include constraints from the
unitarity required in scattering amplitudes, the perturbativity of the
Higgs self--coupling and considerations about the stability of the
vacuum of the theory. We summarize and combine all the limits in the
last subsubsection.

\subsubsection{Unitarity constraint}

The SM is embedded within the framework of quantum field
theory. As for any consistent quantum theory, where the description is
probabilistic in the sense that any physical observable has a
probability to be measured by an observer, the total probability is
conserved and equal to the unity. In the context of the scattering
amplitudes within the SM, it means that we require that the
$S$--matrix, which encodes all the necessary information between the
initial and final states, is unitary. We thus want that the SM to be
unitary in that sense, which will translate into an upper bound for
the SM Higgs boson mass.

It was actually one of the main arguments to abandon the Fermi theory
of the weak interaction, together with the requirement of accomodate
within a gauge theory the fact that experimentally the weak bosons
are to have a mass. To take an example the weak process $\nu_\mu e \to
\nu_e \mu$ is proportional to the Mandelstam variable $s$, which then
violates unitarity at a certain point as we should instead have
$\sigma(\nu_\mu e\to \nu_e \mu) = \mathcal{O}(s^{-1})$. If we go further
the unitarity constraint applied to all SM processes does even lead to
the requirement of the Higgs field itself, as presented in
section~\ref{section:UnitarityWW}\footnote{To be more rigourous this
  leads to the requirement of a new scalar degree of freedom which
  exactly looks like the Higgs 
  field of the SM. This does not mean that this new degree of freedom
  is elementary and it may well be a strongly coupled composite bound
  state, as proposed in many Higgsless
  theories~\cite{Kaplan:1983sm,Agashe:2004rs}.}. In some sense, we can
think about the SM as the theory which cures the unitarity problem at
the Fermi scale $\Lambda \sim 300$ GeV.

However there is still some potential unitarity concern in the SM, at
much higher energies than the Fermi scale, in particular if it is
believed that the SM is valid up to the Planck scale\footnote{That is
  not the current belief of the community, though. If SM were indeed
  valid up to the Planck scale we would stop working on particle
  physics! That would be a shame for young particle physicists and we
  hope that Nature does not trick us with that\dots.}. We take as an
example the $ZZ$ scattering, looking in particular
at the longitudinal component $Z_L Z_L$ scattering. This scattering is
(one of) the focus point of the spontaneously electroweak symmetry
breaking as the $Z_L$ is one of the degrees of freedom of the unbroken
Higgs field. The situation is that of the Fermi theory: the $Z_L
Z_L$ interaction grows with the momenta of the ingoing particles,
which then may lead to a violation of the unitarity
requirement. Considering that at high center--of--mass energies we can
work directly with the corresponding Goldstone bosons for the
scattering of the longitudinal component, we obtain very easily
\beq
\mathcal{A}(Z_L^0 Z_L^0\to Z_L^0 Z_L^0 ) = - \left[ 3 \frac{M_H^2}{v^2}
  +  \left( \frac{M_H^2}{v} \right)^2 \frac{1}{s-M_H^2} + \left(
    \frac{M_H^2}{v} \right)^2\frac{1}{t-M_H^2} \right]
\label{wlwlamp}
\eeq
where $s,t$ are the Mandelstam variables (the center--of--mass energy
$s$ is the square of the sum of the momenta of the initial or final
states, while $t$ is the square of the difference between the momenta
of one initial and one final state).

In order to study the unitarity of this amplitude we use the partial
wave decomposition of the scalar amplitude on the Legendre polynomials
basis:

\beq 
\mathcal{A}= 16 \pi \sum_{k=0}^{\infty} (2k+1) a_k P_k (\cos \theta)
\eeq
with $a_k$ being the partial waves along the angular momentum $k$ and
$\theta$ the scattering angle in the center--of-mass frame. We
recall the reader that the Legendre polynomials is an orthogonal basis
of the vector space $\mathbb{R}[X]$. In particular we have the
orthonormal condition
\beq
\int_{-1}^1 d\cos\theta P_k(\cos \theta) P_l(\cos\theta) =
\frac{2}{2k+1} \delta_{kl}
\eeq
where $\delta_{kl}$ is the Kronecker symbol. Since for a $2\to 2$
process, the cross section is given by ${\rm d} \sigma /{\rm d} \Omega
= |\mathcal{A}|^2 /(64 \pi^2 s)$ with d$\Omega=2\pi$d$\cos\theta$, the
total cross section is
\beq
\sigma &=& \frac{8\pi}{s} \sum_{k=0}^{\infty} 
\sum_{l=0}^{\infty} (2k+1) (2l+1) a_k a_{l}^{*} \int_{-1}^1
d\cos \theta  P_k (\cos \theta )  P_{l} ( \cos \theta ) \nonumber \\
&=& \frac{16 \pi}{s} \sum_{k=0}^{\infty}  (2k +1) |a_k|^2
\eeq

We now use the optical theorem which relates the imaginary part of the
amplitude taken on the beam line $\mathcal{A}(\theta=0)$ and the total
cross section $\sigma$:

\beq
\sigma = \frac{1}{s} {\rm Im} \left[\mathcal{A}(\theta=0)\right] =
\frac{16 \pi}{s} \sum_{k=0}^{\infty}  (2k +1) |a_k|^2 
\eeq

This leads to the unitary
conditions~\cite{Luscher:1988gc,Luscher:1988uq}
\beq
|a_k|^2 = {\rm Im} (a_k) & \Longleftrightarrow & \left[{\rm Re} (a_k)\right]^2 +
\left[{\rm Im}(a_k) \right]^2 = {\rm Im}(a_k) \nonumber \\
&\Longleftrightarrow & \left[{\rm Re} (a_k) \right]^2 + 
\left[{\rm Im}(a_k) -\frac{1}{2}\right]^2 = \frac{1}{4}
\eeq

We thus obtain the equation of a circle of radius $\frac{1}{2}$
and center $(0, \frac{1}{2})$ in the plane $[{\rm Re} (a_\ell), {\rm Im}(a_\ell)
]$. The real part lies between $-\frac{1}{2}$ and $\frac{1}{2}$, we
obtain the unitarity condition which follows:

\beq
|{\rm Re} (a_\ell)| < \frac{1}{2} 
\label{eq:unitaritycondition} 
\eeq

The kinematics of the process gives $\displaystyle
t=\frac{-s}{2}(1-\cos\theta)$. If we use the orthonormality relation
of the Legendre polynomials and knowing that $P_0(x)=1$ we easily
obtain the $J=0$ partial wave $\displaystyle a_0 =
\frac{1}{32\pi}\int_{-1}^1
\mathcal{A}(\cos\theta) d\cos\theta$. Using the $t$ variable this
transforms in
\beq
a_0=\frac{1}{16\pi s} \int_{-s}^0 dt \mathcal{A}(t)
\eeq

We then obtain for the $\mathcal{A}(Z_L^0Z_L^0\to Z_L^0Z_L^0)$:

\beq
a_0 = - \frac{M_H^2}{16 \pi v^2} \left[3 + \frac{M_H^2}{s-M_H^2} -
  \frac{M_H^2}{s} \ln \left( 1 + \frac{s}{M_H^2} \right) \right]
\label{eq:unitarityeq1}
\eeq

If we now assume that the Higgs boson mass to be much smaller than
$\sqrt{s}$ and then use the unitary
condition~\ref{eq:unitaritycondition} we obtain
\beq
M_H \leq \frac{8\pi v^2}{3}\Rightarrow M_H\lsim 710~{\rm GeV}
\label{eq:unitaritylimit}
\eeq

The same analysis has to be done for any channel of the theory:
$W_L^+W_L^-$, $HH$, $Z_LH$, $W_L^{\pm}H$, $W^{\pm}_L Z_L$, etc. The
condition obtained in Eq.~\ref{eq:unitaritylimit} is formally valid only
at tree--level, and since the Higgs boson self--coupling becomes
strong for large Higgs masses, $\lambda = M_H^2/(2v^2)$, the argument
could be destroyed if radiative corrections are
taken into account. This is then only a perturbative tree--level
unitarity limit. The complete perturbative unitarity argument should
then be given within the context of a perturbative expansion analysis
and assuring that the SM remains a perturbative theory where the
radiative corrections are not too large. Taking this into account and
with all the channel in the theory, the unitarity constraint is
(incidentally!) still the one given by
Eq.~\ref{eq:unitaritylimit}~\cite{Marciano:1987un,Dawson:1988va,Marciano:1989ns}.

We could adopt a different point of view and look the other way around
where $M_H$ is taken very large, and we would then obtain a bound on
the possible compatible $\sqrt{s}$
energies~\cite{Chanowitz:1984ne}. We take the example of the
$W_L^+ W_L^-$ channel. We take the limit $s\gg M_H^2$ in
Eq.~\ref{eq:unitarityeq1} where the number 3 is replaced by the number
2, and then apply the unitarity condition~\ref{eq:unitaritycondition}
to obtain
\beq
\sqrt{s} \leq v\sqrt{16\pi}\Rightarrow \sqrt{s}\lsim 1.7~{\rm TeV}
\eeq

Considering all the possible channels this reduces to
\beq
\sqrt{s} \lsim 1.2~{\rm TeV}
\eeq

Therefore we face two possibilities: either some new physics should
manifest at the TeV scale range if the Higgs boson is very massive (or
not existing at all), or the unitarity breakdown is canceled by large
high--order terms which signal the failure of perturbation theory and
the loss of the predictive power of the SM. The last possibility is of
course a nightmare that we would avoid, and then we have two final
reasonable conclusions:

\begin{enumerate}[i)]
\item{the SM Higgs boson exists and its mass should be bounded,
    $M_H\lsim 710$ GeV, in order to retain unitarity.}
\item{the SM Higgs boson is very massive or does not exist: then new
    physics effects are to emerge at the TeV scale.} 
\end{enumerate}

\subsubsection{Constraint from the perturbativity of the self--Higgs
  coupling}

Even if we forget about the unitarity issue, the requirement of having
a perturbative theory in particular with processes involving the Higgs
self--coupling will induce an upper bound on the Higgs boson
mass. Indeed it is known that for large values of the Higgs boson mass
the perturbation theory is lost. If we take the SM Higgs decay into
massive gauge bosons, we obtain (see Ref.~\cite{Djouadi:2005gi}):

\beq
\Gamma(H\to ZZ) = \frac{M_H^3}{32\pi v^2} \left[ 1 +  3 \hat
  \lambda +62 \hat \lambda^2 + \mathcal{O}(\hat \lambda^3) \right]
\eeq

with $\hat \lambda= \lambda/(16 \pi^2)$. We recall that we have
$M_H^2=2\lambda v^2$. Thus if we require that the perturbativity of
the calculation remains, that is each term in the expansion is smaller
that its predecessor, we can derive an upper bound on the Higgs boson
mass. Indeed if we have $M_H \simeq 2.5$ TeV the 1--loop term is of
order 1, $3\hat \lambda =1$, breaking the perturbativity. We can
reduce this bound even more with the limit $3\hat \lambda = 62 \hat
\lambda^2$ where the 1--loop term is of order the 2--loop term, we
then have $M_H \simeq 960$ GeV in this limit. Thus we have to impose
$M_H \ll 1$ TeV to retain the perturbativity of the expansion.

The jeopardy of perturbation theory at large Higgs masses can also be
seen in the scattering of longitudinal gauge bosons from which were
derived above the upper bound on $M_H$ using perturbative unitarity
argument. In the case of the $W^+_L W^-_L \to W^+_L W^-_L$
scattering, the radiative corrections have been calculated at one and
two loops in
Refs.~\cite{Dawson:1989up,Durand:1993vn,Riesselmann:1995en,Nierste:1995zx}.
The logarithmic scale
dependence found in these corrections can be absorbed in a running
$\lambda$ coupling through Renormalization Group analysis and we find
(see Ref.~\cite{Djouadi:2005gi}):

\beq
\sigma (W^+_L W^-_L \to W^+_L W^-_L) \sim \frac{1}{s} \hat \lambda (s)
\left( 1- 48.64 \hat \lambda +333.21  \hat \lambda^2 \right)
\eeq

Here, the coefficients of the corrections are much larger than in
Higgs decays and in fact, the one--loop correction become of order
unity already for $\lambda (s)$ values close to 3, that transforms
into $M_H\simeq 700$ GeV.

The perturbativity of the self--Higgs coupling in the end implies that
$M_H\lsim 700$ GeV in order to keep a calculable pertubative
theory. We then obtain a bound that is very close to that obtained
using the unitarity argument.

\subsubsection{Triviality constraint}

The triviality constraint is closely related to the vacuum stability
that is the next constraint that will be discussed. An interacting
theory is said to be trivial when the coupling is indeed null, which
means that it is actually a non--interacting theory. The scalar sector
of the SM is a $\phi^4$--theory, and for these theories to remain
perturbative at all scales it is required that the theory is trivial:
$\lambda=0$, and thus the Higgs boson is
massless~\cite{Wilson:1973jj,Cabibbo:1979ay,Dashen:1983ts,Callaway:1983zd,Hasenfratz:1987tk,Kuti:1987nr,Chivukula:1996sn,
 Luscher:1988gc,Luscher:1988uq}.

\begin{figure}[!t]
\begin{center}
\includegraphics[scale=0.9]{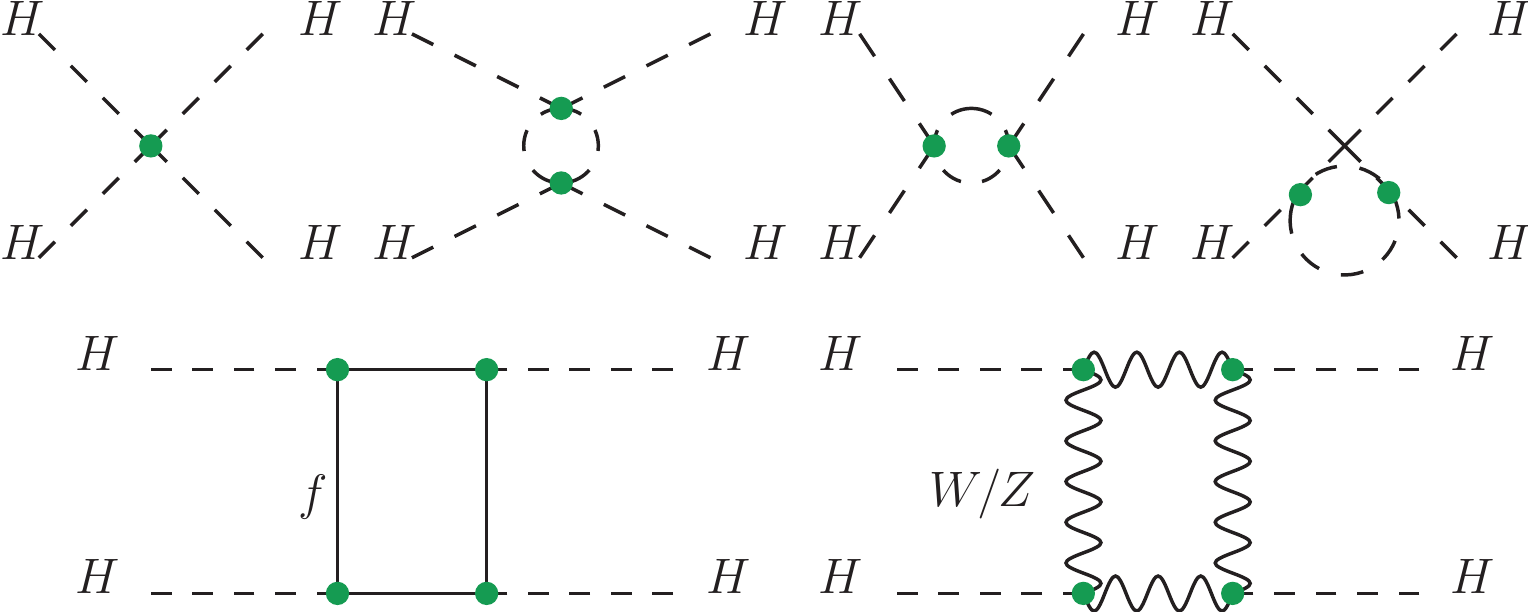} 
\end{center}
\vspace*{-.5cm}
\caption[Feynman diagrams up to one--loop correction for the Higgs
self--coupling]{Feynman diagrams up to one--loop correction for the
  quartic Higgs self--coupling.}
\label{fig:Higgsselfcoupling} 
\end{figure}

We will take another point of view, and look at the Renormalization
Group Equation (RGE) for the quartic Higgs coupling $\lambda$ and
require that it remains finite and the theory perturbative. It will
define a range for the scale where the SM is valid and thus we will
obtain an upper bound for the Higgs
mass. Fig.~\ref{fig:Higgsselfcoupling} shows the tree--level Higgs
self--coupling together with its one--loop corrections. These
contribute to the $\beta$--function for the $\lambda$--coupling; as we
are interested in the high $\lambda$ regime the leading terms will
only deal with the Higgs--Higgs couplings. In this limit the RGE
equation is written at the one--loop order
as~\cite{Cheng:1973nv,Pendleton:1980as,Hill:1980sq,Bagger:1984gn,Beg:1983tu,Duncan:1985ab,Babu:1985ut}:

\beq
\frac{d \lambda (Q^2)} { d\ln Q^2 } = \frac{3}{4\pi^2}
\lambda^2 (Q^2)
\label{eq:RGElambda}
\eeq

When choosing the electroweak symmetry breaking scale $Q_0=v$ as the
reference scale we then obtain
\beq
\lambda (Q^2) = \frac{\lambda (v^2)}{\displaystyle 1 -
  \frac{3}{4\pi^2} \lambda (v^2) \ln \frac{Q^2}{v^2}}  
\eeq

The triviality is recovered at small energy scale $Q \ll  v$ where
$\lambda \to 0$. The other limit will give us the bound that we are
looking for. Indeed at a certain point we see that $\lambda$ becomes
infinite if we increase the energy scale $Q$. We thus reach a Landau
pole, at the scale
\beq
\Lambda_P= v \exp \left(\frac{2\pi^2}{3\lambda(v^2)}\right) = v \exp
\left(\frac{4\pi^2 v^2}{3 M_H^2}\right)
\eeq

If we would like that the scale $\Lambda_P$ be large, the Higgs mass
should be small. If we assume the SM to be valid up to the Planck
scale and then take $\Lambda_P \simeq 10^{19}$ GeV this requires a
light Higgs boson mass $M_H \lsim 145$ GeV. If we take a small
cut--off $\Lambda_P \simeq 10^3$ GeV, the Higgs boson can be heavier,
$M_H \lsim 750$ GeV. In particular, if the cut--off is set at the
Higgs boson mass itself, $\Lambda_P = M_H$, this implies that  $M_H
\lsim  700~{\rm GeV}$. Nevertheless we should bear in mind that all
the calculations presented so far use the perturbation theory and if
$\lambda$ becomes too large the calculation becomes
non--consistent. According to Ref.~\cite{Djouadi:2005gi}, simulations of gauge
theories on the lattice save the argument as the rigorous bound that
is obtained is $M_H \lsim 640$
GeV~\cite{Hasenfratz:1992,Gockeler:1992zj}, in a good agreement
with the perturbative results obtained in this paragraph.

\subsubsection{Stability requirement and lower bound on Higgs mass}

To obtain a lower bound on the Higgs mass we retain in the RGE
equation~\ref{eq:RGElambda} all the terms at the one--loop order, in
particular the fermionic corrections depicted in
Fig.~\ref{fig:Higgsselfcoupling} as the Higgs boson couplings is
proportionnal to the fermion masses. The one--loop RGE equation for
the Higgs self--coupling~\ref{eq:RGElambda} becomes, including all weak
and fermion masses:

\beq
\hspace{-4mm}\frac{d \lambda}{d\ln Q^2} = \frac{1}{16\pi^2} \left[
12 \lambda^2 + 12\frac{m_t^2}{v^2} \lambda - 12\frac{m_t^4}{v^4}  -
\frac{3}{2}\lambda (3g_2^2+ g_1^2) + \frac{3}{16} \left(2 g_2^4+
  (g_2^2+g_1^{2})^2 \right) \right]
\eeq

where $g_1,g_2$ are respectively the hypercharge and weak coupling
constants. The triviality bounds are marginally affected, but
in an important way as the scale of physics beyond the SM will now
depend on the precise value of the top quark mass, which explains why
the top sector is of such great importance in the quest for new
physics. The large impact of the additional fermionic and bosonic
contributions to the Higgs self--coupling $\beta$--function is for
very small values of $\lambda$. For $\lambda \ll \frac{m_t}{v},
g_{1},g_{2}$, the RGE can be approximated by
\beq
\frac{d \lambda}{d\ln Q^2} \simeq \frac{1}{16\pi^2} \left[
12 \lambda^2 - 12 \frac{m_{t}^4}{v^4} + \frac{3}{16} \left(2 g_2^4+ 
(g_2^2+g_1^{2})^2 \right) \right] 
\eeq

This translates into the solution
\beq
\lambda(Q^2)=\lambda(v^2)+ \frac{1}{16 \pi^2} \left[- 12
  \frac{m_{t}^4}{v^4} + \frac{3}{16} \left(2 g_2^4+ (g_2^2+g_1^{2})^2
  \right) \right] \ln \frac{Q^2}{v^2}
\eeq

The top quark contribution becomes dominant for $\lambda\to 0$ and
indeed drives the self--coupling to negative values, which then imply
that the vacuum is not stable anymore. If we want to have a bounded
scalar potential and thus a vacuum we should keep
$\lambda(Q^2)>0$~\cite{Lindner:1988ww,Sher:1993mf,Altarelli:1994rb,Casas:1994qy,
 Sher:1988mj} and
then the Higgs boson mass should have the lower bound
\beq 
M_H^2 >  \frac{v^2}{8 \pi^2} \left[-12 \frac{m_{t}^4}{v^4} +
  \frac{3}{16} \left(2 g_2^4+ (g_2^2+g_1^{2})^2 \right) \right] \ln
\frac{Q^2}{v^2}
\label{eq:Higgsmasslower}
\eeq

Thus we have a lower bound limit dependent on the cut--off $\Lambda_P$
that is the appearance of the new physics. We thus obtain 
\beq
\Lambda_P \sim 10^{3}~{\rm GeV} &\Rightarrow& M_H \gsim 70~{\rm GeV}
\nonumber \\ 
\Lambda_P \sim 10^{16}~{\rm GeV} &\Rightarrow& M_H \gsim 130~{\rm GeV}
\eeq

\subsubsection{Combination of the theoretical bounds}

As stated in Ref.~\cite{Djouadi:2005gi} (nearly) all the bounds presented up
until now were derived in a perturbative approach at the one--loop
level at most. The use of the most advanced results helps to improve
the limits~\cite{Sher:1988mj,Lindner:1985uk,Grzadkowski:1986zw,Hambye:1996wb}.

In particular the $\beta$ functions of all SM couplings have been
calculated up to two loops and can then be included in the analysis of
the Higgs boson mass. From Ref.~\cite{Djouadi:2005gi} we extract for the
$\lambda$ Higgs self--coupling
\beq
\frac{d \lambda}{d\ln Q^2} \equiv 
\beta_\lambda = 24 \frac{\lambda^2}{(16 \pi^2)} - 312 \frac{\lambda^3}
{(16 \pi^2)^2} 
\eeq

At the two--loop level the $\lambda$ coupling can reach an
ultraviolet fixed point where $\beta_\lambda=0$. This means that the
value of the cut--off scale can be lower that obtained above. The
stability bound is again obtained by the requirement that
$\lambda(Q^2)>0$ and we use a two--loop $\beta$ function retaining all
the Higgs, weak boson and fermion corrections, including matching
conditions, that is the precise relation between the physical masses
of the gauge bosons and the top quark and their corresponding
couplings.

\begin{figure}[!t]
\begin{center}
\includegraphics[scale=0.7]{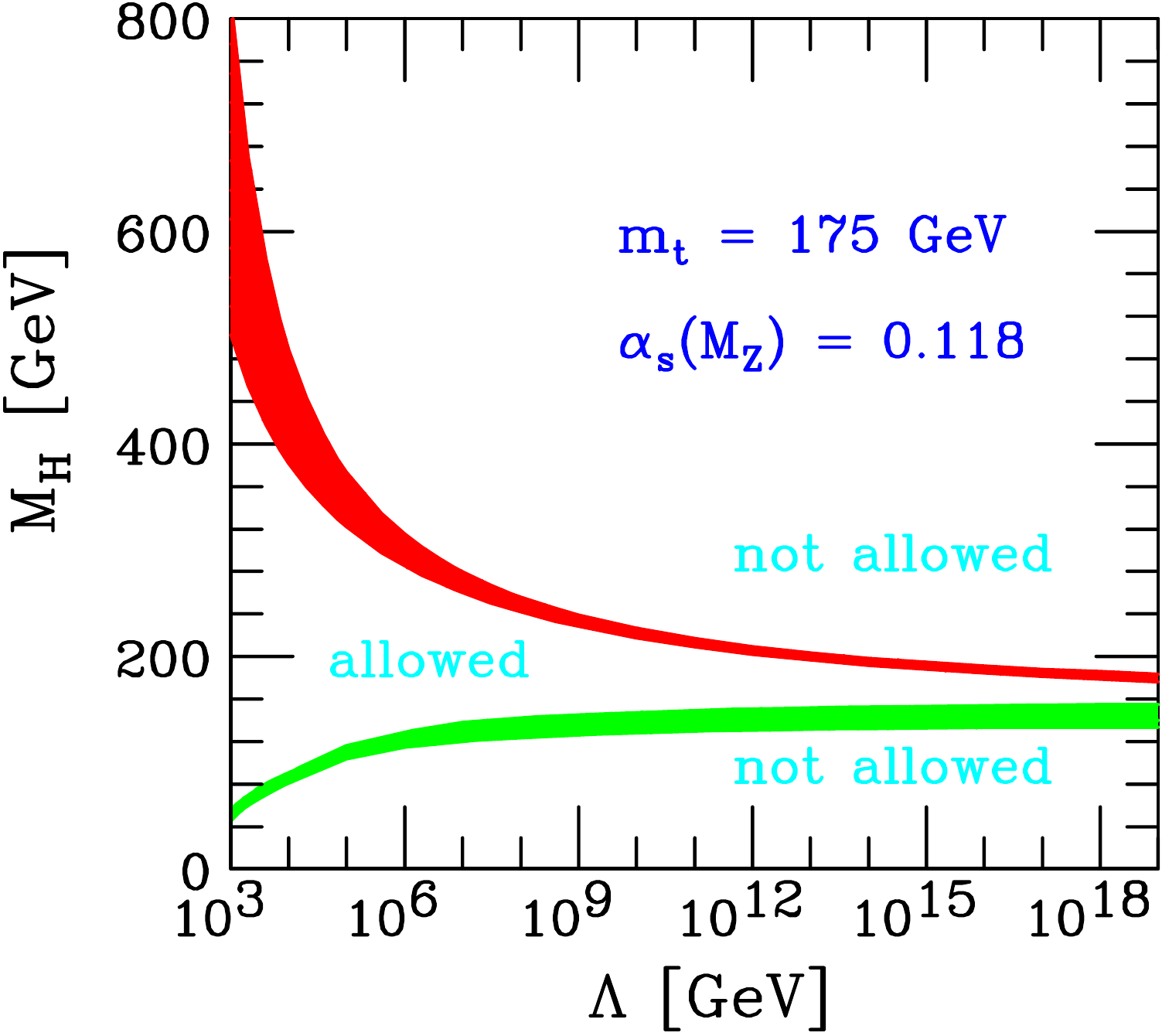} 
\end{center}
\vspace*{-.5cm}
\caption[Theoretical bounds on the Higgs mass in function of the scale
of new physics beyond the SM]{The triviality (upper) bound and the
  vacuum stability (lower) bound on the Higgs boson mass as a function
  of the New Physics or cut--off scale $\Lambda$ for a top quark mass
  $m_t=175 \pm 6$ GeV and $\alpha_s (M_Z) =0.118 \pm 0.002$; the
  allowed region lies between the bands and the colored/shaded bands
  illustrate the impact of various uncertainties. Taken from
  Ref.~\cite{Hambye:1996wb} according to Ref.~\cite{Djouadi:2005gi}.}
\label{fig:Higgstheorybound} 
\end{figure}

We obtain~\cite{Hambye:1996wb} the Roman plot shown in
Fig.~\ref{fig:Higgstheorybound} for 
the stability and triviality bounds on the Higgs boson mass. The graph
displays the allowed values for the Higgs boson mass in function of
the scale $\Lambda$ for new physics effects. The upper band is the
triviality limit and the lower band is the stability limit. These are
bands instead of lines in order to take into account some theory and
experimental uncertainties in the determination (e.g. scale
dependence, imput top quark mass, etc.).

In the end the Higgs boson mass is allowed to be in the range
\beq  
50~{\rm GeV}~ \lsim M_H \lsim ~750~{\rm GeV}  
\eeq 
when new physics is to arise at the TeV scale, taking also into
account the unitarity constraint. If we require that the
SM be valid up to the Grand Unification scale, 
$\Lambda_{\rm GUT} \simeq 10^{16}$ GeV, the Higgs boson mass should
lie in the range
\beq  
130~{\rm GeV}~ \lsim M_H \lsim ~180~{\rm GeV}  
\eeq

\subsection{Experimental bounds on the Higgs
  mass \label{section:HiggsboundsExp}}

The experimental quest for the evidence of the SM Higgs boson has
started more than twenty years ago, at the CERN Large Electron Positron
(LEP) collider. As the Higgs boson has sizeable contributions
to radiative corrections to many processes in the SM, its existence
can be probed through its effects on electroweak observables. Hence we
can obtain indirect constraints on its mass. We will also present of
course the constraints obtained with its direct search both at the LEP
collider and the Fermilab Tevatron collider. We note that the last
constraint, obtained at the Tevatron, will be challenged in the
following chapters of this thesis.

\subsubsection{Indirect searches through precision data}

The SM has been tested in many experiments for decades and the results
have reached a very high precision. It then imposes stringent
conditions on new physics phenomena as well as on the Higgs boson
itself, the only yet unknown SM particle to be discovered.

We will summarize the electroweak precision data used to constraint
indirectly the SM Higgs boson mass, classified by the different experiments:

\begin{enumerate}[$\bullet$]

\item{{\it LEP1 searches}: LEP1 was the first era of running for the
    LEP collider, when the designed center--of--mass energy was 110
    GeV (the actual energy used was 45.6 GeV per beam near the $Z$
    peak). This has given us much information 
    on the $Z$ lineshape: the $Z$ mass, the total $Z$ width
    $\Gamma_Z$, the peak
    hadronic cross section $\sigma^{\rm had} \equiv \sigma(e^+ e^- \to
    {\rm hadrons}$, the ratio $R_{\ell, c,b}$  between the $Z$ partial
    decay width into leptons, charm and bottom quarks, and the total
    hadronic $Z$ decay width, the forward--backward asymmetries
    $A_{FB}^f$ for leptons and heavy $c,b$ quarks, the $\tau$
    polarization asymmetry $A^\tau_{\rm pol}$, etc.}

\item{{\it LEP2 searches}: LEP2 was the second step for the LEP
    collider when the designed energy did reach 209 GeV
    center--of--mass energy in 2000, the year of the end of the
    collider operations. It has given valuable information on the
    $W$ boson: its mass $M_W$, the total decay width $\Gamma_W$, again
    the different asymmetries that are used to measure $\sin^2\theta_W$
    through the lepton channel, the hadronic contribution
    $\Delta^{(5)}_{\rm had} \alpha$ to the value of $\alpha$, etc. As
    discussed in the next subsubsection the Higgs boson was also
    hunted directly at LEP2.}

\item{{\it SLC searches}: The Stanford Linear Collider (SLC)
    experiments in the beginning of the nineties have given unique
    results at the high precision level in electroweak physics thanks
    to its very polarized beams. The longitudinal polarization
    asymmetry $A_{LR}^f$ which has been measured at the SLC thus gives
    the best individual measurement of $\sin^2\theta_W$. It has also
    measured the left--right forward--backward asymmetries for the
    heavy $b,c$ quarks,  $A^{b,c}_{LR,FB}$.}

\item{{\it Tevatron searches}: The Fermilab Tevatron collider has
    started to operate in 1987 and will end running in the end of
    2011. It has given precise measurements of the $W$ mass, the total
    decay width $\Gamma_W$ and in particular of the top quark mass
    $m_t$, which were discovered at the Tevatron in 1995. Being a
    proton--antiproton collider it is also a good place to measure
    asymmetries. Its most important experimental programm is the
    search for the Higgs boson, and as discussed later it has given
    direct constraints on its mass.}

\item{In addition there are high--precision measurements at low
    energies, for example the $\nu_\mu$-- and $\bar{\nu}_\mu$--nucleon
    deep--inelastic scattering cross sections, from which we can
    extract a determination of $\sin^2\theta_W$ through the measure of
    the $Z$ couplings to fermions, both the right--handed and the
    left--handed.}

\end{enumerate}

\begin{figure}[!t]
\begin{center}
\includegraphics[scale=0.6]{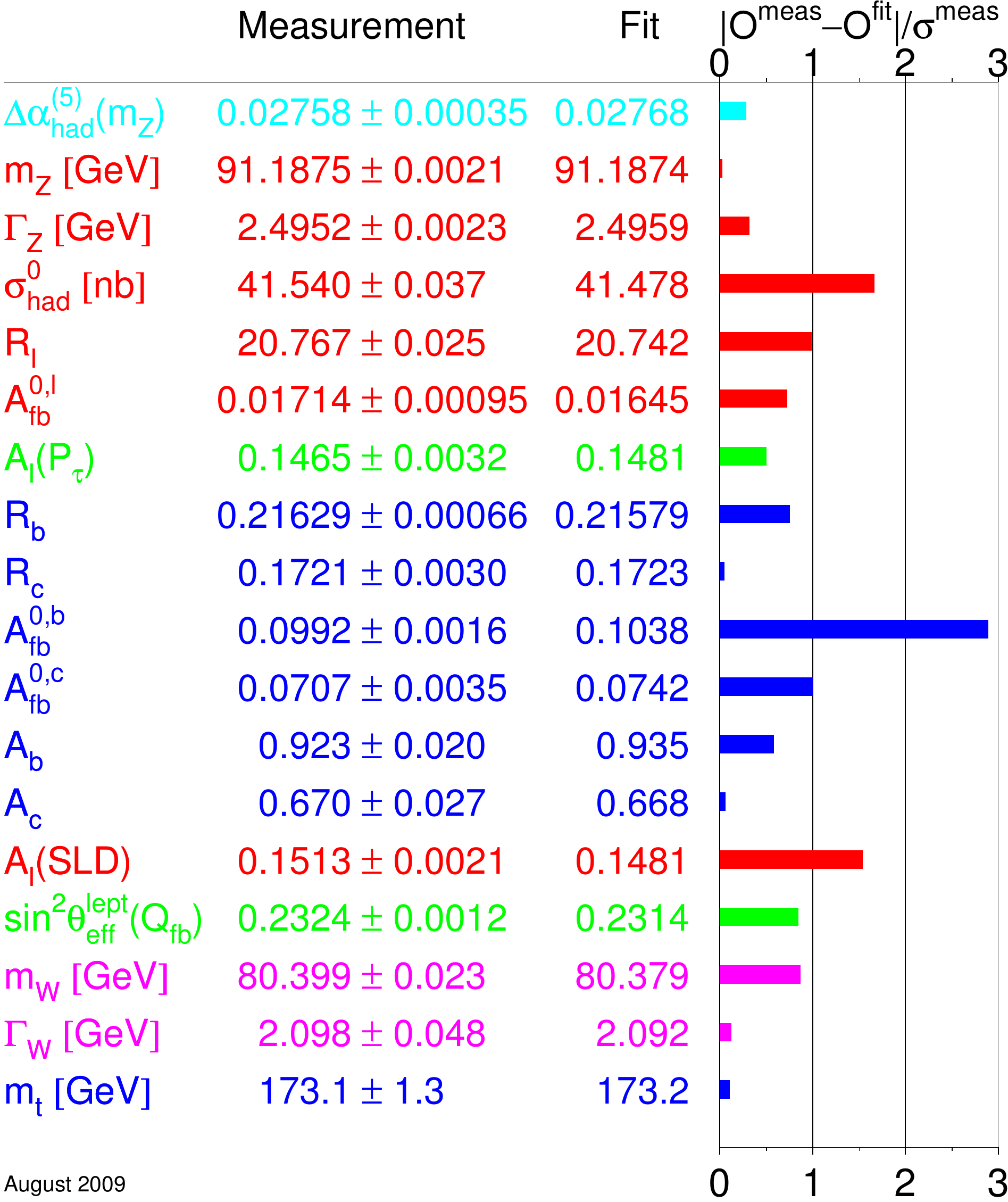} 
\end{center}
\vspace*{-.5cm}
\caption[Electroweak precision data]{The SM fits results compared to
  the data for electroweak precision observables, together with the
  standard deviations. Figure taken from the LEP EW Working Group
  webpage, dated on August 2009.}
\label{fig:EWdata} 
\end{figure}

The comparison with the SM calculation of all these precision data is
displayed in Fig.~\ref{fig:EWdata} which has been taken from LEP
ElectroWeak Working Group webpage. The theoretical predictions are in
a very good agreement with the data, apart from a tension in the
measure of the forward--backward asymmetry for the $b$--quark which is
nearly 3$\sigma$ away from the SM value.

\begin{figure}[!t]
\begin{center}
\includegraphics[scale=0.6]{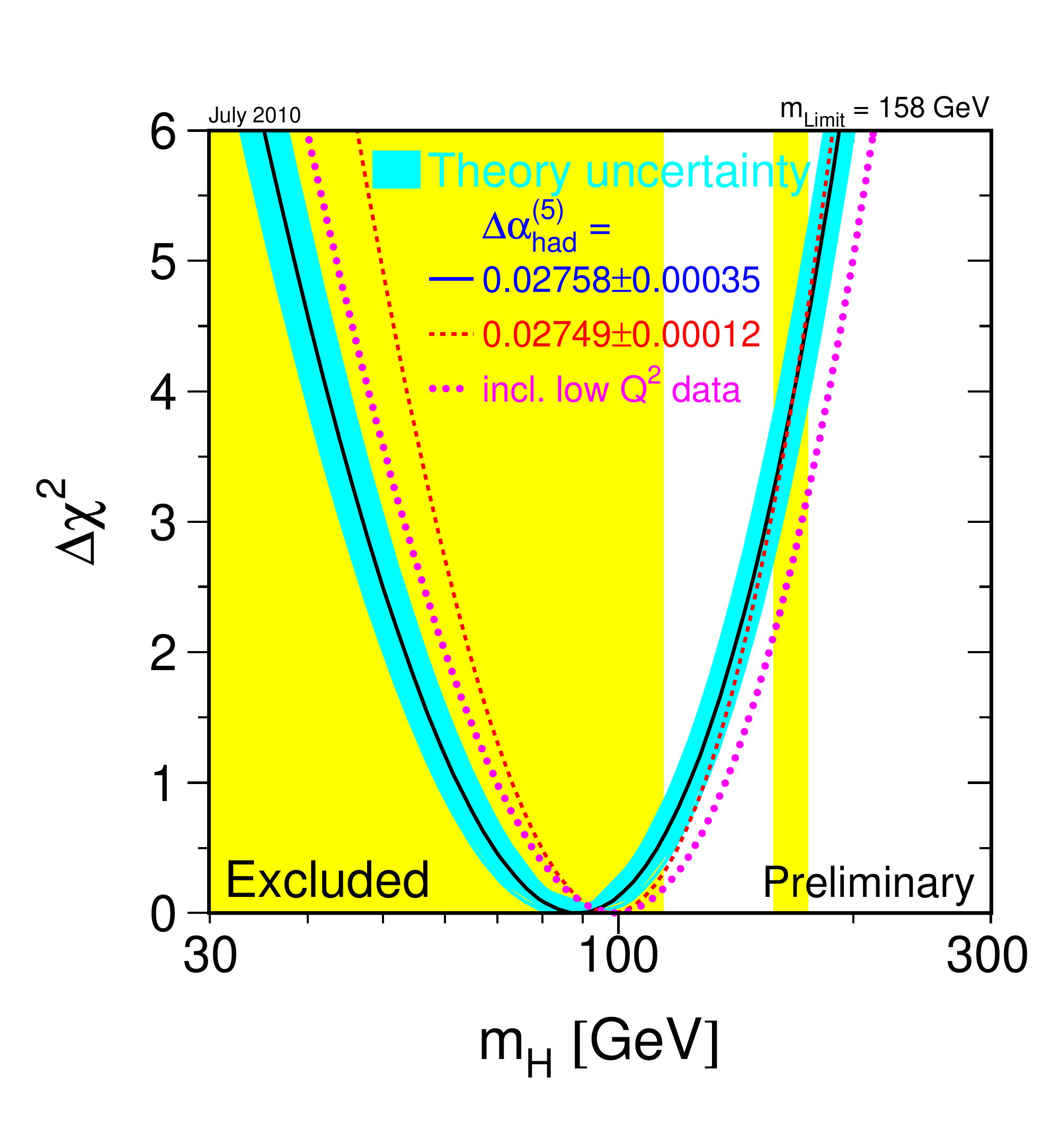} 
\end{center}
\vspace*{-.5cm}
\caption[Indirect constraints on the SM Higgs boson mass]{The
  $\Delta\chi^2$ of the fit of the electroweak precision data to the
  predictions of the SM as a function of the SM Higgs mass value. The
  solid line include all data but the low energies data, and the
  effect of varying the $\Delta^{(5)}_{\rm had}\alpha$ is also shown.}
\label{fig:indirectHiggsconstraint} 
\end{figure}

When combining all these precision data measurements, stringent
constraints on SM the Higgs boson mass can be obtained. When one
performs a fit on all the data, the $\Delta \chi^2=\chi^2-\chi_{\rm
  min}^2$ of the fit
displayed in Fig.~\ref{fig:indirectHiggsconstraint} is obtained,
depending on the Higgs boson mass. 

On the left of the plot the area depicted in yellow (shaded), close to
the minimum value of the fit, is the excluded limit by the LEP2
experiments as discussed just below. The small yellow (shaded) band on
the right of the plot displays the limits set by the Tevatron run II
experiments, again discussed later on. The fitted SM Higgs boson mass
is then
\beq
M_H = 89^{+35}_{-26}~{\rm GeV}
\eeq
and the 95\%CL upper bound is
\beq
M_H \leq 158~{\rm GeV~at}~95\%{\rm CL}
\eeq

This last limit can be a bit relaxed is the direct search limits at
the LEP and the Tevatron colliders are taken into account in the fit,
leading to $M_H\leq 185$ GeV at 95\%CL. The data thus strongly
disfavour a very heavy SM Higgs boson $M_H\lsim 700$ GeV which anyway
is (nearly) outside the theoretical bounds presented above.

It should be worth mentioning that the SM fit is not as good because
of the nearly 3$\sigma$ deviation of the forward--backward asymmetry
of the $b$--quark $A^{0,b}_{\rm fb}$. Nevertheless if this quantity were
removed from the fit, the obtained central value for the Higgs boson
mass together with the uncertainties would fall below the direct
search limit, increasing the tension between the central value from
the EW fit and the bounds obtained by the direct searches.

\subsubsection{Direct searches at the LEP collider}

Direct searches of the SM Higgs boson started at the LEP collider in
the LEP1 era, but the interesting results have been obtained during
the LEP2 era when $\sqrt s = 209$ GeV. The dominant production channel
is the Higgs--strahlung $e^+e^-\to ZH$, see
Refs.~\cite{Ellis:1975ap, Lee:1977eg, Jones:1979bq,
  Ioffe:1976sd, Ma:1979id, Finjord:1979br, Berends:1984dw}. The search
has been conducted in several final states topologies:

\begin{enumerate}[$\bullet$]
\item{{\it Final state $b\bar b\nu \bar \nu$}: this final state is
    obtained with $H\to b\bar b$ together with $Z\to \nu \bar \nu$.}
\item{{\it Final state $b\bar b \ell^+ \ell^-$}: with $\ell=e,\mu$
    this is obtained with the same SM Higgs decay as above but with a
    leptonic $Z$ decay $Z\to \ell^+ \ell^-$.}
\item{{\it Final state $b\bar b \tau^+ \tau^-$}: this topology is
    obtained through two different channels, $e^+ e^-\to (H\to b\bar
    b)~(Z\to \tau^+ \tau^-)$ and $e^+ e^-\to (H\to \tau^+
    \tau^-)~(Z\to b\bar b)$ as the two Higgs decay channel are
    sizeable, see section~\ref{section:SMHiggsDecay}.}
\end{enumerate}

\begin{figure}[!t]
\begin{center}
\includegraphics[scale=0.50]{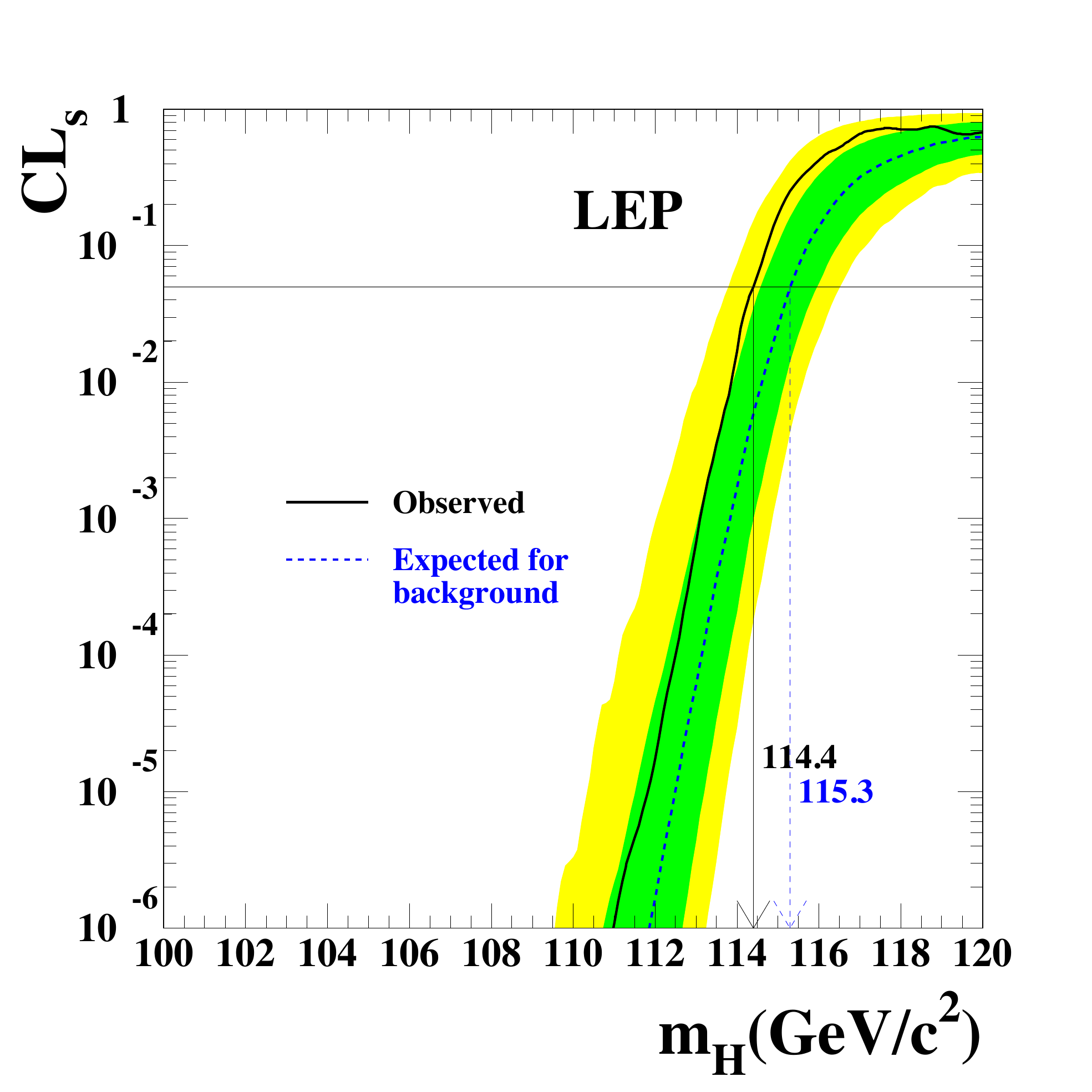} 
\end{center}
\vspace*{-.5cm}
\caption[95\%CL exclusion limit on the SM Higgs boson mass at the LEP
collider]{Confidence level CL$_s$ for the signal plus background
  hypothesis in the Higgs production at LEP2. The black (solid) line
  is for the observed value, the blue (dashed) line is for the
  expected value. The intersection between the horizontal line CL$_s =
0.05$ with the data lines define the lower 95\%CL limit on the SM
Higgs boson mass. Figure taken from Ref.~\cite{Barate:2003sz}.}
\label{fig:DirectLEPHiggsconstraint} 
\end{figure}

Even if there were tantalizing results in the end of year 2000 of the
observation of a SM Higgs boson around 115 GeV, the excess of
$1.7\sigma$ (after a re--evaluation as the first claimed observation
was of $2.9\sigma$ evidence) was not significant enough to claim a
discovery~\cite{Barate:2003sz}. The combination of the results from
the OPAL, L3, DELPHI and ALEPH experiments at LEP has set an exclusion limit
\beq 
M_H > 114.4~{\rm GeV~at}~95\%{\rm CL}
\label{eq:HiggsMassLEP}
\eeq 
This is displayed in Fig.~\ref{fig:DirectLEPHiggsconstraint}, which
shows the statistical CL$_s$ for the signal plus background hypothesis
as a function of the Higgs boson mass. If the CL$_s$ is below 0.05 the
Higgs boson mass is excluded. The expected exclusion is found to be
$M_H> 115.3$ GeV, the discrepency with the observed exclusion being
due to the excess discussed above.

It is worth mentioning that this direct search limit is rather robust
as it has been obtained in pure electroweak processes (at LO). In
particular the theoretical uncertainties on the predicted cross
section times branching ratios used for the comparison with the data
are very small; this has to be compared with the direct search limit
discussed in the following at the Tevatron collider which is a hadron
collider.

\subsubsection{Direct searches at the Tevatron collider}

The direct search for the SM Higgs boson at the Fermilab Tevatron
collider is its major experimental program and started more than a
decade ago. The production channel will be discussed in details in the
next section, but we can already mention briefly that there are two
dominant channels, the Higgs--strahlung processes $p\bar p\to H V$
with $V=W,Z$ for low Higgs mass searches $M_H \lsim 135$ GeV, and the
gluon--gluon fusion channel $gg\to H$for high Higgs mass searches
$M_H\gsim 135$ GeV. The most important decay channels used at the
Tevatron are the hadronic $H\to b\bar b$ decay for low Higgs mass
searches and the bosonic $H\to W^{(*)}W^{(*)}$ for high Higgs mass
searches. Again, this will be discussed in details in the following,
see section~\ref{section:SMHiggsDecayIntro}.

\begin{figure}[!t]
\begin{center}
\includegraphics[scale=0.65]{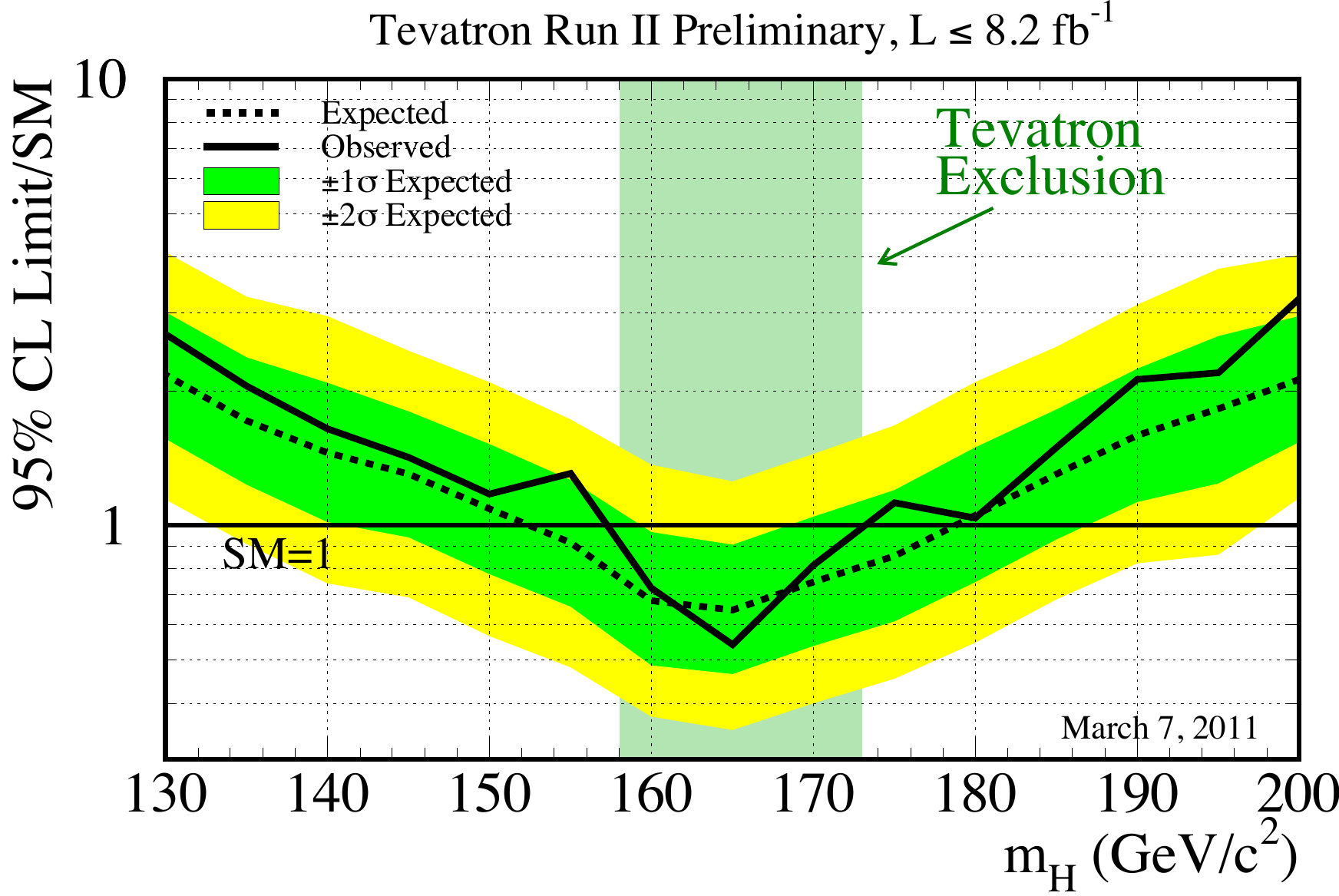} 
\end{center}
\vspace*{-.5cm}
\caption[95\%CL exclusion limit on the SM Higgs boson mass at the
Tevatron collider]{Ratio $R$ between the 95\%CL exclusion cross
  section and the predicted SM cross section as a function of the SM
  Higgs boson mass at the Tevatron run II. The solid black line shows
  the observed value and the dashed black lines shows the expected
  value. When the lines fall below the $R=1$ the Higgs boson mass is
  excluded at 95\%CL. This exclusion band is displayed by the light
  green (shaded) band. Figure taken from Ref.~\cite{Aaltonen:2011gs}.}
\label{fig:DirectTEVHiggsconstraint} 
\end{figure}

The Tevatron has collected enough luminosity to be sensitive
to Higgs signal for $M_H \simeq 165$ GeV, but has not yet seen the
elusive particle. For the first time since the LEP experiments it was
able to produce a new exclusion limit in the year
2008~\cite{Tevatron:2009je} which then was extended in the next
years~\cite{Aaltonen:2010yv,Tevatron:2010ar} to reach the current
exclusion limit~\cite{Aaltonen:2011gs} at 95\%CL:
\beq
M_H \leq 158~{\rm GeV~or}~M_H\geq 173~{\rm GeV~at}~95\%{\rm CL}
\eeq
This is displayed in Fig.~\ref{fig:DirectTEVHiggsconstraint} where the
  ratio between the 95\%CL exclusion limit on the Higgs cross section
  and the SM prediction is shown as a function of the SM Higgs boson
  mass. When the ratio falls below the unity the SM Higgs boson mass
  in excluded.

We remind the reader that this impressive result has been obtain at a
hadron collider. Contrary to the result~\ref{eq:HiggsMassLEP} obtained
at the LEP collider, which is rather robust and theoretically well
behaved, the exclusion limit at the Tevatron is based upon a
comparison between data and SM prediction at a hadron collider, which
are well known to be plagued with various theoretical uncertainties as
discussed in this thesis. We will see that these limits are put into
question, see Refs.~\cite{Baglio:2010um, Baglio:2011wn}, and that will be
discussed in particular in section~\ref{section:SMHiggsTevExclusion}.

\vfill

\pagebreak

\section{Higgs production at the Tevatron}

\label{section:SMHiggsTev}

The Fermilab Tevatron collider has started its operation back in 1987
and has been the home of major particle physics discoveries, e.g. the
discovery of the top quark in 1995 which leads to a Nobel prize. The
Higgs search program is currently one of the most important research
plan for the collider. In this context, the CDF and D0 experiments at
the Tevatron have collected enough data to be sensitive to the
Standard Model Higgs boson. The two collaborations recently performed
a combined analysis on the search for this particle and excluded at
the 95\% confidence level the possibility of a Higgs boson in the mass
range between 158 and 173
GeV~\cite{Tevatron:2009je,Aaltonen:2010yv,Tevatron:2010ar,Aaltonen:2011gs}
(see section~\ref{section:HiggsboundsExp}). We are thus entering a new
era in the quest of the Higgs particle as this is the first time that
the mass range excluded by the Large Electron--Positron (LEP)
collaborations in the late 1990s, $M_H \geq 114.4$
GeV~\cite{Barate:2003sz}, is extended.

Although current hadronic colliders have a center--of--mass energy
that is greater than the LEP energy, which is good to produce
new physics events, QCD interactions are dominant and plague the
theoretical prediction with various sources of uncertainties in contrast
with the predictions obtained at LEP where the production cross sections
were mainly sensitive to (small) electroweak effects which are well
under control. Among these are the contributions of yet uncalculated
higher order corrections which can be important as the strong
coupling constant $\alpha_s$ is rather large; the errors
due to the folding of the partonic cross sections with the parton distribution
functions (PDFs) to obtain the production rates at the hadronic level; and the
errors on some important input parameters such as $\alpha_s$. An
accurate calculation of the estimation of these uncertainties is then
mandatory in order to have a complete theoretical prediction for the
production rates and eventually the production cross section times
branching ratio prediction. It would allow for a consistent
comparison with the experimental data available at the Tevatron which
are described in term of exclusion bounds\footnote{An example of such
  a situation is the $p\bar p \to b\bar b$  production cross section
  that has been measured at the Tevatron (and elsewhere) and which was
  a factor of two to three larger than the theoretical prediction,
  before higher order effects and various uncertainties were
  included~\cite{Cacciari:2003uh}.}. 

At the Tevatron, only two production channels are important for the  Standard
Model Higgs boson\footnote{The CDF/D0 exclusion limits
 ~\cite{Tevatron:2009je,Tevatron:2010ar} have
been obtained by considering a large variety of Higgs production and decay
channels (36 and 54 exclusive final states for, respectively, the CDF and D0
collaborations) and combining them using artificial neural network techniques.
However, as will be seen later, only a few channels play a significant role  in
practice.}. In the moderate to high mass range, 140 GeV$\lsim M_H \lsim 200$
GeV, the Higgs boson decays dominantly into $W$ boson pairs (with one $W$ state
being possibly off
mass--shell)~\cite{Djouadi:1997yw,Butterworth:2010ym} and the main production channel
is the gluon--gluon fusion mechanism  $gg \to H$~\cite{Georgi:1977gs} which proceeds
through heavy (mainly top and, to a lesser extent, bottom) quark triangular
loops. The Higgs particle is then detected through the  leptonic decays of the
$W$ bosons, $H\to WW^{(*)}\to \ell^+ \nu \ell^- \bar \nu$ with $\ell=e,\mu$,
which exhibits different properties than the $p\bar p\to W^+ W^- \to \ell \ell$
plus missing energy continuum background~\cite{Dittmar:1996ss}. 

It is well known that the $gg\to H$ production process is subject to extremely
large QCD radiative corrections
~\cite{Djouadi:1991tka, Dawson:1990zj, Graudenz:1992pv, Spira:1993bb,
  Spira:1995rr, Spira:1997dg, Harlander:2002wh, Anastasiou:2002yz,
  Ravindran:2003um, Catani:2003zt, deFlorian:2009hc,Anastasiou:2009bt}.
In contrast, the electroweak radiative corrections are much smaller, being at
the level of a few percent~\cite{Djouadi:1994ge, Aglietti:2004nj,
  Degrassi:2004mx, Actis:2008ug,Actis:2008ts,Anastasiou:2008tj} as in the
case of Higgs production at the LEP collider.
For the corrections due to the strong interactions, the $K$--factor defined as
the ratio of the higher order (HO) to the lowest order (LO) cross 
sections, consistently evaluated with the $\alpha_s$ value and the PDF sets
at the chosen order, 
\beq
K_{\rm HO}= \sigma^{\rm HO} |_{ (\alpha_s^{\rm HO} \, , \, {\rm PDF^{HO} )} } 
\; / \; 
           \sigma^{\rm LO} |_{ (\alpha_s^{\rm LO} \, , \,  {\rm PDF^{LO}}) } \, , 
\label{eq:kfactor}
\eeq
is about a factor of two at next-to-leading order
(NLO)~\cite{Djouadi:1991tka, Dawson:1990zj, Graudenz:1992pv,
  Spira:1993bb,Spira:1995rr} and about a factor of three at the
next-to-next-to-leading  order
(NNLO)~\cite{Harlander:2002wh,Anastasiou:2002yz,Ravindran:2003um}. In
fact, this exceptionally large $K$--factor is what allows
a sensitivity on the Higgs boson at the Tevatron with the presently collected
data. Nevertheless, the $K$--factor is so large that one may question the
reliability of the perturbative series, despite of the fact that there seems to be
kind of a convergence of the series as the NNLO correction is smaller
than the NLO correction\footnote{At LHC energies, the problem of the
convergence of the perturbative series is less severe as the QCD
$K$--factor is only $\sim 1.7$ at NLO and $\sim 2$ at NNLO in the relevant
Higgs mass range.}. 

In the low mass range, $M_H \lsim 140$ GeV, the main Higgs decay channel is 
$H\to b\bar b$~\cite{Djouadi:1997yw} and the $gg$ fusion mechanism cannot be used
anymore as the $gg \to H \to b\bar b$ signal is swamped by the huge QCD jet
background. The Higgs particle has then to be detected through its associated
production with a $W$ boson  $q \bar q \to WH$~\cite{Glashow:1978ab} which leads to
cleaner $\ell \nu b \bar b$ final states~\cite{Stange:1994bb}. Additional topologies
that can also be considered in this context are  $q \bar q \to  WH$ with $H \to
WW^* \to \ell \ell \nu \nu$ or the twin production process    $q \bar q \to 
ZH$ with the subsequent decays $H \to b\bar b$ and $Z \to \nu \bar \nu$ or
$\ell ^+\ell^- $. Other production/decay channels are expected to lead to very
low rates and/or to be affected with too large QCD backgrounds. 

At the Tevatron, the Higgs--strahlung processes $q \bar q \to VH$  with
$V=W,Z$ receive only moderate higher order corrections: the QCD corrections
increase the cross sections by about 40\% at
NLO~\cite{Altarelli:1979ub, KubarAndre:1978uy, Han:1991ia,
  Ohnemus:1992bd, Djouadi:1999ht} and 10\% at
NNLO~\cite{Brein:2003wg}, while the impact of the one--loop
electroweak corrections is small, leading to a $\approx 5\%$ decrease
of the cross sections~\cite{Ciccolini:2003jy}. 
Thus, in contrast to the gluon--gluon fusion process, the production cross
sections in the Higgs--strahlung processes should be well under control. 
The purpose of this section is to study the Higgs boson production at
the Tevatron in these two main search channels, including all relevant
higher order QCD and electroweak corrections, using the latest MSTW
2008 set of PDFs~\cite{Martin:2009iq}\footnote{The PDFs are available
  in Ref.~\cite{MSTWpage}.} and comparing
with other PDFs set available on the market. The calculation has been
performed in various recent
analyses~\cite{deFlorian:2009hc,Anastasiou:2008tj} for the 
gluon--gluon fusion process and CDF/D0 collaborations for example used
their normalized cross sections to produce the Higgs mass 95\% CL
limits~\cite{Tevatron:2009je,Tevatron:2010ar}
in their combined analysis. The results were published
in~\cite{Baglio:2010um} and were the first update for the
Higgs--strahlung production channels $q\bar q \to VH$ with the newest
PDFs available on the market. The Tevatron analyses before July 2010
used the normalised cross sections available in
Refs.~\cite{Brein:2004ue, Assamagan:2004mu} 
which make use of the old MRST2002 set of PDFs~\cite{Martin:2002dr}, a
parametrisation that was approximate as it did not include the full
set of evolved PDFs at NNLO. For completeness,
an update of the cross sections for the two other single Higgs production
channels at hadron colliders will also be presented briefly: the weak
boson fusion $p \bar p \to qqH$~\cite{Jones:1979bq, Cahn:1983ip,
  Harn:1985iy, Altarelli:1987ue,Han:1992hr, Figy:2003nv} and the
associated production with top quark pairs  $p\bar p \to t \bar t
H$~\cite{Raitio:1978pt, Kunszt:1984ri, Ng:1983jm,Beenakker:2001rj,
  Beenakker:2002nc, Reina:2001sf, Dawson:2002tg}. These channels play
only a minor role at the Tevatron but have also been included in the
CDF/D0 analysis~\cite{Tevatron:2009je,Tevatron:2010ar}.

The study of the Higgs boson production will also include a
comprehensive investigation of all possible sources of uncertainties
mentioned earlier that have a significant impact on the central
prediction for the two main search channels. We will first discuss the
choice of the central renormalization scale $\mu_R$ and factorization
scale $\mu_F$ in the gluon--gluon fusion mechanism and explain why
some choices have been made about the order of calculation. We will
then begin the study of the uncertainties with the analysis of the
unknown higher order effects, which are usually estimated by exploring
the cross sections dependence on the renormalization scale $\mu_R$ and the 
factorization scale $\mu_F$. In most recent analyses, the two scales are varied
within a factor of two from a median scale which is considered as the most
natural one. We show that this choice slightly underestimates the higher order
effects and we use a criterion that allows a more reasonable estimate of the
latter: the range of variation of the two scales $\mu_R$ and $\mu_F$  should be
the one which allows the uncertainty band of the LO/NLO cross section
to match the central value of the cross section at the highest
calculated order. In the case of $gg \to H$, for the uncertainty band
of the LO cross section to reach the central result of the NNLO cross
section, a variation of  $\mu_R$ and $\mu_F$ within a factor of $\sim
3$ from the central value $\mu_R= \mu_F =\frac12 M_H$ is required;
this choice of central scale will also be discussed. When the scales
are varied within the latter range, one obtains an uncertainty on the
NNLO cross section of $\approx 18\%$, which is slightly larger than
what is usually assumed in inclusive calculation. We will see that we
obtain result that is comparable with what is used in current Tevatron
combined analysis and which comes from 0,1,2 jets bin
analysis~\cite{Anastasiou:2009bt}.

We then discuss the errors resulting from the folding of the partonic cross
sections with the parton densities, considering not only the recent MSTW set 
of PDFs as in
Refs.~\cite{deFlorian:2009hc,Anastasiou:2009bt,Anastasiou:2008tj}, but
also two other PDF sets that are available in the literature:
CTEQ~\cite{Nadolsky:2008zw}\footnote{The PDFs are available in
  Ref.~\cite{CTEQpage}.} and ABKM~\cite{Alekhin:2009ni}\footnote{The
  PDFs are available in Ref.~\cite{ABKMpage}.}. We will also make a
comprehensive comparison of all the NNLO available PDFs set on the
market: apart from MSTW and ABKM we will also consider
HERAPDF~\cite{HERApage} and JR09~\cite{JimenezDelgado:2009tv}. In the
case of the cross section for the $gg \to H$ process at the Tevatron,
we find that while the PDF uncertainties evaluated within the same
scheme are moderate, as also shown in
Refs.~\cite{deFlorian:2009hc,Anastasiou:2009bt,Anastasiou:2008tj}, the
central values of the cross sections obtained using the three schemes
can be widely different. We show that it is only when the experimental
as well as the theoretical errors on the strong coupling constant
$\alpha_s$ are accounted for that one obtains results that are
consistent when using the MSTW/CTEQ and ABKM schemes. As a result, the
sum of the PDF+$\Delta^{\rm exp} \alpha_s$ and $\Delta^{\rm th}
\alpha_s$ uncertainties, that we evaluate using a set--up recently
proposed by the MSTW collaboration to determine simultaneously the
errors due to the PDFs and to $\alpha_s$, is estimated to be at least
a factor of two larger than what was generally assumed in earlier
analyses before the publication of Ref.~\cite{Baglio:2010um}. We will
also show that the error we obtain is actually comparable to what
should be taken when using the PDF4LHC recommendation~\cite{PDF4LHC}.

Finally, a third source of potential errors is considered in the $gg$
fusion mechanism: the one resulting from the use of an effective field theory
approach, in which the loop particle masses are assumed to be much larger than
the Higgs boson mass, to evaluate the NNLO contributions. While this error is
very small in the case of the top--quark contribution at the Tevatron,
it is at the percent 
level in the case of the $b$--quark loop contribution at NNLO QCD where the
limit $M_H \ll m_b$ cannot be applied and where the $b$--loop is
indeed totally absent as there are no $b$--loop NNLO calculation
available up until now. This is also the case of the three--loop
mixed QCD--electroweak radiative corrections that have obtained in the
effective limit $M_H \ll M_W$, which lead to a few percent uncertainty. In
addition, an uncertainty of about 1\% originates from the freedom in the
choice  of the input $b$--quark mass in the $Hgg$ amplitude. The total
uncertainty in this context is thus not negligible and amounts to a few
percent.

We then address the important issue of how to combine the theoretical errors
originating from  these different sources. Since using the usually adopted
procedures of adding these errors either in quadrature, as is done  by the
experimental collaborations for instance, or linearly as is generally the case
for theoretical errors, lead to either an underestimate or to an overestimate
of the total error, we propose a procedure that seems to be more
adequate. One first determines the maximal and minimal values of the cross
sections obtained from the variation of the renormalization and factorization
scales, and then estimate directly on these extrema cross sections the 
combined uncertainties due to the PDFs and to the experimental and theoretical
errors on $\alpha_s$. The other smaller theoretical uncertainties, such as
those coming from the use of the effective approach in $gg\to H$, can be then
added linearly to this scale, PDF and $\alpha_s$ combined error.

We will in the end show that the total theoretical error that we
obtain is nearly twice the one often quoted in the literature and used
for the Tevatron analyses. In particular, in the
case of the most sensitive Higgs production channel at the Tevatron,  $gg \to H
\to \ell \ell \nu \nu$,  the overall uncertainty on the NNLO total cross
section is found to be of the order of $\pm 40\%$. This is
significantly larger than the uncertainty of $\approx \pm 20\%$ assumed
by the CDF/D0 combined Higgs search analysis. As
a result, we believe that the exclusion range  given by the Tevatron
experiments for the Higgs mass in the Standard Model, 158 GeV $\le M_H \le 173$
GeV, should be discussed in the light of these results and that 
will be discussed in section~\ref{section:SMHiggsTevExclusion}.

\subsection{The main production
  channels \label{section:SMHiggsTevIntro}}

We present the procedure used to obtain the central or ``best'' values
of the total cross sections for SM Higgs production at the
Tevatron. We recall the reader that we mainly concentrate on the two
main search channels, that is the gluon--gluon fusion and
Higgs--strahlung processes, but we also mention very briefly the two
other productions channels for single Higgs production: the vector
boson fusion and the associated Higgs production with top quark pairs.

The Higgs production at hadron colliders, as for any hadronic process,
requires that the partonic elements which actually collide have to be
extracted from the initial (anti)protons. This is summarized by the
factorization procedure which is behind the concept of parton
distribution functions (PDFs): a PDF for a parton Y evaluated at the scale $Q$ and
for a momentum fraction $x$ is the probability to extract from the
(anti)proton the parton Y at the scale $Q$ with a linear momentum $x
\times \vec{P}$ where $\vec{P}$ is the initial linear momentum of the
(anti)proton. We then have to factorize the PDF with the partonic production
cross section. 

If we call $S$ for the center--of--mass energy and
$\shat$ for the partonic center--of--mass energy, $x_1$ for the
momentum fraction of the first parton and $x_2$ for the momentum
fraction of the second parton, we have $\shat = x_1 x_2 S$ in the
massless limit for the two incoming partons, which is always the case
at current hadronic colliders (we just have to compare the mass of
the (anti)proton which is nearly 1 GeV to the Tevatron 1.96 TeV 
center--of--mass energy). We then have

\begin{eqnarray}
\displaystyle \sigma(p\bar p\to A B) = & ~ \nonumber\\ 
 ~& \hspace{-15mm}\displaystyle \sum_{i,j} \int_0^1 f_i(x_1) dx_1 \int_0^1
 f_j(x_2) dx_2 \displaystyle \hat{\sigma}_{i j}(\shat=x_1 x_2 S) \Theta\left(\shat
\geq (M_A+M_B)^2\right)
\end{eqnarray}

where $\hat{\sigma}_{i,j}$ is the partonic cross section $i j \to A
B$ and $\Theta$ is the usual step function. The dependence of the PDFs
over the scale $Q$ called the factorization scale $\mu_F$ has been
made implicit in the equation above for simplification. We sum over
all possible initial partonic states and we could also include
additional jets in the production denoted collectively by $X$. If we
make the following variable replacement $\displaystyle
x_2=\frac{\shat}{x_1 S}$ we obtain

\beq
\sigma(p\bar p\to A B) = \sum_{i,j} \int_0^1 dx_1 f_i(x_1)
\int_{(M_A+M_B)^2}^{x_1 S} \frac{d\shat}{S} f_j\left(\frac{\shat}{x_1 S}\right)
\hat{\sigma}_{i j}(\shat)
\eeq

then followed by the reordering of the interval of variation of
$\shat$ and $x_1$ which then gives

\beq
\sigma(p\bar p\to A B) = \sum_{i,j} \int_{(M_A+M_B)^2}^{S} d\shat
\frac{\hat{\sigma}_{i j}(\shat)}{S} \int_{\shat/S}^1 f_i(x_1)
  f_j\left(\frac{\shat}{x_1 S}\right) \frac{dx_1}{x_1}
\eeq

We finally use the following variables: $\displaystyle \tau_{AB} =
\frac{(M_A+M_B)^2}{S}$, $\displaystyle \tau=\frac{\shat}{S}$ and obtain

\beq
\sigma(p\bar p\to A B) = \sum_{i,j} \int_{\tau_{AB}}^{1} d\tau
\hat{\sigma}_{i j}(\shat=\tau S) \int_{\tau}^1 f_i(x)
  f_j\left(\frac{\tau}{x}\right) \frac{dx}{x}
\label{eq:PDFcross}
\eeq

The quantity $\displaystyle \frac{d \mathcal{L}^{i
j}}{d\tau} (\tau)=\int_{\tau}^1 f_i(x)
f_j\left(\frac{\tau}{x}\right) \frac{dx}{x}$ is called the $i j$
luminosity. This quantity encodes all the necessary information that
is stored in the PDFs.

We are now ready to describe the calculation of the production cross
sections of the two main channels at the Tevatron.

\subsubsection{The gluon--gluon fusion}

The production rate for the $gg\to H+X$ process, where X denotes the
additional jets that appear at higher orders in QCD, is evaluated
following the Eq.~\ref{eq:PDFcross}. We then define the variable $\displaystyle
\tau_H = \frac{M_H^2}{S}$ which is our $\tau_{AB}$, and we also define
a new variable $\displaystyle z = \frac{\tau_H}{\tau}$ wich quantify
the departure from the soft--gluon limit $z\to 1$ where $\shat \to S$
with no jets production.

\begin{figure}[!t]
\begin{center}
\includegraphics[scale=0.75]{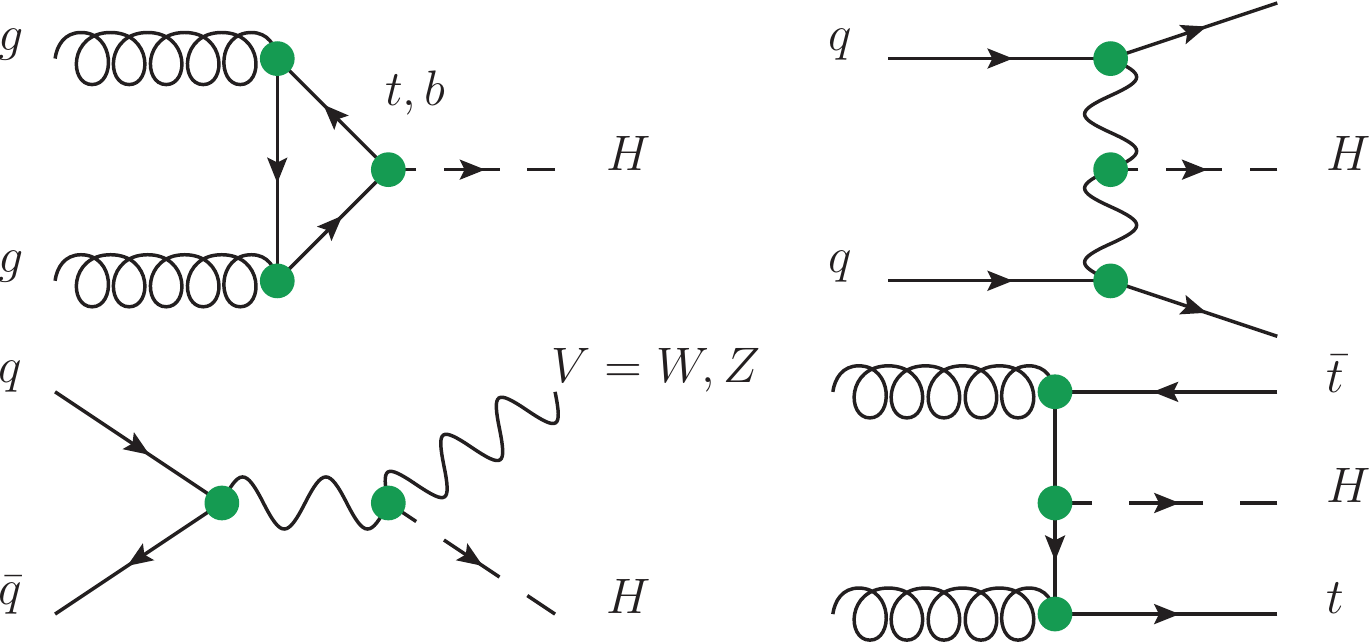} 
\end{center}
\vspace*{-.5cm}
\caption[Feynman diagrams of the four main SM Higgs production channel]{
The Feynman diagrams for the four main channels for SM Higgs
production at hadron colliders as discussed in the text: the
gluon--gluon fusion $gg\to H$ (upper left), the vector boson fusion
$q_1 q_2\to q_1 q_2 H$ (upper right), the Higgs--strahlung channels
$q\bar q\to VH$ with $V=W,Z$ (lower left) and the associated top quark
production $p\bar p\to t\bar t H$ (lower right).}
\label{fig:FeynmanMain} 
\end{figure}

This cross section is already a loop calculation at leading order (LO)
in QCD, as it involves in the SM triangular quark loops, mainly the
top quark and in a lesser extent the bottom quark, see
Fig.~\ref{fig:FeynmanMain}. We write the total cross section in
$\alpha_s$ expansion as

\begin{eqnarray}
\displaystyle \sigma(p\bar p\to H) = & ~ \nonumber\\
~ & \hspace{-25mm} \displaystyle \sum_{(ij)} \int_{\tau_H}^1 d\tau
\frac{d \mathcal{L}^{i j}}{d\tau} (\tau)
\alpha_s^2 \sigma^0 \left(\delta^{(0)}_{ij}(z)+\frac{\alpha_s}{\pi} \delta^{(1)}_{ij}(z)
  + \left(\frac{\alpha_s}{\pi}\right)^2 \delta^{(2)}_{ij}(z) + \cdots \right)
\end{eqnarray}

\begin{figure}[!t]
\begin{center}
\includegraphics[scale=0.75]{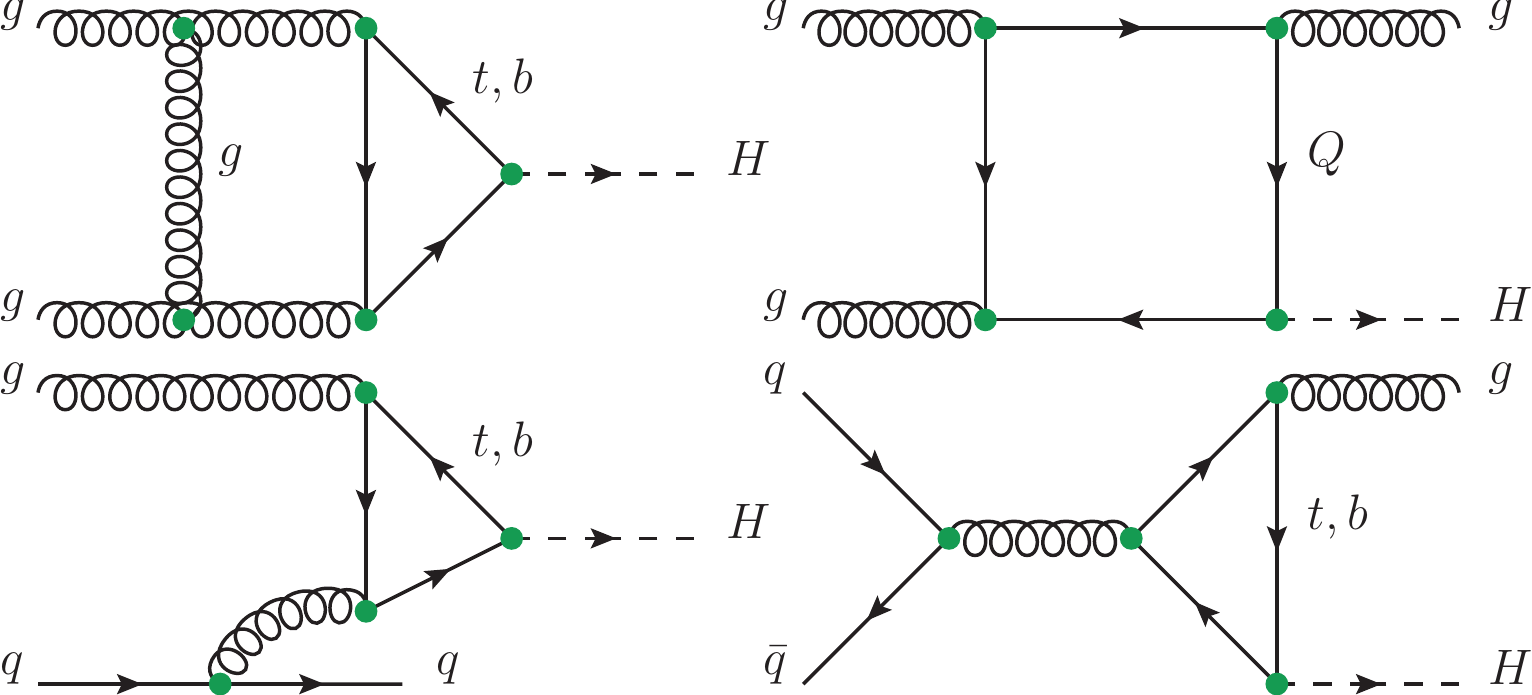} 
\end{center}
\vspace*{-.5cm}
\caption[Some Feynman diagrams for NLO SM $gg\to H$ production]{
Some Feynman diagrams for the NLO corrections to $gg\to H$ SM Higgs
production channel.}
\label{fig:NLOggH} 
\end{figure}

$(ij)$ means either $gg$ which occurs already at LO, or $qg$ and
$q\bar q$/$qq$/$qq'$ pairs which occur from the NLO, see
Fig.~\ref{fig:NLOggH} for typical NLO diagrams. $\sigma^0$ is the LO
kernel and $\delta^{(K)}_{ij}$ is the $K$th order correction to the
total cross section induced by a $(ij)$ partonic initial state. We
have 

\beq
\sigma^0 = \frac{G_F}{288 \pi \sqrt{2}} \left| \frac 3 4 \sum_q
  A(\tau_q) \right|^2
\label{eq:LOkernel}
\eeq

with $G_F=1.16637\times 10^-5$ GeV$^{-2}$ as the Fermi constant. In
Eq.~\ref{eq:LOkernel} we have $\displaystyle
\tau_q=\frac{M_H^2}{4m_q^2}$ and 

\begin{eqnarray}
A(\tau)=2 \left(\frac 1 \tau + \frac{\tau-1}{\tau^2}f(\tau) \right) \nonumber\\
f(\tau)= \left\{
\begin{matrix}
\arcsin^{2} \sqrt{\tau} && \tau \leq 1 \\
\displaystyle -\frac{1}{4} \left[ \ln
        \frac{1+\sqrt{1-\tau^{-1}}}{1-\sqrt{1-\tau^{-1}}} - \imath
        \pi\right]^2 && \tau \geq 1\\
\end{matrix}
\right.
\label{eq:LOfunction}
\end{eqnarray}

\begin{figure}[!t]
\begin{center}
\includegraphics[scale=0.75]{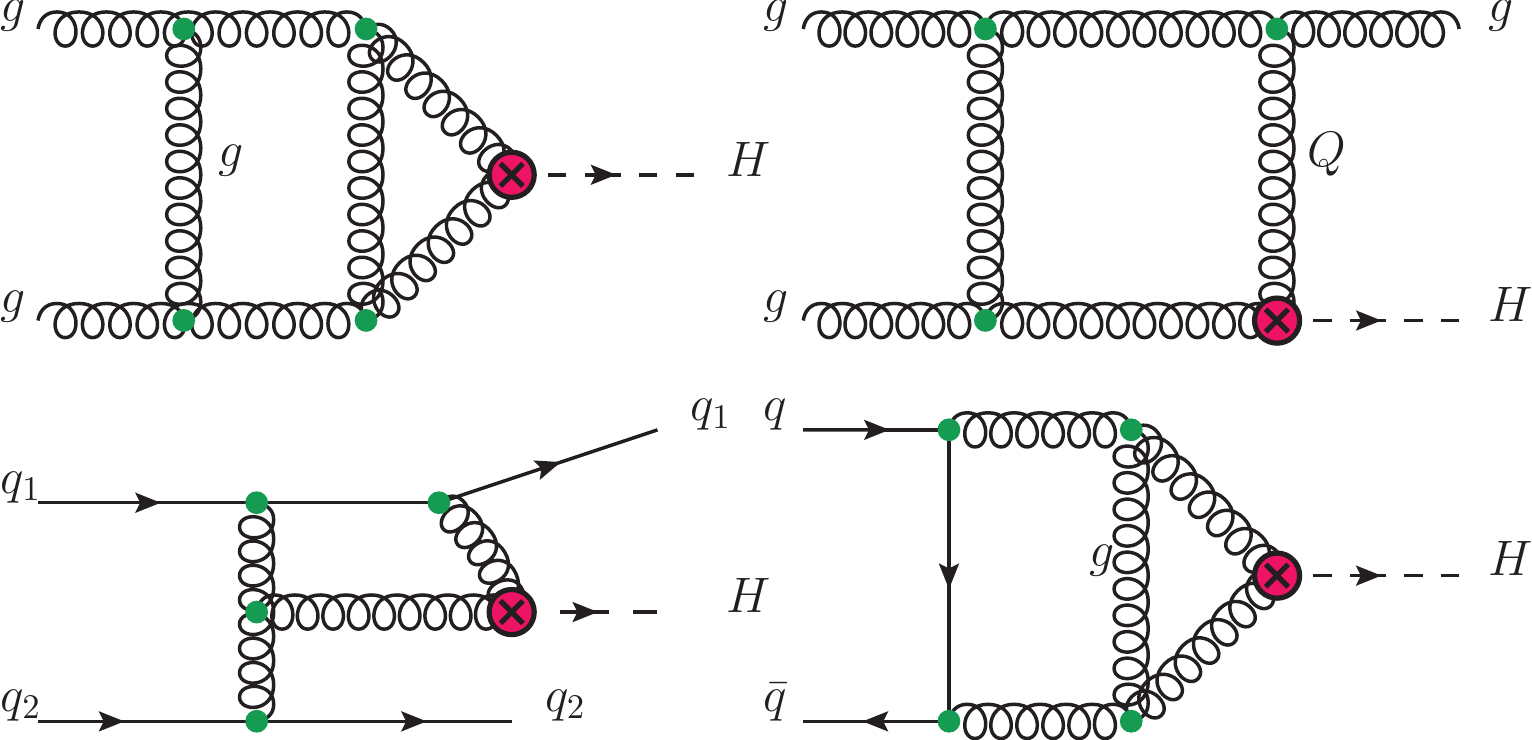} 
\end{center}
\vspace*{-.5cm}
\caption[Some Feynman diagrams for NNLO SM $gg\to H$ production]{
Some Feynman diagrams for the NNLO corrections to $gg\to H$ SM Higgs
production channel in the infinite top mass limit. The red circle with
a cross represents the effective $Hgg$ vertex where the top quark has
been integrated out.}
\label{fig:NNLOggH} 
\end{figure}

The cross section for SM Higgs production in gluon--gluon fusion is then
evaluated in the following way. Up to NLO in QCD we use the Fortran
code {\tt HIGLU}~\cite{Spira:1995mt}\footnote{The public code is
  available in Ref.~\cite{Spirapage}.} which includes the complete set of radiative
corrections at this order, taking into account the full dependence on the top
and bottom quark masses~\cite{Spira:1995rr}. We also want to take into account
the NNLO corrections that are known in an effective approach where
only the dominant top quark contribution is included in the limit where the top
mass in taken as
infinite~\cite{Harlander:2002wh,Anastasiou:2002yz,Ravindran:2003um},
see Fig.~\ref{fig:NNLOggH} for some NNLO diagrams. These contribution are
implemented into the code taking the analytical expression given in
Ref.~\cite{Anastasiou:2002yz}. We rescale the NNLO correction by the full
$m_t$ dependent Born cross section, an approximation which at NLO is
accurate at the level of a few percent for Higgs masses below the
$t\bar t$  kinematical threshold, $M_H \lsim 300$
GeV~\cite{Spira:1995rr,Spira:1997dg}.

The dependence on the renormalization scale $\mu_R$
and the factorization scale $\mu_F$ of the partonic NNLO cross sections has
been reconstructed from the scale independent expressions of
Ref.~\cite{Anastasiou:2002yz} using the fact that the full hadronic cross sections do
not depend on them and the $\alpha_s$ running between the $\mu_F$ and $\mu_R$
scales\footnote{The analytical expressions for the scale dependence have only
been given in  Ref.~\cite{Ravindran:2003um} in the limit $\mu_F\!=\!\mu_R$ from which
one can straightforwardly obtain the case $\mu_F\!\neq \!\mu_R$ as
stated in the text (see also Ref.~\cite{Kramer:1996iq}).}. Indeed, the
complete cross section taking into account all orders in perturbation
theory should not depend on the (unphysical) renormalization and
factorization scales; we first take $\mu_F=\mu_R=\mu$ and we then
have $\displaystyle \frac{d\sigma}{d\ln \mu^2} = 0$. This equation is
then expressed in term of the evolution of the partonic cross section
convoluted with DGLAP evolution for the PDFs, which expressed order by
order in $\alpha_s$ gives the dependence on the scale $\mu$ of
$\hat{\sigma}_{i j}$ (see also Ref.~\cite{Anastasiou:2002yz}).

In the end we can obtain the final $\mu_R$--dependent terms using the
evolution of the strong coupling constant $\alpha_s$ between
$\alpha_s(\mu=\mu_F)$ and $\alpha_S(\mu_R)$. For NNLO terms the LO
evolution is nearly enough: 
\beq
\alpha_s(\mu_F)=\alpha_s(\mu_R)\left(
  1+\alpha_s(\mu_R) b_0\ln\left(\frac{\mu_R^2}{\mu_F^2}\right)\right)
\eeq
with $\displaystyle b_0=\frac{33-2n_f}{12\pi}$ as the first term in the
$\beta$--function of QCD strong coupling constant $\alpha_s$, where
$n_f$ stands for the number of active flavors in the running (here
$n_f=5$). The NLO running of $\alpha_s$ with $\displaystyle
b_1=\frac{153-19n_f}{24\pi^2}$ is needed only for the NNLO
$gg$ virtual correction coming from the LO $gg$ diagram. The complete
$\mu_R$--dependent terms at NNLO are available in the
appendix~\ref{ggH:appendix} of this section.

What will be the choice for the central scale $\mu_0=\mu_R=\mu_F$? The
usual choice would be to use $\mu_0=M_H$ as it seems to be the most
natural scale fixed by the dynamics of the process. Some results using
this scale choice can be found in the first pages of
Ref.~\cite{Baglio:2010um} and have been derived before some criticisms, made by the
members of the Tevatron New Physics and Higgs working group (TEVNPHWG) of the
CDF and D0 collaborations~\cite{Tevatron:response} concerning the
theoretical modeling of the $gg\to H$ production cross section
proposed in the beginning of Ref.~\cite{Baglio:2010um} appeared on
the web in May 2010. A new analysis was then derived with a new
central scale $\mu_0=\frac{1}{2} M_H$ after ICHEP 2010 (in July 2010) as we got aware
of the criticisms only during ICHEP\footnote{The results with
  $\mu_0=m_H$ were also presented in ICHEP, see
  Ref.~\cite{Baglio:2010zz}.} where the new combined analysis of CDF and
D0 for the Higgs search at the Tevatron was
released~\cite{Kilminster:2010, Aaltonen:2010yv}. This
new analysis addressed the criticisms and reinforced the conclusion of
the first pages of Ref.~\cite{Baglio:2010um}\footnote{This was presented in
  the Higgs Hunting workshop in Orsay which followed
  ICHEP~\cite{Baglio:2010hh}.}. We will then use in the rest of the
thesis the scale choice $\mu_0=\frac{1}{2} M_H$ as explained below.

Although soft--gluon resumation contributions for the total cross
section have been calculated up to next--to--next--to--leading
logarithm (NNLL)~\cite{Catani:2003zt} we do not include them in our
calculation. There are also other additional estimated small
contributions at N$^3$LO~\cite{Moch:2005ky} as well as beyond the soft
NNLL approximation~\cite{Ravindran:2006cg, Idilbi:2005ni,
  Laenen:2005uz, Ahrens:2008nc} that we do not include. There are some
reason to do so, that we explain in three points:
\begin{itemize}
\item{these corrections are known only for the inclusive total cross
    section and not for the cross sections when experimental cuts are
    incorporated; this is also the case for the differential cross
    sections~\cite{Ravindran:2003ia, Anastasiou:2005qj, Catani:2007vq,
      Anastasiou:2009kn} and many distributions that are used
    experimentally, which have been evaluated only at NNLO at
    most. Even if we concentrate on the total cross section in this
    work, we should have in mind that these results are then used in
    experimental analyses which also take differential cross
    sections. In order to be fully consistent in the whole analysis,
    our total cross section should be calculated at NNLO. This
    interpretation is strengthened by the CDF/D0 analysis
    itself~\cite{Aaltonen:2010yv,Tevatron:2010ar,Aaltonen:2011gs}, as
    the $gg\to H$ cross section in this analysis has been broken into
    three pieces which yield different final state signal topologies
    for the main decay $H \to WW^{(*)} \to \ell \ell \nu \nu$, namely
    $\ell \ell \nu \nu$+0\,jet,  $\ell\ell \nu \nu$+1\,jet and $\ell
    \ell \nu \nu$+2\,jets or more: 
\beq
\sigma^{\rm NNLO}_{\rm gg\to H}= \sigma^{\rm 0jet}_{\rm gg\to H}+
\sigma^{\rm 1jet}_{\rm gg\to H}+
\sigma^{\rm \geq2jets}_{\rm gg\to H}
\eeq
 These channels have been analyzed separately and these individual
 components, with $\sigma^{\rm 0jet}_{\rm gg\to H}$ evaluated at NNLO,
 $\sigma^{\rm 1jet}_{\rm gg\to H}$ evaluated at NLO and  $\sigma^{\rm
   \geq2jets}_{\rm gg\to H}$ evaluated at LO, represent respectively
 $\approx 60\%$, $\approx 30\%$ and  $\approx 10\%$ of the  total
 $gg\to H$ cross section at NNLO. Since these three pieces add up to
 $\sigma^{\rm NNLO}_{\rm gg\to H}$, why using different normalisation
 for these jet cross sections and the total sum, and include
 soft--gluon resumation in the latter and not in the former?}
\item{it is theoretically not very consistent to fold a resumed cross
    section with PDF sets which do not involve any resumation, as is
    the case for the presently available PDF sets which at at most at
    NNLO. Even if the effects of the resumation on the PDFs might be
    rather small in practice, this has also been discussed in details in
    Ref.~\cite{Corcella:2005us}.}
\item{it is well known that the NNLL contributions increase the NNLO
    result by a factor of $\sim 10$--15\% at the
    Tevatron~\cite{Catani:2003zt}. This increase can be very closely
    approached by evaluating the NNLO cross section at
    $\mu_0=\frac{1}{2}M_H$~\cite{Anastasiou:2008tj,Anastasiou:2010hh}
    which is one of the reasons to use this central scale choice,
    which is then fully consistent with the
    point b) while taking into account this $\sim 10$--15\% increase
    in the total cross section\footnote{This choice of ignoring the
      contributions beyond NNLO has also been adopted  in
      Ref.~\cite{Anastasiou:2009bt} in which the theoretical predictions have
      been confronted to the CDF/D0 results, the focus being the
      comparison between the distributions obtained from the matrix
      elements calculation with those given by the event generators
      and Monte-Carlo programs used by the experiments. They find
      excellent agreement with NNLL calculation at a central scale
      $\mu_0=M_H$}.}
\end{itemize}

We see here that a central scale $\mu_0=\frac{1}{2}M_H$ seems more
appropriate while sticking at NNLO order. In fact there is also
another reason why this scale choice is particularly appropriate for
gluon--gluon fusion. As pointed out by Anastasiou and collaborators
some time ago~\cite{Anastasiou:2008tj,Anastasiou:2010hh} (see also
Ref.~\cite{Djouadi:2005gi}), lowering the central value of the
renormalization and factorization scales from $\mu_0=M_H$ to
$\mu_0=\frac12 M_H$ improves the convergence of the perturbative
series and is more appropriate to 
describe the kinematics of the process. Recalling that if the scale
value $\mu_0= \frac12 M_H$ is chosen, the central value of $\sigma^{\rm
NNLO}_{\rm gg\to H}$ increases by more than $10\%$, we then end with a
NNLO calculation with no difference between  $\sigma^{\rm NNLO}_{\rm
  gg\to H}(\mu_0= \frac12 M_H)$ and $\sigma^{\rm NNLL}_{\rm gg\to
  H}(\mu_0=M_H)$ as calculated for instance in~\cite{deFlorian:2009hc}, which is
fully consistent with the PDF order, and which improves the
convergence of the perturbative series. All of these reasons explain
this central scale choice used in the rest of our analysis.

For the electroweak part, we include the complete one--loop  corrections to the
$gg\to H$ amplitude which have been calculated in Ref.~\cite{Actis:2008ts} taking
into account the full  dependence on the top/bottom quark and the $W/Z$ boson
masses. These corrections are implemented in the so--called partial
factorization scheme in which the electroweak correction $\delta_{\rm EW}$ is
simply added to the QCD corrected cross section at NNLO, $\sigma^{\rm tot}=
\sigma^{\rm NNLO} + \sigma^{\rm LO} (1+\delta_{\rm EW})$.  In the alternative
complete factorization scheme discussed  in Ref.~\cite{Actis:2008ts}, the
electroweak correction $1+\delta_{EW}$ is multiplied by the fully QCD corrected
cross section, $\sigma^{\rm tot}=\sigma^{\rm NNLO}(1+ \delta_{EW})$ and, thus,
formally involves terms of  ${\cal O} (\alpha_s^3 \alpha)$ and ${\cal O}
(\alpha_s^4 \alpha)$ which have not been fully calculated. Since the QCD
$K$--factor is large, $K_{\rm NNLO} \approx 3$, the electroweak corrections
might be overestimated by the same factor. We have also included the mixed
QCD--electroweak corrections at NNLO due to light-quark
loops~\cite{Anastasiou:2008tj}.
These are only part of the three--loop ${\cal O}(\alpha \alpha_s)$ corrections
and have been calculated in an effective approach that is valid only when $M_H
\lsim M_W$ and which cannot be so easily extrapolated to $M_H$ values above this
threshold; this will be discussed in more details in the next section. In
Ref.~\cite{Anastasiou:2008tj}, it has been pointed out that this procedure, i.e. adding
the NLO full result and the mixed QCD--electroweak correction  in the partial
factorization scheme, is equivalent to simply including only the NLO
electroweak correction in the complete factorization scheme. 

We finally fold the partonic cross section with the MSTW PDF
set~\cite{Martin:2009iq}, setting as mentioned above the renormalization and
factorization scales at $\mu_R\!=\!\mu_F\!=\frac{1}{2} M_H$. We
obtain for the Tevatron energy $\sqrt s = 1.96$ TeV the central values
displayed in Fig.~\ref{fig:Tev_fig1} for the gluon--gluon fusion
production cross sections as a function of the Higgs mass on the
entire interesting range for the Tevatron experiment. Our results for
the total cross sections are in an excellent agreement with those given in
Refs.~\cite{Tevatron:2009je,Aaltonen:2010yv,Tevatron:2010ar,Aaltonen:2011gs,deFlorian:2009hc}. For
instance, for $M_H=160$ GeV, we obtain with 
our procedure a total $p \bar p  \to H+X$ cross section of  $\sigma^{\rm
tot}=427$ fb, compared to  the value $\sigma^{\rm tot}=434$ fb quoted in 
Refs.~\cite{Tevatron:2009je,Aaltonen:2010yv,Tevatron:2010ar,Aaltonen:2011gs,deFlorian:2009hc}. The
small difference comes from the 
different treatment of the electroweak radiative corrections 
(partial factorization plus mixed QCD--electroweak contributions in our case
versus complete factorization in Ref.~\cite{deFlorian:2009hc}) and another one percent
discrepancy can be attributed to the numerical uncertainties in the various
integrations of the partonic sections\footnote{We have explicitly
  verified, using the program {\tt HRESUM}~\cite{Grazzini:2008tf,
    Grazzinipage} which led to the results of
  Ref.~\cite{deFlorian:2009hc}, that our NNLO cross section is in
  excellent agreement with those available in the literature. In
  particular, for $M_H=160$  GeV and scales $\mu_R=\mu_F= M_H$, one
  obtains $\sigma^{\rm NNLO}=380$ fb with  {\tt HRESUM} compared to
  $\sigma^{\rm NNLO}=374$ fb in our case; the 1.5\% discrepancy being
  due to the different treatment of the electroweak corrections and
  the integration errors. Furthermore, setting the renormalization and
  factorization scales to $\mu_R=\mu_F=\frac12 M_H$, we find
  $\sigma^{\rm NNLO}=427$ fb which is in excellent agreement with the
  value $\sigma^{\rm NNLO}=434$ fb obtained in
  Ref.~\cite{Anastasiou:2008tj} and with {\tt HRESUM}, as well as the
  value in the NNLL approximation when the scales are set at their
  central values $\mu_R=\mu_F= M_H$. This gives us confidence that our
  implementation of the NNLO contributions in the NLO code {\tt
    HIGLU},  including the scale dependence, is correct.}.

We should also note that for the Higgs mass value $M_H=160$ GeV, we obtain $K
\simeq 2.15$ for the QCD $K$--factor at NLO and $K \simeq 2.8$ at NNLO. These
numbers are slightly different from those presented in Ref.~\cite{Anastasiou:2009bt},
$K \simeq 2.4$ and $K \simeq 3.3$, respectively. The reason is that the
$b$--quark loop contribution, for which the $K$--factor at NLO is
significantly smaller than the one for the top quark contribution~\cite{Spira:1995rr}
has been ignored for simplicity in the latter paper; this difference will be
discussed in section~\ref{section:SMHiggsTevEFT}.

\subsubsection{The Higgs--strahlung production channels}

The Higgs--strahlung processes $q\bar q\to WH$ and $q\bar q\to ZH$ are
also evaluated using Eq.~\ref{eq:PDFcross}. The variable $\tau_{A B}$
is $\displaystyle \tau_{H V}= \frac{(M_H+M_V)^2}{S}$. It has been
shown (see Ref.~\cite{Djouadi:2005gi}) that this process can be interpretated
as the Drell--Yan production $p\bar p\to V^*$ of a virtual vector boson
which then splits into a real vector boson $V$ and a Higgs boson
$H$~\cite{Drell:1970wh}. We could then write $\displaystyle \hat{\sigma}_{i j}=
\hat{\sigma}(i j\to V^*) \times \frac{d\Gamma}{dk^2}$ where $k^2$
is the mass of the virtual vector boson $V^*$, which is subject to
vary between $\tau_{H V}$ and $S$. If we rewrite the equality we can
obtain $\displaystyle \frac{d\Gamma}{d k^2} = \frac{\hat{\sigma}^{\rm
    LO}_{i j}}{\hat{\sigma}^{\rm LO}( i j\to V^*)}$. In the end, using
Eq.~\ref{eq:PDFcross} and the fact that we have to integrate the
variable $\displaystyle Z=\frac{k^2}{S \tau}$ between $\displaystyle
\frac{\tau_{H V}}{\tau}$ and 1:

\beq
\sigma(p\bar p\to H V) = \int_{\tau_{H V}}^1 \sum_{(i j)} \frac{d
  \mathcal{L}^{i j}}{d\tau} (\tau) \int_{\tau_{H V}/\tau}^1 dZ~
\hat{\sigma}_{i j}^{\rm LO}(\tau Z S)~\Delta_{i j}( i j\to V^*)
\label{eq:HVcross}
\eeq

In Eq.~\ref{eq:HVcross} the quantity $\Delta_{i j}$ describes the
corrections to the partonic Drell--Yan process $i j\to V^*$. At LO we thus
have $\Delta_{i j} = \delta_{i q} \delta_{j \bar q}$ (with $\delta$ as
the Kronecker symbol) as only the
$q\bar q$ partonic initial state contributes with no QCD correction,
Higgs--strahlung processes being pure electroweak processes at LO. We
can calculate $\hat{\sigma}_{i j}$ directly and use the factorization
with the Drell--Yan process for HO corrections. We obtain

\beq
\hat{\sigma}_{i j} = \delta_{i q} \delta_{i \bar q} \frac{G_F^2
  M_V^4}{288 \pi \shat}\left(\hat{v}_q^2+\hat{a}_q^2\right)
\sqrt{\beta}~\frac{\beta+12M_V^2/\shat}{\left(1-M_V^2/\shat\right)}
\eeq

where $\displaystyle \beta= \left( 1 -
  \frac{M_V^2}{\shat}-\frac{M_H^2}{\shat}\right)^2-4
\frac{M_V^2M_H^2}{\shat^2}$ is the phase--space two bodies
function. $\hat{a}_q$ and $\hat{v}_q$ stand for the reduced axial and
vector fermion--gauge boson couplings: $\hat{v}_q=\hat{a}_q=\sqrt{2}$
for $V=W$, $\hat{v}_q=2 I_3(q)-4 Q_q \sin\theta_W$ and
$\hat{a}_q=2I_3(q)$ for $V=W$ where $I_3$ is the weak isospin
$z$--projection, $\sin\theta_W$ stands for the Weinberg angle and
$Q_q$ is the electric charge of the quark $q$.

\begin{figure}[!t]
\begin{center}
\includegraphics[scale=0.9]{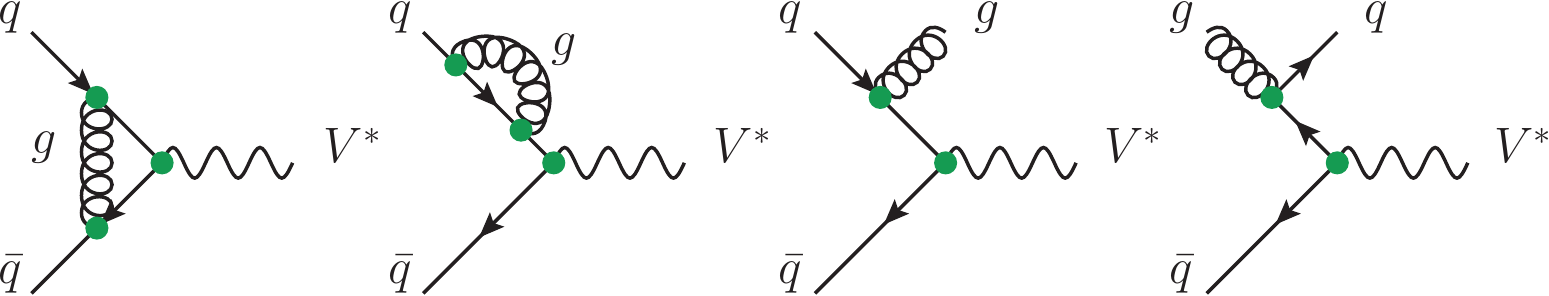} 
\end{center}
\vspace*{-.5cm}
\caption[NLO QCD corrections to $p\bar p\to V^*$]{NLO QCD corrections
  to virtual vector boson production $p\bar p\to V^*$.}
\label{fig:NLOqqVH} 
\end{figure}

\begin{figure}[!t]
\begin{center}
\includegraphics[scale=0.9]{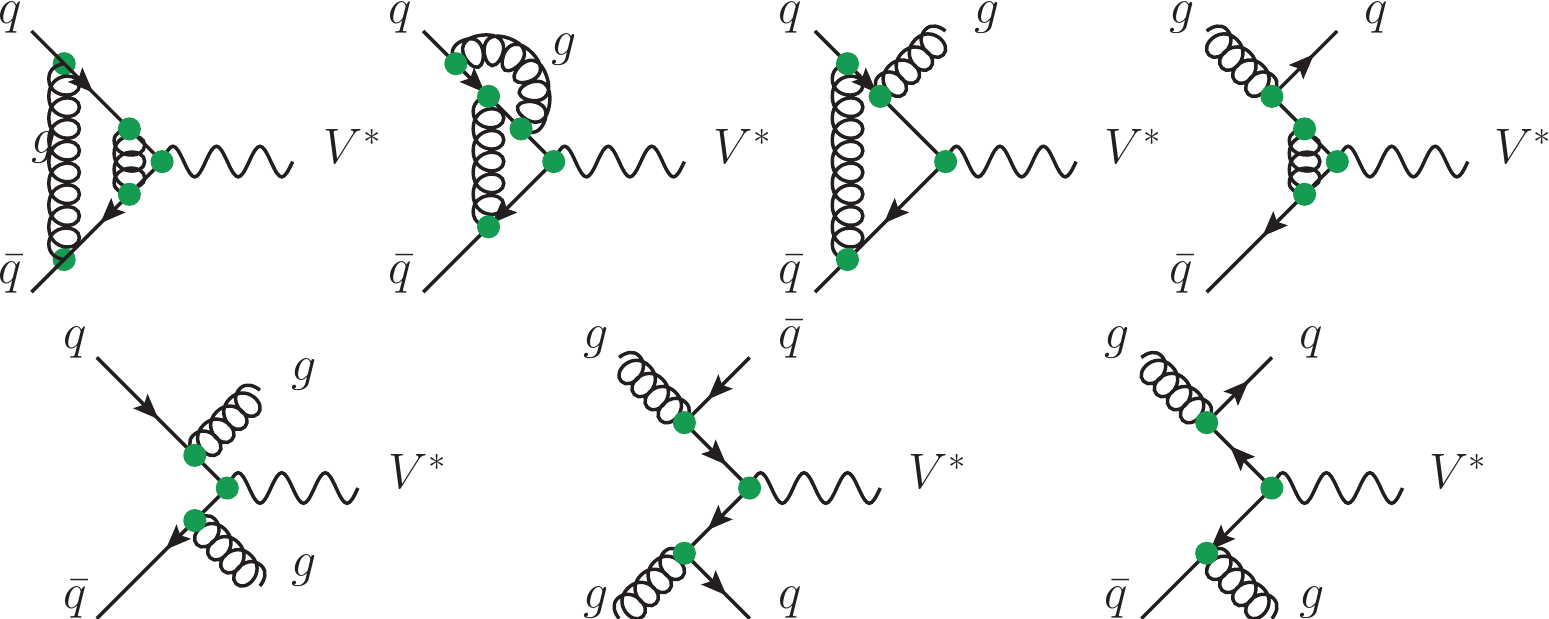} 
\end{center}
\vspace*{-.5cm}
\caption[NNLO QCD corrections to $p\bar p\to V^*$]{Some Feynman
  diagrams for NNLO QCD corrections to virtual vector boson production
  $p\bar p\to V^*$.}
\label{fig:NNLOqqVH} 
\end{figure}

We then add the NLO corrections in the Drell--Yan process, see
Fig.~\ref{fig:NLOqqVH} for the Feynman diagrams. The exact
expressions for the NLO QCD corrections can be found in
Refs.~\cite{Altarelli:1979ub, KubarAndre:1978uy, Han:1991ia,
  Ohnemus:1992bd, Djouadi:1999ht, Djouadi:2005gi}. The NNLO
corrections for $p\bar p\to HV$ have been presented in
Ref.~\cite{Brein:2003wg} and use the NNLO Drell-Yan
corrections~\cite{Hamberg:1990np, Hamberg:2002np,Harlander:2002wh},
see Fig.~\ref{fig:NNLOqqVH} for some Feynman diagrams. It is worth
mentioning that there are other specific corrections for
Higgs--strahlung processes which occur at NNLO and that have nothing
to do with the Drell--Yan process $p\bar p\to V^*$, such as the $gg\to
ZH$ channel. They have been reviewed in Ref.~\cite{Djouadi:2005gi} and
what comes out is that we can neglect them at the Tevatron both for
$W$ and $Z$ bosons (the $gg\to ZH$ channel does not exist for $W$
channel for example and is at the pemille level at the Tevatron for the $Z$ channel).

The total Higgs--strahlung cross sections are then calculated in the
following way. We use the Fortran code {\tt V2HV}~\cite{Spirapage} which 
evaluates the full cross sections at NLO in QCD. The NNLO QCD contributions to
the cross sections~\cite{Brein:2003wg}, as well as the one--loop  electroweak corrections
evaluated in Ref.~\cite{Ciccolini:2003jy}, are incorporated in the program {\tt
  V2HV} by us. The central scale adopted in this case is the invariant mass of the $HV$
system, $\mu_R=\mu_F = M_{HV} = (p_H+p_V)^2$.

Folding the partonic cross sections with the MSTW parton
distribution functions~\cite{Martin:2009iq} and setting the renormalization and
factorization scales at the most natural values
$\mu_R\!=\!\mu_F\!=\!M_{HV}$, we obtain for the Tevatron energy $\sqrt
s=1.96$ TeV, the central values displayed in Fig.~\ref{fig:Tev_fig1} for the $p\bar
p\to VH$ Higgs production cross sections as a function of the Higgs
mass. Note that including the combined HERA data and the Tevatron
$W\to \ell \nu$ charge asymmetry data in
the MSTW2008 PDF set~\cite{Thorne:2010kj} might lead to an increase of
the $p \bar p \to  (H+)Z/W$ cross sections by $\approx 3$\%; a small
change in $\sigma (gg \to H)$ is also expected.

The central values of the cross sections of Higgs--strahlung from  $W$
and $Z$ bosons that we obtain are comparable to those given in 
Refs.~\cite{Tevatron:2009je,Brein:2004ue, Assamagan:2004mu},
with at most a $\sim 2\%$ decrease in the low Higgs mass range, $M_H
\lsim 140$ GeV. The reason is that the quark and 
antiquark densities, which are the most relevant in these processes and are
more under control than the gluon densities, are approximately the same in the
new MSTW2008 and old  MRST2002 sets of PDFs (although the updated set includes
a new fit to run II Tevatron and HERA inclusive jet data). We should note that
for $M_{H}=115$ GeV for which the production cross sections are the largest,
$\sigma^{\rm WH}= 175$ fb and $\sigma^{\rm ZH}=104$ fb, the QCD $K$--factors are
$\sim 1.2 \; (1.3)$ at NLO (NNLO), while the electroweak corrections decrease
the LO cross sections by $\approx -5$\%. The correcting factors do not change
significantly for increasing $M_H$ values for the Higgs mass range relevant at
the Tevatron. 

We make here a small comment which will simplify the analysis of the
theoretical uncertainties affecting the central predictions. Since in
this case, the NNLO QCD corrections and the one--loop electroweak
corrections have been obtained exactly and no effective 
approach was used, only the scale variation and the PDF+$\alpha_s$
uncertainties have to be discussed. In addition, since the NNLO gluon--gluon
fusion contribution to the cross section in the $p \bar p \to ZH$ case, which
is absent in $p \bar p \to WH$, is very small at the Tevatron and
because the scales and phase space are only slightly different for the $p \bar p
\to WH$ and $ZH$ processes, as the difference $(M_Z^2-M_W^2)/ \hat s$ is tiny,
the kinematics and the $K$--factors for these two processes are very similar.
We will thus restrict our analysis to the $WH$ channel but the same
results will hold for the $ZH$ channel. 

\subsubsection{The VBF and associated heavy quarks channels}

\begin{figure}[!t]
\begin{center}
\includegraphics[scale=0.9]{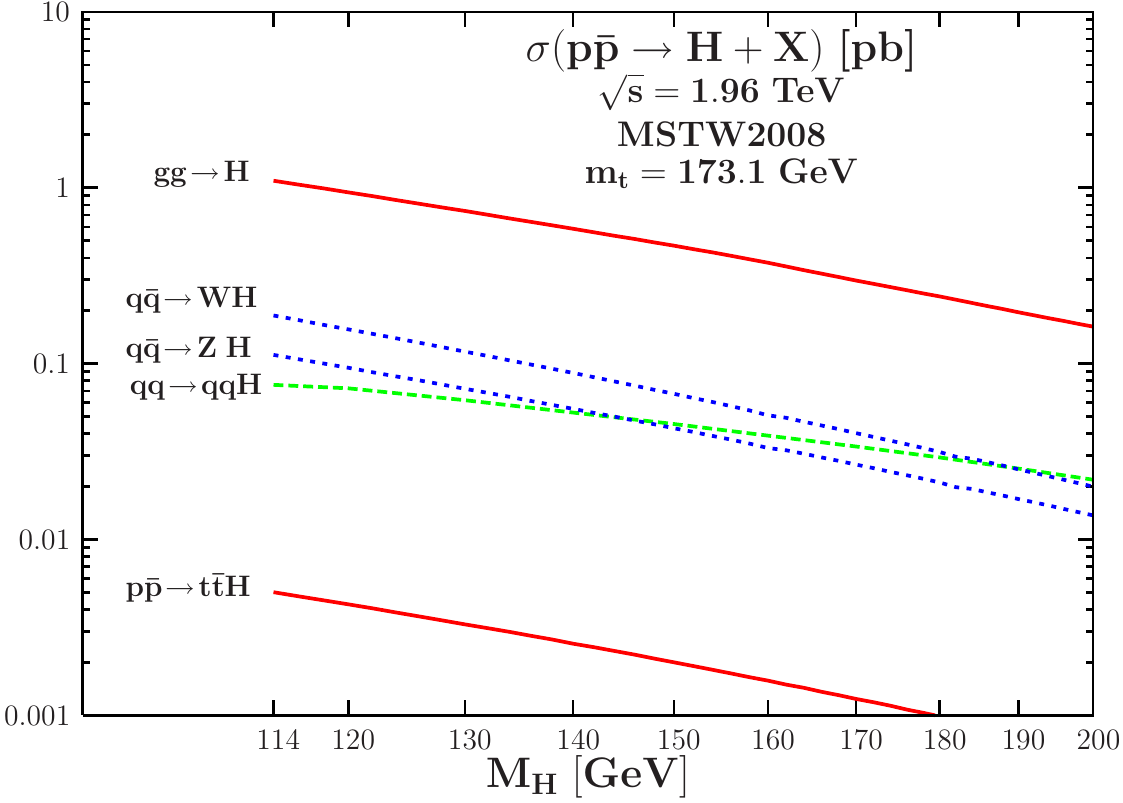} 
\end{center}
\vspace*{-.5cm}
\caption[Total cross sections for Higgs production at the Tevatron in
the four main channels]{The total cross sections for Higgs production
 at the Tevatron as a function of the Higgs mass. The MSTW set of PDFs
 has been used and the higher order corrections are included as
 discussed in the text.} 
\label{fig:Tev_fig1}
\end{figure}

We have also evaluated the cross sections of the two sub-leading
processes $qq \to V^* V^* qq \to Hqq$ and $q \bar q/gg \to t\bar t H$
that we also include in Fig.~\ref{fig:Tev_fig1} for completeness, we
have not entered into very sophisticated considerations. We
have simply followed the procedure outlined in Ref.~\cite{Djouadi:2005gi} and used the
public Fortran codes again given in Ref.~\cite{Spirapage}. The vector boson total
cross section is evaluated at NLO in QCD~\cite{Han:1992hr, Figy:2003nv} at a scale
$\mu_R=\mu_F=Q_V$ (where $Q_V$ is the momentum transfer at the gauge boson
leg), while the presumably small electroweak corrections, known for the LHC
~\cite{Ciccolini:2007ec}, are omitted. We have not included the recent partial
NNLO QCD corrections presented in Refs.~\cite{Bolzoni:2010xr,
  Bolzoni:2010as}, as they do not modify greatly the central cross
section that interest us in this small update. We have not estimated
the uncertainties affecting these calculations.

In the case of associated $t\bar tH$  production, the LO cross section
is evaluated at  scales $\mu_R=\mu_F=\frac12(M_H+2m_t)$ but is
multiplied by a factor $K \sim 0.8$ over the entire Higgs mass range
to account for the bulk of the NLO QCD
corrections~\cite{Beenakker:2001rj, Beenakker:2002nc, Reina:2001sf,
  Dawson:2002tg}. In the latter case, we use the updated value
$m_t=173.1$ GeV for the top quark mass~\cite{Tevatron:2009ec}. The
only other update compared to the cross section values given in
Ref.~\cite{Djouadi:2005gi} is thus the use of the recent MSTW set of PDFs.

The cross sections for the vector boson fusion channel in which the
recent MSTW set of PDFs is used agree well with those given in
Refs.~\cite{Tevatron:2009je,Aaltonen:2010yv,Aglietti:2006ne,Aaltonen:2011gs}\footnote{The
  values from Ref.~\cite{Aglietti:2006ne} can be found
  in~\cite{Tev4LHCpage}.}. In the case of the $t\bar t H$ associated
production process, a small difference is observed compared to
Ref.~\cite{Djouadi:2005gi} in which the 2005 $m_t=178$ GeV value is
used: we have a few percent increase of the rate due the presently
smaller $m_t$ value which provides more phase space for the process,
overcompensating the decrease due to the smaller top--quark Yukawa
coupling. This is shown in Fig.~\ref{fig:Tev_fig1} as a function of
the Higgs mass.

Before closing this subsection, let us make a few remarks on the Higgs decay 
branching ratios and on the rates for the various individual channels that are
used to detect the Higgs signal at the Tevatron, bearing in mind that
this will be discussed in more details in
section~\ref{section:SMHiggsDecay}. In the interesting range 160 GeV
$\le M_H \le 175$ for which the Tevatron 
experiments are most sensitive, the branching ratio for the $H
\to WW$ is largely dominant, being above 90\%. In addition, in this mass range,
the $gg \to H$ cross section is one order of magnitude larger than the cross
sections for the  $q \bar q \to WH,ZH$ and $qq \to qqH$ processes as for  $M_H
\sim 160$ GeV for instance, one has $\sigma( gg\to H)=425$ fb compared to
$\sigma(WH) \simeq 49$ fb, $\sigma(ZH) \simeq 31$ fb and $\sigma(qqH) \simeq
40$ fb.  Thus, the channel $gg \to H \to W^*W^* $ represents, even before
selection cuts are applied,  the bulk of the events leading to $\ell \ell \nu
\nu +X$ final states, where here $X$ stands for additional jets or leptons
coming from $W,Z$ decays as well as for jets due to the higher order
corrections to the $gg\to H$ process. In the lower Higgs mass range, $M_H
\lsim 150$ GeV, all the production channels above,  with the exception of the
vector boson  $qq \to qqH$ channel which can be selected using specific
kinematical cuts, should be taken into account but with the process $q\bar q
\to WH \to \ell \nu b\bar b$ being dominant for $M_H \lsim 130$ GeV. This
justifies the fact that we concentrate mainly on the gluon--gluon fusion and
Higgs--strahlung production channels for the study of the theoretical
predictions both for the central cross sections and the theoretical
uncertainties; the inclusion of the other channels would marginally
affect our results, especially in the Higgs mass region around $M_H =
160$ GeV where the Tevatron is the most sensitive: the effect would be
below the percent level due to the huge impact of the $H\to WW$
branching fraction.

\subsection{Scale variation and higher order
  terms \label{section:SMHiggsTevScale}}

The effect of unknown (yet to be calculated) higher order
contributions to production cross sections and differential
distributions at hadron colliders is usually estimated by studying the
variation of these observables, evaluated at the highest known
perturbative order, with the renormalization scale defining the strong
coupling constant $\alpha_s$ scale and the factorization scale $\mu_F$
at which the matching between the perturbative calculation of the
matrix elements, that is the partonic cross section, and the
non--perturbative part which is described in the parton distribution
functions, is performed. These two scales are unphysical and should be
viewed as a technical artefact; indeed the dependence of the final
result on these two scales should be null, as when all orders of the
perturbative series are summed, the observables should be scale
independent. This scale dependence appears only because the
perturbative series are truncated, as only its few first orders are
evaluated in practice. This is the reason this scale dependence is
taken as as a guess of the impact of the higher order contributions.

Starting from a median scale $\mu_0$ which, with a smart guess, is
considered as the most ``natural'' scale of the process and absorbs potentially
large logarithmic corrections, the current convention is  to vary these two
scales within the range 
\beq
\mu_0/\kappa \le \mu_R, \mu_F \le \kappa \mu_0 \, .  
\label{eq:scalek}
\eeq
with the constant factor $\kappa$ to be determined. One then uses the following
equations to calculate the deviation of, for instance, a cross section
$\sigma(\mu_R, \mu_F)$  from the central value evaluated at scales
$\mu_R=\mu_F=\mu_0$,

\beq
\Delta\sigma_{\mu}^{+} &= &  
\max_{(\mu_{R},\mu_{F})} \sigma(\mu_{R},\mu_{F}) -\sigma(\mu_{R}=\mu_{F}=\mu_0) 
\, , \nonumber \\
\Delta\sigma_{\mu}^{-} &=& 
\sigma(\mu_{R}=\mu_{F}=\mu_0)-\min_{(\mu_{R},\mu_{F})} \sigma(\mu_{R},\mu_{F}) 
\, .  
\label{eq:scaleminus}
\eeq

This procedure is by no means a true measure of the higher order effects and
should be viewed only as providing a guess of the lower limit on the scale
uncertainty. The variation of the scales in the range of Eq.~\ref{eq:scalek} can
be individual with $\mu_R$ and $\mu_F$ varying independently in this domain,
with possibly some constraints such as $ 1/\kappa \le  \mu_R/\mu_F \le \kappa$
in order not to generate ``artificially large logarithms'', or collective when,
for example, keeping one of the two scales fixed, say to $\mu_0$,
and vary the other scale in the chosen domain.  Another possibility which is
often adopted, is to equate the two scales, $\mu_0/\kappa \le \mu_R= \mu_F \le
\kappa \mu_0$,  a procedure that is possibly more consistent as most PDF sets
are determined and evolved according to $\mu_R=\mu_F$, but which has no
theoretical ground as the two scales enter different parts of the calculation
(renormalization versus factorization). 

The choice of the variation domain for a given process, hence the
constant factor $\kappa$ in Eq.~\ref{eq:scalek}, is somewhat arbitrary and a
matter of taste: depending on whether one is optimistic or
pessimistic, that is wether one believes or not that the higher order 
corrections to the process are under control, it can range from
$\kappa\!=\!2$ to much higher values.

In most recent analyses of production cross sections at hadron colliders, a
kind of consensus has emerged and the domain,
\beq  \frac12 \mu_0 \le \mu_R,
\mu_F \le 2 \mu_0 \ ,  \ \     \frac12 \le \mu_R/\mu_F \le 2 \, , 
\label{eq:scale2}
\eeq
has been generally adopted for the scale variation. A
first remark is that the condition $ \frac12 \le \mu_R/\mu_F \le 2$ to avoid
the appearance of large logarithms might seem too restrictive: after all,
these possible large logarithms can be viewed as nothing else than the
logarithms involving the scales and if they are large, it is simply  a
reflection of a large scale dependence. A second remark is that in the case of
processes in which the calculated higher order contributions are small to
moderate and the perturbative series appears to be well
behaved\footnote{This is indeed the case for some important production
  processes at the Tevatron, such as  the Drell--Yan process $p\bar p
  \to V$ ~\cite{Hamberg:1990np, Hamberg:2002np,Melnikov:2006di,
    Catani:2009sm}, weak boson pair production~\cite{Ohnemus:1991kk,
    Frixione:1992pj, Frixione:1993yp, Baur:1993ir, Dixon:1999di} and
  even top quark pair production~\cite{Cacciari:2008zb,
    Kidonakis:2008mu, Moch:2008qy} once the central scale is taken to
  be $\mu_0=m_t$, which have moderate QCD corrections.}, the choice of
such a narrow domain for the scale variation with  $\kappa=2$, appears
reasonable, as will be seen when we will study the scale dependence of
the Higgs--strahlung processes. However, this might be a complete
different story when dealing with processes in which the calculated radiative
corrections turn out to be extremely large. As the higher order contributions
might also be significant in this case, the variation domain of the
renormalization and factorization scales might have to be extended and
a range with a factor $\kappa$ substantially larger than two may seem more
appropriate\footnote{This would have been the case, for instance, in
  top--quark pair production at the Tevatron if the central scale were
  fixed to the more ``natural'' value $\mu_0=2m_t$ (instead of the
  value $\mu_0={m_t}$ usually taken~\cite{Cacciari:2008zb,
    Kidonakis:2008mu, Moch:2008qy}) and a scale variation within
  $\frac14 M_H \le \mu_R,\mu_F \le 4 M_H$ were adopted. Another well
  known example is Higgs production in association with $b$--quark
  pairs in which the cross section can be determined by evaluating the
  mechanism $gg/q\bar q  \to b\bar b H$~\cite{Dittmaier:2003ej} or
  $b\bar b$ annihilation, $b\bar b\to H$~\cite{Campbell:2002zm,
    Harlander:2003ai, Maltoni:2003pn}. The two calculations performed
  at NLO for the former process and NNLO for the later one, are
  consistent only if the central scale is taken to be $\mu_0 \approx
  \frac14 M_H$ instead of the more ''natural" value $\mu_0 \approx
  M_H$, see Ref.~\cite{Assamagan:2004mu} (page 5). Again, without
  prior knowledge of the higher order corrections, it would have been
  wiser, if the central scale $\mu_0=M_H$ had been  adopted, to assume
  a wide domain, e.g.  $\frac14 M_H \le \mu_R,\mu_F \le 4 M_H$,  for
  the scale variation. Note that even for the scale choice  $\mu_0
  \approx \frac14 M_H$, the $K$--factor for the $gg \to b\bar b H$
  process remains very large, $K_{\rm NLO} \approx 2$ at the
  Tevatron. In addition, here, it is the factorization scale $\mu_F$
  which generates the large contributions $\propto \ln(\mu_F^2/m_b^2)$
  in a 5 flavour scheme and not the renormalization scale which can be
  thus kept at the initial value $\mu_R \approx M_H$. We will study
  this process in the section~\ref{section:MSSMHiggsTevCross}
  and~\ref{section:MSSMHiggsLHCCross} in part~\ref{part:four} when
  dealing with MSSM Higgs production.}.

\subsubsection{The case of $gg\to H$ production}

In the case of the $gg \to H$ production process, the most natural
value for the median scale would be the Higgs mass itself,
$\mu_0=M_H$, and if the effects of the higher order contributions to the
cross section is again usually estimated by 
varying $\mu_R$ and $\mu_F$ as in Eq.~\ref{eq:scale2}, i.e. with the choice
$\frac1\kappa \le \mu_R/\mu_F\le \kappa$ and $\kappa=2$, we obtain at
the Tevatron a variation of approximately $\pm 15\%$  of the NNLO cross section
with this specific choice~\cite{Harlander:2002wh,Anastasiou:2002yz} and the uncertainty drops
to the level of $\approx \pm 10\%$ in the NNLL approximation.

As we have stated in the former section that it is wiser to choose
$\mu_0=\frac12 M_H$ as the central scale, e.g. as
in. Ref.~\cite{Anastasiou:2008tj}, the variation domain  $\frac14 M_H \le \mu_R=\mu_F
\le  M_H$ is then adopted, leading also to a $\approx 15\%$ uncertainty.

Nevertheless, as the $K$--factor is extraordinarily large in the $gg \to H$
process, $K_{\rm NNLO} \approx 3$, the domain of Eq.~\ref{eq:scale2} for the
scale variation seems too narrow. If this scale domain was chosen for the LO
cross section for instance, the maximal value of $\sigma(gg \to H)$ at LO would
have never caught, and by far, the value of $\sigma(gg \to H)$ at NNLO,  as it
should be the case if the uncertainty band with $\kappa=2$ were indeed the
correct  ``measure" of the higher order effects. Only for a much larger value
of $\kappa$ that this would have been the case. 

Here, we will use a criterion which allows an empirical evaluation of the
effects of the still unknown high orders of the perturbative series and, hence,
the choice of the variation domain of the factorization and renormalization
scales in a production cross section (or distribution). This is done in two
steps: 

\begin{enumerate}
\item{The domain of scale variation, $\mu_0/{\kappa} \le \mu_R, \mu_F \le
\kappa \mu_0$, is derived by calculating the factor $\kappa$ which allows the
uncertainty band of the lower order cross section resulting from the
variation of  $\mu_R$ and $\mu_F$, to reach the central value (i.e. with
$\mu_R$ and $\mu_F$ set to $\mu_0$), of the cross section that has been
obtained at the higher perturbative order.}
\item{The scale uncertainty on the cross section at the higher perturbative
order is then taken to be the band obtained for a variation of the scales
$\mu_R$ and $\mu_F$ within the same range and, hence, using the same $\kappa$
value.}
\end{enumerate}

\begin{figure}[!h]
\begin{center}
\mbox{
\hspace*{-1.cm}
\includegraphics[scale=0.7]{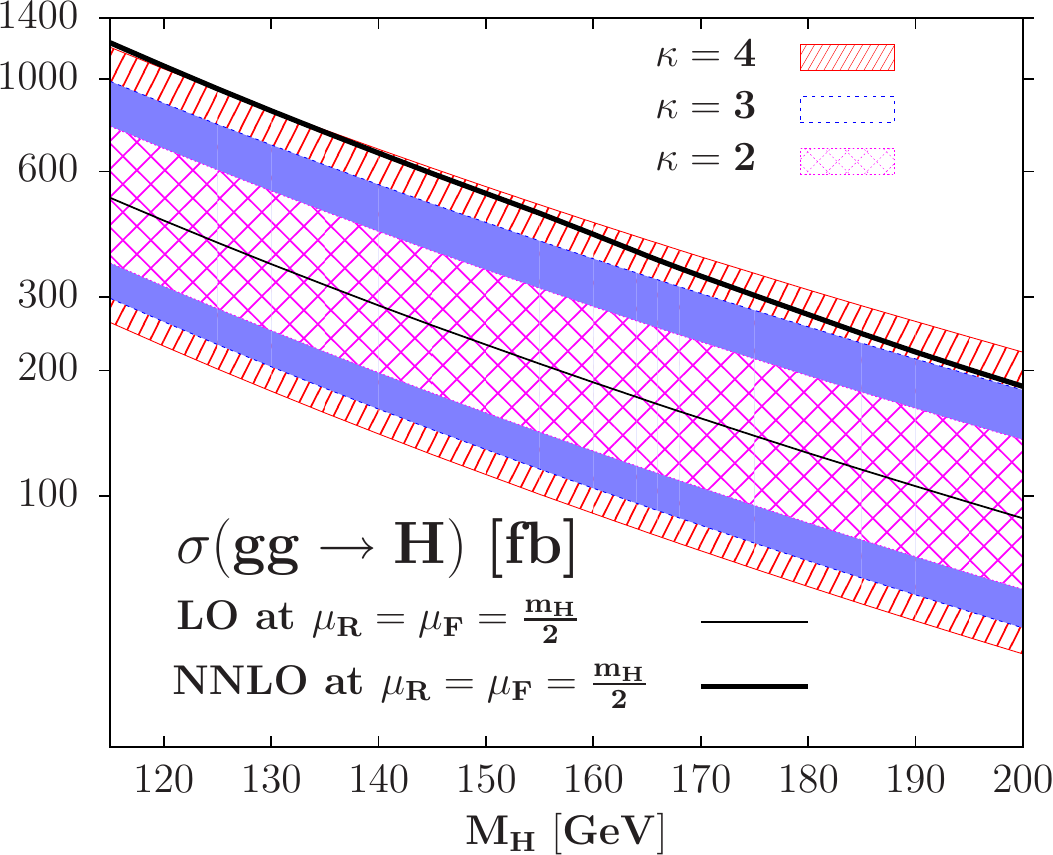} \hspace{-0.1cm}
\includegraphics[scale=0.7]{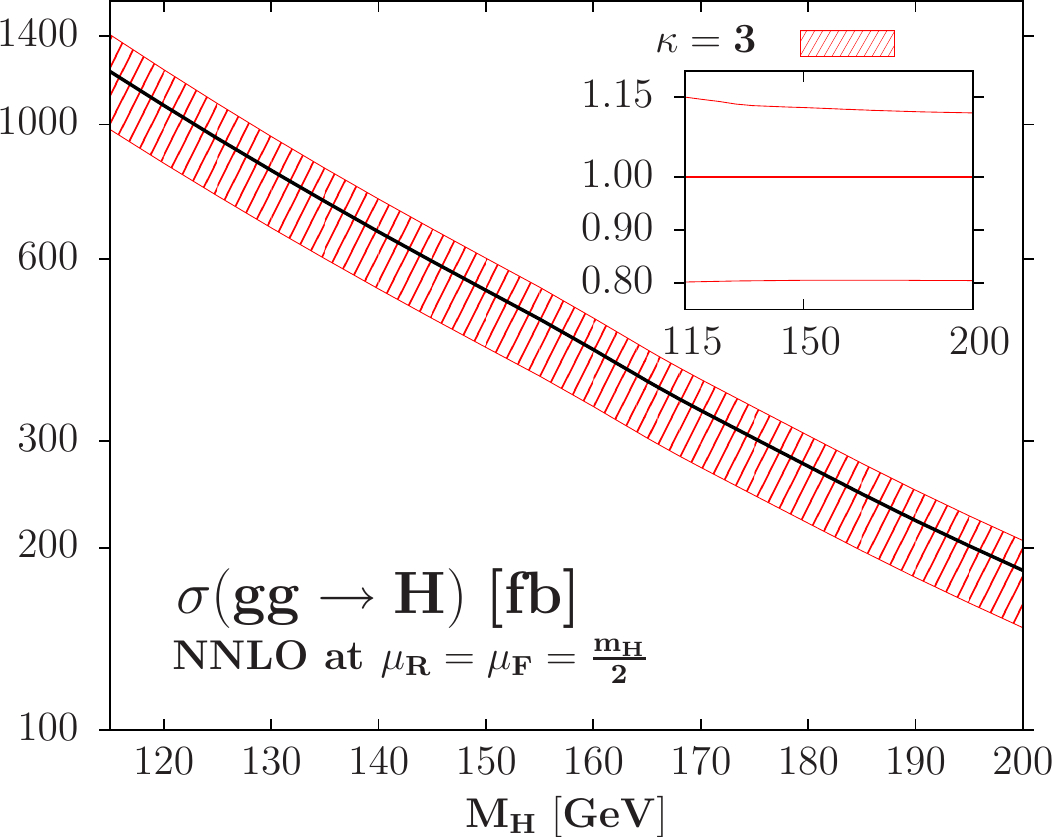}
}
\end{center}
\vspace*{-5mm}
\caption[Scale variation in the $gg\to H$ process at the Tevatron]{Left:
  the scale variation of $\sigma^{\rm LO}_{gg\to H}$  as a 
function of $M_H$ in the domain $ \mu_0/\kappa \leq \mu_R=\mu_F \leq \kappa
\mu_0$  for $\mu_0= \frac12 M_H$ with $\kappa=2,3$ and $4$ compared to
$\sigma^{\rm NNLO}_{gg\to H} (\mu_R=\mu_F=\frac12 M_H)$. Right: the uncertainty
band of  $\sigma^{\rm NNLO}_{gg\to H}$ as  a function of $M_H$ for a scale
variation   $\mu_0/\kappa \leq  \mu_R= \mu_F \leq \kappa \mu_0$ with 
$\mu_0=\frac12 M_H$ and $\kappa=3$. In the inserts shown are the relative 
deviations.}
\vspace*{-3mm}
\label{fig:ggHTeV_scale}
\end{figure}

In the case of the $gg \to H$ process at the Tevatron, if the lower order 
cross section is taken to be $\sigma^{\rm LO}$ and the higher order one 
$\sigma^{\rm NNLO}$, this is shown in the left--hand side of
Fig.~\ref{fig:ggHTeV_scale}. The
figure shows the uncertainty band  of $\sigma^{\rm LO}$ resulting from a scale
variation in the domain 

$$\frac12 M_H/ {\kappa} \le \mu_R,\mu_F \le
\kappa \frac12 M_H$$

with $\kappa=2,3,4$, which is then compared to $\sigma^{\rm NNLO}$  evaluated at
the central scale $\mu_R=\mu_F= \frac12 M_H$.  We first observe that, as expected,
the uncertainty bands are larger with increasing values of $\kappa$.

The important observation that one can draw from this figure is that it is only for
$\kappa\!=\!4$, i.e. a variation of the scales in a range that is much wider
than the one given in Eq.~\ref{eq:scale2}  that the uncertainty band of the LO
cross section becomes very close to (and still does not yet reach for low  Higgs
mass values) the curve giving the NNLO result. Thus, as the scale uncertainty
band of  $\sigma^{\rm LO} (gg \to H)$ is supposed to provide an estimate of
the resulting cross section at NNLO and beyond, the range within which the two
scales $\mu_R$ and $\mu_F$ should be varied must be significantly larger than
$\frac12\mu_0 \le \mu_R, \mu_F \le 2 \mu_0$.

\vspace{2mm}

Nevertheless, one might be rightfully reluctant to use $\sigma^{\rm LO}$  as a
starting point for estimating the higher order effects, as it is well known
that it is only after including at least the next--order QCD corrections that a
cross section is somewhat stabilized and, in the particular case of the $gg\to
H$ process, the LO cross section does not describe correctly the kinematics as,
for instance, the Higgs transverse momentum is zero at this order. We thus
explore also the scale variation of the  NLO cross section $\sigma^{\rm NLO}$ 
instead of that of $\sigma^{\rm LO}$ and compare the resulting uncertainty band
to the central value of the cross  section again at NNLO (we refrain here from
adding the $\sim 15\%$ contribution at NNLL as well as those arising from
higher order corrections, such as the estimated ${\rm N^3LO}$ correction
~\cite{Moch:2005ky}). On the other hand, the choice of $\mu_0=\frac12 M_H$
incorporate much of the higher order correction already at LO as it
lowers the $K$--factors drastically.

If we had chosen to compare NLO
band with NNLO central predictions with the central scale choice
$\mu_0=\frac12 M_H$, we would have been within the band
$\kappa=2$ and touching the lower limit of the $\kappa=3$
band\footnote{see here the Ref.~\cite{Baglio:2010um} where this point is
  mentioned in a footnote. C. Anastasiou is thanked for a
  discussion on this point.}. In addition, this is only for low Higgs
masses that we cannot reach the $\kappa=3$ LO band with NNLO central
prediction. Thus, to attain the (N)LO values of the $gg \to H$ cross section at the
Tevatron with the scale variation of the NLO cross section, when both cross
sections are taken at the central scale choice, we believe that the
right choice is the value $\kappa=3$, and
hence a domain of scale variation that is wider than that given in
Eq.~\ref{eq:scale2}. This choice of the domains of scale variation might seem
somewhat conservative at first sight.  However, we emphasise again that in
view of the huge QCD corrections which affect the cross section of this
particular process, and which almost spoil completely the convergence of the
perturbative series, this choice appears to be justified, even if the
choice of the central scale adopted here helps to reduce these giant
$K$--factors. In fact, this scale choice is not so unusual and in
Refs.~\cite{Anastasiou:2002yz,Ravindran:2003um,Catani:2003zt,Ravindran:2004cx,
  Harlander:2002vv, Cafarella:2005nb} for instance, scale variation
domains comparable to those discussed here, and sometimes even wider,
have been used for illustration.

In addition, we have mentioned that that the scales $\mu_R$ and $\mu_F$ are varied 
independently and with no restriction such as $\frac13 \leq \mu_R/\mu_F \leq 3$
for instance.  In fact, this is only a general statement and this requirement
has absolutely no impact on the NNLO analysis as the minimal and maximal values of
$\sigma^{\rm NNLO}_{\rm gg\to H}$ due to scale variation are obtained for
equal  $\mu_R$ and $\mu_F$ values: for a central scale $\mu_0=M_H$, we have
$\sigma_{\rm min}^{\rm NNLO}$ for $\mu_R=\mu_F=3 \mu_0$ and $\sigma_{\rm
max}^{\rm NNLO}$ for $\mu_R=\mu_F=\frac13 \mu_0$. We then choose for
simplicity to take $\mu_R$ and $\mu_F$ as equal so that there is no
more discussion about the possibility of generating artificially large
logarithms if we take two widely different $\mu_R/\mu_F$ scales.  

Thus, in our analysis, rather than taking the usual choice for the scale domain
of variation with $\kappa=2$ given in Eq.~\ref{eq:scale2}, we will adopt the
slightly more conservative possibility given by the wider variation 
domain\footnote{One might argue that since in the case of $\sigma (gg
  \to H)$, the  NLO and NNLO contributions are both positive and
  increase the LO rate, one should expect a positive contribution from
  higher orders (as is the case for the re-summed NNLL contribution)
  and, thus, varying the scales using $\kappa=2$  is more
  conservative, as the obtained maximal value of the cross section
  would be smaller than the value that one would obtain for
  e.g. $\kappa=3$. However, one should not assume that the higher
  order contributions always increase the lower order cross
  sections. Indeed, had we taken the central scales at
  $\mu_R=\mu_F=\frac15 M_H$, the NNLO (and even NNLL) corrections
  would have  reduced the total cross section evaluated at NLO, see a
  remark made in Ref.~\cite{Baglio:2010um}. Hence, the higher order
  contributions to $\sigma(gg\to H)$ could well be negative beyond
  NNLO and could bring the value of the production cross section close
  to the lower range of the scale uncertainty band of $\sigma^{\rm
    NNLO}$. Another good counter-example of a cross section that is
  reduced by the higher order contributions is the process of
  associated Higgs production with top quark pairs at the Tevatron
  where the NLO QCD corrections decrease the LO cross section by $\sim
  20\%$ ~\cite{Beenakker:2001rj, Beenakker:2002nc, Reina:2001sf,
    Dawson:2002tg} once the central scale is chosen to be
  $\mu_0=\frac12(2m_t+M_H)$.}

\beq
\frac13 \mu_0 \leq \mu_R = \mu_F \leq 3 \mu_0 \ , \ \mu_0= \frac12 M_H
\label{eq:scale3}
\eeq

Having made this choice for the factor $\kappa$, we now 
estimate the higher order effects of $\sigma ({ gg \to H})$ evaluated at
the highest perturbative order that we take to be NNLO, ignoring again the 
known small contributions beyond this fixed order.  

The uncertainty bands resulting from scale variation  of  $\sigma^{\rm NNLO}(gg
\to H)$  at NNLO in the domain given by Eq.~\ref{eq:scale3}
is shown in the right--hand side of Fig.~\ref{fig:ggHTeV_scale} as a
function of $M_H$. As expected, the scale uncertainty
is slightly larger for $\kappa=3$ than for $\kappa=2$. Averaged over the
entire Higgs mass range, the final scale uncertainty is about $\simeq +15\%,
-20\%$. If we had chosen the usual domain $\frac12 \mu_0  \leq \mu_R =
\mu_F \leq 2 \mu_0$, the scale variation would have been of about
$\approx + 10\%, -12\%$ for $M_H \approx 160$ GeV. The minimal cross
section is obtained for the largest values of the two scales as stated
above, $\mu_F=\mu_R= \kappa \frac12 M_H$, while the maximal
value is obtained for the lowest value of the renormalization scale,
$\mu_R=\frac1\kappa \frac12 M_H$, almost independently of the
factorization scale $\mu_F$, but with a slight preference for the
lowest $\mu_F$ values, $\mu_F=\frac1\kappa \frac 12 M_H $.

We should note that the $\approx 10\%$ scale uncertainty obtained in 
Ref.~\cite{deFlorian:2009hc} and adopted in earlier publications by the CDF/D0
collaborations~\cite{Tevatron:2009je} is
even smaller than the ones discussed above. The reason is that it is the
resummed NNLL cross section, again with $\kappa\!=\!2$ and $\frac12\! \le\!
\mu_R/\mu_F\! \le\! 2$, that was considered, and the scale variation of
$\sigma^{\rm NNLL}$ is reduced compared to that of  $\sigma^{\rm NNLO}$ in this
case. As one might wonder if this milder dependence also occurs for our adopted
$\kappa$ value, we have explored the scale variation of  $\sigma^{\rm NNLL}$ in
the case of  $\kappa=3$, without the restriction $\frac13\! \le\!
\mu_R/\mu_F\!  \le\! 3$. Using again the program {\tt
  HRESUM}~\cite{Grazzinipage}, we find that the difference between the
maximal value of the NNLL cross section, obtained for $\mu_R \approx
M_H$ and $\mu_F \approx 3M_H$, and its minimal value, obtained for
$\mu_F \approx  \frac13 M_H$ and $\mu_R \approx 3M_H$, is as large as
in the NNLO case (this is also true for larger $\kappa$ values).
The maximal decrease and maximal increase of $\sigma^{\rm NNLL}$ from the
central value are still of about $\pm 20\%$ in this case. Hence, the relative
stability of the NNLL cross section against scale variation, compared to the
NNLO case,  occurs only  for $\kappa=2$  and may appear as accidentally due to 
a restrictive choice of the variation domain. However, if the additional
constraint $1/\kappa \le \mu_F/\mu_R \le \kappa$ is implemented, the situation
would improve in the NNLL case, as the possibility $\mu_F \approx 
\frac1\kappa M_H$ and $ \mu_R \approx \kappa M_H$ which minimizes  $\sigma^{\rm
NNLL}$  would be absent and the scale variation reduced. Nevertheless, even in
this case, the variation  of $\sigma^{\rm  NNLL}$ for $\kappa=3$ is of the
order of $\approx \pm 15\%$ and, hence, the scale uncertainty is larger than 
what is obtained in  the domain of Eq.~\ref{eq:scale2}.

It is important to notice that if the NNLO $gg\to H$ cross section, evaluated  
at $\mu_0=M_H$,  is broken into the three pieces with 0,1 and 2 jets, and one
applies a scale variation for the individual pieces in the range $\frac12 \mu_0
\leq \mu_R, \mu_F \leq 2 \mu_0$, one obtains with selection cuts similar to
those adopted by the CDF/D0 collaborations~\cite{Anastasiou:2009bt}: 
\beq
\left.  
\Delta \sigma/\sigma 
\right|_{\rm scale} =  
60\% \cdot \left({^{+5\%}_{-9\%}} \right) 
+29\% \cdot \left({^{+24\%}_{-23\%}} \right) 
+11\% \cdot \left({^{+91\%}_{-44\%}} \right) = \left({^{+20.0\%}_{-16.9\%}} 
\right)
\label{eq:jetscale}
\eeq
Averaged over the various final states with their corresponding weights, an
error on the ``inclusive"  cross section which is about $+20\%,-17\%$ is
derived\footnote{The error might be reduced when including higher--order
corrections in the 1\,jet and 2\,jet cross sections.}. This is very close to
the result obtained in the CDF/D0
analysis~\cite{Aaltonen:2010yv,Tevatron:2010ar,Aaltonen:2011gs} which
quotes a scale uncertainty of $\approx \pm 17.5\%$ on the total cross section,
when the weighted uncertainties for the various jet cross sections are
added. Thus, our alleged conservative choice $\frac13 \mu_0 \leq
\mu_R=\mu_F \leq 3 mu_0$ for the scale variation of the total
inclusive cross section $\sigma^{\rm NNLO}_{\rm gg\to H}$,  leads to
a scale uncertainty that is very close to that obtained when one adds
the scale uncertainties of the various jet cross sections for a
variation around the more consensual range $\frac12 \mu_0 \leq  \mu_R,
\mu_F \leq 2 \mu_0$.

We also note that when breaking $\sigma^{\rm NNLO}_{\rm gg\to H}$ into jet
cross sections, an additional error due to the acceptance of jets is
introduced; the CDF and D0 collaborations, after weighting, have estimated it
to be $\pm 7.5\%$. We do not know if this weighted acceptance error should be
considered as a theoretical or an experimental uncertainty. But this error,
combined with the weighted uncertainty for scale variation, will certainly
increase the total scale error in the CDF/D0 analysis, possibly (and depending
on how the errors should be added) to the level where it almost reaches or even
exceeds our own supposedly ``conservative" estimate. 

We could finally give another reason for a more conservative choice of
the scale variation 
domain for $\sigma^{\rm NNLO}$, beyond the minimal $\frac 12 M_H \le
\mu_R,\mu_F \le 2 M_H$ range. It is that it is well known that the QCD
corrections are significantly larger for the total inclusive cross section than
for that on which basic selection cuts are applied; see e.g.
Refs.~\cite{Ravindran:2003ia, Anastasiou:2005qj, Catani:2007vq,
  Anastasiou:2009kn}. This can be seen from the recent analysis of
Ref.~\cite{Anastasiou:2009bt}, in which the higher order corrections
to the inclusive cross section for the  main Tevatron Higgs signal,
$gg \to H \to \ell \ell \nu \nu$, have been compared  to those
affecting the cross section when selection 
cuts, that are very similar to those adopted by the CDF and D0 collaborations 
in their analysis (namely lepton selection and isolation, a minimum
requirement for the missing transverse energy due to the neutrinos, and a veto
on hard jets to suppress the $t\bar t$ background), are applied.  The output of
this study is that the $K$--factor
for the cross section after cuts  is $\sim 20$--30\% smaller than the
$K$--factor for the inclusive total cross section (albeit with a reduced scale
dependence).  For instance,  one has $K^{\rm NNLO}_{\rm cuts}=2.6$ and $K^{\rm
NNLO}_{\rm total}= 3.3$ for  $M_H=160$ GeV and scales set to
$\mu_F=\mu_R=M_H$. 

Naively, one would expect that this $\sim 20$--30\% reduction of the higher
order QCD corrections when selection cuts are applied,  if not implemented from
the very beginning in the normalisation of the cross section after cuts that is
actually used by the experiments (which would then reduce the acceptance of the
signal events,  defined as $\sigma^{\rm NNLO}_{\rm cuts}/\sigma^{\rm NNLO}_{\rm
total}$), to be at least  reflected in the scale variation of the inclusive
cross section and, thus, accounted for in the theoretical uncertainty. This
would be partly the case for scale variation within a factor $\kappa=3$ from
the central scale, which leads to a maximal reduction of the $gg \to H \to \ell
\ell \nu \nu$ cross section by about $20\% $, but not with the choice
$\kappa=2$  made in
Refs.~\cite{Tevatron:2009je,Aaltonen:2010yv,Tevatron:2010ar,Aaltonen:2011gs}
which would have led to a possible reduction of the cross section by
$\approx 10$\% only\footnote{The discussion is, however, more involved
  as one has to consider the efficiencies obtained with the NNLO
  calculation compared to that obtained with the Monte--Carlo used by
  the experiments; see  Ref.~\cite{Anastasiou:2009bt}.}.

\subsubsection{The scale variation uncertainty in Higgs--strahlung
  processes}

We finish this subsection by the discussion of the theoretical uncertainties in the Higgs
strahlung mechanism $q\bar q \to VH$, following the same line of arguments as
in the previous subsection.

To evaluate the uncertainties due to the variation of the renormalisation and 
factorisation scales in the Higgs--strahlung processes, the choice of the 
variation domain is in a sense simpler than for the $gg\to H$ mechanism.
Indeed, as the process at leading order is mediated only by massive gauge
boson exchange and is then a pure electroweak process at the partonic
level, only the factorisation scale $\mu_F$ appears when the partonic cross
section is folded with the $q$ and $\bar q$ luminosities and there is no
dependence on the renormalisation scale $\mu_R$ at this order. It is only at
NLO, when gluons are exchanged between or radiated from the $q,\bar q$ initial
states, that both scales $\mu_R$ and $\mu_F$ appear explicitly.

\begin{figure}[!h]
\begin{center}
\includegraphics[scale=0.7]{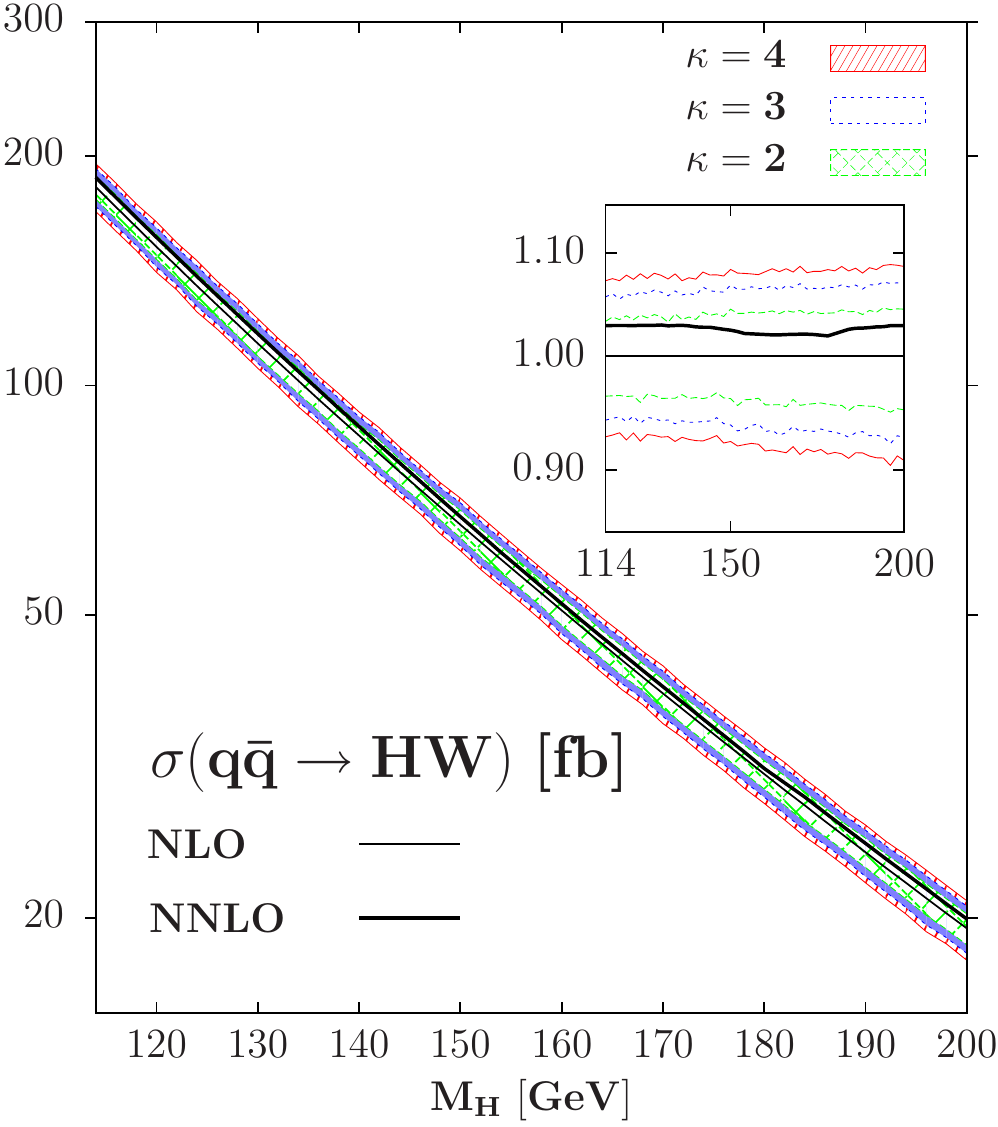}
\includegraphics[scale=0.7]{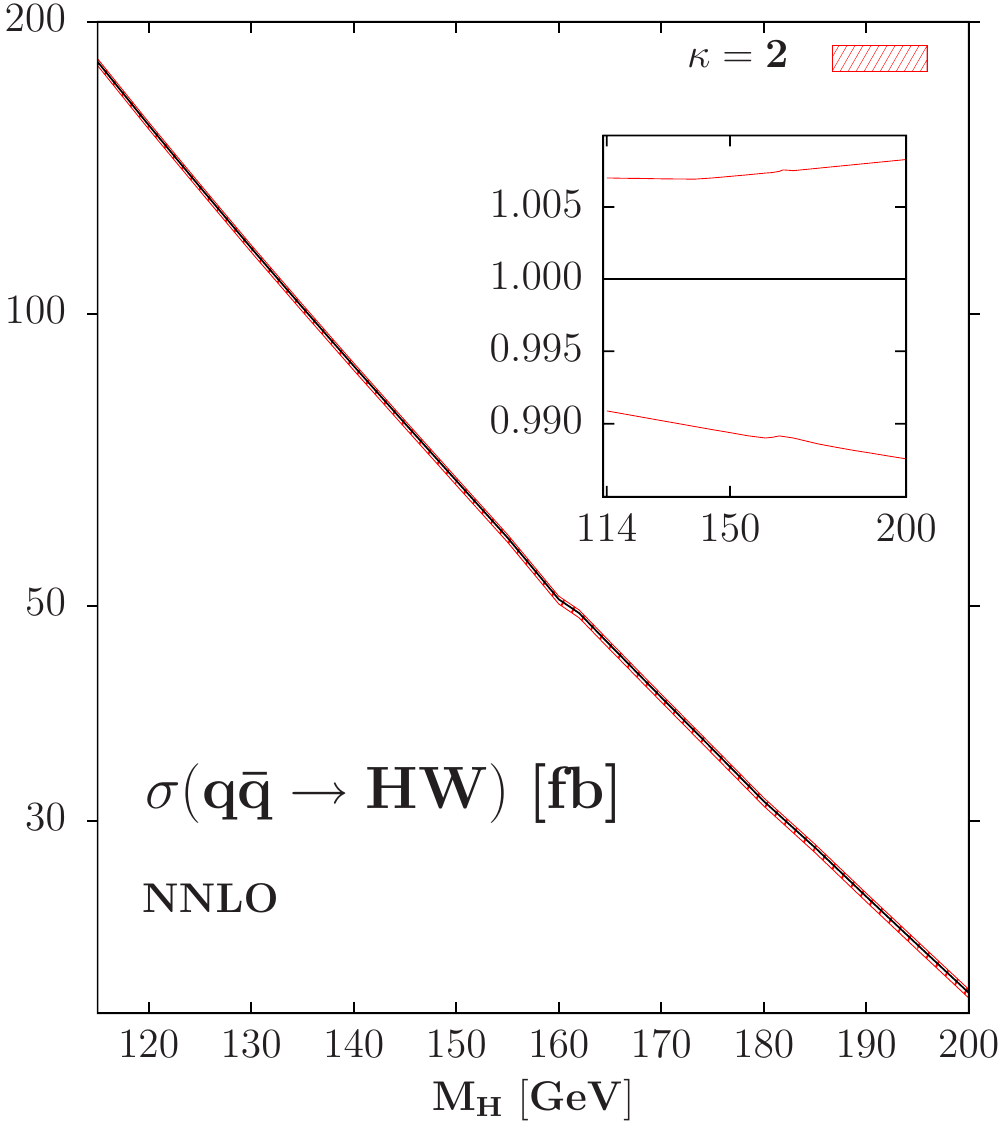}
\end{center}
\vspace*{-3mm}
\caption[Scale variation in the $p\bar p\to WH$ process at the Tevatron] {Left:
  the scale dependence of $\sigma(p \bar p \to WH)$ at NLO for
  variations $M_{HV}/{\kappa} \le \mu_R, \mu_F \le \kappa M_{HV}$ with 
$\kappa=2,3$ and 4, compared to the NNLO value; in the insert, shown are the 
variations in percentage  and where the NNLO cross section is normalized to the NLO
one. Right: the scale  dependence of $\sigma( p\bar p \to WH)$ at NNLO for a
variation in the domains  $M_{HV}/2 \le \mu_R= \mu_F \le 2 M_{HV}$; the relative
deviations from the  central value are shown in the insert.}
\vspace*{-2mm}
\label{fig:pphvTeV_scale}
\end{figure}

 Using the proposed criterion for the estimate of the perturbative higher order
effects, we thus choose again to consider the variation domain of the scales
from their central values, $\mu_0/\kappa \le \mu_R, \mu_F \le \kappa \mu_0$
with $\mu_0=M_{HW}$, of the NLO cross section to determine the value
of the factor $\kappa$ to be used at NNLO\footnote{It is clear that
  in this case the comparison with the LO prediction would be no use
  as the scale uncertainty is a pure QCD uncertainty; the LO partonic
  cross section being a pure electroweak process, we would not be
  consistent to compare a full NNLO calculation with a scale variation
at LO induced only by the PDF through the factorization scale
$\mu_F$.}. We display in the left-hand side of
Fig.~\ref{fig:pphvTeV_scale} the variation of the NLO cross section
$\sigma^{\rm NLO} (p \bar p \to WH)$ at the Tevatron as a function of $M_H$ for
three values of the constant $\kappa$  which defines the range spanned by the
scales, $M_{HW}/\kappa \le \mu_R, \mu_F \le \kappa M_{HW}$. One sees that, in
this case, a value $\kappa=2$ is sufficient (if the scales $\mu_R$ and $\mu_F$
are varied independently in the chosen domain) in order that the uncertainty
band at NLO reaches the central value of the cross section at NNLO. In fact,
the NLO uncertainty band would have been only marginally  affected if one had
chosen the values $\kappa=3$, $4$ or even 5. This demonstrates than the cross
sections for the Higgs--strahlung processes, in contrast to $gg \to H$, are
very stable against scale variation, a result that is presumably due to the
smaller $q\bar q$ color charges compared to gluons, $\approx  C_F/C_A$,  that
lead to more moderate QCD corrections. 

In the right--hand side of Fig.~\ref{fig:pphvTeV_scale}, the NNLO
$p\bar p \to WH$ total cross section is displayed as a function of
$M_H$ for a  scale variation $\frac12 M_{HW} \le \mu_R, \mu_F  \le 2
M_{HW}$. Contrary to the $gg \to H$ mechanism, the scale 
variation within the chosen range is rather mild and only  a $\sim 0.7\%$ (at
low $M_H$) to $1.2\%$ (at high $M_H$) uncertainty is observed for the relevant 
Higgs mass range at the Tevatron. This had to be expected as the $K$--factors in
the Higgs--strahlung processes, $K_{\rm NLO} \approx 1.4$ and  $K_{\rm NNLO}
\approx 1.5$,  are substantially smaller than those affecting the $gg$ fusion
mechanism and one expects perturbation theory to have a better behavior in the
former case. This provides more confidence that the Higgs--strahlung cross
section is stable against scale variation and, thus, that higher order effects
should be small.

\subsection[The PDF puzzle]{The parton distribution functions
  puzzle \label{section:SMHiggsTevPDF}}

The second (and sometimes the most important) source of theoretical
uncertainties on production cross sections and distributions at hadron
colliders is due to the still imperfect parametrisation of the parton
distribution functions (PDFs). Within a given parametrisation, for
example the one in the MSTW scheme, these uncertainties
are estimated as
follows~\cite{Martin:2002dr,Martin:2009bu,Djouadi:2003jg}. The scheme
is based on a Hessian matrix method which enables a characterization of a
parton parametrization in the neighborhood of the global $\chi^2$
minimum fit and gives an access to the uncertainty estimation through
a set of PDFs that describes this neighborhood. The corresponding PDFs
are constructed by: $i)$ performing  a global fit of the data using
$N_{\rm PDF}$ free parameters ($N_{\rm PDF}=15$ or  20, depending on
the scheme); this provides the nominal PDF or reference set denoted by
$S_0$; $(ii)$ the global $\chi^2$ of the fit is increased to a given
value $\Delta \chi^2$ to obtain the error matrix, this value is
subject to vary between different collaborations; $(iii)$ the error 
matrix is diagonalized to obtain $N_{\rm PDF}$ eigenvectors corresponding to
$N_{\rm PDF}$ independent directions in the parameter space; $(iv)$ for each
eigenvector, up and down excursions are performed in the tolerance gap,
$T=\sqrt{\Delta \chi^2_{\rm global}}$, leading to $2N_{\rm PDF}$ sets of new
parameters, denoted by $S_i$, with $i=1, 2N_{\rm PDF}$. It is worth
mentioning that the NNPDF collaboration~\cite{Ball:2009mk} does not use this
method and rely on Monte--Carlo replication sample to generate its
uncertainty PDF sets.

These sets of PDFs can be used to calculate the uncertainty on a cross section
$\sigma$ in the general following way: one first evaluates the cross section with the
nominal PDF $S_0$ to obtain  the central value $\sigma_0$, and then calculates 
the cross section with  the $S_i$ PDFs, giving $2N_{\rm PDF}$ values $\sigma_i$,
and defines, for each $\sigma_i$ value, the deviations 
\beq
\sigma_i^\pm =\mid \sigma_i -\sigma_0\mid \ \ {\rm when} \  
\sigma_i \ ^{>}_{<}  \sigma_0
\eeq
The uncertainties are summed quadratically to calculate the cross section, 
including the error from the PDFs that are given at the 68\% or 90\%
confidence level (CL),
\beq
\sigma_0|^{+\Delta \sigma^+_{\rm PDF} }_{- \Delta \sigma^-_{\rm PDF}} \ \ 
{\rm with} \ \Delta \sigma^\pm_{\rm PDF}  = \left( \sum_i \sigma_i^{\pm 2} 
\right)^{1/2}
\eeq

The procedure outlined above has been applied to estimate the PDF
uncertainties in the Higgs production cross sections in the
gluon--gluon fusion mechanism at the Tevatron in
Refs.~\cite{deFlorian:2009hc,Anastasiou:2008tj}. This has led to a
90\% CL uncertainty of $\approx 6\%$ for the low mass range $M_H
\approx 120$ GeV to $\approx  10\%$ in the high mass range, $M_H
\approx 200$ GeV. These uncertainties was adopted in the early CDF/D0
combined Higgs search and represented the second largest source of
errors after the scale variation. This has been shown since that at
least in the case of the gluon--gluon fusion mechanism it
underestimated the PDF uncertainties for at least the first reason
that will be mentioned below, if not for that of the second that
concerns the strong coupling constant $\alpha_s$ itself, see in the
next pages.

First of all, the MSTW collaboration~\cite{Martin:2009iq} is not the
only one which uses the above scheme for PDF error estimates, as the
CTEQ~\cite{Nadolsky:2008zw} and ABKM~\cite{Alekhin:2009ni}
collaborations, for instance, also provide similar schemes, as well as
NNLO JR set~\cite{JimenezDelgado:2009tv} appeared in 2010 or
HERAPDF~\cite{HERApage} (NNPDF set~\cite{Ball:2009mk} also allows for
error estimates even if their method is a bit different). It is thus
more appropriate to compare the results given by the different sets
and take into account the possibly different errors that are
obtained. In addition, as the parameterisations of the PDFs are
different in the various PDF schemes, we might obtain different
central values for the cross sections and the impact of this
difference should also be addressed\footnote{This difference should
  not come as a surprise as, even within the same scheme, there are
  large differences when the PDF sets are updated. For instance, as
  also pointed out in Refs.~\cite{deFlorian:2009hc,Anastasiou:2008tj},
  $\sigma^{\rm NNLO} (gg \to H)$ evaluated with the MRST2004 set is
  different by more than 10\% compared to the current value obtained
  with the MSTW2008 set, as a result of a corrected treatment of the
  $b,c$ densities among other improvements.}.

These two aspects will be taken into account in the following. We will
begin the analysis by looking at the PDF uncertainties given
separately by MSTW, ABKM and CTEQ. Even if in the case of CTEQ
collaboration there is no NNLO PDFs set yet available, we can still use
the available NLO sets, evaluate the PDF errors on the NLO cross
sections and take these errors as approximately available at NNLO once
we rescale the cross sections by including the NNLO corrections. This
constitutes a good approximation for error estimates that is our goal
in this first analysis.

We then will analyze the results obtained with the current four NNLO
PDFs sets available on the market, that are the MSTW collaboration,
the ABKM PDFs set, the HERA PDF 1.0 and the JR 09 PDF set, comparing
only the central values. This will be full of information and we will
see that there are too many differences between the predictions to
discard the problem. Note that this PDF ``puzzle'' has been the
subject of a recommendation by the PDF4LHC group, see~\cite{PDF4LHC}.
This will also be discussed.

\subsubsection{PDF uncertainties in gluon--gluon fusion}

\begin{figure}[!t]
\begin{center}
\includegraphics[scale=0.7]{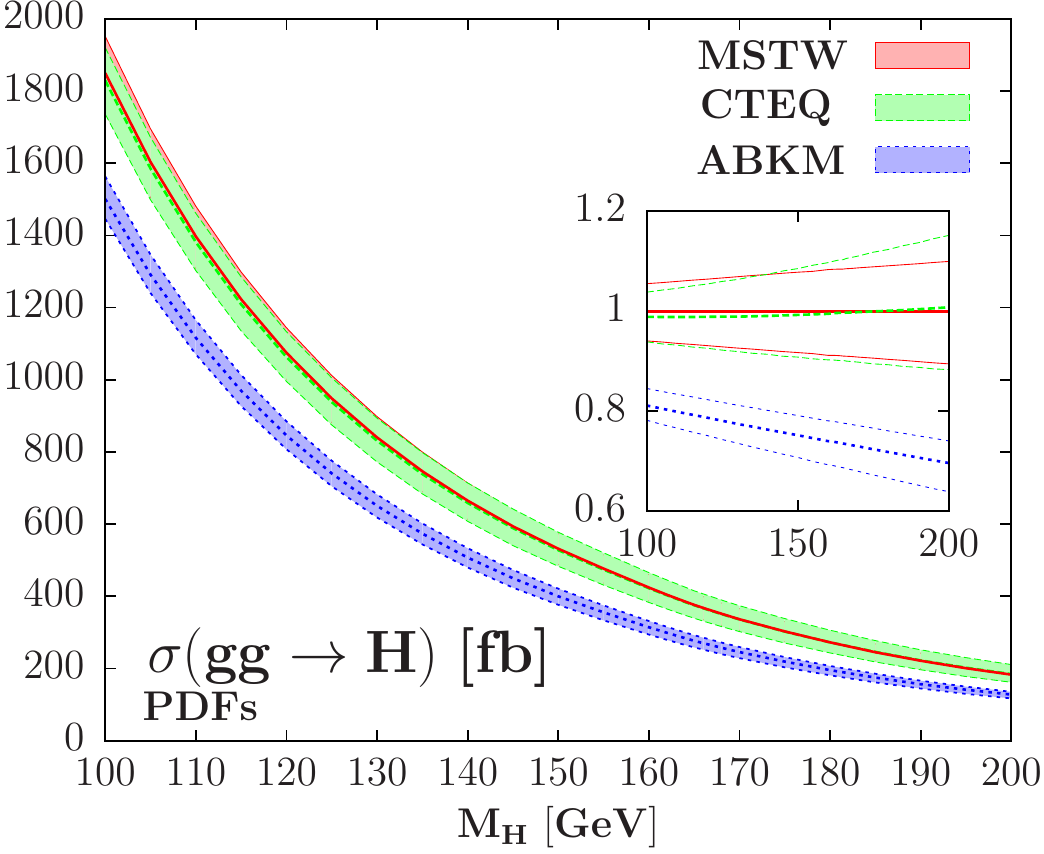}
\includegraphics[scale=0.7]{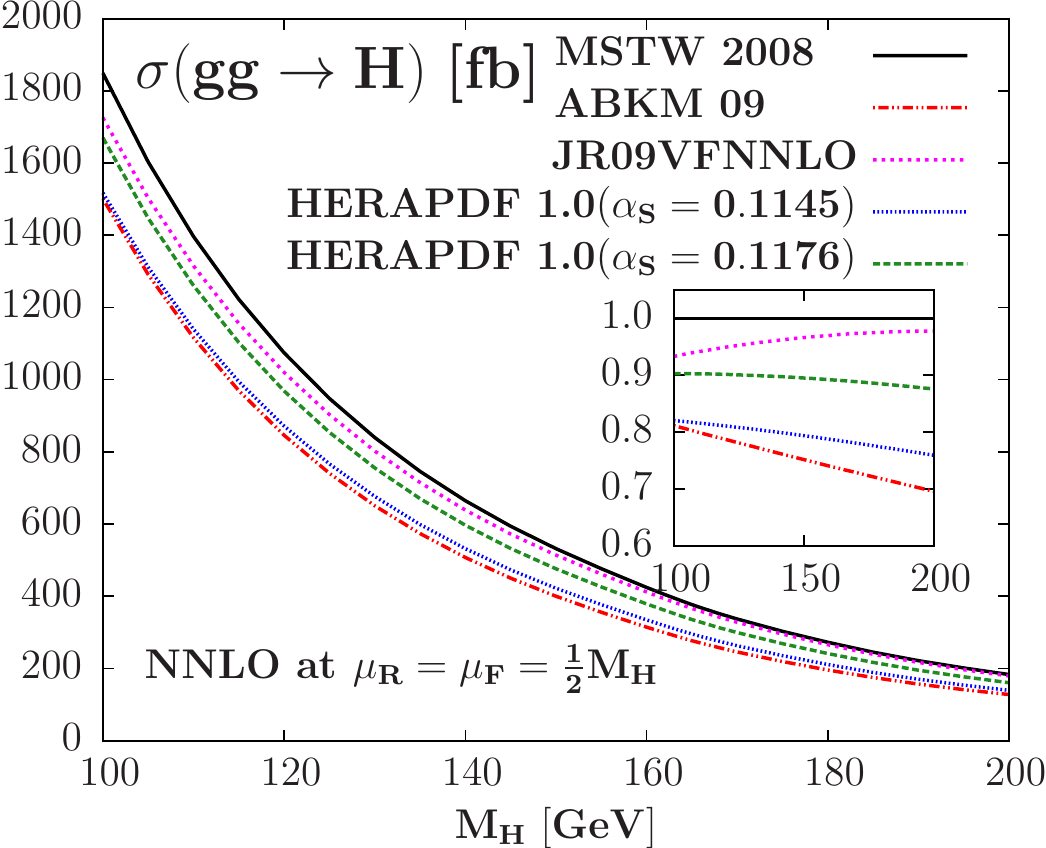}
\end{center}
\vspace*{-5mm}
\caption[Comparison between different PDFs sets in $gg\to H$ at the
Tevatron using CTEQ/ABKM/MSTW
PDF sets for 90\%CL uncertainties and MSTW/ABKM/HERA/JR for central
predictions comparison]{Left: the central values and the 90\%CL PDF
  uncertainty bands in the NNLO cross section $\sigma(gg \to H+X)$ at
  the Tevatron when evaluated within the MSTW, CTEQ and ABKM
  schemes. In the insert, shown in percentage are the deviations
  within a given scheme and the CTEQ and ABKM central values when the
  cross sections are normalized to the MSTW central value. Right: the
  $gg\to H$ cross section as a function of $M_H$ when the four NNLO
  PDF sets, MSTW, ABKM, JR and HERAPDF, are used. In the inserts,
  shown are the deviations with respect to the central MSTW value} 
\vspace*{-2mm}
\label{fig:ggHTeV_pdf1}
\end{figure}

In the case of the  $gg \to H$ cross section at the Tevatron, the
90\% CL PDF errors using the first three schemes discussed above (namely MSTW,
CTEQ, ABKM) are shown in the left--hand side of
Fig.~\ref{fig:ggHTeV_pdf1} as a function of $M_H$? The spread of the
cross section due to the PDF errors is approximately 
the same in the MSTW and CTEQ schemes, leading to an uncertainty band
of less than $10$\% in both cases. For instance, in the MSTW scheme and in 
agreement with Refs.~\cite{deFlorian:2009hc,Anastasiou:2008tj}, we obtain a $\sim \pm 6\%$ error
for $M_H=120$ GeV and $\sim \pm 9\%$ for $M_H=180$ GeV; the errors are only
slightly asymmetric and for $M_H=160$ GeV, one has  $\Delta \sigma^+_{\rm
PDF}/\sigma=+8.1\%$ and $\Delta \sigma^-_{\rm PDF}/\sigma=-8.6\%$.  The errors
are relatively smaller in the ABKM case in the entire Higgs mass range and, for
instance, we obtain a $\Delta \sigma^\pm_{\rm PDF}/\sigma \approx \pm 5\%$
(7\%) error for $M_H=120~(180)$ GeV. 

The left--hand side of Fig.~\ref{fig:ggHTeV_pdf1} opens a very
important issue: the (very!) large discrepency between the central
values of the cross sections calculated with the differents PDFs sets,
with MSTW/CTEQ on the one hand and ABKM on the other hand, even when
taking into account their individual uncertainties. This problem has been
discussed in~\cite{Martin:2009bu,Alekhin:2009ni,Grazzini:2010zc} and
is a long--standing issue still pending in the community. Indeed, the
use of the ABKM parametrisation results in a cross section that is $\sim
25\%$ smaller than the cross section evaluated with the MSTW or CTEQ
PDFs. Thus, even if the PDF uncertainties evaluated within a given
scheme turn out to be relatively small and apparently well under
control, the spread of the cross sections due to the different
parameterisations can be much more important. 

If one uses the old way of estimating the PDF uncertainties (i.e. before the
advent of the PDF error estimates within a given scheme) by comparing  the
results given by different PDF parameterisations, one arrives at an uncertainty 
defined as
\beq 
\Delta \sigma^+_{\rm PDF}  &=& {\rm max} (  \sigma^0_{\rm MSTW} ,
\sigma^{0}_{\rm 
CTEQ}, \sigma^0_{\rm ABKM}) -\sigma^0_{\rm MSTW} \nonumber \\
\Delta \sigma^-_{\rm PDF}  &=& \sigma^0_{\rm MSTW} - {\rm min} 
(\sigma^0_{\rm MSTW} , \sigma^{0}_{\rm CTEQ}, \sigma^0_{\rm ABKM})
\label{eq:conservative-PDF} 
\eeq
where the central value of the $gg\to H$ cross section is taken to be that
given by the MSTW nominal set $S_0$\footnote{We do no include in this
  comparison the uncertainties obtain within a given PDF set, which
  would add 5\% to 7\% to the uncertainty quoted.}. Hence, for
$M_H=160$  GeV for instance, one would have $\Delta \sigma^+_{\rm PDF}
\approx 1\%$ given by the small difference between the CTEQ and MSTW
central values of the cross section and $\Delta \sigma^-_{\rm PDF}
\approx -25\%$ given by the large difference between the ABKM and MSTW
central values.

To make the discussion solid and consistent, we have compared the
central predictions obtained when calculating the NNLO hadronic $gg\to H$
cross section using the various available NNLO sets on the market:
besides the already mentioned ABKM~\cite{Alekhin:2009ni} and
MSTW~\cite{Martin:2009iq} sets, we have also included the HERA PDF
set~\cite{HERApage} and the JR 09
set~\cite{JimenezDelgado:2009tv}. The result is displayed in the
right--hand side of Fig.~\ref{fig:ggHTeV_pdf1}. As can be seen, there
is a very large spread in the four predictions, in particular at large $M_H$ values
where the poorly constrained gluon densities at high--$x$ are involved. The
largest rate is obtained with MSTW, but the cross section using the
ABKM set is $\approx\! 30\%$ lower for $M_H\!\approx\! 160$ GeV. That
would mean that using the naive way to estimate this spread would
end--up with an approximate $-30\%$ uncertainty only due to PDF
uncertainty! Nearly similar results would have been obtained using
HERAPDF set\footnote{It is often argued against the HERAPDF scheme,
  which uses consistently only HERA data to determine the flavour
  decomposition, that it does not use any jet (Tevatron or DIS) data
  which is in principle important in the determination of the gluon
  densities. However, HERAPDF describes well not only the Tevatron jet
  data but also the $W,Z$ data. Since this is a prediction beyond
  leading order, it has also  the contributions of the gluon
  included. This gives an indirect test that the  gluon densities are
  predicted in a satisfactory way. See also Ref.~\cite{Alekhin:2010dd}.}
with the smaller $\alpha_s(M_Z)$ value, with an approximate $22\%$
decrease of the cross section for $M_H\!\approx\! 160$ GeV.

Nevertheless we would like to keep considering the MSTW scheme for at
least two reasons: first because it is the widest used PDF scheme and
then allows for comparison with other calculation and especially with
the numbers used
in~\cite{Tevatron:2009je,Aaltonen:2010yv,Tevatron:2010ar,Aaltonen:2011gs},
and then because the MSTW scheme include the di--jet Tevatron run II
data which are of utmost importance in this context. We will of course
not discard the discrepency outlined above, and try to solve it within
the MSTW scheme.

Indeed, the spread of the predictions does not only result from the
different gluon densities used (and it is well known that these
densities are less severely constrained by experimental data than
light quark densities, especially at high Bjorken--$x$ values), but
also from the difference in the value of the 
strong coupling constant $\alpha_s$ which is fitted altogether with
the PDF sets. The $gg\to H$ process is very special in light of this
issue, as this process is already a one--loop calculation at LO which
is proportionnal to the square of $\alpha_s$ and the large NLO and
NNLO QCD contributions are, respectively, of ${\cal O}(\alpha_s^3)$
and ${\cal O}(\alpha_s^4)$. The $\alpha_s$ value
used in the ABKM set, $\alpha_s(M_Z^2)=0.1129 \pm 0.0014$ at NLO in
the BMSM scheme~\cite{Buza:1996wv},  is $\approx 3 \sigma$ smaller than the
one in the MSTW set $\alpha_s(M_Z^2)=0.12018\pm 0.014$ at NLO. Similar
difference arise with the HERA PDF set ($\alpha_s(M_Z^2)=0.1135$ or
$0.1176$ at NNLO depending on the PDF set) or with JR 09 set
($\alpha_s(M_Z^2)=0.1124\pm 0.0020$ at NNLO) when compared to MSTW
value displayed in Eq.~\ref{eq:alphas} \footnote{The significant
  difference between the world average $\alpha_s$
  value~\cite{Nakamura:2010zzi} and the one from deep-inelastic
  scattering (DIS) data used in the PDFs is connected to that
  issue.}. Since the $K$--factors 
for the gluon--gluon fusion process are very large at the
Tevatron, $K_{\rm NLO} \sim 2$ and $K_{\rm NNLO} \sim 3$, a one percent
uncertainty in the input value of $\alpha_s$ will generate a $\approx 3\%$
uncertainty in $\sigma^{\rm NNLO} (gg \to H)$.

\beq
\alpha_s(M_Z^2)&=& 0.11707~^{+0.0014}_{-0.0014}~{\rm (68\%CL)}~~
                      ^{+0.0032}_{-0.0032}~{\rm (90\%CL)}~~{\rm
                        at~NNLO} 
\label{eq:alphas}
\eeq

Experimental collaborations are now aware of this issue and have
started in 2010 to take into account related ``$\alpha_s$''
uncertainties in the quoted number they use for their analysis, see
~\cite{Tevatron:2010ar,Aaltonen:2011gs}. This
uncertainty was certainly lacking in earlier Tevatron
analysis~\cite{Tevatron:2009je,Aaltonen:2010yv} and that was after the
output of Ref.~\cite{Baglio:2010um} that the incorporation of such an
uncertainty was widely accepted.

In general, $\alpha_s$ is fitted together with the PDFs:  the PDF sets are only
defined for the special value of $\alpha_s$ obtained with the best fit and, to
be consistent,  this best value of $\alpha_s$ that we denote $\alpha_s^0$,  
should also be used for the partonic part of the cross section. Indeed
there is an interplay between the PDFs and the value of
$\alpha_s$ and, for instance, a larger value of $\alpha_s$ would lead to a
smaller gluon density at low $x$~\cite{Martin:2009bu}.  The MSTW
collaboration released last year a new set--up which allows for a
simultaneous evaluation of the errors due to the PDFs and those due to
the experimental uncertainties on $\alpha_s$ of Eq.~\ref{eq:alphas},
taking into account the possible correlations~\cite{Martin:2009bu}.  The
procedure to obtain the different PDFs and their associated errors is
similar to what was presented in the introduction of this subsection,
but provided is a collection of five PDF+error sets for different
$\alpha_{s}$ values: the best fit value $\alpha_s^0$ and its 68\% CL
and 90\% CL maximal and minimal values.

Using the following equations to calculate the PDF error for a fixed value
of $\alpha_{s}$,
\beq
\left(\Delta\sigma_{{\rm PDF}}^{\alpha_{s}}\right)_{+} =\sqrt{\sum_{i}
\left\{\max\left[
 \sigma (\alpha_{s}, S_{i}^{+})-\sigma (\alpha_{s}^0,S_{0}),
 \sigma (\alpha_{s}, S_{i}^{-})-\sigma (\alpha_{s}^0,S_{0}),0 \right] 
 \right\}^2}\, ,
 \label{eq:pdferrorplus} \nonumber \\
\left(\Delta\sigma_{{\rm PDF}}^{\alpha_{s}}\right)_{-} =\sqrt{\sum_{i}
\left\{\max\left[
 \sigma (\alpha_{s}^0,S_{0})-\sigma (\alpha_{s},S_{i}^{+}),
 \sigma (\alpha_{s}^0,S_{0})-\sigma (\alpha_{s},S_{i}^{-}),0 \right]
 \right\}^2}\, ,
 \label{eq:pdferrorminus}
    \eeq
we then compare these five different values and finally arrives, with
$\alpha_{s}^0$ as the best--fit value of $\alpha_{s}$ given by the central 
values of Eq.~\ref{eq:alphas} and $S_0$ the nominal PDF set with this 
$\alpha_s$ value, at the 90\% CL PDF+$\Delta^{\rm exp} \alpha_s$
errors given by~\cite{Martin:2009bu}
\beq
\Delta\sigma^+_{{\rm PDF}+\alpha^{\rm exp}_{s}} = \max_{\alpha_{s}}\left( 
\left\{\sigma (\alpha_{s}^0,S_{0})+
\left(\Delta\sigma_{{\rm PDF}}^{\alpha_{s}}\right)_{+}  \right\}\right)
-\sigma (\alpha_{s}^{0},S_{0}) \, ,
 \label{eq:pdfaserrorplus} \nonumber \\
\Delta\sigma^-_{{\rm PDF}+\alpha^{\rm exp}_{s}} =\sigma (\alpha_{s}^{0},S_{0})-
\min_{\alpha_{s}}\left( \left\{\sigma (\alpha_{s}^0,S_{0})-
\left(\Delta\sigma_{{\rm PDF}}^{\alpha_{s}}\right)_{-}  \right\}\right)\, .
 \label{eq:pdfaserrorminus}
    \eeq

\begin{figure}[!h]
\begin{center}
\vspace*{3mm}
\includegraphics[scale=0.7]{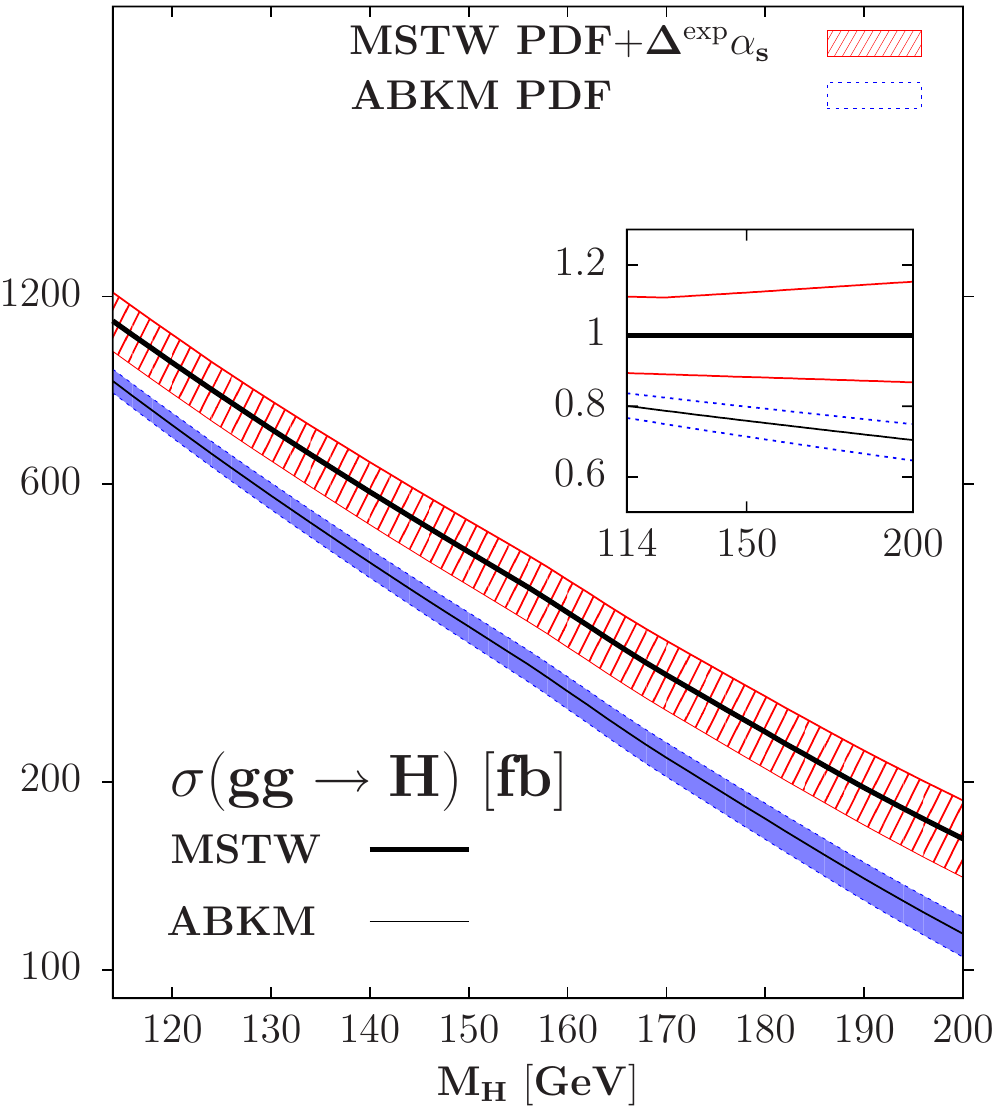}
\includegraphics[scale=0.7]{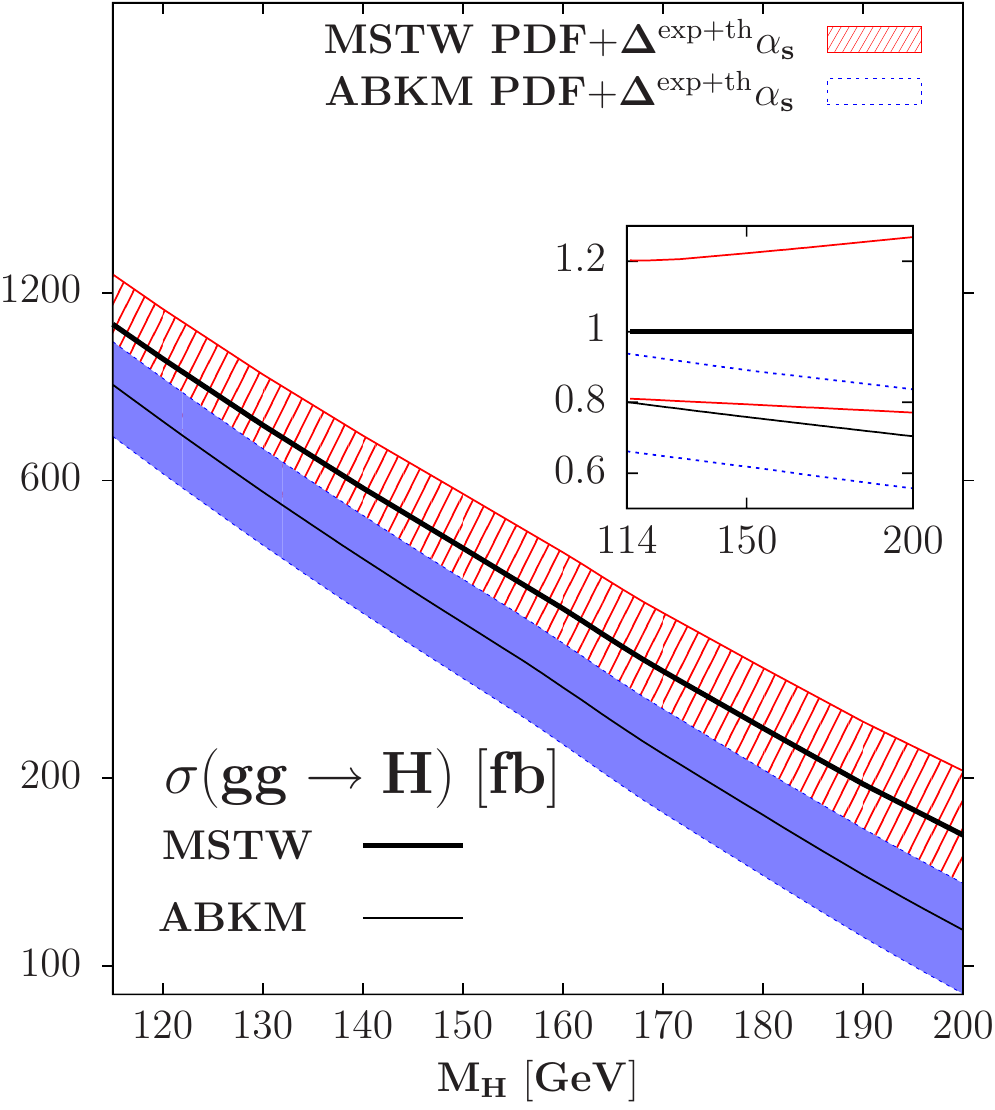}
\end{center}
\vspace*{-3mm}
\caption[Comparison between MSTW PDFs set and ABKM PDFs set
predictions in $gg\to H$ channel at the Tevatron as for the
uncertainties related to 
PDF+$\Delta\alpha_s$]{Left: the PDF+$\Delta^{\rm exp}\alpha_s$ uncertainties 
in the MSTW scheme and the PDF uncertainties in the ABKM schemes on
the $gg \to H$ cross section at the Tevatron as a function of
$M_H$. Right: the  PDF+$\Delta^{\rm exp}\alpha_s +\Delta^{\rm
  th}\alpha_s$ uncertainties in the MSTW scheme using the new set--up
and the PDF+$\Delta^{\rm exp}\alpha_s$+$\Delta^{\rm th}\alpha_s$ error
in the ABKM scheme using a naive procedure. In the inserts, shown are
the same but with the cross sections normalized to the MSTW central
cross section.}
\label{fig:ggHTev_PDF_2}
\end{figure}

Using this procedure, we have evaluated the PDF+$\Delta^{\rm exp} \alpha_s$
uncertainty on the NNLO $gg \to H$ total cross section at the Tevatron and the
result is displayed in the left--hand side of
Fig.~\ref{fig:ggHTev_PDF_2} as a function of $M_H$.
The PDF+$\Delta^{\rm exp} \alpha_s$ error ranges from $\approx \pm 11\%$ for
$M_H=120$ GeV to $ \approx \pm 14\%$ for $M_H=180$ GeV with, again, a slight
asymmetry between the upper and lower values; for a Higgs mass  $M_H=160$ GeV,
one has $\Delta\sigma^\pm_{ {\rm PDF}+\alpha_{s}} /\sigma = ^{+12.8\%}_{-
12.0\%}$.   That is, the experimental uncertainty on $\alpha_s$  adds a
$\approx 5\%$ error to the PDF error alone over the entire $M_H$ range relevant
at the Tevatron.

Nevertheless, this larger PDF+$\Delta^{\rm exp} \alpha_s$ uncertainty compared
to the PDF uncertainty alone does not yet reconcile the evaluation of MSTW and
ABKM (in this last scheme the $\Delta^{\rm exp}\alpha_s$ uncertainty has not
been included since no PDF set with an error on $\alpha_s$ is provided) 
of the $gg\to H$ cross section at the Tevatron, the difference between the
lowest MSTW value and the highest ABKM value being still at the level of
$\approx  10\%$. 

In order to (at least partially) reconcile the MSTW and ABKM/HERA
predictions which differ drastically, we introduce the impact of the
theoretical uncertainy on the value of $\alpha_s$. So far, only the
impact of the experimental errors on $\alpha_s$ has been discussed,
while it is well known that the strong coupling constant is also
plagued by theoretical uncertainties due to scale variation or ambiguities in
heavy quark flavor scheme definition. This theoretical uncertainty has
been estimated in Ref.~\cite{Martin:2009iq} to be at least $\Delta^{\rm th}
\alpha_s= \pm 0.003$ at NLO which turns out to be  $\pm 0.002$ at
NNLO. The estimate of Refs.~\cite{Alekhin:2002fv,Alekhin:2002rk} leads
to a slightly larger uncertainty, $\Delta^{\rm th} \alpha_s= \pm
0.0033$. We will begin by using the MSTW uncertainty, that is
\beq
\Delta^{\rm th}_{\rm NNLO} \alpha_s= 0.002 \,
\label{eq:alphas_th}
\eeq

To calculate in a consistent way the uncertainty due to this
$\alpha_s$ variation, we use the fixed--$\alpha_s$ NNLO PDF
grid also provided by the MSTW collaboration, which is a set of central PDFs
but at  fixed values of $\alpha_s$ different from the best--fit value. Values
of $\alpha_s$ in a range comprised between $0.107$ and $0.127$ in steps of
0.001 are selected, which then include the values $\alpha_s^0\pm
0.002$ that we will use for our estimation. We also point out that
this also allows for a broader range of $\Delta^{\rm th} \alpha_s$
value, that we will use in the following. Using this PDF grid with
the theoretical error on $\alpha_s$ of Eq.~\ref{eq:alphas_th}
implemented, the upper and lower values of the cross sections will be  given by
\beq
\Delta \sigma^{+}_{\rm PDF+\alpha_s^{th}} = \sigma (\alpha_s^{0}+ \Delta^{\rm th} 
\alpha_s, S_0(\alpha_s^{0}+\Delta^{\rm th} \alpha_s)) - \sigma (\alpha_s^0, 
S_0(\alpha_s^{0})) \nonumber \\
\Delta \sigma^-_{\rm PDF+\alpha_s^{ th}} = \sigma (\alpha_s^{0}, S_0(\alpha_s^0 )) 
-\sigma (\alpha_s^{0}- \Delta^{\rm th} \alpha_s, S_0(\alpha_s^{0}- \Delta^{\rm 
th} \alpha_s))
\label{eq:as-naive}
\eeq
with again $S_0(\alpha_s)$ being the MSTW best--fit PDF set at the fixed
$\alpha_s$ value which is either $\alpha_{s}^0$ or $\alpha_{s}^0
\pm\Delta^{\rm  th}\alpha_s$. For the calculation with ABKM set, we
use a naive estimate, that is use the central PDF set at the central
$\alpha_s(M_Z^2)$ but with different $\alpha_S(M_Z^2)$ values, as this
PDFs set does not provide different PDFs set according to a variation
in the value of the strong coupling constant.

Nevertheless with this procedure and the value adopted in
Eq.~\ref{eq:alphas_th} the uncertainty is still far from solving the
issue. Indeed, as pointed in Ref.~\cite{Baglio:2010um}, taking this
additional uncertainty with the value of $\Delta^{\rm th}\alpha_s$ in
Eq.~\ref{eq:alphas_th}  would have reconciled the MSTW/CTEQ and
ABKM/HERA prediction only in the case where $\mu_0=M_H$. When
switching to $\mu_0=\frac12 M_H$, it seems that this increases the
spread between the differents PDFs sets predictions. We will then take
as $\Delta^{\rm th}\alpha_s$ the difference between the strong
coupling constant value in the MSTW set and in the ABKM set. This
gives in the end
\beq
\Delta^{\rm th}_{(2)} \alpha_s(M_Z^2) = 0.004
\label{eq:alphas_th2}
\eeq

Note that despite of the fact that the uncertainty on
$\alpha_s$ is that of a theoretical and is not at the 90\% CL, we will take the
PDF+$\Delta^{\rm th}\alpha_s$ error that one obtains using
Eq.~\ref{eq:as-naive} to be at the 90\% CL. In this way we will be able
to add the uncertainty to the rest: to obtain the total
PDF+$\alpha_s$ uncertainty, we then  
add in quadrature the PDF+$\Delta^{\rm exp}\alpha_s$ and PDF+$\Delta^{\rm th}
\alpha_s$ uncertainties, 
\beq
\Delta \sigma^\pm_{\rm PDF+\alpha_s^{exp}+\alpha_s^{th}}&=&
\left( (\Delta \sigma^\pm_{\rm PDF+\alpha_s^{exp}})^2+
       (\Delta \sigma^\pm_{\rm PDF+\alpha_s^{th} })^2 \right)^{1/2} . 
\label{eq:as-combined}
\eeq

The result for the total PDF+$\alpha_s$ 90\% CL uncertainty on $\sigma^{\rm
NNLO}(gg\to H)$ in the MSTW scheme using the procedure outlined above is
shown in the right--hand side of Fig.~\ref{fig:ggHTev_PDF_2} as a
function of $M_H$ and compared to the result when the PDF error in the
ABKM scheme is combined with the $\Delta^{\rm  exp}\alpha_s$ and
$\Delta^{\rm  th}\alpha_s$ uncertainties using the naive procedure
briefly mentioned previously as, in this case, no PDF with an
$\alpha_s$ value different from that obtained with the best--fit is
provided. Stretching the range of $\Delta^{\rm th}\alpha_s$ to
$\Delta^{\rm th}\alpha_s=0.004$ thus helps to reconcile MSTW and ABKM
predictions as can be seen on the graph, the two uncertainty bands overlaping.

\begin{figure}[!h]
\begin{center}
\vspace*{3mm}
\includegraphics[scale=0.8]{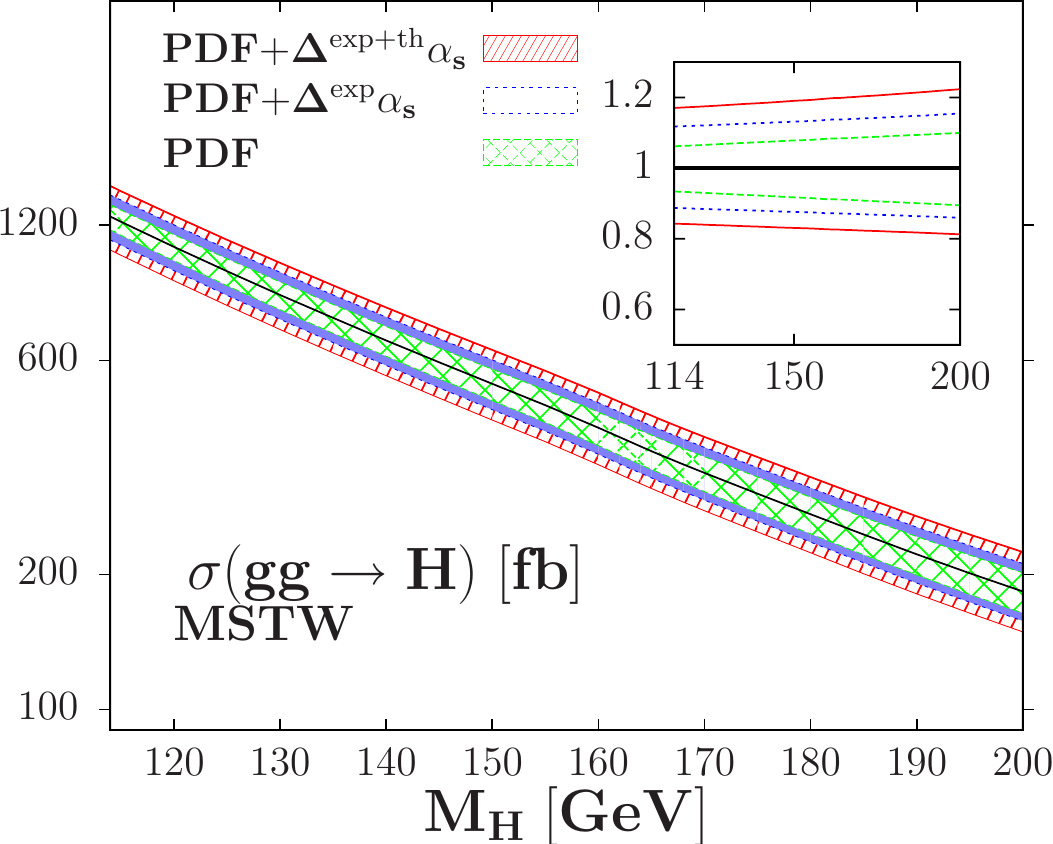}
\end{center}
\vspace*{-3mm}
\caption[The total PDF, PDF+$\Delta^{\rm exp}\alpha_s$ and
PDF+$\Delta^{\rm exp+th}\alpha_s$ uncertainties in $gg\to H$ at the
Tevatron using the MSTW PDFs set.]{The 90\% CL PDF,
  PDF,PDF+$\Delta^{\rm exp}\alpha_s$ and PDF+$\Delta^{\rm
    exp+th}\alpha_s$ uncertainties on $\sigma^{\rm   NNLO}_{gg\to H}$
  in the MSTW parametrisation. In the inserts, shown in \% are the
  deviations with respect to the central MSTW value.}
\label{fig:ggHTeV_PDFfinal}
\end{figure}

We then compare MSTW PDF, PDF+$\Delta^{\rm exp}\alpha_s$ and the total
PDF+$\Delta^{\rm exp+th}\alpha_s$ uncertainties in
Fig.~\ref{fig:ggHTeV_PDFfinal}. The total uncertainty due to the PDF
and theoretical and experimental errors on the strong coupling
constant $\alpha_s$ is now significant, and amounts to $\pm 15\%$ to
$20\%$ on the central cross section depending on
the $M_H$ value. This is still a bit larger than the 12.5\% error
which has been assumed in the most recent CDF/D0 combined analyses
~\cite{Tevatron:2010ar,Aaltonen:2011gs}, and of course even larger
than the $\approx \pm 8\%$ assumed in the earliest Tevatron analysis
~\cite{Tevatron:2009je}) where the strong coupling constant
uncertainties were lacking.  

In any case we face the situation where we have large differences in
the predictions obtained using the various PDFs set on the
market. This is what we call the ``pdf puzzle''; this is also
discussed in Ref.~\cite{Alekhin:2011ey} which analyzes the question of
the discrepency between global fits and DIS fits predictions,
assessing that this can be relied to NMC data handling in the PDFs.

\subsubsection{The case of Higgs--strahlung processes}

In the case of the associated Higgs boson production with a $W$ boson,
$p\bar p\to WH$, we follow the same line of arguments developed above
in the case of the gluon--gluon fusion. We recall that the case of the
associated production with a $Z$ boson is absolutely similar to that
of the $W$ boson.

We thus calculate the PDF, PDF+$\Delta^{\rm exp}\alpha_s$ and
PDF+$\Delta^{\rm exp+th}\alpha_s$ 90\% CL uncertainties in the MSTW
scheme and compare it to what could be obtained using the ABKM PDFs
set. In this case we use the $\Delta^{\rm th}\alpha_s$ value of
Eq.~\ref{eq:alphas_th2} as it is sufficient to reconcile the different
predictions as seen in the results below.

\begin{figure}[!h]
\begin{center}
\includegraphics[scale=0.8]{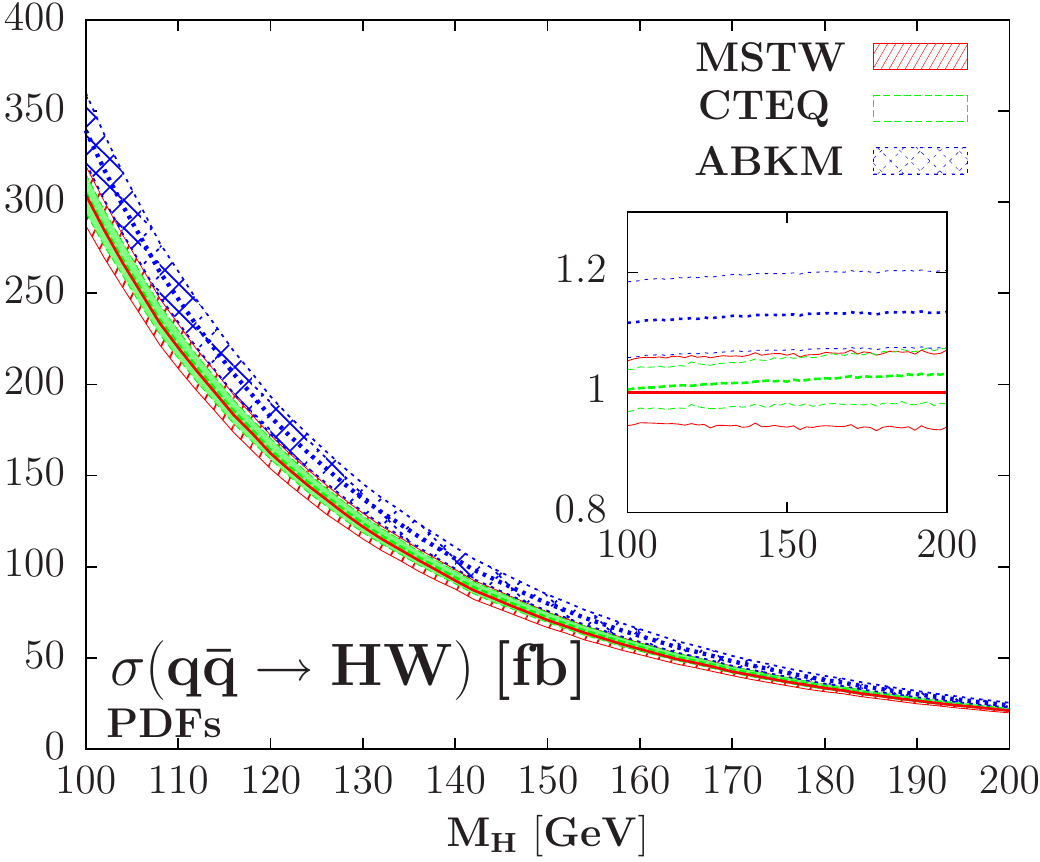}
\end{center}
\vspace*{-2mm}
\caption[Central predictions for NNLO $p\bar p\to WH$ at the Tevatron
using the MSTW, CTEQ and ABKM PDFs sets, together with their 90\% CL
PDF uncertainty]{The central values and the PDF uncertainties in the cross 
section $\sigma( p \bar p \to WH)$ at the Tevatron when evaluated within the 
MSTW, CTEQ and ABKM schemes. In the insert, the relative deviations from the 
central MSTW value are shown.} 
\label{fig:pphvTeV_pdf1}
\vspace*{-2mm}
\end{figure}

\begin{figure}[!h]
\vspace*{5mm}
\begin{center}
\includegraphics[scale=0.7]{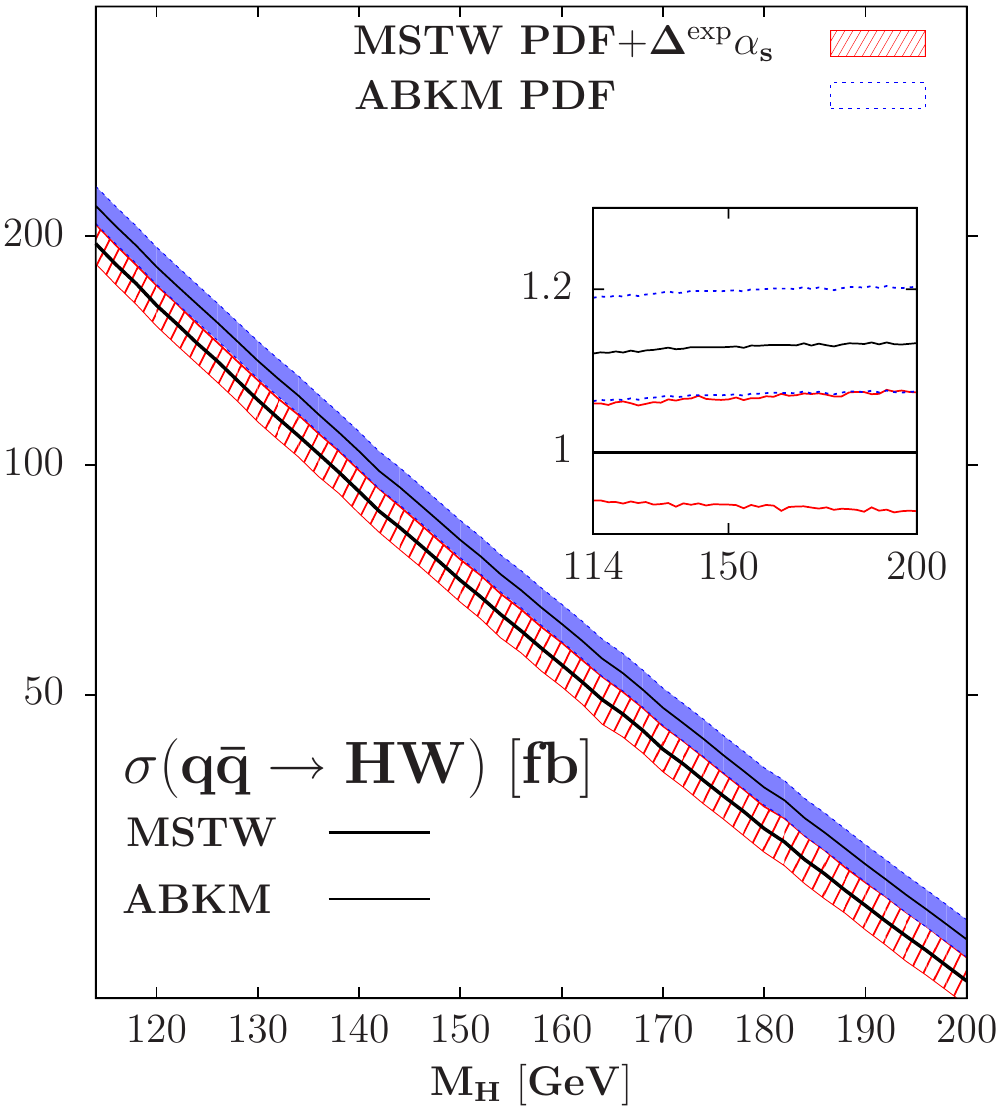}
\includegraphics[scale=0.7]{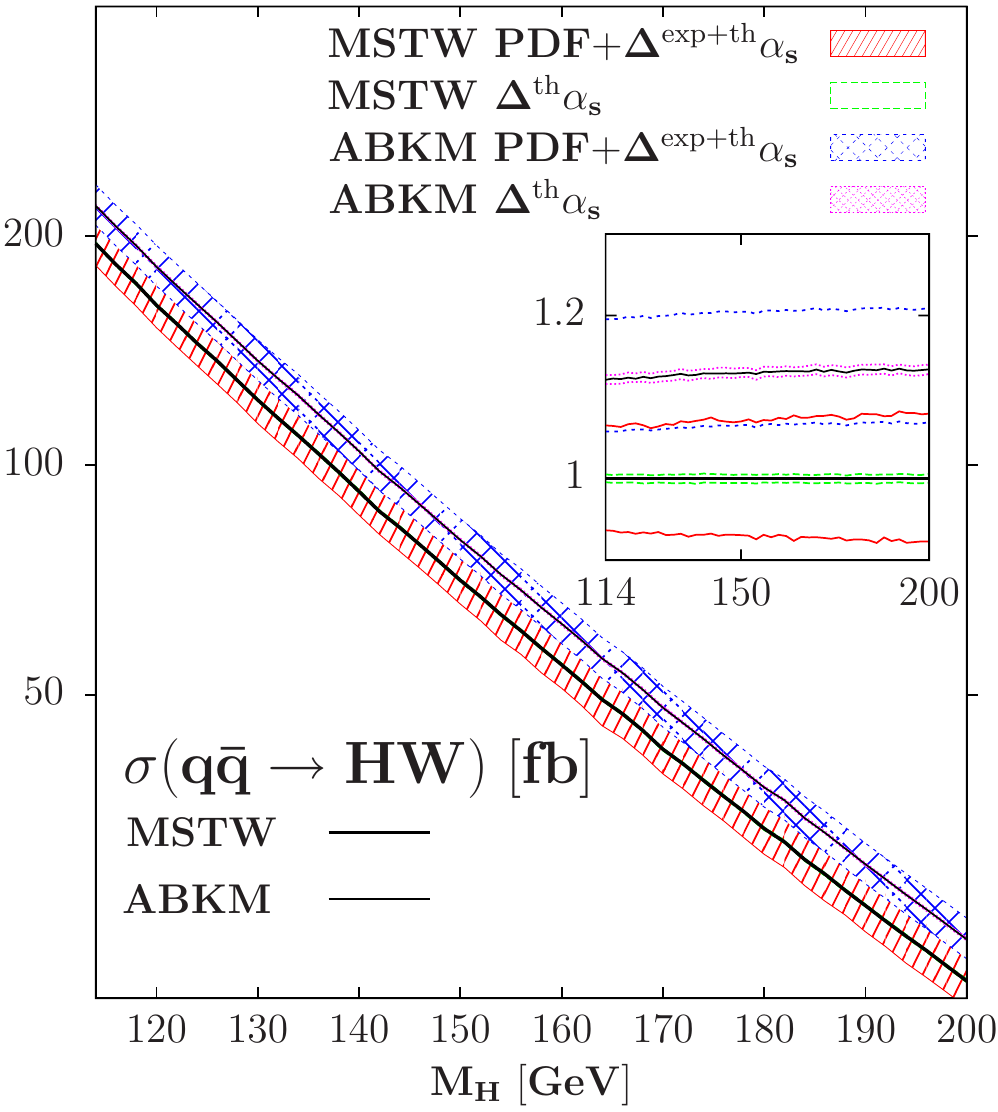}
\end{center}
\vspace*{-5mm}
\caption[Comparison between MSTW PDFs set and ABKM PDFs set
predictions in $p\bar p\to WH$
channel at the Tevatron as for the uncertainties related to
PDF+$\Delta\alpha_s$]{Left: the  PDF uncertainties in the MSTW and ABKM 
schemes when the additional experimental errors on $\alpha_s$ is 
included in MSTW  as discussed in  the text; in the insert, the
relative deviations from the  central MSTW value are shown. Right: the
same as in a) but when the  theoretical error on $\alpha_s$ is added
in both the MSTW and ABKM cases.}
\label{fig:pphvTeV_pdf2}
\vspace*{-2mm}
\end{figure}

 The results are displayed in Fig.~\ref{fig:pphvTeV_pdf1} and
Fig.~\ref{fig:pphvTeV_pdf2} for the Tevatron center--of--mass energy 1.96
TeV and for Higgs mass relevant in the Tevatron SM Higgs searches. The
spread of the MSTW/CTEQ/ABKM schemes is displayed in
Fig.~\ref{fig:pphvTeV_pdf1}. The uncertainty bands are of order $\pm
4\%$ in CTEQ/MSTW schemes and slightly larger in ABKM PDFs
set. The discrepancy between central predictions is of order $10\%$
which is nearly three times less than compared to the gluon--gluon
fusion case. This does not constitute a surprise as at LO the
Higgs--strahlung processes use quark densities which are better known
than gluon density. However, contrary to the gluon--gluon fusion case,
the MSTW/CTEQ and ABKM bands almost touch each other.

In the left--hand side of Fig.~\ref{fig:pphvTeV_pdf2}, we show the
bands resulting from the PDF+$\Delta^{\rm exp} \alpha_s$ uncertainty
in the MSTW mixed scheme, while the right--hand side of the figure
shows the uncertainty bands when the
additional theoretical error $\Delta^{\rm th} \alpha_s$ is included in both
the MSTW scheme using Eq.~\ref{eq:as-combined} and ABKM scheme using the 
naive estimate and Eq.~\ref{eq:as-naive}. As expected, the errors due to the
imprecise value of $\alpha_s$ are much smaller than in the $gg \to H$
mechanism, as in Higgs--strahlung, the process does not involve $\alpha_s$ in
the Born approximation and the $K$--factors are reasonably small, $K_{\rm NNLO}
\lsim 1.5$. Hence, $\Delta^{\rm exp}\alpha_s$ generates an additional error
that is about $\approx  2\%$ when included in the PDF fits, while the error
due to $\Delta^{\rm th}\alpha_s$ is about one to two percent.

In any case, the total PDF+$\Delta^{\rm exp}\alpha_s$+$\Delta^{\rm
th}\alpha_s$ uncertainty is at the level of $\approx 7$--8\% in the MSTW
scheme, that is slightly larger than the errors due to the PDFs alone, and
we again have a significant overlap between the MSTW and ABKM
uncertainty band but with the use of MSTW theoretical error estimate
of $\Delta^{\rm th}\alpha_s=0.002$.\bigskip

To conclude this subsection we would like to insist on the fact that
the procedure that we have used is only one particular way of handling
the PDF puzzle which is still currently pending and on active
investigation by theorists, especially in the view of the LHC
predictions. In particular we could also have used the difference
between the central predictions given by the different NNLO PDFs set
on the market, that is the MSTW, ABKM, HERA and JR sets. The PDF
uncertainty, when compared to the central MSTW prediction, would have
ended being of order $-30\%,+0\%$ in the critical range $155\leq
M_H\leq 180$ GeV, as seen in the right--hand side of
Fig.~\ref{fig:ggHTeV_pdf1} .

We also note that there is another recipe that has been suggested by the
PDF4LHC working group for evaluating PDF uncertainties for NNLO cross sections
(besides taking the envelope of the predicted values obtained using several PDF
sets)~\cite{Thorne:2010hh}:  take the 68\% CL MSTW PDF+$\Delta^{\rm
  exp} \alpha_s$ error and multiply it by a factor of
two~\cite{Dittmaier:2011ti}. In our case, this would lead to an
uncertainty of $\approx \pm 25\%$ which, for the minimal value, is
close to the recipe discussed just above, and is 
larger than what we obtain when considering the
PDF+$\Delta^{\rm exp+th} \alpha_s$ uncertainty given by MSTW. This
gives some credits to our way of handling the PDF issue.

\subsection[EFT and its uncertainties]{Effective field theory and its
  uncertainties \label{section:SMHiggsTevEFT}}

This subsection is intended to cover a specific issue in the
gluon--gluon fusion production channel and will not concern the
Higgs--strahlung mechanism. Indeed, while both the QCD and electroweak
radiative corrections to the process $gg \to H$ have been calculated
exactly at NLO taking into account both the top quark and the bottom quarks
loop with the exact finite mass of both quark flavours, these
corrections are derived at NNLO only in an effective field theory
(EFT) approach in which the loop particles are assumed to be very
massive, $m \gg M_H$, and integrated out. Thus these corrections have
only been calculated for the top loop and currently there are no NNLO
corrections for the $b$--quark loop available.

\subsubsection{The $b$--loop uncertainty}

At the Born level, taking into account only the dominant contribution
of the top quark loop and  working in the limit $m_t \to \infty$
provides an approximation~\cite{Spira:1995rr,Spira:1997dg} that is
only good at the 10\% level for Higgs masses below the $t\bar t$
kinematical threshold, $M_H \lsim 350$ GeV. The difference from the
exact result is mainly due to the absence of the contribution of the
$b$--quark loop: although the $b$--quark mass is small, the $gg \to H$
amplitude exhibits a dependence  $\propto m_b^2/M_H^2 \times \ln^2
(m_b^2/M_H^2)$ which, for relatively low values of the Higgs mass,
generates a non--negligible contribution that interferes destructively
with the dominant top--quark loop contribution. In turn, when
considering only the top quark loop in the $Hgg$ amplitude, the
approximation $m_t  \to \infty$ is extremely good for Higgs masses
below $2m_t$, compared to the amplitude with the exact top quark mass
dependence as shown in
Refs.~\cite{Harlander:2002wh,Anastasiou:2002yz,Ravindran:2003um}.

In the NLO approximation for the QCD radiative corrections, it has
been shown ~\cite{Spira:1995rr} that the exact $K$--factor when the full
dependence on the top and bottom quark masses is taken into account,
$K^{\rm exact}_{\rm NLO}$, is smaller than the $K$ factor obtained in
the approximation in which only the top quark contribution is included
and the asymptotic limit $m_t \to \infty$ is taken,
$K^{m_t\!\to\!\infty}_{\rm NLO}$. The reason is that when only the
$b$--quark loop contribution is considered in the $Hgg$ amplitude (as
in the case of supersymmetric theories in which the $b$--quark Yukawa
coupling is strongly enhanced compared to its Standard Model
value~\cite{Djouadi:2005gj}, see section~\ref{section:MSSMHiggsTev}),
the $K$--factor for the $gg\to H$ cross section at the Tevatron is
about $K \sim 1.2$ to  1.5, instead of $K \sim 2.4$ when only the top
quark is included in the loop. The approximation of infinite loop
particle mass significantly improves when the full $t,b$ mass
dependence is included in the LO order cross section and $\sigma_{\rm
  NLO}^{m_t\! \to \! \infty} = K^{m_t\!\to\! \infty}_{\rm NLO} \times
\sigma_{\rm LO} (m_t,m_b)$ gets closer to the cross section
$\sigma^{\rm exact}_{\rm NLO}$ in which the exact $m_t, m_b$
dependence is taken into account.

\begin{figure}[!t]
\begin{center}
\includegraphics[scale=0.8]{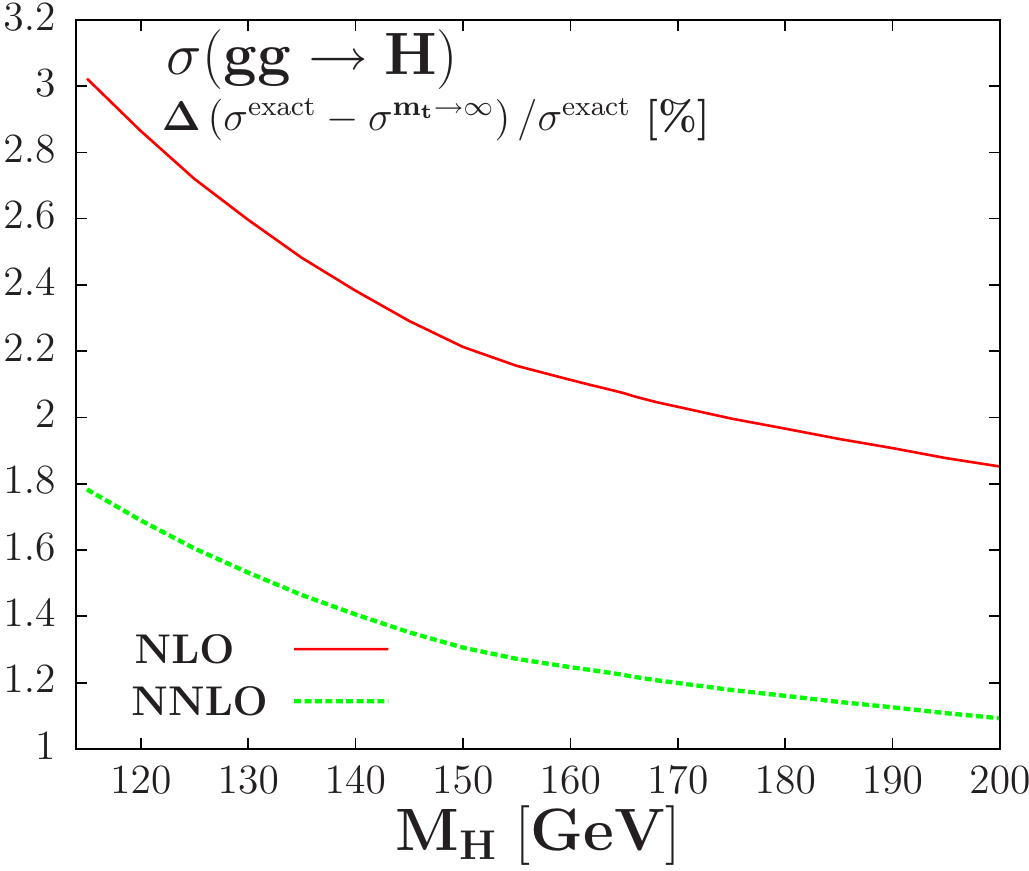}
\end{center}
\vspace*{-5mm}
\caption[$b$--loop uncertainty in $gg\to H$ at the Tevatron]{Relative
  difference (in \%) at Tevatron energies and as a function 
of $M_H$ between the exact NLO and NNLO $gg\to H$ cross sections
$\sigma^{\rm exact}_{\rm NLO/NNLO}$ and the cross section in the
effective approach with an infinite top quark mass $\sigma^{m_t \! \to
  \! \infty}_{\rm NLO/NNLO}$.}
\vspace*{-2mm}
\label{fig:ggHTeV_bloop}
\end{figure}

 The difference between $\sigma^{\rm exact}_{\rm NLO}$ and $\sigma^{m_t \! \to
\! \infty}_{\rm NLO}$ at Tevatron energies is shown in
Fig.~\ref{fig:ggHTeV_bloop} as a function of 
the Higgs mass we show that there is a few percent discrepancy between
the two cross sections. As mentioned previously, in the Higgs mass range 115
GeV$\lsim M_H \lsim 200$ GeV relevant at Tevatron energies, this difference is
only due to the absence of the $b$--quark loop contribution and its
interference with the top quark loop in the $Hgg$ amplitude and not to the
fact that the limit $m_t \gg M_H$ is taken. 

At NNLO, because of the complexity of the calculation, only the result in the
effective approach in which the loop particle masses are assumed to be infinite
is available. In the case of the NNLO QCD
corrections~\cite{Harlander:2002wh,Anastasiou:2002yz,Ravindran:2003um},
the $b$--quark loop contribution and its interference with the
contribution of $t$--quark loop is therefore missing. Since the NNLO
correction increases the cross section by $\sim 30\%$, the question
wether this missing piece might lead to an overestimate of the total
$K$--factor should be asked. We will assume that it might be indeed
the case and assign an error on the NNLO QCD result  which is
approximately the difference between the exact result $\sigma^{\rm
  exact}_{\rm NLO}$ and the approximate result $\sigma^{m_t \! \to \!
  \infty}_{\rm NLO}$ obtained at NLO while taking the exact LO result
and shown in Fig.~\ref{fig:ggHTeV_bloop}, but rescaled with the
relative magnitude of the $K$--factors that is obtained at NLO and
NNLO, i.e. $K_{\rm  NLO}^{m_t \to \infty}/K_{\rm NNLO}^{m_t \to
  \infty}$. This leads to an uncertainty on the NNLO cross section
which ranges from $\sim \pm 2\%$  for low Higgs values $M_H  \sim 120$
GeV at which the $b$--quark loop contribution is significant at LO, to
the level of $\sim \pm 1\%$ for Higgs masses above $M_H \sim 180$ GeV
for which the $b$--quark loop contribution is much smaller. This
uncertainty is not really taken into account in the scale uncertainty,
as we are not only missing the $b$--loop contribution but also, and
mainly if we compare to what happens at NLO, the (likely negative)
top--bottom interference at NNLO which can be significant. This
interference term cannot be really well estimated through NNLO scale
variation as the latter will mainly estimate missing N$^3$LO terms.

In addition we should also assign to the $b$--quark contribution an error
originating from the freedom in choosing the input value of the $b$--quark mass
in the loop amplitude and the scheme in which it is defined\footnote{We thank
M. Spira for reminding us of this point.}. Indeed, apart from the
difference obtained when using the $b$--quark pole mass, $M_b^{\rm
pole}\approx 4.7$ GeV, as we have done here, or the running $\overline{ \rm MS}$
mass evaluated at the scale of the $b$--quark mass,  $\bar m_b^{\rm MS} (M_b)
\sim 4.2$ GeV, there is an additional $\frac43 \frac{\alpha_s}{\pi}$ factor
which enters the cross section when switching from the on--shell to the
$\overline{\rm MS}$ scheme. This leads to an error of approximately 1\% on the
total cross section, over the $M_H$ range that is relevant at the Tevatron. In
contrast, according to very recent
calculations~\cite{Harlander:2009mq, Pak:2009dg, Harlander:2009my,
  Marzani:2008az}, the $m_t \to \infty$ limit is a very good
approximation for the top--quark loop contribution to $\sigma(gg \to
H)$ at NNLO as the higher order terms, when expanding the amplitude in
power series of $M_H^2/(4m_t^2)$, lead to a difference that is smaller
than one percent for $M_H \lsim  300$ GeV.

\subsubsection{The electroweak uncertainty}

The other part of the calculation that is concerned by an effective
theory approach deals with the electroweak corrections in $gg\to H$
production channel. As mentioned previously, while the ${\cal
O}(\alpha)$ NLO corrections have been calculated with the exact dependence on
the loop particle masses~\cite{Actis:2008ts}, the mixed QCD--electroweak
corrections due to light quark loops at ${\cal O}(\alpha \alpha_s)$ have been
evaluated~\cite{Anastasiou:2008tj} in the effective theory approach where the $W,Z$
bosons have been integrated out and which is only valid for $M_H \ll M_W$.
These contributions are approximately equal to the difference between the exact
NLO electroweak corrections when evaluated in the complete factorization and
partial factorization schemes~\cite{Anastasiou:2008tj}. 

\begin{figure}[!t]
\begin{center}
\vspace*{3mm}
\includegraphics[scale=0.8]{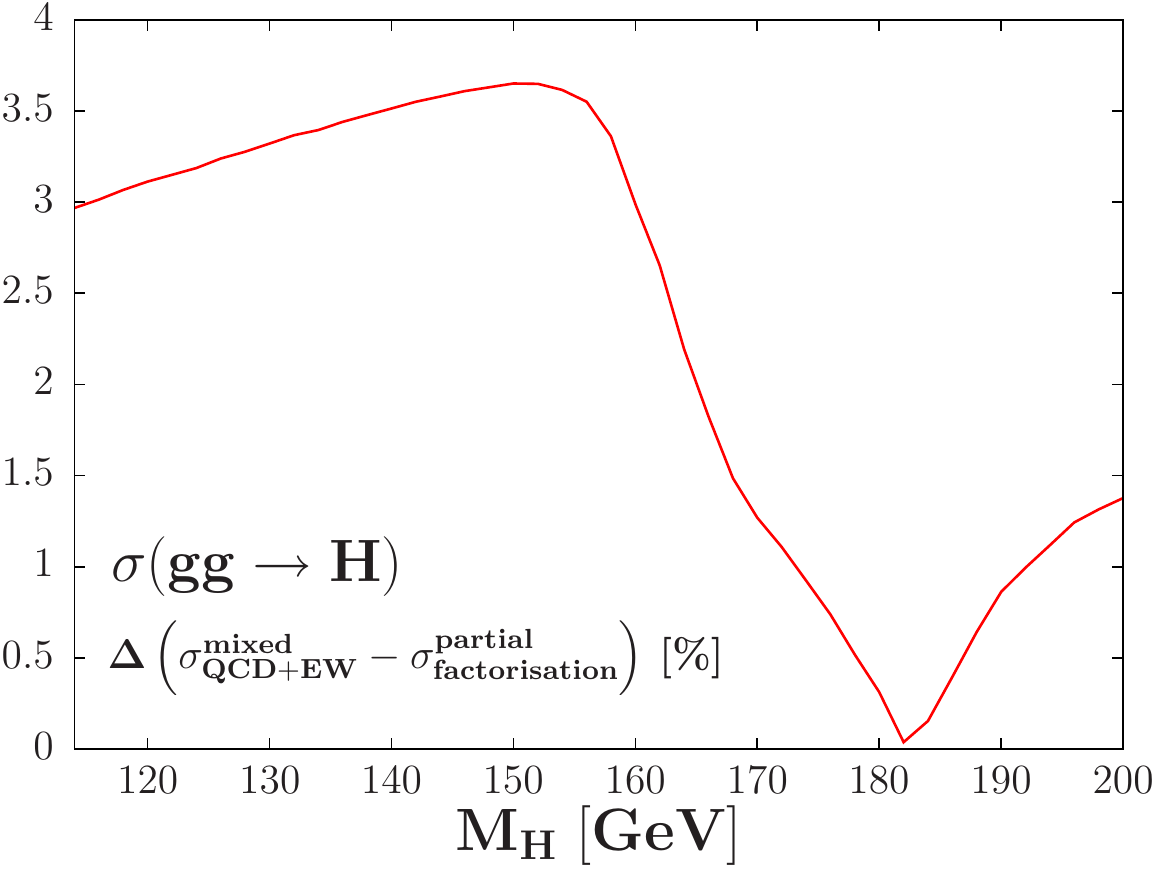}
\end{center}
\vspace*{-5mm}
\caption[EW uncertainties in $gg\to H$ at the Tevatron]{Relative
  difference (in \%) between the complete factorization and partial
  factorization approaches for the electroweak radiative corrections
  to the NLO $gg \to H$ cross section at the Tevatron  as a function of $M_H$.}
\vspace*{-2mm}
\label{fig:ggHTeV_EW}
\end{figure}

However, as the results for the mixed corrections are only valid at most for
$M_H < M_W$ and given the fact that the $\delta_{\rm EW}$
electroweak correction at ${\cal O}(\alpha)$ exhibits a completely different
behavior below and above the $2M_W$ threshold\footnote{Indeed, the NLO
electroweak correction $\delta_{EW}$  of Ref.~\cite{Actis:2008ts} is positive
below the $WW$ threshold $M_H \lsim 2M_W$ for which the effective approach is
valid in this case and turns to negative for $M_H \gsim 2M_Z$ for which the
effective approach cannot be applied and the amplitude develops imaginary 
parts. This behavior can also be seen in Fig.~\ref{fig:ggHTeV_EW}
which, up to the overall 
normalisation, is to a very good approximation  the $\delta_{\rm EW}$
correction factor given in  Fig.~1 of Ref.~\cite{Actis:2008ts} for $M_H \lsim
2M_Z$.},  one should be cautious and assign an uncertainty to this mixed
QCD--electroweak correction, even if this may overestimate the
uncertainty due to EFT effects. Conservatively, we then have chosen in
Ref.~\cite{Baglio:2010um} to assign an error that is of the same size as the
${\cal O}(\alpha \alpha_s)$ contribution itself. This is equivalent to
assigning an error to the full  ${\cal O}(\alpha)$ contribution that
amounts to the difference between the correction obtained in the
complete factorization and partial factorization schemes as done in
Ref.~\cite{Actis:2008ts}. As pointed out in the latter reference, this
reduces to adopting the usual and well--established procedure that has
been used at LEP for attributing uncertainties due to unknown higher
order effects. We then obtain an uncertainty ranging  from 1.5\% to
3.5\%  for Higgs masses below $M_H \lsim 2M_W$ and below 1.5\% for
larger Higgs masses as is shown in Fig.~\ref{fig:ggHTeV_EW}.

Finally, we should note that we do not address here the issue of the threshold
effects from virtual $W$ and $Z$ bosons which lead to spurious spikes in the
${\cal O}(\alpha)$ electroweak correction in the mass range $M_H=160$--190 GeV
which includes the Higgs mass domain that is most relevant at the Tevatron (the
same problem occurs in the case of the $p\bar p \to HV$ cross sections once the
electroweak corrections are included). These singularities are smoothened by
including the finite widths of the $W/Z$ bosons, a procedure which might
introduce potential additional theoretical ambiguities that has been
ignored in the analysis of Ref.~\cite{Baglio:2010um} upon which is
constructed a large part of this thesis.

\subsection{Combination and total
  uncertainty \label{section:SMHiggsTevTotal}}

We present in this subsection the tables which summarize the
individual sources of theoretical uncertainties, and we discuss how to
combine them to obtain the overall theoretical uncertainty. We start
with the gluon--gluon fusion process and then turn our attention to
the Higgs--strahlung channels.

The central cross section in the $gg \to H$  process at the Tevatron,
as well as the various associated theoretical uncertainties, are
summarized in Table~\ref{table:ggHTeV}. For a set of Higgs mass values
that is relevant at the Tevatron (we choose a step of 5 GeV as done by
the CDF and D0
experiments~\cite{Tevatron:2009je,Aaltonen:2010yv,Tevatron:2010ar,Aaltonen:2011gs}
except in the critical range 160--170 GeV where a 2 GeV step 
has been adopted), the second column of the table gives the central values of the
total cross section at NNLO (in fb) for the renormalization and factorization
scale choice $\mu_R=\mu_F=\frac12 M_H$,  when the partonic cross sections are folded
with the MSTW parton densities. The following columns give the errors on the
central value of the cross section originating from the various sources
discussed in
subsection~\ref{section:SMHiggsTevScale}/\ref{section:SMHiggsTevPDF}/\ref{section:SMHiggsTevEFT},
namely, the uncertainties due to the scale variation in the adopted
range $\frac13 \mu_0 \le  \mu_R, \mu_F \le 3 \mu_0$ with
$\mu_0=\frac12 M_H$, the 90\% CL errors due to the MSTW PDF,
PDF+$\Delta^{\rm  exp}\alpha_s$ and PDF+$\Delta^{\rm
  exp}\alpha_s$+$\Delta^{\rm th}\alpha_s$ uncertainties as well as
the estimated uncertainties from the use of the effective approach in
the calculation of the NNLO QCD (the $b$--quark loop contribution and its
interference with the top--quark loop) and electroweak (difference between the
complete and partial factorization approaches) radiative corrections. 

The largest of these errors, $\sim +15,-20\%$, is due to the scale variation,
followed by the PDF+$\Delta^{\rm exp} \alpha_s +\Delta^{\rm th}\alpha_s$
uncertainties which are at the level of $\approx 15\%$; the errors due to the
effective theory approach (including that due to the definition of the
$b$--quark mass) are much smaller, being of the order of a few percent for both
the QCD and electroweak parts.

The very important issue which remains to be solved is how to combine
these various uncertainties. We present here the procedure that was
developed in Ref.~\cite{Baglio:2010um} and then used for the following
analyses at the LHC and in the MSSM (see the following sections). In
accord with Refs.~\cite{Anastasiou:2009bt,Dittmaier:2011ti}, there is
currently a vivid debate whether to 
add these errors in quadrature as done, for instance, by the CDF and D0
collaborations\footnote{In earlier analyses, the CDF
  collaboration~\cite{Aaltonen:2010cm,Bernardi:2008ee} adds in
  quadrature the 10.9\% scale uncertainty obtained at NNLL with a
  scale variation in the range $\frac12 M_H \le  \mu_R, \mu_F \le 2
  M_H$ with a 5.1\% uncertainty due the errors on the MSTW PDFs (not
  including the errors from $\alpha_s$), resulting in a 12\% total
  uncertainty. The D0
  collaboration~\cite{Abazov:2010ct,Bernardi:2008ee}  assigns an even
  smaller total error, 10\%, to the production cross section. This has
  changed a lot in the most recent analyses where they quote an
  approximate 20\% total uncertainty using PDF+$\alpha_s$
  uncertainties and jet--analysis for scale uncertainties which
  amounts to $\simeq
  18\%$.}~\cite{Aaltonen:2010cm,Abazov:2010ct,Bernardi:2008ee} or not,
and we believe that quadratic addition should be the accurate
procedure when dealing with theoretical uncertainties.

To begin with, the uncertainties associated to the PDF
parameterisations should be viewed as theoretical errors and they have
been considered as such since a long time. Indeed, although the PDF
sets use various experimental data which have intrinsic errors which
then induce ``might-be'' statistical grounds, the main uncertainty is
due to the theoretical assumptions which go into the different
parameterizations. This uncertainty cannot be easily quantified within
one given parametrization but it is reflected in the spread of the
central values given by the various PDF parameterizations that are
available. If the PDF uncertainty is defined as to be the difference in the
cross sections when using the different available PDF sets, as
recommanded for example by the PDF4LHC recommendation~\cite{PDF4LHC},
this uncertainty has no statistical ground.

For the scale uncertainty, the situation is of course clear: it has no
statistical ground at all and any value of the cross section in the
uncertainty band is as respectable as another\footnote{In statistical
  language, both the scale and PDF uncertainties have a flat
  prior. See Refs.~\cite{Baglio:2011wn, Dittmaier:2011ti} for a more elaborated
  discussion on this subject. It is worth mentioning that
  Ref.~\cite{Dittmaier:2011ti} recommands a linear combination.}. The same is also
true for the uncertainties due to the use of an effective approach:
they should be viewed as purely theoretical uncertainties and thus be
addded linearly with each others.

As a result, the scale and PDF uncertainties, cannot be combined in
quadrature as done, for instance, by the CDF and D0 collaborations.  This is
especially true as in the $gg\to H$ process, a strong correlation between the 
renormalization and factorization scales that are involved (and that we have
equated here for simplicity, $\mu_R\!=\!\mu_F$), the value of $\alpha_s$ and
the $gg$ densities is present. For instance, decreasing (increasing)  the 
scales will increase (decrease) the $gg \to H$ cross section not only because
of the lower (higher)  $\alpha_s (\mu_R^2)$ value that is obtained and which
decreases (increases) the magnitude of the matrix element squared (that is
proportional to $\alpha_s^2$ at leading order  and  the cross section is
minimal/maximal for the highest/lowest $\mu_R=\mu_F$ values), but also
because
at the same time,  the $gg$ densities become smaller (larger) for higher
(smaller) $\mu_F\!=\!\sqrt{Q^2}$ values. Some of these issues have
been discussed in details in Ref.~\cite{Thorne:2010hh}. On the other hand, if
we add everything linearly it might appear too conservative and also
because of the aforementioned reasons.

We then have proposed a rather simple procedure to combine at least
the two largest uncertainties, those due to the scale variation and to
the PDF+$\alpha_s$ uncertainties that may avoid the drawbacks of the
two other possibilities mentioned above and have been presented for
the first time in Ref.~\cite{Baglio:2010um}. We review this procedure below.

One first derives the maximal and
minimal values of the production cross sections when the
renormalization and factorization scales are varied in the adopted
domain, that is, $\sigma_0 \pm \Delta \sigma^\pm_\mu$ with  $\sigma_0$
being the cross section evaluated for
the central scales $\mu_R=\mu_F=\mu_0$ and the deviations  $\Delta
\sigma^\pm_\mu$  given in Eq.~\ref{eq:scaleminus}. We then evaluate, on
these maximal and 
minimal cross sections from scale variation, the PDF+$\Delta^{\rm exp}
\alpha_s$ as well as the PDF+$\Delta^{\rm th} \alpha_s$ uncertainties (combined
in quadrature) using the new MSTW set-up, i.e  as in Eq.~(\ref{eq:as-combined})
but with $\sigma_0$ replaced by $\sigma_0 \pm \Delta \sigma_\mu^\pm$.  

\begin{table}[!h]
{\small%
\let\lbr\{\def\{{\char'173}%
\let\rbr\}\def\}{\char'175}%
\renewcommand{\arraystretch}{1.20}
\vspace*{2mm}
\begin{center}
\begin{tabular}{|c|c||c|ccc|cc||cc|}\hline
$~~M_H~~$ & $\sigma^{\rm NNLO}_{\rm gg \to H}$ [fb]&~~scale~~&PDF& PDF+$
\alpha_s^{\rm exp}$&$\alpha_s^{\rm th}$& EW & b--loop & total & \% total \\ \hline
$100$ & 1849 & $^{+318}_{-371}$ & $^{+102}_{-109}$ & $^{+210}_{-201}$ &
$^{+219}_{-199}$ & $^{+45}_{-45}$ & $^{+42}_{-42}$ & $^{+817}_{-648}$ & $^{+44.2\%}_{-35.0\%}$ \\ \hline
$105$ & 1603 & $^{+262}_{-320}$ & $^{+91}_{-98}$ & $^{+184}_{-176}$ &
$^{+192}_{-174}$ & $^{+41}_{-41}$ & $^{+39}_{-39}$ & $^{+700}_{-565}$ & $^{+43.7\%}_{-35.3\%}$ \\ \hline
$110$ & 1397 & $^{+219}_{-277}$ & $^{+83}_{-89}$ & $^{+163}_{-156}$ &
$^{+170}_{-152}$ & $^{+37}_{-37}$ & $^{+35}_{-35}$ & $^{+602}_{-496}$ & $^{+43.1\%}_{-35.5\%}$ \\ \hline
$115$ & 1222 & $^{+183}_{-242}$ & $^{+75}_{-81}$ & $^{+144}_{-138}$ &
$^{+151}_{-134}$ & $^{+33}_{-33}$ & $^{+32}_{-32}$ & $^{+521}_{-437}$ & $^{+42.6\%}_{-35.7\%}$ \\ \hline
$120$ & 1074 & $^{+156}_{-211}$ & $^{+69}_{-73}$ & $^{+129}_{-123}$ &
$^{+135}_{-119}$ & $^{+30}_{-30}$ & $^{+29}_{-29}$ & $^{+454}_{-386}$ & $^{+42.2\%}_{-36.0\%}$ \\ \hline
$125$ & 948 & $^{+134}_{-186}$ & $^{+63}_{-67}$ & $^{+115}_{-110}$ &
$^{+121}_{-106}$ & $^{+28}_{-28}$ & $^{+24}_{-24}$ & $^{+397}_{-342}$ & $^{+41.9\%}_{-36.1\%}$ \\ \hline
$130$ & 839 & $^{+115}_{-164}$ & $^{+57}_{-61}$ & $^{+104}_{-99}$ &
$^{+108}_{-94}$ & $^{+25}_{-25}$ & $^{+21}_{-21}$ & $^{+349}_{-304}$ & $^{+41.5\%}_{-36.2\%}$ \\ \hline
$135$ & 746 & $^{+100}_{-145}$ & $^{+53}_{-56}$ & $^{+94}_{-89}$ &
$^{+98}_{-84}$ & $^{+23}_{-23}$ & $^{+18}_{-18}$ & $^{+309}_{-272}$ & $^{+41.4\%}_{-36.5\%}$ \\ \hline
$140$ & 665 & $^{+88}_{-129}$ & $^{+48}_{-51}$ & $^{+85}_{-80}$ &
$^{+88}_{-76}$ & $^{+21}_{-21}$ & $^{+16}_{-16}$ & $^{+275}_{-243}$ & $^{+41.4\%}_{-36.6\%}$ \\ \hline
$145$ & 594 & $^{+78}_{-115}$ & $^{+45}_{-47}$ & $^{+77}_{-73}$ &
$^{+80}_{-68}$ & $^{+19}_{-19}$ & $^{+14}_{-14}$ & $^{+246}_{-218}$ & $^{+41.4\%}_{-36.8\%}$ \\ \hline
$150$ & 532 & $^{+69}_{-103}$ & $^{+41}_{-44}$ & $^{+70}_{-66}$ &
$^{+73}_{-61}$ & $^{+17}_{-17}$ & $^{+13}_{-13}$ & $^{+221}_{-197}$ & $^{+41.6\%}_{-37.0\%}$ \\ \hline
$155$ & 477 & $^{+61}_{-92}$ & $^{+38}_{-40}$ & $^{+64}_{-60}$ &
$^{+67}_{-55}$ & $^{+15}_{-15}$ & $^{+10}_{-10}$ & $^{+198}_{-176}$ & $^{+41.5\%}_{-37.0\%}$ \\ \hline
$160$ & 425 & $^{+54}_{-82}$ & $^{+35}_{-37}$ & $^{+58}_{-54}$ &
$^{+60}_{-50}$ & $^{+11}_{-11}$ & $^{+9}_{-9}$ & $^{+175}_{-155}$ & $^{+41.3\%}_{-36.6\%}$ \\ \hline
$162$ & 405 & $^{+51}_{-78}$ & $^{+33}_{-35}$ & $^{+56}_{-52}$ &
$^{+58}_{-48}$ & $^{+9}_{-9}$ & $^{+8}_{-8}$ & $^{+166}_{-146}$ & $^{+40.9\%}_{-36.2\%}$ \\ \hline
$164$ & 386 & $^{+48}_{-75}$ & $^{+32}_{-34}$ & $^{+53}_{-50}$ &
$^{+55}_{-45}$ & $^{+8}_{-8}$ & $^{+8}_{-8}$ & $^{+158}_{-139}$ & $^{+40.9\%}_{-36.0\%}$ \\ \hline
$165$ & 377 & $^{+47}_{-73}$ & $^{+31}_{-33}$ & $^{+52}_{-48}$ &
$^{+54}_{-44}$ & $^{+7}_{-7}$ & $^{+8}_{-8}$ & $^{+154}_{-135}$ & $^{+40.8\%}_{-35.9\%}$ \\ \hline
$166$ & 368 & $^{+46}_{-71}$ & $^{+31}_{-33}$ & $^{+51}_{-47}$ &
$^{+53}_{-44}$ & $^{+6}_{-6}$ & $^{+8}_{-8}$ & $^{+150}_{-132}$ & $^{+40.9\%}_{-35.8\%}$ \\ \hline
$168$ & 352 & $^{+44}_{-68}$ & $^{+30}_{-31}$ & $^{+49}_{-46}$ &
$^{+51}_{-42}$ & $^{+5}_{-5}$ & $^{+8}_{-8}$ & $^{+144}_{-126}$ & $^{+40.9\%}_{-35.7\%}$ \\ \hline
$170$ & 337 & $^{+42}_{-65}$ & $^{+29}_{-30}$ & $^{+47}_{-44}$ &
$^{+49}_{-40}$ & $^{+4}_{-4}$ & $^{+7}_{-7}$ & $^{+137}_{-119}$ & $^{+40.6\%}_{-35.4\%}$ \\ \hline
$175$ & 303 & $^{+37}_{-59}$ & $^{+26}_{-28}$ & $^{+43}_{-40}$ &
$^{+45}_{-36}$ & $^{+2}_{-2}$ & $^{+6}_{-6}$ & $^{+122}_{-106}$ & $^{+40.4\%}_{-35.1\%}$ \\ \hline
$180$ & 273 & $^{+33}_{-53}$ & $^{+24}_{-26}$ & $^{+39}_{-36}$ &
$^{+41}_{-33}$ & $^{+1}_{-1}$ & $^{+6}_{-6}$ & $^{+111}_{-95}$ & $^{+40.6\%}_{-34.9\%}$ \\ \hline
$185$ & 245 & $^{+30}_{-47}$ & $^{+22}_{-24}$ & $^{+36}_{-33}$ &
$^{+38}_{-30}$ & $^{+1}_{-1}$ & $^{+6}_{-6}$ & $^{+101}_{-87}$ & $^{+41.1\%}_{-35.3\%}$ \\ \hline
$190$ & 222 & $^{+27}_{-43}$ & $^{+21}_{-22}$ & $^{+33}_{-30}$ &
$^{+35}_{-27}$ & $^{+2}_{-2}$ & $^{+5}_{-5}$ & $^{+92}_{-79}$ & $^{+41.4\%}_{-35.7\%}$ \\ \hline
$195$ & 201 & $^{+24}_{-39}$ & $^{+19}_{-20}$ & $^{+31}_{-28}$ &
$^{+32}_{-25}$ & $^{+2}_{-2}$ & $^{+3}_{-3}$ & $^{+83}_{-72}$ & $^{+41.4\%}_{-35.8\%}$ \\ \hline
$200$ & 183 & $^{+22}_{-35}$ & $^{+18}_{-19}$ & $^{+28}_{-26}$ &
$^{+30}_{-23}$ & $^{+2}_{-2}$ & $^{+3}_{-3}$ & $^{+77}_{-67}$ & $^{+42.0\%}_{-36.3\%}$ \\ \hline
\end{tabular} 
\end{center} 
\caption[The NNLO total Higgs production cross sections in the
$\protect{gg\to H}$  process at the Tevatron together with the
detailed theoretical uncertainties as well as the total
uncertainty]{The NNLO total Higgs production cross sections in the
  $\protect{gg\to H}$ process at the Tevatron (in fb) for given Higgs
  mass values (in GeV) at a central scale $\mu_F=\mu_R=\frac12
  M_H$. Shown also are the corresponding shifts due to the theoretical
  uncertainties from the various sources discussed, as well as the
  total uncertainty when all errors are added using the  procedure
  described in the text.}
\label{table:ggHTeV}
\vspace*{-2mm}
}
\end{table}

 We then obtain the maximal and minimal values of the cross
section when scale, PDF and $\alpha_s$ (both experimental and theoretical)
uncertainties are included, 
\beq 
\sigma_{\rm max}^{\rm \mu+PDF+\alpha_s} &=& (\sigma_0 + \Delta
\sigma^+_\mu)+ \Delta  (\sigma_0 + \Delta \sigma^+_\mu)^+_{\rm PDF+
 \alpha_s^{\rm exp}+ \alpha_s^{\rm th}} 
\, , \nonumber \\
\sigma_{\rm min}^{\rm \mu+PDF+\alpha_s} &=& (\sigma_0 - \Delta \sigma^-_\mu)-
\Delta  (\sigma_0 - \Delta \sigma^-_\mu)^-_{\rm PDF+ \alpha_s^{\rm exp}
+ \alpha_s^{\rm th}}   \, . 
\label{eq:combined}
\eeq 

To these new maximal and minimal cross sections, we finally add in the
case of gluon--gluon fusion the smaller EFT uncertainty, that is the
one resulting from the missing $b$--loop at NNLO and the small EW
uncertainties. This last addition is linear as these errors have no
correlations with the scale and PDF uncertainties\footnote{In
the case of the $b$--loop contribution, the $K$--factor when  varying the scale
from the central value $\frac12 M_H$ to the values $\approx \frac16 M_H$ or $\approx
\frac32 M_H$.which  maximise and minimise the cross section, might be slightly
different and thus, the error will not be exactly that given in Table~\ref{table:ggHTeV}.
However, since the entire effect is very small, we will ignore this tiny
complication here.}. 

The two last columns of Table~\ref{table:ggHTeV} display the maximal
and minimal deviations of the $gg \to H$ cross section at the Tevatron
when all uncertainties are added, as 
well as the percentage deviations of the cross section from the central
value. We should note that the actual PDF+$\alpha_s$ uncertainty and
the uncertainty from the use of the  effective theory approach are
different from those of Table~\ref{table:ggHTeV}, which are given for
the best value of the cross section, obtained for the central scale
choice $\mu_F=\mu_R=\frac12 M_H$; nevertheless, the relative or 
percentage errors are approximately the same for $\sigma_0$ and $\sigma_0 \pm
\Delta \sigma^\pm_\mu$.

\begin{figure}[!h]
\vspace*{1mm}
\begin{center}
\includegraphics[scale=0.8]{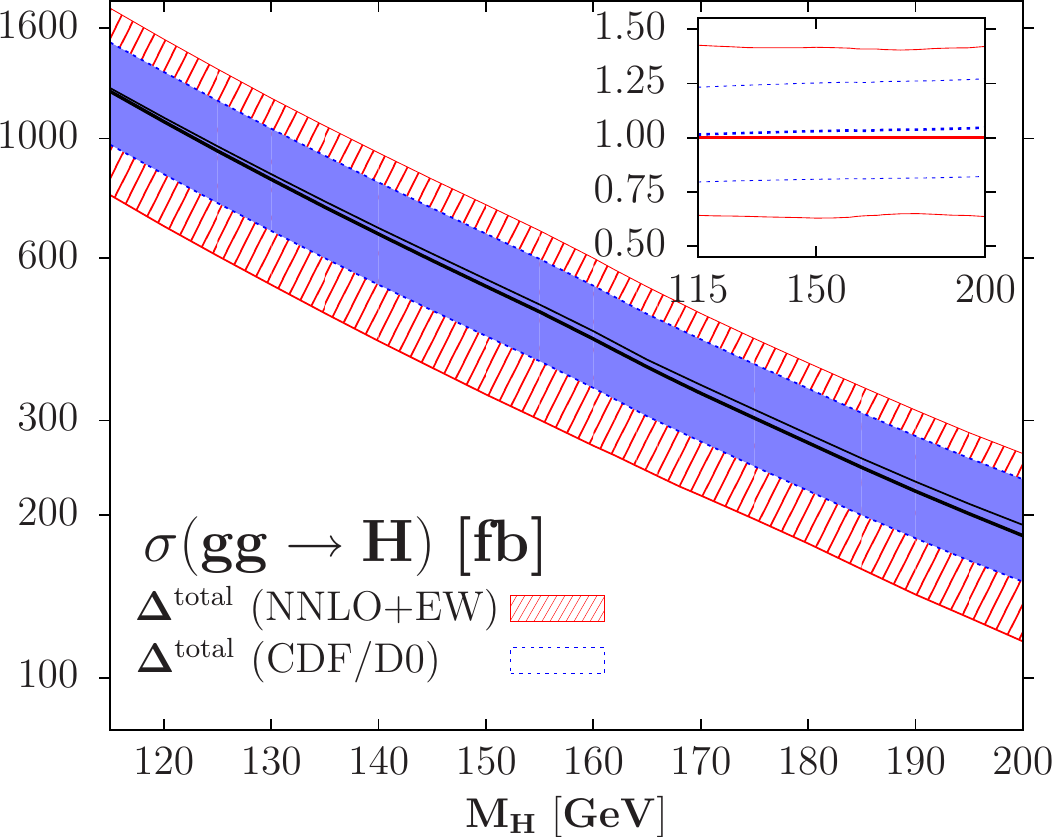}
\end{center}
\vspace*{-5mm}
\caption[Production cross sections for $gg\to H$ at the Tevatron
together with the total theoretical uncertainties]{The production
  cross section $\sigma(gg\to H)$ at NNLO for the QCD 
and NLO for the electroweak corrections at the Tevatron at a central scale 
$\mu_F=\mu_R= \frac12 M_H$ with the uncertainty band when all theoretical
uncertainties are added using our procedure. It is compared to $\sigma(gg 
\to H)$ at NNLL~\cite{deFlorian:2009hc} with the errors quoted by the CDF/D0 
collaboration~\cite{Aaltonen:2010yv,Tevatron:2010ar,Aaltonen:2011gs}. In
the insert, the relative deviations compared to the central value are shown.} 
\label{fig:ggHtotalTeV}
\vspace*{-2mm}
\end{figure}

Doing so for the $gg\to H$ NNLO cross section with a central scale
$\mu_0=\frac12 M_H$, we obtain the total error shown in Fig.~\ref{fig:ggHtotalTeV},
that we compare to the $ \approx \pm 22\%$ error assumed in the CDF/D0
analysis. For $M_H= 160$ GeV for example, we obtain $\Delta \sigma/\sigma
\approx +41\%, - 37\%$, that is a spread from the central value
$\sigma_0=425$ fb which is up to $\sigma^{\rm max}=600$ fb and down to
$\sigma^{\rm min}=270$ fb. This is also shown in Table~\ref{table:ggHTeV}
in the last columns where all theoretical uncertainties are
combined. This large variation of approximately $\pm 40\%$ is seen on
the entire Higgs mass range. We should also note that the central
cross section displayed in Fig.~\ref{fig:ggHtotalTeV} is in excellent
agreement with Ref.~\cite{deFlorian:2009hc}; the striking new result is then the
large uncertainties that we have found.

Hence, our procedure for the combination does not reduce to a linear sum of all
uncertainties. If we had added linearly all errors, we would have 
had, for the negative part at $M_H=160$ GeV,  a total uncertainty of $\Delta
\sigma/\sigma \approx - 42\%$, compared to  the value $- 37\%$ with our
procedure. On the other hand, one has an error of $\approx - 30\%$, i.e. close
to the total error assumed by CDF/D0 if the scale and PDF+$\alpha_s$
uncertainties were added in quadrature and the EFT approach error linearly 
(the latter being ignored by the CDF/D0 collaborations). We also want
to point out that the Ref.~\cite{Dittmaier:2011ti} in its last section
recommands a procedure of combination which is linear, as the prior on
the different theoretical uncertainties are not gaussian. Thus our
combination is far from being too conservative or exaggerated.

\begin{table}[!h]
\small%
\let\lbr\{\def\{{\char'173}%
\let\rbr\}\def\}{\char'175}%
\renewcommand{\arraystretch}{1.30}
\begin{center}
\begin{tabular}{|c|cc||c|ccc|cc|}\hline 
$~M_H~$ & $~\sigma_{HW}$  & $\sigma_{HZ}$~ &~~~scale~~~&PDF&PDF+$\alpha_s
^{\rm exp}$ & $~~\alpha_s^{\rm th}~~$ &  total & \% total \\ \hline 
$115$ & 174.5 & 103.9 & $^{+1.3}_{-1.6}$ & $^{+10.5}_{-9.1}$ & $^{+10.7}_{-10.7}$ &
$^{+1.3}_{-0.9}$ & $^{+12.1}_{-12.3}$ & $^{+7\%}_{-7\%}$ \\  \hline 
$120$ & 150.1 & 90.2 & $^{+1.1}_{-1.4}$ & $^{+9.2}_{-8.1}$ & $^{+9.6}_{-9.4}$ &
$^{+1.2}_{-0.9}$ & $^{+10.7}_{-10.9}$ & $^{+7\%}_{-7\%}$ \\  \hline
$125$ & 129.5 & 78.5 & $^{+0.9}_{-1.3}$ & $^{+7.5}_{-6.8}$ & $^{+8.6}_{-8.7}$ &
$^{+1.1}_{-0.8}$ & $^{+9.6}_{-10.0}$ & $^{+7\%}_{-8\%}$ \\  \hline  
$130$ & 112.0 & 68.5 & $^{+0.8}_{-1.1}$ & $^{+6.8}_{-6.4}$ & $^{+7.2}_{-7.5}$ &
$^{+1.1}_{-0.8}$ & $^{+8.0}_{-8.6}$ & $^{+7\%}_{-8\%}$ \\  \hline 
$135$ & 97.2 & 60.0 & $^{+0.7}_{-1.0}$ & $^{+5.6}_{-5.5}$ & $^{+6.7}_{-6.6}$ &
$^{+1.0}_{-0.7}$ & $^{+7.4}_{-7.6}$ & $^{+8\%}_{-8\%}$ \\  \hline 
$140$ & 84.6 & 52.7 & $^{+0.6}_{-0.9}$ & $^{+5.6}_{-4.5}$ & $^{+5.8}_{-5.7}$ &
$^{+0.9}_{-0.7}$ & $^{+6.5}_{-6.6}$ & $^{+8\%}_{-8\%}$ \\  \hline 
$145$ & 73.7 & 46.3 & $^{+0.5}_{-0.8}$ & $^{+4.4}_{-4.1}$ & $^{+5.4}_{-5.2}$ &
$^{+0.9}_{-0.7}$ & $^{+5.9}_{-6.0}$ & $^{+8\%}_{-8\%}$ \\  \hline 
$150$ & 64.4 & 40.8 & $^{+0.5}_{-0.7}$ & $^{+4.2}_{-3.9}$ & $^{+4.4}_{-4.3}$ &
$^{+0.8}_{-0.6}$ & $^{+5.0}_{-5.0}$ & $^{+8\%}_{-8\%}$ \\  \hline 
$155$ & 56.2 & 35.9 & $^{+0.4}_{-0.6}$ & $^{+3.4}_{-3.1}$ & $^{+4.2}_{-4.1}$ &
$^{+0.7}_{-0.6}$ & $^{+4.6}_{-4.7}$ & $^{+8\%}_{-8\%}$ \\  \hline 
$160$ & 48.5 & 31.4 & $^{+0.4}_{-0.6}$ & $^{+3.3}_{-3.0}$ & $^{+3.6}_{-3.3}$ &
$^{+0.7}_{-0.5}$ & $^{+4.1}_{-4.0}$ & $^{+8\%}_{-8\%}$ \\  \hline 
$162$ & 47.0 & 30.6 & $^{+0.4}_{-0.5}$ & $^{+3.4}_{-2.8}$ & $^{+3.5}_{-3.3}$ &
$^{+0.7}_{-0.5}$ & $^{+3.9}_{-3.8}$ & $^{+8\%}_{-8\%}$ \\  \hline 
$164$ & 44.7 & 29.1 & $^{+0.3}_{-0.5}$ & $^{+3.1}_{-2.7}$ & $^{+3.4}_{-3.4}$ &
$^{+0.6}_{-0.5}$ & $^{+3.7}_{-3.9}$ & $^{+8\%}_{-9\%}$ \\  \hline 
$165$ & 43.6 & 28.4 & $^{+0.3}_{-0.5}$ & $^{+2.8}_{-2.4}$ & $^{+3.4}_{-3.3}$ &
$^{+0.6}_{-0.5}$ & $^{+3.8}_{-3.8}$ & $^{+8\%}_{-8\%}$ \\  \hline 
$166$ & 42.5 & 27.8 & $^{+0.3}_{-0.5}$ & $^{+3.0}_{-2.6}$ & $^{+3.1}_{-3.0}$ &
$^{+0.6}_{-0.5}$ & $^{+3.4}_{-3.5}$ & $^{+8\%}_{-8\%}$ \\  \hline 
$168$ & 40.4 & 26.5 & $^{+0.3}_{-0.5}$ & $^{+2.8}_{-2.4}$ & $^{+3.1}_{-2.9}$ &
$^{+0.6}_{-0.5}$ & $^{+3.4}_{-3.4}$ & $^{+9\%}_{-8\%}$ \\  \hline 
$170$ & 38.5 & 25.3 & $^{+0.3}_{-0.4}$ & $^{+2.9}_{-2.2}$ & $^{+3.0}_{-2.7}$ &
$^{+0.6}_{-0.5}$ & $^{+3.3}_{-3.1}$ & $^{+9\%}_{-8\%}$ \\  \hline 
$175$ & 34.0 & 22.5 & $^{+0.3}_{-0.4}$ & $^{+2.2}_{-1.9}$ & $^{+2.7}_{-2.6}$ &
$^{+0.5}_{-0.4}$ & $^{+3.0}_{-3.0}$ & $^{+9\%}_{-9\%}$ \\  \hline 
$180$ & 30.1 & 20.0 & $^{+0.2}_{-0.4}$ & $^{+2.1}_{-1.8}$ & $^{+2.2}_{-2.2}$ &
$^{+0.5}_{-0.4}$ & $^{+2.5}_{-2.6}$ & $^{+8\%}_{-9\%}$ \\  \hline 
$185$ & 26.9 & 17.9 & $^{+0.2}_{-0.3}$ & $^{+1.8}_{-1.5}$ & $^{+2.1}_{-2.1}$ &
$^{+0.5}_{-0.4}$ & $^{+2.3}_{-2.4}$ & $^{+9\%}_{-9\%}$ \\  \hline 
$190$ & 24.0 & 16.1 & $^{+0.2}_{-0.3}$ & $^{+1.6}_{-1.6}$ & $^{+1.8}_{-1.8}$ &
$^{+0.4}_{-0.3}$ & $^{+2.1}_{-2.1}$ & $^{+9\%}_{-9\%}$ \\  \hline 
$195$ & 21.4 & 14.4 & $^{+0.2}_{-0.3}$ & $^{+1.3}_{-1.2}$ & $^{+1.8}_{-1.7}$ &
$^{+0.4}_{-0.3}$ & $^{+2.1}_{-2.0}$ & $^{+10\%}_{-10\%}$ \\  \hline 
$200$ & 19.1 & 13.0 & $^{+0.2}_{-0.2}$ & $^{+1.4}_{-1.2}$ & $^{+1.5}_{-1.4}$ &
$^{+0.4}_{-0.3}$ & $^{+1.8}_{-1.7}$ & $^{+9\%}_{-9\%}$ \\  \hline 
\end{tabular} 
\end{center}
\vspace*{1mm}
\caption[The NNLO total cross section for Higgs--strahlung processes
at the Tevatron together with the detailed theoretical uncertainties
and the total uncertainty]{The central values of the cross sections
  for the $p\bar p\to WH$ and $ZH$ processes at the Tevatron  (in fb)
  for given Higgs mass values (in  GeV) with, in the case of the $WH$
  channel, the uncertainties from scale variation, PDF,
  PDF+$\Delta^{\rm exp} \alpha_s$ and $\Delta^{\rm th}\alpha_s$, as
  well as the total uncertainty when all errors are added using the
  procedure described in the text.}
\vspace*{-1mm}
\label{table:ppHV}
\end{table}

The consequences on the exclusion limits quoted by CDF and D0
collaborations on the Higgs mass range between 158 and 173 GeV
~\cite{Aaltonen:2010yv,Tevatron:2010ar,Aaltonen:2011gs} will be
discussed in the section~\ref{section:SMHiggsTevExclusion}. We can
already tell than they should be taken with caution in the light of
our uncertainties.\bigskip

We now discuss the results in the Higgs--strahlung processes and
similarly to the $gg \to H$ case we display in Table~\ref{table:ppHV}
the central values of the cross sections for $p \bar p \to WH$ and $p
\bar p \to ZH$ at the Tevatron, evaluated at scales
$\mu_R=\mu_F=M_{HV}$ with the MSTW set of PDFs (second and third
columns). In the remaining columns, we display only the $WH$ channel
results about the uncertainties coming from the scale variation (with
$\kappa=2$), the PDF, mixed PDF+$\Delta^{\rm exp} \alpha_s$ and
PDF+$\Delta^{\rm exp} \alpha_s$+$\Delta^{\rm th} \alpha_s$
uncertainties in the MSTW scheme. In the last columns, we 
give the total error and its percentage; this percentage error is, to a very
good approximation, the same in the $p \bar p \to ZH$ channel which
explains why we have chosen to study the $W$ boson channel only, recalling
that the result for the $Z$ boson channel are in a very good
approximation obtained by rescaling the absolute error by the correct
central value.

 In contrast to the $gg\to H$ mechanism, since the errors due to scale
 variation are rather moderate in this case, there is no large
 difference between the central cross section $\sigma_0$ and the
 cross sections $\sigma_0 \pm \Delta \sigma^\pm_\mu$ and, hence, the
 PDF, PDF+$\Delta^{\rm exp} \alpha_s$ and PDF+$\Delta^{\rm exp}
 \alpha_s$+$\Delta^ {\rm th} \alpha_s$ errors on $\sigma^0$ are, to
 a good  approximation, the same as the errors on $\sigma_0 \pm \Delta
\sigma^\pm_\mu$ displayed in Table~\ref{table:ppHV}.  

\begin{figure}[h]
\vspace*{-1mm}
\begin{center}
\includegraphics[scale=0.8]{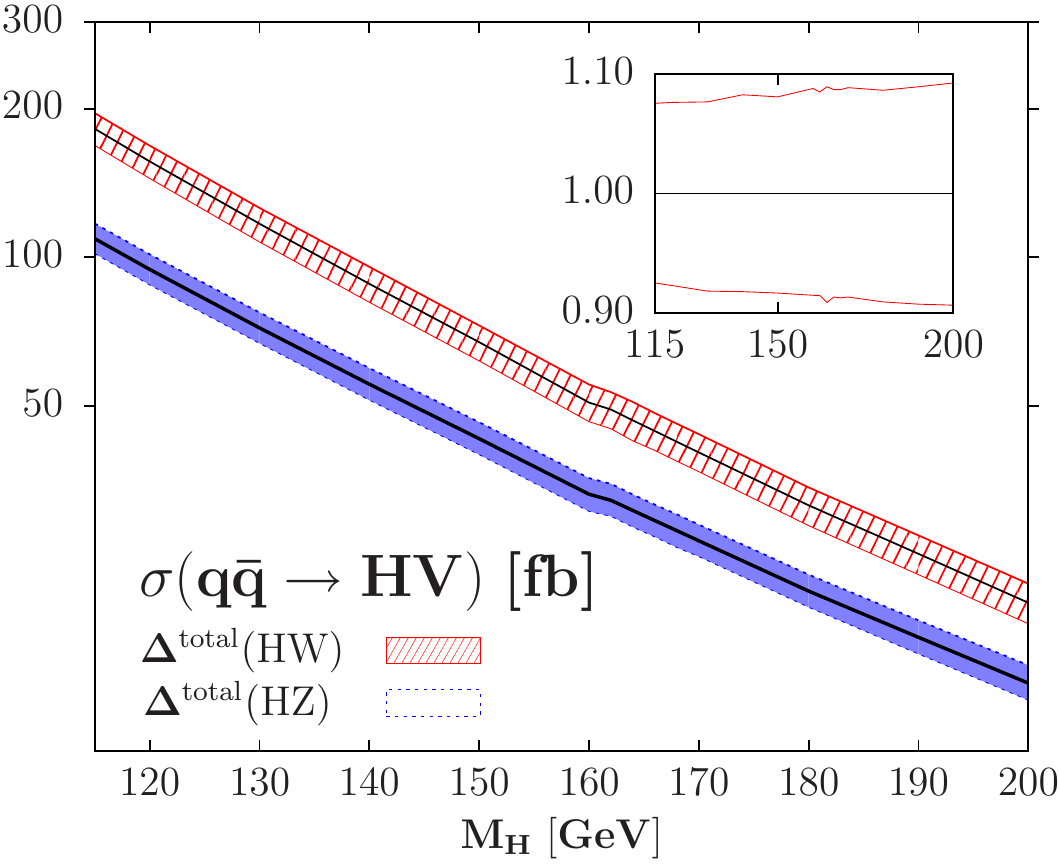}
\end{center}
\vspace*{-6mm}
\caption[Production cross sections for $p\bar p\to WH$ and $p\bar p\to
ZH$ at the Tevatron together with the total theoretical uncertainties]
{The production cross section $\sigma(p \bar p \to WH)$ and 
$\sigma(p \bar p \to ZH)$ at  NNLO in QCD and electroweak NLO at the Tevatron
evaluated with the MSTW set  of PDFs, together with the uncertainty bands when
all the theoretical errors  are added. In the insert, the relative deviations
from the central MSTW value are shown in the case of $\sigma(p\bar p \to WH)$.} 
\vspace*{-1mm}
\label{fig:pphvTotal}
\end{figure}

 The total uncertainty is once more summarized in
 Fig.~\ref{fig:pphvTotal}, where the cross
sections for $WH$ and  $ZH$ associated production at the Tevatron, together
with the total uncertainty bands (in absolute values in the main frame and in
percentage in the insert), are displayed as a function of the Higgs
mass. The total error on the cross sections in the Higgs--strahlung 
processes is about $\pm 9\%$ in the entire Higgs mass range, possibly 1\% to
2\% smaller for low $M_H$ values and $\sim 1\%$ larger for high $M_H$
values, as displayed in the figure. Thus, the theoretical errors are
much smaller than in the case of the $gg \to H$ process and the cross
sections for the Higgs--strahlung processes are well under
control. Nevertheless, it is worth mentionning that the total
uncertainty that is obtained in our analysis is almost twice as large
as the total 5\% uncertainty assumed by the CDF and D0 
collaborations in their combined analysis of this channel
~\cite{Tevatron:2009je,Aaltonen:2010yv,Tevatron:2010ar,Aaltonen:2011gs}.

Before closing this section, we point out that the uncertainties in the 
Higgs--strahlung processes can be significantly reduced by using the
Drell--Yan processes of massive gauge boson production as standard
candles, as first suggested in  Ref.~\cite{Dittmar:1997md}. Indeed,
normalizing the cross sections of associated $WH$ and $ZH$ production
to the cross sections of single $W$ and $Z$ production, respectively,
allows for a cancellation of several experimental errors such as the
error on the luminosity measurement, as well as the partial
cancellation (since the scales that are involved in the $p\bar p \to
V$ and $HV$  processes are different) theoretical errors such as those
due to the PDFs, $\alpha_s$  and the higher order radiative
corrections.

\subsection{Summary and outlook} 

In this section were presented the theoretical predictions for the
Standard Model Higgs boson production total cross section at the
Tevatron in the two main channels,
the gluon--gluon fusion $gg\to H$ and the Higgs--strahlung processes
$p\bar p\to VH$, where $V$ stands for either the $W$ or the $Z$
boson. We have assumed a central scale $\mu_F=\mu_R=\mu_0$ to be
$\mu_0=\frac12 M_H$ for the gluon--gluon fusion process and
$\mu_0=M_{HV}$ (the invariant mass of the pair $HV$) for the
Higgs--strahlung processes, and calculated the cross section up to
NNLO in QCD and EW corrections.

We have then estimated the theoretical uncertainties associated to the
prediction: the scale uncertainty, the uncertainties from the PDF
parametrisation and the associated error on $\alpha_s$, as well as
uncertainties due to the use of the EFT approach for the mixed
QCD-electroweak radiative corrections and the $b$-quark loop 
contribution in the gluon--gluon fusion case. The major result is that
we obtain in the critical region for Higgs searches at the Tevatron,
that is the region $M_H=158$--173 GeV, an overall theoretical
uncertainty of order $\pm 40\%$ when using our procedure of
combination of the uncertainties. That is twice the uncertainty used
in the experimental analyses of the CDF/D0 collaborations
~\cite{Aaltonen:2010yv,Tevatron:2010ar,Aaltonen:2011gs}. This is a mere
consequence of the different ways to combine the individual scale and
PDF+$\alpha_s$ uncertainties and, to a lesser extent, the impact on the
theoretical uncertainty on $\alpha_s$ and the EFT uncertainties which have not
been considered by the CDF/D0 collaborations. The uncertainty obtained
in the Higgs--strahlung mechanism is of order $\pm 8\%$, which implies
that this process is much more under control than the gluon--gluon
fusion mechanism. The uncertainty is also twice the one assumed in low
Higgs mass searches at the
Tevatron~\cite{Aaltonen:2010yv,Tevatron:2010ar,Aaltonen:2011gs}.

The consequence of this doubling of the errors will be analyzed in
section~\ref{section:SMHiggsTevExclusion}. Before that we now turn our
attention on the Higgs production at the other hadron colliders which
has started to run in 2010 in the TeV energy range: the CERN Large
Hadron Collider (LHC). Indeed, it is of crucial importance to analyze
the theoretical uncertainties affecting the Higgs production at the
LHC as it is the designed machine to discover the Higgs boson:
theoretical uncertainties have an impact not only on the discovery of
the elusive boson but also on its couplings to fermions and gauge
bosons~\cite{Zeppenfeld:2000td, Duhrssen:2004cv, Anastasiou:2005pd}.

\pagebreak

\begin{subappendices}
\subsection[Appendix: analytical expressions for $\mu_R$--NNLO terms
in $gg\to H$]{Appendix: analytical terms for NNLO $\mu_R$--dependent
  corrections to $gg\to H$ total cross section \label{ggH:appendix}}

We present in this appendix the exact analytical expressions for the
NNLO corrections which include $\mu_R$ terms. Even if this expressions
can be recovered in a straightforward way, they are not presented in
the current literature and it may be of use to present them in their
all extent.

We write $\displaystyle \hat{\sigma}_{ij} = \sigma_{0}
\alpha_{s}^2(\mu_{R})\left(\eta_{ij}^{(0)}+\frac{\alpha_{s}(\mu_{R})}{\pi}\eta_{ij}^{(1)}+\left(\frac{\alpha_{s}
      (\mu_{R})}{\pi}\right)^2 \eta_{ij}^{(2)}\right)$ with
$$\displaystyle \sigma_{0} = \frac{G_F}{288 \pi \sqrt{2}} \left| \frac 3 4 \sum_q
  A(\tau_q) \right|^2$$
 and $z$ is the variable
defined as $\displaystyle z=\frac{\tau_H}{\tau}$. We separate the
expression of $\hat{\sigma}_{gg}$ in two pieces as done in the literature:
the soft+virtual part containing the virtual corrections as well as
the soft--gluon terms with $\D_i$ plus--distributions, and the hard
part containing the polylogarithms. We then have:\bigskip

\begin{itemize}
\item{For $\hat{\sigma}_{ij}$ with
    $ij=qq,\bar{q}\bar{q},qq',q\bar{q'},\bar{q}q',\bar{q}\bar{q'}$
    there are no $\displaystyle \frac{\mu_{R}}{\mu_{F}}$--terms in
    $\eta_{ij}^{(2)}$.}

\item{For $\hat{\sigma}_{q\bar{q}}$ we have $\delta \eta_{q\bar q}^{(1)} = 0$ and :
\beq
\delta\eta_{q\bar{q}}^{(2)} =
(33-2n_{f})\ln\left(\frac{\mu_{R}^2}{\mu_{F}^2}\right)\frac{8}{27}(1-z)^3
\eeq}
\item{For $\hat{\sigma}_{gq}$ we have $\delta \eta_{gq}^{(1)}=0$ and :
\beq
\begin{array}{cl}\delta \eta_{gq}^{(2)} = & \displaystyle
\frac{33-2n_{f}}{4}\ln\left(\frac{\mu_{R}^2}{\mu_{F}^2}\right)\left[
  \frac{2}{3}\ln\left(\frac{m_{H}^2}{\mu_{F}^2}\right)-1+2z-
  \frac{1}{3}z^2\right.\\
& \\
&
\displaystyle
\left. -\frac{2}{3}\left(1+(1-z)^2\right)\ln\left(\frac{z}{(1-z)^2}\right)\right]
\end{array}
\eeq}

\item{For $\hat{\sigma}_{gg}^{\text{soft+virtual}}$ we have $\displaystyle
    \delta \eta_{gg}^{(1),s+v} = \frac{33-2n_{f}}{6}
    \ln\left(\frac{\mu_{R}^2}{\mu_{F}^2}\right)$ and :
\beq
\begin{array}{cl}\delta \eta_{gg}^{(2),s+v} = & \displaystyle
  \frac{(33-2n_{f})^2}{48}\ln^2\left(\frac{\mu_{R}^2}{\mu_{F}^2}\right)\delta(1-z)+3(33-2n_{f})
  \D_{1}(z)\ln\left(\frac{\mu_{R}^2}{\mu_{F}^2}\right)\\
 & \\
 &
 \displaystyle+\frac{3}{2}\left(33-2n_{f}\right)\ln\left(\frac{m_{H}^2}{\mu_{F}^2}\right)
 \ln\left(\frac{\mu_{R}^2}
   {\mu_{F}^2}\right)\D_{0}(z)\\
& \\
&
\displaystyle
+\ln\left(\frac{\mu_{R}^2}{\mu_{F}^2}\right)\left(3\zeta(2)
  \frac{33-2n_f}{2} +
  \frac{1395-104n_f}{24}\right)\delta(1-z)\end{array}
\eeq}

\item{For $\hat{\sigma}_{gg}^{\text{hard}}$ we have $\delta
    \eta_{gg}^{(1),h}=0$ and :
\beq
\begin{array}{cl}\delta \eta_{gg}^{(2),h} = & \displaystyle
  \frac{3}{2}(33-2n_{f})\ln
  \left(\frac{\mu_{R}^2}{\mu_{F}^2}\right)\ln\left(\frac{\mu_{F}^2}{m_{H}^2}\right)
  z\left(2-z(1-z)\right)\\ 
& \\
& \displaystyle - \frac{3}{2}(33-2n_{f})\frac{\ln
    z}{1-z}\ln\left(\frac{\mu_{R}^2}{\mu_{F}^2}\right)\left(1-z(1-z)\right)^2\\
 & \\
 & \displaystyle
 -3(33-2n_{f})\ln\left(\frac{\mu_{R}^2}{\mu_{F}^2}\right)z\ln(1-z)\left(2-z(1-z)\right)\\
& \\
& \displaystyle  -\frac{33}{8}(33-2n_{f})\ln\left(\frac{\mu_{R}^2}{\mu_{F}^2}\right)\left(1-z\right)^3
\end{array}
\eeq}
\end{itemize}

\end{subappendices}
\vfill
\pagebreak

\section{Higgs production at the LHC}

\label{section:SMHiggsLHC}

The CERN proton--proton collider successfully started its operations
in 2009 at a center--of--mass energy of 900 GeV and then really in
2010 but at a reduced center of mass energy of 7
TeV~\cite{Heuer:2010}. It has been running in physics mode for more
than one year up until now and has collected enough data to produce
already interesting results. Indeed the quest for the SM Higgs boson
has began and some results have been presented in
Refs.~\cite{Chatrchyan:2011tz, Aad:2011qi}.

In order to distinguish between the current early run at 7 TeV and the
designed run at an energy close to 14 TeV, we will call the early 7
TeV experiment as the $\lhc$ for littler Hadron Collider. The $\lhc$
will also be sensitive to the SM Higgs particle and in particular be
competitive with the Tevatron experiment once the accumulated
luminosity will be at least of order 1
fb$^{-1}$~\cite{Heuer:2010,Dittmaier:2011ti,ATLAS:1303604,CMS-NOTE-2010-008}. However,
with the present expectations, the Tevatron and presumably also the
$\lhc$ will only be able to exclude the existence of the SM Higgs
particle in some given mass range. We will present in this section
some results for the $\lhc$ energy and also for SM Higgs boson
production at the designed $\sqrt s = 14$ TeV LHC.

The situation at the $\lhc$ is much in the same way as what has to be
carefully done at the Tevatron: both are hadron colliders which then
imply that the predictions are plagued with various theoretical
uncertainties. The exclusion of Higgs mass regions relies crucially on
the theoretical predictions for the production cross sections for the
Higgs signal as well as for the relevant SM backgrounds, and this was
discussed in detail in the last section. We remind the reader that the
main production channel at the Tevatron which were presented are the
top and bottom quark loops mediated  gluon--gluon fusion mechanism
$gg\to H$~\cite{Georgi:1977gs} with the Higgs decaying into $WW$ pairs
which lead to $\ell \nu \ell \bar \nu$ (with $\ell\!=\!e,\mu$) final
states~\cite{Dittmar:1996ss} and the Higgs--strahlung processes $q
\bar q\! \to\! VH$ (with $V\!=\!W,Z$)~\cite{Glashow:1978ab} with the
subsequent $H\!\to \!b\bar b$ and $V \to \ell\!+\!X$ decays of the
Higgs and the associated gauge bosons.

At the $\lhc$, the Higgs--strahlung processes, as well as as other
production channels such as weak vector--boson fusion and associated
Higgs production with top quark pairs, have too small cross sections
and/or are plagued with too large QCD backgrounds. This implies that
the gluon--gluon fusion process with the Higgs boson decaying into
$H\!\to\! WW\!\to\! \ell \ell \nu \bar \nu$,  $H\!\to\! ZZ\!
\to\!2\ell\!+\!X$, where $X$ stands for charged leptons, neutrinos and
eventually also jets including $b$--quark jets, and to a lesser extent
$H \to \gamma \gamma$ final states, will be mostly relevant.  At a
center of mass energy $\sqrt{s}\!=\!7$ TeV and with 1 fb$^{-1}$ of
data, recent studies by the ATLAS and CMS collaborations have shown
that the mass range $M_H\!\approx\! 150$--190 GeV can be excluded at
95\%CL if no Higgs signal is
observed~\cite{ATLAS:1303604,CMS-NOTE-2010-008}. This is the reason
why we will only concentrate in this section in the theoretical
predictions in the gluon--gluon fusion production channel.

We will reproduce the same line of arguments developed in
section~\ref{section:SMHiggsTev} and presented in
Ref.~\cite{Baglio:2010um}.  We present an analysis of the gluon--gluon
fusion production channel for SM Higgs production $gg\to H$  at the
$\lhc$, beginning with the theoretical prediction for the Higgs
production cross sections, including all the relevant higher order QCD
and electroweak corrections. We then analyze the various uncertainties
that affect them. We show that the scale and PDF uncertainties, as
well as the non--negligible uncertainty due to the use of an effective
approach in the calculation of QCD and electroweak higher order
corrections beyond next--to--leading order, add up to $\approx
25$--30\% depending on the considered Higgs mass range. If we compare
with what has been obtained in section~\ref{section:SMHiggsTevTotal}
for the Tevatron case, the total uncertainty at the $\lhc$ is
significantly smaller than that obtained at the
Tevatron, as a result of smaller QCD radiative corrections and a
better knowledge of the gluon distribution function at the energies
relevant at the $\lhc$.

We will also give some predictions for different center--of--mass
energies. Indeed the LHC commissionning group did think about running
the LHC at intermediate center--of--mass energies of 8, 9, 10 TeV
before the long shut--down of the accelerator in order to complete
design operations needed for the full--flegded 14 TeV
LHC~\cite{Heuer:2010}. This has been finally abandonned and the CERN
accelerator will run at 7 TeV until the one year shut--down for the
design of the 14 TeV run~\cite{ATLASnews:2010}. Nevertheless we will
give the central cross sections predictions for these intermediate
energies, as well as a detailed analysis of the gluon--gluon fusion
production channel for the full--fledged LHC with $\sqrt s =14$
TeV. The main result that we obtain is that
while the production cross sections for the SM Higgs particle are higher at
energies $\sqrt s=8$--14 TeV compared to $\sqrt s= 7$ TeV, the overall
theoretical uncertainties affecting the predictions do not change
much.

\subsection[The main channel at the lHC]{The main channel at the lHC
  (or LHC at 7 TeV) \label{section:SMHiggsLHCIntro}}

The hierarchy in the production channels for the SM Higgs boson at the
$\lhc$ does not substantially change when compared to the situation at
the Tevatron collider presented in
section~\ref{section:SMHiggsTev}. The main production channel remains
the gluon--gluon fusion channel $gg\to H$ proceeding through
triangular top and bottom quark loops~\cite{Georgi:1977gs} and as stated in
the introduction of this section we will discard the analysis of the
other sub--dominant production channels, due to the fact that the
$\lhc$ analysis is mostly relevant only in the gluon--gluon fusion
channel.  As stated in the Tevatron analysis this process is known to
be subject to extremely large QCD radiative
corrections~\cite{Djouadi:1991tka, Dawson:1990zj,Graudenz:1992pv,
  Spira:1993bb, Spira:1995rr,Harlander:2002wh, Anastasiou:2002yz,
  Ravindran:2003um,Catani:2003zt,Moch:2005ky, Ravindran:2006cg,
  Ahrens:2008nc,Anastasiou:2008tj,deFlorian:2009hc} that can be
described by an associated $K$--factor defined as the ratio of the
higher order (HO) to the LO cross sections, consistently evaluated 
with the value of the strong coupling $\alpha_s$ and the PDF taken at
the considered order, see Eq.~\ref{eq:kfactor}.

The NLO corrections in QCD are known both for
infinite~\cite{Djouadi:1991tka, Dawson:1990zj, Graudenz:1992pv,
  Spira:1993bb} and finite~\cite{Spira:1995rr} loop quark masses and,
at $\sqrt{s}=7$ TeV, lead to a $K$--factor $K_{\rm NLO}\sim 1.8$ in
the low Higgs mass range, if the central scale of the cross section is
chosen to be the Higgs mass. We remind the reader that it has been
shown in Ref.~\cite{Spira:1995rr} that working in an effective field
theory (EFT) approach in which the top quark mass is assumed to be
infinite is a very good approximation for Higgs mass values below the
$t\bar t$ threshold $M_H \lsim 2m_t$, provided that the leading order
cross section contains the full  $m_t$ and $m_b$ dependence. The
calculation of the NNLO contribution has then been
done~\cite{Harlander:2002wh, Anastasiou:2002yz, Ravindran:2003um} only
in the EFT approach where $M_H \ll 2m_t$ and, at $\sqrt s=7$ TeV, it
leads to a $\approx 25\%$ increase of the cross section with $K_{\rm
  NNLO}\sim 2.5$. The comparison with the Tevatroon K--factors shows
that the QCD corrections to $gg\to H$ at $\sqrt s = 7$ TeV are thus
smaller than the corresponding ones at the Tevatron as the
$K$--factors in this case are $K_{\rm NLO} \approx 2$ and $K_{\rm
  NNLO} \approx 3$ (with a central scale equal to $M_H$). At the LHC
with $\sqrt s=14$ TeV, the $K$--factors are even smaller, $K_{\rm NLO}
\approx 1.7$ and $K_{\rm NNLO} \approx 2$.

The NNLL resummation of soft gluons increases the cross section by
slightly less than 10\%~\cite{Catani:2003zt,deFlorian:2009hc} and as
in the Tevatron case and for the same reasons, we do not include this
effect and mimick the $10\%$ increase by using as the central scale
\beq
\mu_R=\mu_F=\mu_0=\frac12 M_H
\label{eq:central-scale_lhc}
\eeq

We recall the reader that this central scale choice improves the
convergence of the perturbative series and is more appropriate to describe the
kinematics of the process~\cite{Anastasiou:2010hh}. We again also include the
electroweak corrections known exactly up to
NLO~\cite{Djouadi:1994ge, Aglietti:2004nj, Degrassi:2004mx,
  Actis:2008ug, Actis:2008ts} and which contribute at the level of a
few percent; there are also small mixed NNLO QCD--electroweak effects
which have been calculated~\cite{Anastasiou:2008tj} in an effective
approach valid for $M_{H}\ll M_W$ that are also included in our
calculation.\bigskip

\begin{figure}[!t]
\begin{center}
\includegraphics[scale=0.75]{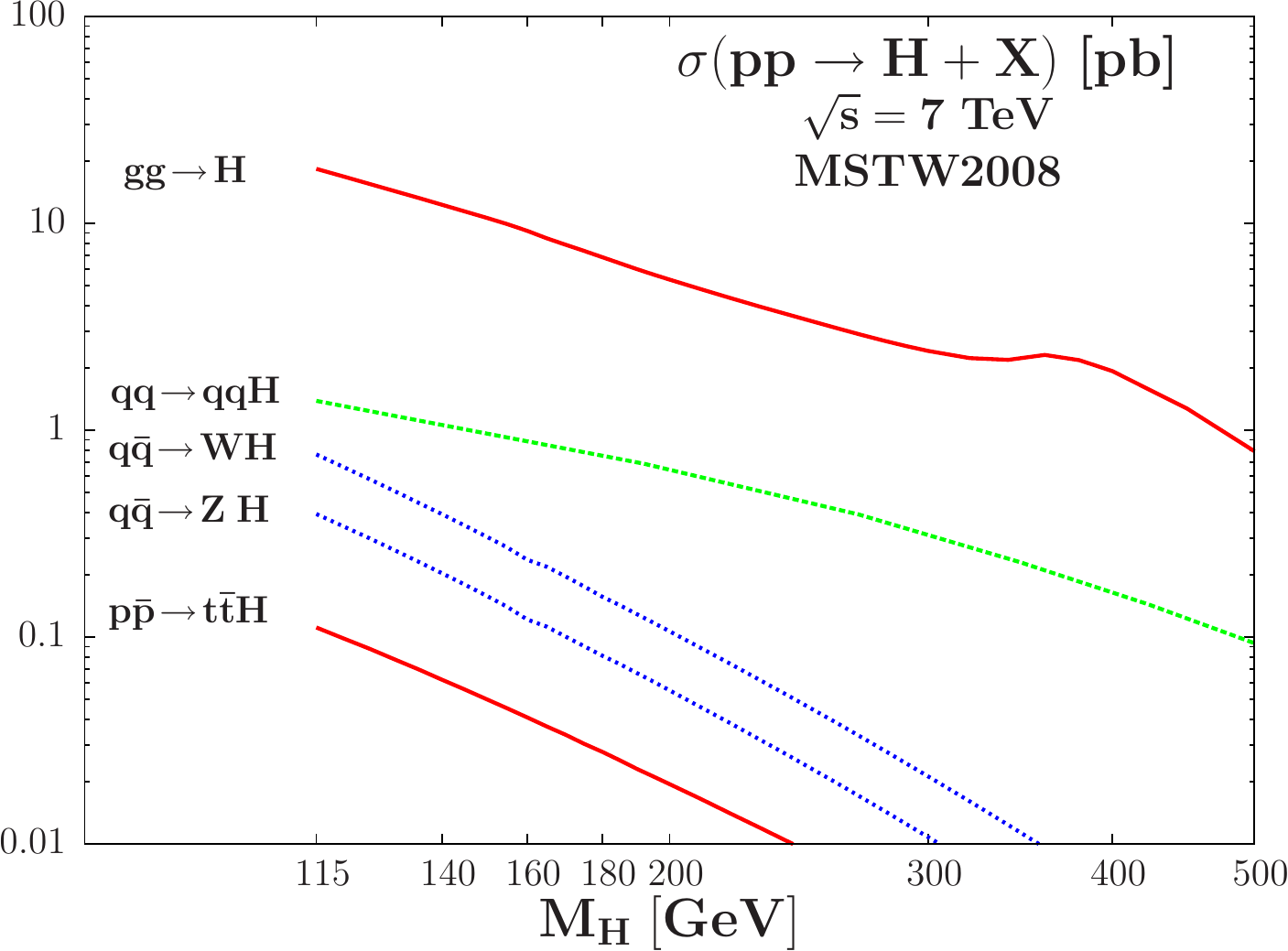}
\end{center}
\vspace*{-4mm}
\caption[Total cross sections for SM Higgs production at the
$\lhc$]{The total cross sections for Higgs production at the $\lhc$ 
with $\sqrt s=7$ TeV as a function of the Higgs mass. The MSTW set of PDFs 
has been used and the higher order corrections are included as discussed in 
the text.}
\vspace*{-2mm}
\label{fig:pp-H-LHC}
\end{figure}

We use the same procedure described in~\ref{section:SMHiggsTevIntro}
for the calculation of the production cross sections\footnote{Other
  updates of the $gg\to H$ cross section at the LHC can be found in
  Refs.~\cite{Anastasiou:2008tj,deFlorian:2009hc, Berger:2010nc,
    Demartin:2010er, Ahrens:2010rs, Alekhin:2010dd}.}: we use the Fortran
code {\tt HIGLU}~\cite{Spirapage} with our modification in order to
include the complete NLO corrections for the top and bottom quark
loops and the NNLO corrections for the top loop in the infinite mass
limit; grids for the electroweak corrections are provided within the
code and we also implement the NNLO mixed QCD--EW corrections in our
calculation. The production cross sections are shown at the $\lhc$
with $\sqrt s=7$ TeV in Fig.~\ref{fig:pp-H-LHC} for an updated value of the
top  quark mass $M_t=172.5$ GeV\footnote{in accord with most predictions at
the $\lhc$, see Ref.~\cite{Dittmaier:2011ti}.} and when the partonic
cross section is folded with the NNLO MSTW2008 public set of
PDFs~\cite{Martin:2009iq}. The renormalization and
factorization scales are fixed to the central values of Eq.~\ref{eq:central-scale_lhc}.
Our results published in Ref.~\cite{Baglio:2010ae} agree with those given
in Refs.~\cite{Anastasiou:2008tj,deFlorian:2009hc} and updated in
Ref.~\cite{Dittmaier:2011ti} within a few percent.

As in the Tevatron calculation we also display in
Fig.~\ref{fig:pp-H-LHC} for completeness the cross sections for the three
other Higgs production channels at hadron colliders that we evaluate using the
programs of Ref.~\cite{Spirapage}:

\vspace{-2mm}
\begin{enumerate}[{\it i)}]
\itemsep-3pt
\item{the Higgs--strahlung processes $q\bar q \to HV$ with $V=W,Z$
    that are known exactly up to NNLO in QCD~\cite{Altarelli:1979ub,
  KubarAndre:1978uy, Han:1991ia, Ohnemus:1992bd, Djouadi:1999ht,
  Spira:1997dg, Hamberg:1990np, Hamberg:2002np, Brein:2003wg} and up
to NLO for the electroweak corrections~\cite{Ciccolini:2003jy}; they
are evaluated at  $\mu_0= M_{HV}$ (the invariant mass of the $HV$
system) for the central scale\footnote{Here, the QCD $K$--factors are 
  moderate, $K_{\rm NNLO} \sim 1.5$ and the electroweak  corrections
  reduce the cross section by an amount of $\approx 3-8\%$. For the
  evaluation of the cross section, we have used the NLO code {\tt
    V2HV}~\cite{Spirapage} in which we implemented  these higher order
  contributions. Our results agree with those of
  Ref.~\cite{Dittmaier:2011ti}.};}

\item{the weak vector boson fusion channel $qq \to Hqq$ evaluated at
    the scale $\mu_0=Q_V$ (the momentum transfer at the gauge boson
    leg) in which only the NLO QCD corrections~\cite{Han:1992hr,
      Figy:2003nv, Spira:1997dg} have been included; the NNLO
    corrections have been found to be very small~\cite{Bolzoni:2010xr,
      Bolzoni:2010as} and we omit the electroweak
    corrections~\cite{Ciccolini:2007ec}, as was done in the Tevatron
    calculation earlier;} 

\item{associated Higgs production with top quark pairs in which only
    the leading order order contribution  is implemented but at a
    central scale $\mu_0\!=\!\frac12(M_H+2m_t)$, which is a good
    approximation  at these energies as the NLO $K$--factor is very
    close to unity~\cite{Beenakker:2001rj, Beenakker:2002nc,
      Reina:2001sf, Dawson:2002tg}, see also
    Ref.~\cite{Dittmaier:2011ti}.}
\end{enumerate}

\begin{table}[!h]{\small%
\let\lbr\{\def\{{\char'173}%
\let\rbr\}\def\}{\char'175}%
\renewcommand{\arraystretch}{1.36}
\vspace*{2mm}
\begin{center}
\begin{tabular}{|c|ccccc|}\hline
$M_H$ & $\sigma^{\rm NNLO}_{g g\to H}~$ & $\sigma^{\rm NLO}_{q q\to H q q}~$ &
$\sigma^{\rm NNLO}_{q \bar q\to HW}~$ & $\sigma^{\rm NNLO}_{q \bar q\to HZ}~$ 
& $\sigma^{\rm LO}_{pp\to t\bar t H}~$ \\ \hline
$115$ & $18347.4$ & $1386.1$ & $764.1$ & $394.0$ & $111.5$ \\ \hline
$120$ & $16844.6$ & $1313.2$ & $664.8$ & $343.2$ & $98.7$ \\ \hline
$125$ & $15509.2$ & $1259.5$ & $580.0$ & $300.3$ & $87.9$ \\ \hline
$130$ & $14322.6$ & $1192.1$ & $507.5$ & $263.2$ & $78.1$ \\ \hline
$135$ & $13260.6$ & $1148.6$ & $446.0$ & $231.4$ & $69.8$ \\ \hline
$140$ & $12305.6$ & $1087.0$ & $392.9$ & $203.6$ & $62.3$ \\ \hline
$145$ & $11446.4$ & $1051.5$ & $347.3$ & $180.0$ & $56.0$ \\ \hline
$150$ & $10665.9$ & $1006.2$ & $307.1$ & $159.1$ & $50.2$ \\ \hline
$155$ & $9936.4$ & $964.6$ & $272.0$ & $140.4$ & $45.3$ \\ \hline
$160$ & $9205.9$ & $908.4$ & $236.8$ & $121.7$ & $40.9$ \\ \hline
$165$ & $8470.8$ & $875.3$ & $218.9$ & $112.3$ & $37.0$ \\ \hline
$170$ & $7872.3$ & $842.5$ & $196.0$ & $100.6$ & $33.7$ \\ \hline
$175$ & $7345.3$ & $796.6$ & $175.6$ & $90.2$ & $30.5$ \\ \hline
$180$ & $6861.2$ & $768.0$ & $156.8$ & $81.2$ & $27.9$ \\ \hline
$185$ & $6416.3$ & $732.8$ & $142.7$ & $74.0$ & $25.4$ \\ \hline
$190$ & $6010.0$ & $705.0$ & $129.3$ & $67.1$ & $23.1$ \\ \hline
$195$ & $5654.8$ & $683.8$ & $117.5$ & $60.9$ & $21.2$ \\ \hline
$200$ & $5344.1$ & $651.5$ & $106.8$ & $55.3$ & $19.5$ \\ \hline
$220$ & $4357.0$ & $556.7$ & $74.4$ & $38.4$ & $14.0$ \\ \hline
$240$ & $3646.4$ & $479.6$ & $52.9$ & $27.2$ & $10.4$ \\ \hline
$260$ & $3110.7$ & $411.6$ & $38.4$ & $19.7$ & $7.8$ \\ \hline
$280$ & $2706.4$ & $360.7$ & $28.3$ & $14.5$ & $6.1$ \\ \hline
$300$ & $2415.4$ & $312.6$ & $21.2$ & $10.8$ & $4.8$ \\ \hline
\end{tabular}
\vspace*{-2mm}
\end{center}
\caption[The total Higgs production cross sections in the four main
production channels at the $\lhc$ with $\sqrt{s}=7$ TeV]{The total
  Higgs production cross sections (in fb) in the processes 
$\protect{g g\to H}$, vector--boson fusion  $\protect{q q\to H q q}$,
Higgs--strahlung  $\protect{q\bar q\to HW.HZ}$  and  associated 
production $\protect{pp \to t\bar t H}$  at the $\lhc$  with $\sqrt s=7$ TeV 
for given Higgs mass values (in GeV) with the corresponding central scales 
described in the main text. The MSTW sets of PDFs have been used at the 
relevant  order.} 
\vspace*{-1mm}
\label{table:allLHC}
}
\end{table} 

The hierarchy shows a small difference compared to that of the Tevatron,
as the vector boson fusion production channel becomes the second most
important production channel at the $\lhc$ whereas it was next to the
Higgs--strahlung processes at the Tevatron. It is also demonstrated in
Fig.~\ref{fig:pp-H-LHC} that the $gg\to H$ process is by far
dominating in the entire Higgs mass range, with a cross section that
is one to two orders of magnitudes larger than in the other production
channels. Table~\ref{table:allLHC} displays the values of the cross
sections for the five Higgs production processes for a selection of
Higgs masses relevant at the $\lhc$.

\subsection{The scale uncertainty \label{section:SMHiggsLHCScale}}

In the calculation of production cross sections and kinematical distributions
at hadron colliders, as the perturbative series are truncated and the results
are available only at a given perturbative order, there is a residual
dependence of the observables on the renormalization scale $\mu_{R}$ which
defines the strong coupling constant $\alpha_{s}$ and on the factorization
scale $\mu_{F}$ at which the matching between the perturbative matrix elements
calculation and the non--perturbative parton distribution functions is
performed. The  uncertainty due to the variation of these two scales
is then viewed
as an estimate of the unknown (not yet calculated) higher--order terms and is
rather often the dominant source of theoretical uncertainties.

This picture has been sketched in
section~\ref{section:SMHiggsTevScale} and we will use the same
techniques for the estimation of this scale uncertainty induced by the
missing higher order terms. We start at the median scale
$\mu_R=\mu_F=\mu_0=\frac12 M_H$ for which the central or ``best''
value of the cross section is obtained, and we vary the two scales
$\mu_R$ and $\mu_F$ within the intervall defined
in~\ref{section:SMHiggsTevScale}:

$${\mu_0}/{\kappa} \le \mu_R, \mu_F \le {\kappa}\! \times \! {\mu_0}$$

with the value of the constant $\kappa=2,3,4,\ldots,$ to be
chosen. The constant $\kappa$ is again chosen as to catch the NNLO
central calculation with the (N)LO scale uncertainty bandd\footnote{In
  the case of the $gg\to H$ process, the maximal (minimal) cross
  sections at a given fixed order in perturbation theory (in
  particular at NNLO) is obtained for the scale choices $\mu_R=\mu_F=
  \mu_0/\kappa~(\kappa \mu_0)$; there is thus no relevant restriction
  to put on the intervall of variation in practice and we will usually
  use the simplification $\mu_R=\mu_F$.}. At the $\lhc$ with $\sqrt s
= 7$ TeV, at $\mu_0=\frac12 M_H$, the value $\kappa=2$ is enough as to
catch the NNLO central prediction with the NLO uncertainty band of
$\sigma^{\rm NLO} (gg\to H)$. This result still nearly holds if
we use the LO bands to catch the NNLO central prediction, as the LO
bands nearly touch the NNLO central prediction.

\begin{figure}[!h]
\vspace*{3mm}
\begin{center}
\mbox{
\includegraphics[width=9cm]{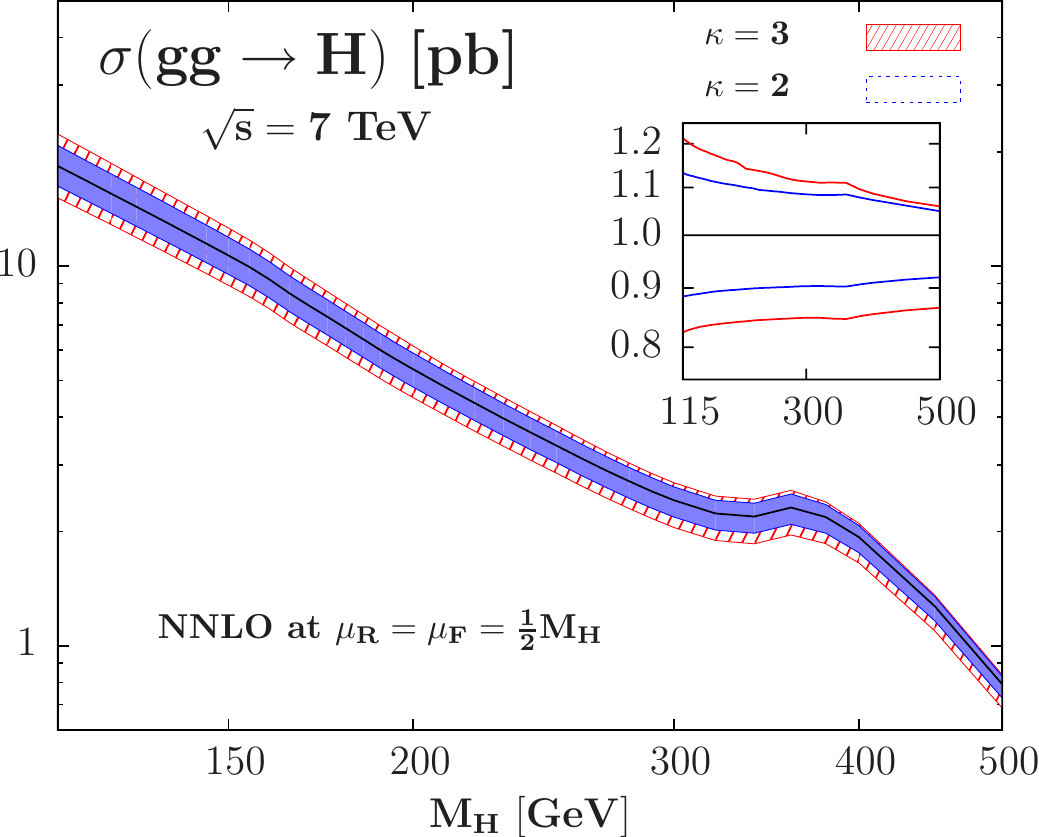} }
\end{center}
\vspace*{-4mm}
\caption[Scale uncertainty at the $\lhc$ in $gg\to H$ at NNLO]{
The scale uncertainty band of $\sigma^{\rm NNLO} (gg\to H)$ at the $\lhc$
as a function of $M_H$ for a scale variation in the domain $M_H/(2\kappa) 
\le \mu_R, \mu_F  \le \kappa\!\times\! \frac12 M_H$ for $\kappa=2$ and 3; 
in the insert, shown are the relative deviations from the central value with a 
scale $\mu_R=\mu_F=\frac12 M_H$.}
\label{fig:scale-SM-lhc}
\vspace*{-.1mm}
\end{figure}

Adopting the range
\beq
\frac12 \mu_0 \le \mu_R,\mu_F \le 2\mu_0
\label{eq:scale_intervall_lhc}
\eeq
with $\mu_0=\frac12 M_H$ for the cross section $\sigma^{\rm
  NNLO}(gg\to H)$, the scale variation displayed in
Fig.~\ref{fig:scale-SM-lhc} as a function of
$M_H$ is obtained; it is compared to a variation with a factor
$\kappa=3$.  In the insert, we
present the maximal and minimal variations of $\sigma^{\rm NNLO} (gg\to H)$ 
compared to the central value. We can see that for $\kappa=2$, a scale
uncertainty of $\approx \pm 10 \%$ is obtained in the low mass range, $M_H
\approx 120$ GeV, which decreases to the level of $\approx -8\%, +4\%$ at high
masses, $M_H \approx 500$ GeV. If the domain for scale variation were
extended to $\kappa=3$, the uncertainty band would have increased to
$\approx \pm 17\%$ in the low Higgs mass range as shown in
Fig.~\ref{fig:scale-SM-lhc}. Using at the $\lhc$ the value $\kappa=2$
instead of $\kappa=3$ can be related to the fact that the higher order
QCD corrections in the gluon--gluon fusion process at the $\lhc$ are
smaller than that of the Tevatron case where we have then used
$\kappa=3$ and hence a larger domain for the scale variation.

\subsection{The PDF+$\alpha_S$
  uncertainty \label{section:SMHiggsLHCPDF}}

The second major source of uncertainty in the Higgs production in the
gluon--gluon fusion at the $\lhc$ is again related to the imprecise
determination of the parton distribution functions (PDFs), much in the
same way at the Tevatron. This uncertainty, as shown later, is even
more important at the $\lhc$ as for high Higgs mass values it becomes
the largest uncertainty.

As discussed in the Tevatron case, see~\ref{section:SMHiggsTevPDF},
there exist many different PDFs parametrizations on the market. We
thus have different way to estimate the PDF uncertainty on the total
production cross section and again we will summarize our lines of
argument as in the Tevatron case.\bigskip

\begin{figure}[!h]
\vspace*{-1mm}
\begin{center}
\mbox{
\includegraphics[width=9cm]{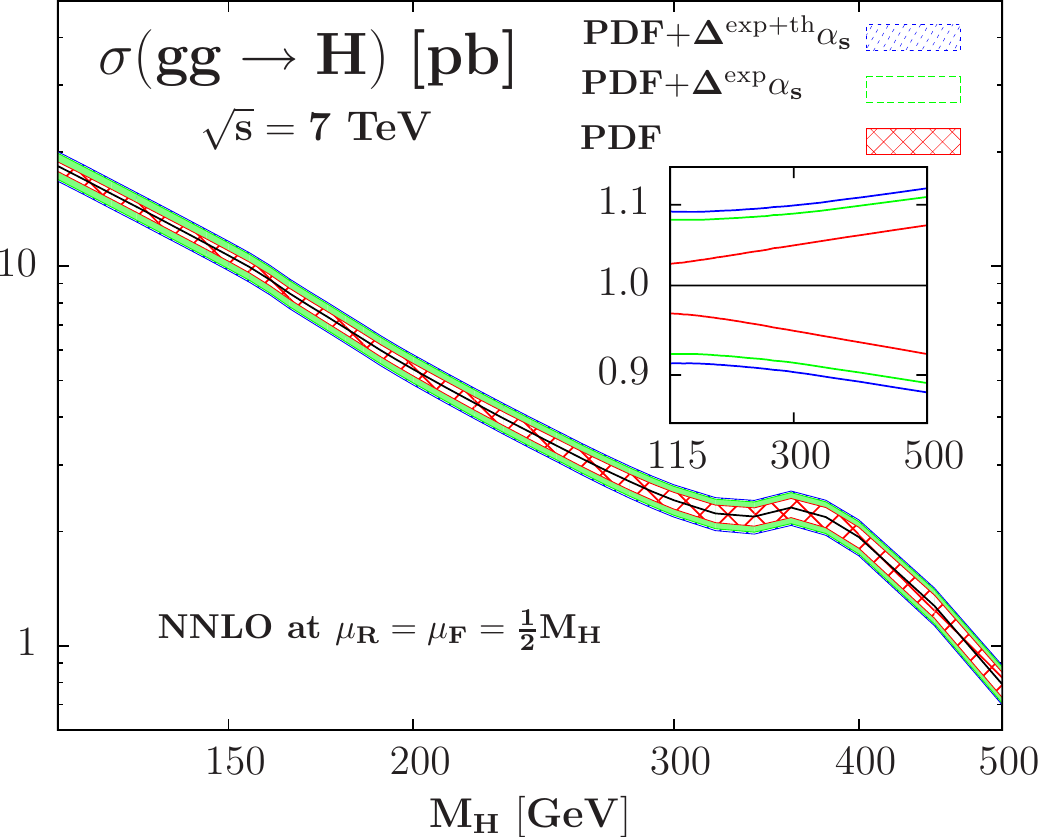} 
}
\end{center}
\vspace*{-4mm}
\caption[PDF and $\Delta^{\rm exp, th}\alpha_s$ uncertainties in
$gg\to H$ at the $\lhc$]{The central values and the 90\% CL PDF, PDF+$\Delta^{\rm
exp}\alpha_s$ and PDF+$\Delta^{\rm exp}\alpha_s+\Delta^{\rm th}\alpha_s$
uncertainty bands in  $\sigma^{\rm NNLO} (gg \to H+X)$ at the 
$\lhc$ when evaluated within the MSTW scheme. In the inserts, shown are the 
same but with the cross sections normalized to the central cross section.}
\label{fig:ggH-pdf1-lhc}
\end{figure}

The first method is the so--called Hessian method where, besides the best fit PDF
with which the central values of the cross sections are evaluated,  a set of
$2N_{\rm PDF}$ PDF parameterizations is provided that reflect the $\pm
1\sigma$ variation of all ($N_F$) parameters that enter into the global fit.
These uncertainties are thus mostly due to the experimental errors in the
various data that are used in the fits. Taking the NNLO public set
provided by the MSTW collaboration~\cite{Martin:2009iq,Martin:2009bu} for
instance, the PDF uncertainty at the 90\% CL that is
obtained at the $\lhc$ for the NNLO $gg\to H$ cross section is shown (by the
red lines and red band) in Fig.~\ref{fig:ggH-pdf1-lhc} as a function of $M_H$.
For low Higgs masses, the uncertainty is at the level of 5\% but it increases
to reach the level of $\approx 8\%$ at high Higgs masses, $M_H\approx 500$ GeV.
If only the (more optimistic) 68\%CL errors are to be considered, the previous
numbers have to be divided by $\approx 1.6$.

The thorough analysis conducted for the gluon--gluon Higgs production
at the Tevatron has demonstrated that these (Hessian) uncertainties
are not sufficient as they do not account for the theoretical
assumptions entering into the parametrizations of the PDFs and which
explain why there are so many different PDFs sets on the market. It is
therefore very difficult to estimate within a given set the
uncertainties behind these different assumptions. Nevertheless, an
accurate determination of the PDFs uncertainty requires that we take
into account this spread observed in the theoretical predictions using
the different NNLO PDFs sets that are available. 

\begin{figure}[!h]
\vspace*{-1mm}
\begin{center}
\mbox{
\includegraphics[width=9cm]{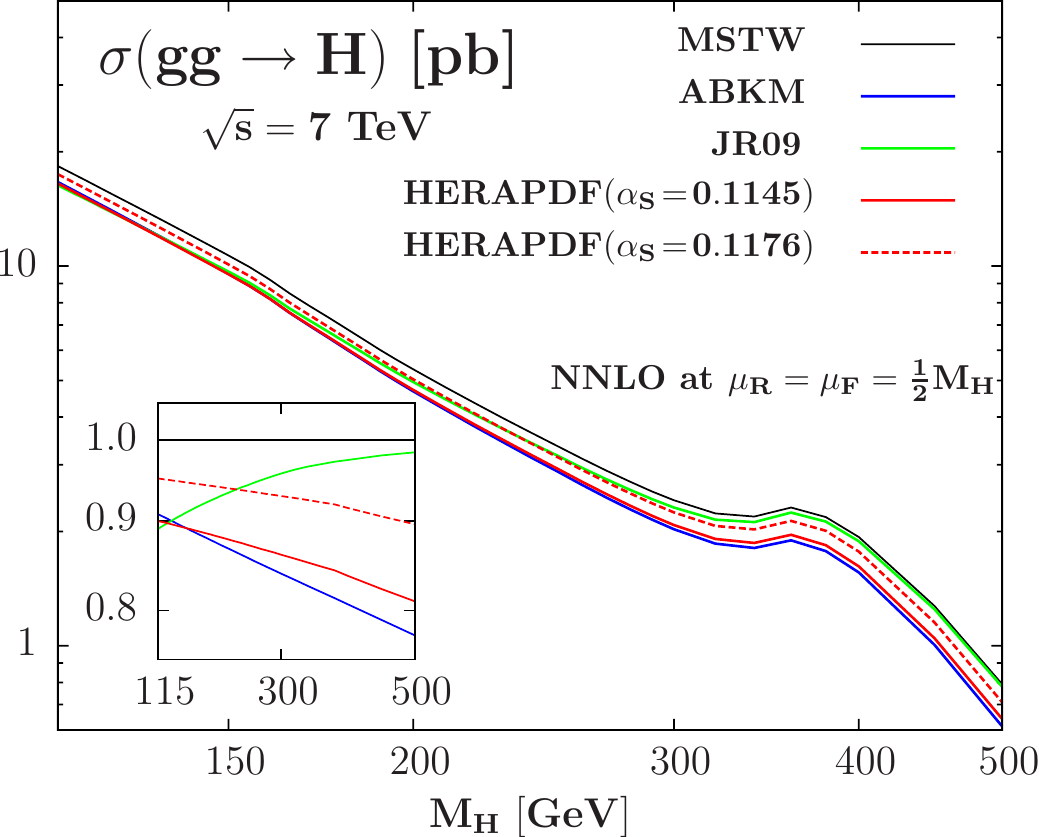}
}
\end{center}
\vspace*{-4mm}
\caption[Comparison between the predictions given by the four NNLO
PDF sets for $gg\to H$ at the $\lhc$]{The central values of the  NNLO
  cross section $\sigma(gg \to H+X)$ 
at the  $\lhc$ as a function of $M_H$ when evaluated in the four schemes 
which provide NNLO PDFs. In the inserts, shown are the same but with the 
cross sections normalized to the MSTW central cross section.}
\label{fig:ggH-pdf2-lhc}
\end{figure}

As stated in section~\ref{section:SMHiggsTevPDF}, one possible way to
estimate this more accurate uncertainty is to compare the results
for the central values of the cross section (and hence, using the best
fit PDFs only) when using the different NNLO sets of PDFs which
involve, in principle, different assumptions. We display in
Fig.~\ref{fig:ggH-pdf2-lhc} the values of $\sigma^{\rm NNLO}(gg\to H)$
calculated with our procedure when folding the partonic cross section with
the gluon densities that are predicted by the four PDF
sets\footnote{We discard in this analysis the two other major PDF
  collaborations on the market, the CTEQ
  collaboration~\cite{Nadolsky:2008zw} and the NNPDF
  collaboration~\cite{Ball:2009mk} as their PDF sets were only 
  provided at the NLO order in QCD when this analysis has been
  done. We also point out that the
  analysis conducted in section~\ref{section:SMHiggsTevPDF} has shown
  that the MSTW and CTEQ parametrizations are very similar, hence we
  could easily believe that this should be in much the same way at the
  $\lhc$. It is worth mentioning that the NNPDF collaboration has
  released an update during the final stage of the writing of the
  thesis, which includes a NNLO PDFs set~\cite{Ball:2011uy}. Their PDF
  fit is close to that of the MSTW collaboration.} that have
parameterizations at NNLO: MSTW~\cite{Martin:2009iq}, JR
09~\cite{JimenezDelgado:2009tv}, ABKM~\cite{Alekhin:2009ni} and
HERAPDF~\cite{HERApage}. In the latter case, two sets are provided: one
with the value $\alpha_s(M_Z^2)=0.1176$ that is close to the world
average value~\cite{Nakamura:2010zzi} and the MSTW best--fit
$\alpha_s(M_Z^2)=0.1171$~\cite{Martin:2009iq}, and one with the value
$\alpha_s(M_Z^2)=0.1145$ used also in the ABKM
set~\cite{Alekhin:2009ni}, that is close to the preferred values that
one obtains using deep--inelastic scattering data alone.

The Fig.~\ref{fig:ggH-pdf2-lhc} shows that there are significant
differences bewteen the various NNLO PDFs sets predictions. Indeed,
while the differences in the cross sections are moderate in the low
Higgs mass range, being of the order of 10\% or less, there is a
significant difference at higher Higgs masses, ${\cal O}(25\%$) for
$M_H\approx 500$ GeV. Even with the same $\alpha_s$ input or best--fit
values, HERAPDF with $\alpha_s(M_Z^2)=0.1176$ and MSTW results are
strikingly different. For the large $M_H$ values, this discrepancy is
mainly due to the gluon densities at moderate to high Bjorken--$x$
values which are less constrained by the data (in particular if the
Tevatron high $E_T$ jet data are not included as is the case of the
ABKM and HERAPDF sets).

We again encounter the same situation developed at the Tevatron and
discussed earlier. The comparison between the estimate using only one
PDF set with the usual Hessian method displayed in
Fig.~\ref{fig:ggH-pdf1-lhc} and the estimate comparing the central
predictions with various NNLO PDF sets, displayed in
Fig.~\ref{fig:ggH-pdf2-lhc}, leads to very different uncertainty
estimations. That is why the inclusion of the uncertainty due to the
value of $\alpha_S$ is crucial as pointed out in Refs.~\cite{Baglio:2010um,
  Baglio:2010ae} and explained in details in
section~\ref{section:SMHiggsTevPDF}. If the reader reads carefully the
legend of Fig.~\ref{fig:ggH-pdf2-lhc} he will see that even within one
PDF set, say the HERAPDF set, the value of $\alpha_s$ could change
drastically depending on the PDF fit that is done. Indeed this value
is fitted altogether with the PDF, but the difference in the value of
$\alpha_s(M_Z^2)$ alone is not enough to explain the striking
differences in the various predictions and we should think about the
differences in the shape of the gluon densities. We will then also use this
additionnal source of uncertainty in the case of the Higgs production
in gluon--gluon fusion at the $\lhc$.

We recall that we use the MSTW scheme~\cite{Martin:2009bu} with
\beq
\label{eq:alphas-lhc}
\alpha_s(M_Z^2)&=& 0.1171~^{+0.0014}_{-0.0014}~{\rm (68\%~CL)}~{\rm or}~
                      ^{+0.0032}_{-0.0032}~{\rm (90\%~CL)}~{\rm at~NNLO} 
\eeq

We have evaluated  the 90\% CL correlated PDF+$\Delta^{\rm exp} \alpha_{s}$
uncertainties and the result is shown in  Fig.~\ref{fig:ggH-pdf1-lhc}
(green band and green lines) as a function of $M_H$. The uncertainties
that we have obtained, as displayed in Fig.~\ref{fig:ggH-pdf1-lhc},
are much larger than the uncertainties solely due to the PDF Hessian
errors. At low Higgs masses we even double the
uncertainty. Nevertheless, in much the same was as in the case of the
Tevatron predictions, we cannot yet reconcile MSTW and ABKM/HERAPDF
predictions, particularly at high Higgs masses. We then miss something
that would be able to explain the spread of the different PDF sets
predictions using only one PDF scheme.

We will thus consider, in addition to the correlated PDF+$\Delta^{\rm
  exp}\alpha_s$ uncertainty, the one that comes from the theoretical
uncertainty on the value of the strong coupling constant, stemming
from the truncation of the perturbation series, the different heavy
flavour scheme in the various PDF collaborations, and so on. If we
compare the world average value for $\alpha_s(M_Z^2)$ and the value
obtained using only deep--inelastic scattering data, the difference is
significant. Considering this additionnal source of uncertainty then
helps to reconcile the different Higgs production cross section predictions.

The MSTW collaboration estimates the theoretical
uncertainty to be  $\Delta^{\rm th} \alpha_{s}  =  0.003~{\rm
  at~NLO}$~\cite{Martin:2009bu}, which gives $\Delta^{\rm th}
\alpha_{s} = 0.002$ at most at NNLO.
Using a fixed $\alpha_s$ NNLO grid with central PDFs given by the MSTW
collaboration, with $\alpha_s$ values different from the best--fit value (in the
range 0.107--0.127 with steps of 0.001 and which thus include the values
$\alpha_s(M_Z^2) =  0.1171\pm 0.002$ at NNLO), we have evaluated this
uncertainty. Adding it in quadrature to the PDF+$\Delta^{\rm exp} \alpha_{s}$
uncertainty, we obtain at the $\lhc$ a total  PDF+$\Delta^{\rm
  exp}\alpha_s+\Delta^{\rm th}\alpha_s$ uncertainty of  $\approx
11\%-15\%$ depending on the Higgs mass. At least for not too heavy
Higgs bosons, this larger uncertainty reconciles the MSTW and
ABKM/HERAPDF predictions\footnote{In order to be as much thorough as
  possible, we could also take into account the effects due to the
  bottom and charm quark masses on the PDFs. Indeed, a change in the
  fitted masses for these quarks may affect the gluon splitting which
  in turn alters the shape of the gluon--gluon luminosity in
  $gg\to H$. We have estimated quantitatively this effect by using the
  $m_b$ and $m_c$ dependent MSTW PDFs~\cite{Martin:2010db} with $m_c = 1.40
  \pm 0.15$ GeV and $m_b=4.75 \pm 0.05$ GeV. In the case of the charm
  quark, we obtain an approximate $0.2\%$  change at $M_H=115$ GeV and at
  most $\approx 1.5\%$ change at $M_H=500$ GeV. In the case of the
  $b$--quark, the change is below the percent level. As we add in
  quadrature these uncertainties to obtain the total PDF+$\Delta^{\rm
    exp+th}\alpha_s$ uncertainty, these additionnal sources are
  totally negligible and will not be included in the end.}.  

If we compare the situation at the $\lhc$ with the results we obtained
in the Tevatron study in section~\ref{section:SMHiggsTevPDF} we see
that for $M_H \lsim 200$ GeV we obtain smaller uncertainties,
approximately $\pm 11\%$ compared to
the $\approx \pm 15$--20\% uncertainty of $\sigma^{\rm NNLO}(gg\to H)$ at the
Tevatron. This is due to the better control on the  behavior of the gluon
density at moderate--$x$ values which are relevant for the $\lhc$
compared to high--$x$ values relevant for the Tevatron. In addition to
this remark we point out the fact that we required to extend
$\Delta^{\rm th}\alpha_s$ to the value of $0.004$ in the Tevatron
study in order to reconcile MSTW prediction with that of ABKM. This is
nearly done at the $\lhc$ with the much smaller value of $\Delta^{\rm
  th}\alpha_s=0.002$ as estimated by the MSTW collaboration.

We again stress that this procedure is by no mean universal and
unique. If we compare to the results obtained with the use of the
PDF4LHC recommendation~\cite{PDF4LHC}, that is to take the 68\%CL MSTW
PDF+$\Delta^{\rm exp}\alpha_s$ band and multiply it by a factor of
two\footnote{More precisely the procedure advocated by the PDF4LHC
  group is to use the envelop of MSTW,CTEQ and NNPDF prediction at the
NLO level, rescaled to the central NNLO MSTW prediction. As argued
in~\cite{Dittmaier:2011ti} this is equivalent to the procedure outlined in
the main text.} we obtain a slightly larger uncertainty while we
obtained a comparable uncertainty in the Tevatron analysis. This is
not much a suprise as Fig.~\ref{fig:ggH-pdf2-lhc} displays a very
large discrepency between the various NNLO predictions that cannot be
handled in a satisfactory way by the PDF4LHC
recommendation. Nevertheless we obtain similar uncertainties, a
situation which gives us some confidence in our way of trying to
handle the still pending PDF puzzle.

\subsection[EFT]{Effective field theory
  approximation \label{section:SMHiggsLHCEFT}}

The last source of theoretical uncertainties that we consider is the
one specific to the gluon--gluon fusion production channel, and that
was already discussed in section~\ref{section:SMHiggsTevEFT} in the
case of the Tevatron collider.

Indeed, although NLO QCD and EW corrections for the $gg\to H$ cross
section are known precisely with no approximation, the calculation at
the NNLO order is done in an effective theory (EFT) approach where the
particles running in the loop are assumed to have an infinite mass,
or more precisely to have a much heavier mass than that of the
produced Higgs boson. We will thus take into account an uncertainty
related to that approximation, both for QCD and EW corrections.\bigskip

At NLO in QCD, the approximation  $m_t \gg M_{H}$ for the contribution of the
top quark in the loop is rather good for Higgs masses below the $t \bar t$
threshold,  $M_H \lsim 340$ GeV, in particular when the full quark mass
dependence of the leading order cross section $\sigma^{\rm LO}_{\rm exact}$ is
taken into account~\cite{Spira:1995rr}. At NNLO,  this approximation for the top quark
contribution seems also to be accurate as studies of the effect of a finite
$m_t$ value in expansions  of $M_H/(2m_t)$ have shown a difference below the
percent level with respect to the EFT calculation for $M_H \lesssim
300$ GeV~\cite{Harlander:2009mq, Pak:2009dg, Harlander:2009my,
  Marzani:2008az}. At the Tevatron this result was enough to keep us
looking at the uncertainty related to the finiteness of the top quark
mass as the results were derived in the interesting Higgs mass range
of $115\leq M_H\leq 200$ GeV for the Tevatron experiments.

The situation at the $\lhc$ is quite different: the ATLAS/CMS
experiments have been designed to look for the Higgs boson in its
entire mass range $115\leq M_H\leq 1000$ GeV\footnote{the upper bound
  should be in fact approximately 800 GeV, related to the theoretical
  bounds on the SM Higgs mass, see
  section~\ref{section:HiggsboundsTheory}.} which reduces at
$\sqrt s = 7$ TeV to the mass range $115\leq M_H\leq 600$ GeV. The EFT
approach should then definitely not be valid for Higgs masses beyond the
$t\bar t$ threshold, that is beyond $M_H \geq 350$ GeV, where the
$gg\to H$ amplitude develops imaginary parts. This can be seen at NLO
where both the exact and the approximate results are known. We will
thus include an uncertainty related to the use of the EFT approach for
$M_H \gsim 2m_t$ which is taken as the difference between $\sigma^{\rm
  NLO}_{\rm exact}$ and $\sigma^{\rm NLO}_{m_t \to  \infty}$ when the
exact top quark mass dependence is included in the LO cross section
and when this difference is rescaled with the relative magnitudes of
the NLO and NNLO $K$--factors, i.e. $K^{\rm NLO}_{m_t \to
  \infty}/K^{\rm NNLO}_{m_t \to \infty}$. 

\begin{figure}[!h]
\vspace*{3mm}
\begin{center}
\mbox{
\includegraphics[scale=0.7]{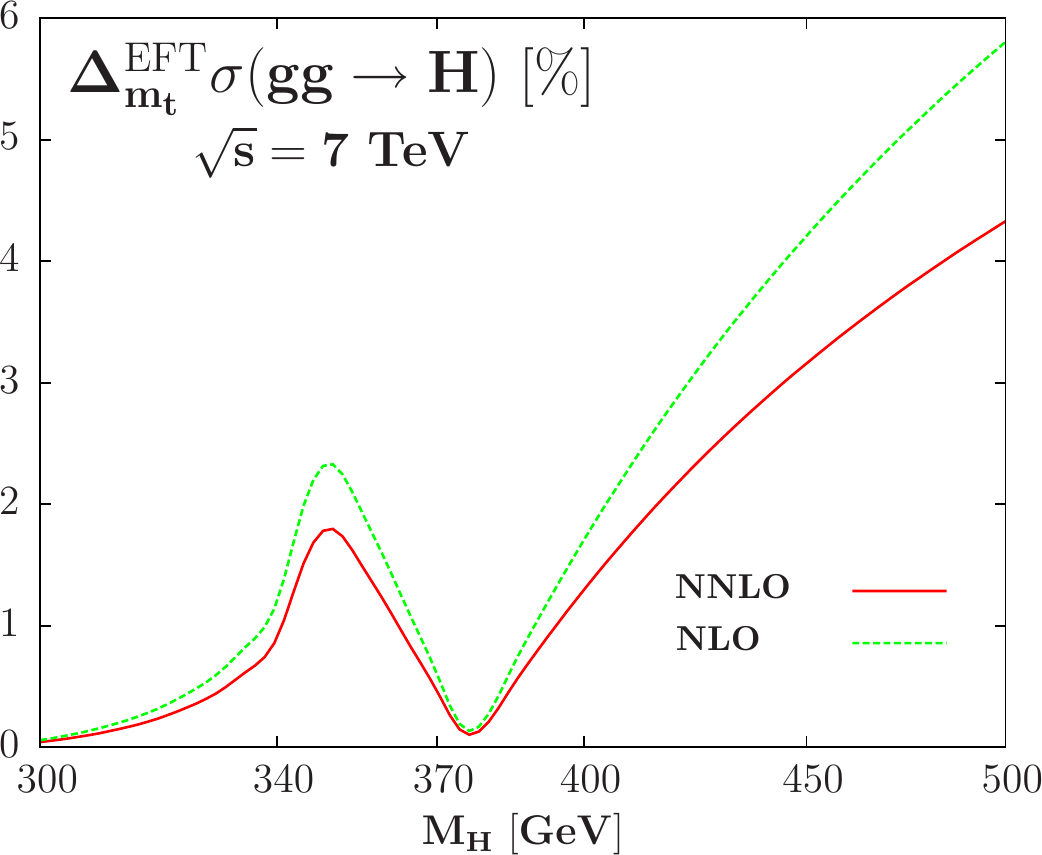}\hspace*{2mm} 
\includegraphics[scale=0.7]{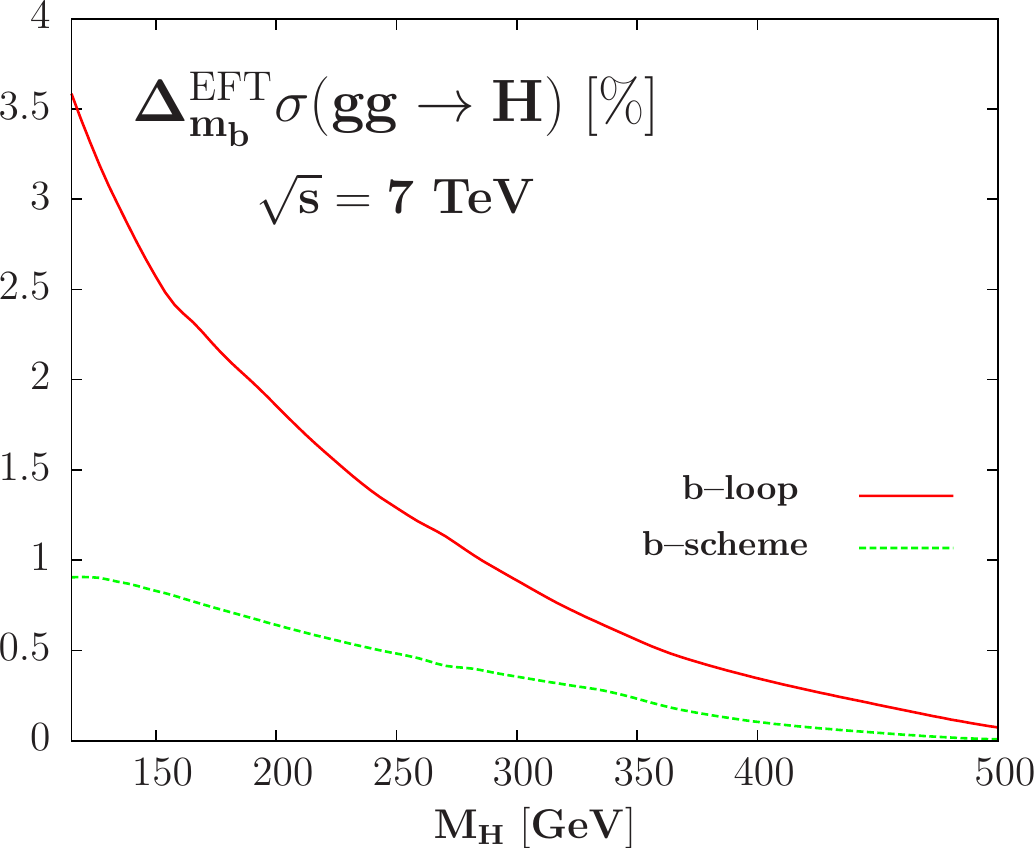} 
}
\end{center}
\vspace*{-3mm}
\caption[Uncertainties due to EFT in the top quark and bottom quark
loops of $gg\to H$ at NNLO at the $\lhc$]{The estimated uncertainties
 (in \%) due to the use of the EFT approach in the evaluation of
 $\sigma(gg \to H)$ at NNLO for the top quark loop contribution for
 Higgs masses beyond  $2m_t$ (left) and for the bottom--quark loop
 contributions (right).}
\label{fig:EFT-mq-lhc}
\vspace*{-3mm}
\end{figure}

In the left--hand side of Fig.~\ref{fig:EFT-mq-lhc} we display the
difference between the exact calculation and the EFT calculation, both
at NLO and NNLO QCD order, for $\sqrt s = 7$ TeV. As expected the
difference is very small for $M_H\leq 350$ GeV where the EFT approach
is very well motivated, and then oscillate around the 2\% level up to
400 GeV from which the error increases very fast. For $M_H\gsim 500$
GeV we obtain an uncertainty larger that 5\% and thus we cannot treat
this uncertainty as being negligible in contrast with the Tevatron
analysis~\cite{Baglio:2010um}.

The two other sources of uncertainty that enter in the EFT approach
have already been encountered in the Tevatron analysis. Indeed it is
well known that of course the EFT approximation is not valid at all
for the bottom quark loop in the gluon--gluon fusion as the bottom
quark mass is small. The omission of this contribution at LO leads to a
$\approx 10 \%$ difference compared to the exact case. The inclusion
of the bottom quark loop actually decreases the cross section because
of the significant negative interference between the top and the
bottom quark loops, the contribution of the bottom quark loop itself
being rather small. Neglecting this contribution then actually overestimates
the NLO (and also the NNLO) cross section and this approximation
cannot be taken into account in the scale variation performed at NNLO
because of the absence of the top--bottom interference at NNLO.

In order to estimate the uncertainty due the missing $b$--loop contribution at
NNLO, we simply follow the previous procedure for the top quark: we assign an
error on the NNLO QCD result  which is approximately the difference between the
exact result $\sigma_{\rm exact}^{\rm NLO}$ and the approximate result
$\sigma_{m_t \! \to \! \infty}^{\rm NLO}$  obtained at NLO  but rescaled with
the relative magnitude of the $K$--factors that one obtains at NLO and NNLO for
the top--loop, i.e. $K^{\rm NLO}_{m_t \to \infty}/K^{\rm NNLO}_{m_t \to
\infty}$. This procedure is the one that has been used in
section~\ref{section:SMHiggsTevEFT} in the study of the EFT approach
at the Tevatron collider. This leads to the uncertainty on the
$\sigma^{\rm NNLO}(gg \to H)$ at $\sqrt  s=7$ TeV that is shown in the
right--hand side of Fig.~\ref{fig:EFT-mq-lhc}. Below $M_H \sim 120$
GeV where the $b$--loop plays an important role the uncertainty is
not negligible, reaching the order of $\pm 3\%$. For Higgs mass above
$M_H \sim 300$ GeV the uncertainty is below the percent level and thus
completely negligible especially in the view of the EFT uncertainty
due to the top--loop foreseen in the previous lines.

In addition to this missing loop uncertainty, there is some freedom in
the choice of the renormalization scheme for the $b$--quark mass in
the $gg\to H$ amplitude: either the the on--shell scheme in which the
pole mass is $m_b \approx  4.7$ GeV or the $\overline{\rm MS}$ scheme
in which the mass $\overline{m}_{b}(\overline{m}_{b}) \approx 4.2$ GeV
is adopted. This leads to a difference of $\approx 1\%$ in the 
$b$--quark loop contribution at NLO that we will take as an additional 
uncertainty due to the scheme dependence. This scheme dependence will
also be discussed in part~\ref{part:four} when dealing with MSSM
$gg\to$ Higgs production at the Tevatron and the $\lhc$. Note that
this cannot be taken into account with the scheme dependence as these
are two widely different scales: the scheme dependence is probed
around $\mu_0 =\frac12 M_H$ while the scheme dependence can be
actually probed around $\mu=\overline{m}_b(\overline{m}_b)$.

The last source of EFT uncertainties is the mixed QCD--EW corrections
that have been calculated at NNLO~\cite{Anastasiou:2008tj} in an EFT
approach with $M_{W/Z} \gg M_H$. Obviously this limit is not valid in
practice as Higgs mass above 115 GeV are probed at hadron
colliders. Some caution should be taken when including this correction
and we have made the choice to assign an uncertainty that is of the
same size as the contribution of this correction itself, as in
Refs.~\cite{Baglio:2010um, Baglio:2010ae} and in
section~\ref{section:SMHiggsTevEFT}\footnote{According to a discussion
  with C. Anastasiou we tend to somewhat underestimate the quality of
  the approximation which may be valid, up to a certain point,
  above the $M_{W/Z}$ threesold. Nevertheless, in the view of the
  smallness of this (still non--negligible) uncertainty, we decide to
  stick to our procedure for the entire Higgs mass range.}. This
uncertainty is comparable in size to the difference between the
electroweak correction calculated exactly at NLO~\cite{Actis:2008ug,
  Actis:2008ts} evaluated in the partial factorization scheme, where
the correction $\sigma^{\rm LO} \Delta_{\rm EW}$ is added to the QCD
corrected cross section, and the EW corrections calculated in the
complete factorization scheme where the NNLO cross section is
multiplied by $1+\Delta_{\rm EW}$. This generates an additional
uncertainty of $\approx 3\%$ at most, nearly exactly the same as
discussed in section~\ref{section:SMHiggsTevEFT} for the Tevatron.

\begin{figure}[!h]
\vspace*{-1mm}
\begin{center}
\mbox{
\includegraphics[scale=0.78]{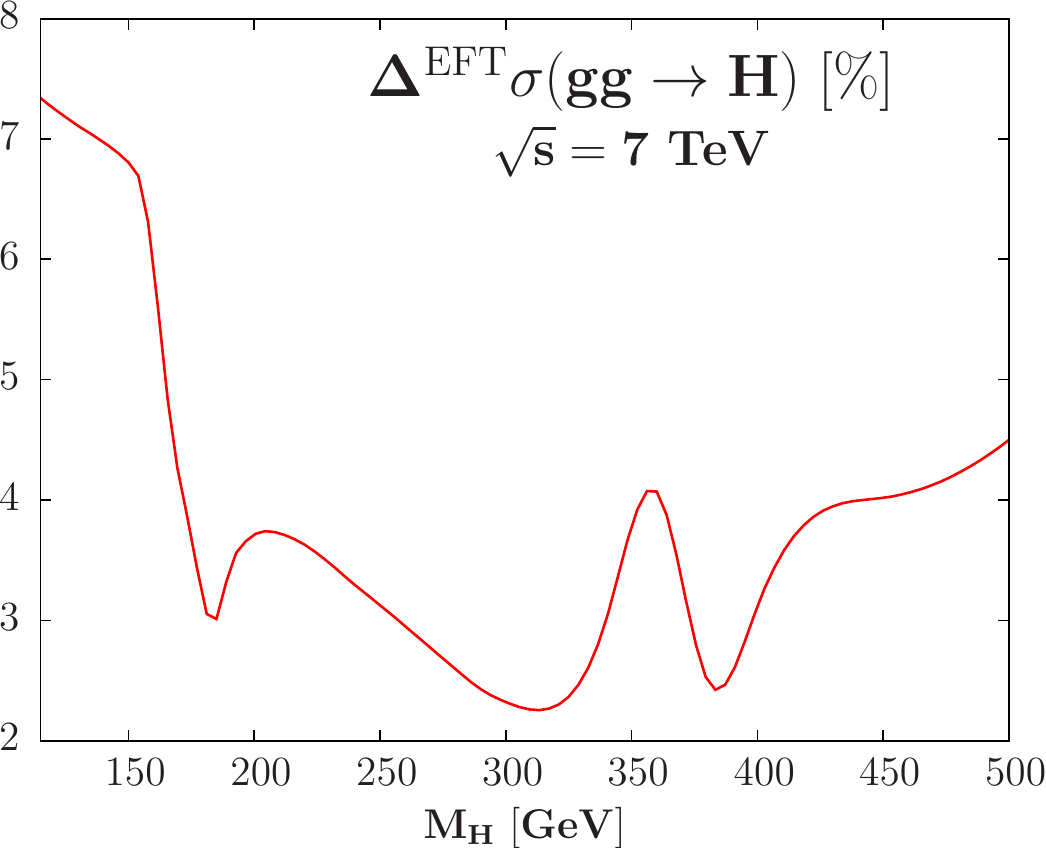}
}
\end{center}
\vspace*{-4mm}
\caption[Total uncertainty due to the EFT approach in $gg\to H$ at
NNLO at the $\lhc$]{The estimated total uncertainty (in \%)  at $\lhc$ energies 
and as a function of $M_H$ from  the use of the EFT approach for the 
calculation of the $gg\to H$ amplitude at NNLO and from the scheme dependence
of the $b$--quark mass.}
\label{fig:ggH-eft-lhc}
\vspace*{-2mm}
\end{figure}

Adding the uncertainties from these three sources linearly, the resulting
overall uncertainty is displayed in  Fig.~\ref{fig:ggH-eft-lhc}  as a function of $M_H$.
It amounts to  $\approx 7\%$ in the low Higgs mass range, drops to the level of
$\approx 4\%$ in the mass range $M_H \approx 200$--400 GeV, and then increases
to reach the level of $\approx 6\%$ at  600 GeV, as a result of the bad $M_H
\ll 2 m_t$ approximation for the top loop contribution. Since this is a pure
theoretical uncertainty with no statistical grounds, it should be
added linearly to the uncertainty from scale variation, as will be
done in the next subsection.

\subsection{Total uncertainy at 7 TeV \label{section:SMHiggsLHCTotal}}

We have described in the previous subsection the different sources of
theoretical uncertainties affecting our prediction. The last task
which remains to be done is the final combination in order to obtain
the overall theoretical uncertainty on the $gg\to H$ production cross
section at the $\lhc$.

As discussed in section~\ref{section:SMHiggsTevTotal} and also
in~\cite{Baglio:2011wn}, the uncertainty associated to the
PDFs(+$\alpha_s$) should be considered as a pure theoretical
uncertainty despite the fact that the fit is done on experimental data
using $\chi^2$ techniques. Indeed this is the only reasonable way to
handle the PDF puzzle, that is the very large discrepency between the
various predictions with the different NNLO PDFs sets on the
market. In this way we consider the different theoretical assumptions
in the determination of the parton densities, and consider the
PDF(+$\alpha_s$) uncertainty to have no statistical ground\footnote{In
  statistical language, the PDF uncertainties should be considered as
  having a flat prior, exactly like the scale uncertainty. A more
  elaborated discussion can be found in~\cite{Dittmaier:2011ti} where the
  recommended combination is that of the linear type.} and not added
in quadrature with the scale uncertainty and the EFT uncertainty which
are purely theoretical beyond any doubts.

We will thus summarize the three possible ways to combine the various
uncertainties, that have been presented in Ref.~\cite{Baglio:2010ae}.

\vspace{-2mm}
\begin{enumerate}[A)]
\itemsep-3pt
\item{The first procedure has been presented for the first time
    in~\cite{Baglio:2010um} and has been used in the
    section~\ref{section:SMHiggsTevTotal} in order to obtain the total
    uncertainty at the Tevatron in the gluon--gluon fusion channel,
    taking into account the possible correlation between the scale and
    the PDFs (that are evaluated at a given factorization scale). We
    calculate the extremal cross sections regarding to the scale
    variation and then apply on this minimal/maximal cross section our
    evaluation of the PDF+$\Delta^{\rm exp+th}\alpha_s$ uncertainty,
    with the factorization and renormalization scales fixed at the
    values corresponding to the minimal and maximal cross sections
    with respect to scale variation\footnote{This procedure has been
      in fact already proposed, together with other possibilities
      which give similar results, in Ref.~\cite{Cacciari:2008zb} where
      top quark pair production at hadron colliders was discussed.}. We
    then add linearly the smaller EFT uncertainty to obtain the
    overall theoretical uncertainty on the central prediction. This
    procedure labeled as procedure ``A'' is our preferred way of
    handling with the total uncertainty.} 

\item{The second procedure, labeled as procedure ``B'',  is simply the
    linear addition of the scale, 
    EFT/scheme and PDF+$\alpha_s$ uncertainties, as all of them are
    considered as pure theoretical uncertainties. This procedure is
    the one advocated by Ref.~\cite{Dittmaier:2011ti}. The final uncertainty that
    is obtained is in general slightly larger that what can be
    obtained with the procedure A described above.} 

\item{The third and last procedure labeled as procedure ``C'' is to
    consider the PDF uncertainty as being the spread of the four NNLO
    central predictions with the four NNLO PDFs sets, using the MSTW
    parametrization as the central set. Since the maximal cross
    section for $gg\to H$ is obtained with the MSTW parametrization
    and the minimal one with the ABKM set, as displayed in
    Fig.~\ref{fig:ggH-pdf2-lhc}, the total uncertainty on the
    production cross section will be on the one hand the linear
    addition of the scale and EFT uncertainties evaluated with the
    MSTW PDF set for the upper uncertainty, on the other hand the
    linear addition of the scale and EFT uncertainties evaluated with
    the ABKM PDF set for the lower uncertainty, adding the difference
    between the MSTW and the ABKM predictions.} 
\end{enumerate}

\begin{figure}[!h]
\vspace*{-1mm}
\begin{center}
\mbox{
\includegraphics[scale=.89]{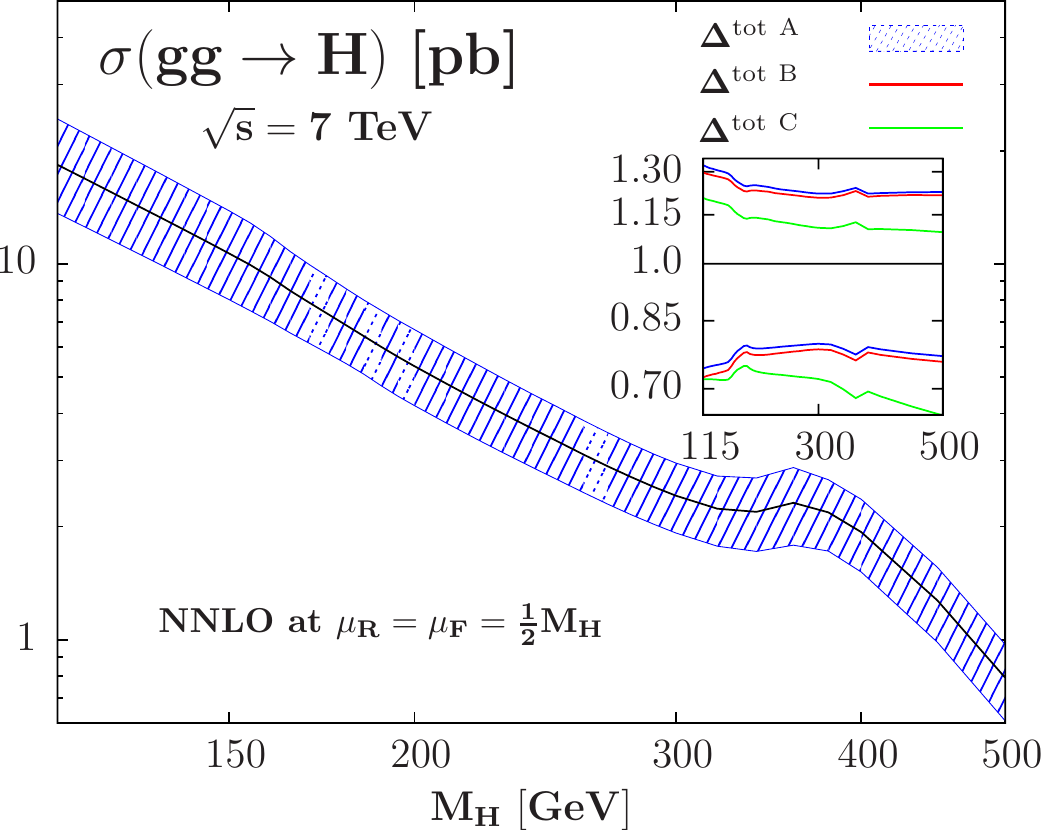} 
}
\end{center}
\vspace*{-4mm}
\caption[Central prediction with its total uncertainty for $gg\to H$
at NNLO at the $\lhc$]{The production cross section $\sigma(gg\to H)$ at NNLO at the
$\lhc$ with $\sqrt s=7$ TeV,  including the total theoretical uncertainty band 
when all the individual uncertainties are combined using the three procedures
A, B and C described in the text. In the inserts, the relative deviations from 
the central cross section value are shown.}
\label{fig:ggH_all_lhc7}
\vspace*{-1mm}
\end{figure}

The overall theoretical uncertainties on $gg\to H$ cross section at
the 7 TeV$\lhc$ that we obtain when using the three ways A, B, C of
handling with the final combination are displayed in
Fig.\ref{fig:ggH_all_lhc7} as a function of the Higgs mass. With the
procedure A, the uncertainty amounts to $\approx -23\%,+25\%$ for $M_H
\lsim 160$ GeV, reduces to $\approx -21\%, +22\%$ for $M_H \gsim 200$
GeV to reach the value $\approx \pm 22\%$ at $M_H \approx 500$
GeV\footnote{ Note that the overall 
  uncertainty on $\sigma^{\rm NNLO} (gg\to H)$ at the $\lhc$ is
  significantly smaller than what has been obtained at the Tevatron in
  section~\ref{section:SMHiggsTevTotal} and published
  in~\cite{Baglio:2010um}, where the procedure A for the
  combination has been used ($\approx -35\%, +40\%$). This is a result
  of the reduction of both the scale and the PDF uncertainties at
  $\sqrt s=7$ TeV compared to $\sqrt s=1.96$ TeV, because of the
  reduction of the QCD $K$--factors at the $\lhc$ compared to the
  Tevatron and also because of the gluon density itself which has a
  less uncertain behavior for not too heavy Higgs bosons.}. With the
procedure B, that is when the various sources of uncertainties are
added linearly, the total uncertainty that is obtained is nearly the
same as in the case of the procedure A, but slightly smaller (2 to
3\%) for the upper uncertainty and slightly higher for the lower
uncertainty. We remind the reader that this procedure is advocated by
Ref.~\cite{Dittmaier:2011ti}; the results obtained with the procedure A are then
reasonable. In the case of procedure C, as we use for the central
prediction of $\sigma^{\rm NNLO} (gg\to H)$ the MSTW PDF set, the
upper uncertainty is simply the sum of the scale and scheme
uncertainties which add to less than 20\% in the entire Higgs mass and
are thus much smaller (in particular in the low  Higgs mass range)
than in procedures A and B. This is a different story for the lower
uncertainty as in addition to the scale and EFT uncertainties we have
to take into account the difference between the MSTW and ABKM
predictions.In this case, the overall uncertainty is comparable to
what is obtained using the procedures A and B in the low Higgs mass
range where the difference between MSTW and ABKM is of the order of
10\% and, hence, the Hessian MSTW PDF+$\alpha_s$ uncertainty alone. It
becomes much larger at higher $M_H$ values where the difference
between the MSTW and ABKM predictions becomes significant. A
significant example can be taken at $M_H=500$ GeV where an uncertainty
of $\approx -35\%$ is obtained in procedure C, compared to $\approx
-23\%$ using procedures A and B.

\begin{table}
   \begin{bigcenter}
    \small
    \begin{tabular}{|c|ccccccc|}\hline
    $M_H$ & $\sigma$ & Scale [\%] &
    PDF+$\Delta^{\rm exp+th}_{\alpha_s}$  [\%]  & ${\rm EFT}$  [\%] & A
    [\%] & B [\%] & C [\%]  \\ \hline
$115$ & $18.35$ & ${+9.1}\;{-10.2}$ & ${+9.1}\;{ -8.8}$ & ${\pm 7.3}
$ & ${+27.8}\;{-24.6}$ & ${+25.6}\;{-26.3}$ & ${+16.4}\;{-26.8}$  \\  
$120$ & $16.84$ & ${+8.9}\;{-10.2 }$ & ${+9.1}\;{ -8.8}$ & ${\pm 7.3}
$ & ${+27.5}\;{-24.5}$ & ${+25.3}\;{-26.2}$ & ${+16.2}\;{-26.8}$  \\  
$125$ & $15.51$ & ${ +8.8}\;{-9.9 }$ & ${+9.1}\;{ -8.8}$ & ${\pm 7.2}
$ & ${+27.1}\;{-24.1}$ & ${+25.1}\;{-25.8}$ & ${+16.0}\;{-26.7}$  \\  
$130$ & $14.32$ & ${ +8.5}\;{-9.8 }$ & ${+9.1}\;{ -8.8}$ & ${\pm 7.1}
$ & ${+26.7}\;{-24.0}$ & ${+24.7}\;{-25.6}$ & ${+15.6}\;{-26.7}$  \\  
$135$ & $13.26$ & ${ +8.4}\;{-9.6 }$ & ${+9.1}\;{ -8.8}$ & ${\pm 7.0}
$ & ${+26.4}\;{-23.7}$ & ${+24.5}\;{-25.4}$ & ${+15.4}\;{-26.6}$ \\  
$140$ & $12.31$ & ${ +8.3}\;{-9.5 }$ & ${+9.1}\;{ -8.8}$ & ${\pm 7.0}
$ & ${+26.2}\;{-23.6}$ & ${+24.3}\;{-25.2}$ & ${+15.3}\;{-26.7}$ \\  
$145$ & $11.45$ & ${ +8.2}\;{-9.5 }$ & ${+9.1}\;{ -8.8}$ & ${\pm 6.9}
$ & ${+26.0}\;{-23.6}$ & ${+24.2}\;{-25.2}$ & ${+15.1}\;{-26.8}$  \\  
$150$ & $10.67$ & ${ +8.1}\;{-9.5 }$ & ${+9.1}\;{ -8.8}$ & ${\pm 6.8}
$ & ${+25.8}\;{-23.5}$ & ${+24.0}\;{-25.1}$ & ${+14.9}\;{-26.9}$  \\  
$155$ & $9.94$  & ${ +7.9}\;{-9.4 }$ & ${+9.1}\;{ -8.8}$ & ${\pm 6.6}
$ & ${+25.4}\;{-23.2}$ & ${+23.6}\;{-24.8}$ & ${+14.5}\;{-26.8}$  \\  
$160$ & $9.21$  & ${ +7.8}\;{-9.4 }$ & ${+9.1}\;{ -8.8}$ & ${\pm 5.9}
$ & ${+24.4}\;{-22.7}$ & ${+22.8}\;{-24.2}$ & ${+13.7}\;{-26.4}$  \\  
$165$ & $8.47$  & ${ +7.7}\;{-9.4 }$ & ${+9.1}\;{ -8.8}$ & ${\pm 4.9}
$ & ${+23.4}\;{-21.7}$ & ${+21.7}\;{-23.1}$ & ${+12.6}\;{-25.5}$ \\  
$170$ & $7.87$  & ${ +7.7}\;{-9.4 }$ & ${+9.1}\;{ -8.8}$ & ${\pm 4.2}
$ & ${+22.7}\;{-21.0}$ & ${+21.1}\;{-22.5}$ & ${+11.9}\;{-25.0}$  \\  
$175$ & $7.35$  & ${ +7.6}\;{-9.4 }$ & ${+9.1}\;{ -8.9}$ & ${\pm 3.7}
$ & ${+22.0}\;{-20.5}$ & ${+20.4}\;{-21.9}$ & ${+11.3}\;{-24.6}$  \\  
$180$ & $6.86$  & ${ +7.5}\;{-9.3 }$ & ${+9.2}\;{ -8.9}$ & ${\pm 3.1}
$ & ${+21.4}\;{-19.9}$ & ${+19.8}\;{-21.3}$ & ${+10.6}\;{-24.2}$  \\  
$185$ & $6.42$  & ${ +7.4}\;{-9.3 }$ & ${+9.2}\;{ -8.9}$ & ${\pm 3.0}
$ & ${+21.1}\;{-19.8}$ & ${+19.6}\;{-21.2}$ & ${+10.4}\;{-24.2}$  \\  
$190$ & $6.01$  & ${ +7.4}\;{-9.3 }$ & ${+9.2}\;{ -8.9}$ & ${\pm 3.4}
$ & ${+21.5}\;{-20.3}$ & ${+20.0}\;{-21.7}$ & ${+10.8}\;{-24.9}$ \\  
$195$ & $5.65$  & ${ +7.4}\;{-9.3 }$ & ${+9.2}\;{ -8.9}$ & ${\pm 3.6}
$ & ${+21.8}\;{-20.5}$ & ${+20.3}\;{-21.9}$ & ${+11.0}\;{-25.2}$  \\  
$200$ & $5.34$  & ${ +7.3}\;{-9.3 }$ & ${+9.3}\;{ -9.0}$ & ${\pm 3.7}
$ & ${+21.8}\;{-20.6}$ & ${+20.3}\;{-21.9}$ & ${+11.0}\;{-25.4}$  \\  
$210$ & $4.81$  & ${ +7.2}\;{-9.3 }$ & ${+9.3}\;{ -9.0}$ & ${\pm 3.7}
$ & ${+21.7}\;{-20.6}$ & ${+20.2}\;{-22.0}$ & ${+10.9}\;{-25.8}$  \\  
$220$ & $4.36$  & ${ +7.2}\;{-9.2 }$ & ${+9.3}\;{ -9.1}$ & ${\pm 3.6}
$ & ${+21.6}\;{-20.6}$ & ${+20.2}\;{-21.9}$ & ${+10.9}\;{-26.0}$  \\  
$230$ & $3.97$  & ${ +7.0}\;{-9.2 }$ & ${+9.4}\;{ -9.2}$ & ${\pm 3.5}
$ & ${+21.4}\;{-20.5}$ & ${+19.9}\;{-21.8}$ & ${+10.5}\;{-26.2}$  \\  
$240$ & $3.65$  & ${ +7.0}\;{-9.2 }$ & ${+9.5}\;{ -9.2}$ & ${\pm 3.3}
$ & ${+21.1}\;{-20.5}$ & ${+19.8}\;{-21.8}$ & ${+10.3}\;{-26.5}$  \\  
$250$ & $3.37$  & ${ +6.9}\;{-9.2 }$ & ${+9.5}\;{ -9.3}$ & ${\pm 3.1}
$ & ${+20.9}\;{-20.4}$ & ${+19.5}\;{-21.6}$ & ${+10.0}\;{-26.6}$  \\  
$260$ & $3.11$  & ${ +6.8}\;{-9.2 }$ & ${+9.6}\;{ -9.4}$ & ${\pm 3.0}
$ & ${+20.6}\;{-20.3}$ & ${+19.3}\;{-21.6}$ & ${+9.7}\;{-26.8}$  \\  
$270$ & $2.89$  & ${ +6.7}\;{-9.2 }$ & ${+9.7}\;{ -9.5}$ & ${\pm 2.8}
$ & ${+20.5}\;{-20.2}$ & ${+19.1}\;{-21.4}$ & ${+9.4}\;{-27.0}$  \\  
$280$ & $2.71$  & ${ +6.8}\;{-9.2 }$ & ${+9.8}\;{ -9.5}$ & ${\pm 2.6}
$ & ${+20.5}\;{-20.1}$ & ${+19.2}\;{-21.4}$ & ${+9.4}\;{-27.2}$  \\  
$290$ & $2.55$  & ${ +6.8}\;{-9.1 }$ & ${+9.8}\;{ -9.6}$ & ${\pm 2.4}
$ & ${+20.3}\;{-20.0}$ & ${+19.1}\;{-21.1}$ & ${+9.2}\;{-27.2}$ \\  
$300$ & $2.42$  & ${ +6.7}\;{-9.1 }$ & ${+9.9}\;{ -9.7}$ & ${\pm 2.3}
$ & ${+20.2}\;{-19.9}$ & ${+18.9}\;{-21.1}$ & ${+9.0}\;{-27.4}$  \\  
$320$ & $2.23$  & ${ +6.7}\;{-9.2 }$ & ${+10.1}\;{-9.9}$ & ${\pm 2.3}
$ & ${+20.2}\;{-20.1}$ & ${+19.0}\;{-21.4}$ & ${+9.0}\;{-28.2}$  \\  
$340$ & $2.19$  & ${ +6.9}\;{-9.2 }$ & ${+10.3}\;{-10.1}$& ${\pm 3.0}
$ & ${+21.4}\;{-21.1}$ & ${+20.2}\;{-22.3}$ & ${+10.0}\;{-29.6}$ \\  
$360$ & $2.31$  & ${ +7.0}\;{-9.2 }$ & ${+10.5}\;{-10.3}$& ${\pm 4.1}
$ & ${+22.5}\;{-22.3}$ & ${+21.6}\;{-23.6}$ & ${+11.0}\;{-31.3}$ \\  
$380$ & $2.18$  & ${ +6.3}\;{-9.1 }$ & ${+10.7}\;{-10.5}$& ${\pm 2.5}
$ & ${+20.4}\;{-20.9}$ & ${+19.6}\;{-22.2}$ & ${+8.8}\;{-30.3}$  \\  
$400$ & $1.93$  & ${ +5.9}\;{-8.8 }$ & ${+11.0}\;{-10.7}$& ${\pm 3.1}
$ & ${+20.7}\;{-21.5}$ & ${+20.0}\;{-22.7}$ & ${+9.0}\;{-31.3}$  \\  
$450$ & $1.27$  & ${ +5.0}\;{-8.4 }$ & ${+11.6}\;{-11.3}$& ${\pm 4.0}
$ & ${+21.4}\;{-22.6}$ & ${+20.5}\;{-23.8}$ & ${+9.0}\;{-33.4}$  \\  
$500$ & $0.79$  & ${ +4.4}\;{-8.1 }$ & ${+12.2}\;{-11.9}$& ${\pm 4.5}
$ & ${+22.1}\;{-23.3}$ & ${+21.1}\;{-24.5}$ & ${+8.9}\;{-35.2}$  \\  
$600$ & $0.31$  & ${ +3.7}\;{-7.7 }$ & ${+13.3}\;{-13.0}$& ${\pm 6.6}
$ & ${+24.5}\;{-26.1}$ & ${+23.7}\;{-27.3}$ & ${+10.4}\;{-40.0}$  \\ \hline  
\end{tabular} 
\caption[The NNLO total Higgs production cross sections in the $gg\to H$
process at the $\lhc$ with $\sqrt{s}=7$ TeV together with the
associated theoretical uncertainties]{The NNLO total Higgs production
 cross sections in the $\protect{gg\to
H}$ process at the $\lhc$ with $\sqrt s=7$ TeV  (in pb) for given Higgs mass
values (in GeV) at a central scale $\mu_F=\mu_R=\frac12 M_H$. Shown also are 
the corresponding shifts due to the theoretical uncertainties from the various
sources discussed (first from scale, then from PDF+$\Delta^{\rm \exp+th}\alpha_s$ 
at 90\%CL and from EFT), as well as the total uncertainty when all errors are
added using the procedures A, B and C  described in the text.}
\label{table:ggH_lhc7_sm}
\end{bigcenter} 
\end{table}
\clearpage

The results obtained in this section are summarized in
Table~\ref{table:ggH_lhc7_sm}\footnote{An extended table for Higgs
  masses up to 1 TeV can be found in the $gg$--fusion section of
  Ref.~\cite{Dittmaier:2011ti}. The full table has not been displayed in this
  thesis as we want to insist on the fact that predictions for Higgs
  mass above 600 GeV are irrelevant at the $\lhc$, as the
  center--of--mass energy and the luminosity are too small to obtain a
  viable signal.}. The $gg\to H$ production cross section for values of
the Higgs mass relevant at the  $\lhc$ with $\sqrt s=7$ TeV are given
together with the uncertainties from scale variations, the
PDF+$\Delta^{\rm exp+th} \alpha_{s}$ uncertainty in the MSTW scheme
and the uncertainty due the use of the EFT approach beyond the NLO
order. The combined uncertainties obtained using the three 
procedures A, B and C proposed above are also given.

\subsection{LHC results at different center--of--mass
  energies \label{section:SMHiggsLHC14TeV}}

As stated in the introduction of this section, the LHC commissionning
group had thought about the idea of raising gradually the
center--of--mass energy at the LHC, by going from 7 TeV to 14 TeV
through intermediate 8, 9, 10 TeV center--of--mass
energies~\cite{Heuer:2010} in order to increase significantly the luminosity
beyond the 1 fb$^{-1}$ level without having to change much the
accelerator design. Even if this idea was abandoned in the beginning
of year 2011~\cite{ATLASnews:2010} we will present some results
concerning these intermediate energies and then give a detailed
analysis of the case of the full--fledged LHC at $\sqrt s = 14$ TeV.

\subsubsection{The case of the $\lhc$ with $\sqrt s = 8, 9, 10$ TeV}

\begin{figure}[!h]
\vspace*{-1mm}
\begin{center}
\mbox{
\includegraphics[scale=.8]{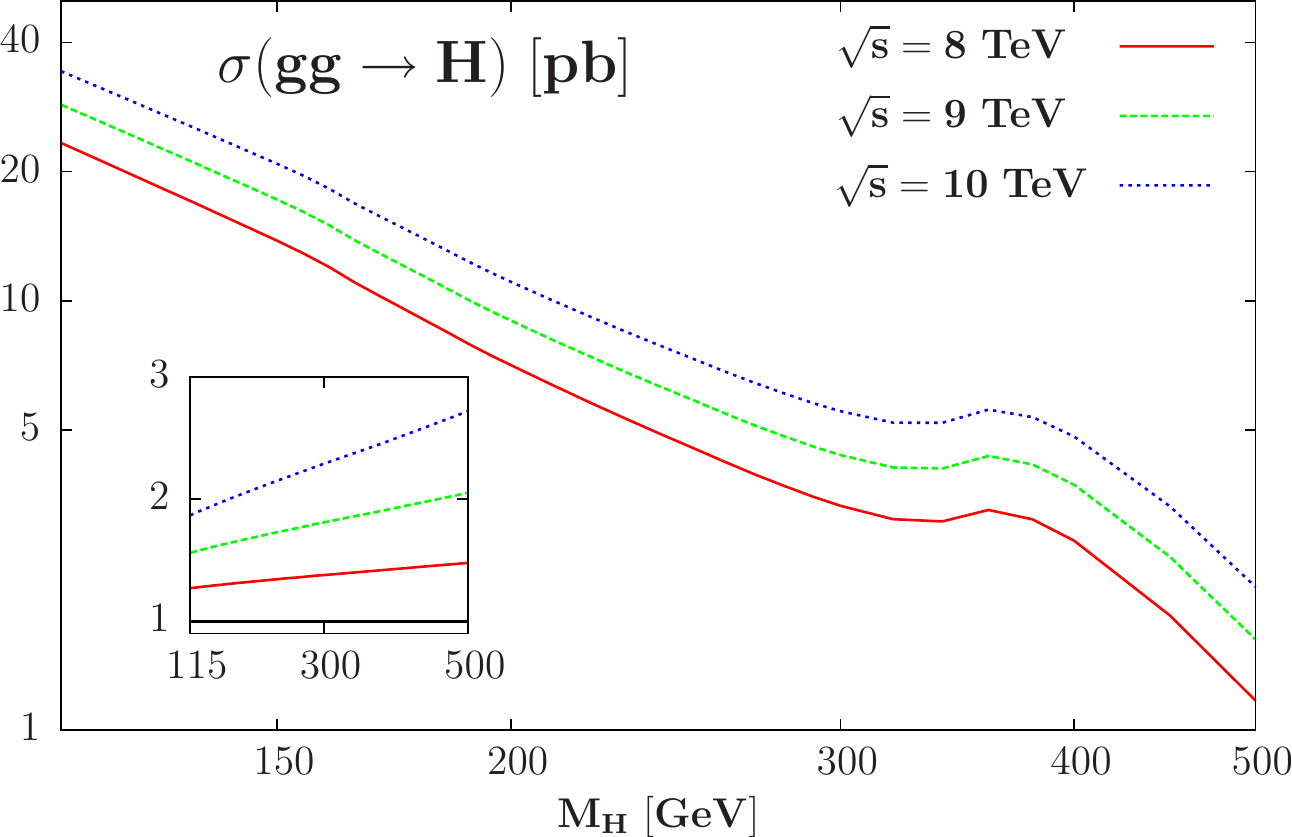} 
}
\end{center}
\vspace*{-4mm}
\caption[Central predictions for $gg\to H$ at NNLO at the $\lhc$ with
$\sqrt{s} = 8,9,10$ TeV]{The production cross section $\sigma(gg\to
  H)$ at NNLO as a function of $M_H$ at the $\lhc$ with center of mass
  energies of $\sqrt s=8,9$ and 10 TeV. In the inserts, the relative
  increase compared to the cross section at $\sqrt s=7$ TeV are shown.}
\label{fig:7-10}
\vspace*{-2mm}
\end{figure}

We use the same recipe for the calculation of the central
predictions, namely a calculation at NNLO order in QCD and EW
corrections, with a central scale $\mu_0=\mu_R=\mu_F=\frac12 M_H$,
using the MSTW 2008 NNLO PDFs set. The results for $\sigma^{\rm NNLO}
(gg\to H)$ are displayed for $\sqrt s=8,9$ and 10 TeV in
Fig.~\ref{fig:7-10} as a function of $M_H$. They are also available
numerically in Table~\ref{table:lhc8} for the relevant Higgs mass values . 

As shown in Fig.~\ref{fig:7-10}, in the low Higgs mass range $M_H
\approx 120$ GeV the cross section $\sigma(gg\to H)$ is approximately
$20\%, 40\%$ and  100\% higher at, respectively, $\sqrt s=8,9$ and 10
TeV, compared to $\sqrt s=7$ TeV. At higher Higgs masses, $M_H \gsim
300$ GeV, the increase of the cross section is slightly larger as the
phase space available for the Higgs production is reduced at lower
energies.

It has been verified that the expectations for the theoretical
uncertainies does not change significantly between the $\lhc$ at 7 TeV
and the intermediate 8, 9, 10 TeV center--of--mass energies. At most
we obtain a 2\% decrease due to the better behaviour of the gluon
density at high Higgs masses. To a very good approximation, we can
therefore view the results of the scale, EFT and PDF+$\Delta^{\rm
  exp+th}\alpha_s$ uncertainties as well as the final combination (in
either procedure A, B or C) all given in Table~\ref{table:ggH_lhc7_sm}
for the $\sqrt s = 7$ TeV case as being the same at the $\lhc$ for the
intermediate $\sqrt s = 8, 9, 10$ TeV.

\begin{table}[!t]{\small%
\let\lbr\{\def\{{\char'173}%
\let\rbr\}\def\}{\char'175}%
\renewcommand{\arraystretch}{1.1}
\vspace*{-1mm}
\begin{center}
\begin{tabular}{|c|ccc||c|ccc|}\hline
$M_H$ & $\sigma^{\rm 8~TeV}_{g g\to H}$ & $\sigma^{\rm 9~TeV}_{g g\to H}$ 
& $\sigma^{\rm 10~TeV}_{g g\to H}$
& $M_H$ & $\sigma^{\rm 8~TeV}_{g g\to H}$ & $\sigma^{\rm 9~TeV}_{g g\to H}$ 
& $\sigma^{\rm 10~TeV}_{g g\to H}$ \\ \hline
$115$ & $23.31$ & $28.63$ & $34.26$ &
$210$ & $6.39$ & $8.15$ & $10.07$ \\ \hline 
$120$ & $21.46$ & $26.42$ & $31.68$ &
$220$ & $5.82$ & $7.44$ & $9.22$ \\ \hline 
$125$ & $19.81$ & $24.44$ & $29.37$ &
$230$ & $5.33$ & $6.84$ & $8.50$ \\ \hline 
$130$ & $18.34$ & $22.68$ & $27.30$ &
$240$ & $4.91$ & $6.32$ & $7.88$ \\ \hline 
$135$ & $17.03$ & $21.10$ & $25.44$ &
$250$ & $4.55$ & $5.88$ & $7.34$ \\ \hline 
$140$ & $15.84$ & $19.67$ & $23.76$ &
$260$ & $4.22$ & $5.47$ & $6.85$ \\ \hline 
$145$ & $14.77$ & $18.38$ & $22.24$ &
$270$ & $3.94$ & $5.13$ & $6.44$ \\ \hline 
$150$ & $13.80$ & $17.21$ & $20.86$ &
$280$ & $3.70$ & $4.83$ & $6.08$ \\ \hline 
$155$ & $12.89$ & $16.10$ & $19.55$ &
$290$ & $3.50$ & $4.58$ & $5.78$ \\ \hline 
$160$ & $11.97$ & $14.98$ & $18.22$ &
$300$ & $3.33$ & $4.37$ & $5.53$ \\ \hline 
$165$ & $11.04$ & $13.85$ & $16.87$ &
$320$ & $3.10$ & $4.09$ & $5.20$ \\ \hline 
$170$ & $10.28$ & $12.92$ & $15.77$ &
$340$ & $3.06$ & $4.07$ & $5.19$ \\ \hline 
$175$ & $9.62$ & $12.11$ & $14.80$ &
$360$ & $3.26$ & $4.35$ & $5.58$ \\ \hline 
$180$ & $9.00$ & $11.36$ & $13.90$ &
$380$ & $3.09$ & $4.15$ & $5.36$ \\ \hline 
$185$ & $8.44$ & $10.66$ & $13.08$ &
$400$ & $2.76$ & $3.73$ & $4.83$ \\ \hline 
$190$ & $7.92$ & $10.03$ & $12.32$ &
$450$ & $1.85$ & $2.53$ & $3.32$ \\ \hline 
$195$ & $7.47$ & $9.47$ & $11.65$ &
$500$ & $1.17$ & $1.62$ & $2.15$ \\ \hline 
$200$ & $7.07$ & $8.99$ & $11.07$ &
$600$ & $0.47$ & $0.68$ & $0.92$ \\ \hline
\end{tabular}
\vspace*{-2mm}
\end{center}
\caption[The NNLO total production cross section in the $gg\to H$ channel
at the LHC with $\sqrt{s}=8,9,10$ TeV]{The cross sections in the
  $\protect{ \sigma^{\rm NNLO}( gg\to H)}$ at the $\lhc$ with $\sqrt
  s=8,9,10$ TeV (in pb) for given Higgs mass values (in GeV) at a
  central scale $\mu_F=\mu_R=\frac12 M_H$ using  the MSTW PDF set.}
\label{table:lhc8}
\vspace*{-4mm}
}
\end{table}

\subsubsection{The case of the designed LHC at $\sqrt s = 14$ TeV}

We end this subsection with the case of the designed LHC, with $\sqrt
s = 14$ TeV. It is expected to collect at least $\sim 30$ fb$^{-1}$ of
data, a luminosity that should allow either to discover the SM Higgs 
boson or to exclude its existence at 95\%CL in its entire mass range
$115 \leq\! M_H \!\lsim \!  800$ GeV.

\begin{figure}[!h]
\begin{center}
\vspace*{-.1mm}
\includegraphics[scale=0.65]{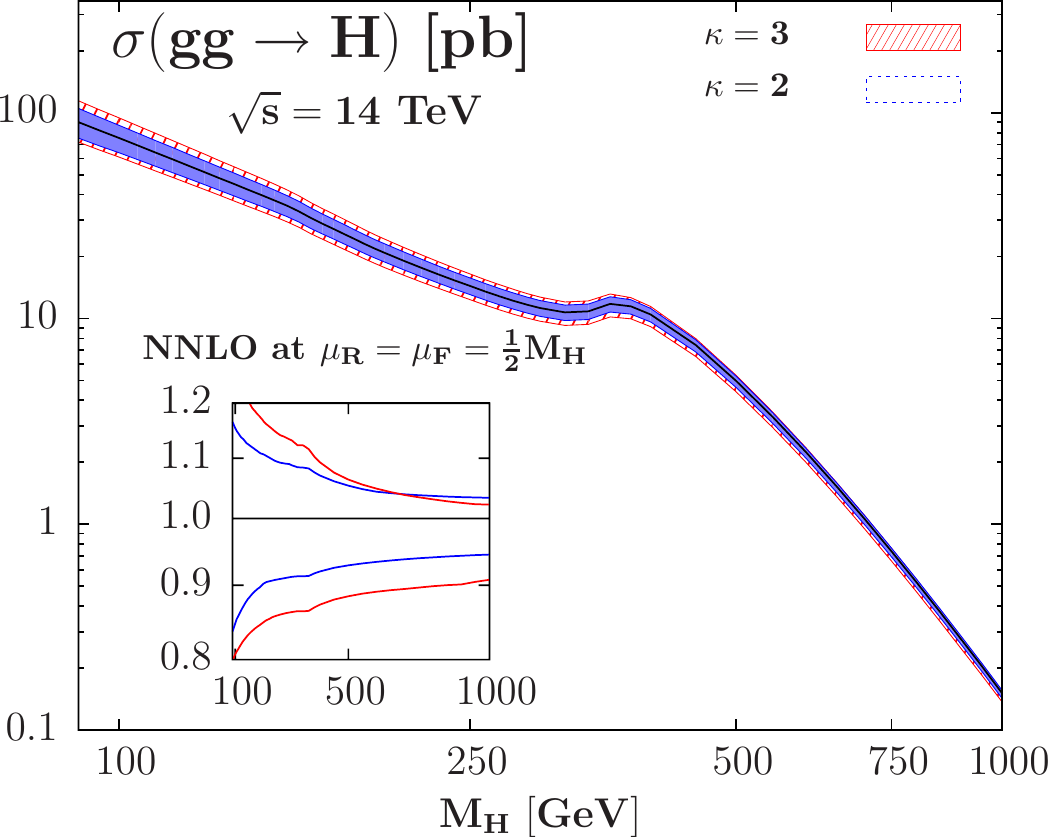}
\includegraphics[scale=0.65]{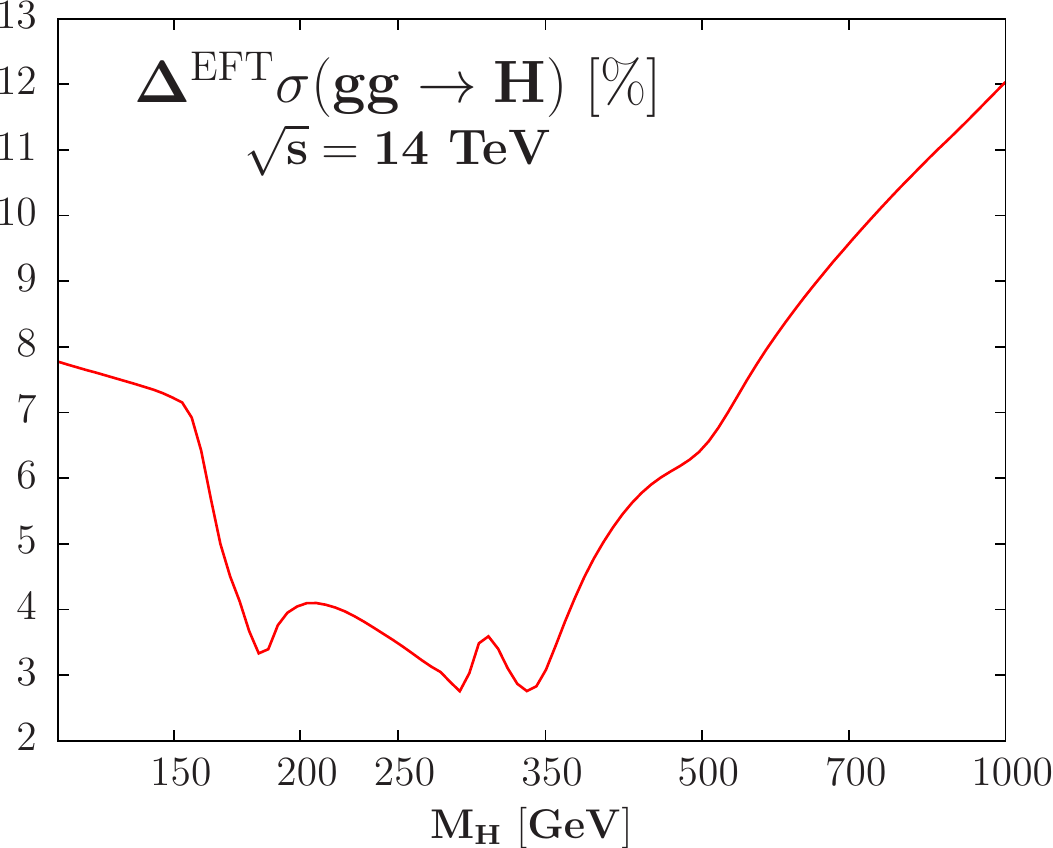}
\end{center}
\vspace*{-7mm}
\caption[Scale and total EFT uncertainties in $gg\to H$ at the LHC
with $\sqrt{s}=14$ TeV]{The uncertainty bands of $\sigma^{\rm NNLO}_{gg \to H}$ from
scale variation with $\kappa=2$ (left) and the total EFT uncertainty (right) 
at $\sqrt s=14$ TeV  as a function of $M_H$.}
\label{fig:scale_ggH_lhc14}
\end{figure}

\begin{figure}[!h]
\begin{bigcenter}
\vspace*{-.1mm}
\mbox{
\includegraphics[scale=0.65]{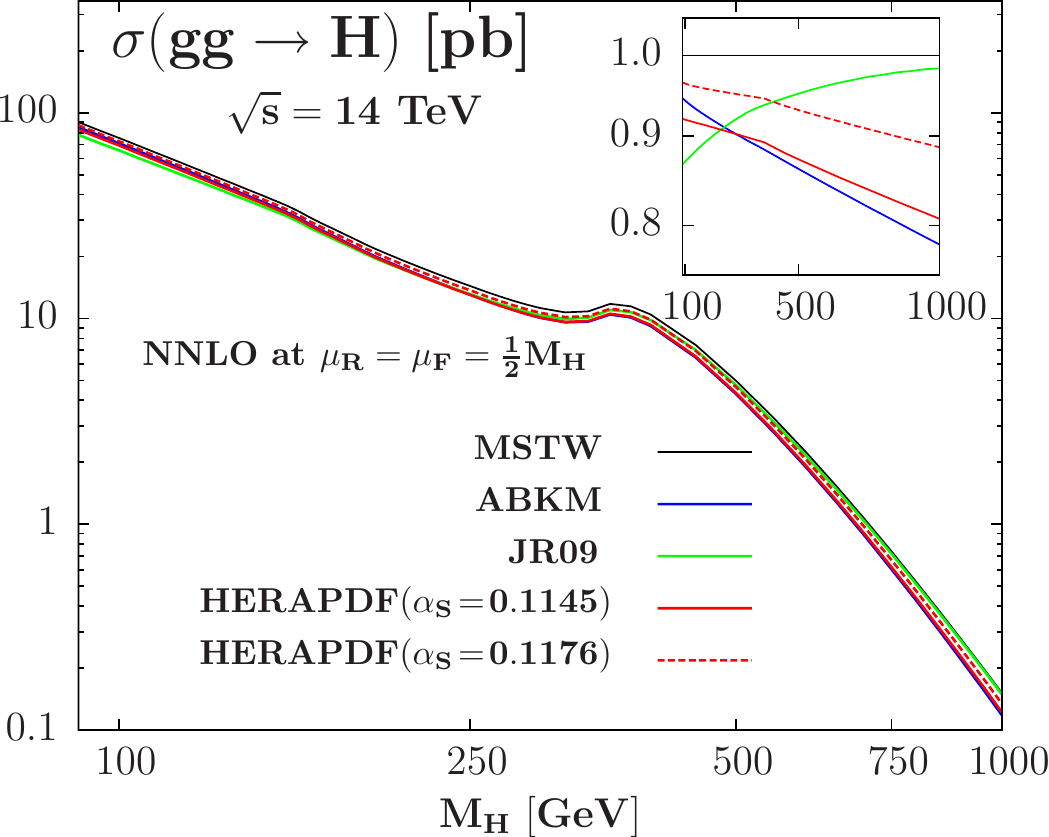}\hspace*{-0.1cm}
\includegraphics[scale=0.65]{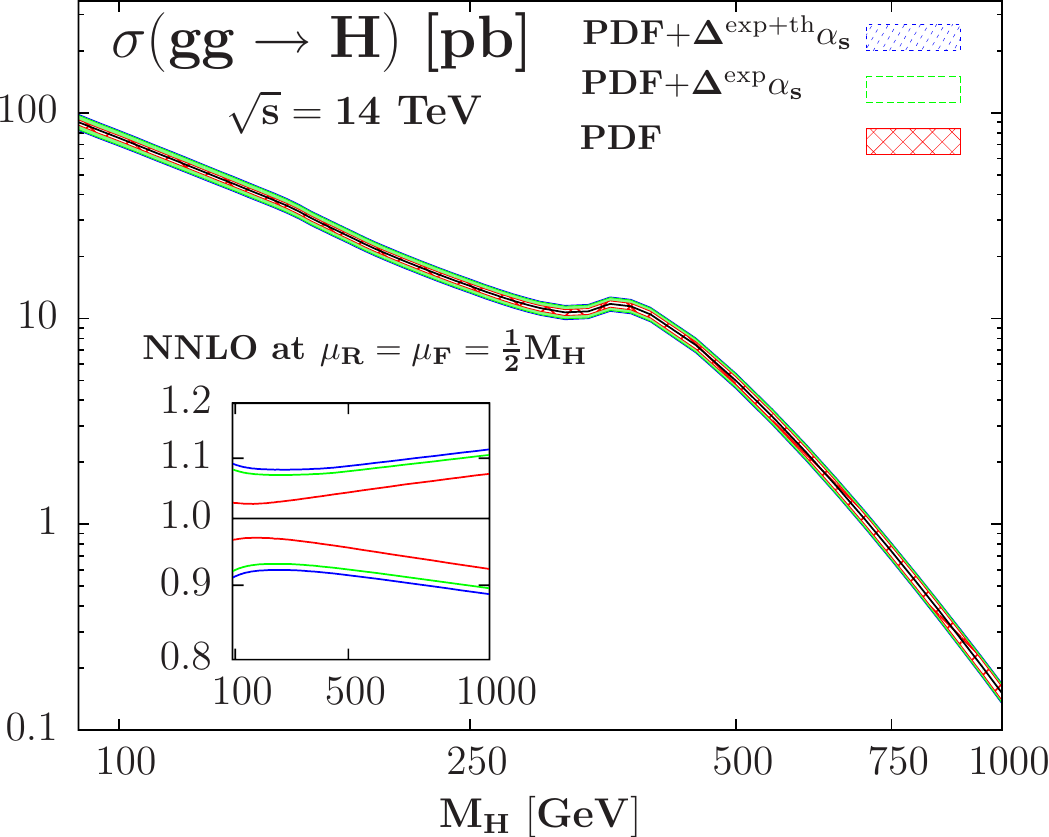}
}
\end{bigcenter}
\vspace*{-7mm}
\caption[PDF+$\Delta^{\rm exp,th}\alpha_s$ uncertainties and the
comparison between the 4 NNLO PDF sets in $gg\to H$ at the LHC with
$\sqrt{s}=14$ TeV]{The PDF uncertainties in $\sigma^{\rm NNLO}(g g\to H)$ at the LHC
with $\sqrt s= 14$ TeV as a function of $M_H$. Left: the central  values when
using the four NNLO PDFs and right:  the 90\% CL PDF, PDF+$\Delta^{\rm exp}
\alpha_s$ and PDF+$\Delta^{\rm exp}\alpha_s +\Delta^{\rm th}\alpha_s$ 
uncertainties in the MSTW scheme.}
\label{fig:pdf_ggH_lhc14}
\end{figure}

\begin{figure}[!h]
\begin{center}
\vspace*{-.01mm}
\includegraphics[scale=0.65]{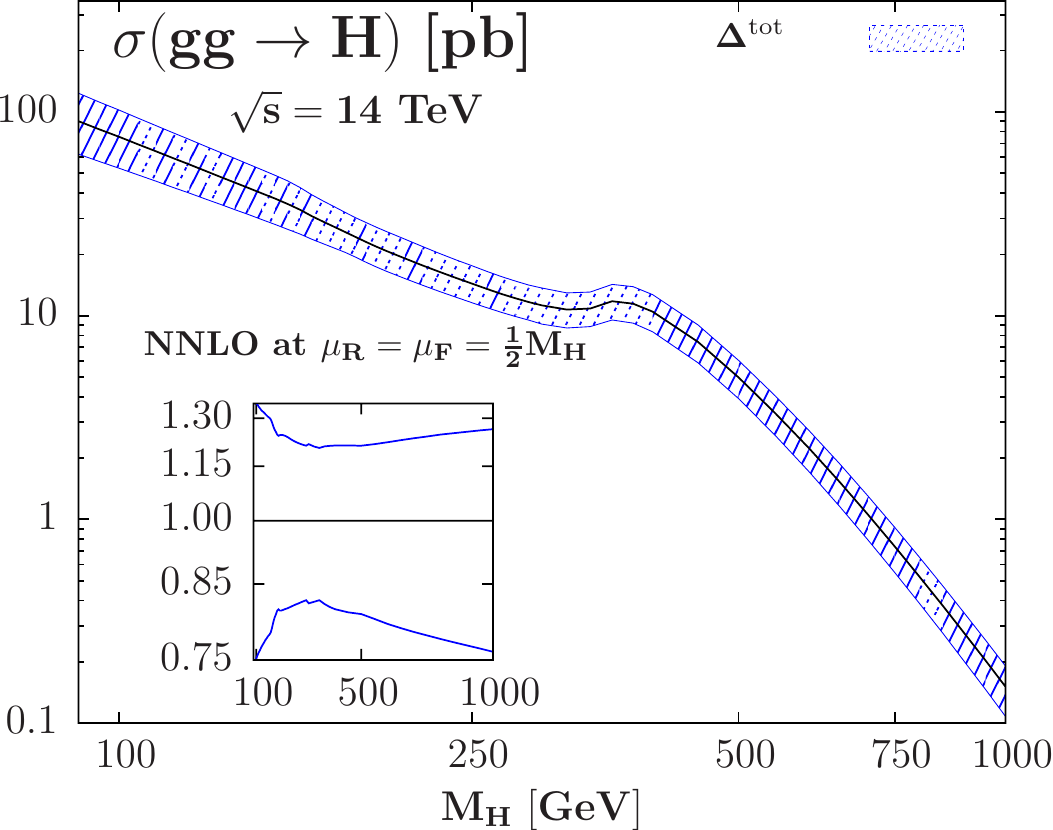}
\end{center}
\vspace*{-5mm}
\caption[Central prediction and total uncertainty in $gg\to H$ at NNLO
at the LHC with $\sqrt{s}=14$ TeV]{The cross section $\sigma^{\rm
    NNLO} (gg\to H)$ at the LHC with the uncertainty band when all
  theoretical uncertainties are added using our procedure A described
  in the text.}
\label{fig:ggH_all_lhc14}
\vspace*{-7mm}
\end{figure}

We use exactly the same outlines that were developed troughout this
section for the case of the $\lhc$ at $\sqrt s = 7$ TeV. The results
for the gluon--gluon fusion Higgs production cross section at the LHC
with $\sqrt s = 14$ TeV are displayed in Figs.~\ref{fig:scale_ggH_lhc14}
(for the scale uncertainty), \ref{fig:pdf_ggH_lhc14} for the
PDF+$\Delta^{\rm exp+th}\alpha_s$ uncertainty and \ref{fig:ggH_all_lhc14}
for the overall combination. The numerical results can also be found
in Table~\ref{table:lhc14}.

The main differences are highlighted in the next points:

\begin{itemize}

\item{The scale uncertainty is estimated by varying $\mu_R$ and
    $\mu_F$ in the domain $\frac14 M_H \le \mu_R=\mu_F \le M_H$, as
    was the case for the $\lhc$ at 7 TeV. The result does not
    significantly change when comparing to the $\lhc$ case: we obtain
    a variation of approximately $\pm 10\%$ at low Higgs masses and
    nearly $\pm 5\%$ at high Higgs masses, as displayed in the left
    part of Fig.~\ref{fig:scale_ggH_lhc14}.}

\item{The EFT/scheme uncertainty is almost exactly the same than at
    $\sqrt s=7$ TeV, as its most important component enters as a
    multiplicative factor in the $gg\!\to\! H$ amplitude but is larger
    starting from $M_H \gsim 500$ GeV. This is exemplified in the
    right part of Fig.~\ref{fig:scale_ggH_lhc14}.}

\item{The left part of Fig.~\ref{fig:pdf_ggH_lhc14} demonstrates that
    there is still a very large spread in the different predictions
    when folding the partonic $gg\to H$ cross section with the gluon
    luminosities given by the four NNLO PDFs sets available. The
    results obtained within the MSTW scheme for the 90\%CL PDF,
    PDF+$\Delta^{\rm exp}\alpha_s$ and the combined PDF+$\Delta^{\rm
      exp+th}\alpha_s$ uncertainties are displayed in the right part
    of Fig.~\ref{fig:pdf_ggH_lhc14}. A slightly smaller PDF+$\alpha_s$
    uncertainty than at the $\lhc$ is obtained, of order 1 to $2\%$,
    since lower Bjorken $x$ values are probed.}

\item{The overall uncertainty on $\sigma^{\rm NNLO} (gg\to H)$
    displayed in Fig.~\ref{fig:ggH_all_lhc14} is more or less the same
    at 14 TeV than at 7 TeV in the low Higgs mass range, but is
    slightly smaller for heavier Higgs bosons. As an example, we
    obtain for $M_H=500$ GeV a total $\approx \pm 21\%$ uncertainty at
    14 TeV compared to $\approx \pm 23\%$ at 7 TeV. We note that we
    have restricted to procedure A as we believe it is the most
    reasonable procedure as discussed in
    section~\ref{section:SMHiggsLHCTotal}.}

\end{itemize}\pagebreak

\vspace{-2mm}

\vspace{-2mm}
\clearpage

\begin{table}[!h]{\small%
\let\lbr\{\def\{{\char'173}%
\let\rbr\}\def\}{\char'175}%
\renewcommand{\arraystretch}{1.37}
\vspace*{1mm}
\begin{bigcenter}
\begin{tabular}{|c|ccc||c|ccc|}\hline
$\!M_H\!$ & $\sigma_{g g\to H}^{\pm \Delta_{\mu}\pm \Delta_{\rm PDF}\pm
\Delta_{\rm EFT}}$ & A & B &
$\!M_H\!$ & $\sigma_{g g\to H}^{\pm \Delta_{\mu}\pm \Delta_{\rm PDF}\pm 
\Delta_{\rm EFT}}$ & A & B\\ \hline 
$115$ & $59.37^{+9.4\% +8.7\% +7.8\%}_{-12.2\% -8.5\% -7.8\%} $ & $^{+28.1\%}_{-26.6\%} $ & $^{+25.8\%}_{-28.5\%}$ &
$240$ & $15.30^{+7.0\% +8.0\% +3.7\%}_{-8.2\% -7.8\% -3.7\%} $ & $^{+20.0\%}_{-18.6\%} $ & $^{+18.7\%}_{-19.6\%}$ \\ \hline
$120$ & $55.20^{+9.2\% +8.6\% +7.7\%}_{-11.9\% -8.4\% -7.7\%} $ & $^{+27.8\%}_{-26.1\%} $ & $^{+25.5\%}_{-28.0\%}$ &
$250$ & $14.38^{+6.9\% +8.0\% +3.5\%}_{-8.2\% -7.8\% -3.5\%} $ & $^{+19.7\%}_{-18.4\%} $ & $^{+18.4\%}_{-19.5\%}$ \\ \hline
$125$ & $51.45^{+9.0\% +8.5\% +7.6\%}_{-11.6\% -8.4\% -7.6\%} $ & $^{+27.4\%}_{-25.7\%} $ & $^{+25.1\%}_{-27.5\%}$ &
$260$ & $13.52^{+6.8\% +8.0\% +3.3\%}_{-8.1\% -7.8\% -3.3\%} $ & $^{+19.4\%}_{-18.2\%} $ & $^{+18.1\%}_{-19.2\%}$ \\ \hline
$130$ & $48.09^{+8.9\% +8.5\% +7.5\%}_{-11.4\% -8.3\% -7.5\%} $ & $^{+27.2\%}_{-25.5\%} $ & $^{+25.0\%}_{-27.3\%}$ &
$270$ & $12.79^{+6.7\% +8.0\% +3.1\%}_{-8.1\% -7.8\% -3.1\%} $ & $^{+19.1\%}_{-18.1\%} $ & $^{+17.9\%}_{-19.1\%}$ \\ \hline
$135$ & $45.06^{+8.7\% +8.4\% +7.5\%}_{-11.1\% -8.2\% -7.5\%} $ & $^{+26.8\%}_{-25.1\%} $ & $^{+24.6\%}_{-26.8\%}$ &
$280$ & $12.17^{+6.7\% +8.0\% +2.9\%}_{-8.1\% -7.8\% -2.9\%} $ & $^{+18.9\%}_{-17.8\%} $ & $^{+17.7\%}_{-18.8\%}$ \\ \hline
$140$ & $42.30^{+8.5\% +8.4\% +7.4\%}_{-10.8\% -8.2\% -7.4\%} $ & $^{+26.4\%}_{-24.7\%} $ & $^{+24.3\%}_{-26.4\%}$ &
$290$ & $11.65^{+6.6\% +8.0\% +2.8\%}_{-8.0\% -7.8\% -2.8\%} $ & $^{+18.6\%}_{-17.6\%} $ & $^{+17.4\%}_{-18.6\%}$ \\ \hline
$145$ & $39.80^{+8.4\% +8.3\% +7.3\%}_{-10.6\% -8.1\% -7.3\%} $ & $^{+26.0\%}_{-24.4\%} $ & $^{+24.0\%}_{-26.1\%}$ &
$300$ & $11.22^{+6.5\% +8.0\% +3.4\%}_{-8.0\% -7.8\% -3.4\%} $ & $^{+19.2\%}_{-18.4\%} $ & $^{+18.0\%}_{-19.3\%}$ \\ \hline
$150$ & $37.50^{+8.3\% +8.3\% +7.2\%}_{-10.4\% -8.1\% -7.2\%} $ & $^{+25.8\%}_{-24.1\%} $ & $^{+23.8\%}_{-25.7\%}$ &
$320$ & $10.70^{+6.5\% +8.1\% +3.1\%}_{-8.0\% -7.9\% -3.1\%} $ & $^{+18.8\%}_{-18.1\%} $ & $^{+17.8\%}_{-19.1\%}$ \\ \hline
$155$ & $35.32^{+8.1\% +8.3\% +7.0\%}_{-10.2\% -8.1\% -7.0\%} $ & $^{+25.3\%}_{-23.7\%} $ & $^{+23.4\%}_{-25.3\%}$ &
$340$ & $10.83^{+6.5\% +8.1\% +2.8\%}_{-8.0\% -7.9\% -2.8\%} $ & $^{+18.4\%}_{-17.8\%} $ & $^{+17.4\%}_{-18.7\%}$ \\ \hline
$160$ & $33.08^{+8.0\% +8.2\% +6.3\%}_{-10.0\% -8.0\% -6.3\%} $ & $^{+24.4\%}_{-22.9\%} $ & $^{+22.6\%}_{-24.4\%}$ &
$360$ & $11.77^{+6.4\% +8.1\% +3.5\%}_{-8.0\% -8.0\% -3.5\%} $ & $^{+19.0\%}_{-18.6\%} $ & $^{+18.1\%}_{-19.5\%}$ \\ \hline
$165$ & $30.77^{+7.9\% +8.2\% +5.3\%}_{-9.8\% -8.0\% -5.3\%} $ & $^{+23.1\%}_{-21.8\%} $ & $^{+21.4\%}_{-23.1\%}$ &
$380$ & $11.46^{+6.0\% +8.2\% +4.4\%}_{-7.7\% -8.1\% -4.4\%} $ & $^{+19.5\%}_{-19.2\%} $ & $^{+18.6\%}_{-20.2\%}$ \\ \hline
$170$ & $28.89^{+7.8\% +8.2\% +4.5\%}_{-9.7\% -8.0\% -4.5\%} $ & $^{+22.3\%}_{-21.0\%} $ & $^{+20.6\%}_{-22.3\%}$ &
$400$ & $10.46^{+5.6\% +8.2\% +5.0\%}_{-7.4\% -8.1\% -5.0\%} $ & $^{+19.7\%}_{-19.7\%} $ & $^{+18.9\%}_{-20.6\%}$ \\ \hline
$175$ & $27.24^{+7.8\% +8.2\% +4.0\%}_{-9.5\% -7.9\% -4.0\%} $ & $^{+21.6\%}_{-20.2\%} $ & $^{+20.0\%}_{-21.4\%}$ &
$450$ & $7.42^{+5.0\% +8.4\% +6.0\%}_{-7.0\% -8.3\% -6.0\%} $ & $^{+20.0\%}_{-20.4\%} $ & $^{+19.3\%}_{-21.3\%}$ \\ \hline
$180$ & $25.71^{+7.7\% +8.2\% +3.5\%}_{-9.4\% -7.9\% -3.5\%} $ & $^{+20.9\%}_{-19.6\%} $ & $^{+19.3\%}_{-20.8\%}$ &
$500$ & $4.97^{+4.6\% +8.6\% +6.4\%}_{-6.7\% -8.6\% -6.4\%} $ & $^{+20.4\%}_{-20.9\%} $ & $^{+19.7\%}_{-21.7\%}$ \\ \hline
$185$ & $24.28^{+7.6\% +8.1\% +3.3\%}_{-9.1\% -7.9\% -3.3\%} $ & $^{+20.6\%}_{-19.2\%} $ & $^{+19.0\%}_{-20.4\%}$ &
$550$ & $3.32^{+4.3\% +8.9\% +7.4\%}_{-6.5\% -8.8\% -7.4\%} $ & $^{+21.3\%}_{-21.9\%} $ & $^{+20.7\%}_{-22.7\%}$ \\ \hline
$190$ & $22.97^{+7.6\% +8.1\% +3.8\%}_{-9.1\% -7.9\% -3.8\%} $ & $^{+21.0\%}_{-19.6\%} $ & $^{+19.5\%}_{-20.7\%}$ &
$600$ & $2.24^{+4.1\% +9.2\% +8.3\%}_{-6.3\% -9.1\% -8.3\%} $ & $^{+22.1\%}_{-22.9\%} $ & $^{+21.6\%}_{-23.7\%}$ \\ \hline
$195$ & $21.83^{+7.5\% +8.1\% +4.0\%}_{-9.0\% -7.9\% -4.0\%} $ & $^{+21.1\%}_{-19.7\%} $ & $^{+19.6\%}_{-20.8\%}$ &
$650$ & $1.53^{+3.9\% +9.5\% +9.0\%}_{-6.2\% -9.4\% -9.0\%} $ & $^{+22.7\%}_{-23.9\%} $ & $^{+22.4\%}_{-24.6\%}$ \\ \hline
$200$ & $20.83^{+7.4\% +8.1\% +4.1\%}_{-8.8\% -7.9\% -4.1\%} $ & $^{+21.0\%}_{-19.6\%} $ & $^{+19.5\%}_{-20.7\%}$ &
$700$ & $1.05^{+3.8\% +9.8\% +9.6\%}_{-6.1\% -9.6\% -9.6\%} $ & $^{+23.4\%}_{-24.6\%} $ & $^{+23.2\%}_{-25.3\%}$ \\ \hline
$210$ & $19.10^{+7.3\% +8.1\% +4.1\%}_{-8.6\% -7.8\% -4.1\%} $ & $^{+20.9\%}_{-19.4\%} $ & $^{+19.5\%}_{-20.5\%}$ &
$800$ & $0.52^{+3.5\% +10.4\% +10.5\%}_{-6.0\% -10.2\% -10.5\%} $ & $^{+24.6\%}_{-26.1\%} $ & $^{+24.4\%}_{-26.7\%}$ \\ \hline
$220$ & $17.64^{+7.2\% +8.1\% +4.0\%}_{-8.4\% -7.8\% -4.0\%} $ & $^{+20.6\%}_{-19.2\%} $ & $^{+19.3\%}_{-20.3\%}$ &
$900$ & $0.27^{+3.4\% +11.0\% +11.3\%}_{-5.9\% -10.7\% -11.3\%} $ & $^{+25.6\%}_{-27.5\%} $ & $^{+25.7\%}_{-28.0\%}$ \\ \hline
$230$ & $16.38^{+7.1\% +8.0\% +3.8\%}_{-8.3\% -7.8\% -3.8\%} $ & $^{+20.4\%}_{-18.9\%} $ & $^{+19.0\%}_{-20.0\%}$ &
$1000$ & $0.15^{+3.3\% +11.5\% +12.0\%}_{-5.8\% -11.3\% -12.0\%} $ & $^{+26.4\%}_{-28.8\%} $ & $^{+26.9\%}_{-29.1\%}$ \\ \hline
\end{tabular}
\end{bigcenter} 
\vspace*{-3mm}
\caption[The NNLO total Higgs production cross section in the $gg\to H$
process at the LHC with $\sqrt{s} = 14$ TeV together with the
associated theoretical uncertainties]{The NNLO total Higgs production
  cross sections in the $\protect{gg\to H}$ process at the LHC with
  $\sqrt s=14$ TeV (in pb) for given Higgs mass values (in GeV) at a
  central scale $\mu_F=\mu_R=\frac12 M_H$. We also display the
  corresponding shifts due to the theoretical uncertainties from the
  various sources discussed (first from scale, then from
  PDF+$\Delta^{\rm \exp+th}\alpha_s$ at 90\%CL and from EFT), as well
  as the total uncertainty when all errors are added  using the
  procedures A and B  described in the text.} 
\label{table:lhc14}
\vspace*{-1mm}
}
\end{table}

Table~\ref{table:lhc14} which displays the cross sections together with the
individual and overall theoretical uncertainties for the Higgs masses relevant
at the LHC summarizes the results obtained at the full--fledged LHC
with $\sqrt s = 14$ TeV.

\subsection{Summary and outlook}

In this section were presented the theoretical predictions for the
Standard Model Higgs boson production total cross section at the early
$\lhc$, that is the LHC at 7 TeV, in the gluon--gluon fusion $gg\to H$
that is the main channel. We have assumed a central scale
$\mu_F=\mu_R=\mu_0$ to be $\mu_0=\frac12 M_H$ and calculated the cross
section up to NNLO in QCD and EW corrections.

We have then estimated the theoretical uncertainties associated to the
prediction: the scale uncertainty, the uncertainties from the PDF
parametrisation and the associated error on $\alpha_s$, as well as
uncertainties due to the use of the EFT approach for the mixed
QCD-electroweak radiative corrections and the $b$-quark loop 
contribution in the gluon--gluon fusion case. We have followed the
recipe proposed for the first time in the case of the Tevatron
collider and presented in section~\ref{section:SMHiggsTevTotal} for
the combination of the different sources of uncertainties. The results
that we obtain in the case of the $\lhc$ are smaller than at the
Tevatron, mainly because the QCD corrections are smaller at the
$\lhc$, the $K$--factor beeing smaller. There is also a better
behaviour of the gluon density at the high Bjorken $x$ values probed
at the $\lhc$.

We have also given some predictions for the intermediate
center--of--mass energies $\sqrt s = 8,9, 10$ TeV as well as for the
full--fledged $\sqrt s= 14$ TeV LHC. The total uncertainty is in all
cases nearly the same as for the $\lhc$ with $\sqrt s = 7$ TeV.

We have finished the analysis of the production cross section. Before
the investigation of the consequences on the experimental results we
have to analyze the crucial Higgs decay branching ratios in the next
section.

\vfill
\pagebreak

\section[Higgs decay and the implications for Higgs searches]{Higgs decay
  branching ratios and the implication on Higgs searches at hadron
  colliders}

\label{section:SMHiggsDecay}

The last two sections have been devoted to the Higgs production itself,
following the main production channels at the Tevatron and $\lhc$
colliders. The focus has been on the total cross section regardless of
the Higgs decay channel. This section is then devoted to the crucial
study of the Higgs decay branching ratios, as the experiments are
looking for the traces of the Higgs boson in their detectors, which
then require that the Higgs decay chain has to be well understood.

We will first review the most important channels for experimental
searches at the Tevatron and the LHC. The main reference for this
subsection is Ref.~\cite{Djouadi:2005gi}. We then discuss the theoretical
uncertainties affecting the Higgs decay branching ratios, that have
not been considerered in the experimental
analyses~\cite{Aaltonen:2011gs,ATLAS:1303604,CMS-NOTE-2010-008} up
until now and is under active investigation in
Ref.~\cite{Dittmaier:2011ti, Baglio:2010ae}. Indeed, while the Higgs
decays into lepton and gauge boson pairs are well under control (as 
mainly small electroweak effects are involved), the partial decays 
widths into quark pairs and gluons are plagued with uncertainties that
are mainly due to the imperfect knowledge of the bottom and charm quark
masses and the value of the strong coupling constant $\alpha_s$. This
was first studied in Ref.~\cite{Djouadi:1995gt} and the analysis presented in 
Ref.~\cite{Baglio:2010ae} was an update of this previous result, and will be
the subject of this section. We show that at least in the intermediate
mass range, $M_H\approx 120$--150 GeV, where the SM Higgs decay rates
into $b\bar b$ and $W^+W^-$ final states have the same order of
magnitude, the parametric uncertainties on these two main Higgs decay
branching ratios are non--negligible, being of the order of 3 to 10\%
at the $1\sigma$ level.

\subsection{Important channels for experimental
  search \label{section:SMHiggsDecayIntro}}

\begin{figure}[!t]
\begin{center}
\includegraphics[scale=1.5]{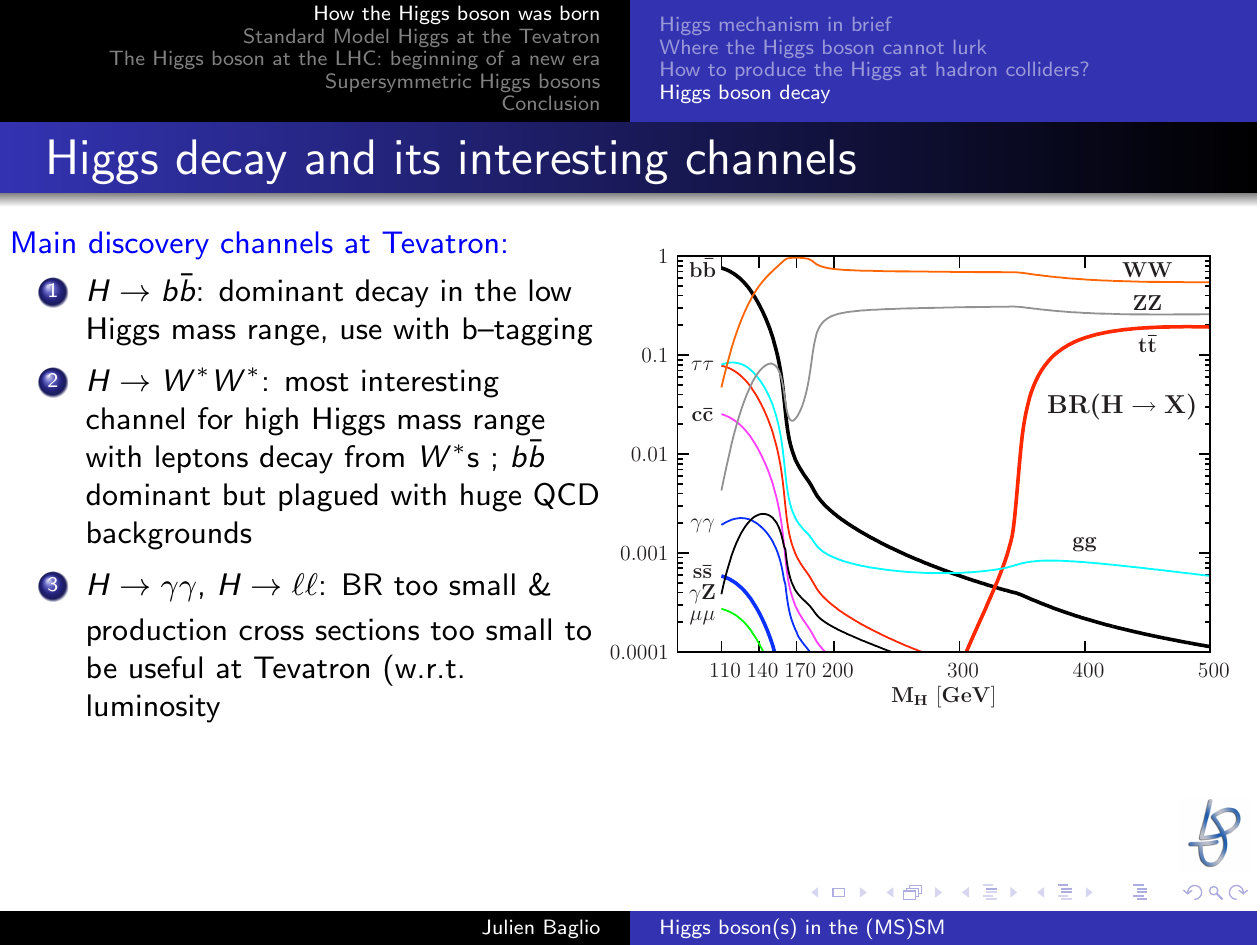}
\end{center}
\vspace*{-4mm}
\caption[SM Higgs decay channels on the interesting Higgs mass range]{SM
  Higgs boson decay branching ratios on the Higgs mass range relevant
  at the Tevatron and the $\lhc$, calculated with the programm {\tt
    HDECAY}~\cite{Djouadi:1997yw}.}
\vspace*{-2mm}
\label{fig:Hdecay_intro}
\end{figure}

We present in this subsection the main channels used at the Tevatron
and the LHC for the experimental searches of the SM Higgs boson. We
display in Fig.~\ref{fig:Hdecay_intro} the decay pattern on the Higgs
mass range that is relevant for the current search at the Tevatron and
in particular at the LHC, which will illustrate the comments that follow for
each important decay mode. We already point out that the reader can
find most of the information relative to the LHC case in
Refs.~\cite{Ball:2007zza, Aad:2008aa, Aad:2009wy} which summurize all
the efforts made by the two major ATLAS and CMS collaborations.

\subsubsection{$H\to b\bar b$ channel}

The partial width of the decay of the Higgs boson into quark pairs
is given at LO in QCD by
\beq
\Gamma(H\to q\bar q)=\frac{3 G_F}{4\pi \sqrt 2} M_H m_q^2
\left(1-4\frac{m_q^2}{M_H^2}\right)^{3/2}
\label{eq:Hdecay_bbar}
\eeq

Eq.~\ref{eq:Hdecay_bbar} then shows that the partial width is
proportionnal to the square of the quark mass for not too light Higgs
masses, which means that the dominant channel will be for the bottom
quark. This partial width receives large NLO QCD corrections, and the
quark mass has to be taken in the $\overline{\rm MS}$
renormalization scheme in order to absorb large logarithms
corrections. We refer the reader to Ref.~\cite{Djouadi:2005gi}, page 39. In
Refs.~\cite{Kniehl:1993jc, Kniehl:1994ph} the NNLO QCD corrections have
been calculated as well as the three-loops $\mathcal{O}(\alpha_s^2 G_F m_t^2)$ in
Refs.~\cite{Chetyrkin:1996ke, Chetyrkin:1996wr}.

The decay of the SM Higgs boson in quark pairs is one of the most
important channels for experimental searches in particular for the
bottom quark. This is the dominant mode for light Higgs masses $M_H
\lsim 135$ GeV at the Tevatron through the Higgs production in the
Higgs--strahlung channel $p(p/\bar p) \to VH$ with $V$ being either a
$W$ or a $Z$ boson~\cite{Carena:2000yx}.

This mode is the golden mode at the Tevatron for the low Higgs mass
searches, with $\ell \nu b \bar b$, $\ell \ell b \bar b$ and $\nu \bar
\nu b \bar b$ final states where the requirements of isolated leptons
and missing energy help to reduce the background. The use of
$b$--tagging techniques crucially improves the signal--over--background
ratio and is used along with neural network techniques for the signal
reconstruction and the rejection of $p\bar p\to VV$, $p\bar p\to t
\bar t$ and $p\bar p\to t + X$ backgrounds, see
Ref.~\cite{Bhat:2000dr} and Ref.~\cite{Bhat:2000dr} (page 69).

The situation at the LHC is nos as good as the QCD background is far
too large. Only with a luminosity of order 30 fb$^{-1}$ at the
full--fledged LHC we may obtain a 3$\sigma$ evidence at $M_H = 120$
GeV~\cite{RichterWas:2000uy, RichterWas:2000ux}. This decay mode also
exists in association with a SM
Higgs production through associated top quark production $p p\to t\bar
t H$ and can improve the LHC search, but this require a very challenging
$b$--tagging with four bottom quarks in the final state. A clear
evidence is beyond the reach of the $\lhc$ at 7 TeV and require a
luminosity greater than 100 fb$^{-1}$ at the LHC to claim a discovery
with only this channel~\cite{RichterWas:1999sa, Drollinger:2001ym,
  CMS:2001039, Cammin:685523}.

\subsubsection{$H\to V^* V^*$ channel}

As displayed in Fig.~\ref{fig:Hdecay_intro} the Higgs decay channels
involving a pair of weak bosons are the dominant channels for Higgs
mass above $M_H=135$ GeV. These channels are thus of utmost importance
both at the Tevatron and the LHC colliders.

We should distinguish between two regimes: either above the
$WW$ and $ZZ$ threesold where both weak bosons are real and then
follow their own decay chains, or below the threesold where we have
off--shell weak bosons and thus three and four--body decays. The
partial width for for the $H\to VV$ with boths $V$ bosons being
on--shell is given at LO by
\beq
\Gamma(H\to VV) = \frac{G_F M_H^3}{16\sqrt{2}\pi} A_V \sqrt{1-4z}
\left(12z^2-4z+1\right),~~x=\frac{M_V^2}{M_H^2}
\label{eq:Hdecay_VV}
\eeq

The symmetry factor $A_V$ is either $A_W=2$ or $A_Z=1$. We see that
for very heavy Higgs bosons we have $\Gamma(H\to VV)\propto M_H^3$
which means, this channel being largely dominant, that $\Gamma^{\rm
  tot}_H \propto M_H^3$ for very heavy Higgs bosons. It implies that
considering the Higgs boson still as a particule is far from obvious
given that its width is comparable to its mass. 

In order to obtain the decay width where the two weak bosons are
off--shell we have to take into account the total decay width of the
weak bosons through a Breit--Wigner modelisation and use
Eq.~\ref{eq:Hdecay_VV} with general $p^{V}_1$ and $p^{V}_2$ momentum. It
then has the expression given in Ref.\cite{Grau:1990uu}:

\begin{eqnarray}
\Gamma(H\to V^* V^*)= & \\ \nonumber
 & \hspace{-30mm}\displaystyle \frac{1}{\pi^2} \int_0^{M_H^2} dp_1^2 \frac{M_V
  \Gamma_V}{(M_V-p_1^2)^2+M_V^2\Gamma_V^2} \int_0^{(M_H-p_1)^2} dp_2^2
\frac{M_V \Gamma_V}{(M_V-p_2^2)^2+M_V^2\Gamma_V^2}
\Gamma(p_1^2,p_2^2,M_H^2)
\label{eq:Hdecay_VVoffshell}
\end{eqnarray}

with $\Gamma(p_1^2,p_2^2,M_H^2)$ being given by Eq.~\ref{eq:Hdecay_VV}
but with $p_1^2 \neq M_V^2$ and $p_2^2\neq M_V^2$. NLO corrections
have been calculated in Refs.~\cite{Chanowitz:1978uj, Chanowitz:1978mv,
  Fleischer:1980ub, Dawson:1988th, Hioki:1989eu, Kniehl:1990mq,
  Kniehl:1991xe} as well as higher order corrections up to
NNLO~\cite{Bredenstein:2006ha}  that are summarized in
Ref.~\cite{Djouadi:2005gi} page 58.

Higgs seaches through these decay channels are very powerful, using
either the gluon--gluon fusion production mode, the vector boson
fusion mode or the Higgs--strahlung processes:

\begin{enumerate}[-]
\item{{\it $H\to W^{(*)} W^{(*)}$}: the Higgs decaying into a pair of
    $W$ bosons is very promising when the Higgs has been produded in
    the gluon--gluon fusion channel or the Higgs--strahlung process
    $p(p/\bar p)\to HW$. The channel $gg\to H\to WW\to \ell \nu \ell
    \nu$ is the most powerful channel at the Tevatron for high Higgs
    masses $M_H \gsim 135$ GeV searches and very promising at the LHC
    even at low Higgs masses around $M_H=120$ GeV, see
    Refs.~\cite{Glover:1988fn, Aad:2009wy}. We have to know
    very precisely the $WW$ and $ZZ$ backgrounds as well as top quark
    pair production in order to observe a clear excess of events in
    the distributions. Spin correlation and suitable cuts help to
    improve the significance. The process $p\bar p\to HW\to WWW,WWZ$
    at the Tevatron is not very useful as the production cross section
    is too small. It is more promising at the LHC for Higgs masses
    between $M_H=160$ GeV and $M_H=180$ GeV, see Ref.~\cite{Djouadi:2005gi}
    page 122. Finally the WW/ZZ fusion followed by the $H\to WW^{(*)}$
    decay is very useful at the LHC with the jet veto techniques. A
    $3\sigma$ evidence in this channel can be obtained with a
    luminosity of order 30 fb$^{-1}$, see Ref.~\cite{Asai:2004ws}.}

\item{{\it $H\to Z^{(*)} Z^{(*)}$}: in association with the Higgs
    boson production through gluon--gluon fusion, this is the golden
    channel for high Higgs mass searches $M_H\gsim 2 M_Z$ both at the
    Tevatron and the LHC~\cite{Stirling:1985bi, Gunion:1985vt,
      Gunion:1986cc}, where in the latter case a discovery can already
    be claimed for a Higg mass $M_H\simeq 150$ GeV with a 5 fb$^{-1}$
    luminosity~\cite{Aad:2009wy}, and with a 8 fb$^{-1}$ luminosity in the
    CMS detector on the entire high mass range (except for $M_H\simeq
    170$ GeV where 100 fb$^{-1}$ in this channel is needed), see
    Ref.~\cite{Ball:2007zza}. The main background is the
    $ZZ$ production which can be removed with a side--bands study and
    the interpolated in the signal region. This channel is not useful
    with other production channels because of the smallness of the
    production cross section and the very high backgrounds.} 
\end{enumerate}

\subsubsection{$H\to \gamma \gamma$ channel}

The $H\to \gamma \gamma$ channel as well as the $H\to gg$ decay
process that will be discussed below are loop--induced
processes. These processes are known up to NNLO in EW corrections and
NNLO in QCD corrections in the infinite quark mass limit in the loops,
see page 58 of Ref.~\cite{Djouadi:2005gi}. The expression for the LO partial
decay widths are similar to what has been obtained for the $gg\to
H$ production channel with the same functions $A$ and $f$ involved,
see Eq.~\ref{eq:LOfunction}:
\beq
\Gamma(H\to \gamma \gamma) = \frac{G_F \alpha_{\rm EM}^2
  M_H^3}{128\pi^3\sqrt 2} \left| \sum_{\rm fermions} N_c Q_f^2
  A\left(\frac{M_H^2}{m_f^2}\right) +
  B\left(\frac{M_H^2}{M_W^2}\right)\right|^2
\label{eq:Hdecay_gamgam}
\eeq
with $N_c=3$ for the quarks, $N_c=1$ for the leptons, $Q_f$ being the
electric charge of the fermion $f$ and $B$ as the form factor

$$B(\tau) = -\left( 2+ \frac{3}{\tau} + 3 \frac{(2\tau -1)}{\tau^2}
  f(\tau) \right) \, .$$

This decay channel is the best mode for light Higgs mass searches $M_H
\lsim 150$ GeV at the LHC~\cite{Fayard:682120, CMS:19940289,
  Froidevaux:682442, Gianotti:682459, Wielers:684265}, and has been used in
Tevatron experimental analyses for $M_H \leq 130$ GeV but as a
sub--dominant channel and only to improve the efficiency of the
experimental programm with multivariate techniques. Even if the
branching fraction is small as can be seen in
Fig.~\ref{fig:Hdecay_intro}, its very clean final state compensate the
trickiness of the analyses. Both $pp\to VH$
and $gg\to H$ production channels are used even if the former is not
as useful as the latter. The backgrounds are similar in both
cases. The QCD background from jets that fake photons is huge and
require a very efficient triggering of the electromagnetic
calorimeters of ATLAS and CMS detectors. These backgrounds can be
determined by the measurements of the side--bands in the invariant
di--photon mass distribution. Nevertheless this channel require a
large amount of data as a small bump in the di--photon mass
distribution is searched.

At low luminosities the combination of all production channels is
required: $gg\to H\to \gamma \gamma$, $gg\to H\to \gamma \gamma + {\rm
jets}$, $pp\to HW\to \gamma \gamma \ell \nu$ and $pp\to t\bar t H\to
b\ell \nu \bar b \bar \ell \bar \nu \gamma \gamma$ will enhance the
statistics and we may then have a $5\sigma$ discovery though this
decay mode with a luminosity of order 30 fb$^{-1}$ and already a
3$\sigma$ evidence for $\mathcal{L} = 10$ fb$^{-1}$~\cite{Aad:2009wy}.

\subsubsection{A comment on the $H\to gg$ channel}

The $H\to gg$ channel will be discussed in the following subsection,
even if it adds little to the experimental searches. We just mention
that the theoretical predictions follow a very similar recipe
compared to that of the di--photon decay channel presented above. We
have at LO:
\beq
\Gamma(H\to gg) = \frac{G_F\alpha_s^2 M_H^3}{36\pi^3\sqrt 2} \left|
  \frac34 \sum_{b,t} A\left(\frac{M_H^2}{m_q^2}\right)\right|^2
\eeq

We will discuss in brief this channel when dealing with the
uncertainties that affect its branching fraction. The QCD corrections
are known up to $\mathcal{O}(\alpha_s^2)$ order in the infinite top
mass approximation, see Ref.~\cite{Chetyrkin:1997iv}.

\subsubsection{$H\to \tau^+ \tau^-$ channel}

The last important decay channel for experimental searches is the only
leptonic channel that is not swamped by the other decay channels. The
partial width at LO is given in Eq.~\ref{eq:Hdecay_bbar} where we
replace $m_q$ by $m_\tau$ and divide the result by the factor of 3, as
the tau lepton has no color factor. The NLO and NNLO corrections
follow what has been presented in the $H\to b\bar b$ subsubsection
above.

The production channel which allows for a search in this decay mode is
at first look only the vector boson fusion $pp\to qq H$ process. It
has been considered in particular in LHC analyses for the intermediate
Higgs mass $M_H \in [120~{\rm GeV}~;~140~{\rm GeV}]$
range~\cite{Cavalli:2002vs} (page 56), where for a luminosity of order
30 fb$^{-1}$ a $6\sigma$ combined statistical significance can be
obtained when considering all possible decay products of the
$\tau$'s. The major backgrounds are found to be QCD production $pp\to
Zjj$ with the $Z$ boson decaying into a di--$\tau$ pair. This can be
measured in the side--bands of the invariant mass distribution of the
lepton pair and then extrapolated in the signal region.

In the section~\ref{section:SMHiggsTauTau} of this thesis we will
present a possible new way for the use of this decay channel in
association with gluon--gluon fusion Higgs production.


\subsection{Uncertainties on the branching
  ratios \label{section:SMHiggsDecayResult}}

The sections~\ref{section:SMHiggsTev} and~\ref{section:SMHiggsLHC}
have introduced the detailed analysis of the theoretical uncertainties
affecting the production channels for the SM Higgs boson. We need to
perform the same investigation when dealing with the Higgs decay
branching ratios in order to obtain in the end a combined cross section
times branching ratios analysis with all sources of theoretical
uncertainties taken into account.

\subsubsection{The parametric uncertainties}

An analysis of the uncertainties affecting the Higgs branching ratios has been 
performed long ago in Ref.~\cite{Djouadi:1995gt}. This section, based on the
results published in Ref.~\cite{Baglio:2010ae}, update the earlier results
by taking into account the parametric uncertainties on the value of
the strong coupling constant $\alpha_s(M_Z)^2$ as well as the value of the
bottom quark and charm quark masses. These uncertainties are taken as
being experimental which means that we will not take into account the
theoretical uncertainty on the value of $\alpha_s$ in this analysis,
mainly because the impact would be negligible in the view of the final
combination of the uncertainties on the branching fractions. As we
consider these uncertainties as experimental they have a statistical
ground, and we will choose to stick at the 68\%CL (i.e. 1$\sigma$
level) for the choice of the error interval.

We will then use this following recipe:

\begin{enumerate}[a)]
\item{The strong coupling constant $\alpha_s$ plays a very important
    role either at LO for the $H\to gg$ channel or beyond LO when
    dealing with the QCD corrections; the uncertainty will be thus
    significant on the branching ratios. We adopt for consistency
    reasons the same value that we have used for the determination of
    the central predictions for the production cross section in the
    former sections: $\alpha_s(M_Z^2)=0.1171\pm 0.0014$ at NNLO at the
    68\%CL.}

\item{In Ref.~\cite{Baglio:2010ae} we used $\overline{\rm MS}$ masses
    evaluated at the mass itself as starting points for the $b$ and
    $c$ quarks, with the central values and the uncertainties taken
    directly from Ref.~\cite{Nakamura:2010zzi}\footnote{with these starting inputs
      and the value of $\alpha_s(M_Z^2)$ given in point a), the
      central values of the $b,c$ pole masses are $M_b=4.71$ GeV and
      $M_c=1.54$ GeV using a 2--loops calculation with variation
      ranges of, respectively, 4.64--4.90 GeV and 1.42--1.63 GeV. For
      the pole bottom--quark mass, the central value is rather close
      to the one adopted in the MSTW scheme for parameterizing the
      parton densities~\cite{Martin:2009iq}, $M_b=4.75$ GeV, and it was
      expected that the difference would have no practical
      impact}. Their uncertainties were larger than those considered
    in Ref.~\cite{Chetyrkin:2009fv} with quite the same central values. In a
    recent discussion within the LHC Higgs Cross Section Working Group
    a new agreement has been reached and the results presented in the
    following will use $\overline{m}_b(\overline{m}_b)=4.16\pm 0.06$ GeV and
    $\overline{m}_c(\overline{m}_c)=1.28\pm 0.03$ GeV, which then
    transform into pole masses $M_b=4.49\pm 0.06$ GeV and
    $M_c=1.41\pm 0.03$ GeV with a 1--loop calculation\footnote{We
      greatly thank A. Singer for having clarified the discussion
      about the final values to be taken.}, as the
    parameters in Ref.~\cite{Nakamura:2010zzi} were seen to be too
    conservative on 
    the one side, the parameters in Ref.~\cite{Chetyrkin:2009fv} being seen
    a little too optimistic on the other side. When we will quote separately
    the impact of these uncertainties on the Higgs branching ratio, we
    will assume the central value $\alpha_s(M_Z^2)=0.1171$ at NNLO
    for the strong coupling constant\footnote{We thank M. Spira for having
    clarified the link between the order at which the transformation
    from $\overline{\rm MS}$ masses to pole masses is done and the
    correlation between $\alpha_s$ and the uncertainties on the quark
    masses.}.}

\item{For the $H\to gg$ channel we will consider only the parametric
    uncertainties quoted above. At NNLO, the pure theoretical scale
    uncertainty is of the order 10\% but since the branching ratio is
    small it migrates into an error of less than  1\%. In addition,
    the N$^3$LO contribution would reduce this scale variation to the
    level of a few percent at most and this uncertainty can be safely
    neglected. The same analysis can be done for the other channel
    that we will consider in the following, meaning that the scale
    variation uncertainties will be neglected in the rest of the section.}
\end{enumerate}

All these three points mentioned above and in particular the last
point c) are dependant on the interval of variation for the Higgs
mass. Indeed for $M_H\gsim 350$ GeV the $H\to t \bar t$ opens up, as
displayed in Fig.~\ref{fig:Hdecay_intro}. We should then perform a
detailed analysis of this channel for these high mass regime and in
particular take into account the scale variation. Nevertheless this
channel might play a role only for Higgs masses greater than $M_H
\simeq 400-450$ GeV, where we reach the possibility of the $\lhc$ at 7
TeV, not to mention the Tevatron experiments. This is why its analysis
has been totally discarded in this thesis.

\subsubsection{The final uncertainties on the branching ratios}

In order to calculate the central values for the branching ratios
together with the associated uncertainties we use the latest version
3.80 of the program {\tt HDECAY}~\cite{Djouadi:1997yw} available, which
calculates the partial and total Higgs decay widths including the
relevant QCD and electroweak higher order corrections. It is worth
mentioning that the version used in Ref.~\cite{Baglio:2010ae} was older: the
results presented in this section are then an update of the previous
results published in this latter reference. We will use the parameters
discussed above and add in quadrature the final errors that we obtain
in order to present the total uncertainties on the different branching
ratios considered. We stress that the result we obtain are very
similar to those presented in Ref.~\cite{Baglio:2010ae} in the crucial decay
channels that are the $H\to b \bar b$ and $H\to W^{(*)} W^{(*)}$
modes. Thus the conclusion of Ref.~\cite{Baglio:2010ae} remains globally
unchanged.

\begin{figure}[!h]
\vspace*{.1cm}
\begin{center}
\hspace*{-.2cm}
\includegraphics[scale=0.9]{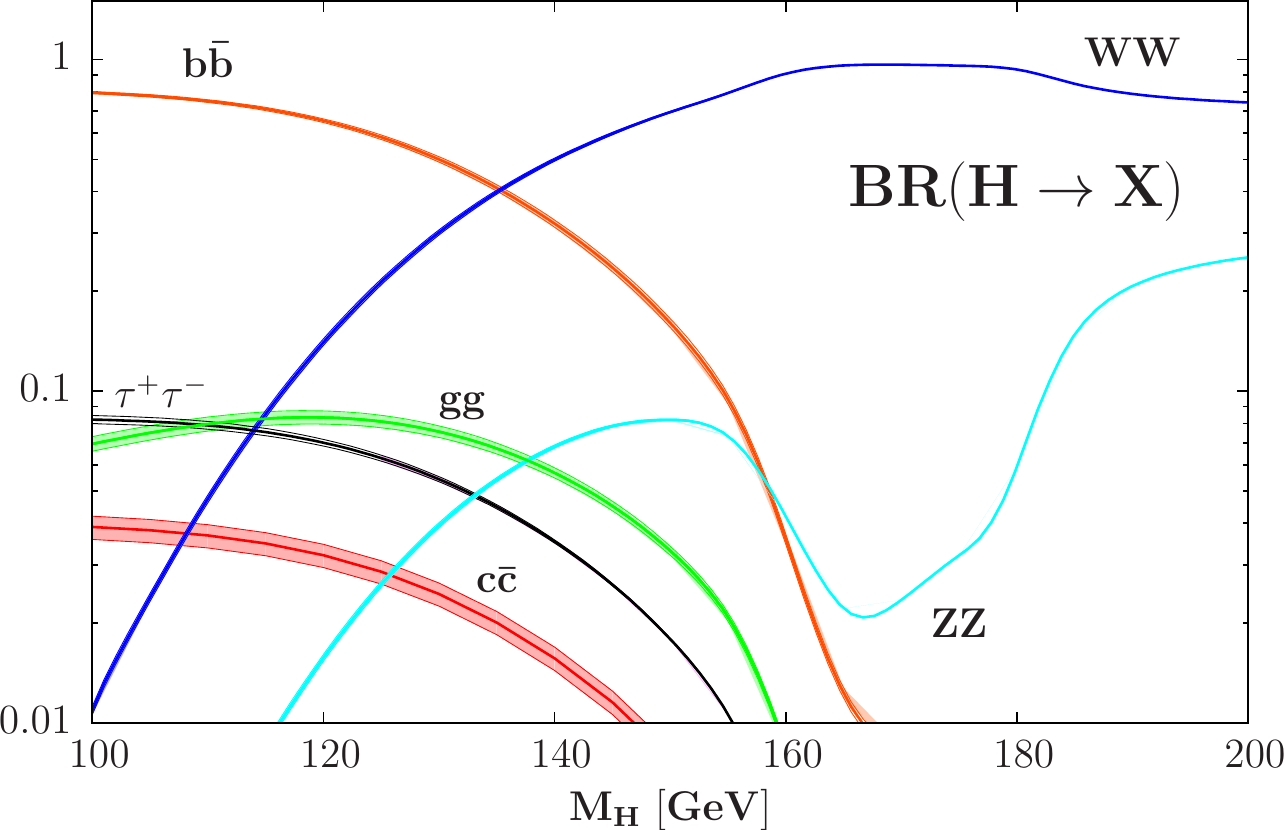}
\end{center}
\vspace*{-6mm}
\caption[The Higgs decays branching ratios together with the total
uncertainty bands]{The Higgs decay branching ratios as a function of
  $M_H$ including the total uncertainty bands from the 1$\sigma$
  errors on the input quark masses and the coupling $\alpha_s$ (the
  individual errors have been added in quadrature).} 
\vspace*{-2mm}
\label{fig:Hdecay_uncertainties}
\end{figure}

We obtain the branching ratios (BR) shown in
Fig.~\ref{fig:Hdecay_uncertainties} as a function of the
Higgs mass, including the total uncertainty bands. We present in
Table~\ref{table:Hdecay1} the branching ratios for the main decay
modes $H\to b\bar b$ and $H\to W^{(*)}W^{(*)}$, for selected values of
the Higgs mass, $M_H=120,135$ and 150 GeV, together with the
individual and total uncertainties. The uncertainties (in percentage)
on the branching ratios for the $H\to \tau^+\tau^-$ and $H\to
ZZ,\gamma \gamma$ channels are the same as those affecting the $H\to
WW$ mode. In Table~\ref{table:Hdecay2}, we display the branching
ratios for the various decays as well as their corresponding total
uncertainties for a selection of Higgs masses that are relevant for
the $\lhc$ and the Tevatron
colliders. Fig.~\ref{fig:Hdecay_uncertainties} displays the branching
ratios for Higgs mass up to $M_H=200$ GeV. The numbers above this mass
are not useful as the uncertainties for the only relevant channels,
that are the $ZZ$ and $WW$ modes, drop to the level of 0\%. We will
thus not include the values for Higgs mass above 200 GeV and up to
$M_H=500$ GeV in Table~\ref{table:Hdecay2} but they are of course
available on demand.

The largest total errors are by far the ones that affect BR($H \to c\bar c)$ 
which are of the order of 8\%. In contrast the errors on BR$(H\to
gg)$ are at the level of a few to 5\% at most and the errors in the
$WW$ channel do not exceed 3\%. In the case of the $H\to c\bar c$
channel, it is mainly due to the uncertainty on the input charm--quark mass
$\overline{m}_c$ and, to a lesser extent, to the uncertainty on $\alpha_s$ ; their
combination leads to a very strong variation of the charm--quark mass at the
high scale $\mu$, $\overline{m}_c(\mu) \propto [\alpha_s
(\mu)]^{12/13}$. For the $H\to g g$ channel the uncertainty is mainly
due to the uncertainty on $\alpha_s$, and in both cases, the error on
the input $b$--quark mass leads to a 3\% uncertainty at
most. Nevertheless, since the branching ratios for these two decays
are small, at most a few percent in the Higgs mass range of interest,
the associated uncertainties will affect the Higgs total width, and
hence the branching ratios for the other decay channels, in a less
significant way.

\begin{table}[!h]{\small%
\let\lbr\{\def\{{\char'173}%
\let\rbr\}\def\}{\char'175}%
\renewcommand{\arraystretch}{1.4}
\vspace*{-1mm}
\begin{center}
\begin{tabular}{|c|c|c|ccc|c|}\hline
channel & $M_H$ & BR(\%) & $\Delta m_c$ & $\Delta m_b$
& $\Delta\alpha_s$ & $\Delta$BR    \\ \hline
& $120$ & 65.40 & $^{+0.2\%}_{-0.2\%}$ & $^{+1.1\%}_{-1.1\%}$ & 
$^{+0.6\%}_{-0.6\%}$ & $^{+1.3\%}_{-1.3\%}$ 
\\ 
$H\to b\bar b$ & $135$ & 41.01 & $^{+0.1\%}_{-0.1\%}$ & $^{+1.9\%}_{-1.9\%}$ & 
$^{+1.1\%}_{-1.1\%}$ & $^{+2.2\%}_{-2.2\%}$ 
\\ 
& $150$ & 16.07 & $^{+0.1\%}_{-0.1\%}$ & $^{+2.7\%}_{-2.7\%}$ & 
$^{+1.5\%}_{-1.5\%}$ & $^{+3.1\%}_{-3.1\%}$ 
\\ \hline
& $120$ & 14.06 & $^{+0.2\%}_{-0.1\%}$ & $^{+2.1\%}_{-2.0\%}$ & 
$^{+1.1\%}_{-1.0\%}$ & $^{+2.4\%}_{-2.2\%}$ 
\\ 
$H\to WW$ & $135$ & 39.86 & $^{+0.1\%}_{-0.1\%}$ & $^{+1.3\%}_{-1.3\%}$ & 
$^{+0.6\%}_{-0.6\%}$ & $^{+1.5\%}_{-1.4\%}$ 
\\ 
& $150$ & 69.45 & $^{+0.0\%}_{-0.1\%}$ & $^{+0.5\%}_{-0.5\%}$ & 
$^{+0.2\%}_{-0.2\%}$ & $^{+0.6\%}_{-0.5\%}$ 
\\ \hline
\end{tabular}
\end{center}
\vspace*{-3mm}
\caption[The SM Higgs decay branching ratios in the $b\bar b$ and $WW$
modes for representatives Higgs masses together with the different
sources of uncertainties as well as the total uncertainty.]{The Higgs
  decay branching ratio  BR($H\rightarrow b\bar b)$ and
  BR($H\rightarrow WW)$ (in \%) for given Higgs mass values (in GeV)
  with the corresponding uncertainties from the various sources
  discussed in the text; the total uncertainties (adding the
  individual ones in quadrature) are also shown.}
\label{table:Hdecay1}
\vspace*{-2mm}
}
\end{table}

The uncertainty on the $H\to b\bar b$ decay channel that is the most
important channel at the Tevatron for low Higgs mass searches, depends
strongly on the considered Higgs mass range. For $M_H  \lsim 120$ GeV
where the branching ratio is BR$(H\to b\bar b) \gsim 65\%)$ and thus
largely the dominant channel, the uncertainty is less than a percent,
reaching the percent level at $M_H=110$ GeV. This is mainly due to the
fact that as the channel is dominant, the $b\bar b$ partial width is
the major component of the total width and the errors partly cancel in
the branching ratio. There is, however, a residual error coming from
the input bottom mass and to a lesser extent $\alpha_s$ and the charm
mass errors. This is also exemplified by the comparison between the
results presented in this section and the former results of
Ref.~\cite{Baglio:2010ae} where the errors on the input bottom mass where
larger: we see that the errors on the branching fractions where also
correspondingly larger.

In the mass range $M_H\! \approx\! 120$--150 GeV, the partial widths
for $H\!\to\! b\bar b$ and $WW$ decays have the same magnitude (we
reach the crossing region of the two decay modes) and the two
branching ratios have larger uncertainties. The uncertainty on
BR$(H\!\to\! b\bar b)$ increases from 1\% at $M_H\! \approx\! 120$ GeV
to approximately 3\%, that is the triple, at $M_H\!\approx \!150$
GeV. At the same time, the uncertainty on BR$(H\!\to \!WW)$, which is
the same as the one on BR($H\!\to\! ZZ^*)$ and BR($H\! \to
\!\gamma\gamma)$, drops from $\pm 2.5\%$ at $M_H \! \approx \! 120$
GeV to $\pm 0.5\%$ at $M_H\! \approx \!150$ GeV. In this case
$\overline{m}_c$ has nearly no impact as BR$(H\!\to \!c\bar c)$ is
small. For higher Higgs masses, $M_H \gsim 160$ GeV, the Higgs width is mainly
controlled by the $H\to WW$ decay and then jointly with $H \to ZZ$ for
$M_H \gsim 2M_Z$. The uncertainty on the branching ratio BR$(H\to WW)$
drops to nearly 0\%. The uncertainty on BR$(H\to b\bar b)$ remains at
the 8\% level but this has no impact as the branching ratio is far too
small to be relevant.

For Higgs masses beyond $M_H \gsim 350$ GeV, the decay channel $H\to t\bar t$
opens up and has a branching ratio of $\approx 20\%$ at $M_H \approx 500$ GeV  
dropping to less than $10\%$ for $M_H \lsim 400$ GeV or $M_H \gsim 700$ GeV. It
will be affected by the uncertainties on $m_t, \alpha_s$ as well as by some
electroweak contributions from the Higgs self--couplings. This leads to a 
total uncertainty that is estimated to be below the 5\% level~\cite{Dittmaier:2011ti},
which translates to an uncertainty of less than 1\% in the branching ratios for
the important decays $H \to WW,ZZ$. 

Thus, the uncertainties on the important Higgs branching ratios
BR($H\to WW,ZZ,\gamma \gamma)$ as well as BR($H\to b\bar b)$  can be
significant in the intermediate mass range $M_H \approx 120$--150 GeV where
the $b\bar b$ and $WW$ decays are competing with each other. Even if
the uncertainty remains reasonable in this new analysis, at most 8\%,
this should be taken into account in experimental analyses. We also
see that the conclusions remain the same as in Ref.~\cite{Baglio:2010ae}
even if the uncertainties have been reduced by a factor of 2.

\begin{table}
\renewcommand{\arraystretch}{1.3}
\small
\begin{center}
\begin{tabular}{|c||cc|cc|cc|ccccc|}\hline
$M_H$ & $b\bar b $ & $\Delta^{\rm tot}$
& $c\bar c$ & $\Delta^{\rm tot}$ & gg & $\Delta^{\rm tot}$
&WW & ZZ & $\tau \tau$ & $\gamma \gamma$ & $\Delta^{\rm tot}$\\ \hline
$100$ & 79.48 & $^{+0.8\%}_{-0.8\%}$ &
3.90 & $^{+7.8\%}_{-8.3\%}$ & 6.94 & $^{+5.2\%}_{-5.0\%}$ &
1.09 & 0.11 & 8.22 & 0.16 & $^{+2.9\%}_{-2.8\%}$ \\ \hline
$105$ & 77.69 & $^{+0.8\%}_{-0.9\%}$ &
3.81 & $^{+7.8\%}_{-8.3\%}$ & 7.50 & $^{+5.1\%}_{-4.9\%}$ &
2.39 & 0.21 & 8.11 & 0.18 & $^{+2.8\%}_{-2.7\%}$ \\ \hline
$110$ & 74.96 & $^{+1.0\%}_{-1.0\%}$ &
3.68 & $^{+7.8\%}_{-8.3\%}$ & 7.97 & $^{+5.0\%}_{-4.8\%}$ &
4.75 & 0.43 & 7.89 & 0.19 & $^{+2.7\%}_{-2.6\%}$ \\ \hline
$115$ & 70.95 & $^{+1.1\%}_{-1.1\%}$ &
3.48 & $^{+7.8\%}_{-8.2\%}$ & 8.27 & $^{+4.8\%}_{-4.7\%}$ &
8.54 & 0.86 & 7.53 & 0.21 & $^{+2.6\%}_{-2.5\%}$ \\ \hline
$120$ & 65.40 & $^{+1.3\%}_{-1.3\%}$ &
3.21 & $^{+7.9\%}_{-8.2\%}$ & 8.34 & $^{+4.6\%}_{-4.5\%}$ &
14.06 & 1.57 & 7.00 & 0.22 & $^{+2.4\%}_{-2.2\%}$ \\ \hline
$125$ & 58.32 & $^{+1.6\%}_{-1.6\%}$ &
2.86 & $^{+7.9\%}_{-8.1\%}$ & 8.11 & $^{+4.4\%}_{-4.2\%}$ &
21.35 & 2.62 & 6.29 & 0.23 & $^{+2.1\%}_{-2.0\%}$ \\ \hline
$130$ & 50.00 & $^{+1.9\%}_{-1.9\%}$ &
2.45 & $^{+7.9\%}_{-8.1\%}$ & 7.56 & $^{+4.1\%}_{-4.0\%}$ &
30.13 & 3.95 & 5.43 & 0.22 & $^{+1.8\%}_{-1.8\%}$ \\ \hline
$135$ & 41.01 & $^{+2.2\%}_{-2.2\%}$ &
2.01 & $^{+8.0\%}_{-8.1\%}$ & 6.72 & $^{+3.8\%}_{-3.7\%}$ &
39.86 & 5.42 & 4.49 & 0.21 & $^{+1.5\%}_{-1.4\%}$ \\ \hline
$140$ & 32.02 & $^{+2.5\%}_{-2.5\%}$ &
1.57 & $^{+8.1\%}_{-8.1\%}$ & 5.68 & $^{+3.6\%}_{-3.5\%}$ &
49.92 & 6.82 & 3.53 & 0.19 & $^{+1.1\%}_{-1.1\%}$ \\ \hline
$145$ & 23.60 & $^{+2.8\%}_{-2.8\%}$ &
1.16 & $^{+8.2\%}_{-8.1\%}$ & 4.51 & $^{+3.4\%}_{-3.3\%}$ &
59.82 & 7.86 & 2.62 & 0.17 & $^{+0.8\%}_{-0.8\%}$ \\ \hline
$150$ & 16.07 & $^{+3.1\%}_{-3.1\%}$ &
0.79 & $^{+8.3\%}_{-8.2\%}$ & 3.31 & $^{+3.2\%}_{-3.1\%}$ &
69.45 & 8.21 & 1.79 & 0.14 & $^{+0.5\%}_{-0.6\%}$ \\ \hline
$155$ & 9.47 & $^{+3.3\%}_{-3.3\%}$ &
0.46 & $^{+8.3\%}_{-8.2\%}$ & 2.09 & $^{+3.0\%}_{-3.0\%}$ &
79.28 & 7.33 & 1.06 & 0.10 & $^{+0.3\%}_{-0.3\%}$ \\ \hline
$160$ & 3.58 & $^{+3.6\%}_{-3.5\%}$ &
0.18 & $^{+8.3\%}_{-8.3\%}$ & 0.84 & $^{+2.9\%}_{-2.8\%}$ &
90.63 & 4.19 & 0.40 & 0.05 & $^{+0.1\%}_{-0.1\%}$ \\ \hline
$165$ & 1.23 & $^{+3.6\%}_{-3.6\%}$ &
0.06 & $^{+8.4\%}_{-8.3\%}$ & 0.30 & $^{+2.8\%}_{-2.8\%}$ &
95.97 & 2.22 & 0.14 & 0.02 & $^{+0.0\%}_{-0.0\%}$ \\ \hline
$170$ & 0.82 & $^{+3.7\%}_{-3.6\%}$ &
0.04 & $^{+8.5\%}_{-8.3\%}$ & 0.21 & $^{+2.8\%}_{-2.7\%}$ &
96.43 & 2.35 & 0.09 & 0.02 & $^{+0.0\%}_{-0.0\%}$ \\ \hline
$175$ & 0.63 & $^{+3.7\%}_{-3.6\%}$ &
0.03 & $^{+8.5\%}_{-8.3\%}$ & 0.17 & $^{+2.8\%}_{-2.8\%}$ &
95.84 & 3.21 & 0.07 & 0.01 & $^{+0.0\%}_{-0.0\%}$ \\ \hline
$180$ & 0.51 & $^{+3.7\%}_{-3.6\%}$ &
0.02 & $^{+8.4\%}_{-8.4\%}$ & 0.15 & $^{+2.8\%}_{-2.7\%}$ &
93.28 & 5.94 & 0.06 & 0.01 & $^{+0.0\%}_{-0.0\%}$ \\ \hline
$185$ & 0.40 & $^{+3.7\%}_{-3.6\%}$ &
0.02 & $^{+8.5\%}_{-8.4\%}$ & 0.12 & $^{+2.8\%}_{-2.7\%}$ &
84.52 & 14.86 & 0.05 & 0.01 & $^{+0.0\%}_{-0.0\%}$ \\ \hline
$190$ & 0.32 & $^{+3.7\%}_{-3.6\%}$ &
0.02 & $^{+8.5\%}_{-8.4\%}$ & 0.10 & $^{+2.8\%}_{-2.7\%}$ &
78.72 & 20.77 & 0.04 & 0.01 & $^{+0.0\%}_{-0.0\%}$ \\ \hline
$195$ & 0.28 & $^{+3.7\%}_{-3.7\%}$ &
0.01 & $^{+8.4\%}_{-8.4\%}$ & 0.10 & $^{+2.7\%}_{-2.7\%}$ &
75.89 & 23.66 & 0.03 & 0.01 & $^{+0.0\%}_{-0.0\%}$ \\ \hline
$200$ & 0.25 & $^{+3.7\%}_{-3.6\%}$ &
0.01 & $^{+8.4\%}_{-8.4\%}$ & 0.09 & $^{+2.7\%}_{-2.7\%}$ &
74.26 & 25.34 & 0.03 & 0.01 & $^{+0.0\%}_{-0.0\%}$ \\ \hline
\end{tabular}
\end{center}
\caption[The SM Higgs decay branching ratios together with the total
uncertainty for the most important decay channels]{The Higgs decay
  branching ratios (in \% ) for given Higgs mass values (in GeV) with
  the corresponding total uncertainties from the various sources
  discussed in the text.}
\label{table:Hdecay2}
\end{table}


\subsection{Combination at the
  Tevatron \label{section:SMHiggsFinalTev}}

The section~\ref{section:SMHiggsDecayResult} above has introduced the
branching ratios together with their uncertainties that are of
parametric nature and have thus be combined with a quadratic
addition. We need in the end to combine both the production cross
section and the branching fractions in order to obtain a complete
prediction for each specific search channel.

This combination is in general complicated to perform when several
production and decay channels are involved. Unfortunately that is the
case at the Tevatron where both $p\bar p\to HV$ and $gg\to H$
production channels on the one hand, both $H\to b\bar b$ and $H\to
W^{(*)}W^{(*)}$ on the other hand have to be considered especially in
the intermediate mass range $M_H \simeq 135$ GeV. In addition to this
difficulty we should also take into account two important issues:

\begin{enumerate}[$i)$]
\item{The total Higgs decay width becomes large at high Higgs masses
    and thus should be taken into account in the processes $p (p/\bar
    p)\to H+X\to X' +X$ where the narrow width approximation for the
    Higgs boson propagator fails.}
\item{We should take into account interferences between the signal and
    background, as for example in $gg\to H\to VV$ and $gg\to VV$.}
\end{enumerate}

These two issues will not be addressed in this thesis, and we refer
the reader to a work in progress~\cite{Djouadi}. We also note that there
are anti--correlations between the different decay rates as the sum of
all branching ratios is equal to unity.

Combining the uncertainty in the production rate and the uncertainty
in the Higgs decay branching ratios is then not obvious at all. We
should also think about the diffence between the pure theoretical
uncertainties considered in the case of the production cross sections
and the uncertainties on the branching ratios which (mainly) come from
experimental errors on the input parameters $\alpha_s$ and the quark
masses. In statistical language this means that we should make a
combination of a flat prior uncertainty and a gaussian prior
uncertainty.  In addition, there is one parameter which is common in
the calculation of the branching ratios and the production cross
sections: the coupling $\alpha_s$. The uncertainty on $\alpha_s$ will
affect at the same time $\sigma(gg\to H)$ and BR($H\to X)$; it occurs
that in both the Higgs production and in Higgs decays the minimal
(maximal) values are obtained with the minimal (maximal) value of
$\alpha_s$ when the error $\Delta^{\rm exp} \alpha_s$ is included.

We will then adopt a simple method that is a linear addition of the
errors on the Higgs branching ratios to the theoretical uncertainties
on the production cross section, keeping in mind that if there are
anti--correlations between the production and the decay chain the
related uncertainty should nearly vanish. For consistency reason we
should increase the uncertainty due to the experimental errors on the
strong coupling constant $\alpha_s$ in the branching ratios as in the
production cross sections we have taken the $90\%$CL PDF+$\Delta^{\rm
  exp} \alpha_s$ uncertainty. We will present results for the two main
search channels $p\bar p\to HV$ followed by $H\to b\bar b$ decay and
$p\bar p\to H$ through gluon--gluon fusion followed by $H\to W^+W^-$
decay. In the intermediate Higgs mass range the two decay compete each
other. The combination of the two channels should then take into
account the correlations between the two decay channels on the one
hand, and should also take into account the relative importance of the
two production channels on the other hand. We thus have a mean with
anti--correlated uncertainties on the $b$--quark mass between the two
decay chains. The final result will be used in
section~\ref{section:SMHiggsTevExclusion} in order to compare with the
experimental
results~\cite{Aaltonen:2010yv,Tevatron:2010ar,Aaltonen:2011gs}.

\begin{figure}[!h]
\begin{bigcenter}
\hspace{-4mm}
\includegraphics[scale=0.65]{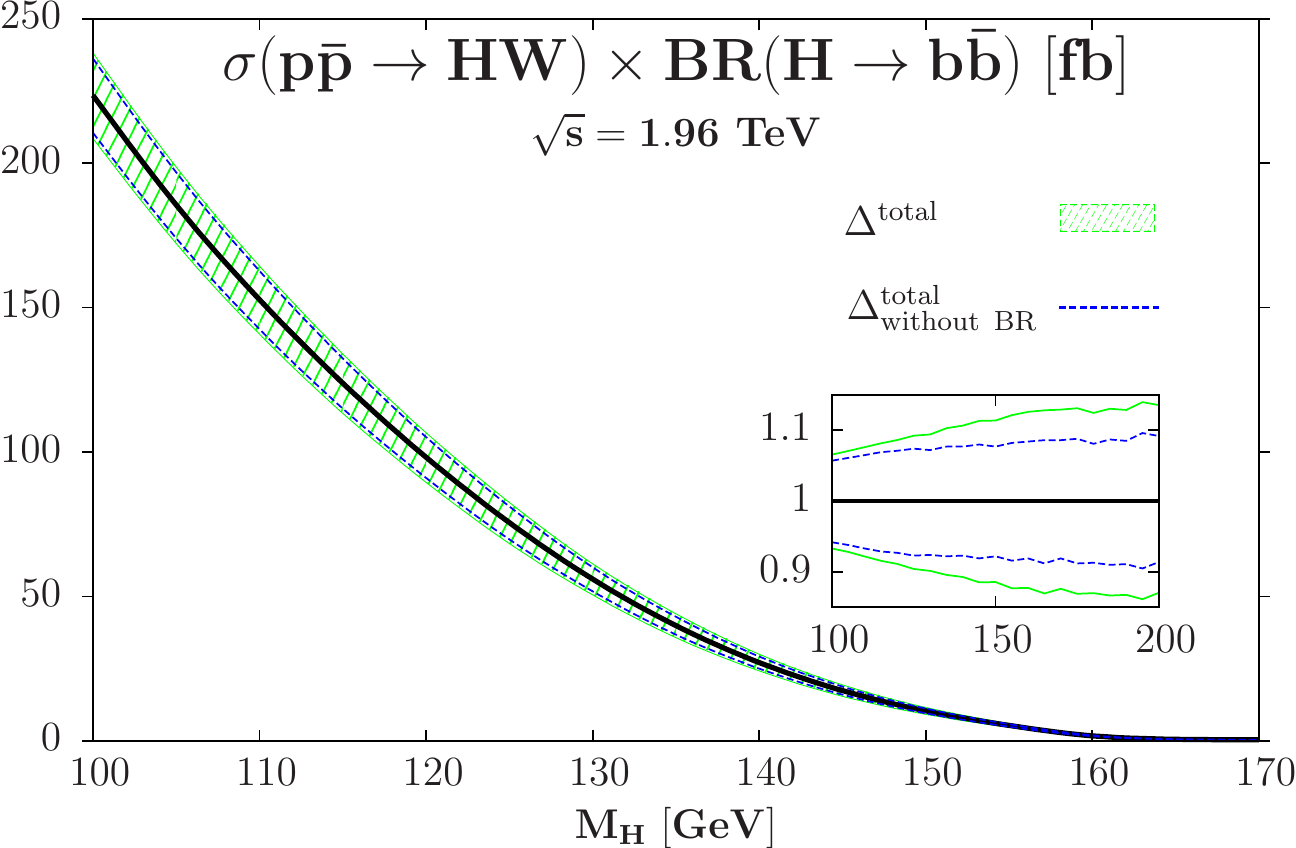}
\includegraphics[scale=0.65]{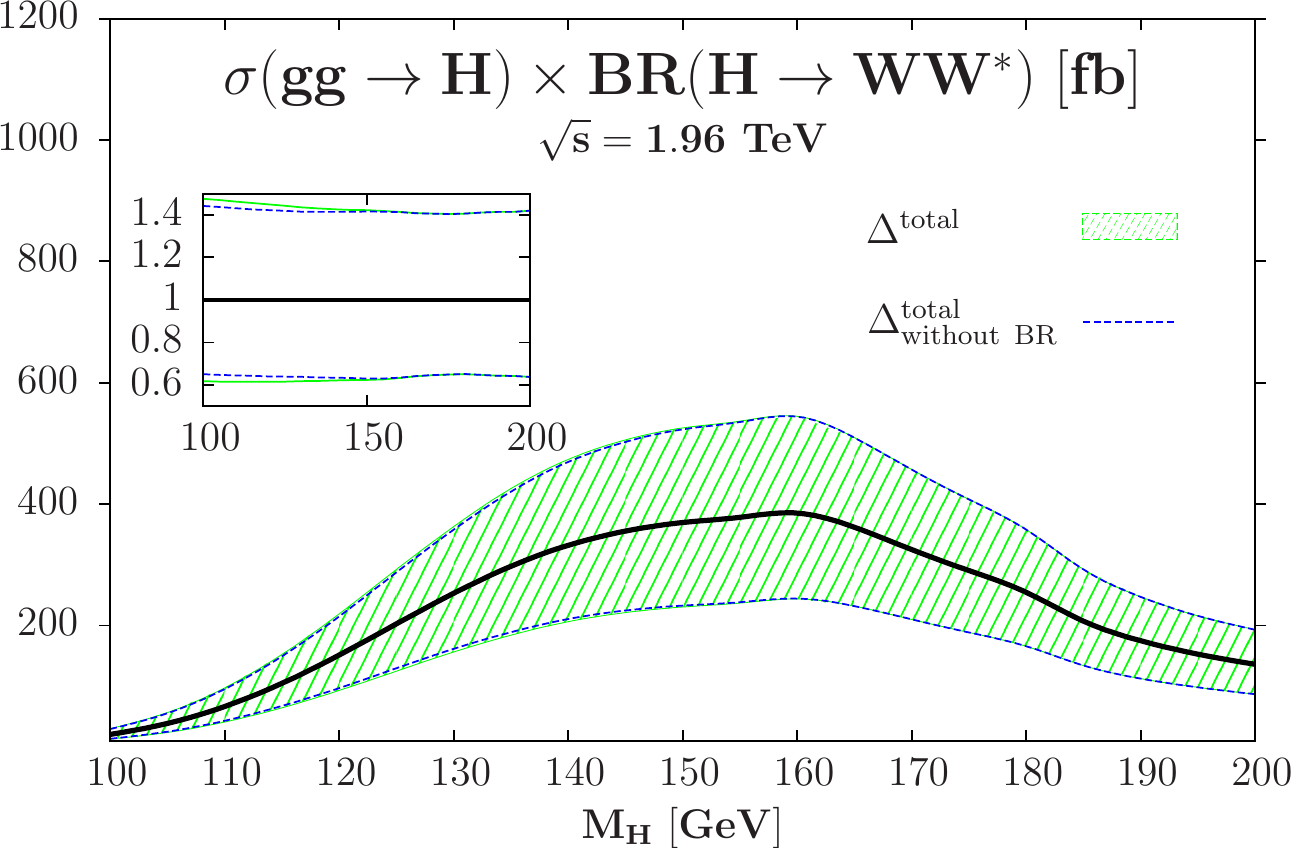}
\end{bigcenter}
\vspace*{-4mm}
\caption[The production cross section times branching ratio for SM
$p\bar p\to WH\to W b\bar b$ and $gg\to H\to W^+W^-$ at the Tevatron
together with the total uncertainty]{The production cross section
  times branching ratio for the process $p\bar p\to H\to b\bar b$
  (left) and $gg\to H \to WW$ (right) at the Tevatron with
  $\sqrt s=1.96$ TeV, including the total theoretical uncertainty band
  when the errors in the decay branching ratios are taken into
  account. In the inserts the relative deviations are shown.}
\label{fig:ggHBR_tev}
\vspace*{-1mm}
\end{figure}

The results for $\sigma(p\bar p\to HW)\times {\rm BR}(H\to b\bar b)$
and $\sigma(gg\to H)\times {\rm BR}(H\to W^{(*)}W^{(*)}$ are displayed
in Fig.~\ref{fig:ggHBR_tev} as a function of the relevant Higgs masses at
the Tevatron, together with the total uncertainty when taking into
account that on the branching ratio. The left of Fig.~\ref{fig:ggHBR_tev}
shows the result for the dominant search channel for low Higgs masses
in the $H\to b\bar b$ decay channel. It is then demonstrated that the
uncertainties on the branching ratio are really sizeable and have to
be taken into account. They add $\sim 5\%$ total uncertainty to be
compared to the $\pm 7\%$ uncertainty on the production cross section
alone. We note that the total $\sigma(p\bar p\to HW\to b\bar b)$ plays
no role at all above $M_H\gsim 160$ GeV as the total cross section
times branching ratio tends to the null value.

The right of Fig.~\ref{fig:ggHBR_tev} shows the
main channel for high Higgs masses searches in the $H\to
W^{(*)}W^{(*)}$ channel, including the total uncertainty with or
without those of the branching ratio included. The effect of the total
uncertainty on the branching ratio is only visible in the region $M_H
\approx 100$--150 GeV where the errors on BR$(H\to WW)$, dominated in
 practice by the errors in the decay channel $H\to b\bar b$, is
 significant. Above the $M_H \gsim 2M_W$ threshold, the branching
 ratio uncertainty is nearly null which means that it has no effect in
 the Higgs mass range $M_H\approx 150$--180 GeV to which both the
 Tevatron and the $\lhc$ are most sensitive. 

The net result is that even if the uncertainties on the branching
ratios play no role in the range where the Tevatron experiment is the
most sensitive, that is the range $M_H \in [150~;~180]$ GeV, they have
to be taken into account for $M_H \lsim 150$ GeV in both the $H \to
b\bar b$ and $WW$ channels.

\subsection{Combination at the LHC \label{section:SMHiggsFinalLHC}}

The situation at the $\lhc$ and even at the LHC with $\sqrt s = 14$
TeV the situation is much simpler compared to that of the
Tevatron. Indeed only the $gg\to H$ production channel is to be
considered in practice, only the decays $H\to WW, ZZ$ and to a lesser
extent $H\to \gamma \gamma$ are relevant. Since we have considered
only the dominant QCD uncertainties, these decays are affected by the
same uncertainties as discussed in previous
section~\ref{section:SMHiggsDecayResult} and displayed in the last
column of Table~\ref{table:Hdecay2}.

\begin{figure}[!h]
\begin{center}
\includegraphics[scale=0.8]{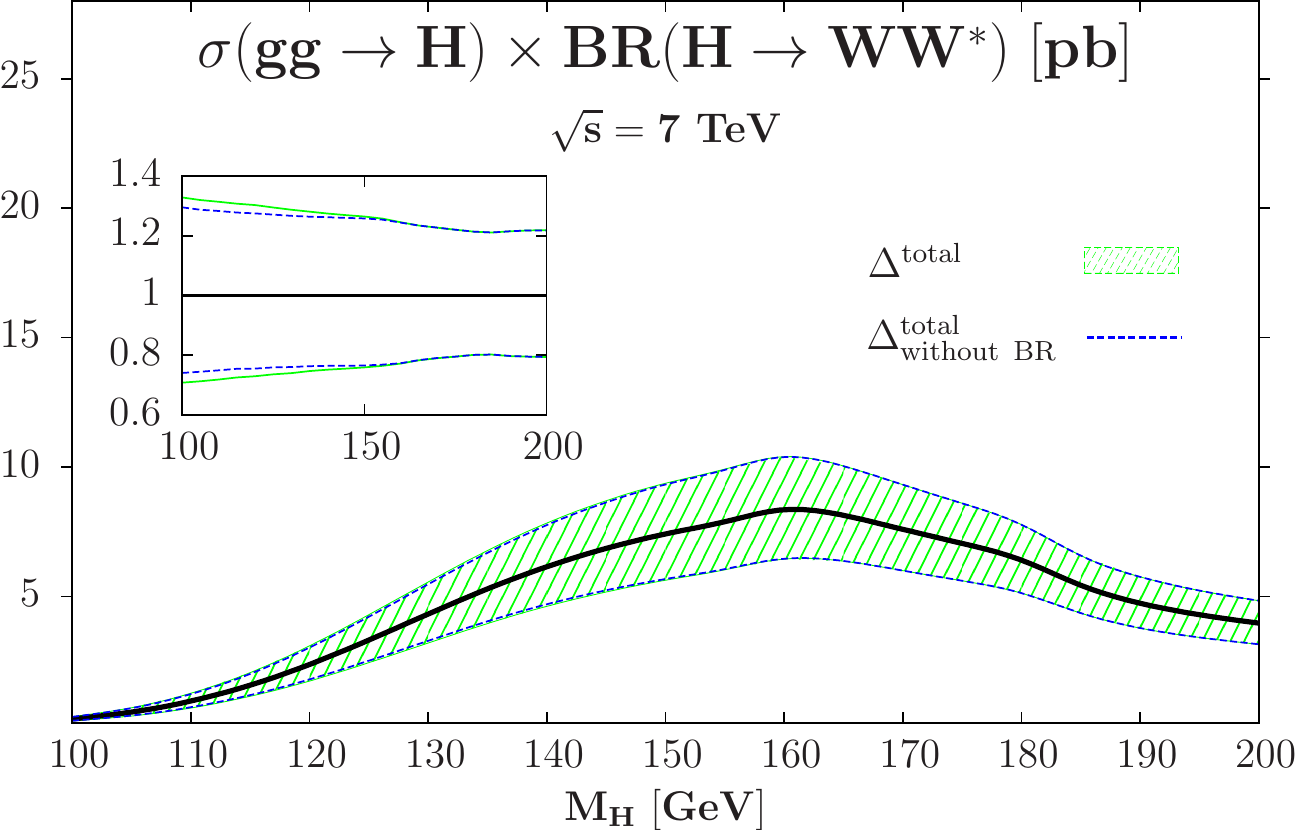}
\end{center}
\vspace*{-4mm}
\caption[The production cross section times branching ratio for SM $gg\to
H\to W^+W^-$ at the $\lhc$ together with the total uncertainty]{The
  production cross section times
  branching ratio for the process $gg\to H \to WW$ at the $\lhc$ with
  $\sqrt s=7$ TeV,  including the total theoretical uncertainty band
  when the errors in the decay branching ratios are taken into
  account. In the inserts the relative deviations are shown.}
\label{fig:ggHWW_lhc}
\vspace*{-1mm}
\end{figure}

In this case, we obtain the combined uncertainty of $\sigma^{\rm
  NNLO}(gg\to H) \times {\rm BR}(H \to VV^*)$ at the $\lhc$ that is
displayed in Fig.~\ref{fig:ggHWW_lhc} in the case of the $H \to WW$
decay as a function of $M_H$. Besides the uncertainty on the cross
section which is shown by the 
dashed lines, we display the effect of adding the error on the $H \to WW$ 
branching ratio as shown by the full lines which slightly increase the
overall uncertainty. As expected this is only visible in the region
where the uncertainties are sizeable for $M_H\leq 150$ GeV where the
uncertainty on the branching fraction is in practice dominated by the
uncertainty on the partial decay width $H\to b\bar b$. Above the WW
threesold the uncertainty drops to the null value, meaning that in
particular there is no need to consider this uncertainty in the high
Higgs mass regime, as seen above in the case of the Tevatron
colliders. However as already stated the total Higgs width as well as
the interference between the signal and the background become
significant at high masses and should be taken into account in both
the production cross section and the decay widths~\cite{Dittmaier:2011ti}.

In the case of the $H\to ZZ, \gamma \gamma$ (and $\tau\tau)$ decays,
since the errors on the branching ratios are the same as those
affecting BR($H\to WW)$, the overall uncertainties in $\sigma \times
{\rm BR}$ can also be seen from Fig.~\ref{fig:ggHWW_lhc}: only the
normalization of the branching ratio is different and can be obtained
for a specific decay mode from Table \ref{table:Hdecay2}.

\subsection{The Tevatron exclusion
  limit \label{section:SMHiggsTevExclusion}}

After having calculated the production cross sections together with
their theoretical uncertainties in section~\ref{section:SMHiggsTev} at
the Tevatron, the branching ratios that are relevant in
section~\ref{section:SMHiggsDecayResult} and made the combination in
section~\ref{section:SMHiggsFinalTev}, we are now ready to compare
with the experimental results given by the CDF and D0
collaborations~\cite{Tevatron:2009je,Aaltonen:2010yv,Tevatron:2010ar,Aaltonen:2011gs}.

The comparison with low mass Higgs seaches results shows that while
the Tevatron collaborations use an approximate 5\% uncertainty on the
main search channel $p\bar p\to WH\to W b\bar b$, we have found a
total uncertainty of nearly 15\% taking into account all possible
sources of uncertainties. In particular the uncertainties on the
branching fractions are sizeable and add up to nearly 5\%, when this
has been overlooked by experimental collaborations. Nevertheless, it
does not change significantly the main result on the 95\%CL exclusion
for the SM Higgs boson mass in the $[115:150]$ GeV mass range.

The situation in the high Higgs mass range is completely
different. This is the key range at the Tevatron as the experiments
are the most sensitive for Higgs mass $M_H \sim 165$ GeV: indeed, the
CDF and D0 experiments claim that they exclude a SM Higgs boson at
95\%CL for $158\leq M_H\leq 173$ GeV. In
section~\ref{section:SMHiggsTevTotal} we have found a total
uncertainty of order $\pm 40\%$, to be compared with the nearly $\pm
20\%$ used by the experimental collaborations. This has to be analyzed
in detail to see wether it has an impact on the exclusions
bounds~\cite{Aaltonen:2010yv,Tevatron:2010ar,Aaltonen:2011gs}. In
addition we should add the uncertainties on the branching ratios, but
as can be seen on Fig.~\ref{fig:ggHBR_tev} they have no impact for
$M_H\gsim 150$ GeV. In the intermediate Higgs mass range $130\lsim M_H
\lsim 150$ GeV where both search channels $p\bar p\to WH\to W b\bar b$
and $gg\to H\to W^{(*)}W^{(*)}$ compete each other, the uncertainties
on the two branching ratios have to be taken into account.

\begin{figure}[!h]
\begin{center}
\includegraphics[scale=0.8]{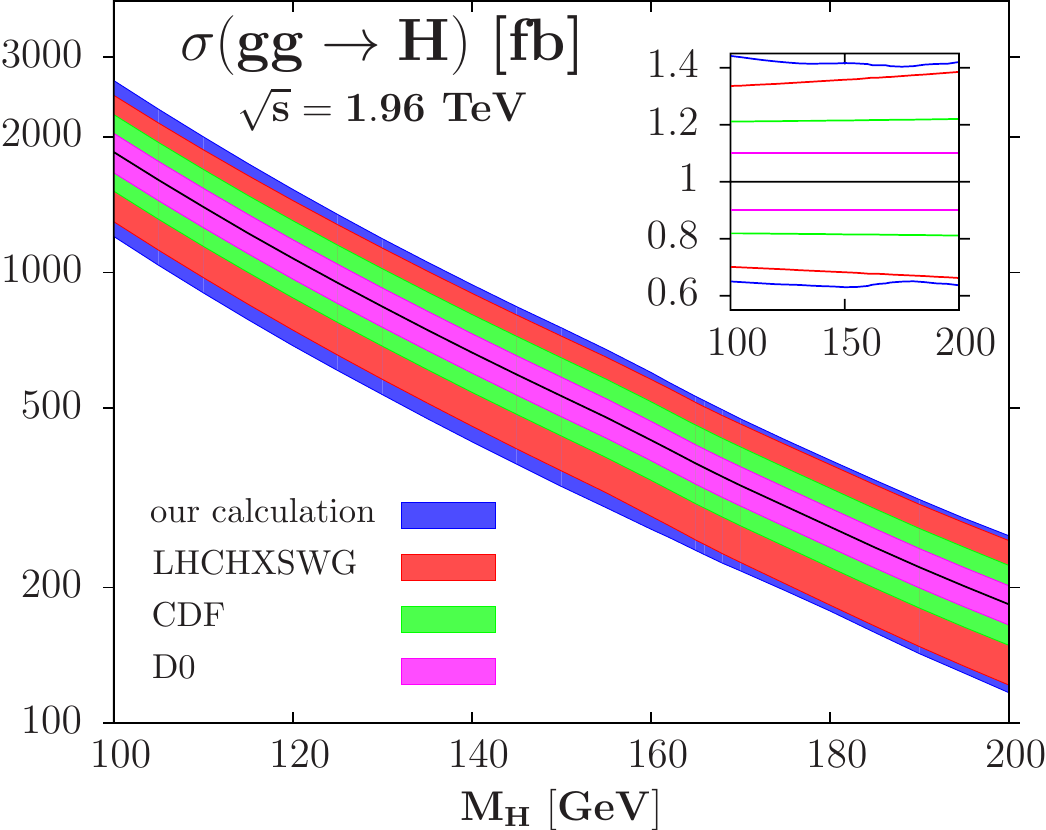}
\end{center}
\vspace*{-4mm}
\caption[The SM Higgs boson production cross section $gg\to H$ at the
Tevatron together with the total uncertainty using 4 different ways of
adding the theoretical uncertainties]{The SM Higgs
  boson production cross section $gg\to H$ at the Tevatron including
  the total theoretical uncertainty band  when the uncertainties are
  combined using our procedure, those of CDF/D0 experiment or the
  recommendation of the LHC Higgs Cross Section Working
  Group~\cite{Dittmaier:2011ti}.}
\label{fig:ggH_all_limit}
\vspace*{-1mm}
\end{figure}

We display in Fig.~\ref{fig:ggH_all_limit} the comparison between our
calculation on the production cross section and the CDF/D0
expectations on the one hand, the prediction using the way of handling
the theoretical uncertainties described in Ref.~\cite{Dittmaier:2011ti} on the
other hand. The CDF and D0 experiments simply add in
quadrature the uncertainties from the scale variation and the
PDF+$\alpha_s$ uncertainties obtained through the Hessian method and
ignore the smaller EFT uncertainty. They obtain an overall uncertainty
of order 20\% on the inclusive cross section. As stated in the former
section, see section~\ref{section:SMHiggsTevTotal}, we believe that
this is not 
reasonable and has little justification. Indeed, the uncertainties
associated  to the PDFs in a given scheme should be viewed as purely
theoretical uncertainties (due to the theoretical assumptions in the
parameterization) despite of the fact that they are presented as the
$1\sigma$ or more departure from the central values of the data
included in the PDF fits. They should be equivalent to the spread of
the different predictions when use the four NNLO PDFs sets available
on the market and thus having no statistical
interpretation\footnote{If we want to use a stasticial analysis we
  then should use a flat prior and not a gaussian prior to handle the
  PDF puzzle.} and thus combined linearly with the uncertainties from
the scale variation and the EFT approach which are pure theoretical
uncertainties. This is the procedure recommended, for instance, by
the LHC Higgs Cross Section Working Group~\cite{Dittmaier:2011ti} that is also
displayed in Fig.~\ref{fig:ggH_all_limit}. Our procedure has been
proposed first in Ref.~\cite{Baglio:2010um} where we apply the combined
PDF--$\alpha_s$ uncertainties directly on the maximal/minimal cross
sections with respect to scale variation\footnote{We remind the reader
  that a similar procedure has also been advocated in
  Ref.~\cite{Cacciari:2008zb} for top quark pair production.}, and then adds
linearly the small uncertainty from the  EFT approach.  This last procedure,
that we have used here, provides an overall uncertainty that is similar (but
slightly smaller) to that obtained with  the linear sum of all
uncertainties. 

In the mass range $M_H \! \approx \! 160$ GeV with almost the best
sensitivity, our procedure gives an approximate $+41\%,-37\%$ total
uncertainty, to be compared to the $\approx\! 10\%$ and $\approx\!
20\%$ uncertainties assumed, respectively, by the CDF and D0
collaborations. The comparison with the LHC Higgs cross section
working group recommendation use the uncertainties from scale
($\!+\!20\%,\!-\!17\%$ on the sum of the jet cross
sections\footnote{An additional uncertainty of $\approx\! 7.5\%$ from
  jet acceptance is introduced when considering the Higgs+jet cross
  sections. We will consider it to be experimental and, when added in
  quadrature to others, it will have little impact.} and PDFs
($+16\%,-15\%$ when the MSTW 68\% CL PDF+$\Delta^{\rm exp} \alpha_s$
error is multiplied by a factor of two following the PDF4LHC
recommendation)~\cite{PDF4LHC}, leading to a total of $\approx \!+\!36\%,
\!-\!32\%$ for $M_H\! \approx\! 160$ GeV. Thus, the uncertainty that we assume
is comparable to the one obtained using the LHC procedure~\cite{Dittmaier:2011ti}, the
difference being simply due to the additional ${\cal O}(5\%)$ uncertainty from
the use of the EFT approach that we also include. In the end this
demonstrates that our total uncertainty is certainly not too
conservative and then has to be taken seriously.

\begin{figure}[!h]
\begin{center}
\includegraphics[scale=0.8]{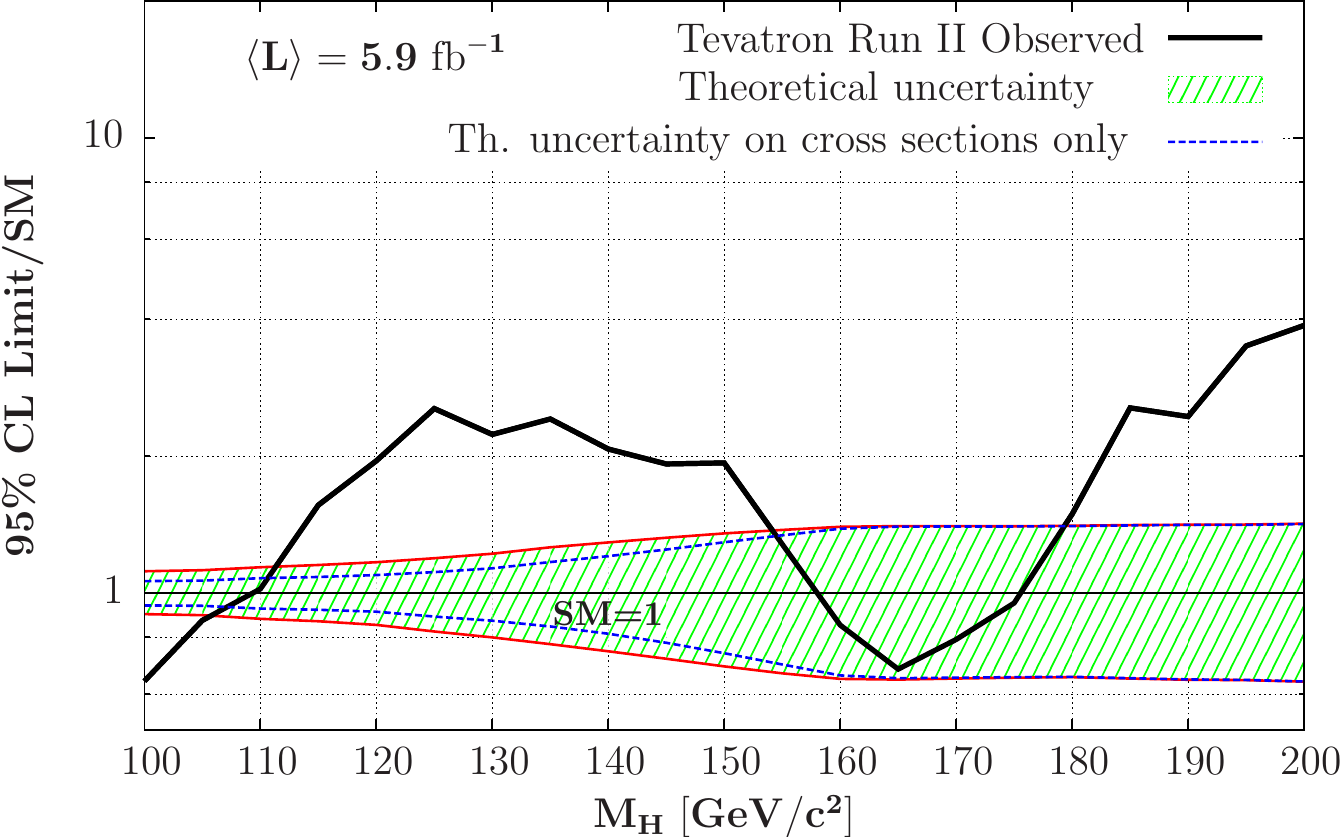}
\end{center}
\vspace*{-4mm}
\caption[The CDF/D0 95\%CL limit on the SM Higgs boson mass confronted
to our theoretical expectations in a naive approach.]{The SM Higgs
  boson production cross section $gg\to H$ at the Tevatron including
  the total theoretical uncertainty band  when the uncertainties are
  combined using our procedure, those of CDF/D0 experiment or the
  recommendation of the LHC Higgs Cross Section Working
  Group~\cite{Dittmaier:2011ti}.}
\label{fig:naive_comb_tev}
\vspace*{-1mm}
\end{figure}

We display in Fig.~\ref{fig:naive_comb_tev} the comparison between the
experimental 
results presented in
Refs.~\cite{Aaltonen:2010yv,Tevatron:2010ar,Aaltonen:2011gs} and our
theoretical expectations. This presentation is of course naive, as the
true combination of the different channel is very difficult, given
the fact that many channels are involved. Nevertheless we know in
practice that two channels only matter at the Tevatron, that are the
$H\to b\bar b$ after Higgs--strahlung production for low Higgs mass
searches and the $H\to W^{(*)}W^{(*)}$ channel after gluon--gluon
fusion Higgs production for high Higgs mass searches. We pay a
particular attention in the intermediate Higgs mass range where both
channel are to be considered, and we take into account the correlation
between the two decay channel.

If we want to claim a 95\%CL exclusion we should require that the
observed line be below the lower band of the SM prediction. That is
not the case when we compare with our prediction, thus we believe that
the Tevatron experiment have not excluded the SM Higgs boson in the
Higgs mass range $158\leq M_H\leq 175$ GeV.\bigskip

The comparison displayed above is indeed not precise. We should plug
directly our prediction in the experimental analysis as some
uncertainties have been already taken into account by the experiments
in the calculation of the SM line of
Fig.~\ref{fig:naive_comb_tev}. Nevertheless we advocate to separate
experimental and theoretical results on such a graph in order to make
the comparison more relevant. We will in the next consider different
scenarios and give the results published in
Ref.~\cite{Baglio:2011wn} where the procedure described below has
been used. The reader should note that this work is based on the
former results of the Tevatron experiments when the luminosity was of
order 5.9 fb$^{-1}$. The new result of the CDF experiment use a
luminosity of 7.1 fb$^{-1}$; nevertheless our conclusions will remain,
especially that the new Tevatron results have induced a slightly
reduction of the exclusion range from 158--175 GeV down to 158--173
GeV.

Before describing the more precise way of handling the comparison
between our results and those of the CDF/D0 collaborations, we stress
again that the comparison between our values and those assumed by the
experimental collaborations becomes even worse when the cross section
is evaluated with another set of PDFs. For instance, with the ABKM PDF
parametrization, there is a reduction of $\approx 25--30\%$ of the
normalisation  compared to the central value adopted in the CDF/D0
combined analysis. We should also take into account the uncertainties
on the backgrounds in order to make complete predictions. The by far
largest background is $p\bar p\! \to\! W^+W^-$ for which CDF/D0
assume the inclusive cross section to be $\sigma  \!=\!  11.34
^{+4.9\%}_{-4.3\%} ({\rm scale})^{+3.1\%}_{-2.5\%} ({\rm PDF})$ pb. We
have reevaluated the rate using {\tt MCFM}~\cite{Campbell:2010ff} and
find $\sigma\!= \!  11.55^{+5\%}_{-6\%} ({\rm scale})^{+5\%}_{-8\%}
({\rm 90\%CL\;PDF})$ pb using the MSTW scheme (the errors due to
$\alpha_s$ are negligible here) which gives $\sigma= 11.55^
{+11\%}_{-14\%}$ pb if the errors are added according to our procedure
developed for the signal cross section. If we adopt the ABKM or
HERAPDF sets, we would obtain a rate of, respectively, 12.35 pb and
11.81 pb. i.e. nearly $9\%$ higher in the maximal case. We will thus
consider that  $\sigma(p\bar p\! \to\! W^+W^-)$ can be $\approx 10\%$
larger/lower than assumed by CDF/D0.

Thus if the $\approx 20\%$ total uncertainty assumed by the CDF
collaboration is adopted, one can consider three scenarios:

\begin{enumerate}
\item{The first scenario is a reduction of $\sigma^{\rm NNLO}_{gg \to
      H}$ by $\approx 20\%$ to account for the difference between the
    quadratic and (almost) linear ways of combining the individual
    uncertainties.}
\item{The second scenario is simply to adopt the normalisation
    obtained using the ABKM PDFs which gives an $\approx 30\%$
    reduction of $\sigma^{\rm NNLO}_{gg \to H}$ in the critical
    region. In both scenarios 1 and 2, the
    remaining $\approx 20\%$ uncertainty due to scale variation and
    the EFT will correspond to the overall theoretical uncertainty
    that has been assumed in the Tevatron analysis.}
\item{The additionnal third scenario will be to consider the same
    assumptions in the first scenario for the signal and change the
    normalization of the $p \bar p\! \to \! WW$ background by $\pm
    10\%$.}
\end{enumerate}

In Ref.~\cite{Baglio:2011wn}, the exploration of the Tevatron
exclusion limit is based on the CDF study published in
Ref.~\cite{CDF:10232} which provides with all the necessary 
details. In the analysis of the $gg \to H \to WW \to \ell \ell \nu \nu
$ signal, the production cross section has been broken into the three
pieces which yield different final state signal topologies, namely
$\ell \ell \nu \nu$+0\,jet,  $\ell\ell \nu \nu$+1\,jet and $\ell \ell
\nu \nu$+2\,jets or more. These channels which represent,
respectively,  $\approx 60\%$, $\approx 30\%$ and $\approx 10\%$ of
the total $\sigma_{gg\to H}^{\rm NNLO}$~\cite{Anastasiou:2009bt}, have been
studied separately. In the $\ell \ell \nu \nu$+0\,jet and +1\,jet
samples, two configurations have been analyzed, one with a high and
one with a low signal over background ratio (depending on the quality
of the lepton identification). In addition, a sample with a low
invariant mass for the two leptons, $M_{\ell \ell} \leq 16$ GeV, has
been  included. Five additional channels resulting from the
contributions of the  Higgs--strahlung processes are also included:
$p\bar p \to VH \to VWW$ leading to same sign dilepton and to
trilepton final states. These channels give rather small signal rates,
though.

One then estimates the necessary relative variation of the integrated
luminosity needed to reproduce the currently quoted sensitivity of the
CDF collaboration, if the normalization of the Higgs signal cross section
(as well as the corresponding backgrounds) is different from the one
assumed to obtain the results. In Ref.~\cite{Baglio:2011wn} it
has first been tried to reproduce as closely as possible the CDF results
for a Higgs mass $M_H=160$ GeV for which the sensitivity is almost the
best. They have used the information given in Ref.~\cite{CDF:10232},
that is the background, signal and data numbers for all the search
channels of Tables I--VIII, and also their neural network outputs for
the 10 search channels (each one for the  signals, backgrounds and
data) presented in Figs.~2,4,$\cdots$,16 to build the background only
and the background plus signal hypotheses, implemented them in the
program {\tt MClimit}~\cite{Junk:1999kv} and used a ratio of
log--likelihood ``\`a la LEP'' as a test--statistic for which we
combined the above channels; this provided the 95 \%CL/$\sigma_{\rm
  SM}$ sensitivity limit on the Higgs boson at the considered mass of
$M_H\!=\!160$ GeV. A median expected 95 \%CL/$\sigma_{\rm SM}$  limit
of $S_0\!=\!1.35$ has been obtained, to be compared to $S_0\!=\!1.05$
in the CDF analysis; for the observed 95 \%CL/$\sigma_{\rm SM}$
limit, the agreement is better as we obtain 1.35 compared to 1.32. As
the CDF and D0 collaboration agree in their methods only within 10\%
accuracy for the same imput Monte Carlo and data~\cite{Peters:2010},
it is believed that the procedure is satisfactory enough to be used in the
three scenarios described above.

The first two scenarios discussed previously which, in
practice, when compared to the CDF/D0 results, reduce the
normalisation of the $gg\! \to\!  H\! \to\!  WW \! \to \! \ell \ell
\nu \nu$ signal cross section by approximatively
$20\%$ when all the uncertainties are added using our procedure,
approximatively $30\%$ when the ABKM set is used to derive the
central value of the cross section, will be considered. The third
scenario adds the effect of the background uncertainties. The authors
of Ref.~\cite{Baglio:2011wn} then estimate
the relative variation of the sensitivity and increase the integrated
luminosity until we recover our initial sensitivity. Finally, it is
assumed that the obtained relative variations of the sensitivity, as
well as the required luminosity to reproduce the initial sensitivity,
would be the same for the CDF experiment.

\begin{figure}[!h]
\begin{center}
\includegraphics[scale=0.7]{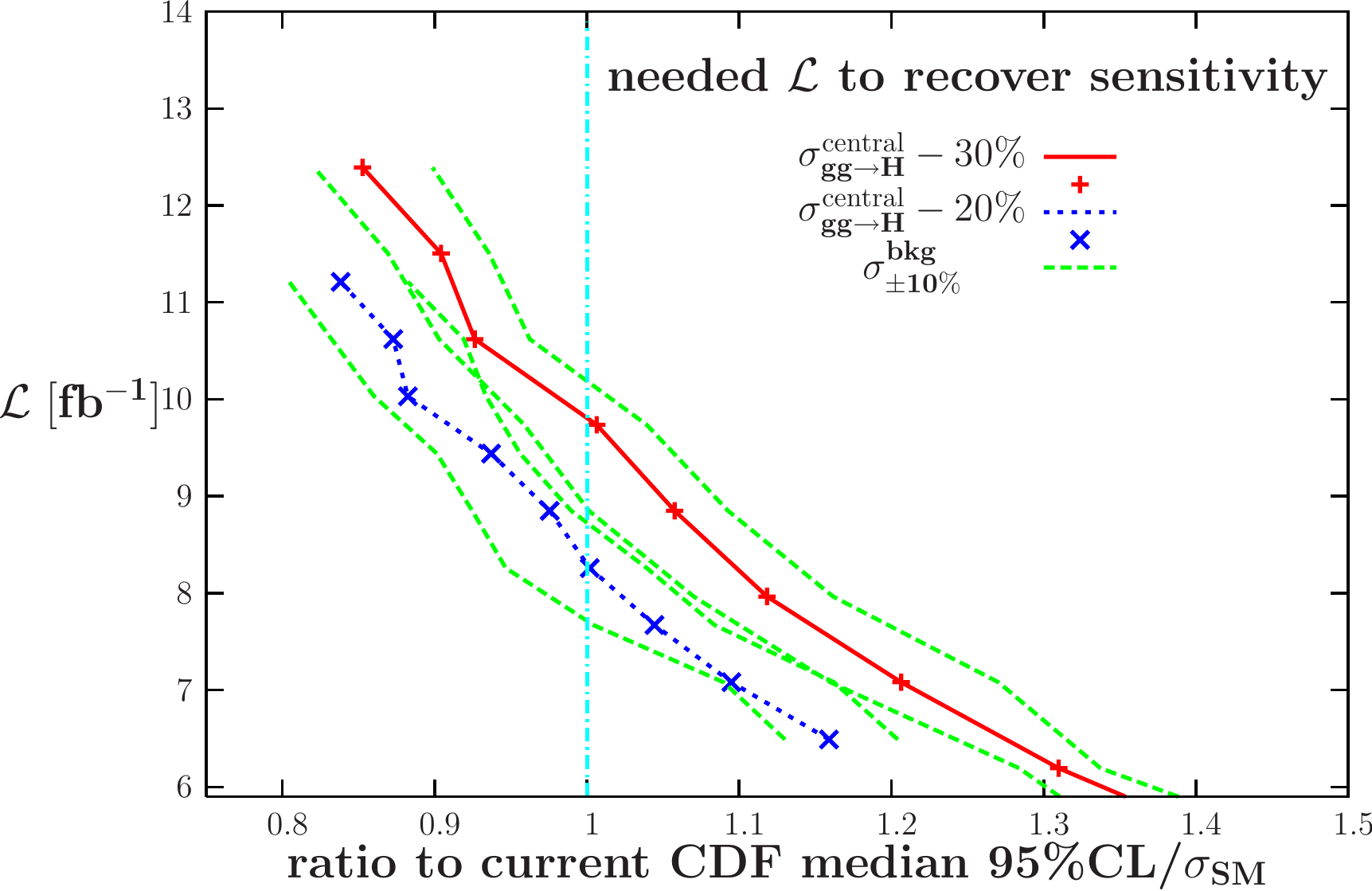}
\end{center}
\caption[The luminosity needed by the CDF experiment to recover their
current claimed sensitivity when compared to our theoretical
expectations for the uncertainty]{The luminosity needed by the CDF
  experiment to recover the current sensitivity (with 5.9 fb$^{-1}$
  data) when the $gg\!\to\! H\! \to \!\ell \ell \nu \nu$ signal rate
  is lowered by 20 and 30\% and with a $\pm 10\%$ change in the $p\bar
  p\! \to\! WW$  dominant background. Figure taken from
  Ref.~\cite{Baglio:2011wn}.}
\vspace*{-5mm}
\label{fig:tev_lumi_results}
\end{figure}

For each scenario, the expected signals and the corresponding
backgrounds at the Tevatron have been multiplied by a luminosity
factor that has been varied. For each value of the luminosity factor,
the corresponding median expected 95\%CL/$\sigma_{\rm SM}$ has been
estimated and normalized to the initial sensitivity $S_0=1.35$
obtained above. We display in Fig.~\ref{fig:tev_lumi_results} the
final result, where the Tevatron luminosity is shown as a function of
the obtained normalized sensitivity. The luminosity needed to recover
the current $S_0$ CDF sensitivity is given by the intersection of the
vertical blue line with the luminosity curves. The third scenario
where the normalization of the background cross sections is changed by
approximatively $\pm 10\%$\footnote{The 
  correlation between signal and background is implicitly taken into
  account as we use the results of Ref.~\cite{CDF:10232}; we assume
  though that it is almost the same when another PDF set is adopted
  for both signal and background.} has also been considered. The final
result is that if $\sigma_{gg\to H}^{\rm NNLO}$ is lowered by $20\%$,
a luminosity of $\approx 8$ fb$^{-1}$, compared to 5.9 fb$^{-1}$ used
in~\cite{CDF:10232} would be required for the same analysis to
obtain the current sensitivity. If the rate is lower by 30\%, the
required luminosity should increase to $\approx\! 10$ fb$^{-1}$,
i.e. nearly a factor of two, to obtain the CDF sensitivity with 5.9
fb$^{-1}$; as the newest CDF analysis with 7.1 fb$^{-1}$ leads to a
slightly worse exclusion limit $158\leq M_H\leq 173$ GeV, we believe
that recovering this last exclusion limit should require $\approx 13$
fb$^{-1}$, that is still a factor of two than the current used
luminosity . In the third scenario, we see that increasing/decreasing
the background will degrade/improve the sensitivity  and an $\approx
10\%$ higher/lower luminosity would be required to recover the
sensitivity.

\subsection{Summary of the results}

After the results of sections~\ref{section:SMHiggsTev} and
\ref{section:SMHiggsLHC} on the production cross sections we have
given in this section our theoretical prediction of the Higgs decay
branching fractions, together with a detailed analysis of the
parametric uncertainties that affect the predictions.

We have found that in the two main decay search channels $H\to b\bar
b$ and $H\to W^{(*)}W^{(*)}$ the uncertainties are sizeable and have
some impact in the intermediate mass range $130\lsim M_H\lsim 150$
GeV. We then have combined the production cross section and decay
branching ratio in section~\ref{section:SMHiggsFinalTev} for the
Tevatron and section~\ref{section:SMHiggsFinalLHC} for the LHC and
found that in particular for the $p\bar p\to WH\to W b\bar b$ channel
at the Tevatron the uncertainties are more important than for the
production cross section alone.

We then have compared our results with those of the CDF and D0
experiments. We have presented a naive approach that has cast some
doubts on the 95\%CL exclusion limit. This is then confirmed by a more
rigorous approach that we have presented in the end of
section~\ref{section:SMHiggsTevExclusion}, where we conclude that the
reduction of the signal by 30\% as would be the case if the ABKM set
were used for its normalization, would reopen the entire mass range
$M_H\!=\!158$--173 GeV excluded by the CDF/D0 analysis with 12.6
fb$^{-1}$ combined data: this means that even if our uncertainties are
viewed as being too conservative (which we believe is certainly not
the case, as demonstrated above) the exclusion limit critically
depends on the considered PDF. That is a non--satisfactory situation
which needs to be cured and in particular the PDF case is still open,
see for example Refs.~\cite{Alekhin:2011cf, CooperSarkar:2011vp,
  Thorne:2011kq} on the comparison between DIS--only and global fits
that is still strongly disputed between the experts. We also note that
the exclusion limit is also disputed in Ref.~\cite{Berger:2010xi} with
arguments different from that presented in this thesis. Henceforth the
Tevatron results on the SM Higgs boson mass is strongly debated within
the community.\bigskip

This results put an end to the first two parts of this thesis that
have been devoted to the study of the Higgs boson within the Standard
Model. However this is not the end of the story, as there are strong
indications that new physics beyond the SM is needed to describe the
sub--atomic world, just to mention the experimental proof of the
neutrinos masses which are not included in the SM\footnote{This can be
  accommodated in some ways without introducing new physics, e.g. with
  a right--handed neutrino yet--to--be--seen put by hand in the
  theory; this is not a satisfactory way as we do not have any (good) 
  explanation why this right--handed neutrino has not yet been
  observed.}. We will then turn our attention to one of the most
popular beyond--the--SM scenario: supersymmetric
theories. Part~\ref{part:three} will introduce the theoretical
concepts of supersymmetry (SUSY) and its application to the minimal
extension of the SM, called the MSSM. Part~\ref{part:four} will give
our results on the SUSY Higgs boson(s) production and decay at the Tevatron
and LHC colliders.

\vfill
\pagebreak


\part{The Minimal Supersymmetric extension of the Standard Model}
\label{part:three}

\section{Why Supersymmetry is appealing}

\label{section:SUSYIntro}

The first two parts of this thesis were devoted to the extensive study
of the Higgs boson phenomenology at hadron colliders within the SM
theory. However we have already stated that this theory is bound to be
considered as an effective model of a more elaborated theory
yet--to--be--discovered: indeed, there are many models on the market
that go beyond the SM, none of them being well established by
experiments.

We will study in the following sections one of the most appealing
scenarios beyond the SM: the minimal extension of the SM in the
context of supersymmetry (SUSY). This theory will be presented in the
following sections but to begin with we remind the reader that SUSY is
a symmetry between bosons and fermions: it relates these two degrees of
freedom through new fermionic operators, and any supersymmetric
lagrangian has to be invariant by the exchange of the bosonic and
fermionic degrees of freedom grouped in one same multiplet. We will
first give some reasons why SUSY is a very appealing theory and then
in the two next sections introduce its formal aspects and application
to the minimal extension of the SM.

\subsection{The hierarchy problem \label{section:IntroHierarchy}}

When the Standard Model was described in section~\ref{section:IntroHistory}, one
of the key arguments in going from the old $V-A$ theory for the weak
interaction to the current electroweak theory was that it enables
calculations beyond the Fermi scale and in particular the weak
electron--neutrino scattering be well behaved at high center--of--mass
energies, that is the unitarity requirement. However, we have seen
that the SM itself is not free from potential unitarity problem and we
have assessed bounds on the SM Higgs boson following from the
unitarity requirement. This exercice, along other ways of obtaining
theoretical bounds on the SM Higgs boson, has shown that the scalar
sector of the SM is in some sense ill--defined: it concentrates most
of the difficulties of the SM, as it is an unstable sector not
protected by any symmetry, and this will be presented in this
subsection.

The major problem of the Higgs boson mass is that quantum corrections
of this fundamental parameter of the theory is of quadratically
divergent type. If we take the one--loop corrections, they are
described by the Feynman diagrams depicted in
Fig.~\ref{fig:SUSYhierarchy} below.

\begin{figure}[!h]
\begin{center}
\includegraphics[scale=0.85]{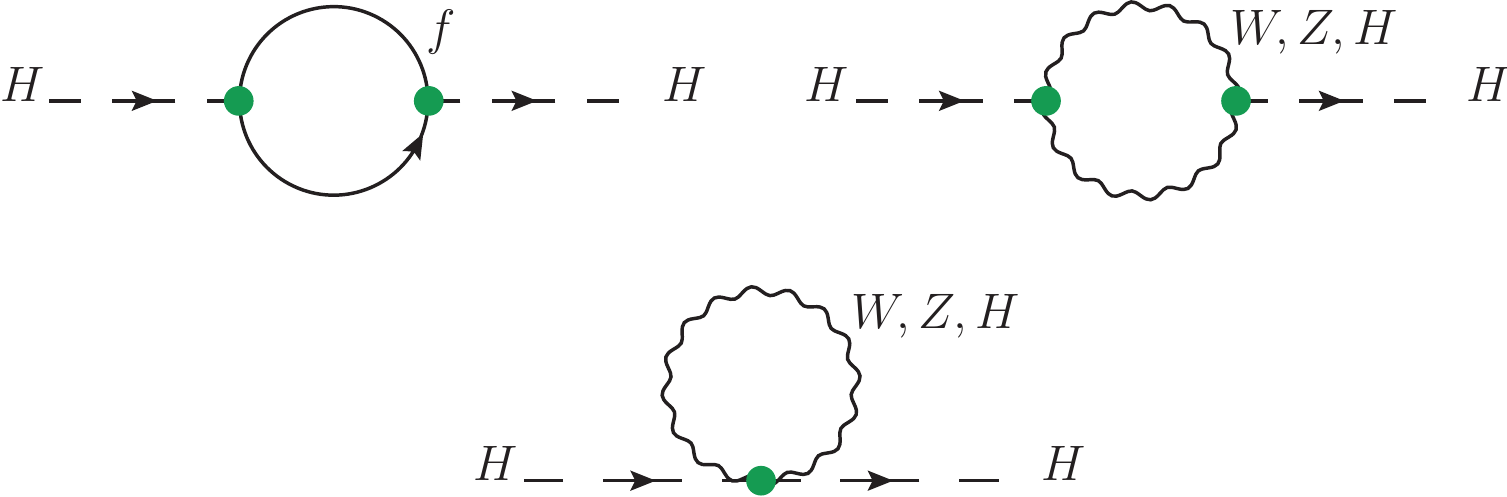} 
\end{center}
\vspace*{-.5cm}
\caption[One--loop corrections to the Higgs boson mass within the
SM]{Feynman diagrams for one--loop corrections to the Higgs boson
  mass within the SM: fermionic loops (upper left), 3--points bosonic
  vertex (upper right) and 4--points bosonic vertex (lower).}
\label{fig:SUSYhierarchy} 
\end{figure}

If we take for example the first fermionic diagram, the typical
contribution is
\beq
\Pi^{f}_{HH}(0) = -4 \imath N_f \left(\frac{\lambda_f}{\sqrt
    2}\right)^2 \frac{1}{8\pi^2}
  \int_{0}^{+\infty} dk~ k^3 \frac{k^2+m_f^2}{(k^2-m_f^2)^2}
\label{eq:Higgscorrection0}
\eeq
were $N_f$ is 3 for a quark, 1 for a lepton. If we rewrite the
integrand it reads
\beq
\frac{k^3(k^2+m_f^2)}{(k^2-m_f^2)^2}  & =  & \frac{k^3}{k^2-m_f^2} +
\frac{2m_f^2 k^3}{(k^2-m_f^2)^2} \nonumber \\
 & = & k + 3 m_f^2 \frac{k}{k^2-m_f^2} + 2 m_f^4
 \frac{k}{(k^2-m_f^2)^2}\nonumber
\eeq
which contains a quadratic divergence as the first term, after the
integration with a cut--off $\Lambda$ at high $k$ momentum, gives a
term proportionnal to $\Lambda^2$. The actual dominant contribution,
taking into account self--Higgs interactions and weak bosons
contributions along that of the fermionic, reads~\cite{Djouadi:2005gi,
Djouadi:2005gj}:
\beq
\Delta M_H^2 = \frac{3 \Lambda^2}{8\pi^2 v^2} \left(
  M_H^2+2M_W^2+M_Z^2-4m_t^2\right)
\label{eq:Higgscorrection}
\eeq

This is a very severe divergence which is retained even after
renormalization. Indeed, if the SM were to be valid up to the Planck
scale $\Lambda_P \simeq 10^{18}$ GeV the Higgs boson mass would be
attracted to that scale and would have a mass of order $\Lambda_P$: we
would need a counterterm that is hugely fined--tuned in order to keep the
Higgs boson mass below the theoretical bounds of order 1 TeV as
presented in section~\ref{section:HiggsboundsTheory}. If we forget
about the bosonic contributions to the Higgs boson mass, we see that
the correction in Eq.~\ref{eq:Higgscorrection} does not depend on
$M_H$: this means that even if we were to set $M_H=0$ we still would
have the same problem. No symmetry then protects the SM Higgs boson to
be driven to high masses, hence making this parameter unnatural. This
is exactly the naturalness problem of the SM, see
Ref.~\cite{Weinberg:1978ym, LlewellynSmith:1981yi}. The related
question, which is the gauge hierarchy problem, is why the new physics
scale that drives the SM Higgs boson mass, is much heavier than say
$M_Z$ which is typical of the electroweak scale, see also
Ref.~\cite{DreesMSSM}.

How to solve this technical (but theoretically unsatisfactory)
problem? One solution would be to introduce new degrees of freedom
which would arrange themselves to cancel out exactly the contributions
of the standard bosons and fermions. We start from
Eq.~\ref{eq:Higgscorrection0} where the leading quadratic divergence
reads $\Delta M_H^2 = -\lambda_f^2 \Lambda^2/8\pi^2$ (we discard in
this analysis the additionnal $N_f$ factor that is irrelevant for the
point discussed). We follow Refs.~\cite{DreesMSSM, Martin:1997ns}; we
introduce $N$ new complex scalar fields $\tilde{f}$ with
\beq
{\cal L}_{\tilde f \tilde f H} = -\frac12 \lambda_{\tilde f} H^2 |\tilde
f|^2 -\lambda_{\tilde f} v H |\tilde f|^2
\eeq

This induces two types of corrections to the Higgs boson mass following
the two bosonic diagrams in Fig.~\ref{fig:SUSYhierarchy}, and again
when writing only the quadratic divergence it reads:
\beq
\Pi^{\tilde f}_{HH}(0) = N \imath \frac{\lambda_{\tilde f}}{16 \pi^2}
\Lambda^2
\label{eq:HiggscorrectionScalar}
\eeq

If we combine Eqs.~\ref{eq:Higgscorrection0}
and~\ref{eq:HiggscorrectionScalar} we see that the cancellation of
quadratic divergences, in other words having $\Pi^{f}_{HH}+\Pi^{\tilde
  f}_{HH} = \mathcal{O}(\ln \Lambda)$, can be obtained provided that
\beq
N = 2~,~\lambda_{\tilde f} = \lambda_f^2
\label{eq:SUSYcondition}
\eeq

This results is of utmost importance: it means that if we introduce a
symmetry between bosonic and fermionic degrees of freedom that
arranges to follow Eq.~\ref{eq:SUSYcondition} we can solve the
naturalness of the SM. Following Ref.~\cite{Djouadi:2005gj, DreesMSSM}
we can even go further and cancel the logarithmic divergence without
renormalization analysis if we impose $m_f = m_{\tilde f}$. We are
even temptated to assign for each fermion $f$, which has a left and
right chirality, two new complex scalar partners $\tilde f_R$ and
$\tilde f_L$ where the index refers to the original fermionic degree
of freedom and not any chirality for the scalar fields. Note that the
total number of degree of freedom is conserved by this assignment.

This is exactly the virtue of SUSY which is a symmetry between bosons
and fermions: each SM boson and each SM fermion has its own
(super)partner, respectively a fermion and a boson. The relationship
between the standard particles and these new degrees of freedom ensures
that we get rid of quadratic divergencies, thus solving the weak
hierarchy problem. This is one particularly elegant reason why SUSY is
a very appealing theory beyond the SM.

\subsection{Coupling constants convergence at high
  energies \label{section:IntroGUT}}

We now turn our attention to a less severe problem as it deals with
speculative questions. For centuries atomic and then sub--atomic
physics have tried to unify the description of fundamental processes,
starting from the unification of electricity and magnetism into a
single electromagnetic interaction, currently described by QED. It is
then strongly believed that the current electroweak\footnote{We want
  to point out in passing that the weak and electromagnetic
  interactions are partially unified in the electroweak framework: the
  gauge algebra $SU(2)_L\times U(1)_Y$ has two couplings constants,
  whereas we should only have a single Lie group $X$ with a single
  coupling constant $g_X$ to really have unification. This is the goal
  of any Grand Unified Theory, with the additionnal strong interaction
  unification to the electroweak interaction.} and strong interactions
unify at a very high energy scale, $\Lambda_{\rm GUT} \simeq 10^{16}$
GeV. Several attempts have been made for fourty years and
the most well--known solution is the long--celebrated $SU(5)$
theory~\cite{Georgi:1974sy} (see also Ref.~\cite{Buras:1977yy}). This
model is nowadays excluded because of the too short proton
life--time and the disagreement between the measured Weinberg
angle $\sin^2\theta_W$ and the predicted value of the $SU(5)$ model;
nevertheless it provides a very useful example of a grand unified
theory.

\begin{figure}[!h]
\begin{center}
\includegraphics[scale=0.65]{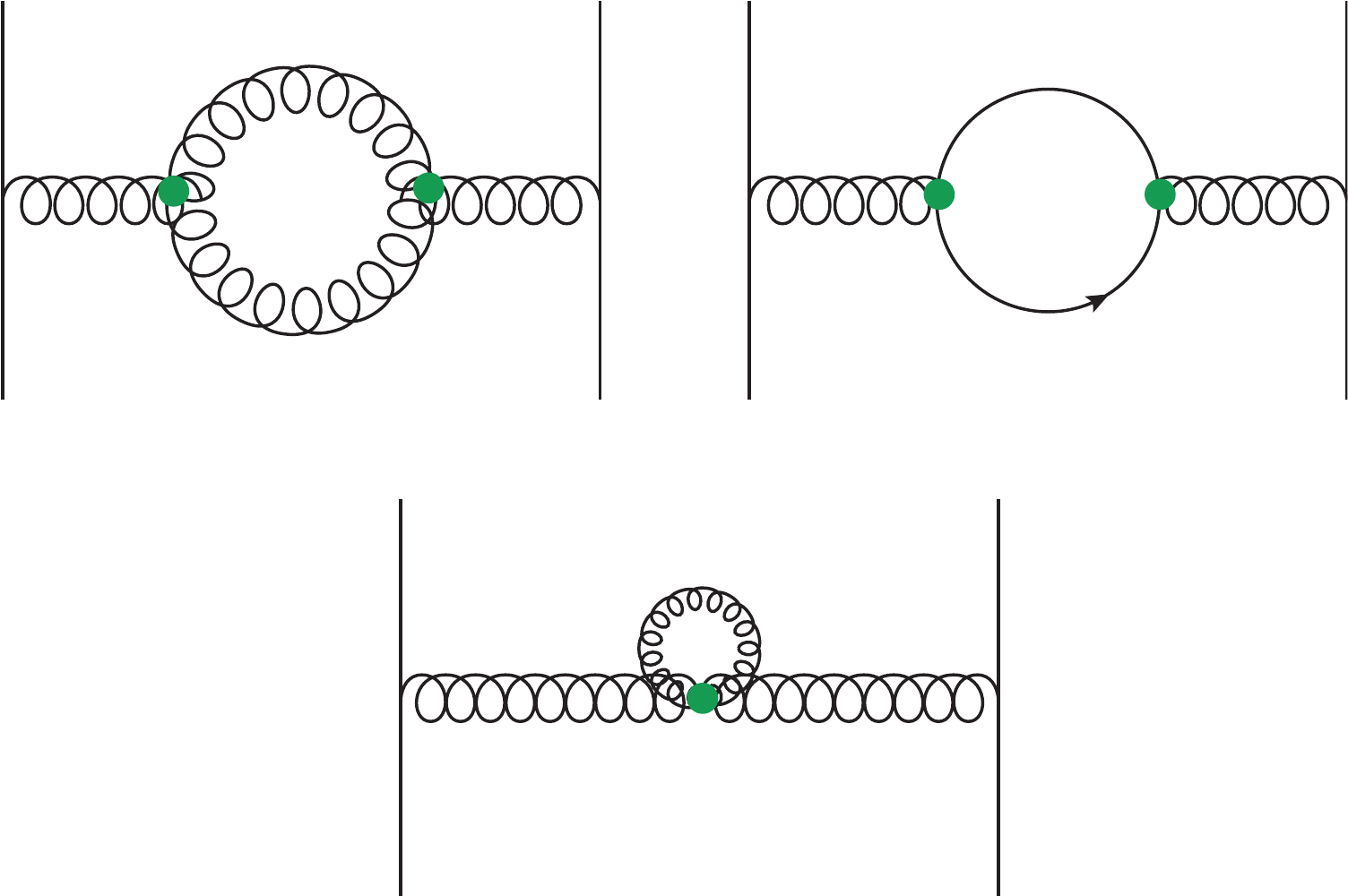} 
\end{center}
\vspace*{-.5cm}
\caption[One--loop corrections to gauge couplings]{Feynman diagrams
  for typical one--loop corrections to a generic gauge coupling. The
  gauge--gauge vertex diagrams are only relevant for non--abelian
  gauge symmetries.}
\label{fig:gaugecorrection} 
\end{figure}

Gauge couplings are renormalized and thus running couplings. The
running proceeds through quantum loop corrections, see
Fig.~\ref{fig:gaugecorrection} for typical diagramms at the 1--loop
order, and is described through renormalization group
equations~\cite{Georgi:1974yf, Langacker:1991an}
\beq
\frac{d\alpha_i}{d\ln Q^2} & = &  -b_i \alpha_i^2\, , \, \alpha_i \equiv
\frac{g_i^2}{4\pi}~(g_L \equiv g)\nonumber\\
b_s & = & \frac{23}{12\pi}~,~b_L =
\frac{19}{64\pi}~,~b_Y=-\frac{41}{40\pi}
\eeq
with the SM coefficients taken at the 1--loop order. The hypercharge
coupling constant $\alpha_Y$ has been rescaled by a factor of 5/3 in
order to match unified gauge description in the covariant derivation.

\begin{figure}[!h]
\begin{center}
\hspace{-4mm}
\includegraphics[scale=0.65]{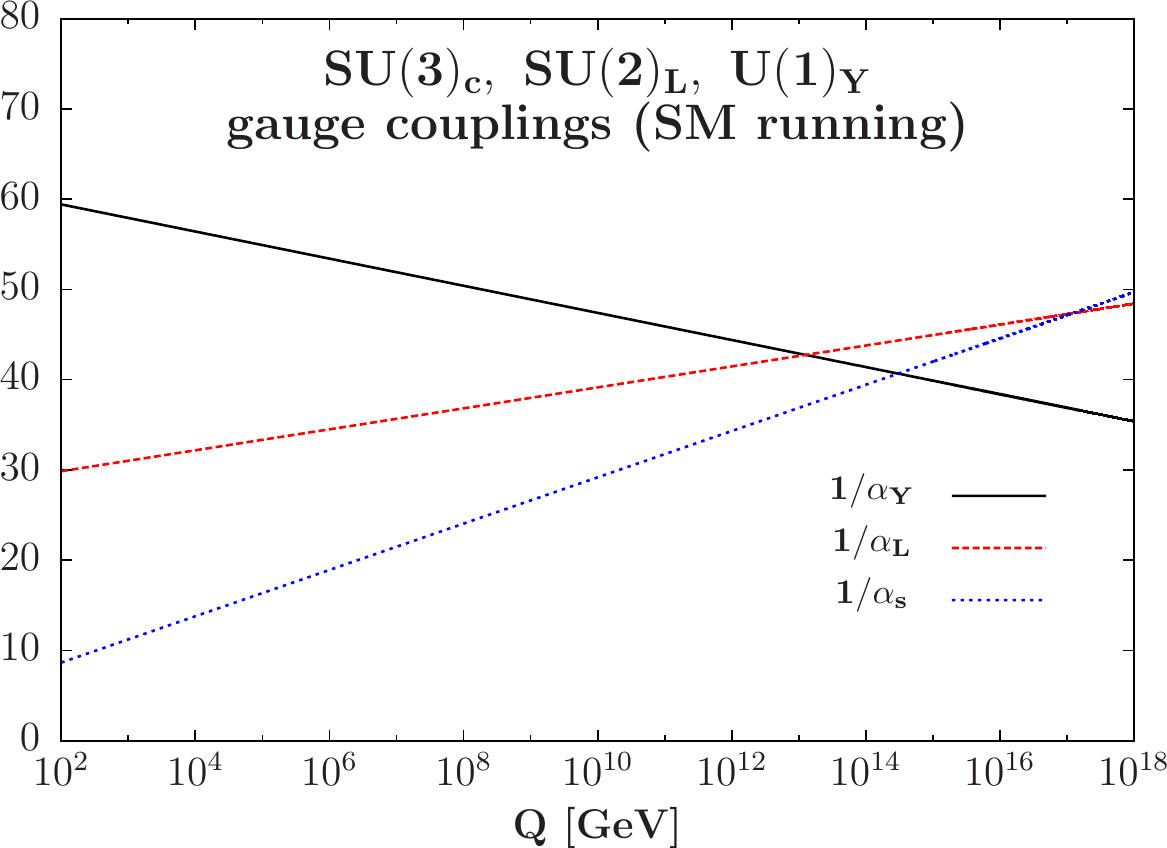}
\includegraphics[scale=0.65]{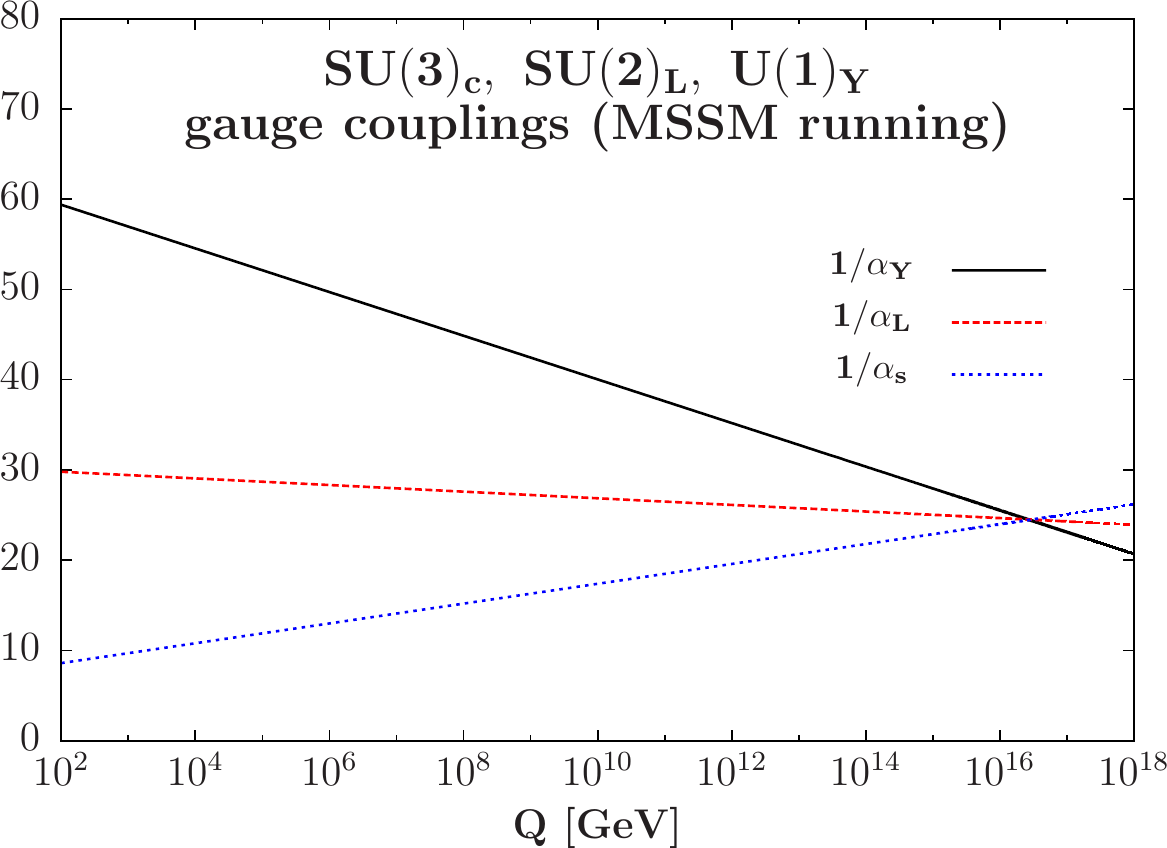}
\end{center}
\vspace*{-4mm}
\caption[$SU(3)_c\times SU(2)_L\times U(1)_Y$ gauge couplings running
from the weak scale up to the GUT scale]{$SU(3)_c\times SU(2)_L\times
  U(1)_Y$ gauge couplings running from the weak scale up to the GUT
  scale, within the SM (left) and within the MSSM (right).}
\label{fig:gauge_unification}
\vspace*{-1mm}
\end{figure}

Fig.~\ref{fig:gauge_unification} shows on the left the resulting
running within the SM, assuming no new particles content between the
weak scale and the GUT scale. The three gauge couplings seem to meet
at high energy but not exactly: unification is not exactly achieved if
we do not assume new particles content between the weak scale and the
GUT scale, assuming that the new degree of freedom required by grand
unification have a mass of order $\Lambda_{\rm GUT}$\footnote{There is no
reason why this should not be the case: if that were the case we would
have to introduce a mechanism explaining why these new particles had a
reduced mass.} and thus not contributing to the
RGE~\cite{Georgi:1974sy}.

The right part of Fig.~\ref{fig:gauge_unification} reproduces the same
exercice but replacing the RGE parameters by those obtained assuming
new particles content due to the additionnal particles within the
minimal supersymmetric extension of the SM, called the MSSM. Indeed
the running of the three gauge couplings is expected to change as
higher--order corrections involve new fermions and bosons (the
so--called superpartners of the standard particles). We thus have
\beq
b_Y \rightarrow -\frac{33}{20\pi}~,~ b_L \rightarrow
-\frac{1}{4\pi} ~,~b_s \rightarrow \frac{3}{4\pi}
\eeq
which now allow the three gauge couplings to perfectly match at
$Q\simeq 10^{16}$ GeV~\cite{Langacker:1991an}. This is a spectacular
reason why the addition of SUSY to the SM is very elegant: it helps to
push toward an actual unification of the three sub--atomic
interactions.

\subsection{SUSY and Dark Matter
  searches \label{section:IntroDarkMatter}}

The two subsections above have presented two technical reasons that
explain why SUSY is an interesting solution to both internal problems
of the SM and external difficulties in the wish to obtain a Grand
Unified Theory (GUT) where all the interactions are unified in a
single Lie algebra, with all the coupling constants merged into a
single one at very high energy. We will in this subsection present
an experimental reason of introducing SUSY, connected to cosmology
puzzles.

Indeed, all the cosmological observations\footnote{We here take the
  (standard) point of view that general relativity still holds at such
large scales. However this is nothing more than a mere assumption,
general relativity not being tested as such large scales. There are
alternative cosmological attempts to explain the cosmological
observations such as the galaxies profiles without the standard
$\Lambda$--CDM model based on general relativity on the cosmological
scales, which does not require a dark matter component, see
Ref.~\cite{Sanders:2002pf, Bekenstein:2004ne} for example. However we
will not come into such unecessary discussions for the rest of the
thesis.} point out to the existence of an unknown matter component of
the Universe which amounts to nearly 23\% of the total energy density
of our current Universe~\cite{Spergel:2003cb, Spergel:2006hy,
  Komatsu:2010fb}. If we take the exact numbers from
Ref.~\cite{Komatsu:2010fb} we have 
\beq
\Omega_{\rm DM} = 0.229~, ~\text{to be compared to}~ \Omega_{\rm
  baryon} = 0.046
\eeq
This matter density is more than four times the density of standard
matter that we know, which means that there is a very large amount of
the Universe that is yet to be understood, not to mention the nearly
73\% of dark energy density that is even more mysterious and drives the
current acceleration of the expansion of the Universe. Particle
physics is then confronted to a critical problem: how to account for
this missing 23\% piece of the Universe, which is absent from the SM?
Any theory that goes beyond the SM ought to provide for realistic dark
matter candidates.

If we come back to SUSY, we remind that we have introduced new
fermions and bosons in the theory, as a consequence of the symmetry
between bosonic and fermionic degrees of freedom. We then have new
vertexes and in particular we should care about the proton
decay. Indeed, the proton lifetime is estimated to be larger than
10$^{32}$ years, thus any theory beyond the SM (where the proton is
stable) must respect this lower bound, which means in practice to
assure that the proton remains stable. Strictly speaking, this is not
possible with the most general SUSY extension of the SM: for example
we could have the decay $p\to e^+ \pi^0$ as depicted in
Fig.~\ref{fig:protondecay} below.

\begin{figure}[!h]
\begin{center}
\includegraphics[scale=1.0]{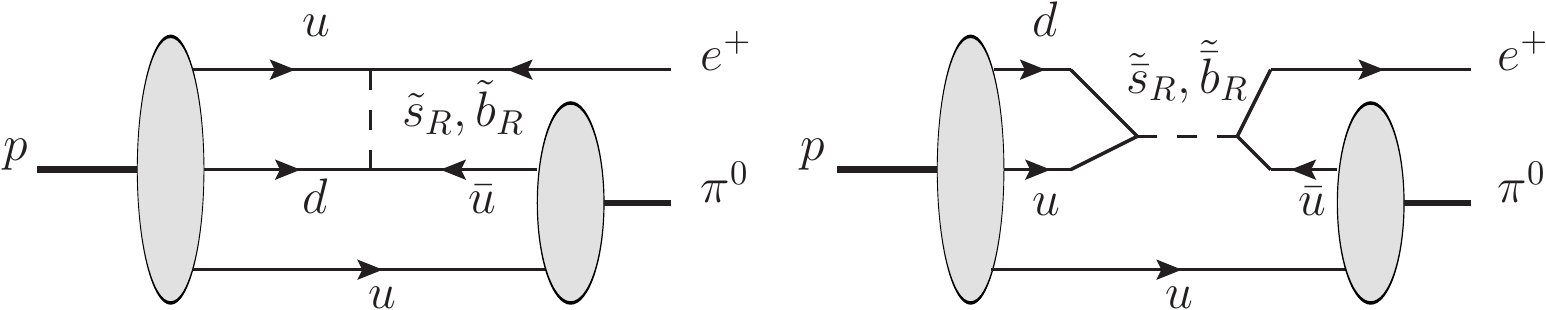} 
\end{center}
\vspace*{-.5cm}
\caption[Possible proton decay in SUSY theories without R--parity
conservation]{Some Feynman diagrams for the proton decay into a pion
  and a positron $p\to e^+ \pi^0$ through SUSY vertexes when
  $R$--parity is violated.}
\label{fig:protondecay} 
\end{figure}

This can be cast away if we impose, as is the case in the minimal
supersymmetric extension of the SM, a new global discrete symmetry
called $R$--parity~\cite{Farrar:1978xj}: we assign a multiplicative
number $R = (-1)^{3B+L+2S}$, where $B$, $L$ and $S$ stand respectively
for the baryon number, the lepton number and the spin. It reduces in
the end to $R=+1$ for standard matter, and $R=-1$ for
superpartners. $R$--parity is connected to the $B-L$ conservation,
indeed we have $(-1)^{3B+L+2S} = (-1)^{3(B-L)+2S}$. Any reaction is
now supposed to conserve this $R$ number. This discrete symmetry can
be seen as a relic of a global $U(1)_R$ symmetry broken down to a
$Z_2$ subgroup, see Ref.~\cite{DreesMSSM} page 359.

We can now evade the problem of the proton decay: as the quarks (u,d)
are standard particles their $R$--number is equal to one, thus it
forbids any vertex with only one superpartner involved such as the
violating vertexes in Fig.~\ref{fig:protondecay}.

How is this statement connected to our cosmological problem? If we
introduce $R$--parity to block the proton decay, it has also a very
interesting consequence: superparticles ought to be produced in pairs,
and if the proton is to be stable on the one hand, we can easily make
the same statement for the lightest supersymmetric particle. Hence we
predict that any supersymmetric spectrum following $R$--parity
conservation must have a stable particle!

The lightest supersymmetric particle (LSP) in the context of the
minimal supersymmetric extension of the SM is a neutral mixture of the
superpartners of the weak bosons and Higgs bosons\footnote{The plural
  form is not accidental: we will see later on that there must be at
  least two Higgs doublets in the MSSM.} called the neutralino
$\chi^{0}_1$. This is a stable, neutral, massive particle, hence an
ideal candidate for the dark matter component of the Universe (see
Ref.~\cite{Feng:2003zu} for a review on this topic). This is worth
mentioning that this LSP is nothing more than just one candidate among
many others. Nevertheless the LSP explication for dark matter is one
of the most popular scenerios that could explain the origin of dark
matter.

\vfill
\pagebreak

\section{Formal SUSY aspects}

Now that we have introduced some reasons to explain our interest in
supersymmetric theories, we are ready to sketch some formal aspects of
supersymmetry. Basically, as stated in the
section~\ref{section:SUSYIntro} above, supersymmetry (SUSY) is a
symmetry between bosons and fermions: each scalar boson has a spin
$\frac{1}{2}$ partner, each spin $\frac{1}{2}$ fermion has a spin 0
partner, each vector boson has a spin $\frac{1}{2}$
partner\footnote{Here we leave aside the case of the graviton and its
  gravitino superpartner of spin $\frac{3}{2}$.}. A theory is said to
be supersymmetric when the exchange of any pair of partners leaves the
action invariant. We will see that it is a very elegant extension of the
usual Poincaré algebra and in some sense ``natural''.

The first subsection will introduce the super--algebra, following the
outlines of a very famous theoretical physics theorem that shapes any
extension of the Poincaré algebra. We will then give some hints of the
superspace formalism that allow for a compact description of the
supermultiplets. We will end by the introduction to SUSY breaking, one
of the most important issue in the study of SUSY theories that is
still to be solved.

\label{section:FormalSUSY}

\subsection[SUSY Algebra]{Graded Lie Algebra and the Coleman--Mandula
  theorem \label{section:FormalAlgebra}}

The mathemetical formalism of SUSY theories is based on the concept of
graded Lie algebras which generalizes the usual Lie algebras to encompass
anticommutators as well as Lie brackets. The actual shape of SUSY
algebra is dictated by very powerful formal theorems: the
Coleman--Mandula theorem~\cite{Coleman:1967ad} and the
Haag--Lopuszanzski --Sohnius (HLS) theorem~\cite{Haag:1974qh} which
generalizes the Coleman--Mandula theorem and leads to SUSY graded Lie
algebra. We will not prove these theorems in this section, but discuss
their consequences.

\subsubsection{The Coleman--Mandula and HLS theorems}

We will follow Refs.~\cite{WessBegger, DreesMSSM} in this
paragraph. The statement of the Coleman--Mandula theorem is:
\begin{quotation}
{\it With the following assumptions:
\begin{enumerate}[$a)$]
\item{The $S$--matrix is based on a local, relativistic quantum field
    theory in four--dimensional spacetime;}
\item{For a given mass, there is only a finite number of different
    particles associated to one--particle states;}
\item{There is an energy gap between the vacuum and the one--particle
    states;}
\end{enumerate}
the most general Lie algebra of symmetries of the $S$--matrix contains
the Poincaré algebra and a finite number of Lorentz scalar operators
$B_l$ which must belong to a compact Lie algebra.}
\end{quotation}

This theorem then puts some restrictions on the form of the symmetries
than can be imposed on the $S$--matrix: any Lie algebra not of the
type of the Poincaré algebra shall be in direct sum with the Poincaré
algebra, that is, if we define the generator as ${T^a}$ and the
structure constants $f_{abc}$ we have:
\beq
[T^a,T^b]=f_{abc} T^{c}~,~ [T^a,P_{\mu}] = [T^a,M_{\mu\nu}]=0
\eeq
where $M_{\mu \nu}$ is the Lorentz generator, $P_\mu$ the translation
generator. We see that this imposes severe restrictions to any {\it
  bosonic} symmetry (represented by a gauge Lie algebra), but what if
we introduce a {\it fermionic} symmetry? This is precisely the spirit
of SUSY, in which graded Lie algebras escape the Coleman--Mandula
theorem by using anticommutators in addition to the usual Lie
commutators. The HLS theorem then states that the only graded Lie
algebra which is compatible with the Coleman--Mandula theorem
assumptions is precisely that of the supersymmetry algebra which will
be described below.

The HLS theorem precisely states the following (see also
Ref.~\cite{DreesMSSM}, page 31):
\begin{quotation}
{\it The Lie algebra which follows the assumptions of the
  Coleman--Mandula theorem is a $Z_2$ graded Lie algebra which odd
  generators belong to the representations $(\mathbf{0,\frac{1}{2}})$
  and $(\mathbf{\frac{1}{2},0})$ of the Lorentz algebra and even
  generators are a direct sum of the Poincaré generators and other
  symmetry generators.}
\end{quotation}

This puts a restriction over the form of the new fermionic symmetry
that we want to introduce through the anticommutators: this must be
represented by a set of two $(Q,\bar{Q})$ generators which are of spin
$\frac{1}{2}$, but not higher. The minimal choice is to take a set of
two Majorana spinors $(Q_{a}, \bar{Q}_{\dot{a}})$ (ie where the spinor
is invariant under charge conjugaison) and we will use Weyl
two--components notation; in four--components notation with Dirac
spinors we would have one Majorana spinor
\beq
Q = \left( \begin{matrix}Q_{a}\\ \bar{Q}^{\dot
      a}\end{matrix}\right)\nonumber
\eeq

We will give the final super--Poincaré algebra relations. If we follow
the HLS theorem, we see that the anticommutator $\{Q_{a},
  \bar{Q}_{\dot{a}}\}$ has no choice but to be proportionnal to $P_\mu$
(if not it would violate the Coleman--Mandula theorem). This precisely
dictates the final shape of the super--Poincaré algebra:
\beq
\left\{ Q_{a},\bar{Q}_{\dot{b}} \right\} & = & 2 \sigma^{\mu}_{a \dot{b}}
P_{\mu}\nonumber\\
\left\{ Q_{a}, Q_{b} \right\} & = & \left\{ \bar{Q}_{\dot{a}},
  \bar{Q}_{\dot{b}} \right\} = 0 \nonumber \\
 \left[ Q_{a}, P_{\mu} \right] & = & \left[ \bar{Q}_{\dot{a}}, P_{\mu} \right]  =
 0\nonumber\\
\left[ P_{\mu}, P_{\nu} \right] & = & 0\nonumber\\
\left[ Q_{a}, M_{\mu \nu} \right] & = &  (\sigma_{\mu\nu})_{a}^{b}
Q_{b}
\label{eq:SUSYalgebra}
\eeq
where we recall that $\displaystyle \sigma^{\mu\nu} \equiv \frac{i}{4}
\sigma^{[\mu}\bar{\sigma}^{\nu]}$, $\sigma^{\mu}$ being the usual set
of Pauli matrices, $\sigma^{0}=\bar{\sigma}^{0}=I$, $\sigma^{1,2,3} =
- \bar{\sigma}^{1,2,3}$. In four--components notation with only one
four--components Majorana generator $Q$ the super--Poincaré
algebra~\ref{eq:SUSYalgebra} reads (see Ref.~\cite{DreesMSSM} page 34)
\beq
\left\{ Q_{a},Q_{b} \right\} & = & -2 \left(\gamma^{\mu} C\right)_{ab}
P_{\mu}\nonumber\\
 \left[ Q_{a}, P_{\mu} \right] & = & 0\nonumber\\
\left[ P_{\mu}, P_{\nu} \right] & = & 0\nonumber\\
\left[ Q_{a}, M_{\mu \nu} \right] & = &  (\Sigma^{\mu \nu})_{a}^{b}
Q_{b}
\label{eq:SUSYalgebra2}
\eeq
where this time we have $\displaystyle \Sigma^{\mu \nu} \equiv \frac{i}{4}
\gamma^{[\mu}\gamma^{\nu]}$ and the charge conjugate operator
$C\equiv\imath \gamma^{2}\gamma^{0}$ in Dirac representation. 

Using the SUSY algebra we can derive two important consequences:
\begin{itemize}
\item{{\it In each supermultiplet spanning a representation of the SUSY
    algebra there is an equal number of bosons and fermions:} if we
  introduce the operator $(-1)^{2S}$ which takes +1 eigenvalue on
  bosonic states and -1 eingenvalue on fermionic states, and take a
  closed supermultiplet $\{|i\rangle\}$ with a definite (non--zero)
  $p_\mu$ momentum, we have
\beq
\text{Tr~} [(-1)^{2S} \left\{Q_a, Q_b\right\}] & = & \text{Tr~}
[(-1)^{2S} Q_a Q_b] + \text{Tr~} [(-1)^{2S} Q_b Q_a]\nonumber\\
 & = & \text{Tr~} [(-1)^{2S} Q_a Q_b] + \text{Tr~} [Q_a (-1)^{2S} Q_b]\nonumber\\
 & = & \text{Tr~} [(-1)^{2S} Q_a Q_b] - \text{Tr~} [(-1)^{2S} Q_a Q_b]
 = 0\nonumber
\eeq
using the cyclicity of the trace operator and the fact that
$\left\{(-1)^{2S},Q_a\right\}=0$, as can be easily verified. If we
replace the anticommutator by the first equality of
Eq.~\ref{eq:SUSYalgebra2} it reads
\beq
\text{Tr~} [(-1)^{2S} P_{\mu}] = p_{\mu} \text{Tr~} (-1)^{2S} =
0\nonumber
\eeq
discarding the constant $\gamma^{\mu}$ factor. Having chosen $p_\mu
\neq 0$ this reduces to
\beq
\text{Tr~} (-1)^{2S} = n_{B} - n_{F} = 0
\eeq
where $n_F$ ($n_B$) stands for the total number of fermionic states
(bosonic states) within the supermultiplet. This is thus exactly the
result we were looking for.}
\item{{\it Each superpartner must have the same mass within a SUSY
      multiplet:} indeed if SUSY is exact, $[Q_a,P_\mu]=0$ implies
    that $[Q_a, P^2]=0$.}
\end{itemize}

We want to note that the super--Poincaré algebra has two important
features: the first is that it is an extension of the usual
space--time symmetry and thus can be considered also as a space--time
symmetry, making SUSY a fundamental symmetry; the second feature is
that as the anticommutator of two supersymmetry generators $Q$
gives the translation generator $P_\mu$, it means that if we gauge
SUSY we also gauge the usual space--time symmetry. Thus there is an
intimate connexion between SUSY and gravity which is described in the
context of supergravity theories~\cite{WessBegger}. It is sometimes
said that gravity is ``the square--root of SUSY''.

\subsubsection[The Wess--Zumino free--field model]{The Wess--Zumino
  free field model as an example of a SUSY lagrangian}

We close this subsection by giving the most simple example of a SUSY
lagrangian, the free field model called the massless non--interacting
Wess--Zumino model~\cite{Wess:1974tw}, following the outlines of
Ref.~\cite{Martin:1997ns}. To parametrize a SUSY transformation we
introduce an infinitesimal $\epsilon$ Grassmann parameter, which
implies using the super--Poincaré algebra of Eq.~\ref{eq:SUSYalgebra}
that $[\epsilon Q, \bar{\epsilon} \bar{Q} ] = 2
\bar{\epsilon} \sigma^{\mu} \bar{\epsilon} P_\mu$.

In the Wess--Zumino model we have one complex
scalar field and one left--handed Weyl spinor field. The former is the
superpartner of the latter; if we count the number of degrees of
freedom we see that we have on--shell 2 bosonic and 2 fermionic degrees
of freedom; nevertheless we would like to describe the most general
lagrangian in the view of the quantum corrections. This means that we
have to introduce an auxiliary scalar field $F$ in order to match the
4 fermionic degree of freedoms off-shell. This constitute the chiral
supermultiplet. The lagrangian is written as:
\beq
{\cal L} = \imath \bar{\psi} \bar{\sigma}^{\mu} \partial_\mu \psi
- \partial^\mu \phi^* \partial_{\mu} \phi + F^* F
\label{eq:WessZumino}
\eeq

In Eq.~\ref{eq:WessZumino} we see explicitely that $F$ is an auxiliary
field, as its equation of motion is $F^*=F=0$. Its has a mass
dimension of 2 which is unusual for a scalar field. A SUSY
transformation is written as
\beq
\delta_{\epsilon} \psi & = &  (\epsilon Q + \bar{\epsilon} \bar{Q})
\psi\nonumber\\
\delta_{\epsilon} \phi & = &  (\epsilon Q + \bar{\epsilon} \bar{Q})
\phi\nonumber\\
\delta_{\epsilon} F & = &  (\epsilon Q + \bar{\epsilon} \bar{Q}) F
\eeq
which reads~\cite{Martin:1997ns, WessBegger}
\beq
\delta_\epsilon \phi & = & \sqrt{2} \epsilon \psi\nonumber\\
\delta_\epsilon \psi & = & \imath \sqrt{2} \sigma^{\mu}
\bar{\epsilon} \partial_\mu \phi + \sqrt{2} \epsilon F\nonumber\\
\delta_\epsilon F & = & \imath \sqrt{2} \bar{\epsilon}
\bar{\sigma}^{\mu} \partial_\mu \psi
\label{eq:WessZuminoSUSY}
\eeq

In Eq.~\ref{eq:WessZuminoSUSY} the role of $F$ is to close the SUSY
algebra, that is $[\epsilon Q, \bar{\epsilon} \bar{Q} ] = 2
\bar{\epsilon} \sigma^{\mu} \bar{\epsilon} P_\mu$, even in the
off--shell case. The lagrangian~\ref{eq:WessZumino} is then invariant
up to a total space--time derivative:
\beq
\delta {\cal L} & = &  \partial_{\mu} \left( i
    F\bar{\psi} \bar{\sigma}^{\mu} \epsilon + \epsilon \sigma^{\nu}
    \bar{\sigma}^{\mu} \psi \partial_{\nu} \phi^* + \epsilon
    \psi \partial^{\mu} \phi^* + \bar{\epsilon}
    \bar{\psi} \partial^{\mu} \phi \right)
\eeq
The action is then fully SUSY invariant. We will see in the next
section how to introduce in a simple way interactions which follow
SUSY invariance, through the concept of superpotential.

\subsection{Superspace, superfields and
  superpotential \label{section:FormalSuperspace}}

We introduce in this subsection the superspace and superfield
formalism which is a very compact and elegant formulation of SUSY,
enlightening its deep connection with space-time symmetries. We will
follow Refs.~\cite{WessBegger, DreesMSSM}.

The first step is to introduce the superspace: this is simply the
usual space--time with the coordinates $x\equiv x_\mu = (c t,X,Y,Z)$ enlarged
to include two additionnal two--components Grassmann (anticommuting) coordinates
$\theta,\bar{\theta}$. We then have $\theta \theta \equiv \theta^a
\theta_a$ and $\bar{\theta} \bar{\theta} \equiv \bar{\theta}^{\dot a}
\bar{\theta}_{\dot a}$, but $\theta_a \theta_b = - \theta_b \theta_a$,
$\theta_a \bar{\theta}_{\dot b} = - \bar{\theta}_{\dot b}
\theta_a$. This means that any product of $\theta,\bar{\theta}$ which
contains more than three $\theta/\bar{\theta}$ terms vanishes.

We introduce fields on this superspace, which we call superfields. By
the virtue of the Grassmann coordinates, any superfield ${\cal F}$ can
be expanded in a finite serie of $\theta,\bar{\theta}$ variables:
\beq
{\cal F}(x,\theta,\bar \theta) & = & \phi(x) + \sqrt{2} \theta \psi(x)
+ \sqrt{2} \bar \theta \bar \chi (x) + \theta \theta M(x) + \bar
\theta \bar \theta N(x) +\nonumber\\
 & & \theta \sigma^{\mu} \bar \theta A_{\mu}(x) + \theta \theta \bar
 \theta \bar \lambda (x) + \bar \theta \bar \theta \theta \zeta (x) +
 \frac{1}{2} \theta \theta \bar \theta \bar \theta D(x)
\label{eq:generalsuperfield}
\eeq

As the operators $\epsilon Q$ and $\bar{\epsilon} \bar Q$ follow an ordinary
Lie algebra, it is temptating to use the following SUSY transformation
representation $\exp(  \imath (x^\mu P_\mu + \epsilon Q + \bar \epsilon
\bar Q))$~\cite{WessBegger}. A linear SUSY transformation then reads
in the superspace as 
\beq
(x^\mu,\theta,\bar \theta) + \delta(x^\mu,\theta,\bar \theta) = (x^\mu
- \imath \theta \sigma^{\mu} \bar \epsilon + \imath \epsilon
\sigma^{\mu} \bar \theta, \theta + \epsilon, \bar \theta + \bar \epsilon)
\eeq
which then induces the following linear representation of the
supersymmetric generators
\beq
Q_a & = & -\imath \left(\frac{\partial}{\partial \theta^a} + \imath
\sigma^{\mu} \bar{\theta} \frac{\partial}{\partial
  x^{\mu}}\right)\nonumber\\
\bar{Q}_a & = & -\imath \left(\frac{\partial}{\partial \bar{\theta}^a}
  + \imath \theta \sigma^{\mu} \frac{\partial}{\partial
    x^{\mu}}\right)\nonumber\\
\delta {\cal F} & = &  \imath (\epsilon Q + \bar{\epsilon} \bar{Q} )
{\cal F}
\eeq

We are now ready to present the chiral and vector superfields that are
used to construct the SUSY extension of the SM.

\subsubsection{The chiral superfield}

The chiral superfield will associate the standard fermionic fields
with their spin 0 superpartners. In order to have a SUSY invariant
definition, we shall construct a SUSY covariant derivative. As $P_\mu
= - \imath \partial_\mu$ commute with $Q$ but not
$\displaystyle \partial_\theta \equiv \frac{\partial}{\partial \theta^a}$ nor
$\partial_{\bar \theta}$. Following Ref.~\cite{DreesMSSM} we define
\beq
{\cal D}_a & \equiv & \partial_{\theta_a} - \imath \sigma^{\mu}_{a
  \dot b} \bar{\theta}^{\dot b} \partial_\mu\nonumber\\
\overline{\cal D}_{\dot a} & \equiv & -\partial_{\bar{\theta}_{\dot a}} +
  \imath \theta^{b} \sigma^{\mu}_{b \dot a} \partial_\mu 
\eeq
with which we define the (left) chiral superfield to follow these two
equivalent conditions:
\beq
{\cal D}_{a} \Phi^{\dagger} = 0~~{\rm or}~~ \overline{\cal D}_{\dot a} \Phi =0
\label{eq:chiralSUSY}
\eeq

This gives this component decomposition
\beq
\Phi(x,\theta,\bar \theta) & = & \phi(x) - \imath \theta
\sigma^{\mu} \bar\theta \partial_\mu \phi(x) -\frac{1}{4} \theta
\theta \bar\theta \bar\theta \partial_\mu \partial^{\mu} \phi(x)
+\nonumber\\
 & & \sqrt{2} \theta \psi(x) + \frac{\imath}{\sqrt 2} \theta
\theta \partial_{\mu} \psi \sigma^{\mu} \bar\theta + \theta \theta
F(x)
\eeq

The SM fermion field will be $\psi(x)$, its scalar superpartner will
be $\phi(x)$ and called a sfermion. The field $F(x)$ will be the
auxiliary field necessary to ensure SUSY invariance in the off--shell
case. Any product of chiral superfields will still be a
superfield~\cite{DreesMSSM, WessBegger}; the $F$--component of a
chiral superfield, that is the $\theta \theta$ component, transforming
as a total spatial derivative under a SUSY transformation, the
$F$--component of any product of chiral superfields is a candidate for
any lagrangian describing a SUSY invariant action.

The product $\Phi^{\dagger}\Phi$ is not a chiral superfield;
nevertheless its highest $\theta \theta \bar \theta \bar\theta$
component is SUSY invariant up to a total derivative like any
$\theta\theta \bar\theta \bar\theta $ component, and reads
\beq
F^* F + \frac12 \partial_{\mu} \phi^* \partial^{\mu} \phi -\frac14
(\phi \partial_\mu \partial^{\mu} \phi^* + ~{\rm h.c.}) + \frac12
\imath \psi \sigma^{\mu} \partial_\mu \bar{\psi} + \imath \frac12
\bar{\psi} \bar{\sigma}^{\mu} \partial_\mu \psi
\eeq

This will constitute the kinetic part of the chiral superfield. We
then can rewrite the Wess--Zumino massless free lagrangian of
Eq.~\ref{eq:WessZumino} simply as
\beq
{\cal L}_{\rm free} = \frac{\partial \Phi^{\dagger}\Phi}{\partial
  \bar\theta\bar\theta \theta \theta} \equiv \Phi^{\dagger}\Phi
|_{\theta\theta \bar\theta\bar\theta}
\eeq

\subsubsection{The vector superfield}

The second type of superfield will yield the gauge vector bosons. We
impose a reality condition which define a vector superfield
\beq
V^{\dagger} = V
\eeq
We can then use the general superfield written in
Eq.~\ref{eq:generalsuperfield} to obtain the expression of a vector
superfield, but we will simplify the task in the next lines.

Following strictly Ref.~\cite{DreesMSSM} it is interesting to note
that if we use a chiral superfield $\Lambda$ the superfield $V =
\imath \Lambda - \imath \Lambda^{\dagger}$  is a vector
superfield. This implies the following abelian super--gauge
transformation $V^{'} = V + \imath \Lambda(x,\theta,\bar\theta) -
\imath \Lambda^{\dagger}(x,\theta,\bar \theta)$. This is indeed a
gauge transformation as $\Lambda$ is a function of the super--space
coordinates. We are free to chose a particular gauge, the Wess-Zumino
gauge in which a vector superfield reduces to
\beq
V = \theta \sigma^{\mu} \bar\theta A_\mu (x) + \theta\theta \bar\theta
\bar{\lambda} (x) + \bar\theta\bar\theta \theta \lambda(x) +\frac12
\theta\theta\bar\theta\bar\theta D(x)
\eeq

The SM gauge boson field will be $A_\mu(x)$, its fermionic
superpartner will be $\lambda(x)$ and called the gaugino. The $D(x)$
field is that of the auxiliary necessary to ensure SUSY invariance
even in the off--shell case, as did the $F(x)$ field in the case of a
chiral superfield. As stated in Ref.~\cite{DreesMSSM} the $D$--term
has the very interesting virtue to remains invariant under a
supergauge transformation, making this term a very good candidate for
writing supergauge and SUSY invariant lagrangian. It is actually very
easy to introduce a $U(1)$ supergauge theory by taking a chiral
superfield $\Phi$ and a vector superfield $V$, with
\beq
\Phi & \rightarrow & \Phi^{'} = e^{-2\imath g Y_{\Phi} \Lambda} \Phi\nonumber\\
\Phi^{\dagger} & \rightarrow & {\Phi^{\dagger}}^{'} = e^{2\imath g
  Y_\Phi \Lambda^{\dagger}} \Phi^{\dagger}\nonumber\\
V & \rightarrow & V^{'} = V + \imath (\Lambda - \Lambda^{\dagger})
\label{eq:SUSYabeliantheory}
\eeq
where $Y_\Phi$ is the $U(1)$ charge of the superfield $\Phi$. A vector
field is of zero mass dimension and thus can be exponentiated, which
means that we can define
\beq
{\cal L}_{\rm kinetic} = \Phi^{\dagger} e^{2 g Y_{\Phi} V}
\Phi|_{\theta\theta\bar\theta\bar\theta}
\eeq
to be the lagrangian part containing both the kinetic term for the
chiral superfield (the first factor in the exponential expansion) and
the gauge invariant interacting term (i.e. the usual covariant
derivative). The extension to an non--abelian supergauge theory
follows the same pattern, that is introducing a gauge multiplet of
chiral superfields $\Phi_I$ and then write down (in the Wess--Zumino
gauge):
\beq
\Phi_I & \rightarrow & \Phi_I^{'} = \left(e^{-2\imath g_2 T_a
    \Lambda^{a}}\right)_{IJ} \Phi_J\nonumber\\
V_a T^{a} & \rightarrow & V^{'}_a T^{a} = V_a T^{a} + \imath
\left(\Lambda_a T^a - (\Lambda_a T^a)^{\dagger}\right) +
\frac{\imath}{2} \left[ V,\Lambda_a T^a + (\Lambda_a
  T^a)^{\dagger}\right]
\eeq
with $T^a$ as the usual generators of the non--abelian gauge
algebra. Note that the $a$ notation is here a non--abelian gauge index
and not a spinorial index.

We now end this paragraph by giving the superfield--strength $W$ which
is a chiral superfield (see Ref.~\cite{DreesMSSM} page 65):
\beq
W_A = -\frac14 \overline{\cal D}_{\dot B} \overline{\cal D}^{\dot B} {\cal D}_A
V ~,~ W_{\dot A}^{\dagger} = -\frac14 {\cal D}_{B} {\cal D}^{B} \overline{\cal
  D}_{\dot A} V
\eeq
with $A$ as a spinorial index. This superfield contains the usual
field--strength tensor $F_{\mu\nu}$. We then define
\beq
{\cal L}_{\rm gauge~kinetic} = \frac{1}{4} \left.\left(W_A W^{A} +
    W_{\dot A}^{\dagger} {W^{\dagger}}^{\dot A}\right)\right|_{\theta
  \theta}
\eeq
which is coherent as we have a sum of product of chiral superfields
which itself is also a chiral superfield, thus its $F$--term is SUSY
invariant.

\subsubsection{The superpotential}

The last remaining piece that is needed to write down the most
complete SUSY (renormalizable) lagrangian is the interaction among
the scalar fields and between the scalars and fermions fields. The
most simple way to introduce these interactions is to use a
superpotential, in much the same spirit as above for the definition of
the covariant derivative.

As the product of chiral superfields is also a superfield, the
definition of a renormalizable superpotential is a superfield ${\cal
  W}$ restricted by the following conditions:

\begin{enumerate}[$1.$]
\item{It is an holomorphic expression of $\Phi_i$ chiral superfield,
    thus not containing any $\Phi^{\dagger}$ superfield;}
\item{It does not contain any product higher that $\Phi_i\Phi_j\Phi_k$
  to ensure renormalizability;}
\end{enumerate}

The same conditions hold for the ${\cal W}^{\dagger}$ superfield where we
replace $\Phi_i$ by $\Phi^{\dagger}_i$ everywhere. The induced term in
the lagrangian will be simply the $F$--term of the superpotential. The
final lagrangian for a $U(1)$ gauge interaction among chiral
superfields $\Phi_i$ mediated by a vector superfield $V$ then writes
\beq
{\cal L} & =  & {\cal L }_{\rm kinetic} + {\cal L}_{\rm
  interaction}\nonumber\\
{\cal L}_{\rm kinetic} & =  & \frac{1}{4} \left.\left(W_A W^{A} +
    W_{\dot A}^{\dagger} {W^{\dagger}}^{\dot A}\right)\right|_{\theta
  \theta} +\left.\left( \sum_i \Phi_i^{\dagger} e^{2 g Y_{\Phi_i} V}
    \Phi_i + \eta V
  \right)\right|_{\theta\theta\bar\theta\bar\theta}\nonumber\\
{\cal L }_{\rm interaction} & = & {\cal W}(\Phi_i)|_{\theta\theta}
+~{\rm h.c.}\nonumber\\
 & = & \left.\left(\sum_i ( h_i \Phi_i) + \frac12 m_{ij} \Phi_i \Phi_j
     +  \frac16 \lambda_{ijk}
     \Phi_i\Phi_j\Phi_k\right)\right|_{\theta\theta} +~\rm{h.c.}
\label{eq:SUSYgeneralLagrangian}
\eeq
in which $h_i$ is a complex number, $m_{ij}$ and $\lambda_{ijk}$ are
complex tensors symmetric in their indices, and we have also included
a $D$--term $\eta V$ with $\eta$ being real, as the $D$--term of any
vector superfield is both SUSY and supergauge invariant. The
generalization to non--abelian gauge interactions is straightforward. 

The use of the Lagrangian equations for the auxiliary fields which can
then be eliminated gives
\beq
F_i = -\overline{\cal W}_i \equiv - \left.\frac{\partial
    {\cal W}^{\dagger}}{\partial
    \Phi^{\dagger}_i}\right|_{\theta=\bar\theta=0} ~,~ F_i^{*} =
-{\cal W}_i \equiv -\left.\frac{\partial {\cal W}}{\partial
    \Phi_i}\right|_{\theta=\bar\theta=0}
\eeq

In the lagrangian~\ref{eq:SUSYgeneralLagrangian} the scalar potential
then reads
\beq
V(\phi_i) = \sum_i |F_i|^2 + \frac12 |D|^2 = \sum_i |{\cal W}_i|^2 +
\frac12 |\eta + g \phi^{*}_i \frac{Y_{\Phi_i}}{2} \phi|^2
\eeq
where the $D$--term comes from the vector superfield and is dictated
by SUSY and gauge invariance. The Yukawa interactions among scalar
fields and fermion fields reads 
\beq
{\cal L}_{\rm Yukawa} = -\frac12 \left( \psi_i \psi_j
  \left.\frac{\partial{\cal W}}{\partial
      \Phi_i\Phi_j}\right|_{\theta=\bar\theta=0} +~{\rm h.c.}\right)
\eeq

The superpotential then controls the shape of the scalar potential
which is of utmost importance when dealing with spontaneous gauge
symmetry breaking. We have finished the description of SUSY lagrangian
in the very compact and elegant superspace description which
generalizes the usual space--time field formalism. If we look at the
lagrangian~\ref{eq:SUSYgeneralLagrangian} the writing of interacting
term is as in the usual field formalism, but with ordinary fields
replaced by superfields and taking the $F$ or $D$ terms in the
end. This is why the superspace description is so powerful.

\subsection{Soft SUSY breaking \label{section:FormalBreaking}}

Up until now we have dealt with exact SUSY lagrangians, that are known
to be renormalizable, see Ref.~\cite{Iliopoulos:1974zv}. However
experiments tells us that SUSY is broken: indeed as seen in
section~\ref{section:FormalAlgebra} we should observe the exact
superpartner of the electron, that is a scalar selectron with the same
mass $m_{\tilde{e}} = 511$ keV. That is obviously not the case, hence
SUSY has to be broken.

\subsubsection{General soft SUSY beaking}

If we recall the virtues of SUSY introduced in
section~\ref{section:SUSYIntro} we do want to break SUSY {\it but
  still keeping these virtues}, and in particular the resolution of
the weak hierarchy problem explained in
section~\ref{section:IntroHierarchy}. We recall that if we take a
Dirac fermion $f$ and its scalar superpartners $\tilde{f}_L$,
$\tilde{f}_R$, exact SUSY which cancels completely the 1--loop
corrections to the Higgs mass leads to $m_f = m_{\tilde{f}_L} =
m_{\tilde{f}_R}$, no trilinear $A_f H \tilde{f}_L \tilde{f}_R^{*}$
coupling and the Yukawa equality $\lambda_{\tilde{f}} =
\lambda_f^2$. We still want to cancel quadratic divergences and thus
keep the equality $\lambda_{\tilde{f}} = \lambda_f^2$, but break the
equality $m_f = m_{\tilde{f}_{L,R}}$ and allow for $A_f\neq 0$, which
means that logarithmic divergences will arise to be renormalized by
the standard procedures. An explicit SUSY breaking that ensures these
conditions is called soft SUSY breaking: only terms of mass dimension
not higher that three are authorized in the soft SUSY breaking part of
the lagrangian~\cite{Iliopoulos:1974zv, Dimopoulos:1981zb,
  Sakai:1981gr}. We thus have
\beq
{\cal L} & = & {\cal L}_{\rm SUSY} + {\cal L}_{\rm soft}\nonumber\\
{\cal L}_{\rm soft} & = & -\phi_i^{*} (m^2)_{ij} \phi_j + \left( \frac16
  f_{ijk}A_{ijk} \phi_i \phi_j\phi_k -\frac12 (B\mu)_{ij}\phi_i\phi_j
  + \sum_i h_iC_i \phi_i +~{\rm h.c.}\right) \nonumber\\
 & & - \frac12\left( \sum_{\alpha} M_{\alpha} \lambda_{\alpha}^a
   \lambda_{\alpha}^a +~{\rm h.c.}\right)
\label{eq:generalsoftSUSY}
\eeq
$M_\alpha$ is the mass term for the gauginos $\lambda_{\alpha}^a$, $a$
being a gauge index for a given algebra $G_\alpha$; $(m^2)_{ij}$ is an
hermitian matrix describing the mass terms for the sfermions, $h_i
C_i$, $(B\mu)_{ij}$ and $f_{ijk}A_{ijk}$ being the tadpole, bilinear
and trilinear scalar couplings. ${\cal L}_{\rm soft}$ obviously breaks
SUSY as there are no fermions nor gauge bosons within it. We note for
consistency that $h_i$, $\mu_{ij}$ and $f_{ijk}$ are SUSY invariant
factors that come from the superpotential, see
Eq.~\ref{eq:SUSYgeneralLagrangian} with $m$ replaced by $\mu$ (thus
reserving $m$ for the SUSY breaking sfermions masses terms).

The tadpole arises only with gauge singlet; as it does not appear in
the minimal supersymmetric extension of the SM, we will discard this
term in the rest of the thesis.

\subsubsection{What is the origin of soft SUSY breaking?}

It is obvious that this way of SUSY breaking is not satisfactory: we
would like to have the same receipt as for the electroweak symmetry
breaking, namely a spontaneous breakdown of supersymmetry that is more
elegant than just explicit breaking and could explain the origin of
the various SUSY breaking terms. This is one of the most tedious
questions in SUSY theoretical studies nowadays and we get several
mechanisms on the market, none of them being convincing in every
situation. We will not come into details but rather present how can
spontaneous SUSY breaking arise and what will be the impact on
phenomenological studies.

In exact SUSY theories the hamiltonian is always positive, as we get,
see Refs~\cite{DreesMSSM, WessBegger, Martin:1997ns},
\beq
H = P_0 =  \frac{1}{4} ( Q_1 \bar{Q}_1 + Q_2 \bar{Q}_2 + \bar{Q}_1
Q_1 + \bar{Q}_2 Q_2) 
\eeq

If the vacuum is SUSY invariant, we should have $\delta|\Omega\rangle
= \imath (\epsilon Q + \bar{\epsilon} \bar{Q}) |\Omega\rangle = 0$
for any pair of $\epsilon,\bar{\epsilon}$ parameters, that is
$Q|\Omega\rangle = \bar{Q} |\Omega\rangle = 0$. It then turns that
{\bf if SUSY is exact the vacuum state $|\Omega\rangle$ must have a
  vanishing energy}. The minimum of the scalar potential must be zero
in exact SUSY theories.

Taking the other point of view, it means that as long as the scalar
potential is non zero at its global minimum, we have spontaneously
broken SUSY: this can be done through auxiliary $F$ or $D$ terms which
are the only non physical fields that can play this role, as the
scalar potential contains these two fields. A SUSY breaking through
$F$--terms is called the O'Raifeartaigh
mechanism~\cite{O'Raifeartaigh:1975pr}, if we use $D$--terms it is
called the Fayet--Iliopoulos mechanism~\cite{Fayet:1974jb}. Only
$F$--terms ensure a spontaneously SUSY breaking with non--abelian
gauge theories, see Ref.~\cite{DreesMSSM} page 140.

In any phenomenological theory which incorporates spontaneous symmetry
breaking, such as the constrained minimal extension of the 
SM,  the spontaneous breakdown of SUSY is assumed to arise in a
``hidden sector'' of the theory at the GUT scale. This hidden sector
has only very small couplings (if any) to the phenomenological sector
(or visible sector) and we need messengers, that is superfields that
share some interactions between the hidden sector and the visible
sector. There are several ways of achieving such a goal (see
Refs.~\cite{DreesMSSM, Martin:1997ns} for more details), one of the
most popular using gravity interactions in the minimal supergravity
theory (mSUGRA). In that case SUSY is broken in the hidden sector by
some vev $\langle F\rangle$ and we get the soft term in the visible
sector through gravitationnal interactions, that is:
\beq
m_{\rm soft} \simeq \frac{\langle F\rangle}{M_P}
\eeq
In this model we have the graviton and its superpartner the
gravitino. The goldstino fermion, relic of the spontaneous breakdown
of SUSY (in much the same way as the Goldstone boson in spontaneously
broken bosonic gauge symmetry), becomes part of the gravitino field
which then has got a mass. We then use renormalization group equations
(RGE) to evolve the couplings from the GUT scale down to the SUSY
scale where the phenomenological SUSY terms, in particular the soft
SUSY beaking terms, are defined. We will present in the next section
the minimal set of parameters at the GUT scale for a mSUGRA model in
the MSSM.

\vfill
\pagebreak

\section{The Minimal Supersymmetric Standard Model}

\label{section:MSSM}

In this part~\ref{part:three} we have already given some reasons why
it is interesting to study supersymmetry by itself and we have given
the SUSY formalism to construct supersymmetric theories. The content
of this section is to apply these theoretical fundations to the
building of a minimal extension of the SM, called the Minimal
Supersymmetric Standard Model or MSSM in short. This will be minimal
in the sense that we incorporate as minimal number of fields as
possible. We will see that it leads to a richer Higgs phenomenology
that will be discussed in the next part~\ref{part:four}.

The first subsection will introduce the MSSM, the second subsection
will explain in more details the Higgs sector; the last subsection
will be an opening to theories beyond the MSSM, explaining that this
latter model is not at all the end of the story for SUSY
phenomenology, by introducing the famous $\mu$--problem in the
MSSM.

\subsection{Fields content: Higgs and SUSY
  sectors of the MSSM \label{section:MSSMContent}}

The sections~\ref{section:FormalSuperspace}
and~\ref{section:FormalBreaking} have introduced all the elements that
are necessary for the minimal supersymmetric extension of the Standard
Model, called the MSSM. We will give in this subsection all the
superfields of the MSSM together with the soft SUSY breaking part of
the lagrangian.

\subsubsection{Superfields and (super)particles content}

The MSSM is still based on the Lie algebra $SU(2)_L\times U(1)_Y\times
SU(3)_c$. We will then introduce twelve vector superfields: the three
$SU(2)_L$ superfields $V^{a}_W$, the hypercharge superfield $V_Y$ and
the eight $SU(3)_c$ superfields $V^{a}_c$. 

All the chiral superfield are left handed in the MSSM. We will then
introduce
\beq
L_i = \left(\begin{matrix}L_{\nu_i}\\L_{e_i}\end{matrix}\right),~
\overline{E}_i
\eeq
where the right--handed singlet leptons are represented by the chiral
superfields $\overline{E}_i$ containing $e_R^{C}$ and
$\tilde{e}_R^{*}$\footnote{The bar notation does not mean any
  (hermitian) conjugation, it is rather a notation in order to remind
  the reader that we deal with the charge conjugate of the original
  spinor, as we deal with left chiral superfield only.}. In this way
all the chiral superfields are left--handed, which will be useful for
the writing of the superpotential. We have the following chiral
superfields for the quarks:
\beq
Q_i = \left(\begin{matrix}Q_{u_i}\\Q_{d_i}\end{matrix}\right),~
\overline{U}_i,~ \overline{D}_i
\eeq
where the same rule presented for the leptonic superfields also apply
for the quark superfields. The index $i$ runs from 1 to 3, labeling
the generations.

We finally introduce the Higgs superfield, but in the context of the
MSSM we have two Higgs superfields $H_1$ and $H_2$. We will review in
the next subsection (some of) the reasons to introduce more than one
Higgs superfield, but it can already be noted that for consistency
reason due to the analyticity of the superpotential, we should have
one Higgs superfield devoted to the coupling to up--type quarks and
another Higgs superfield devoted to the coupling to down--type quarks,
as we cannot take the charge conjugate of only one Higgs field as in
the case of the SM; otherwise we would break the analyticity of the
superpotential. We thus have two $\mathbf{2}$ $SU(2)_L$ superfields
doublets, $H_u$ with hypercharge +1 and $H_d$ with hypercharge -1:
\beq
H_u =\left(\begin{matrix}H_u^+\\H_u^0\end{matrix}\right),~
H_d=\left(\begin{matrix}H_d^0\\H_d^-\end{matrix}\right)
\eeq

We list in the Table~\ref{table:MSSMcontent} below the new
superparticles and Higgs fields in the MSSM in addition to the usual
SM fields. It should be noted that by construction the higgsinos and
gauginos are four--component Majorana spinors, thus their own charge
conjugate.

\begin{table}[!h]
{\small%
\let\lbr\{\def\{{\char'173}%
\let\rbr\}\def\}{\char'175}%
\renewcommand{\arraystretch}{1.35}
\vspace*{2mm}
\begin{center}
\begin{tabular}{|c|c|c|ccc|}\hline
Type & Name & Spin & $SU(3)_c$ rep. & $SU(2)_L$ rep. & 
 $U(1)_Y$ charge \\ \hline
 & $\tilde{\nu}_e, \tilde{e}_L, \tilde{\nu}_\mu, \tilde{\mu}_L,
 \tilde{\nu}_{\tau}, \tilde{\tau}_L$ & $0$ & $\mathbf{1}$ &
 $\mathbf{2}$ & $-1$ \\
SLEPTONS  & $\tilde{e}_R, \tilde{\mu}_R, \tilde{\tau}_R$ & $0$ &
$\mathbf{1}$ & $\mathbf{1}$ & $2$\\ \hline
 & $\tilde{u}_L, \tilde{d}_L, \tilde{s}_L, \tilde{c}_L,
 \tilde{t}_L, \tilde{b}_L$ & $0$ & $\mathbf{3}$ &
 $\mathbf{2}$ & $1/3$ \\
SQUARKS  & $\tilde{u}_R, \tilde{c}_R, \tilde{t}_R$ & $0$ &
$\mathbf{3}$ & $\mathbf{1}$ & $-4/3$\\
& $\tilde{d}_R, \tilde{s}_R, \tilde{b}_R$ & $0$ &
$\mathbf{3}$ & $\mathbf{1}$ & $2/3$\\ \hline
 & $\tilde{B}$ & $1/2$ & $\mathbf{1}$ &
 $\mathbf{1}$ & $1$ \\
GAUGINOS  & $\tilde{W}_1, \tilde{W}_2, \tilde{W}_3$ & $1/2$ &
$\mathbf{1}$ & $\mathbf{3}$ & $0$\\
& $\tilde{g}$ & $1/2$ &
$\mathbf{8}$ & $\mathbf{1}$ & $0$\\ \hline
 & $\tilde{h}_u^+, \tilde{h}_u^0$ & $1/2$ & $\mathbf{1}$ &
 $\mathbf{2}$ & $1$ \\
HIGGSINOS  & $\tilde{h}_d^0, \tilde{h}_d^-$ & $1/2$ &
$\mathbf{1}$ & $\mathbf{2}$ & $-1$\\ \hline
 & $h_u^+, h_u^0$ & $0$ & $\mathbf{1}$ &
 $\mathbf{2}$ & $1$ \\
HIGGS  & $h_d^0, h_d^-$ & $0$ &
$\mathbf{1}$ & $\mathbf{2}$ & $-1$\\ \hline
\end{tabular} 
\end{center} 
\caption[The superparticles and Higgs content of the MSSM before
EWSB]{The superparticles and Higgs content of the MSSM before the
  electroweak symmetry breaking.}
\label{table:MSSMcontent}
\vspace*{-2mm}
}
\end{table}

After electroweak symmetry (and SUSY) breaking the Higgs and the
gauginos/higgsinos sectors are affected by mixing between the
different gauge eigenstates. We finally obtain 2 $CP$--even neutral
Higgs bosons $h, H$, one $CP$--odd neutral Higgs boson $A$ and two
charged Higgs bosons $H^{\pm}$; four neutral gauginos $\chi_i^0$
called the neutralinos and four charged gauginos $\chi_{1,2}^{\pm}$
called the charginos. This is summarized in
Table~\ref{table:MSSMcontent2} below:

\begin{table}[!h]
{\small%
\let\lbr\{\def\{{\char'173}%
\let\rbr\}\def\}{\char'175}%
\renewcommand{\arraystretch}{1.35}
\vspace*{2mm}
\begin{center}
\begin{tabular}{|c|c|c|c|}\hline
Name & Spin & gauge eigenstates& mass eigenstates \\ \hline
Higgs & $0$ & $h_u^+, h_u^0, h_d^0, h_d^-$ & $h, H, A, H^{\pm}$ \\ \hline
Neutralinos & $1/2$ & $\tilde{W}_3, \tilde{B}, \tilde{h}_u^0,
\tilde{h}_d^0$ & $\tilde{\chi}_1^0, \tilde{\chi}_2^0,
\tilde{\chi}_3^0, \tilde{\chi}_4^0$ \\ \hline
Charginos & $1/2$ & $\tilde{W}_{1,2}, \tilde{h}_u^{+},
\tilde{h}_d^{-}$ & $\tilde{\chi}_1^{\pm}, \tilde{\chi}_2^{\pm}$ \\ \hline
\end{tabular} 
\end{center} 
\caption[The neutralinos, charginos and Higgs content of the MSSM
after EWSB]{The neutralinos, charginos and Higgs bosons in the MSSM
  after electroweak symmetry breaking.}
\label{table:MSSMcontent2}
\vspace*{-2mm}
}
\end{table}

The ratio between the two Higgs vacuum expectation values (vev) $v_u$
and $v_d$:
\beq
\tan\beta = \frac{v_u}{v_d}
\eeq
is a fundamental parameter in the MSSM controlling much of its
phenomenology as will be seen later on for example when dealing with
Higgs production in the MSSM. We will discuss in more details the
Higgs sector in the next subsection.

\subsubsection{The MSSM lagrangian}

We are now ready to write down the MSSM lagrangian using all the
superfields listed above, following the formal techniques described in
the previous section~\ref{section:FormalSUSY}. We will separate the
lagrangian in the SUSY invariant part and the soft SUSY breaking part:
\beq
{\cal L}_{\rm MSSM} = {\cal L}^{\rm MSSM}_{\rm SUSY} + {\cal L}^{\rm
  MSSM}_{\rm soft}
\eeq
with
\begin{eqnarray}
{\cal L}^{\rm MSSM}_{\rm SUSY} & = & {\cal L}_{\rm gauge}+{\cal L}_{\rm
  matter} + {\cal L}_{\rm Higgs}\nonumber \\
& & \nonumber\\
& {\cal L}_{\rm gauge}  = & \left.\left(W_c^a W_c^a + W_W^a W_W^a + W_Y
  W_Y\right)\right|_{\theta\theta} +~{\rm h.c.}\nonumber\\
&  {\cal L}_{\rm matter} = & \sum_{i=1}^3 \left( L_i^{\dagger}
   e^{\left(2 g T_a V_W^a + g_Y Y_{L_i} V_Y\right)} L_i +
   \overline{E}_i^{\dagger} e^{g_Y Y_{E_i} V_Y} \overline{E}_i ~+
 \right. \nonumber\\
 & & Q_i^{\dagger} e^{\left(2 g_s V^{a}_c R_a + 2 g T_a V_W^a + g_Y
     Y_{Q_i} V_Y\right)} Q_i + \overline{U}_i^{\dagger} e^{\left(-2
     g_s V^{a}_c \overline{R}_a + g_Y Y_{\overline{U}_i} V_Y\right)}
 \overline{U}_i~+ \nonumber\\
 & & \left.\left. \overline{D}_i^{\dagger} e^{\left(-2
     g_s V^{a}_c \overline{R}_a + g_Y Y_{\overline{U}_i} V_Y\right)}
 \overline{D}_i
\right)\right|_{\theta\theta\bar\theta\bar\theta}\nonumber\\
& {\cal L}_{\rm Higgs} = & \left( H_u^{\dagger} e^{\left(2g
        T_a V^a_W + g_Y Y_{H_u} V_Y\right)} H_u~+ \right.\nonumber\\
 & &  \left. \left. \left. H_d^{\dagger} e^{\left(2g T_a V^a_W + g_Y
           Y_{H_d}
           V_Y\right)} H_d \right)\right|_{\theta\theta\bar\theta\bar\theta}
   + {\cal W}_{\rm MSSM}\right|_{\theta\theta} +~{\rm h.c.}
\label{eq:MSSMSUSYlagrangian}
\end{eqnarray}
where $a$ is a gauge index, $T_a$ are the three $SU(2)_L$ generators
acting on the $\mathbf{2}$ representations, $R_a$ are the eight
$SU(3)_c$ generators acting on the $\mathbf{3}$ representations and
$\overline{R}_a$ their complex conjugates acting on the
$\mathbf{\overline{3}}$ representations. ${\cal W}_{\rm MSSM}$ is the
MSSM superpotential which is governed by gauge invariance and
$R$--parity. Indeed in the SM both $L$ and $B$ are conserved, thus the
natural assumption of the minimal supersymmetric extension of the SM
is to assume $R$--parity which is related to the $B-L$ invariance. The
superpotential contains both the Yukawa interactions and the SUSY
invariant scalar potential, and then reads~\cite{DreesMSSM}:
\beq
{\cal W}_{\rm MSSM} = - \mu H_u\cdot H_d - \lambda^{e}_{ij} (H_d\cdot
L_i) \overline{E}_j - \lambda^{d}_{ij} (H_d \cdot Q_i) \overline{D}_j
- \lambda^{u}_{ij} (Q_i\cdot H_u) \overline{U}_j
\label{eq:MSSMsuperpotential}
\eeq
where $Q\cdot R \equiv \epsilon_{AB}Q^A R^B$ for $A,B$ as $SU(2)_L$
indices and $\epsilon_{01}=- \epsilon_{10} = -1$. From this
superpotential we can extract the SUSY invariant scalar potential that
will be introduced in the next subsection when dealing specifically
with the Higgs sector of the MSSM.

The last piece that remains is the soft SUSY breaking part. We start
from the general lagrangian in Eq.~\ref{eq:generalsoftSUSY}, retaining
only the gauge and $R$--parity invariant terms, in particular we
discard the tadpole terms as the MSSM does not contain any gauge
singlet. It eventually reads
\beq
-{\cal L}_{\rm soft}^{\rm MSSM} & = &  (m_{\tilde q}^2)_{ij}
\tilde{q}_{i L}^* \tilde{q}_{j L} + (m_{\tilde l}^2)_{ij} \tilde{l}_{i
  L}^* \tilde{l}_{j L} + (m_{\tilde e}^2)_{ij} \tilde{e}_{i
  R}^*\tilde{e}_{j R} + (m_{\tilde u}^2)_{ij} \tilde{u}_{i R}^*
\tilde{u}_{j R} + \nonumber\\
 & & (m_{\tilde d}^2)_{ij} \tilde{d}_{i R}^* \tilde{d}_{j R} +
 m_{H_u}^2  h_u^{\dagger} h_u + m_{H_d}^2 h_d^{\dagger} h_d  - \left(
   B\mu h_u\cdot h_d +~{\rm h.c.}\right) +\nonumber \\
 & & \left( \lambda^e_{ij} A^e_{ij}  (h_d \cdot \tilde{l}_{i L}) 
   \tilde{e}_{j R}^* + \lambda^d_{ij} A^d_{ij}  (h_d \cdot
   \tilde{q}_{i L}) \tilde{d}_{j R}^* +\right. \nonumber\\
 & &  \left. \lambda^u_{ij} A^u_{ij} (\tilde{q}_{i L} \cdot h_u)
   \tilde{u}_{j R}^*  +~{\rm h.c.} \right) + \frac12 \left( M_1
   \overline{\tilde{B}} P_L \tilde{B} + M_2^* \overline{\tilde{B}} P_R
 \right) + \nonumber\\
 & & \frac12 \left( M_2 \overline{\tilde{W}}_a P_L \tilde{W}_a + M_2^*
   \overline{\tilde{W}}_a P_R \tilde{W}_a \right) +\nonumber\\
 & & \frac12 \left( M_3 \overline{\tilde{g}}_a P_L \tilde{g}_a + M_3^*
   \overline{\tilde{g}}_a P_R \tilde{g}_a \right)
\label{eq:softMSSMlagrangian}
\eeq
where the fields have been defined in Table.~\ref{table:MSSMcontent}
and with $P_L \equiv \left(1-\gamma_5\right)/2$,
$P_R\equiv \left(1+\gamma_5\right)/2$ being the
left and right chiral projectors. The repeted gauge index $a$ are
summed implicitely. $M_i$ are the complex Majorana gauginos SUSY
breaking mass parameters, $m_{1,2}$ are the real Higgs SUSY breaking
mass parameters. The sleptons and squarks mass parameters are
$3\times3$ hermitian matrices. Finally the trilinear couplings
$A^{e,u,d}_{ij}$ are $3\times 3$ complex matrices of mass dimension;
the $B$ parameter is real and is also of mass dimension.

If one counts all the new parameters, the general (or unconstrained)
MSSM has 134 new parameters. This huge parameter space needs to be
reduced to allow for viable phenomenological studies: that is the case
in the constrained versions of the MSSM (cMSSM) where some assumptions
such as no $CP$--violating mass matrices, mass unification at GUT
scale, universality of the two first generations of squarks and
sleptons, etc. As the origin of many new parameters lies in the soft
SUSY breaking part of the MSSM, many phenomenological studies assume a
definite SUSY breaking scheme at higher scale, such as the mSUGRA
scheme, that induces the soft SUSY breaking parameters through
renormalisation group evolution (RGE) of the parameters down to the
weak scale. In such constrained version of the MSSM, everything is
typically controlled by four parameters:
\beq
m_0,~ M_{1/2},~ A_0,~ \tan\beta,~ {\rm sign}(\mu)
\eeq
where $m_0$ is the common scalar mass, $M_{1/2}$ is the common gaugino
  mass, $A_0$ is the common trilinear coupling, $\tan\beta$ is the vev
  ratio and ${\rm sign}(\mu)$ is the sign of the invariant SUSY
  parameter $\mu$.

\subsubsection{Some experimental constraints on the sparticles masses}

We end this subsection by giving some experimental limits to the mass
of the yet--to--be--observed sparticles. These limits have been
obtained at the LEP, Tevatron\footnote{For the LEP and Tevatron
  searches, see the review in Ref.~\cite{Djouadi:2005gj}. We have
  taken the PDG limits~\cite{Nakamura:2010zzi} for these experiments.}
and very recently LHC colliders in the ATLAS~\cite{Aad:2011hh,
  Aad:2011ks} and CMS~\cite{Chatrchyan:2011wc, Chatrchyan:2011ah}
experiments. The limits are usually given within the cMSSM, that is
assuming $R$--parity, gauginos mass unification at the GUT scale,
first and second generations universality in the squark sector along
with the condition $m_{\tilde{q}_L}=m_{\tilde{q}_R}$. We have at 95\%CL:
\beq
m_{\tilde{\chi}_1^0} > 46~{\rm GeV}, & m_{\tilde{\chi}_2^0} > 62.4~{\rm
      GeV}, & \nonumber\\
m_{\tilde{\chi}_3^0} > 99.9~{\rm GeV}, &  m_{\tilde{\chi}_4^0}
        > 116~{\rm GeV}, &  m_{\tilde{\chi}_1^{\pm}} > 94~{\rm GeV}
\eeq
Apart from the lightest neutralino, these limits assume
$1\leq\tan\beta\leq 40$. We also have for the gluinos
\beq
m_{\tilde g} > 308~{\rm GeV}; & m_{\tilde g} \gsim 500-700~{\rm GeV~ if~
  assuming}~m_{\tilde q} = m_{\tilde g}\nonumber\\
& {\rm and~depending~on~the~mSUGRA~parameters}
\eeq

The squarks limits strongly depends on the parameter space
point. Nevertheless the gross limit either at LEP/Tevatron or at the
LHC points out to
\beq
m_{\tilde q} \gsim 400-500~{\rm GeV}
\eeq
which can be loosen for the stop and sbottom masses:
\beq
m_{\tilde b} > 89~{\rm GeV},~ m_{\tilde t} > 95~{\rm GeV}
\eeq

The limits on the sleptons use the same assumptions as for the
charginos and neutralinos and read:
\beq
m_{\tilde{e}_R} > 107~{\rm GeV},~ m_{\tilde{\mu}_R} > 94~{\rm GeV},~
m_{\tilde{\tau}_R} > 81.9~{\rm GeV}
\eeq

\subsection{The Higgs sector and the number of Higgs
  doublets \label{section:MSSMAnomaly}}

After having presented the full MSSM we move on to a more detailed
analysis of its Higgs sector that will be discussed at hadron
colliders  in the next part~\ref{part:four}. We first remind the
reason to introduce two Higgs doublet instead of the minimal choice in
the SM; we then present the Higgs scalar potential in the MSSM and the
electroweak symmetry breaking, intimately connected to that of SUSY;
we end by presenting the masses and couplings of the MSSM Higgs
bosons.

\subsubsection{The number of Higgs doublet in the MSSM}

We have already presented one reason to introduce two Higgs doublets
within the MSSM. Indeed the mathematical formulation of the MSSM
requires that the superpotential, encoding all the SUSY invariant
interactions between the chiral superfields that represent the
fermions and their sfermions superpartners, to be a holomorphic
expression of these superfields. In particular this means that there
will be no $\Phi^{\dagger}$ term that was used in the case of the SM
to give rise to up--type quark using the $\tilde{\phi}=\imath \tau_2
\phi^*$ field. In other words, $\mathbf{2}$ and $\mathbf{\overline{2}}$
$SU(2)_L$ representations are not equivalent in SUSY theories contrary
to the case of the SM.

Even if we were not to look for a way to give a mass to fermions we
would require two Higgs doublets. Indeed, we know that the SM is free
frome gauge anomalies and in particular the Adler--Bardeen--Jackiw
anomalies originating from triangular fermionic loops with
axial--vector current couplings~\cite{Adler:1969er}. Gauge anomalies
would be disastrous, spoiling the renormalizability and the gauge
identities, but the fact that the sum of the hypercharges of all the
fifteen chiral fermions (or charges, related to hypercharges through
the Gell--Mann--Nishijima relation, see
section~\ref{section:IntroSymmetries}) is zero in the SM explains the
vanishing of these chiral anomalies, ${\rm Tr}(Y_f) = {\rm
  Tr}(Q_f)=0$. In the MSSM, SUSY introduces new fermions in
the theory as superpartners of the standard bosons, in particular in
the higgsino sector. If we were to have only one Higgs doublet with a
definite hypercharge, we would be left with the charged higgsino not
cancelled by anything in the trace sum rule and thus the anomaly
cancellation would be spoiled. With two doublets of Higgs fields with
opposite hypercharge, the cancellation of chiral anomalies still takes
place~\cite{AlvarezGaume:1983ig}.

We now have two solid reasons to introduce two Higgs doublet within
the MSSM, making the minimal supersymmetric extension of the SM a
particular case of 2 Higgs doublet models (2HDM), more precisely a
2HDM of type II, see Ref.~\cite{Gunion:1990} for a review. We will now
sketch the electroweak symmetry breaking within the MSSM.

\subsubsection{The electroweak symmetry breaking in the MSSM}

We recall that we use the following Higgs doublets:
\beq
h_u = \left(\begin{matrix} h_u^+\\h_u^0\end{matrix}\right),~ h_d =
\left(\begin{matrix} h_d^0\\h_d^-\end{matrix}\right)
\eeq

We first need to write down the Higgs scalar potential in the MSSM. We
break the potential into a SUSY invariant part and a soft part
\beq
V = V_{\rm SUSY} + V_{\rm soft}
\eeq
where $V_{\rm SUSY}$ comes from the
superpotential~\ref{eq:MSSMsuperpotential} and the gauge $D$--terms
while $V_{\rm soft}$ comes from the soft SUSY breaking lagrangian in
Eq.~\ref{eq:softMSSMlagrangian}. We then have~\cite{Djouadi:2005gj,
  Martin:1997ns}:
\beq
V_{\rm SUSY} & = & |\mu|^2 \left( |h_u^0|^2 + |h_d^0|^2 +
  |h_u^+|^2+|h_d^-|^2\right) + \nonumber\\
 & & \frac{(g^2+g_Y^2)}{8}\left( |h_d^0|^2+|h_d^-|^2 - |h_u^0|^2 -
   |h_u^+|^2\right)^2 +\nonumber\\
 & &  \frac{g^2}{2} \left| {h_d^0}^* h_u^+ + {h_d^-}^* h_u^0 \right|^2
 \nonumber\\
V_{\rm soft} & = & m_{H_u}^2 \left(|h_u^0|^2 + |h_u^+|^2\right) +
m_{H_d}^2\left( |h_d^0|^2 + |h_d^-|^2\right) +\nonumber\\
 & &  B \mu \left( h_u^+ h_d^- - h_u^0  h_d^0 +~{\rm c.c.} \right)
\label{eq:MSSMscalarpotential1}
\eeq

We want that the minimum of the scalar potential $V$ breaks
$SU(2)_L\times U(1)_Y$ down to $U(1)_{\rm EM}$. Without any loss of
generality we can assume that $\langle \, h_u^+ \, \rangle =0$ at the
minimum using the freedom of a gauge $SU(2)_L$ rotation. Together with
the minimization equation $\partial V / \partial h_u^+ =0$ we obtain
$\langle \, h_d^- \, \rangle=0$, a result that could have been predicted as
the vacuum should not break $U(1)_{\rm EM}$ thus not being charged. We
are left with:
\beq
\langle \, h_u \, \rangle = \frac{1}{\sqrt 2}
\left(\begin{matrix}0\\v_u\end{matrix}\right),~ \langle \, h_d \,
\rangle = \frac{1}{\sqrt 2} \left(\begin{matrix}
    v_d\\0\end{matrix}\right)
\eeq
with $v_u$ and $v_d$ real positive numbers. A $U(1)$ rotation of the
fields $h_u^0$ and $h_d^0$ can make $B\mu$ as a real positive number
too, thus the potential does conserve $CP$ at tree--level. 

The pattern of the vacuum expectation value is more complicated than
compared to the SM case as we now have two vacuum expectation values
(vev) $v_u$ and $v_d$. The coupling of the Higgs doublets to the gauge
bosons remain as in the SM case, thus we can define the weak bosons
masses to be the same as in the SM case, that is:
\beq
v_u^2+v_d^2 & = & v_{\rm SM}^2 \simeq 246~{\rm GeV},\nonumber\\
M_W  & = & \frac12 g\, v_{\rm SM},~ M_Z = \frac12 \sqrt{g^2+g_Y^2} \, v_{\rm SM}
\label{eq:gaugeMSSMmasses}
\eeq

SUSY being broken the scalar potential is not always positive. $V$
must be bounded from below to develop a vev; along the so--called
$D$--flat direction $|h_u^0| = |h_d^0|$ the quartic term vanishes and
the stability requirement thus imposes the following
condition~\cite{DreesMSSM, Djouadi:2005gj, Martin:1997ns}:
\beq
m_{H_u}^2 + m_{H_d}^2 + 2|\mu|^2 > 2 B\mu
\label{eq:EWSBcondition1}
\eeq
This relation must hold at any perturbative order, that is even in the
case where the coefficients become running quantities. In addition we
need to be sure that a linear combination of the two neutral Higgs
fields has a negative mass squared term in order to have spontaneous
electroweak symmetry breaking. The mass matrix is $\partial
V/ \partial h_u^0 \partial h_d^0$ and the quadratic part of $V$ reads
\beq
\left( \begin{matrix} {h_u^0}^* & h_d^0\end{matrix}\right)
\left( \begin{matrix} m_{H_u}^2 + |\mu|^2 & - B\mu\\ - B\mu & m_{H_d}^2
      + |\mu|^2\end{matrix}\right) \left( \begin{matrix}
      h_u^0\\{h_d^0}^*\end{matrix}\right)
\eeq
and its trace being positive because of Eq.~\ref{eq:EWSBcondition1}
its determinant must be negative:
\beq
(B\mu)^2 > (m_{H_u}^2 + |\mu|^2)(m_{H_d}^2 + |\mu|^2)
\label{eq:EWSBcondition2}
\eeq
If not the stable minimum is along the $D$--flat direction $\langle\,
h_u^0\, \rangle = \langle \, h_d^0 \, \rangle = 0$. The very
interesting remark is that the two conditions~\ref{eq:EWSBcondition1}
and~\ref{eq:EWSBcondition2} are not fulfilled if $m_{H_u}^2 =
m_{H_d}^2$ and in particular if these two soft SUSY breaking terms
vanish. This means that the electroweak symmetry breaking in the MSSM
requires that SUSY be broken: there is a deep connection between SUSY
and electroweak symmetry breakings. Furthermore the electroweak
symmetry breaking can be driven radiatively: starting from a very high
energy input scale where $m_{H_u}^2 = m_{H_d}^2$ (even equal to zero
before SUSY breaking in the hidden sector), RGE evolution with
the top/stop and bottom/sbottom contributions does lift the
degeneracy, ending with $m_{H_u}^2 < 0$ or $m_{H_u}^2 \ll
m_{H_d}^2$. This is known as the radiative electroweak symmetry
breaking~\cite{Ibanez:1982fr, AlvarezGaume:1983gj, Ibanez:1984vq}
which is a very elegant way to obtain a gauge symmetry breaking, not
having imposed at first the negative value of the squared mass in the
scalar potential.

\subsubsection{The masses and couplings of the MSSM Higgs bosons}

\paragraph{Charged Higgs, $CP$--even and $CP$--odd neutral Higgs
  masses\newline}

We have seen in the previous subsection the Higgs scalar potential of
the MSSM which controls the electroweak symmetry breaking. We will now
diagonalize the equation~\ref{eq:MSSMscalarpotential1} to obtain the
physical Higgs bosons states.

We will expand the Higgs fields around the minimum $(v_u,v_d)$:
\beq
h_u = \frac{1}{\sqrt 2} \left(\begin{matrix} \sqrt{2}
    h_u^+\\(h_u^0+\imath G_u^0 + v_u)\end{matrix}\right),~ h_d =
\frac{1}{\sqrt 2} \left(\begin{matrix} (h_d^0+\imath G_d^0 + v_d)\\
    \sqrt{2} h_d^-\end{matrix}\right)
\label{eq:MSSMHiggsexpansion}
\eeq
where this time all the fields but those of the charged are reals (we
use the same notation for the full fields and the expanded fields, but
there should be no confusion from now on).

The first step is to expand the scalar potential at the minimum only,
which gives
\beq
V_{\rm min} = \frac12 m_1^2 v_d^2 + \frac12 m_2^2 v_u^2 - B\mu v_u v_d
+ \frac{g^2+g_Y^2}{32}\left(v_d^2-v_u^2\right)^2
\eeq
where we have $m_1^2 \equiv m_{H_d}^2+|\mu|^2$ and $m_2^2 \equiv
m_{H_u}^2+|\mu|^2$. Using $\partial V_{\rm min}/\partial v_u
= \partial V_{\rm min}/\partial v_d =0$ this gives the two consistency
conditions below:
\beq
m_1^2 & = & B\mu \tan\beta -\frac12 M_Z^2 \cos(2\beta)\nonumber\\
m_2^2 & = & B\mu \cotan\beta + \frac12 M_Z^2\cos(2\beta)
\label{eq:MSSMconsistencyHiggs}
\eeq
when we have used $v_u=v\cos\beta$ and $v_d=v\sin\beta$ together with
the expression of $M_Z$ in term of SM vacuum expectation value
$v$. The consistency conditions of Eq.~\ref{eq:MSSMconsistencyHiggs}
implies a reduction of the number of free parameters in the Higgs
sector.\bigskip

We rewrite the scalar potential of Eq.~\ref{eq:MSSMscalarpotential1}
in terms of the real fields:
\beq
V & = & \frac12 m_1^2 \left( (v_d+h_d^0)^2 + {G_d^0}^2 + 2
  |h_d^-|^2\right) + \frac12 m_2^2 \left( (v_u+h_u^0)^2 + {G_u^0}^2 +
  2 |h_u^+|^2\right) - \nonumber\\
 & & \frac12 B\mu \left( (v_u + h_u^0 + \imath G_u^0)(v_d + h_d^0
   +\imath G_d^0) +~{\rm c.c.}~- 2 h_u^+ h_d^- +~{\rm c.c.} \right)
 +\nonumber\\
 & & \frac14 g^2 \left| (v_d + h_d^0 + \imath G_d^0)h_u^+ + (v_u +
 h_u^0+\imath G_u^0) h_d^-\right|^2 + \frac{1}{32} \times \nonumber\\
 & & \left(g^2+g_Y^2\right) \left( (v_d+h_d^0)^2 +
   {G_d^0}^2 + 2 |h_d^+|^2 - (v_u+h_u^0)^2 - {G_u^0}^2 - 2
   |h_u^-|^2\right)^2
\label{eq:MSSMscalarpotential2}
\eeq

We will now read each quadratic part of the potential depending on our
interest in the $CP$--even neutral Higgs bosons $(h_u^0, h_d^0)$, the
$CP$--odd neutral Higgs bosons $(G_u^0,G_d^0)$ or the charged Higgs
bosons $(h_u^+,h_d^-)$.

\begin{enumerate}[$\bullet$]
\item{{\it $CP$--odd Higgs boson $A$:} 

We read the quadratic part in Eq.~\ref{eq:MSSMscalarpotential2}:
\beq
V_{\rm quadratic}^{A} & = & \frac12 m_1^2 {G_d^0}^2 + \frac12 m_2^2
{G_u^0}^2 + \frac12 B\mu (G_d^0 G_u^0 + G_u^0 G_d^0) +\nonumber\\
 & & \frac{1}{32} \left(g^2+g_Y^2\right) \left( 2 v_u^2 {G_u^0}^2 + 2
   v_d^2 {G_d^2}^2 - 2 v_u^2 {G_d^0}^2 - 2 v_d^2 {G_u^0}^2
 \right)\nonumber\\
 & = & \frac12 \left(G_u^0, G_d^0\right) \left( \begin{matrix} m_2^2 +
     \frac18 \left(g^2+g_Y^2\right) (v_u^2-v_d^2) & B\mu\\ B\mu &
     m_1^2 - \frac18 \left( g^2+g_Y^2\right)
     (v_u^2-v_d^2)\end{matrix}\right)
 \left( \begin{matrix}G_u^0\\G_d^0\end{matrix}\right)\nonumber\\
 & = & \frac12 \left(G_u^0, G_d^0\right) \left( \begin{matrix} B\mu
     \cotan\beta & B\mu\\ B\mu & B\mu \tan\beta \end{matrix}\right)
 \left( \begin{matrix}G_u^0\\G_d^0\end{matrix}\right)
\eeq
where we have used in the last line the two consistency conditions in
Eq.~\ref{eq:MSSMconsistencyHiggs}. The vanishing determinant together
with the non--vanishing trace implies a neutral Goldstone boson (which
combines with the $Z$ boson to give its mass) and a massive $CP$--odd
Higgs boson $A$. We then have
\beq
\left(\begin{matrix}G^0\\A\end{matrix}\right) =
\left(\begin{matrix}\cos\beta & -\sin\beta\\ \sin\beta &
    \cos\beta\end{matrix}\right)
\left(\begin{matrix}G_d^0\\G_u^0\end{matrix}\right),~~M_A^2 =
\frac{2B\mu}{\sin(2\beta)}, M_{G^0}^2 = 0
\label{eq:HiggsCPoddmass}
\eeq
In the unitarity gauge the Goldstone boson disappears as in the SM case.}
\item{{\it $CP$--even Higgs bosons $(h,H)$:} 

We do the same exercice from Eq.~\ref{eq:MSSMscalarpotential2} with
the real components of the neutral Higgs fields:
\beq
V_{\rm quadratic}^{H} & = & \frac12 m_1^2 {h_d^0}^2 + \frac12 m_2^2
{h_u^0}^2 - \frac12 B\mu (h_u^0 h_d^0 + h_d^0 h_u^0) +\nonumber\\
 & & \frac{1}{32} (g^2+g_Y^2) \left( 6 v_u^2 {h_u^0}^2 + 6 v_d^2
   {h_d^0}^2 - 2 v_u^2 {h_d^0}^2 - 2 v_d^2 {h_u^0}^2 - 8 v_u v_d h_u^0
   h_d^0\right)\nonumber\\
 & = & \frac12 \left(\begin{matrix}h_u^0 & h_d^0\end{matrix}\right)
 {\cal M}_H \left(\begin{matrix}
      h_u^0\\h_d^0\end{matrix}\right)
\eeq
with ${\cal M}_H$ being the following mass matrix
\beq
{\cal M}_H & = & \left(\begin{matrix} m_2^2 + \frac18
    \left(g^2+g_Y^2\right) \left(3v_u^2-v_d^2\right) & -\frac14
    \left(g^2+g_Y^2\right) v_u v_d - B\mu\\ -\frac14
    \left(g^2+g_Y^2\right) v_u v_d - B\mu &  m_1^2 +\frac18
    \left(g^2+g_Y^2\right) \left(3v_d^2 -
      v_u^2\right)\end{matrix}\right)\nonumber\\
 & = &  \left(\begin{matrix} M_A^2 \cos^2\beta + M_Z^2\sin^2\beta & 
     -(M_A^2+M_Z^2)\sin\beta\cos\beta \\
     -(M_A^2+M_Z^2)\sin\beta\cos\beta & M_A^2\sin^2\beta + 
     M_Z^2\cos^2\beta\end{matrix}\right)
\eeq
where in the last line we have used both the consistency conditions of
Eq.~\ref{eq:MSSMconsistencyHiggs} and the result of
Eq.~\ref{eq:HiggsCPoddmass} on the $CP$--odd Higgs mass $M_A^2$. The
determinant of the mass matrix does not vanish, meaning that the
spectrum contains two physical $CP$--even Higgs bosons conventially
named $h$ for the lightest and $H$ for the heaviest. The rotation
matrix is more complicated than in the $CP$--odd case. Together with
the physical masses we have
\beq
\left(\begin{matrix} h \\ H\end{matrix}\right) & = &
\left(\begin{matrix} \cos\alpha & -\sin\alpha \\ \sin\alpha &
    \cos\alpha \end{matrix}\right) \left(\begin{matrix}h_u^0 \\
    h_d^0 \end{matrix}\right)\nonumber\\
 M_{h,H}^2 & = & \frac12 \left( M_A^2 + M_Z^2 \mp
   \sqrt{(M_A^2+M_Z^2)^2-4M_A^2M_Z^2\cos(2\beta)^2}\right)
\label{eq:HiggsCPevenmass}
\eeq
with the mixing angle $\alpha$ given by
\beq
\cos(2\alpha) & = & -\cos(2\beta)
\frac{M_A^2-M_Z^2}{M_H^2-M_h^2}\nonumber\\
 \sin(2\alpha) & = & - \sin(2\beta) \frac{M_H^2+M_h^2}{M_H^2-M_h^2}
\eeq}
\item{{\it Charged Higgs boson $H^{\pm}$:} 

We end with the charged Higgs boson mass matrix. From
Eq.~\ref{eq:MSSMscalarpotential2} we read:
\beq
V_{\rm quadratic}^{H^{\pm}}  & = & m_1^2 |h_d^-|^2 + m_2^2 |h_u^+|^2 +
B\mu \left( h_u^+ h_d^- + h_u^- h_d^+ \right) + \nonumber\\
  & & \frac14 g^2 \left(v_d^2 |h_u^+|^2 + v_u^2 |h_d^-|^2 + v_u v_d
    \left(h_u^+ h_d^- + h_u^- h_d^+\right) \right) + \nonumber\\
 & &  \frac{1}{16} \left( 2 v_u^2 |h_u^+|^2 + 2 v_d^2 |h_d^-|^2 - 2
   v_d^2 |h_u^+|^2 - 2 v_u^2 |h_d^-|^2\right)\nonumber\\
 & = & \left(\begin{matrix} h_u^+ & h_d^+\end{matrix}\right) {\cal
   M}_{H^{\pm}} \left(\begin{matrix} h_u^- \\ h_d^-\end{matrix}\right)
\eeq
with ${\cal M}_{H^{\pm}}$ being the following charged Higgs matrix
\beq
{\cal M}_{H^{\pm}} & = & \left(\begin{matrix} m_2^2 + \frac14 g^2
    v_d^2 + \frac18 \left(g^2+g_Y^2\right) \left(v_u^2-v_d^2\right) &
    B\mu + \frac14 g^2 v_u v_d \\ B\mu + \frac14 g^2 v_u v_d & m_1^2 + 
      \frac14 g^2 v_u^2 - \frac18 \left( g^2+g_Y^2\right)
      \left(v_u^2-v_d^2\right) \end{matrix}\right)\nonumber\\
 & = & \left(\begin{matrix} \left(M_A^2+M_W^2\right)\cos^2\beta &
     \left(M_A^2+M_W^2\right)\sin\beta\cos\beta \\
     \left(M_A^2+M_W^2\right)\sin\beta\cos\beta &
     \left(M_A^2+M_W^2\right)\sin^2\beta\end{matrix}\right)
\eeq
Again the vanishing determinant and the non--vanishing trace signal a
pair of Goldstone bosons $G^{\pm}$ (which give the mass to the charged
$W^{\pm}$ bosons) and two physical charged Higgs bosons $H^{\pm}$. The
rotation matrix is the same as in the case of the $CP$--odd Higgs
boson and we have
\beq
\left(\begin{matrix}G^{\pm}\\ H^{\pm}\end{matrix}\right) =
\left(\begin{matrix}\cos\beta & -\sin\beta\\ \sin\beta &
    \cos\beta\end{matrix}\right) \left(\begin{matrix} h_d^{\pm} \\
    h_u^{\pm}\end{matrix}\right),~ M_{H^{\pm}}^2 = M_A^2 + M_W^2,
M_{G^{\pm}}^2 = 0
\label{eq:HiggsChargedmass}
\eeq}
\end{enumerate}

At tree--level we then have two parameters only which control the
Higgs sector: the vevs ratio $\tan\beta$ and the $CP$--odd Higgs boson
mass $M_A$. All the other masses and mixing angles (and also
couplings, see below) follow from these two inputs. We also have a
strong hierarchy in the spectrum and in particular $M_h \leq
M_Z$. However it is well  known that the radiative corrections
(hopefully!) enhance the MSSM Higgs bosons masses (see
Ref.~\cite{Djouadi:2005gj} for a review): indeed we have stop/top
contribution which increases the lightest Higgs boson and according to
Ref.~\cite{Martin:1997ns} we have at the one--loop order
\beq
\Delta (M_h^2) = \frac{3}{4\pi^2} \cos^2\! \alpha\, \lambda_t^2
m_t^2\ln\left(\frac{m_{\tilde{t}_1} m_{\tilde{t}_2}}{m_t^2}\right)
\eeq
which helps having the lightest Higgs boson $h$ above the LEP bound
$M_{H_{\rm SM}}>114.4$ GeV as often the Higgs boson $h$ is SM--like,
thus bounded by the experimental SM limits. Including all known
corrections up to three--loops order  (see Refs. 185 to 194 in
Ref.~\cite{Martin:1997ns} for some of them) we
have 
\beq
M_h^2 \lsim 135~{\rm GeV}
\eeq
with $m_t = 173.1$ GeV (see Ref.~\cite{Djouadi:2005gj} page
68). Direct searches at LEP have also put the followings lower
bounds~\cite{Muhlleitner:2011hh}:
\beq
M_{h,H} \gsim 92.6~{\rm GeV},~ M_A \gsim 93.4~{\rm GeV},~M_{H^{\pm}}
\gsim 78.6~{\rm GeV}
\eeq

\paragraph{Higgs bosons couplings to fermions\newline}

Due to the huge number of new states in the MSSM, we will not give the
full list\footnote{The full list is available in standard textbooks
  and reviews, see e.g. Refs.~\cite{DreesMSSM, Djouadi:2005gj}.}. As
we are interested in the Higgs production at hadron colliders in the
main channels, we will simply give the couplings that are of interest
to our goal: the couplings of the Higgs bosons to fermions.

We recall that the Yukawa interactions come directly from the
superpotential ${\cal W}$ through (in four--components notation):
\beq
{\cal L}_{\rm Yukawa} = -\frac12 \sum_{ij} \left( \overline{\psi}_{i L}
\left. \frac{\partial {\cal W}}{\partial \Phi_i \partial
    \Phi_j}\right|_{\theta=\bar\theta=0} \psi_{j L} +~{\rm h.c.}
\right)
\eeq

Using the MSSM superpotential written in
Eq.~\ref{eq:MSSMsuperpotential} and introducing the chiral projector
operators $P_{L,R}$ we have for the first
generation~\cite{Djouadi:2005gj}, the scheme being repeted for the 
other two:
\beq
{\cal L}_{\rm Yukawa}&=& - \lambda_u \left( \bar u P_L u\, h_u^0  - \bar
  u P_L d\, h_u^+ \right) - \lambda_d \left( \bar d P_L d\, h_d^0  - \bar
  d P_L u\, h_d^- \right)
+ {\rm h.c.}
\label{eq:MSSMYukawa-lagrangian}
\eeq
In the lagrangian~\ref{eq:MSSMYukawa-lagrangian} above we have used
the full Higgs fields without any expansion around the minimum. The
fermion masses are generated when the neutral components of the Higgs
fields acquire their vacuum expectation values; they read in terms of
the Yukawa couplings, using the expansion of
Eq.~\ref{eq:MSSMHiggsexpansion} with $v_u = v\sin\beta$ and
$v_d=v\tan\beta$, $v$ being the SM vacuum expectation value: 
\beq
m_u = \lambda_u \sin\beta \frac{v}{\sqrt 2},~ m_d = \lambda_d
\cos\beta \frac{v}{\sqrt 2}
\eeq
\label{eq:YukawaMSSM}

We now expand the Higgs fields around their vevs using the physical
fields expressed with
Eqs.~\ref{eq:HiggsCPoddmass},~\ref{eq:HiggsCPevenmass},~\ref{eq:HiggsChargedmass}:
\beq
{\cal L}_{\rm Yukawa} & = & -\frac{m_u}{v \sin\beta}
\Big(\bar{u}u (H\sin\alpha+ h\cos\alpha) - \imath \bar{u} \gamma_5
  u\,  A \cos\beta \Big) \nonumber \\
 & & -\frac{m_d}{v \cos\beta} \Big(\bar{d} d (H\cos\alpha -
   h\sin\alpha) - \imath \bar{d} \gamma_5 d\,  A \sin\beta \Big)
 \nonumber \\ 
& & +\frac{1}{v \sqrt{2}} V_{ud} \Big[ H^+ \bar{u} \Big(m_d \tan\beta
  (1+\gamma_5) + m_u\cotan\beta (1-\gamma_5)\Big) d \Big]\nonumber\\
& & +\frac{1}{v \sqrt{2}} V_{ud}^* \Big[ H^- \bar{d} \Big(m_d \tan\beta
  (1-\gamma_5) + m_u\cotan\beta (1+\gamma_5)\Big) u \Big]
\label{eq:MSSMYukawa-lagrangian2}
\eeq
with $V_{ud}$ the CKM matrix element which is present in the case of
the quarks. The lagrangian of Eq.~\ref{eq:MSSMYukawa-lagrangian2} is
repeted for the two other generations and has the same structure. The
MSSM Higgs boson couplings to fermions are thus given by
\beq
G_{huu} = \imath \frac{m_u}{v} \frac{\cos\alpha}{\sin\beta} \ , &&
G_{Huu} = \imath \frac{m_u}{v} \frac{\sin\alpha}{\sin\beta} \ , \ \ \ 
G_{Auu} = \frac{m_u}{v} \cot\beta \, \gamma_5 \nonumber \\ 
G_{hdd} = -\imath \frac{m_d}{v} \frac{\sin\alpha}{\cos\beta} \ , & &
G_{Hdd} =  \imath \frac{m_d}{v} \frac{\cos\alpha}{\cos\beta} \ , \ \ \ 
G_{Add} = \frac{m_d}{v} \tan\beta \, \gamma_5 \nonumber 
\eeq
\beq
G_{H^+ \bar u d} & = &  -  \frac{\imath}{\sqrt{2} v}  V_{ud} 
\Big(m_d \tan\beta (1+\gamma_5) + m_u\cotan\beta (1-\gamma_5)\Big)
\nonumber \\
G_{H^- u \bar d} & = &  -  \frac{\imath}{\sqrt{2} v}  V_{ud}^* 
\Big(m_d \tan\beta (1-\gamma_5) + m_u\cotan\beta (1+\gamma_5)\Big)
\label{eq:MSSMGHff}
\eeq

What can be noted from these couplings? First we have to say that
directly for the charged Higgs bosons $H^{\pm}$ and the pseudoscalar
$A$ boson their couplings to up--type fermions ($u$, $c$, $t$ quarks)
are suppressed as $\tan\beta$ grows whereas their couplings to
down--type fermions ($d$, $s$, $b$ quarks and charged leptons) are
greatly enhanced. This has a great inpact on MSSM Higgs production at
colliders as will be seen in the next part~\ref{part:four}, as in the
heavy quarks sector we have a very strong coupling between the
$A$ boson and the bottom quark whereas the $A t$ coupling is strongly
suppressed. This side remark is also true for the $CP$--even Higgs
bosons $h,H$, as if we rescale their couplings to those of the SM we
have:
\beq
g_{hdd} & = & -\frac{\sin\alpha}{\cos\beta} =  \sin(\beta-\alpha) -
\tan\beta \cos(\beta-\alpha) \nonumber \\
g_{huu} & = & \ \frac{\cos\alpha}{\sin\beta} =  \sin(\beta-\alpha) +
\cotan\beta \cos(\beta-\alpha) \nonumber \\
g_{Hdd} & = & \ \frac{\cos\alpha}{\cos\beta} =  \cos(\beta-\alpha) +
\tan\beta \sin(\beta-\alpha) \nonumber \\
g_{Huu} & = & \ \frac{\sin\alpha}{\sin\beta} =  \cos(\beta-\alpha) -
\cotan\beta \sin(\beta-\alpha) 
\label{eq:MSSMgHff}
\eeq
which means that (though depending of the magnitude of
$\cos(\beta-\alpha)$ or $\sin(\beta-\alpha)$) the $dd \, (uu)$
coupling of either the $h$ or $H$ boson are enhanced (suppressed) by a
factor $\tan\beta$.

\subsection{The MSSM is not the end of the
  story \label{section:nMSSM}}

This last subsection will close the presentation of the MSSM and is
intended as a door opened to theories beyond the minimal
supersymmetric extension of the SM. Indeed it is well known that the
MSSM suffers from difficulties even if it solves many problems within
the SM, some having been highlighted in
section~\ref{section:SUSYIntro}. It then should be fair to present the 
limitations of this thesis and to put in perspectives the beauty of the
MSSM.

One of the celebrated issues is the so--called ``$\mu$ problem'' which
can be traced as a naturalness issue. Indeed, let us start with the
two equalities in Eq.~\ref{eq:MSSMconsistencyHiggs} which have to be
fullfilled to obtain a consistent electroweak vacuum. These can be
rewritten as
\beq
B\mu & = & \frac{\left(m_{H_u}^2-m_{H_d}^2\right) \tan(2\beta) -
  M_Z^2\sin(2\beta)}{2}\nonumber\\
|\mu|^2 & = & \frac{\sin^2\beta\, m_{H_u}^2-\cos^2\beta\,
  m_{H_d}^2}{\cos(2\beta)} - \frac12 M_Z^2
\label{eq:muproblem}
\eeq
The two equalities in Eq.~\ref{eq:muproblem} above highlight the
$\mu$--problem: $m_{H_{u,d}}^2$ being soft SUSY breaking terms they
are at most two orders of magnitude of $M_Z^2$, which means that on
the overall $\mu$ has to be of the order of the soft SUSY breaking
terms and cannot be null in order to have a correct electroweak
symmetry breaking. This is unnatural as $\mu$ is a SUSY invariant term
which then is naturally of the order of the highest scale of the
theory, that are either the GUT scale where SUSY is broken or even
the Planck scale. What is the mechanism which forces a SUSY invariant
parameter to be of the same order of the soft terms\footnote{One
  might argue that this is a rather subjective argument to go beyond
  the MSSM, Nature being perhaps ``unnatural''. However one of the
  arguments to introduce supersymmetry {\it is} precisely the
  naturalness argument applied on the Higgs boson mass. It would be
  rather strange to reject the naturalness argument on the one hand
  and to accept it on the other hand.}?

One (elegant) solution is to introduce a new singlet superfield $S$ in
the theory, which then extends to the NMSSM for
Next--to--MSSM~\cite{Fayet:1974pd, Derendinger:1983bz} which was
actually born even before the MSSM. Two new couplings $\lambda$ and
$\kappa$ are introduced in the lagrangian and the superpotential reads
\beq
{\cal W}_{\rm NMSSM} & = & -\lambda\, S H_u\cdot H_d - \frac13 \kappa S^3
- \lambda^{e}_{ij} (H_d\cdot L_i) \overline{E}_j \nonumber\\
 && - \lambda^{d}_{ij} (H_d \cdot Q_i) \overline{D}_j - \lambda^{u}_{ij}
(Q_i\cdot H_u) \overline{U}_j
\eeq
the trilinear $S^3$ term is necessary to prevent the NMSSM lagrangian
to possess a $\mathbf{Z}_3$ symmetry which would lead to domain walls
in cosmology. An effective $\mu$ term is generated when $S$ acquires a
vev:
\beq
\mu_{\rm eff} = \lambda \langle\, S\, \rangle
\eeq
which is then naturally of the order of the soft SUSY breaking terms, thus
solving the $\mu$--problem. The soft SUSY breaking terms read
\beq
-{\cal L}_{\rm soft} = -{\cal L}_{\rm MSSM~soft} + m_{S}^2 s^2
-\frac13 \kappa A_\kappa s^3 - \lambda A_\lambda s h_1\cdot h_2
\eeq

The phenomenology is richer than in the MSSM. Indeed the new singlet
superfield adds two Higgs fields and two Higgsino fields, which mix
with other fields to end up with~\cite{Ellis:1988er,
  Ellwanger:1996gw}:
\begin{enumerate}[$\bullet$]
\item{3 $CP$--even Higgs bosons $h_{1,2,3}$;}
\item{2 $CP$--odd Higgs bosons $a_1$, $a_2$;}
\item{5 neutralinos $\tilde{\chi}^{0}_{1..5}$.}
\end{enumerate}
The lightest neutralino can be singlino--like, affecting the dark
matter searches. The next--to--lightest supersymmetric particle (NLSP)
can be charged, ending up with long--lived $\tilde{\tau}$ tracks thus
being a golden signal for the NMSSM\footnote{This can also happen in
  the MSSM with gauge--mediated SUSY breaking though.}. All these
features are reviewed in Ref.~\cite{Ellwanger:2009dp}; it is worth
mentioning that adding this new singlet in the theory reduces the MSSM
fine--tuning, that is the requirement of having large corrections to
the lightest MSSM Higgs boson in order to cross the $M_Z$
value. Indeed the NMSSM lightest Higgs boson can be very light in the
case of large singlet composition thus evading the LEP bounds, and the
SM--like boson can be heavier than in the case of the MSSM.

We end this perspectives subsection by mentioning that the constrained
version of the NMSSM (cNMSSM) can be very predictive: if all the
experimental constraints are taken into account and in particular the
WMAP constraints on the dark matter density, the cNMSSM can be
described in practice by a single parameter as shown in
Ref.~\cite{Djouadi:2008yj} and depicted in Fig.~\ref{fig:NMSSM}
below.

\begin{figure}[!h]
\begin{center}
\includegraphics[scale=0.85]{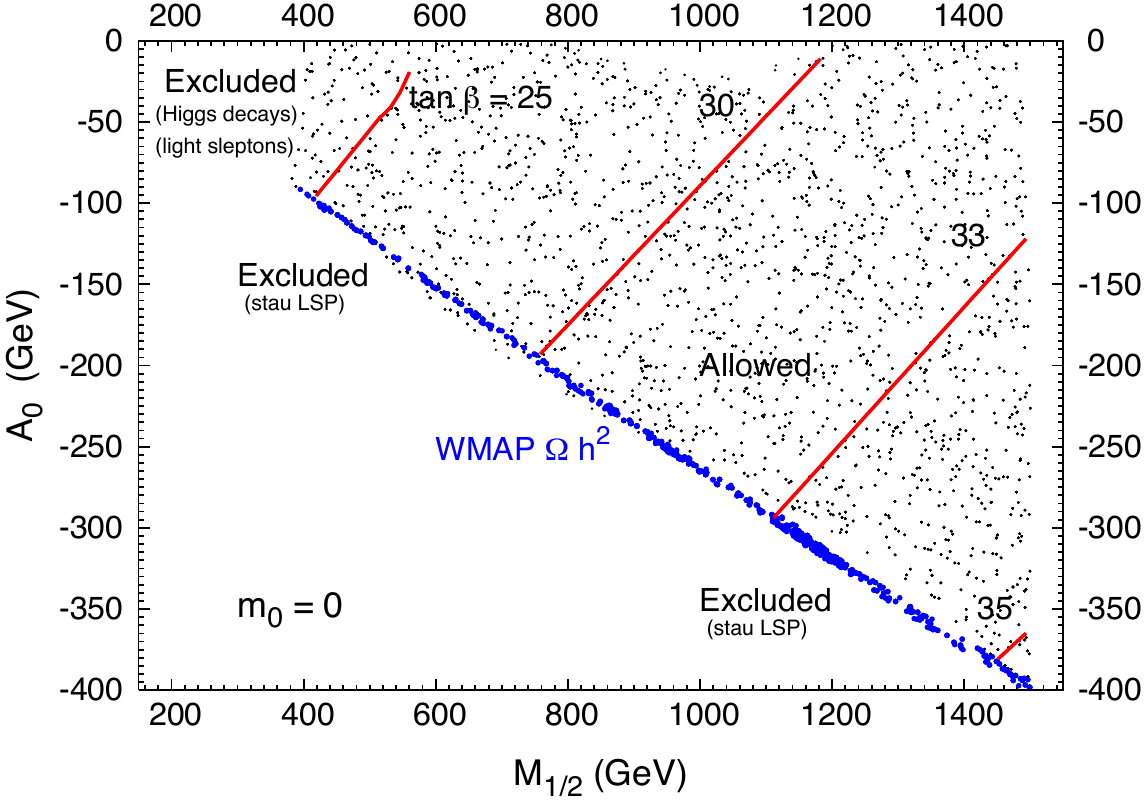} 
\end{center}
\vspace*{-.5cm}
\caption[The constrained NMSSM parameter space]{The viable cNMSSM
  parameter space in the $[M_{1/2}, A_0]$ plane. Figure taken from
  Ref.~\cite{Djouadi:2008yj}.}
\label{fig:NMSSM} 
\end{figure}

We have now ended the sketch of the MSSM and its essential features in
the view of our goal, that is the study of the MSSM Higgs bosons
production and decay at hadron colliders. We will now enter in the
core of this study by moving on part~\ref{part:four}.

\vfill
\pagebreak


\part{MSSM Higgs(es) production and decay}
\label{part:four}

\section{The MSSM Higgs sector at hadron colliders}

\label{section:MSSMHiggsIntro}

Part~\ref{part:three} was devoted to a brief summary of the reasons to
go beyond the SM and in particular why SUSY is appealing. It has
introduced the MSSM, the minimal extension of the SM, in which the
Higgs phenomenology is particularly rich with five Higgs bosons in the
spectrum. The purpose of this last part of the thesis is to study the MSSM
Higgs bosons production and decay at the two major hadron colliders in
activity, which will grossly reproduce the same outlines as done in
part~\ref{part:two}.

We have seen in sections~\ref{section:MSSMContent} and
\ref{section:MSSMAnomaly} the Higgs content of
the MSSM; we need to study in this section how the Higgs couplings to
fermions are affected by specific SUSY corrections which
are in addition to that of the standard QCD and electroweak. We will
then deduce from this analysis that the modelisation that we chose for
the Higgs bosons production and decay will be somewhat model
independent in the sense that it will hardly depend from the GUT scale
parameters, wether it would be that of minimal supergravity, anomaly
mediation or gauge mediation.

\subsection{SUSY corrections to Higgs
  couplings to fermions \label{section:MSSMHiggsIntroCouplings}}

We have presented in section~\ref{section:MSSMAnomaly} the tree--level
analysis of the Higgs sector of the MSSM. The need for quite large
radiative corrections to the Higgs bosons masses is illustrated by the
tree--level upper bound for the lightest $CP$--even Higgs boson $h$,
$M_h \lsim M_Z$. As this scalar boson has not been observed at LEP2
this limit should be crossed with the MSSM radiative corrections that
also affect the Higgs couplings to the other fields. A
thorough review has been done in Ref.~\cite{Djouadi:2005gj} (see also
references herein) from page 45 and we will only reproduce its main
features and in particular list the regimes that we are interested in.

We should first mention that we have already presented some radiative
corrections in section~\ref{section:MSSMAnomaly} in particular in the
case of the lightest Higgs boson $h$ in order to illustrate the impact
of radiative corrections on the tree--level upper bound. Depending on
the value of $\tan\beta$ and the relative value of $M_A$ as compared
to $M_Z$ we may have $h$ approching its maximal value $M_h^{\rm max}
\simeq 135$ GeV. We are mainly interested in the Higgs--fermions
couplings in the view of the Higgs production at hadron colliders in
the main production channels that involve Higgs--fermions
couplings. The one--loop vertex corrections modifying  the tree--level
Lagrangian can be implemented in an effective lagrangian written
as~\cite{Gunion:1990};
\beq
- {\cal L}_{\rm Yukawa} &=& \epsilon_{ij} \left[ (\lambda_b + \delta
  \lambda_b) \bar{b}_R h_d^i Q_L^j + (\lambda_t + \delta \lambda_t)
  \bar{t}_R Q_L^i h_u^j  + (\lambda_\tau + \delta \lambda_\tau)
  \bar{\tau}_R h_d^i L^j \right] \nonumber \\ 
&+&  \Delta \lambda_b \bar{b}_R Q_L^i h_u^{i*} +
     \Delta \lambda_\tau \bar{\tau}_R L^i h_u^{i*} +
     \Delta \lambda_t \bar{t}_R Q_L^i h_d^{i*} + {\rm h.c.} 
\eeq
Thus, at this order, in addition to the expected corrections $\delta
\lambda_{t,b}$ which alter the tree--level Lagrangian, a small
contribution $\Delta \lambda_t \, (\Delta \lambda_b)$ to the top
(bottom) quark will be generated by the doublet $h_u\, (h_d)$. The top
and bottom quark Yukawa couplings, defining $\lambda_b \Delta_b=\delta
\lambda_b+ \Delta \lambda_b \tan\beta$ and $\lambda_t \Delta_t =
\delta \lambda_t + \Delta \lambda_t \cotan\beta$, are then given by 
\beq
\lambda_b = \frac{ \sqrt{2} m_b } {v \cos\beta} \frac{1}{1+ \Delta_b}
\ , \ \ \lambda_t = \frac{\sqrt{2} m_t } {v \sin\beta} \frac{1}{1+
  \Delta_t} 
\label{eq:RC-Yukawas}
\eeq
The (approximate) corrections $\Delta_{t,b}$ are the same affecting
the $b$ and $t$ quarks in the MSSM and is a way, known as the
``$\Delta_b$ approximation'', to resum large corrections into an
effective parameter that gives for example in the case of the $b$
quark
\beq
\bar{m}_b^{\rm MSSM}(M_Z^2) \simeq \frac{\bar{m}_b^{\rm
    SM}(M_Z^2)}{1-\Delta_b}
\eeq
with
\beq
\Delta_b & \simeq & \left(\frac{2\alpha_s}{3\pi} \mu m_{\tilde g}\,
  I(m_{\tilde g}^2, m_{\tilde{b}_1}^2, m_{\tilde{b}_2}^2) +
  \frac{\lambda_t^2}{16\pi^2}A_t \mu\,
  I(\mu^2,m_{\tilde{t}_1}^2,m_{\tilde{t}_2}^2)\right)
\tan\beta\nonumber\\
 I(x,y,z) & = & \frac{x y \ln(x/y) + y x \ln(y/z) + x z
   \ln(z/x)}{(x-y)(y-z)(z-x)}
\eeq
The $b$ quark corrections are then enhanced by $\tb$ factors while
those affecting the top quark are only sizable for large $A_t$ or
$\mu$ values. These corrections will affect the Yukawa couplings and
iIn the case of the neutral Higgs boson couplings to bottom quarks,
we may write
\beq
g_{hbb} &\simeq &- \frac{\sin \bar \alpha}{\cos\beta} \bigg[1-
\frac{\Delta_b} {1+\Delta_b}(1+  \cotan\bar \alpha \cotan\beta )\bigg]
\nonumber \\
g_{Hbb} &\simeq & + \frac{\cos \bar \alpha}{\cos\beta}
\bigg[1-\frac{\Delta_b}
{1+\Delta_b}( 1- \tan \bar \alpha \cotan \beta  )\bigg] \nonumber \\
g_{Abb} &\simeq & \tb \bigg[1-\frac{\Delta_b} {1+\Delta_b} \frac{1}
{\sin^2\beta}\bigg]
\label{eq:ghff:threshold}
\eeq
where $\bar{\alpha}$ is the one--loop corrected $\alpha$ mixing angle
in the $CP$--even Higgs bosons sector; $g_{ijk}$ describe the MSSM
couplings normalized to those of the SM.

As shown in Ref.~\cite{Djouadi:2005gj} there is a strong variation of
the couplings depending on the value of $M_A$ around the critical
value $M_{h}^{\rm max}$. For  $M_A \lsim M_h^{\rm max}$ the lighter
$h$ boson couplings to up--type fermions are suppressed, while the
couplings to down--type fermions are enhanced, with the
suppression/enhancement being stronger at high $\tan\beta$ values. For
$M_A \gsim M_h^{\rm max}$, the normalized $h$ couplings approach the
unity value and reach the values of the SM Higgs couplings,
$g_{hff}=1$, for $M_A \gg M_h^{\rm max}$. The situation of the $H$
boson couplings to fermions is just opposite: they are close to unity
for $M_A \lsim M_h^{\rm max}$, while for $M_A \gsim M_h^{\rm max}$,
the $H$ couplings to up--type (down--type) fermions are strongly
suppressed (enhanced). For $M_H \gg M_h^{\rm max}$, the $H$ boson
couplings become approximately equal to those of the $A$ boson which
couples to down--type and up--type fermions proportionally to,
respectively, $\tan\beta$ and $\cotan\beta$.

We thus have typical regime that are of interest in our study, they
are listed below.
\begin{enumerate}[$1)$]
\item{{\it The decoupling regime}: either for large values of $M_A
    \gsim 300$ GeV at low $\tan\beta$ or for $M_A \gsim M_h^{\rm max}$
    at high $\tan\beta$, the lightest $CP$--even Higgs boson $h$
    becomes SM--like while the heavier $H$ becomes $CP$--odd like and
    decouples~\cite{Haber:1995be}: we have in the spectrum $h$
    SM--like with $M_h \lsim M_{h}^{\rm max} \simeq 135$ GeV,
    $(H,A,H^{\pm})$ with nearly the same mass and couplings to other
    particles. In particular the $H$ couplings to weak bosons become
    suppressed as in the case of the $A$ boson where they are
    forbidden by $CP$ invariance.

    This behaviour is already manifest at tree--level, and the only
    significant change due to radiative corrections is the threesold
    for the $A$ boson mass, which rises from $M_Z$ to $M_h^{\rm
      max}$~\cite{Djouadi:1996pj}. We have for $M_A^2 \gg M_Z^2$ and
    $\tan\beta \gg 1$:
    \beq
    g_{HVV} & = & \cos(\beta-\alpha)  \longrightarrow - \frac{2 M_Z^2}
    {M_A^2 \tan\beta} \,\, \sim 0
    \nonumber\\
    g_{hVV} & = & \sin(\beta-\alpha) \longrightarrow - \frac{2 M_Z^4}
    {M_A^4 \tan^2 \beta} \,\, \sim 1
    \label{eq:gHVVdecoup}
    \eeq
    for the weak bosons couplings and we have
    \beq
    g_{huu} & \longrightarrow & 1 - \frac{2 M_Z^2} {M_A^2 \tan^2
      \beta} \,\, \sim 1\nonumber \\
    g_{hdd} & \longrightarrow & 1 + \frac{2 M_Z^2} {M_A^2} \,\, \sim 1
    \nonumber \\
    g_{Huu} & \longrightarrow & - \cotan\beta \left( 1 + \frac{2
        M_Z^2} {M_A^2} \right) \,\, \sim -\cotan\beta\nonumber \\
    g_{Hdd} & \longrightarrow & \tan\beta \left(1 - \frac{2 M_Z^2}
      {M_A^2\tan^2 \beta} \right) \,\, \sim \tan\beta
    \label{gHff:decoup}
    \eeq
    for the fermions couplings.

    It is then manifest that for $M_A\gg M_Z$, $g_{HVV}$ vanishes
    while $g_{hVV}$ reaches the SM value (that is unity in term of
    normalized couplings). We recover for the $CP$--even Higgs
    couplings to fermions the same behaviour: the couplings of the $h$
    boson approach those of the SM Higgs boson, $g_{huu} = g_{hdd}=1$,
    while the couplings of the $H$ boson reduce, up to a sign, to
    those of the pseudoscalar Higgs boson, $g_{Huu} \simeq g_{Auu} =
    \cotan\beta$ and $g_{Hdd} \simeq g_{Add} = \tan\beta$.}
\item{{\it The anti--decoupling regime}: this regime is exactly the
    opposite of the decoupling regime: for a light pseudoscalar Higgs
    boson $M_A \gg M_{h}^{\rm max}$ the lighter $CP$--even Higgs boson
    mass is given at tree--level by $M_h \simeq M_A |\cos 2\beta|$
    while the heavier $CP$--even Higgs mass is given by $M_H \simeq
    M_Z(1 + M_A^2 \sin^2 2\beta /M_Z^2)$. At large values of
    $\tan\beta$, this is this time the $h$ boson which is degenerate
    in mass with the pseudoscalar Higgs boson $A$ while the $H$ boson
    has a mass $M_H \simeq M_h^{\rm max}$ when taking into account the
    radiative corrections~\cite{Gunion:1995zu}. The role of the $h$
    and $H$ bosons are thus reversed compared to the decoupling
    regime. Again the effect of radiative corrections is in practice
    to level up the $M_A$ threesold from $M_Z$ to $M_h^{\rm max}$. The
    $h$ boson has then the couplings that behave as those of the
    pseudoscalar Higgs boson $A$, while the $H$ boson couplings are
    SM--like:
    \beq
    g_{huu} \longrightarrow \cotan\beta  &,& 
    g_{hdd} \longrightarrow - \tan\beta  \nonumber \\ 
    g_{Huu} \longrightarrow 1  &,& 
    g_{Hdd} \longrightarrow 1  
    \eeq
    and the $H$ couplings to gauge bosons are SM--like, while the
    lighter $h$ boson is degenerate in mass with the pseudoscalar
    Higgs boson, $M_h \simeq M_A$ and has approximately the same
    couplings, that is, very suppressed couplings to gauge bosons.}
\item{{\it The intense coupling regime}: the last interesting regime
    for our stuy is the situation where the mass $M_A$ of the
    pseudoscalar $A$ boson is close to the maximal possible value for
    the lightest $CP$--even Higgs boson $M_h^{\rm max}$ (which
    corresponds to a mass close to the mass of  the $Z$ boson $M_Z$ at
    tree--level). The three neutral Higgs bosons have nearly the same
    mass (also nearly equal to that of the charged Higgs boson), $M_h
    \sim M_H \sim M_A \sim M_h^{\rm max}$. The mass degeneracy is more
    effective when $\tan\beta$ is large; this is the so--called
    intense--coupling regime discussed in details in the case of
    hadron colliders in Ref.~\cite{Boos:2003jt} referenced by
    Ref.~\cite{Djouadi:2005gj}.

    More precisely, this regime is defined as the one where the two
    $CP$--even Higgs bosons $h$ and $H$ are almoste degenerate in mass
    which implies that this degeneracy enlarges to the pseudoscalar
    boson $A$ as well. We have in this regime two possibilities:
    \beq
    M_A \gsim M_{h}^{\rm max} &\Rightarrow& M_H \simeq M_A \quad {\rm
      and} \quad M_h \simeq M_{h}^{\rm max} \nonumber \\
    M_A \lsim M_{h}^{\rm max} &\Rightarrow& M_h \simeq M_A \quad {\rm
      and} \quad M_H \simeq M_{h}^{\rm max}  
    \eeq
    Therefore the $A$ boson is always degenerate in mass with one of
    the $CP$--even Higgs bosons, that we will call $\Phi_A$, while the
    other $CP$--even Higgs particle, called $\Phi_H$, is very close in
    mass with $M_h^{\rm max}$ which in passing is equivalent as being
    the minimal value for the $H$ boson. In addition, the $CP$--even
    $\Phi_A$ boson will have almost the same couplings as $A$, while
    the $\Phi_H$ particle will have almost the couplings of the SM
    Higgs boson. We actually recover in a different situation the two
    regime described in the first two points, depending on the value
    of $M_A$ relative to that of $M_h^{\rm max}$.}
\end{enumerate}

Two other regimes, the intermediate--coupling and the
vanishing--coupling regimes, are not listed above because they do not
scope our study, see Ref.~\cite{Djouadi:2005gj} for more details. In
all cases above the interesting feature is that one of the
two $CP$--even Higgs boson behaves exactly like the $CP$--odd $A$
boson, the other being SM--like. This greatly simplfies the study as 
we can take the SM results of part~\ref{part:two} for the study of the
SM--like $CP$--even Higgs boson and only study the case of the $A$
boson for the two other neutral Higgs bosons.

\subsection{Model independence of the
  results \label{section:MSSMHiggsIntroModel}}

We will now present our set--up for the calculation of the cross
sections and decay branching fractions at hadron colliders. As
presented above three regimes are of great interest: the decoupling,
anti--decoupling and intense coupling regimes. In all of them we are
in the following situation: one of the $CP$--even Higgs boson behaves
as the $CP$--odd Higgs boson, the other is SM--like. We will then
denote the $A$ boson and the $H/h$ $CP$--odd--like boson by $\Phi$,
and study the $\Phi$ production and decay at hadron colliders. As
stated in Ref.~\cite{Baglio:2011xz}, this study is somewhat
model--independent in the following sense that will review the
arguments presented in the mentioned reference.

In both production and decay processes, we will actually assume the
$b\bar b\Phi$ coupling to be SM--like, $\lambda_{\Phi bb}= m_b /v$;
this means that to obtain true cross sections the results have to be
rescaled by a factor of $\tan^2\beta$. What justifies the use of
SM--like reduced couplings (apart from the $\tan\beta$ factor)? This
follow from three reasons:
\begin{enumerate}[$i)$]
\item{our study is conducted in the case of the regimes mentioned in
    the previous subsection; in these cases as explained above one of
    the $CP$--even Higgs boson behaves like the $CP$--odd $A$ boson
    and thus share its production and decay amplitudes. The only
    exception is for the intense coupling regimes where the three
    neutral Higgs bosons have similar enhanced couplings to down--type
    quarks. As the squares of the $CP$--even Higgs couplings add to the
    square of the $CP$--odd Higgs coupling, and since $M_H\! \approx\!
    M_h\!\approx M_A$, we still recover our approximation provided 
    that the cross section times branching ratios for the three
    $h,H,A$ particles are added;}
\item{as the pseudoscalar $A$ boson does not couple to squarks of the
    same flavor ($A \tilde q_i \tilde q_i$ couplings are forbidden by
    $CP$--invariance), there is no superparticle contribution in the
    $gg,b\bar b \to A$ processes (to be discussed later on) at leading
    order and higher order SUSY corrections are
    suppressed. In the $CP$--even Higgs case, there are additional
    superparticle contributions to $gg\to H(h)$ originating mainly
    from stop and sbottom squarks loops. However, these
    contributions are damped by the squark mass squared and are not
    similarly enhanced by $ m_b \tb$ factors; they thus remain small
    so that they can be safely neglected in most cases;}
\item{the most important reason is the last one: the relevant effect
    of supersymmetric contributions appears through the $\Delta_b$
    term in the $\Delta_b$ approximation as stated in the previous
    subsection. This correction can be significant as it grows with
    $\tan\beta$ and is obviously SUSY model dependant. However in the
    case of our study, that is the production times branching
    fraction, this correction almost cancel out between production and
    decay, the remaining part having no practical impact in the view
    of the large QCD uncertainties whatever benchmark scenario to be
    considered. Indeed as shown in Ref.~\cite{Baglio:2011xz} the
    impact of the $\Delta_b$ approximation is
    \beq
    \sigma \times {\rm BR} & \longrightarrow &
    \frac{\sigma}{(1+\Delta_b)^2} \times \frac{\Gamma(\Phi \to
      \tau \tau)}{(1+\Delta_b)^{-2} \Gamma(\Phi\to b\bar b)+
      \Gamma(\Phi \to \tau\tau)}\nonumber\\  
    & & \approx  \sigma \times {\rm BR} \times (1- \frac15 \Delta_b)
    \eeq
    assuming BR$(\Phi \to \tau^+ \tau^- )\approx 10\%$. Thus, unless the
    $\Delta_b$ correction is extremely large, it will lead to only a
    few percent correction at most to the cross section times decay
    branching ratio.}
\end{enumerate}

The impact of the $\Delta_b$ approximation is shown below:

\begin{figure}[!h]
  \begin{center}
    \includegraphics[scale=0.85]{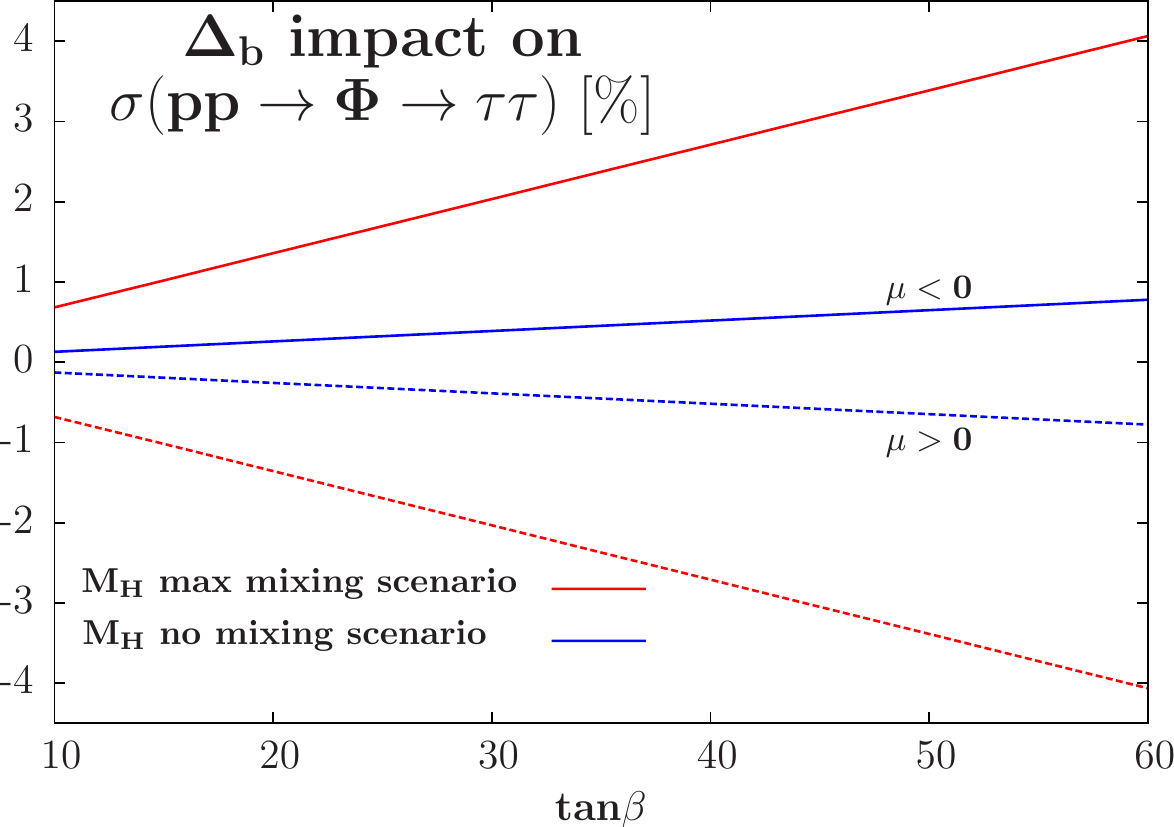}
  \end{center}
  \vspace*{-6mm}
  \caption[The impact of main one--loop SUSY corrections to the $\Phi
  b\bar b$ coupling in the MSSM at hadron colliders]{The impact (in \%) of
    the $\Delta_b$ supersymmetric radiative correction on the cross
    section times branching ratio $\sigma[pp\to A+H(h)] \times{\rm
      BR}[A /H(h) \to \tau^+\tau^-]$ as a function if $\tb$ in two of
    the benchmark scenarios of Ref.~\cite{Carena:2002qg} for both signs of
    $\mu$.}
  \label{fig:MSSMDeltabapproximation}
\end{figure}

We have used in Fig.~\ref{fig:MSSMDeltabapproximation} the program
FeynHiggs~\cite{Heinemeyer:1998yj} to 
evaluate the $\Delta_b$ correction, and we display for a fixed value
of $M_A$ and as a function of $\tan\beta$, the impact of the
$\Delta_b$ correction on $\sigma (gg+b\bar b\! \to \Phi) \times  {\rm
  BR}(\Phi \to \tau \tau)$. This is done in two benchmark scenarios
for the $CP$--conserving  MSSM proposed in Ref.~\cite{Carena:2002qg}:
the maximal $M_h^{\rm max}$ mixing and the $M_h^{\rm min}$ no--mixing
Higgs scenarios\footnote{The gluophobic scenario is now ruled out as
  it leads to light gluinos $m_{\tilde g} =500$ GeV) and squarks
  ($m_{\tilde q} \approx 350$ GeV) which have been  excluded by the
  recent ATLAS and CMS analyses~\cite{daCosta:2011qk,
    Khachatryan:2011tk}. The small $\alpha_{\rm eff}$, which leads to
  $\tilde m_g =500$ GeV and $\tilde m_q =800$ GeV is probably also
  excluded, in particular if the various ATLAS and CMS analyses are
  combined. This latter scenario leads to a huge (and probably rather
  problematic) $\Delta_b$ value but the effect on the cross section
  times branching ratio is again less than 10\% for $\tan\beta \lsim
  30$.}, with the two possible signs of the higgsino parameter $\mu$;
these two benchmark scenarios enter in the regimes described in the
previous subsection.

We see that in both cases the quality of our approximation for the
$pp\!\to\! \Phi\! \to\! \tau^+\tau^-$ cross section is always very
good, the difference with the exact result including the $\Delta_b$
correction being less than 2\% for $\tb \lsim 30$ (and $\lsim 4\%$ for
$\tb \lsim 60$), which is negligible in view of the large QCD
uncertainties that affect the cross section as studied in the
following sections. The quality of the $\Delta_b$ approximation itself
has been discussed in Ref.~\cite{Dawson:2011pe} which states that this
accurately reproduces the exact one--loop SQCD
corrections. Nevertheless in the final stage of the writing of the
thesis some preliminary results presented in
Ref.~\cite{Muhlleitner:2011hh} with new SUSY contributions show that
this might not be the case in every scenarios 
(and in particular in the so--called $\alpha^{\rm eff}$ scenario that
has been ruled out though), thus softening a bit our model
independance argument. Nevertheless it remains quite accurate in
the main benchmark scenarios discussed in this thesis. We will then
use the SM--like couplings to the $A$ 
boson and multiply by a factor of $2\tan^2\beta$ to account for the
degeneracy between the $A$ and one of the $CP$--even Higgs bosons, to
obtain the true $pp\!\to\! \Phi\! \to\! \tau^+\tau^-$ cross
section\footnote{We note in passing that this approximation is very
  useful in practice as it prevents the need of large grids to tackle
  numerically every MSSM scenario as well as CPU time consuming scans
  of the supersymmetric parameter space.}. In fact, our results also
hold in a general two--Higgs doublet model in which two Higgs
particles have the same mass and the same enhanced couplings to
down--type  fermions, as the relevant parameters are again $\tan\beta$
and $M_A$.

We summarize our set--up developed in this subsection: we will study
the production and decay of the pseudoscalar $A$ boson that is similar
to one of the $CP$--even Higgs boson, these two states being
collectively denoted by $\Phi$. We will only present results with
$\tan\beta=1$ and with SM--like couplings, which is an excellent
approximation; the actual results are recovered provided that the
results are multiplied by a factor of $2\tan^2\beta$. We are now ready
to study the MSSM Higgs bosons production at the Tevatron and the
LHC. We will then study the combination with the branching fractions
and finish by the impact of our study on the MSSM parameter space.

\vfill
\pagebreak

\section{MSSM Higgs production at the Tevatron}

\label{section:MSSMHiggsTev}

The section~\ref{section:SMHiggsTev} has been devoted to the
production of the SM Higgs boson at the Tevatron. The search of this
remnant of the spontaneous breaking of the electroweak symmetry is one
of the main goal of the present high energy colliders and in
particular of the Tevatron program. We have seen in
section~\ref{section:MSSMAnomaly} that the Higgs sector is extended in
supersymmetric theories~\cite{DreesMSSM} and in particular in the
minimal extension, the MSSM, two Higgs doublet are required and leads
to the existence of five Higgs bosons: two $CP$--even $h$ and $H$, a
$CP$--odd $A$ and two charged $H^\pm$
particles~\cite{Gunion:1990,Djouadi:2005gj}. We will study the
production and later on the decay of the neutral Higgs bosons at the
Tevatron.

We have seen in section~\ref{section:MSSMAnomaly} that two parameters
are needed to describe the Higgs sector of the MSSM at tree--level:
the mass $M_A$ of the pseudoscalar boson and the ratio of vacuum
expectation values of the two Higgs fields, $\tan\beta$, that is
expected to lie in the range $1 \lsim \tb \lsim 50$ from theoretical
analysis, see Ref.~\cite{Djouadi:2005gj} for a review. The
section~\ref{section:MSSMHiggsIntroCouplings} has presented the main
regimes in which our analysis is conducted, thus we will denote
collectively the $A$ boson and the $CP$--even state which behaves like
the $A$ boson as $\Phi\!=\!A,H(h)$. They are almost degenerate in mass
and have the same properties: no couplings to gauge bosons, while the
couplings to isospin down--type (up--type) quarks and charged leptons
are (inversely) proportional to $\tb$. 

This means that for $\tb\! \gsim\! 10$, the $\Phi$ boson couplings to
bottom quarks and $\tau$--leptons are strongly enhanced while those to
top quarks are suppressed.  As a result, the phenomenology of these
states becomes rather simple. To a very good approximation, the $\Phi$
bosons decay almost exclusively into $b\bar b$ and $\tau^+\tau^-$
pairs with branching ratios of, respectively, $\approx\! 90\%$ and $
\approx\! 10\%$, while the other decay channels are suppressed to a
negligible level~\cite{Djouadi:1997yw}. The main
production mechanisms for these particles are those processes which
involve the couplings to bottom quarks. At hadron colliders,  these
are the gluon--gluon fusion mechanism, $gg \to \Phi$, which dominantly
proceeds through $b$--quark triangular
loops~\cite{Georgi:1977gs,Spira:1995rr} and bottom--quark fusion,
$b\bar b \to \Phi$~\cite{Dicus:1988cx,Campbell:2002zm,
  Maltoni:2003pn,Harlander:2003ai}, in which the bottom quarks are
directly taken from the protons in a five active flavor
scheme~\cite{Dittmaier:2003ej, Dawson:2003kb}. The latter process is
similar to the channel $p\bar p \to b\bar b\Phi$ when no $b$--quarks
are detected in the final state~\cite{Dittmaier:2003ej,
  Dawson:2003kb}. We will study these two processes at the Tevatron
and later on in this thesis at the LHC, leaving the study of the
branching fractions and in particular their combination with
production cross sections for the last section.

With its successful operation in the last years, the Fermilab Tevatron 
collider has now collected a substantial amount of data which allows
the CDF and D0 experiments to be sensitive to the MSSM Higgs
sector. Stringent constraints beyond the well established LEP bounds
$M_{A}, M_h \gsim M_Z$ and $\tb \gsim 3$~\cite{Barate:2003sz}, have
been set on the MSSM parameter space $[M_A, \tb$] using the process
$gg, b\bar b \to \Phi \to \tau^+ \tau^-$. Moderate $A$ masses,  $M_A\!
\approx\! 100$--200 GeV, together with high $\tb$ values, $\tb\!
\gsim\! 30$, have been excluded at the 95\% confidence level
(CL)~\cite{Aaltonen:2009vf, Abazov:2008hu,Benjamin:2010xb}.

However the experimental analyses cited above have not taken into
account the theoretical uncertainties that affect the production and
decay rates, which can be important despite of the fact that some
higher order perturbative corrections to these processes are
known. These are mainly due to the unknown higher order corrections in
perturbation theory as in the SM case, the still not
satisfactory parametrization of the parton distribution functions
(PDFs), as well as the parametric uncertainties stemming from the not
very precisely measured values of the strong coupling constant
$\alpha_s$ and the bottom quark mass $m_b$ which plays a significant
role in the MSSM contrary to the SM case. We will present in the
following the results published in Ref.~\cite{Baglio:2010bn}
concerning the thorough analysis of the theoretical uncertainties of
production cross section of the $\Phi$ boson at the Tevatron, in the
set--up summarized in the end of
section~\ref{section:MSSMHiggsIntroModel}. We will see that they can
affect by nearly $50\%$ the production rates, thus having a great
impact of the combined analysis of the MSSM Higgs bosons by CDF and D0
experiments~\cite{Benjamin:2010xb} that have put high constraints on
the MSSM $[M_A, \tb]$ parameter space.

\subsection{Gluon--gluon fusion and bottom quarks
  fusion \label{section:MSSMHiggsTevCross}}

Our study as said before will be in the context of the regimes where
at least two of the neutral Higgs bosons are almost degenerate. Indeed
it is the case in most benchmark scenarios~\cite{Carena:2002qg} as in
the maximal mixing scenario where $X_t= A_t -\mu/\tb \sim \sqrt 6
M_S$, $M_S$ being the common squark mass and leading to $M_h^{\rm max}
= 135$ GeV~\cite{Heinemeyer:2004gx, Heinemeyer:2004ms,
  Allanach:2004rh}, or the no--mixing scenario where $X_t\approx 0$
which leads to a lower $M_h^{\rm max}$ value.

The Higgs Yukawa couplings to bottom quarks plays  a
major role in the analysis that will be presented in this section and
this explains why we concentrate on the two main production channels
that are the gluon--gluon fusion and the bottom quarks fusion. Indeed
because of $CP$ invariance which forbids $A$ couplings to gaugebosons
at tree--level, the pseudoscalar $A$ boson cannot be produced in the
Higgs-strahlung and vector boson fusion processes; only the $gg\to A$
fusion as well as associated production with  heavy quark pairs,
$q\bar q, gg \to Q\bar Q A$, will be in practice relevant (additional
processes, such as associated production of $CP$--even and $CP$--odd
Higgs particles, have too small cross sections). This will therefore
be also the case of the $CP$--even $H$ and $h$ particles in,
respectively, the decoupling and anti--decoupling  scenario. As
mentioned earlier, in almost the entire parameter space for large
enough $\tb$ values, the couplings of one of the $CP$--even Higgs
particles are SM--like which means that we can use again the results
presented in part~\ref{part:two}, while the couplings of the other
$CP$--even particle are the same as those of the pseudoscalar $A$ boson,
on which we will focus in the rest of our discussion. At the tree
level, this coupling is given in terms of the $b$--quark mass, the SM
vacuum expectation value $v$ and $\tb$, by 
\beq  
\lambda_{\Phi bb} = \frac{\sqrt 2 m_b}{v \cos\beta}
\stackrel{\small \tb \gg 1} \longrightarrow \frac{\sqrt 2 m_b}{v} \tb \ , \ 
\Phi=A, H\;(h)
\label{eq:abb} 
\eeq 
First of all, in the MSSM, one usually uses the modified dimensional
reduction $\overline{\rm DR}$ scheme which, contrary to the
$\overline{\rm MS}$ scheme, preserves Supersymmetry. In the case of
the $b$--quark mass, the relation between the $\overline{\rm DR}$ and
$\overline{\rm MS}$ running masses at a given scale $\mu$
reads~\cite{Pierce:1996zz}
\beq  
{\overline{m}}_{b}^{\overline{\rm DR}} (\mu) =
{\overline{m}}_{b}^{\overline{\rm MS}} (\mu) \, \bigg[ 1- \frac{1}{3}
\frac{\alpha_{s} (\mu^2)}{\pi} - \frac{\alpha_s^2(\mu^2)}{\pi^2} +
\cdots \bigg]
\eeq 
where the strong coupling constant $\alpha_s$ is also evaluated at the
scale $\mu$ and additional but small electroweak contributions are
present. Since the difference between the quark masses in the two
schemes is not very large, $\Delta m_b/m_b \sim 1\%$, to be compared
with an ``experimental" error on $\overline{m}_b(\overline{m}_b)$ of
the order of a few percent, we will neglect the difference as commonly
done at least in unconstrained SUSY models with no RGE evolution of
the parameters from a higher scale. We will thus adopt this
approximation at least when we quote the central values of the cross
sections; nevertheless we will discuss later on the impact of the
$b$--mass renormalization scheme on the theoretical uncertainties. We
will not include SUSY corrections to the Yukawa coupling as explained
and justified in the previous section: the analysis will thus be
somewhat model independent and we will focuse on the standard QCD
uncertainties. However if ones is interested by the genuine SUSY
corrections~\cite{Dawson:1996xz, Harlander:2004tp, Anastasiou:2006hc,
  Aglietti:2006tp, Bonciani:2007ex, Muhlleitner:2006wx,
  Degrassi:2008zj, Dawson:2007ur, Anastasiou:2008rm,
  Muhlleitner:2010nm} they can be evaluated using the program {\tt
  HIGLU} for instance in the case of the gluon--gluon fusion production
channel.

\subsubsection{The gluon--gluon fusion in the MSSM}

The first production channel that we consider is the gluon--gluon
fusion as in the SM case. As we are in moderate to high $\tan\beta$
regime in our set--up, the top--loop is suppressed and only the
$b$--loop is included. As the $b$--quark mass is very small compared
to the Higgs masses, chiral symmetry approximately holds and the cross
sections are approximately the same for the $CP$--even $H\;(h)$ and
$CP$--odd $A$ bosons\footnote{This is only true if the SUSY particle
  loop contributions are not included. In the case of the $CP$--even
  particles, their relative contribution are suppressed at high
  enough $\tb$ and we will ignore them here as stated in the previous
  section. In the case of the $A$ boson, the SUSY contributions appear
  only at two--loops and they can be safely neglected. Indeed these
  additionnal SUSY contributions in $gg\!\to\! H/h$ do not
  appear in $gg\!\to\! A$ but they are very small for a large SUSY
  breaking scale, $M_\Phi \ll M_S$ as seen in
  Ref.~\cite{Muhlleitner:2010nm}. In addition the $\Delta_b$ term is
  again negligible as stated in the previous section, and that can be
  seen from the almost identical tables XI--XIV and Figs.~4 of
  Ref.~\cite{Benjamin:2010xb} that describe four benchmark
  scenarios~\cite{Carena:2002qg}.}. The QCD
corrections are known only to NLO for which the exact calculation with
finite loop quark masses is available~\cite{Spira:1995rr}. Contrary to
the SM case, they increase only moderately the production cross
sections. The calculation of the higher order corrections that have
been made available  for the SM Higgs boson, the NNLO QCD corrections
(performed in the infinite quark mass limit) and the NLO electroweak
corrections (the dominant part of which arises because of the large
Higgs--$t \bar t$ Yukawa coupling)  do not apply here and will be thus
ignored. In order to match the SM calculation which reproduces the
results for the SM--like $CP$--even Higgs boson in order to approach
properly the decoupling and anti--decoupling regime, we will adopt the
central scale $\mu_0=\frac12 M_\Phi$.

We will evaluate the cross section with such a central scale using the
program program {\tt HIGLU}~\cite{Spira:1995mt, Spirapage} with only
the dominant loop contribution of the bottom quark loop included. We
work in the $\overline{\rm MS}$ scheme for the renormalization  of the
$b$--quark mass, that is $\overline{m}_b(\overline{m}_b)$. The
resulting partonic cross sections are then folded with the latest MSTW
sets of PDFs~\cite{Martin:2009iq, Martin:2009bu, Martin:2010db}
consistently at the NLO order in perturbation theory. We will use
$\tb=1$: to obtain the true numbers a factor of $\tan^2\beta$ has to
be included (and then doubled to obtain the full $gg\to A+h/H$ cross
section).

\subsubsection{The bottom quarks fusion in the MSSM}

In the case of the $pp \to b\bar b \Phi$ processes, the NLO QCD
corrections have been calculated in Ref.~\cite{Dittmaier:2003ej,
  Dawson:2003kb} and turn out to be rather large,  in contrast to $pp
\to t\bar t$+Higgs production.  Because of the small $m_b$ value, the
cross sections develop large logarithms $\ln(Q^2/m_b^2)$ with the
scale $Q$ being typically of the order of the factorization scale,
$\mu_F \sim M_\Phi \gg m_b$. These logarithms can be resummed via the
Altarelli--Parisi equations by considering the $b$--quark as a
massless parton and using heavy quark distribution functions at a
scale $\mu_F \sim Q$ in a five active flavor scheme. In this scheme,
the inclusive process where one does not require to observe the $b$
quarks is simply the $2\to 1$ process $b \bar b \to \Phi$ at leading
order~\cite{Dicus:1988cx}. If the observation of a high--$p_T$ final
$b$--quark is required, one has to consider its NLO
corrections~\cite{Campbell:2002zm, Maltoni:2003pn} and in particular
the $2\to 2$ process $gb\to \Phi b$, which indeed generates the $p_T$
of the $b$--quark. Requiring the observation of two $b$ quarks in the
final state, we have to consider the $2 \to 3$ process $gg \to b\bar b
\Phi$ discussed above, which is the leading mechanism at
NNLO~\cite{Harlander:2003ai}. Thus, instead of  $q\bar q, gg \to b\bar
b \Phi$,  we will consider the process $b\bar b \to \Phi$ for which
the cross section is known up to NNLO in QCD \cite{Campbell:2002zm,
  Maltoni:2003pn,Harlander:2003ai}, with corrections that are of
moderate size if: the bottom quark mass in the Yukawa coupling is
defined at the scale $M_\Phi$ to absorb large logarithms
$\ln(\mu_R^2/m_b^2)$; the factorization scale, that we will set here
equal to the renormalization scale, is chosen to be small,
$\mu_F=\mu_R=\mu_0= \frac14 M_\Phi$. Fig.~\ref{fig:bbHFeynman} below
displays some typical Feynman diagrams for bottom quarks fusion in
this set--up.

\begin{figure}[!h]
\begin{center}
\includegraphics[scale=0.70]{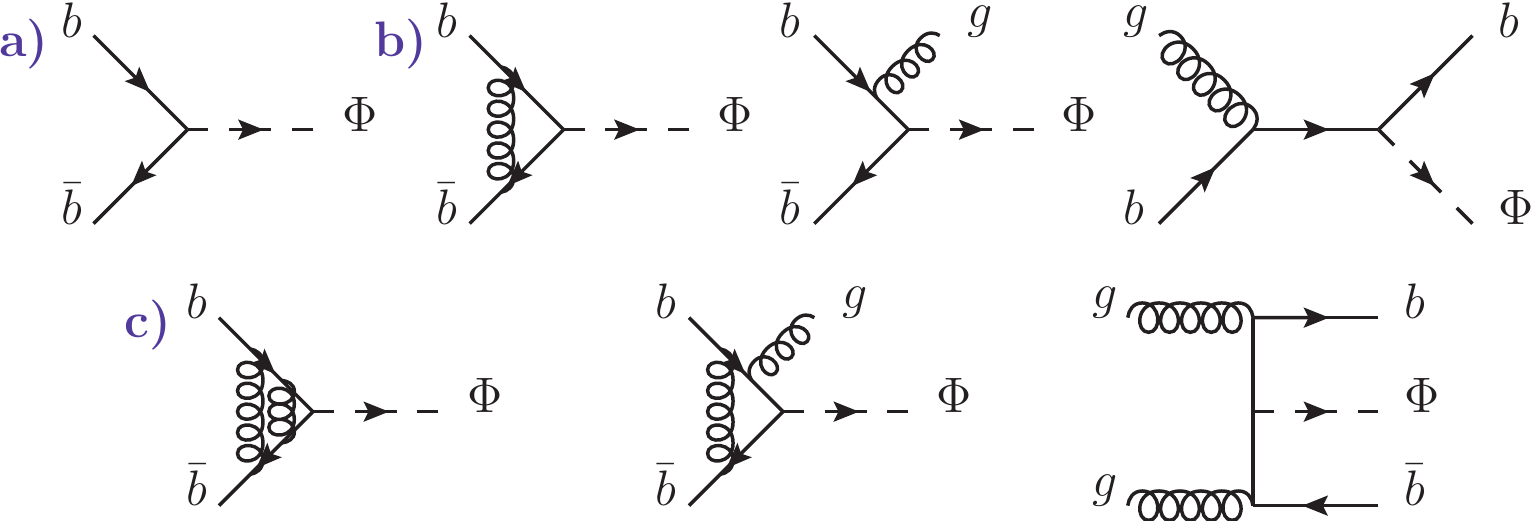} 
\end{center}
\vspace*{-.5cm}
\caption[Feynman diagrams for the bottom quark fusion process in the
MSSM]{Typical Feynman diagrams up to next--to--next--to--leading order
  in QCD in the bottom quark fusion production channel in the
  MSSM. The five flavors scheme has been adopted.}
\label{fig:bbHFeynman} 
\end{figure}

The evaluation of the cross section up to NNLO is done using the
program {\tt bbh@nnlo}\footnote{We thank R. Harlander for providing us
  with the code.} with a central scale $\mu_R\!=\mu_F\!= \!\mu_0\!=\!
\frac14 M_A$ as appropriate for this process according to
Ref.~\cite{Harlander:2003ai}. We work in the $\overline{\rm MS}$
scheme for the bottom quark mass evaluated at the scale of the
process, $\overline{m}_b(\mu_R)$. The resulting partonic cross
sections are then folded again with the MSTW2008 set of PDFs
consistently at the NNLO order in in perturbation theory. We again
assume $\tb=1$, the actual cross sections being obtained after a
multiplication by a factor of $\tan^2\beta$.

\begin{figure}[!h] 
\begin{center} 
\includegraphics[scale=0.80]{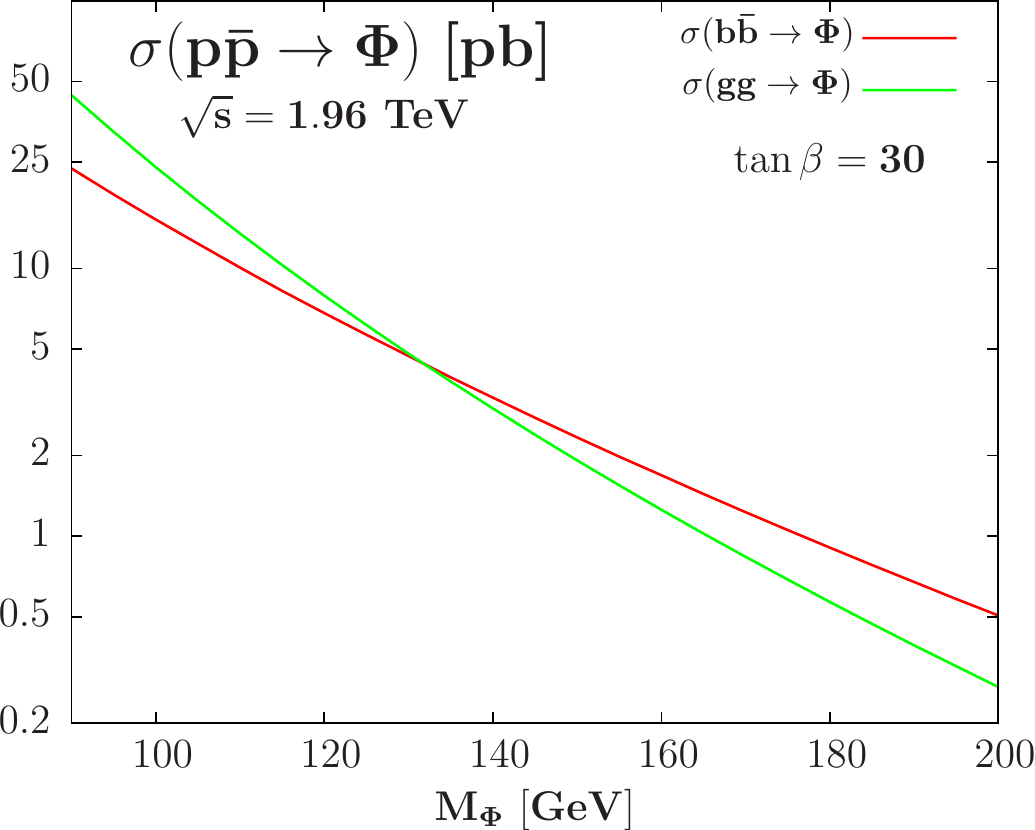} 
\end{center} 
\vspace*{-5mm}
\caption[The NLO $gg\to A$ and NNLO
$b\bar b \!\to\!A$ cross sections at the Tevatron with
$\tb=30$]{The normalization of the cross sections $\sigma^{\rm
    NLO}_{gg\!\to \!\Phi}$ and $\sigma^{\rm NNLO}_{b\bar b \!\to\!\Phi}$ at
  the Tevatron as a function of $M_\Phi$ when using the MSTW PDFs and
  $\tb=30$.}
\vspace*{-3mm}
\label{fig:MSSMproductionTev}
\end{figure}

The results for the cross sections $\sigma(gg\!\to\! \Phi)$ and
$\sigma(b\bar b\! \to\!  \Phi)$ are shown in
Fig.~\ref{fig:MSSMproductionTev} below for the 
Higgs mass range that is relevant at the Tevatron, $M_\Phi=90$--200
GeV and with $\tb=30$ as an example. We have compared our values with
those given by the program that has been used by the CDF and D0
collaborations for their cross section normalization, {\tt
  FeynHiggs}\cite{Heinemeyer:1998yj}. This program, initially supposed
to only provide precise values for the MSSM Higgs masses and
couplings, gives also grids for production cross sections which should
be used with care. For the $b \bar b\!\to\! \Phi$ channel, we obtain
cross sections that are $\approx 30\%$ smaller. The reason is that
{\tt FeynHiggs} simply provides the values given in the original
paper~\cite{Harlander:2003ai} which uses the outdated MRST2002 set of
PDFs which are only partly at NNLO. In the case of $gg\to \Phi$, the
agreement is better as we obtain a cross section that is only $\approx
10\%$ higher, due to the different central scale and renormalization
scheme for $m_b$ that have been used. The comparison between the two
process shows that the dominant process from $M_\Phi \gsim 135$ GeV is 
the bottom quarks fusion. Both production cross sections are far
greater than the similar processes in the SM: this adds up more odds
to find the Higgs boson at the Tevatron.

\subsection{The scale uncertainty \label{section:MSSMHiggsTevScale}}

We will follow the guideline that has been developed in
part~\ref{part:two} in the case of the SM Higgs production. We start
the evaluation of the theoretical uncertainties by looking at the
scale uncertainty. In order to estimate the missing higher orders in
perturbation theory we usually use a variation of the renormalization
and factorization scales in the domains $ \mu_0/\kappa \le \mu_R,\mu_F
\le \kappa \mu_0$ around the central scales $\mu_0$, with sometimes
the additional restriction $1/\kappa \le \mu_R/\mu_F \le \kappa$
imposed.

In the case of the $gg\to \Phi$ process, as for the SM Higgs
boson at the lHC,  the scale uncertainty is evaluated by allowing for
a variation of the renormalization and factorization scales within a
factor of two  around the central scale, $\frac12 \mu_0 \le
\mu_R,\mu_F  \le 2 \mu_0$ with $\mu_0= \frac12 M_\Phi$; this is enough
in the case of the bottom quark loop at NLO in view of our procedure
developed in part~\ref{part:two} where we required that the LO band
catch the NLO central result to define the constant factor $\kappa$.

\begin{figure}[!h]
  \begin{center}
    \vspace*{-2mm}
    \mbox{
      \includegraphics[scale=0.65]{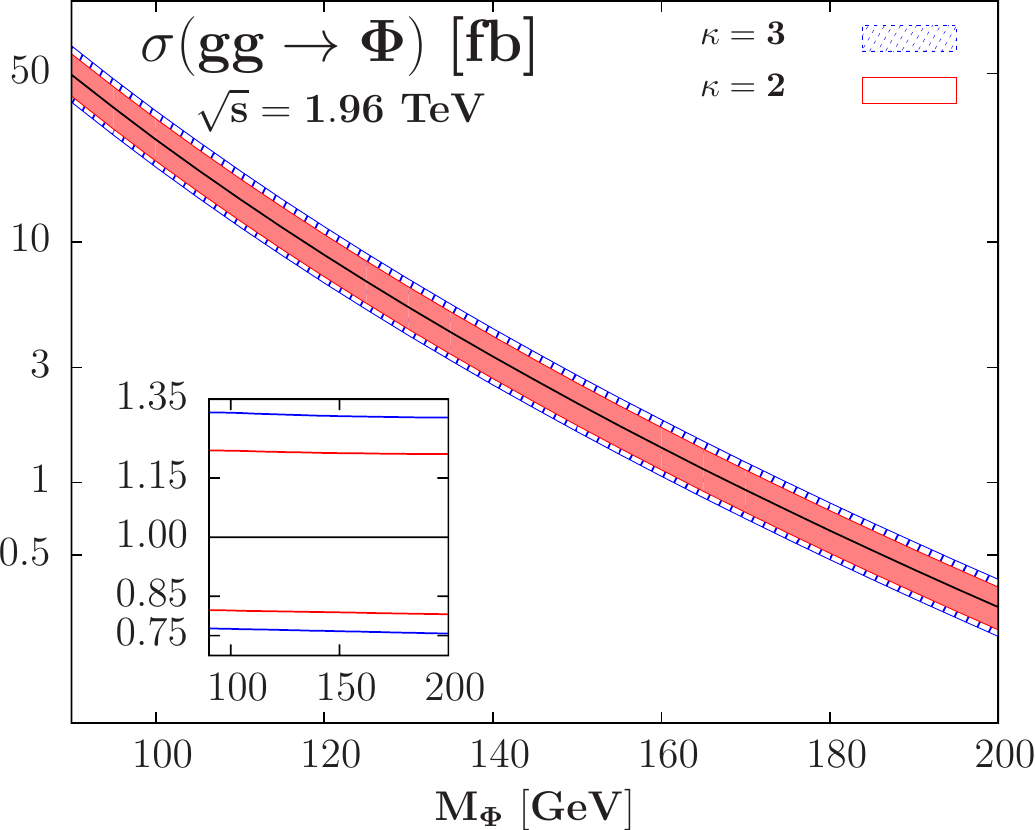}
      \includegraphics[scale=0.65]{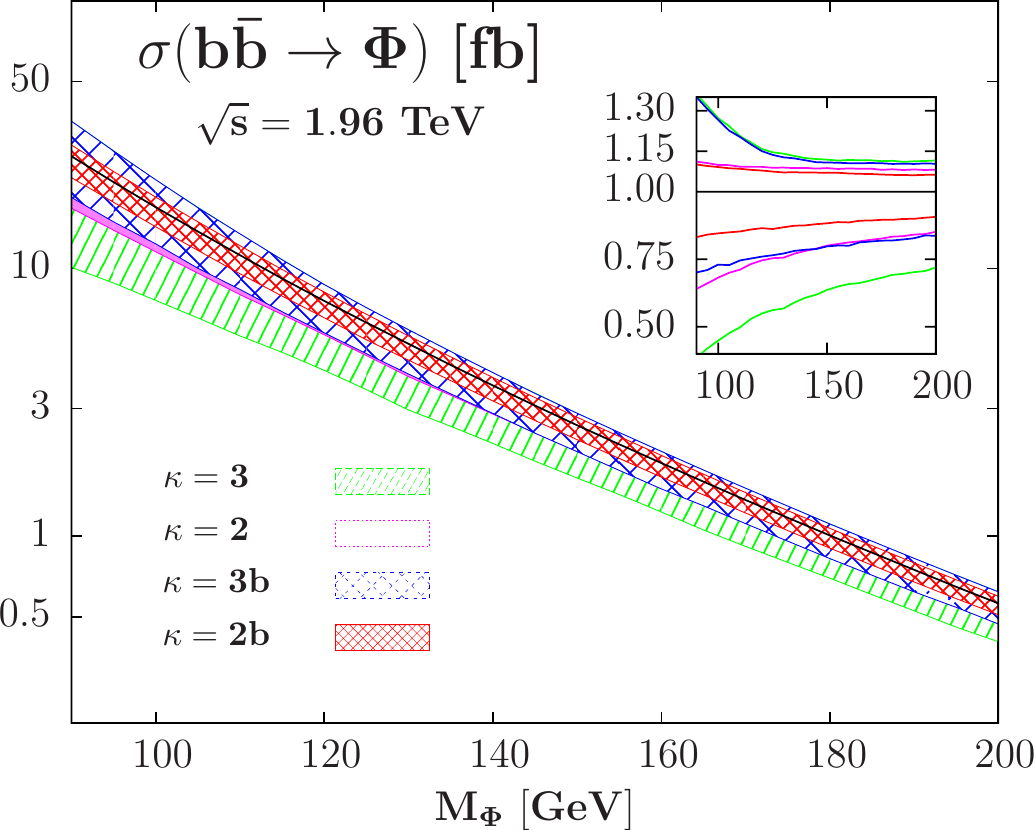}
    }
  \end{center}
  \vspace*{-4mm}
  \caption[Scale uncertainty in the $gg\to \Phi$ and $b\bar b\to \Phi$
  processes at the Tevatron]{The scale uncertainty bands of the NLO $gg \to
    \Phi$ (left) and the NNLO $b\bar b  \to \Phi$ (right) cross
    sections at the Tevatron as a function of $M_\Phi$;
    different values $\kappa=2,3$ are used and the results are shown
    when the additional constraint $1/\kappa \le \mu_R/\mu_F \le
    \kappa$ is imposed or not (marked as $\kappa$b). In the inserts,
    the relative deviations (compared to the central cross section
    values) are shown.}
  \label{fig:MSSM-scaleTev}
  \vspace*{-2mm}
\end{figure}

However for the $b\bar b  \to \Phi$ case we will extend the
domain of scale variation to a factor of three around the central
scale $\mu_0= \frac14 M_\Phi$, $\frac13 \mu_0 \le \mu_R,\mu_F  \le 3
\mu_0$ and impose an additional restriction $1/3 \le \mu_R/\mu_F \le
3$. There are several reasons to do so, the first being that we would
like to include the scheme dependence in the renormalization of the
bottom quark as seen later on; this adds to the scale uncertainty and
as the bottom quark mass in the bottom quark fusion is defined at the
renormalization scale itself, one way to consider this additionnal
source of uncertainty is to include it in the scale uncertainty by
extending the domain of variation for the renormalization and
factorization scales. A second reason is that it is well known that
when the same final states are considered, the cross sections in the
$b\bar b \to \Phi$ process in the five--flavor scheme and in the
$q\bar q, gg \to b\bar b \Phi$ channel in the four--flavor scheme
differ significantly (see Ref.~\cite{Assamagan:2004mu} page 5) and
only by allowing a wider domain for scale variation and, hence, a
larger scale uncertainty that the two results become consistent with
each other. To illustrate the much larger scale uncertainty that is
possible in the $b \bar b \to \Phi$ case, we display the results in
Fig.~\ref{fig:MSSM-scaleTev} below in much the same way as in
Ref.~\cite{Dittmaier:2011ti}.

We obtain in the gluon--gluon fusion process a relatively constant
uncertainty of order $+21\%, -18\%$ in the whole $M_\Phi$ mass range
90--200 GeV relevant at the Tevatron. In the case of the bottom quarks
fusion, the right part of Fig,~\ref{fig:MSSM-scaleTev} shows clearly
that there is an unstability in the choice of the constant factor
$\kappa$ defining the interval of variation of the two scales
$\mu_R,\mu_F$. If one chooses the factor $\kappa=2$, the scale
uncertainty is $+15\%, -9\%$ at $M_\Phi\simeq 100$ GeV with the
additionnal constrain on the ratio $\mu_R/\mu_F$ and reduces to $+9\%,
-6\%$ at $M_\Phi \simeq 200$ GeV. However our final choice leads to a 
scale uncertainty of $\pm 27\%$ at $M_\Phi\simeq 100$ GeV which
reduces to $+15\%, -10\%$ at high $M_\Phi\simeq 200$ GeV.

\subsection{The PDF and $\alpha_S$
  uncertainties \label{section:MSSMHiggsTevPDF}}

\begin{figure}[!h]
  \begin{bigcenter}
    \vspace*{-1mm}
    \mbox{
      \includegraphics[scale=0.65]{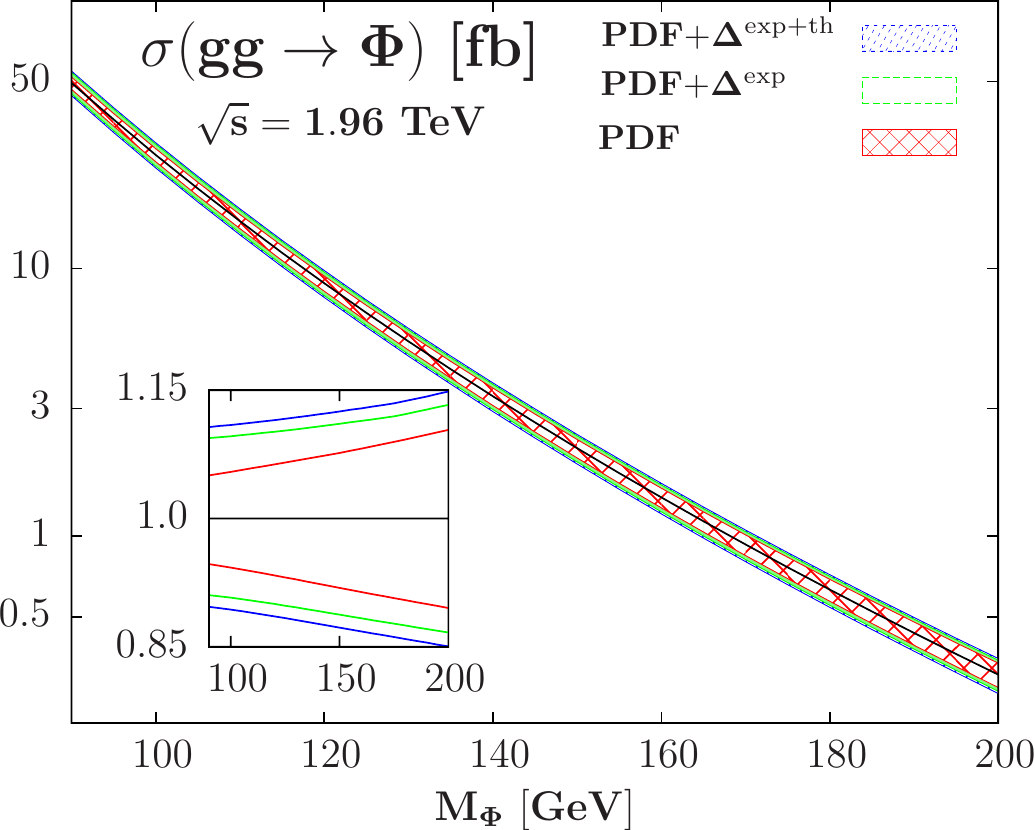}
      \includegraphics[scale=0.65]{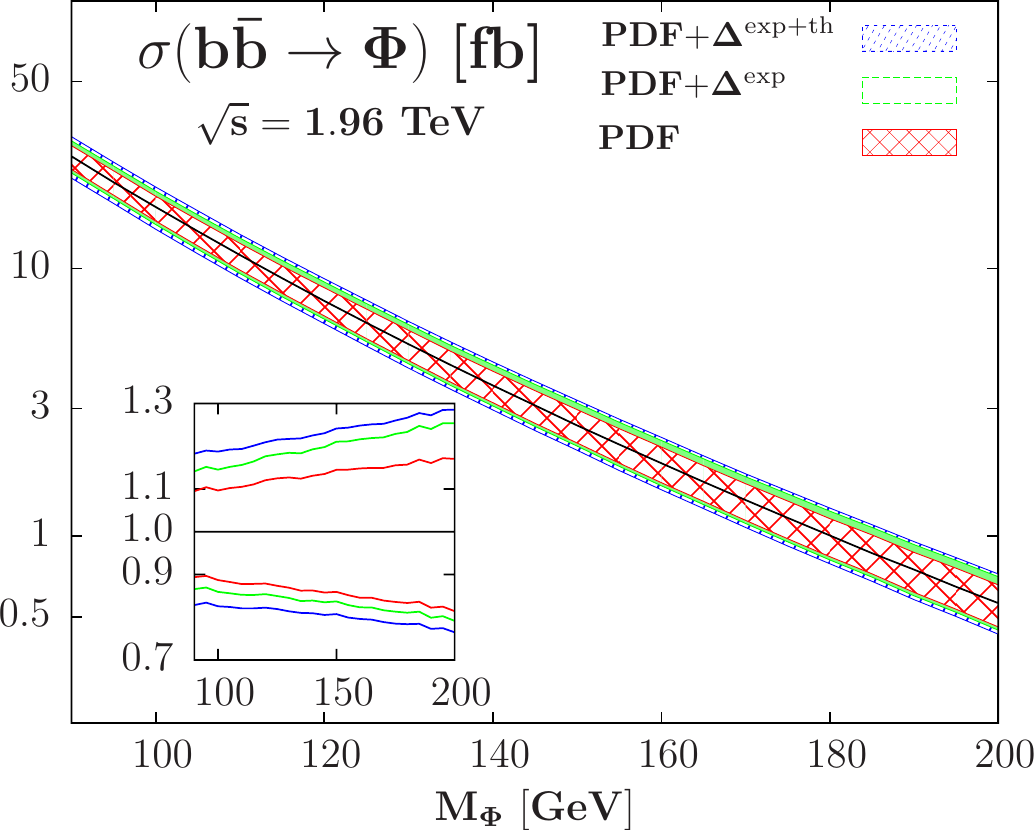}}
  \end{bigcenter}
  \vspace*{-4mm}
  \caption[PDF+$\Delta^{\rm exp,th}\alpha_s$ uncertainty in the $gg\to\Phi$ and
  $bb\to\Phi$ processes at the Tevatron ]{The PDF 90\% CL PDF,
    PDF+$\Delta^{\rm exp} \alpha_s$ and PDF+$\Delta^{\rm exp}\alpha_s
    +\Delta^{\rm th}\alpha_s$ uncertainties in the MSTW scheme in the
    $gg \to \Phi$ (left) and $b\bar b  \to \Phi$ (right) cross
    sections  at the Tevatron as a function of $M_\Phi$. In the
    inserts, the relative deviations are shown.}
  \label{fig:MSSM-PDFTev}
\end{figure}

We now turn our attention to the estimation of the uncertainties from
the parton densities and $\alpha_s$, following stictly the discussion
lead in section~\ref{section:SMHiggsTevPDF} in the case of the SM
Higgs gluon--gluon fusion production channel. The 90\% CL
PDF+$\Delta^{\rm exp}\alpha_s$ uncertainty, with
$\alpha_s(M_Z^2)=0.120 \pm 0.002$ at NLO for $gg \to \Phi$ and
$\alpha_s(M_Z^2)=0.1171 \pm 0.0014$ at 
NNLO for $b\bar b \to \Phi$, is evaluated within the MSTW
parametrization when including the experimental error on
$\alpha_s$~\cite{Martin:2009iq, Martin:2009bu, Martin:2010db}. To
that, we add in quadrature the effect of the
theoretical error on $\alpha_s$, estimated by the MSTW collaboration
to be $\Delta^{\rm th} \alpha_s \approx 0.003$ at NLO and $\Delta^{\rm
  th} \approx 0.002$ at NNLO, using the  MSTW fixed $\alpha_s$ grid
with central PDF sets. The 90\%CL PDF, PDF$+\Delta^{\rm exp}\alpha_s$
and the PDF$+\Delta^{\rm exp+th}\alpha_s$ uncertainties at the Tevatron
are shown in Fig.~\ref{fig:MSSM-PDFTev} as a function of $M_\Phi$. In
the case of the $gg\to\Phi$ process the PDF uncertainty is up to $\pm
10\%$ at high mass $M_\Phi=200$ GeV and the total PDF+$\Delta^{\rm
  exp+th}\alpha_s$ uncertainty is $\pm 11\%$ at low Higgs mass
$M_\Phi\simeq 100$ GeV and $\pm 15\%$ at high Higgs mass $M_\Phi\simeq
200$ GeV. This is then a rather controlled uncertainty, which is quite
different in the case of the $b\bar b\to \Phi$ production
channel. Indeed in the latter case we obtain a total
PDF+$\Delta^{\rm exp+th}\alpha_s$ uncertainty of order $\pm 20\%$ at
low Higgs mass values $M_\Phi\simeq 100$ GeV and up to $+28\%, -24\%$
at $M_\Phi=200$ GeV.

\begin{figure}[!h]
  \begin{bigcenter}
    \vspace*{-1mm}
      \includegraphics[scale=0.75]{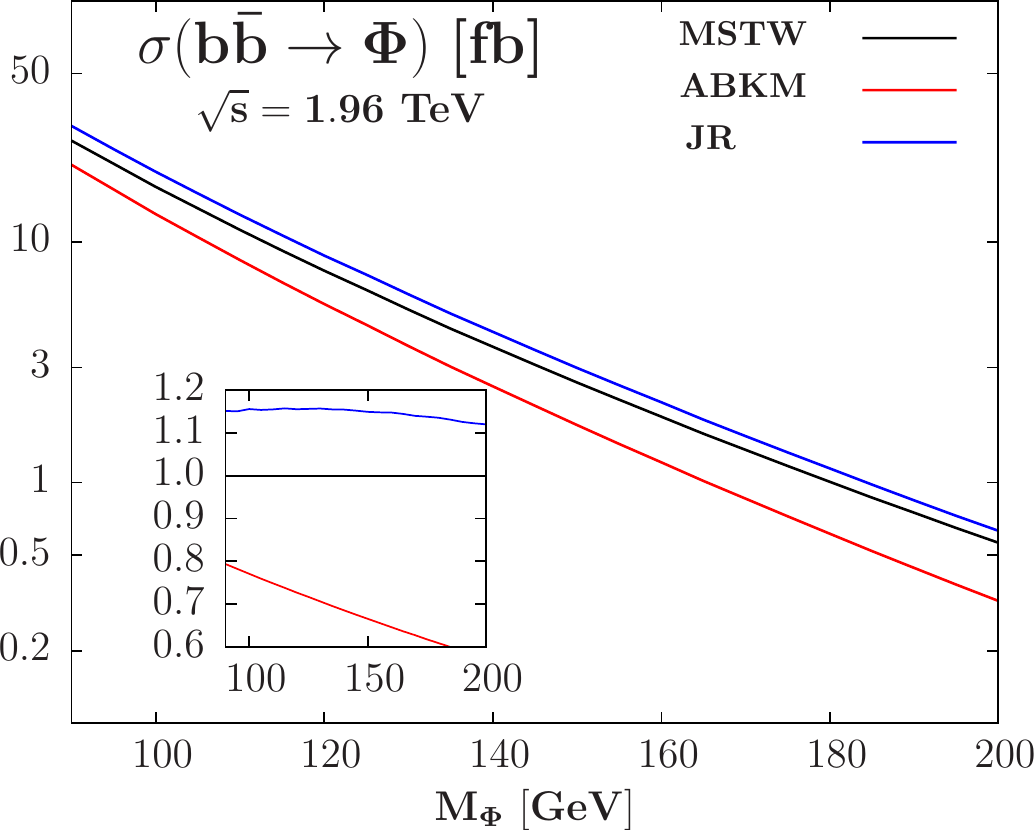}
  \end{bigcenter}
  \vspace*{-4mm}
  \caption[The comparison between the MSTW, ABKM and JR prediction for
  the NNLO bottom quark fusion cross section at the Tevatron]{The
    MSTW, JR09 and ABKM predictions for $\sigma^{\rm NNLO}(b\bar b\to
    \Phi)$ cross section at the Tevatron. In the inserts, the relative
    deviations compared to the central MSTW prediction are shown.}
  \label{fig:MSSM-PDF2Tev}
\end{figure}

This spectacular uncertainty in the case of the bottom quark fusion
channel is in fact not as big if we recall the other way of estimating
the PDF uncertainty, that is the comparison between the central
predictions of the various NNLO PDFs sets on the market. We have
chosen to display in Fig.~\ref{fig:MSSM-PDF2Tev} below the comparison
between the MSTW PDF set~\cite{Martin:2009iq}, the JR09 PDF
set~\cite{JimenezDelgado:2009tv} and the
ABKM PDF set~\cite{Alekhin:2002fv}. The difference between the ABKM
and MSTW predictions is again huge: from $30\%$ at low Higgs mass up
to more than $40\%$ at $M_\Phi = 200$ GeV. In this view the obtained
PDF+$\Delta^{\rm exp+th}\alpha_s$ uncertainty within the MSTW set--up is
acceptable. We have also evalulated the $gg\to \Phi$ sections with
four other PDF sets and found that the maximal  values are obtained
with MSTW PDFs set while some other schemes give $\approx
20\%$ lower rates. The PDF discrepency is then reduced in the case of
the gluon--gluon fusion at the Tevatron in the MSSM.

\subsection{The $b$--quark mass
  uncertainty \label{section:MSSMHiggsTevMb}}

The last piece of uncertainty that remains to be evaluated deals with
the $b$ quark issue. The impact of the $b$ quark mass experimental
error or the choice of the renormalization scheme was only marginal in
the case of the SM Higgs boson both at the Tevatron and the LHC,
mainly because the top loop is by far dominant in this latter
case. There are three different sources of uncertainties related to
the $b$ quark, that are summarized below.

\begin{enumerate}[$\bullet$]
\item{The first one is of purely theoretical nature and is due to the
    choice of the renormalization scheme for the $b$--quark mass. In
    our analysis, we have adopted the $\overline{\rm MS}$
    renormalization scheme for two reasons: first because the natural
    scheme for any SUSY calculation would be the $\overline{\rm DR}$
    scheme which preserves SUSY, a scheme than can be traded in our
    case for the $\overline{\rm MS}$ scheme without too much error as
    stated in the introduction of this section; then because the
    calculation of the $b \bar b \to \Phi$ process is available only
    in this scheme and we to have chosen to treat the same way both
    the $b\bar b \to \Phi$ and  $gg \to \Phi$ channels. Nevertheless,
    one could choose another renormalization scheme such as the
    on--shell scheme as it was discussed in the case of the $b$--loop
    contribution in the $gg$ fusion process for SM Higgs
    production. To estimate this scheme dependence, we evaluate the
    difference of the $gg\to \Phi$ cross section in the cases where
    the $b$--quark mass is defined in the on--shell and in the
    $\overline{\rm  MS}$ schemes. Using the program {\tt HIGLU}, the
    obtained difference amounts to $\approx +5\%$. Bearing in mind
    the fact that the corrections could have been negative if we had
    adopted another scheme (as would have been the case if we have
    used the $\overline{\rm  DR}$ scheme for instance, although the
    difference from the  $\overline{\rm MS}$ result would have been at
    the level of a few percent only), an error $\Delta^{\rm
      scheme}_{m_b} \approx \pm 5\%$ could be assigned to the $gg \to
    \Phi$ cross section. This is the  procedure that we will adopt
    here.

    Note that there is one another way to estimate the renormalization
    scheme dependence of the $b$--quark  mass: it would be to look at
    the differences that one obtains by using $\overline{m}_b(\frac12
    \overline{m}_b)$ and $\overline{m}_b(2\overline{m}_b)$ as inputs
    in the $gg\to \Phi$ cross section\footnote{We should note that the
      scheme dependence actually appears also as a result of the
      truncation of the perturbative series which, in principle,
      should already be accounted for by the scale variation. However,
      the scales which enter in the $b$--quark mass, that is defined
      at $m_b$ itself, and in the rest of the $gg \to \Phi$ matrix
      element, which is the Higgs mass or the scale $\mu_R$, are
      different.  We have checked explictly that the scale
      uncertainties due to the variation of $\mu_R$ (and $\mu_F$)  in
      both schemes are comparable. Adding this scheme dependence to
      the scale variation, as we will do here, is similar in practice
      to increase the domain of scale variation from the central scale
      while sticking to a given mass renormalization scheme. We thank
      Michael Spira for a discussion on this point.}. In this case,
    one obtains an uncertainty that is in fact  much larger than the
    difference between the on--shell and $\overline{\rm  MS}$ schemes
    and which goes both ways, $\Delta^{\rm scheme}_{m_b}  \approx -
    10\%,+30\%$ for $M_\Phi=90$ GeV for instance. 

    In the $b\bar b \to \Phi$ process, we cannot perform this
    exercise  as the cross section, using the program {\tt
      bbh@nnlo} can only be evaluated  in the $\overline{\rm MS}$
    scheme with the Yukawa couplings evaluated at the scale
    $\mu_R$. Nevertheless as stated a few lines above this scheme
    dependance can in this case be related to the scale uncertainty
    itself as in this process we use $\overline{m}_b(\mu_R)$. This is
    the reason why we extended the domain of scale variation in this
    case to $\frac13 \mu_0 \le \mu_R,\mu_F \le 3\mu_0$. The larger
    scale uncertainty obtained this way could be seen as indirectly
    taking care of the scheme dependence.}
\item{The second source of uncertainty is of parametric nature and is
    the same as the one affecting the $H\to b\bar b$ partial decay
    width of the Higgs boson discussed in
    section~\ref{section:SMHiggsDecay}. It is estimated as previously,
    i.e. by evaluating the maximal values of the cross 
    sections when one includes the error on the input $b$--quark mass
    at the scale $\overline{m}_b$, $\overline{m}_b
    (\overline{m}_b)=4.19^{+0.18}_{-0.06}$ GeV\footnote{We use our old
      set--up that was developed before the final stage of the thesis and
      published in Ref.~\cite{Baglio:2010ae}. Note that the final
      uncertainty obtained being quite small, the new choice for $m_b$
      presented in section~\ref{section:SMHiggsDecay} only changes
      marginally the final results, not to mention that this uncertainty
      will cancel out when taking into account the branching fraction, see
      in the following section~\ref{section:MSSMHiggsExp}.}, and in the
    case of the $b\bar b \to \Phi$ process where the Yukawa coupling is
    defined at the high scale, the strong coupling constant,
    $\alpha_s(M_Z^2)=0.1171 \pm 0.0014$ at NNLO, used to run the mass
    $\overline{m}_b (\bar m_b)$ upwards to  $\overline{m}_b (\mu_R)$. In
    the considered Higgs mass range, one obtains an uncertainty of
    $\Delta_{ m_b}^{\rm input} \approx -4\%,+13\%$ and $\approx
    -3\%,+10\%$ in the case of, respectively, the $gg\to \Phi$ and
    $b\bar b\to \Phi$ processes (using the 
    central MSTW PDF set). The difference is mainly due to the fact
    that the bottom quark masses in the two processes are not defined
    at the  same scale and, also, in the case of the $gg \to \Phi$
    process, additional  corrections which involve the $b$--quark
    mass, $\propto \ln(\bar m_b^2/M_\Phi^2)$,  occur at leading order.}
\item{Finally, a third source of uncertainty originates from the
    choice of the $b$--mass value in the $b$--quark densities. The
    MSTW collaboration has released last year a set of PDFs with
    different bottom quark masses \cite{Martin:2010db}: it involves
    six different central PDFs with a range of on--shell $m_b$ values
    between $4.00$ GeV and $5.50$ GeV in $0.25$ GeV steps, in addition
    to  the central best--fit with $m_b=4.75$ GeV. In order to
    distinguish between the parametric uncertainty in $m_b$ and the
    one due to the correlated PDF--$\Delta m_b$ uncertainty, we have
    chosen to calculate the latter uncertainty by taking the minimal
    and maximal values of the production cross sections when using the
    central value $m_b=4.75$ GeV and the two closest ones upwards and
    downwards,  i.e. $m_b=4.5$ GeV and $m_b=5$ GeV in the MSTW PDF
    set\footnote{ Note that when  including the $1\sigma$ errors on
      $\overline{m}_b(\overline{m}_b)$, while the upper value
      corresponds to the pole mass $m_b \approx 5$ GeV, the lower
      value does not correspond to $m_b=4.5$ GeV; we will however
      adopt this smaller value  to estimate the uncertainty as no
      other choice is possible within the MSTW set.}. However, we kept
    in the partonic calculation the central value of
    $\overline{m}_b(\overline{m}_b)=4.19$ GeV which, approximately
    corresponds to the pole mass $m_b=4.75$ GeV. One obtains an
    approximate 7--9\% uncertainty depending on the considered Higgs
    mass range. Note that this uncertainty will not only affect the
    cross section in the $bb\to \Phi$ process in which the
    $b$--densities play the major role, but also the one of the
    $gg\to \Phi$ channel; however, in this case, the change is below
    the percent level and can be safely neglected.}
\end{enumerate}

\begin{figure}[!h]
  \begin{center}
    \vspace*{1mm}
    \mbox{
      \includegraphics[scale=0.65]{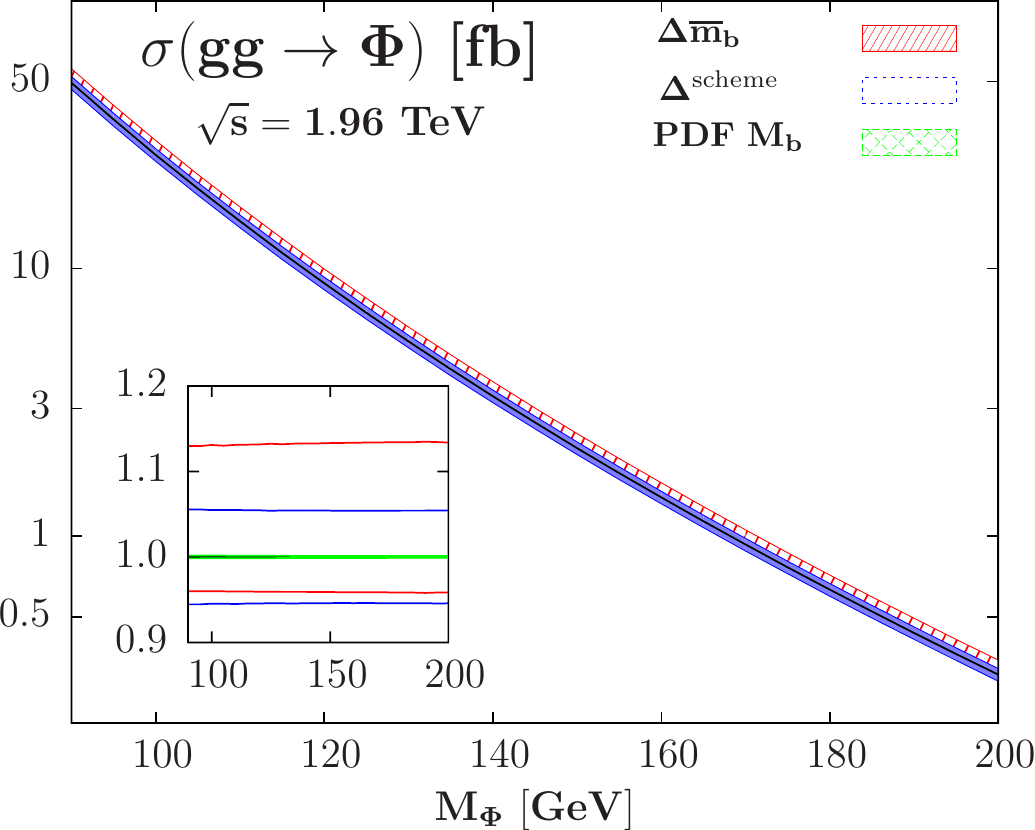}
      \includegraphics[scale=0.65]{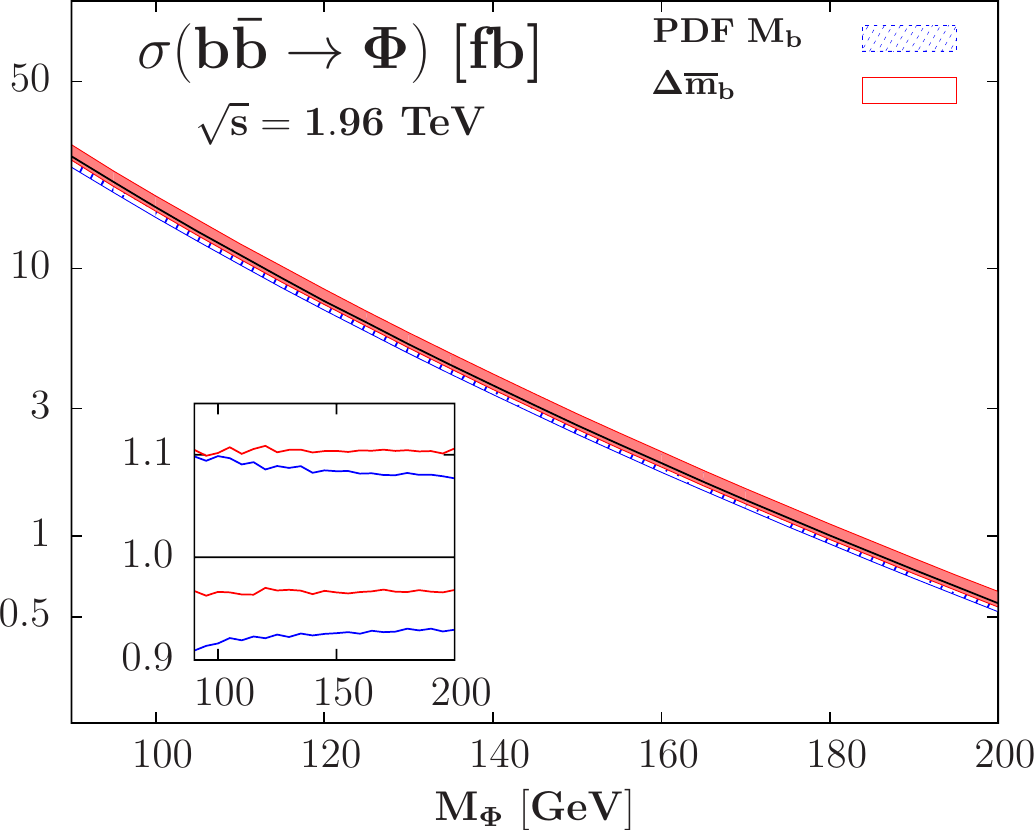}
    }
  \end{center}
  \vspace*{-4mm}
  \caption[Specific $b$--quark mass uncertainties in the $gg\to\Phi$ and
  $b\bar b\to \Phi$ processes at the Tevatron]{The scheme, parametric and PDF
    uncertainties due to the $b$-quark mass in the $gg \to \Phi$
    (left) and $b\bar b  \to \Phi$ (right) cross sections at the
    Tevatron as a function of $M_\Phi$. In the inserts, the
    relative deviations are shown.}
  \label{fig:MSSM-mbTev}
  \vspace*{-3mm}
\end{figure}

The effect of these three sources of uncertainties is displayed  in
Fig.~\ref{fig:MSSM-mbTev} for $gg\!\to\! \Phi$ and $b\bar b\!\to\!
\Phi$ as a function of $M_\Phi$. As can be seen, quite large
uncertainties occur, in particular in $gg\! \to\! \Phi$ where the
scheme uncertainty that is absent in $b \bar  b\! \to\! \Phi$ is of
the same order than the specific $b$ quark mass uncertainty.

\subsection{Summary and combination of the different sources of
  uncertainties\label{section:MSSMHiggsTevTotal}}

\begin{figure}[!h] 
  \begin{center} 
    \mbox{
      \includegraphics[scale=0.65]{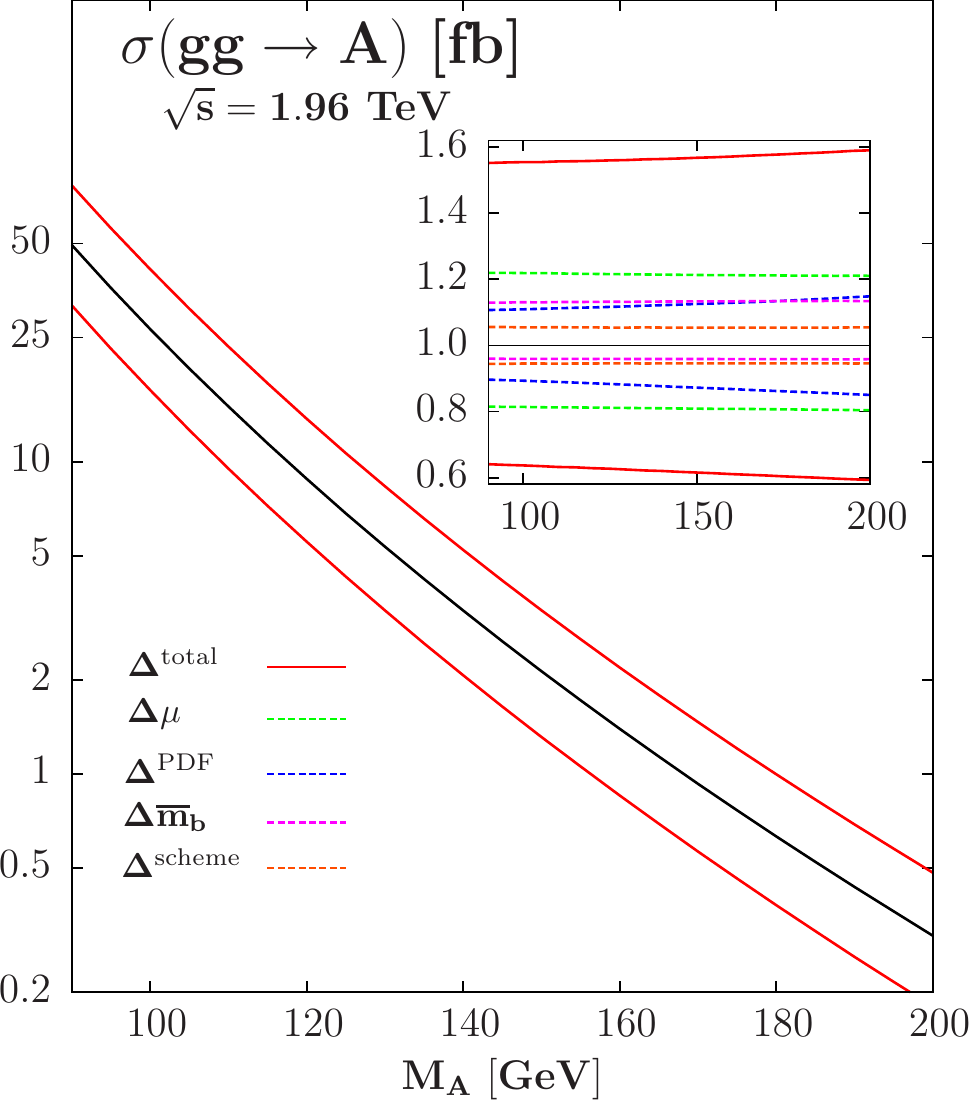}
      \includegraphics[scale=0.65]{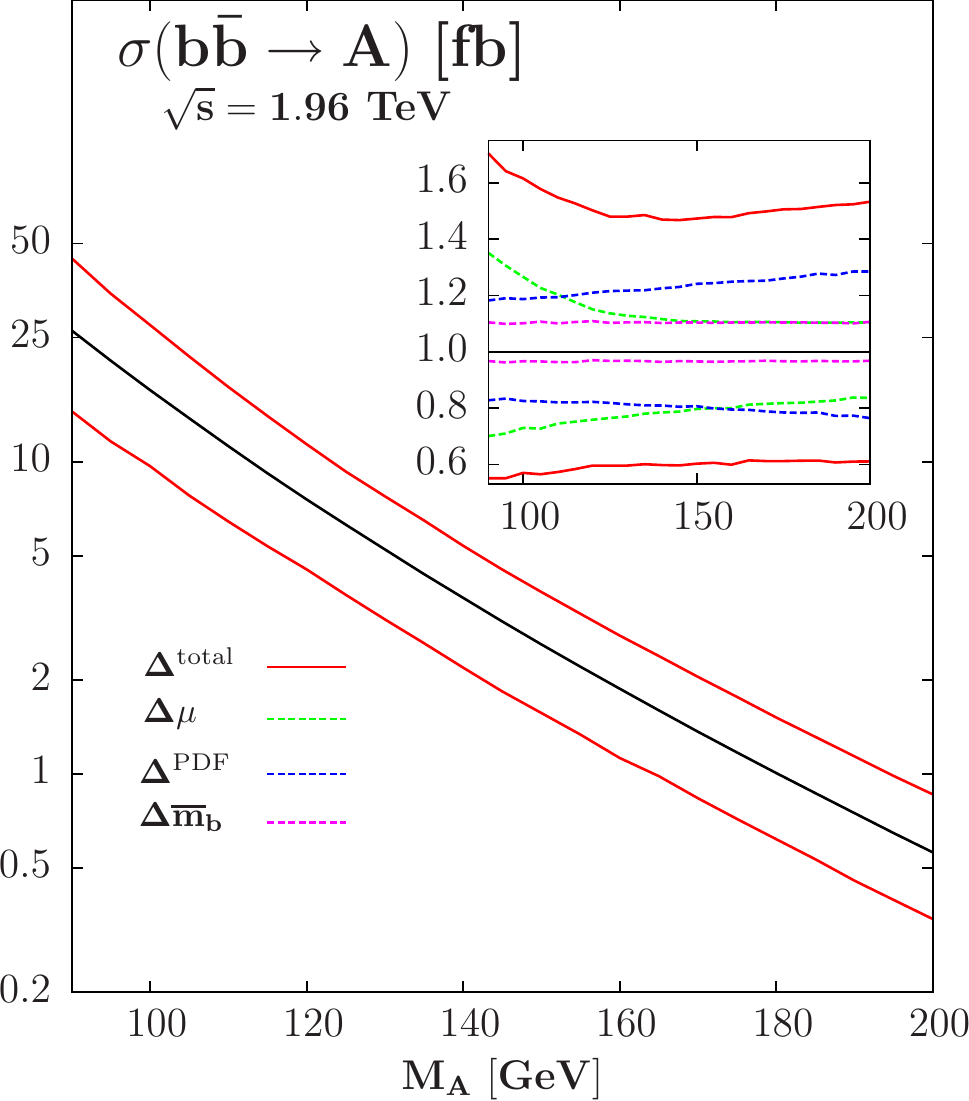}}
  \end{center} 
  \vspace*{-5mm}
  \caption[The $gg\to A$ and $b\bar b\to A$ cross sections at the
  Tevatron together with their different sources of uncertainty and
  the total uncertainties]{The normalization of the cross sections
    $\sigma^{\rm NLO}_{gg\!\to \!A}$ (left) and $\sigma^{\rm
      NNLO}_{b\bar b \!\to\!A}$ (right) at Tevatron energies as a
    function of $M_A$ when using the MSTW PDFs and unit $Ab\bar b$
    couplings together with the total uncertainty. In the inserts,
    shown are the various sources of theoretical uncertainties when
    the rates are normalized to the central values.}
\vspace*{-3mm}
\label{fig:MSSMTotalTev}
\end{figure}

We now summarize and combine the different sources of uncertainties
that have been discussed above and display them in the single
Fig.~\ref{fig:MSSMTotalTev} below where we display the $gg\to\A$ and
$b\bar b\to A$ with $\tb$ set to unity. In the $gg$ case and almost
independently of $M_A$, the scale variation in a domain with
$\kappa=2$ leads to an uncertainty  ${\cal O}(\pm 20\%)$, while the
uncertainty from the scheme dependence in the renormalization of $M_b$
is about $\pm 6\%$; they add up to $\approx 25\%$ that is only
slightly lower than the scale uncertainty in the $b\bar b$  process,
$\approx 30\%$ for low $M_A$,  in which the domain of variation is
extended to $\kappa=3$. In the $gg\to A$ ($b\bar b \to A)$ channel,
the PDF+$\Delta^{\rm exp+th} \alpha_s$ (with $\Delta M_b$ in addition
for $b\bar b \to A$) uncertainties are at the level of $\pm 10\%$
($\pm 20\%$) for $M_A \approx 100$ GeV  and larger ($\pm 30\%$ in
$b\bar b\!\to\! A$) at $M_A \approx 200$ GeV where the more uncertain
high Bjorken--$x$ values for the gluon and bottom quark densities are
probed. The parametric error on $\overline{m}_b$ leads to a $\approx
+13\%, -4\%$ uncertainty in the $gg\to A$ process and slightly less in
the case of $b\bar b \to A$.

How to combine these uncertainties together in the final total
theoretical uncertainty? We will follow the same set--up developed in
part~\ref{part:two} in the case of the SM Higgs production. To begin
with the scheme and scale uncertainties being pure theoretical
uncertainties they add up linearly. We then follow our recipe
developed for the first time in Ref.~\cite{Baglio:2010ae} to deal with
the combination of the scale/scheme and PDF+$\Delta \alpha_s$+$\Delta
m_b$ (the last one refers to the uncertainties related to the choice
of the $b$ quark mass within the PDFs): we would like to view these
uncertainties as a way of handling the theoretical ambiguities due to
the parametrization of the PDF and we will evaluate them on the
minimal and maximal values with respect to the scale variation. We
then add linearly in the end the parametric uncertainties on the $b$
quark mass. This last source of uncertainty will be discussed again in
the following section when dealing with the combination with branching
fractions.

The combined uncertainties on the cross sections, when using this procedure,
are also shown in Fig.~\ref{fig:MSSMTotalTev} for the two production
channels. As shown they are very large: at low Higgs masses,
$M_A\approx 100$ GeV, we obtain $\approx +55\%,-35\%$ for  $\sigma( gg
\to A)$   and $\approx +60\%,-40\%$ for $\sigma(b\bar b \to A)$ which
become at  masses $M_A\approx 200$ GeV, respectively, $\approx
+60\%,-40\%$ and $\approx +50\%,-40\%$.

We have ended the study of the main neutral MSSM Higgs bosons
production channels at the Tevatron. We will now make the same
exercice at the lHC collider, and leave the display of the table of
results for the final section~\ref{section:MSSMHiggsExp} where the
combination with the branching fractions will also be presented.

\newpage

\vfill
\pagebreak

\section{MSSM Higgs production at the LHC}

\label{section:MSSMHiggsLHC}

We now turn our attention to the study of the MSSM Higgs production at
the CERN LHC collider. The LHC started its operations at 7 TeV in 2010
and has already provided interesting results regarding to MSSM
searches, see Ref.~\cite{daCosta:2011qk, Khachatryan:2011tk}. The
quest for MSSM Higgs bosons has also started and some results have
already been published~\cite{Collaboration:2011rv,
  Chatrchyan:2011nx}. In the final stage of the writing of the thesis
new results have been presented at HEP-EPS 2011 conference, but they
will not be commented before section~\ref{section:MSSMHiggsExp} and
the final conclusion of this thesis.

We will reproduce the same outines of
section~\ref{section:MSSMHiggsTev} and use the same set--up, that
is neglecting the genuine SUSY corrections as justified in
section~\ref{section:MSSMHiggsIntroModel}, using SM--like couplings
for the $A$ boson and
conducting our study in the case of the maximal mixing scenario where
the upper bound on the mass of the lightest $h$ boson is shifted from
the tree level value $M_Z$ to the value $M_h^{\rm max} \sim 110$--135
GeV~\cite{Heinemeyer:2004gx, Heinemeyer:2004ms, Allanach:2004rh}, the
$H$ boson being almost degenerate with the $A$ boson and having the
same properties, or the no--mixing scenario where the role of the $h$
boson and the $H$ boson is exchanged. Ref.~\cite{Carena:2002qg} gives
widely used benchmark scenarios that fall into theses two regimes. We
will give the predictions for the bottom quarks fusion and gluon--gluon
fusion production channels at the lHC together with a detailed study
of the various theoretical uncertainties affecting the calculation:
the unknown higher order corrections evaluated by the variation of the
renormalization and factorization scales, the impact of the PDF and
the associated uncertainty related to the value of the strong coupling
constant $\alpha_s$ and finally the uncertainties due to the $b$ quark
mass, its experimental value together with its renormalization scheme
choice. We will see in the next section the impact of these
uncertainties on the $[M_A,\tb]$ parameter space studied by the ATLAS
and CMS experiments. All of the material was published in
Refs.~\cite{Baglio:2010ae, Baglio:2011xz}.

\subsection{Gluon--gluon fusion and bottom quarks
  fusion channels \label{section:MSSMHiggsLHCCross}}

As in the Tevatron case the two most important channels are the bottom
quarks fusion and the gluon--gluon fusion at the LHC. As stated before,
$CP$ invariance forbids $A$ couplings to gauge
bosons  at tree--level: the pseudoscalar $A$ boson cannot be produced in the
Higgs-strahlung and vector boson fusion processes. We concentrate on
the high $\tb$ regime that is relevant at the lHC which means that the
$b$--quark will play the major role as its couplings to the $CP$--odd
like Higgs bosons are enhanced.  

We remind that in the $gg \to \Phi$ processes with only the $b$--quark
loop included, the QCD corrections are known only to  NLO for which
the exact calculation with finite loop quark masses is
available~\cite{Spira:1995rr}. We will again use the central scale
$\mu_R=\mu_F=\mu_0=\frac12 M_H$ to approach the SM case which is valid
for the SM--like $CP$--even Higgs boson, thus having a consistent
set--up for the three neutral Higgs bosons. In the case of the $pp \to
b\bar b \Phi$ processes, the NLO QCD corrections have been calculated
in Ref.~\cite{Dittmaier:2003ej, Dawson:2003kb}. We will again use the
five flavor scheme in which the bottom quarks are directly taken from
the proton sea and the LO partonic process being $b\bar b\to
\Phi$\cite{Dicus:1988cx}. The NNLO corrections are of moderate size
when using the central scale $\mu_F=\mu_R=\mu_0= \frac14 M_\Phi$ as
recommanded in Ref.~\cite{Harlander:2003ai} and the bottom quark mass
in the $\overline{\rm MS}$ renormalization scheme.

In order to evaluate the Higgs production cross sections at LHC energies in
these two main processes, $gg \to \Phi$ and $b\bar b \to \Phi$, we
will proceed as follows. We only evaluate the cross sections  for the
pseudoscalar $A$ boson: in the $gg \to A$ process at NLO using the
program {\tt HIGLU}~\cite{Spira:1995mt, Spirapage} with a central
scale  $\mu_R=\mu_F=\mu_0= \frac12 M_A$ (and where only the bottom
quark loop contribution is included by setting $\lambda_{A tt}=0$),
and in the $b\bar b\to A$ process up to NNLO using again the program
{\tt bbh@nnlo}  \cite{Harlander:2003ai} with a central scale
$\mu_R=\mu_F=\mu_0= \frac14 M_A$. In both cases, we work in the
$\overline{\rm MS}$ scheme for the renormalization of the
bottom--quark mass. However, while $\overline{m}_b(\overline{m}_b)$ is
used in the $gg$ fusion process, $\overline{m}_b(\mu_R)$ is adopted in
the $b\bar b$ fusion channel. In both processes, we assume the $Ab\bar
b$ coupling to be SM--like, that is, we will not include the $\tb$
term and the SUSY corrections to $\Delta m_b$. To obtain the actual
cross section for both $CP$--even and $CP$--odd Higgs production, a
factor of $2\tan^2\beta$ has to be included. As a consequence of
chiral symmetry for $M_\Phi\! \gg\!\overline{m}_b$ and since the
$H\;(h)$ masses and couplings are very close to those of $A$, this
turns out to be an excellent approximation as stated in
section~\ref{section:MSSMHiggsIntroModel}.

\begin{figure}[!h]
  \begin{center}
    \mbox{
      \includegraphics[scale=0.65]{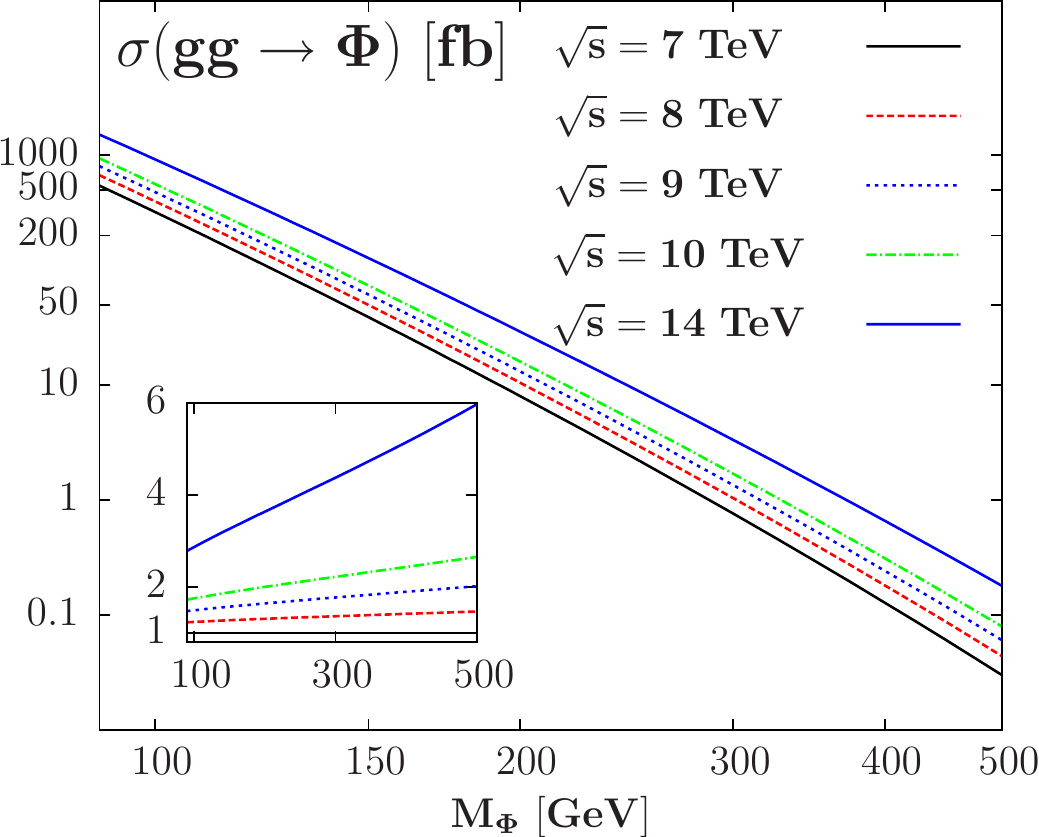}
      \includegraphics[scale=0.65]{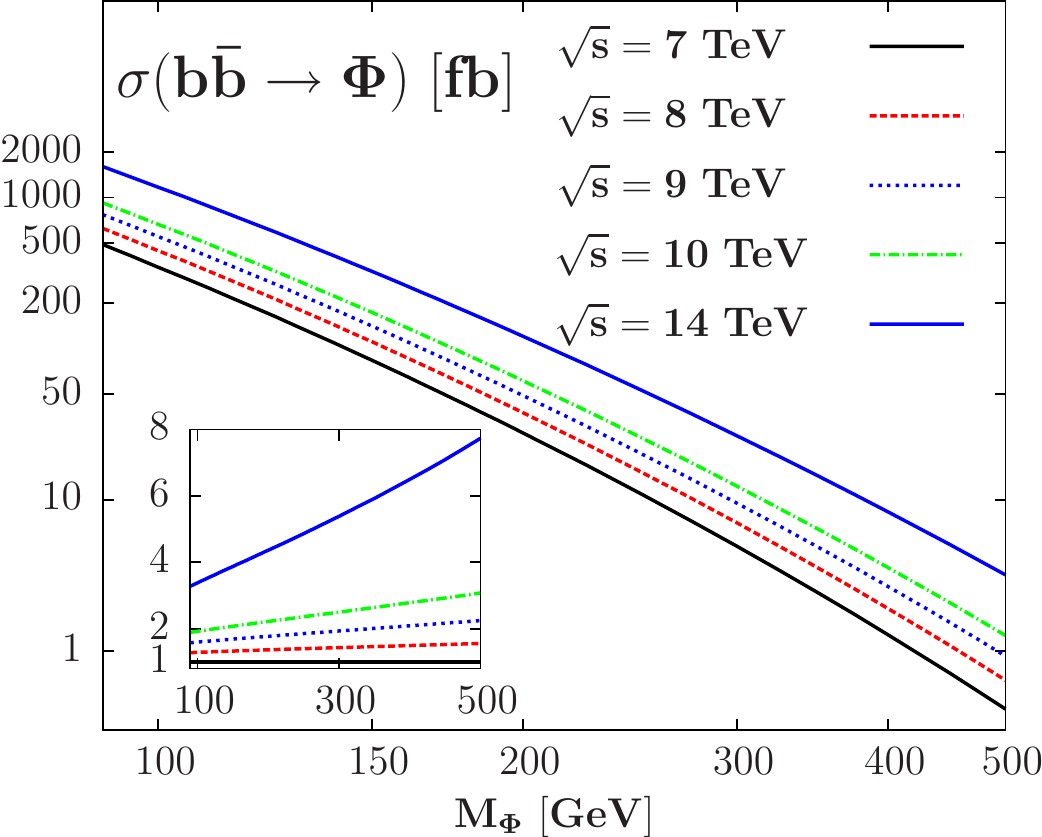}
    }
  \end{center}
  \vspace*{-4mm}
  \caption[The $gg\to \Phi$ and $b\bar b\to \Phi$ at the LHC for
  different center--of--mass energies]{The production cross sections
    in the processes $gg \to \Phi$ (left)  at a central scale
    $\mu_R=\mu_F=\mu_0= \frac12 M_\Phi$ and $b\bar b  \to \Phi$
    (right) at a central scale $\mu_R=\mu_F=\mu_0= \frac14 M_\Phi$ as
    a function of $M_\Phi$ for several center of mass energies
    relevant for the LHC. The MSTW sets of PDFs at the required
    perturbative order have been used. Only the cross section in the
    pseudoscalar $A$ case but with a SM-like Yukawa coupling is
    included.}
  \label{fig:MSSM-xs}
  \vspace*{-.6mm}
\end{figure}

Within this set--up, the best values of the cross sections for $gg \to \Phi$
and  $b\bar b \to \Phi$ are shown in Fig.~\ref{fig:MSSM-xs} as a
function of the Higgs mass $M_\Phi$ when the MSTW sets of (NLO for the
former and NNLO for the latter process) PDFs are used to parametrize
the gluon and bottom--quark densities. Center of mass energies in a
range between $\sqrt s=7$ TeV and 14 TeV relevant for the LHC are
considered. It can be noticed that the cross sections for $gg \to
\Phi$ and  $b\bar b \to \Phi$ are comparable at a given energy and
they significantly increase with increasing center of mass  energies
or decreasing Higgs mass. If, for example, the value $\tb=10$ is
adopted, the numbers in Fig.~\ref{fig:MSSM-xs} have to be multiplied
by a factor $\simeq 200$ to obtain the true cross section for both $A$
and $H(h)$ production. For low to moderate $M_\Phi$ values, the
expected event rates are thus simply huge at the lHC, despite of the
relatively low luminosities that are expected. This explains why the
chances for observing a Higgs boson at the $\lhc$ are much higher in
the MSSM than in the SM, as in the former case the production rates
can be two to three orders of magnitude larger.

We are left to evaluate the theoretical uncertainties on the production cross
sections and, for this purpose, we will follow very closely the
procedure developed for the SM Higgs boson and in particular in the
end use the procedure A) for the total uncertainty. For the numerical
analysis, we will only consider the case of the $\lhc$ at $\sqrt s=7$
TeV: the uncertainties at center of mass  energies slightly above this
value, $\sqrt s=8$--10 TeV, and even for the LHC energy $\sqrt s=14$
TeV are expected to be comparable. We will start by the scale
uncertainty, followed by the estimation of the PDF and $\alpha_s$
effects and end with the $b$ quark issue, strictly following the
procedure developed in the case of the Tevatron\footnote{Note
  that since in our analysis we are not considering the SUSY particle
  contributions and focus only on the standard QCD effects, additional
  uncertainties from the SUSY sector should, in principle, also be
  present. Nevertheless, at high $\tb$, the genuine SUSY contributions
  should not be significant and can be ignored, while the corrections
  entering in $\Delta_b$ will almost cancel out when the $\tau^+
  \tau^-$ decays are considered. See
  section~\ref{section:MSSMHiggsIntroModel} for more details which
  justify this approach.}.

\subsection{The scale uncertainty at the
  lHC \label{section:MSSMHiggsLHCScale}}

In the case of the $gg\to \Phi$ process, as for the SM Higgs boson, the scale
uncertainty is evaluated by allowing for a variation of the renormalization and
factorization scales within a factor of two around the central scale, $\frac12
\mu_0 \le \mu_R,\mu_F  \le 2 \mu_0$ with $\mu_0= \frac12 M_\Phi$.  For the
$b\bar b  \to \Phi$ case, we will extend the domain of scale variation to a
factor of three around the central scale $\mu_0= \frac14 M_\Phi$, $\frac13
\mu_0 \le \mu_R,\mu_F  \le 3 \mu_0$ but with the additional
restriction $1/\kappa \le \mu_R/\mu_F \le \kappa$ imposed. We then
follow strictly the procedure adopted in the case of the Tevatron in
the previous subsection~\ref{section:MSSMHiggsTevScale}. Again, as to
illustrate the much larger scale uncertainty that is possible in the $b \bar b
\to \Phi$ case, we will also show results when this constraint is
relaxed as also displayed in Ref.~\cite{Dittmaier:2011ti}.

\begin{figure}[!h]
  \begin{center}
    \vspace*{-2mm}
    \mbox{
      \includegraphics[scale=0.65]{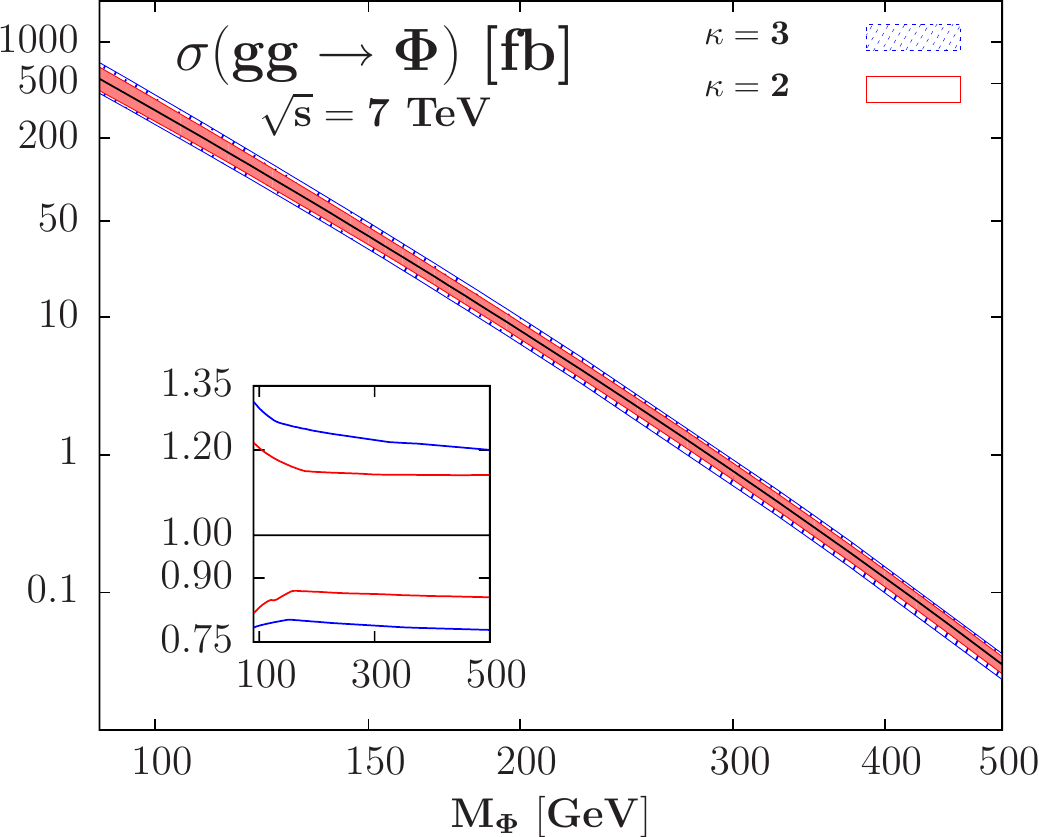}
      \includegraphics[scale=0.65]{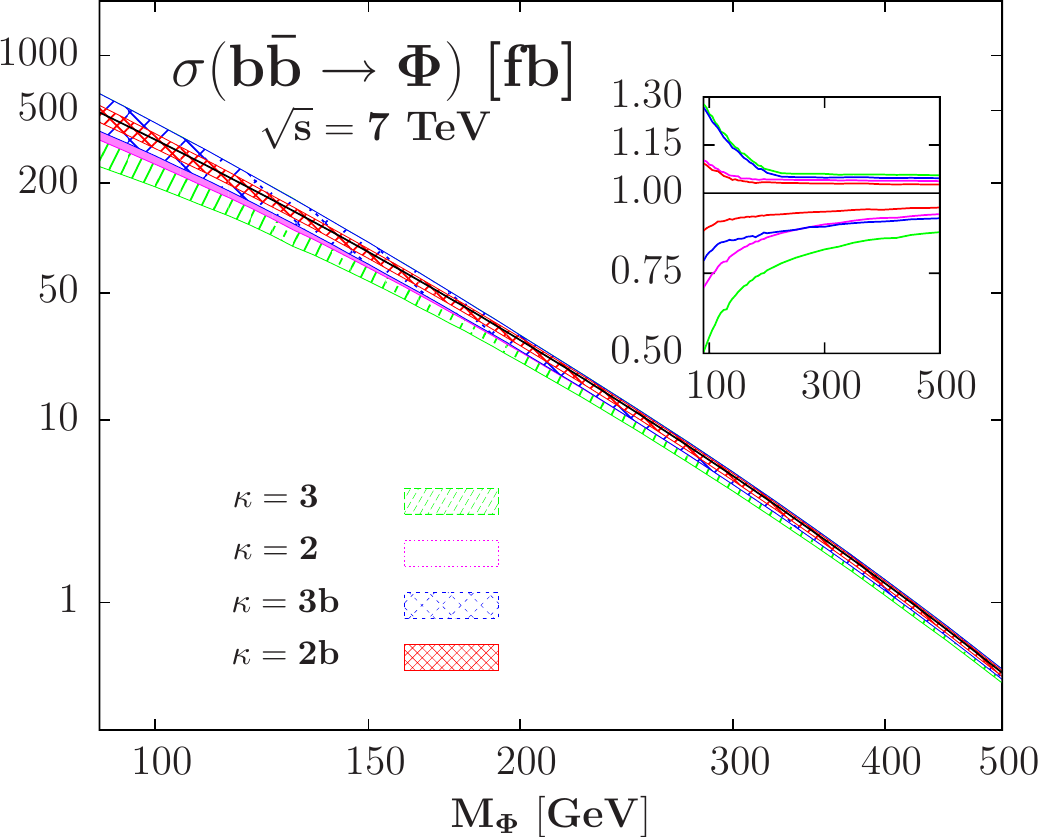}
    }
  \end{center}
  \vspace*{-4mm}
  \caption[Scale uncertainty in the $gg\to \Phi$ and $b\bar b\to \Phi$
  processes at the lHC]{The scale uncertainty bands of the NLO $gg \to
    \Phi$ (left) and the NNLO $b\bar b  \to \Phi$ (right) cross
    sections  at the $\lhc$ at 7 TeV  as a function of $M_\Phi$;
    different values $\kappa=2,3$ are used and the results are shown
    when the additional constraint $1/\kappa \le \mu_R/\mu_F \le
    \kappa$ is imposed or not (marked as $\kappa$b). In the inserts,
    the relative deviations (compared to the central cross section
    values) are shown.}
  \label{fig:MSSM-scaleLHC}
  \vspace*{-2mm}
\end{figure}

The results for $\sigma(gg\to \Phi)$ and $\sigma(b\bar b\to \Phi)$ at
the lHC, for scale variations in the domains $ \mu_0/\kappa \le
\mu_R,\mu_F  \le \kappa \mu_0$ with $\kappa=2$ and 3 are shown in
Fig.~\ref{fig:MSSM-scaleLHC}  as a function of $M_\Phi$. One can see that the
scale variation is moderate for $gg \to \Phi$ with $\kappa=2$, despite
of the fact that the process is known only at NLO, leading to an
uncertainty  of order $\pm 10\%$ in the entire  Higgs mass
range. Extending the variation domain to $\kappa=3$ will increase the
uncertainty by another $\approx 10\%$. As in the SM Higgs case, the
maximal and minimal values of the cross sections are approximately
obtained for $\mu_R \approx \mu_F$ and thus, varying independently the
two scale does not affect the uncertainty. For $\kappa=2$ when the
restriction $\frac12 \le \mu_R/\mu_F  \le 2$ is imposed,  the
uncertainty is also small in the case of  $b \bar b \to \Phi$,
$\approx \pm 10\%$ at low Higgs masses and less for higher masses;
this was to be expected as the process is evaluated at NNLO. However,
the  extension to $\kappa=3$,  while keeping the restriction $\frac13
\le \mu_R/\mu_F  \le 3$, will significantly increase the uncertainty:
one would have $-18\%,+24\%$ at $M_\Phi=100$ GeV and $-13\%,+6\%$ at
$M_\Phi=200$ GeV. If the restriction on  $\mu_R/\mu_F$ is ignored the
uncertainty blows up, especially for very low Higgs masses: one would
have a variation of $-45\%,+25\%$ at $M_\Phi=100$ GeV, as also noticed
in Ref.~\cite{Dittmaier:2011ti}. 

Hence, the cross section for $b \bar b \to \Phi$ is rather unstable against
scale variation and this justifies, a posteriori, the choice of a larger domain
of variation with $\kappa=3$ in this case, a choice that does not  appear  to
be a too extreme  one when looking at Fig.~\ref{fig:MSSM-scaleLHC}.

\subsection{The PDF and $\alpha_S$
  uncertainties at the lHC\label{section:MSSMHiggsLHCPDF}}

\begin{figure}[!h]
  \begin{bigcenter}
    \vspace*{-1mm}
    \mbox{
      \includegraphics[scale=0.65]{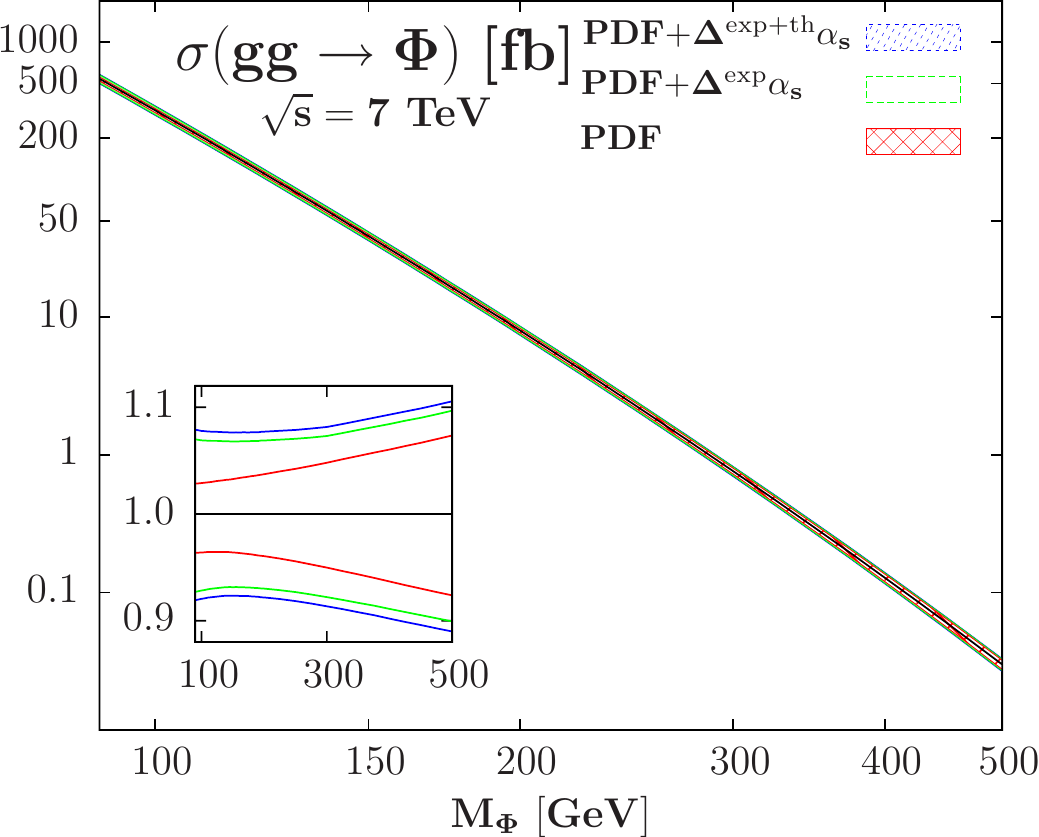}
      \includegraphics[scale=0.65]{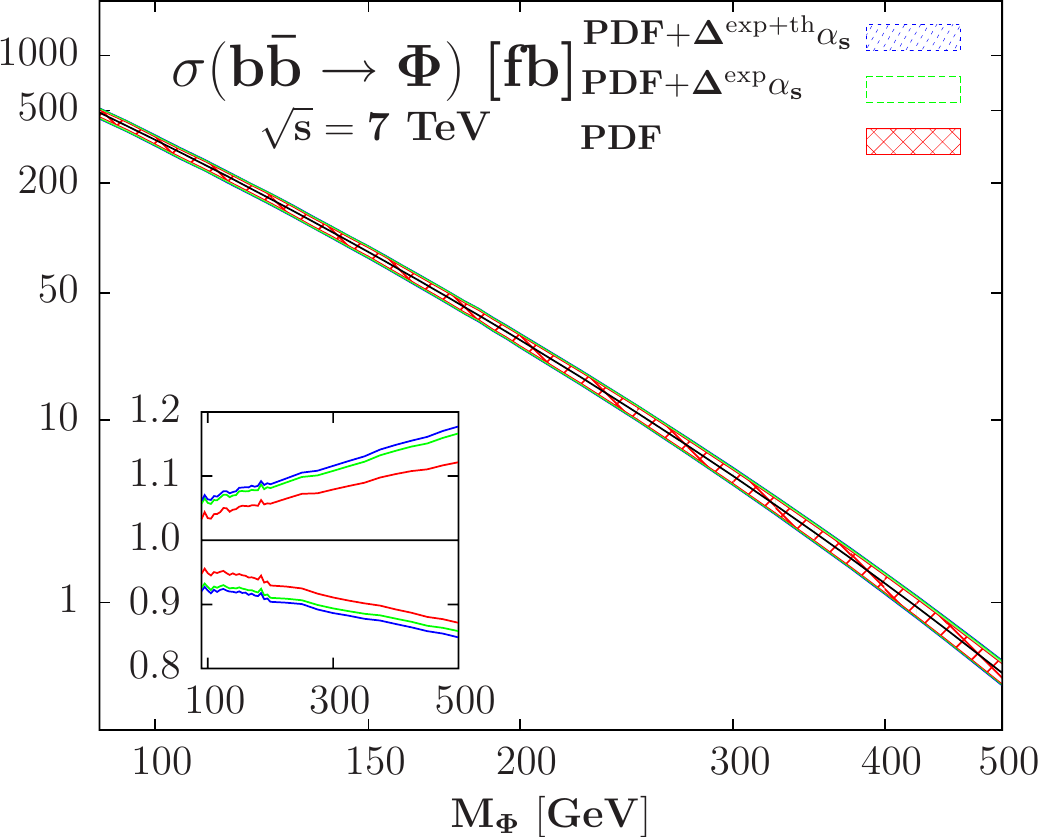}   }
  \end{bigcenter}
  \vspace*{-4mm}
  \caption[PDF+$\Delta\alpha_s$ uncertainty in the $gg\to\Phi$ and
  $bb\to\Phi$ processes at the lHC]{The PDF 90\% CL PDF,
    PDF+$\Delta^{\rm exp} \alpha_s$ and PDF+$\Delta^{\rm exp}\alpha_s
    +\Delta^{\rm th}\alpha_s$ uncertainties in the MSTW scheme in the
    $gg \to \Phi$ (left) and $b\bar b  \to \Phi$ (right) cross
    sections  at the $\lhc$ at 7 TeV as a function of $M_\Phi$. In the
    inserts, the relative deviations are shown.}
  \label{fig:MSSM-PDF}
\end{figure}

Let us now turn to the estimation of the uncertainties from the parton 
densities and $\alpha_s$. The 90\% CL PDF+$\Delta^{\rm exp}\alpha_s$ 
uncertainty, with  $\alpha_s(M_Z^2)=0.120 \pm 0.002$ at NLO for $gg \to \Phi$
and  $\alpha_s(M_Z^2)=0.1171 \pm 0.0014$ at NNLO for $b\bar b \to \Phi$,  is
evaluated within the MSTW parametrization when including the experimental
error on $\alpha_s$. To that, we add in quadrature the effect of the
theoretical error on $\alpha_s$, estimated by the MSTW collaboration to be
$\Delta^{\rm th} \alpha_s \approx 0.003$ at NLO and $\Delta^{\rm th} \approx
0.002$ at NNLO, using the  MSTW fixed $\alpha_s$ grid with central PDF sets. 
The 90\%CL PDF, PDF$+\Delta^{\rm exp}\alpha_s$ and the PDF$+\Delta^{\rm
exp+th}\alpha_s$ uncertainties at the $\lhc$ are shown in  Fig.~\ref{fig:MSSM-PDF}
as a function of $M_\Phi$. In both the $gg\to \Phi$ and $b\bar b\to \Phi$
processes, the total  uncertainty is below  $\pm 10\%$ for $M_\Phi \lsim 200$
GeV but increases at higher masses, in particular in the $b\bar b\to \Phi$ case.

\begin{figure}[!h]
  \begin{bigcenter}
    \vspace*{-1mm}
    \mbox{
      \includegraphics[scale=0.65]{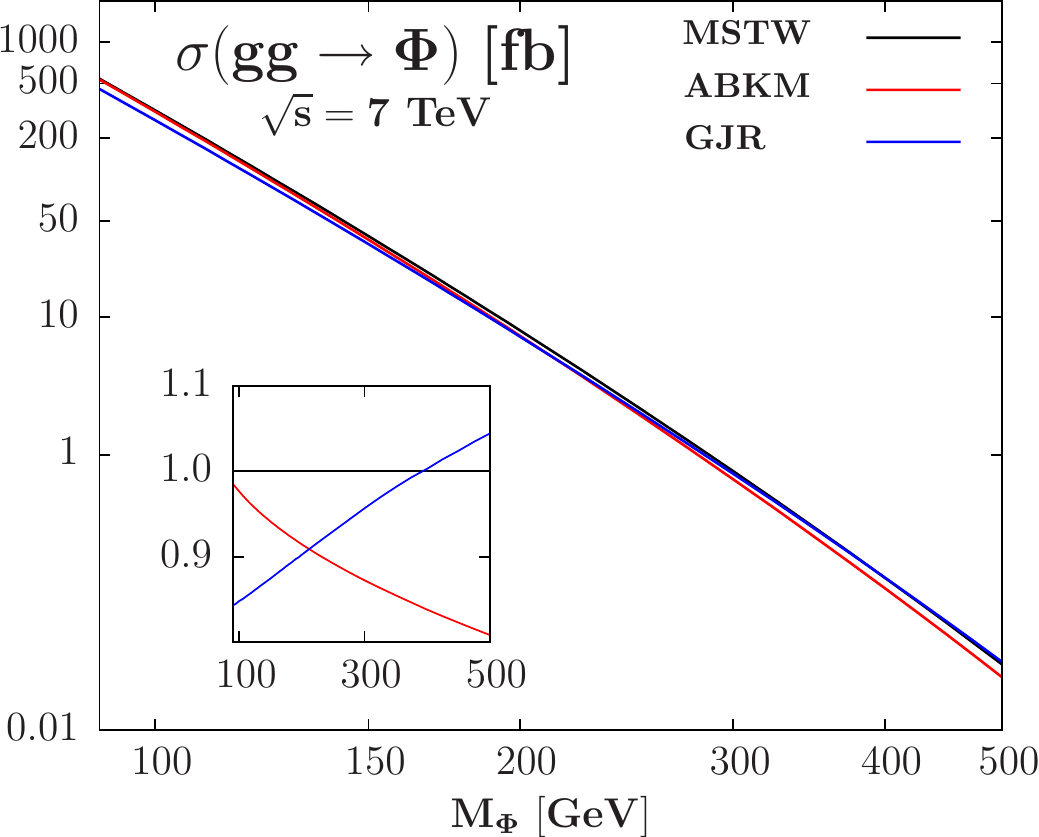}
      \includegraphics[scale=0.65]{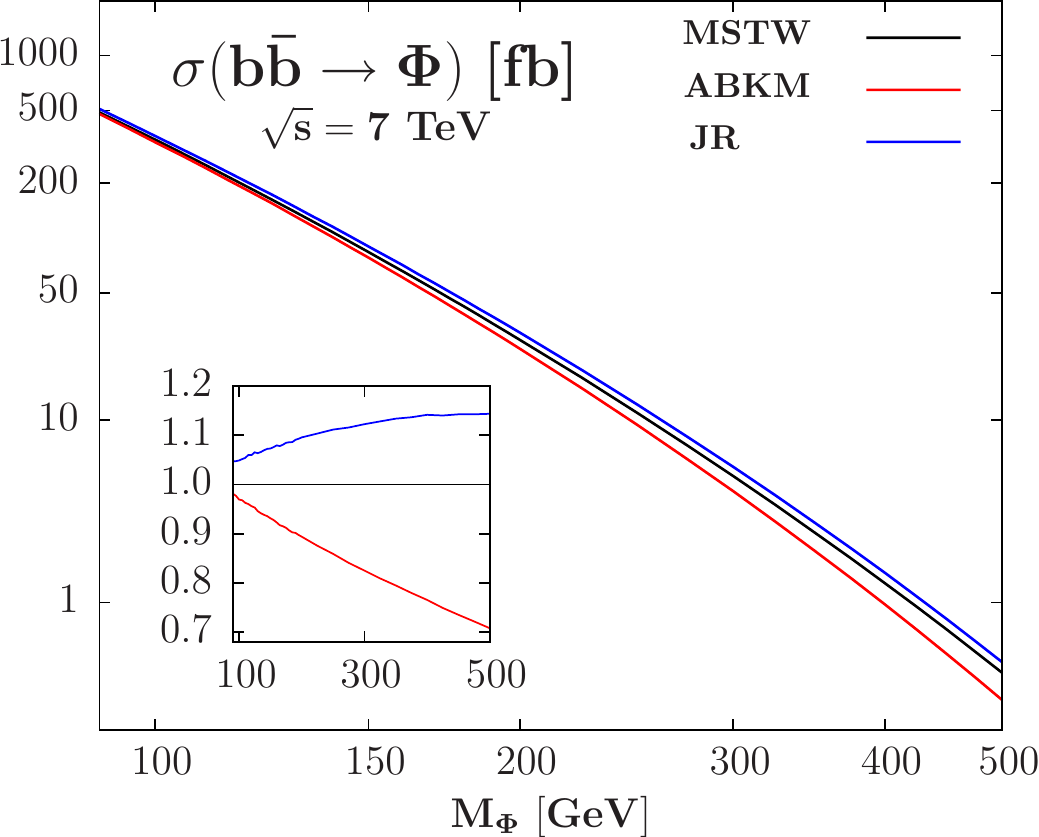}
    }
  \end{bigcenter}
  \vspace*{-4mm}
  \caption[Comparison between the different PDFs sets in the
  $gg\to\Phi$ and $b\bar b\to \Phi$ processes at the lHC]{The cross
    sections in the $gg \to \Phi$ (left) and $b\bar b \to \Phi$
    (right)   at the $\lhc$ at 7 TeV as a function of $M_\Phi$
    evaluated when using the ABKM and (G)JR PDF sets. In the inserts,
    the relative deviations from the value in the MSTW scheme are
    shown.}
  \label{fig:MSSM-PDF2}
  \vspace*{-3mm}
\end{figure}

For completeness, we also display the two cross sections when ones adopts 
two other PDF sets, ABKM and (G)JR, and compare
the results with that of MSTW. As can be seen in
Fig.~\ref{fig:MSSM-PDF2}, the deviations from the MSTW values are
moderate in the case of $b\bar b\to \Phi$ for the JR scheme: a few
percent at $M_\Phi=100$ GeV, increasing to $\approx 10\%$ at
$M_\Phi=500$ GeV. The ABKM scheme leads to substantial deviations at
high masses, $\approx 30\%$ at $M_\Phi = 500$ GeV. In the $gg$ fusion
process, the prediction with the GJR PDF set at $M_\Phi=100$ GeV is
$\approx 20\%$ lower than in the MSTW and ABKM  cases which give
comparable results. The GJR and ABKM parameterizations cross at
$\approx 200$ GeV and at higher masses, the $gg$ cross sections with
the ABKM parametrization are $\approx 20\%$ lower than for MSTW results
while the predictions with the GJR set are $\approx 10\%$ higher than
for MSTW.

\subsection{The $b$--quark mass issue \label{section:MSSMHiggsLHCMb}}

Finally, there is the effect of the uncertainty on the $b$--quark mass
that was discussed in the case of the Tevatron in
subsection~\ref{section:MSSMHiggsTevMb}. We reproduce the same
outlines in this subsection and summarize the three different sources
of uncertainties related to the $b$ quark mass.

The first one is of purely theoretical nature and is due to the choice of the
renormalization scheme for the $b$--quark mass. As in the Tevatron
case we have adopted the $\overline{\rm  MS}$ renormalization scheme
while other scheme could have been chosen. In order to estimate the
effect of the other scheme we calculate the difference between
on--shell and $\overline{\rm MS}$ scheme predictions in the case of
the gluon--gluon fusion process, and we include this scheme dependence
in the scale uncertainty in the case of the bottom quarks fusion
production chanel. Using the program {\tt HIGLU}, we obtain a $\approx
+15\%$ difference in the $gg\to\Phi$ cross section\footnote{Using the
  second way of estimating this scheme uncertainty developed in
  subsection~\ref{section:MSSMHiggsTevMb} we obtain a bigger
  uncertainty than the difference between the on--shell and
  $\overline{\rm  MS}$ schemes and which goes both ways, $\Delta^{\rm
    scheme}_{m_b}  \approx - 20\%,+40\%$ for $M_\Phi=90$ GeV for
  instance.}. Bearing in mind the fact that the corrections could have
been negative if we had adopted another scheme we assign an
uncertainty $\Delta^{\rm scheme}_{m_b} \approx \pm 15\%$  to the $gg
\to \Phi$ cross section.

The second source of uncertainty is due to the experimental errors on
the value of the $b$ quark mass $\overline{m}_b$, $\overline{m}_b
(\overline{m}_b)=4.19^{+0.18}_{-0.06}$ GeV\footnote{The same remark
  developed in the Tevatron study also applies in the LHC case, see
  subsection~\ref{section:MSSMHiggsTevMb}.}, and  in the case of the
$b\bar b \to \Phi$ process where the Yukawa coupling is defined at the
high scale, the strong  coupling constant, $\alpha_s(M_Z^2)=0.1171 \pm
0.0014$ at NNLO, used to run the mass  $\overline{m}_b (\bar m_b)$
upwards to  $\overline{m}_b (\mu_R)$. In the considered Higgs mass
range, one obtains an uncertainty of $\Delta_{ m_b}^{\rm input}
\approx -4\%,+14\%$ and $\approx -3\%,+10\%$ in the case of,
respectively, the $gg\to \Phi$ and $b\bar b\to \Phi$ processes (using
the central MSTW PDF set). The difference is mainly due to the fact
that the bottom quark masses in the two processes are not defined at
the  same scale and, also, in the case of the $gg \to \Phi$ process,
additional corrections which involve the $b$--quark mass, $\propto
\ln(\bar m_b^2/M_\Phi^2)$,  occur at leading order. It is interesting
to note that this uncertainty is rather non sensible to
center--of--mass energies as we obtain the uncertainty calculated at
the Tevatron.

The last source of uncertainty originates from the choice of the
$b$--mass value in the $b$--quark densities. Using the dedicated MSTW
PDFs set with different bottom quark masses~\cite{Martin:2010db} as
already done in the case of the Tevatron study,we have calculated the
correlated PDF--$\Delta m_b$ uncertainty by taking the minimal and
maximal values of the production cross sections when using the central
value $m_b=4.75$ GeV and the two closest ones upwards and downwards,
i.e.  $m_b=4.5$ GeV and $m_b=5$ GeV in the MSTW PDF set while keeping
in the partonic calculation the central value of
$\overline{m}_b(\overline{m}_b)=4.19$ GeV which approximately
corresponds to the pole mass $m_b=4.75$ GeV. We obtain at the lHC a
$\approx 3$--5\% uncertainty depending on the considered Higgs mass
range in the $b\bar b\to \Phi$ channel, that is smaller than the
uncertainty obtained at the Tevatron. As already mentioned in the
Tevatron study the gluon--gluon fusion process is marginally affected,
the uncertainty being below the percent level.

\begin{figure}[!h]
  \begin{center}
    \vspace*{1mm}
    \mbox{
      \includegraphics[scale=0.65]{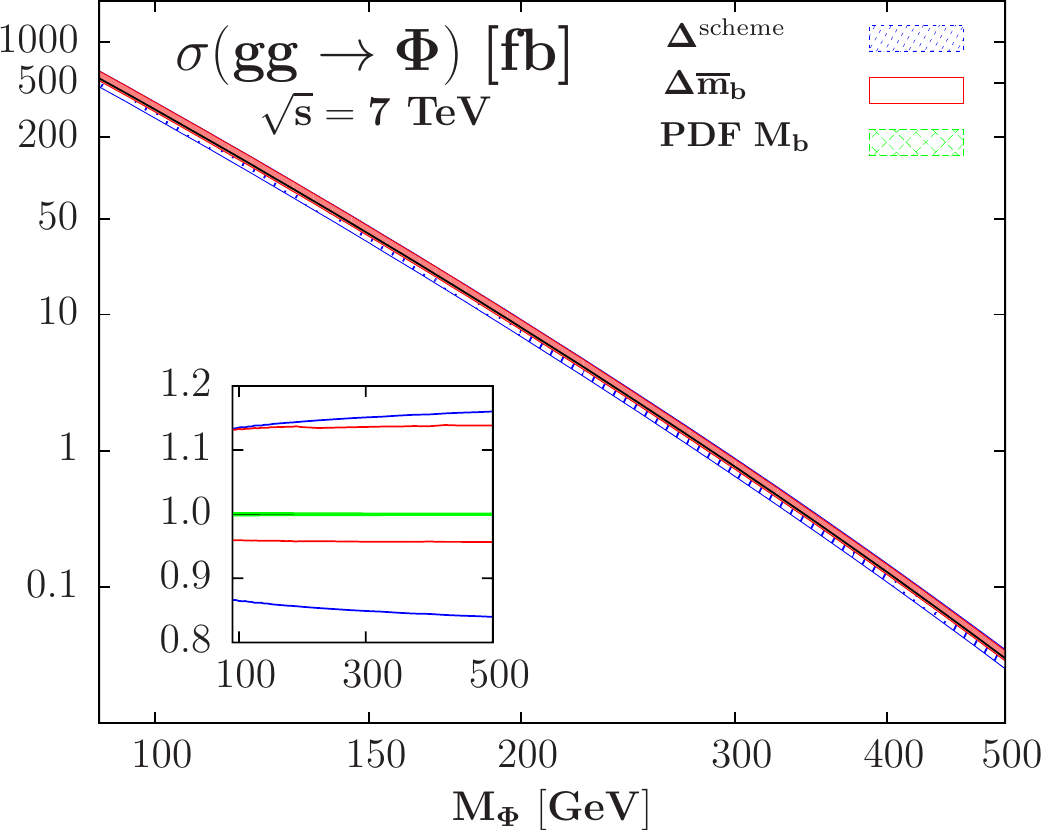}
      \includegraphics[scale=0.65]{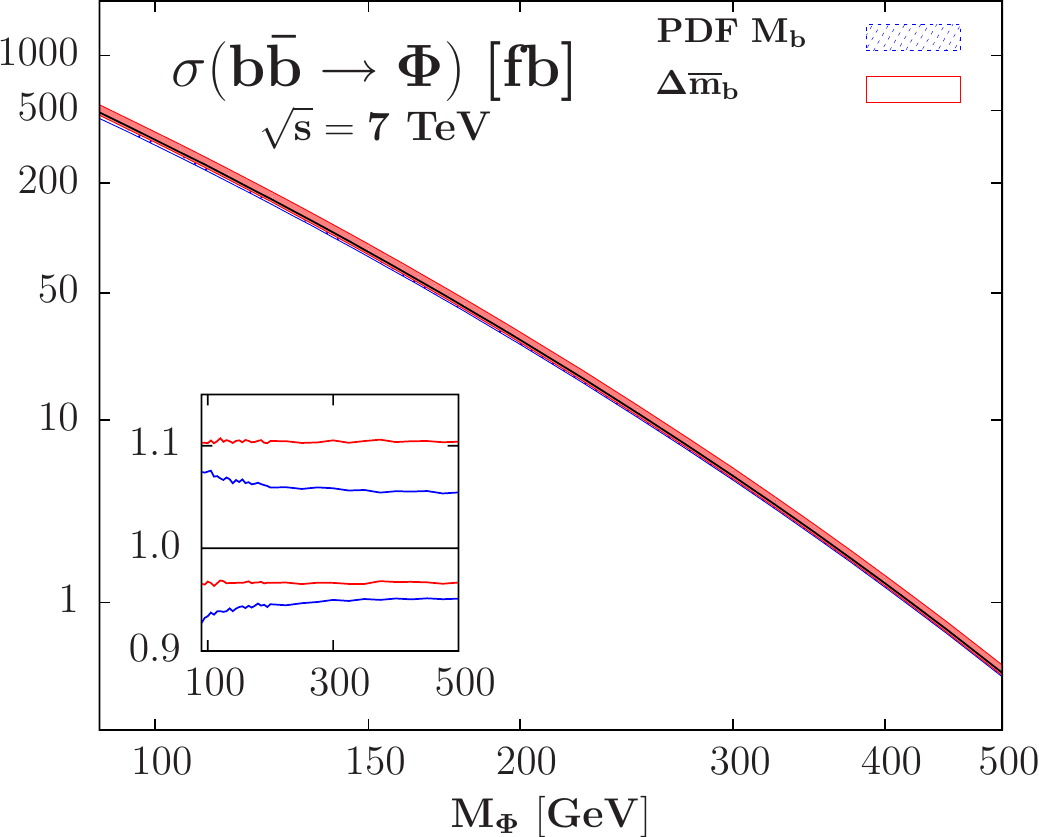}
    }
  \end{center}
  \vspace*{-4mm}
  \caption[Specific $b$--quark mass uncertainties in the $gg\to\Phi$ and
  $b\bar b\to \Phi$ processes at the lHC]{The scheme, parametric and PDF
    uncertainties due to the $b$-quark mass in the $gg \to \Phi$
    (left) and $b\bar b  \to \Phi$ (right) cross sections at the
    $\lhc$ at 7 TeV as a function of $M_\Phi$. In the inserts, the
    relative deviations are shown.}
  \label{MSSM-mb}
  \vspace*{-3mm}
\end{figure}

The effect of these three sources of uncertainties is displayed  in
Fig.~\ref{MSSM-mb} for $gg\!\to\! \Phi$ and $b\bar b\!\to\! \Phi$ as a
function of $M_\Phi$. As can be seen, large uncertainties occur, in
particular in $gg\! \to\! \Phi$ where the $\approx 15\%$ scheme
uncertainty that is absent in $b \bar  b\! \to\! \Phi$ dominates.

\subsection{Combination and total
  uncertainty \label{section:MSSMHiggsLHCTotal}}

\begin{figure}[!h] 
  \begin{center} 
    \mbox{
      \includegraphics[scale=0.65]{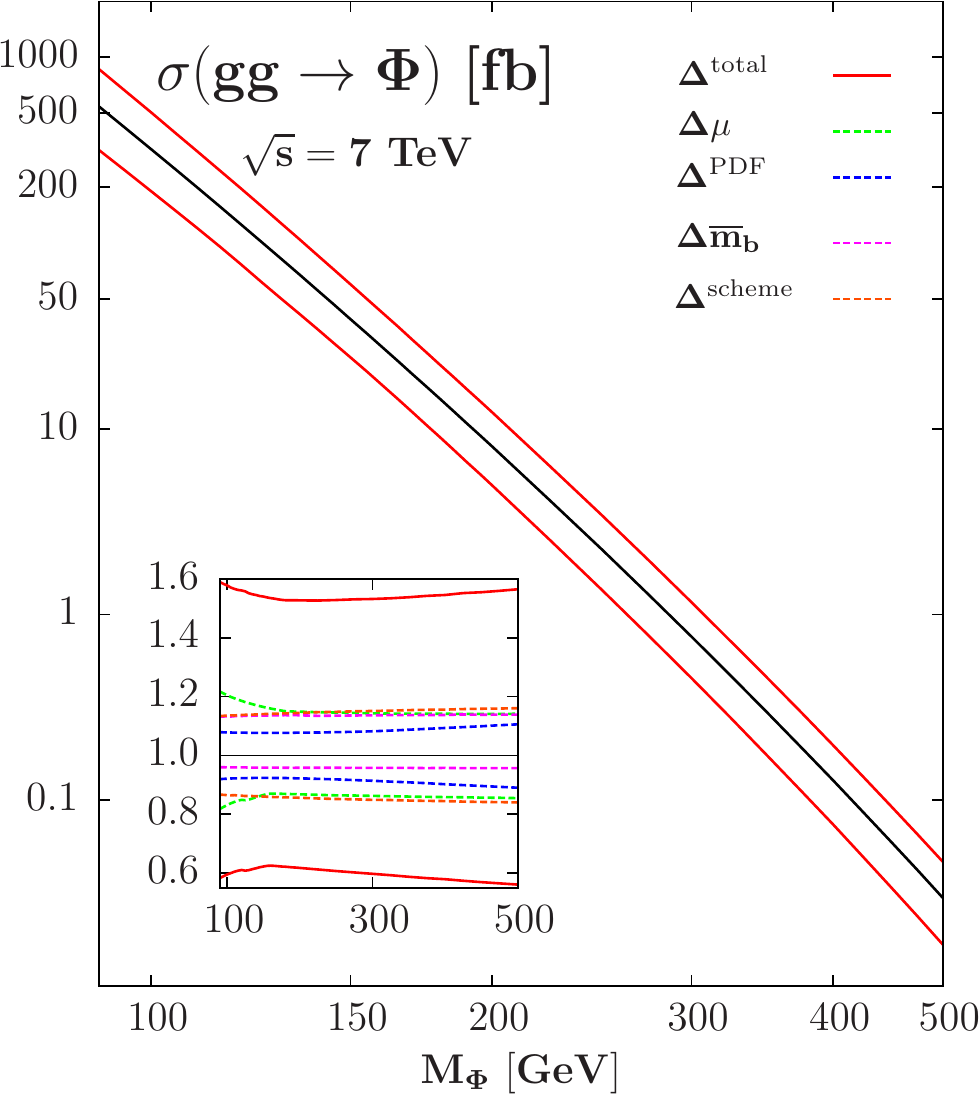}
      \includegraphics[scale=0.65]{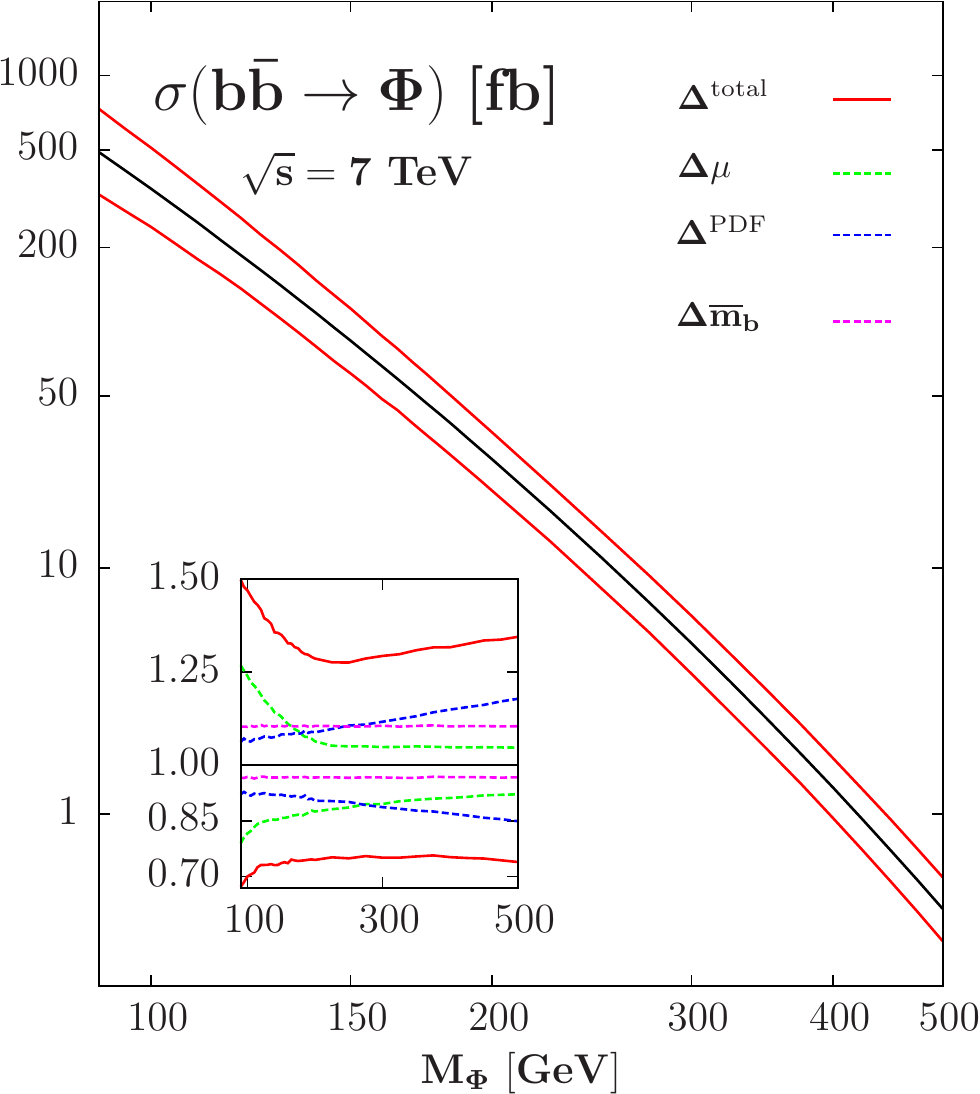}}
  \end{center} 
  \vspace*{-5mm}
  \caption[The $gg\to \Phi$ and $b\bar b\to \Phi$ cross sections at the
  lHC together with their different sources of uncertainty and
  the total uncertainties]{The normalization of the cross sections
    $\sigma^{\rm NLO}_{gg\!\to \!\Phi}$ (left) and $\sigma^{\rm
      NNLO}_{b\bar b \!\to\!\Phi}$ (right) at the lHC with $\sqrt s =
    7$ TeV as a function of $M_\Phi$ when using the MSTW PDFs and unit
    $\Phi b\bar b$ couplings together with the total uncertainty. In
    the inserts, shown are the various sources of theoretical
    uncertainties when the rates are normalized to the central values.}
\vspace*{-3mm}
\label{fig:MSSMTotalLHC}
\end{figure}

We end this section by the summary of the various sources of
uncertainty discussed above then we give the final total
uncertainty. This is displayed in Fig.~\ref{fig:MSSMTotalLHC} below
for the two processes $gg\to \Phi$ and $b\bar b\to \Phi$ at the lHC;
we recall again that $\tb$ has been set to unity. The scale
uncertainty has been taken in the $gg\to\Phi$ process with $\kappa=2$
whereas $\kappa=3$ has been used in the bottom quarks fusion process,
with the additionnal restriction $1/3 \leq \mu_R/\mu_F \leq 3$. We
have obtained almost independently of $M_\Phi$ an uncertainty of order
$\pm 15\%$ for the gluon--gluon fusion process (a little higher for
low Higgs masses $M_\Phi\simeq 90$ GeV) and of order $\pm 10\%$ in the
bottom quarks fusion process nearly in the entire range; more
precisely it is $-18\%,+24\%$ at $M_\Phi=100$ GeV and $-13\%,+6\%$ at
$M_\Phi=200$ GeV. In the $gg\to A$ ($b\bar b \to A)$ channel,
the PDF+$\Delta^{\rm exp+th} \alpha_s$ (with $\Delta M_b$ in addition
for $b\bar b \to A$) uncertainties start from $\pm 10\%$ at low masses
and reach the level of $\pm 15\%$ for $M_\Phi \approx 200$ GeV; in
$b\bar b\!\to\! \Phi$) the situation is almost the same, with a bit
higher uncertainty for high masses than in the gluon--gluon fusion
process, where the more uncertain high Bjorken--$x$ values for the
gluon and bottom quark densities are probed. The parametric error on
$\overline{m}_b$ leads to a $+14\%, -3\%$ uncertainty in the $gg\to
\Phi$ process and slightly less in the case of $b\bar b \to \Phi$,
$+10\%, -4\%$. We also have an additionnal scheme uncertainty in the
gluon--gluon fusion process that amounts to approximately $\pm 15\%$;
this has been taken into account in the bottom quarks fusion channel
when extending the scale interval from $\kappa=2$ to $\kappa=3$.

We now combine the uncertainties following strictly what has been done
in section~\ref{section:MSSMHiggsTevTotal} to handle this delicate
issue, applying the procedure developed in Ref.~\cite{Baglio:2010um}
to obtain the results published in Ref.~\cite{Baglio:2010ae}. The
overall uncertainty on the production cross section is obtained by
applying the PDF+$\Delta^{\rm exp}\alpha_s$+$\Delta^{\rm th}
\alpha_s$ uncertainty on the maximal and minimal cross sections from
the scale variation, adding linearly the scheme uncertainty.

We detail again the procedure developed to handle the parametric $b$
quark mass issue. The uncertainty from the scheme dependence in the
$gg \to \Phi$ process should be simply, that is linearly, added to the
scale uncertainty as both emerge as a result of the truncation of the
perturbative series and are thus of pure theoretical nature.  The
uncertainties of the input $b$--quark mass which appears in the parton
densities should be added (in quadrature, a reflect of its
experimental nature) to the other PDF+$\alpha_s$ uncertainties. Since
this uncertainty is much smaller than that of the PDF+$\Delta^{\rm
  exp+th}\alpha_s$ it will have practically no impact on
the total PDF+$\alpha_s$ error\footnote{Note that one may also
  consider an uncertainty related to the charm--quark mass dependence
  of the PDFs. This can be estimated again as done in the SM case and
  we find at most $\simeq \pm2.6\%$ at $M_\Phi=500$ GeV in both
  channels. We then neglect this uncertainty added in quadrature to
  that of the other PDF--related.}. Finally, the parametric
uncertainty, which has a special status as it appears also in the
Higgs branching ratios as seen later on, can be simply added linearly
to the combined scale+scheme+PDF uncertainty; we will see in the next
section that it will have no impact in practice.

Applying the procedure above for combining the uncertainties,  the results for
the processes $gg \to \Phi$ and $b\bar b\to \Phi$ at the $\lhc$ with $\sqrt
s= 7$ TeV are displayed in Fig.~\ref{fig:MSSMTotalLHC} as a function
of $M_\Phi$; the numerical values will be displayed in the next
section when we will also quote the numbers from the cross section
times branching fractions study. A total uncertainty of approximately
$+60\%,-40\%$ for $\sigma(gg\to \Phi)$ and $+50\%,-35\%$ for
$\sigma(b\bar b \to \Phi)$ is found in the low Higgs mass range and 
slightly less at higher masses.  As mentioned previously, we expect
that these numbers approximately hold at least for slightly higher
energies, $\sqrt s=8$--10 TeV, and probably also at the designed LHC
with $\sqrt s=14$ TeV.

This completes the analysis of the MSSM neutral Higgs bosons
production at the lHC in the two main channels $gg\to\Phi$ and $b\bar
b\to\Phi$. The next step that remains to be done before using these
results to constrain the MSSM parameter space is to study the decay
branching fractions and the combination of the latter with the
production cross sections. This will be the subject of the next-to-next -- and last
-- section of this thesis.

\vfill
\pagebreak

\subsection{The case of the charged Higgs production in association with
  top quark at the LHC}

Before getting to the comparison with experimental results on the MSSM
neutral Higgs bosons production, we wish to present the usefulness of
the study of the charged Higgs production in association with one top
quark at the LHC. We will see that a quite accurate measure of the
fundamental parameter $\tb$ can be done thanks to the left--right
asymmetry in the channel $g b\to t H^{-}$ and in particular for some
values of $\tb$ we can distinguish between type II 2 Higgs Doublet
Models (2HDM) which MSSM is an example, and type I 2HDM where only one
Higgs field couples to all fermions whereas both couple to the weak
and hypercharge bosons.

Once the Higgs bosons have been produced, their mass can be measured 
looking at the kinematical distributions of the decay
products~\cite{Aad:2008aa, Ball:2007zza}. In the MSSM the parameter
$\tan\beta$ can be
determined  looking at the total cross section of processes involving
Higgs bosons. For instance in the MSSM the total cross sections $p
p(\bar p) \to H,\, A$ are proportional to $\tan^2\beta$ as seen in
previous sections. A measurement of the relevant production cross
sections at the LHC allows for a determination of
$\tb$~\cite{Assamagan:2004mu} with an uncertainty of the order of $30
\%$.

Another interesting process  is the production of the charged Higgs
boson in association with  a top quark in bottom--gluon fusion  at
hadron colliders~\cite{Gunion:1986pe,
  Borzumati:1999th,Plehn:2002vy,Beccaria:2004xk,Huitu:2010ad}
\beq 
bg \to t H^-, \qquad  \bar bg \to \bar t H^+,
\label{eq:myprocesses}
\eeq
in which the bottom quark is directly taken from the proton in a five 
flavor scheme. The cross section of this process is proportional to the
square of the Yukawa coupling $g_{H^\pm tb}$. In type II 2HDMs, such
as in the MSSM, $g_{H^\pm tb}$ reads as follows~\cite{Gunion:1990},
\beq
g_{\rm H^\pm tb}&=& \frac{g}{\sqrt{2}M_W} V_{tb} \, \left\{ H^+ \bar{t} 
\left [m_b \tb\, P_R + m_t {\rm cot}\beta\, P_L \right ] b + {\rm h.c.} 
\right\},   
\label{eq:mycoupling}
\eeq
where $g=e/\sin(\theta_W)$ is the $SU(2)_L$ coupling and $P_{L/R}=
(1\mp \gamma_5) / 2$  are the chiral projectors. The CKM matrix
element $V_{tb}$ can be set, to a good approximation, to
unity~\cite{Nakamura:2010zzi}. At tree-level the total production
cross sections of the processes in Eq.~(\ref{eq:myprocesses}) are
equal and proportional to $(m_t^2 \cot^2\beta+
m_b^2\tan^2\beta)$. They are significant both in the  $\tb \le 1$ and
in the $\tb \gg 1$ regions\footnote{The total cross section exhibits a
  minimum at $\tan\beta = \sqrt{m_t /m_b} \approx 7$.}. In type I
2HDMs, all fermions couple to only one Higgs field; thus the $g_{\rm
  H^\pm tb}$ coupling is modified with the substitution $m_b \tb \to
m_b \cot \beta$ in Eq.~(\ref{eq:mycoupling}). The sum of the total
cross section of the two processes in Eq.~(\ref{eq:myprocesses}) is
proportional to ${\rm   cot}^2\beta$ and is enhanced for small $\tb$
values only\footnote{We recall that in the MSSM  the lower bound of
  the mass of $h$ requires that $\tb \gsim
  2$--$3$~\cite{Djouadi:2005gj,Nakamura:2010zzi}. In a general 2HDM
  $\tb$ is less constrained: the region $0.2 \lsim  \tb \lsim 50$ is
  not ruled out and preserves the perturbativity of the Higgs Yukawa
  coupling in Eq.~\ref{eq:mycoupling}.}.

Besides the experimental uncertainties, the cross section measurement
is plagued with various theoretical uncertainties as seen in previous
section~\ref{section:MSSMHiggsLHC}. The most important uncertainties
are related to the dependence of the observables on the
renormalisation and factorisation scales, as   well as the dependence
on the choice of the parton distribution functions (PDFs), and the
related errors on the strong coupling constant $\alpha_s$. These
theoretical uncertainties can be of the  order of
$20-30\%$~\cite{Assamagan:2004mu} and are a  major source of error in
the determination of $\tb$ directly from the Higgs production  cross
section. The study of the left--right top asymmetry in the $gb\to b
H^{-}$ production , followed by the clean and detectable $H^\pm  \to
\tau^\pm \nu$ channel, will be proved to be nearly free of these
theoretical uncertainties. The polarisation asymmetry $A_{LR}^t $ is
defined as the difference of cross sections for the production of
left--handed  and right--handed top quarks divided by their sum
\beq 
A_{LR}^{t}
& \equiv & \frac{\sigma_L -\sigma_R} 
{\sigma_L+\sigma_R},
\label{Eq:ALRt}
\eeq  
where $\sigma_{L/R}$ is the total hadronic cross section of the
process of $t_{L/R}H^-$ associated production. The asymmetry is a
ratio of observables of similar nature. Compared to the cross section,
the asymmetry is then significantly less affected by the scale and PDF
uncertainties. We are then mainly left only with the experimental
uncertainties in the determination of the cross sections and with the
measurement of the polarisation of the top  quarks\footnote{This
  section will not address the issue of the experimental determination
  of the top quark polarization; we refer to
  Refs.~\cite{Beneke:2000hk, Bernreuther:2008ju, Godbole:2010kr} for
  this matter.}. In the MSSM, the asymmetry will nevertheless remain
sensitive to the electroweak and strong radiative corrections from
supersymmetric particles which also strongly affect the cross sections
at high $\tb$
values~\cite{Plehn:2002vy,Beccaria:2009my,Carena:1999py}. All the work
that will be presented in this section is taken from
Ref.~\cite{Baglio:2011ap} and follow the works cited herein that
originate from a first study of the asymmetry in the case of
associated top--charged slepton production in the
MSSM~\cite{Arai:2010ci}. A detailed analysis of the top polarisation
in $bg \to tH^-$ production has also been given in
Ref.~\cite{Huitu:2010ad} which provides material that partly overlaps
with the one presented here.

We will first discuss the asymmetry in the tree--level approximation
and show its dependence on $\tb$. We then move on to the demonstration
that it nearly independent of the scale and PDF distribution choice,
but still remaining dependent on the important MSSM radiative
corrections that indeed may help to distinguish between SUSY and
non--SUSY scenarios.

\paragraph{The $\mathbf{A_{LR}^t}$ asymmetry at tree--level\newline}

We will fix the notation, and begin with the process
\beq
b(p_b, \lambda_b) \; g(p_g, \lambda_g) \; \to \;  t(p_t,\lambda_t) \; H^-(p_H).
\label{Eq:process2}
\eeq
The momentum (helicity) of the particle $i$ is marked as $p_i$ ($\lambda_i$).
In the tree--level approximation the process is mediated by two
Feynman diagrams, one with $s$--channel bottom quark exchange and
another with $u$--channel top  quark exchange. In the case of type II
2HDM couplings the helicity amplitude
$F_{\lambda_b\lambda_g\lambda_t}$ reads as
follows~\cite{Beccaria:2004xk}
\beq
F_{\lambda_b\lambda_g\lambda_t} &=&
{g g_s \lambda^l \sqrt{x_+} \over2  M_W} \Bigg \{  
\frac{\delta_{\lambda_b\lambda_g}}{\sqrt{\hat s}} 
\left [ \lambda(1-r_t)s_{\theta / 2} \delta_{\lambda_b\lambda_t}
+{1+r_t\over2} c_{\theta/2}
\delta_{\lambda_b-\lambda_t} \right ]  \nonumber \\
&+&  
\frac{m_t\delta_{\lambda_b\lambda_g}}{\hat u-m^2_t} \left [ (1+r_t)
  s_{\theta/2}
\lambda\delta_{\lambda_b\lambda_t} + {1-r_t\over2}
c_{\theta/2}\delta_{\lambda_b-\lambda_t} \right] \nonumber\\
&+& \frac{(1-r_t) s_{\theta / 2} \lambda 
\delta_{\lambda_b\lambda_t} }{ \hat u-m^2_t}\left [-p(1+c_\theta)
\delta_{\lambda_b-\lambda_g}+ d_t
\delta_{\lambda_b\lambda_g} \right ]\nonumber\\
&+&\frac{(1+r_t)  c_{\theta/2}
\delta_{\lambda_b-\lambda_t}}{2 (\hat u-m^2_t)}\left [p (1-c_\theta)
\delta_{\lambda_b-\lambda_g}+ d_t
\delta_{\lambda_b\lambda_g}\right ]\Bigg\}
\left [m_t \cot\beta \delta_{\lambda_tL}+m_b \tan\beta
  \delta_{\lambda_tR} \right].
\eeq
We will define the partonic Mandelstam variables as $\hat s =
(p_b+p_g)^2$ and $\hat u=(p_b-p_H)^2$. The angle $\theta$ is the
azimuthal angle in the center-of-mass frame, $g_s$ is the strong
coupling constant. The abbreviations $d_t$, $r_t$,  $x_{\pm}$ and
$\lambda$ read as follows
\beq
d_t = \sqrt{\hat s} -E_t+p \cos\theta, \qquad
r_t =  \sqrt{  \frac{x_-}{x_+} }. \qquad
x_{\pm} = \left (\sqrt{\hat s} \pm m_t \right ) ^2-M^2_{H^\pm},
\eeq
\beq
\lambda\!=\! \sqrt{
\left (1\!-\!(x_t\!+\!x_h  )^2 \right ) \left (1\!-\!(x_t\! - \!x_h)^2
\right )}, \nonumber
\eeq
while $p \equiv |{\bf p}_t|$, $c_\alpha \equiv \cos \alpha$, and
$s_\alpha \equiv \sin \alpha$.

The partonic cross sections for L/R  polarized top quarks in the final
state is then 
\beq 
\hat \sigma_{L/R} =  \frac{p} {384 \pi \hat s^{3/2}}
\int_{-1}^{+1} {\rm d}\!\cos \theta  \sum_{\lambda_b,\lambda_g} \big |
F_{\lambda_b\lambda_g L/R} \big |^2.
\eeq
The integration over the angle $\theta$ finally leads to
\beq
\hat \sigma_{L} & = & \displaystyle \frac{G_F\alpha_s}{24\sqrt{2}
  \hat{s} \lambda}  \Bigg  \{   \lambda  \left [  m_t^2 \cotan^2\beta 
   \left(\frac{7}{2} \lambda x_{ht}^2+2 x_{ht}^2+2
   \left(1-x_{ht}^2\right)^2+\frac{3}{2} (\lambda -1) \lambda
   \right) \right.   \nonumber \\
& & \displaystyle \left. - m_b^2 \tan^2\beta
    \left(-\frac{7}{2} \lambda x_{ht}^2+2 x_{ht}^2+2
   \left(1-x_{ht}^2\right)^2+\frac{3}{2} \lambda  (\lambda
   +1)\right) \right] + \nonumber \\ 
& & \displaystyle \Lambda  \Bigg[m_t^2 \cotan^2\beta
\left(\left(x_{ht}^2+2 \lambda
    \right) \left(1-x_{ht}^2\right)^2+(\lambda +1) \left(x_{ht}^2
      (\lambda
   +1)-1\right)\right)  \nonumber\\
& & \displaystyle + m_b^2 \tan^2\beta \left(\left(2 \lambda
    -x_{ht}^2\right)
   \left(1-x_{ht}^2\right)^2+\left((\lambda -1) x_{ht}^2+1\right)
   (1-\lambda )\right) \Bigg] \Bigg \}, \nonumber\\
\hat \sigma_{R} & = & \displaystyle \frac{G_F\alpha_s}{24\sqrt{2}
  \hat{s} \lambda} \left \{ \lambda  \left[ m_b^2 \tan^2\beta
   \left(\frac{7}{2} \lambda x_{ht}^2+2 x_{ht}^2+2
   \left(1-x_{ht}^2\right)^2+\frac{3}{2} (\lambda -1) \lambda
   \right) \right. \right. \nonumber \\
& & \displaystyle \left. - m_t^2 \cotan^2\beta
    \left(-\frac{7}{2} \lambda x_{ht}^2+2 x_{ht}^2+2
   \left(1-x_{ht}^2\right)^2+\frac{3}{2} \lambda  (\lambda
   +1)\right) \right] + \nonumber \\ 
& & \displaystyle \Lambda  \Bigg  [ m_b^2 \tan^2\beta
\left(\left(x_{ht}^2+2 \lambda
    \right) \left(1-x_{ht}^2\right)^2+(\lambda +1) \left(x_{ht}^2
      (\lambda
   +1)-1\right)\right)  \nonumber\\
& & \displaystyle  + m_t^2 \cotan^2\beta \left(\left(2 \lambda
    -x_{ht}^2\right)
   \left(1-x_{ht}^2\right)^2+\left((\lambda -1) x_{ht}^2+1\right)
   (1-\lambda )\right) \Bigg ] \Bigg \}.
\eeq
where $x_i = m_i/\sqrt{\hat s}$ and $x^2_{ht}=x^2_h  -x^2_t$ and
$\Lambda$ is defined as
\beq
\Lambda = \ln \left (  \frac{1-x_{ht}^2+\lambda}{1-x_{ht}^2-\lambda
    } \right ).
\eeq
The total partonic cross section is then simply the sum of the cross
sections $\hat \sigma_L$ and $\hat \sigma_R$
\beq
\hat \sigma_{\rm tot} = \frac{G_F \alpha_s}{24 \sqrt{2} \hat s}
 \left (m_t^2 \cot^2\beta\! +\! m_b^2 \tan^2 \beta \right ) 
\bigg\{  2 \left[1\!-\!2 x^2_{ht} 
(1\!-\!x^2_{ht})\right]\Lambda \!-\!  (3\!-\!7 x^2_{ht})  \lambda \bigg\}.  \ \ 
\label{siggbH+}
\eeq
The final hadronic cross sections $\sigma_{L,R}$ are the convolution
of these partonic cross sections with the bottom--quark and gluon
parton distribution functions. The results for the type I 2HDM are
obtained once we perform the switch $m_b \tb \to m_b {\rm cot} \beta$.

\begin{figure}[!h]
  \begin{center}
    \mbox{
      \includegraphics[scale=0.65]{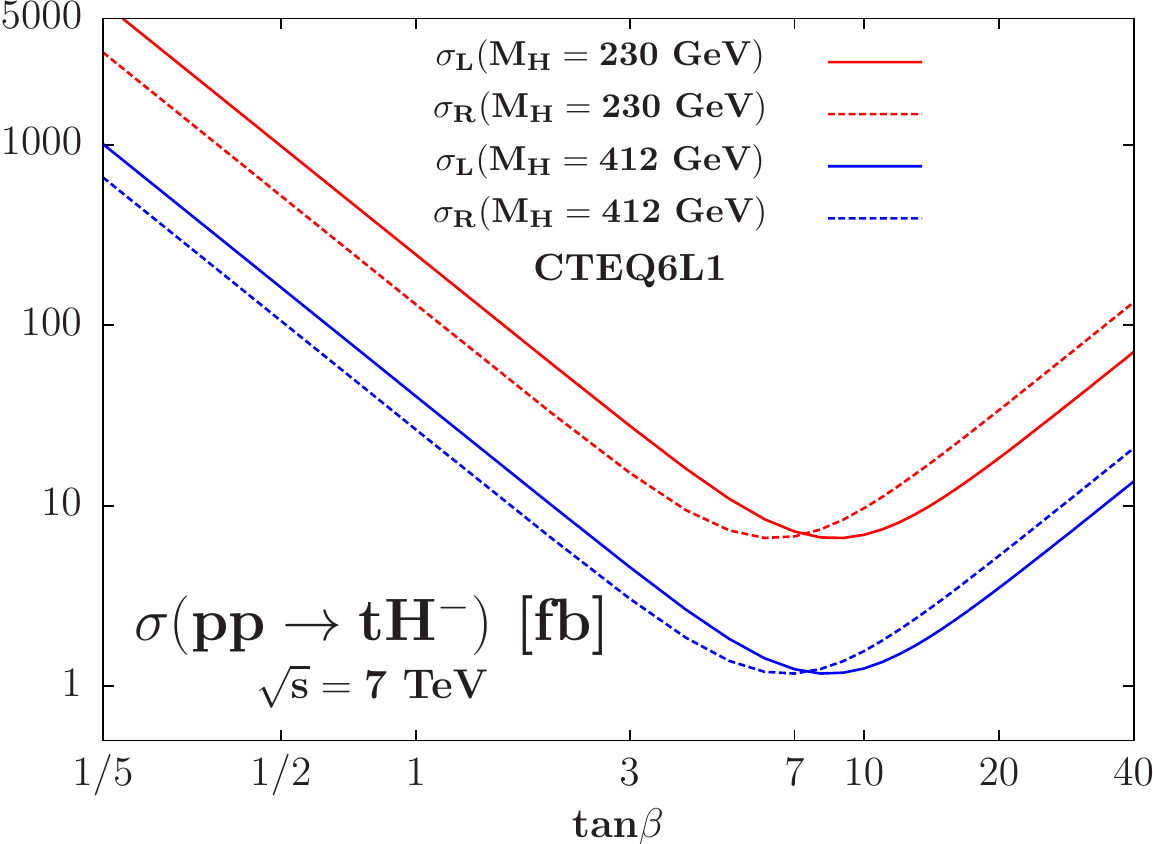}
      \includegraphics[scale=0.65]{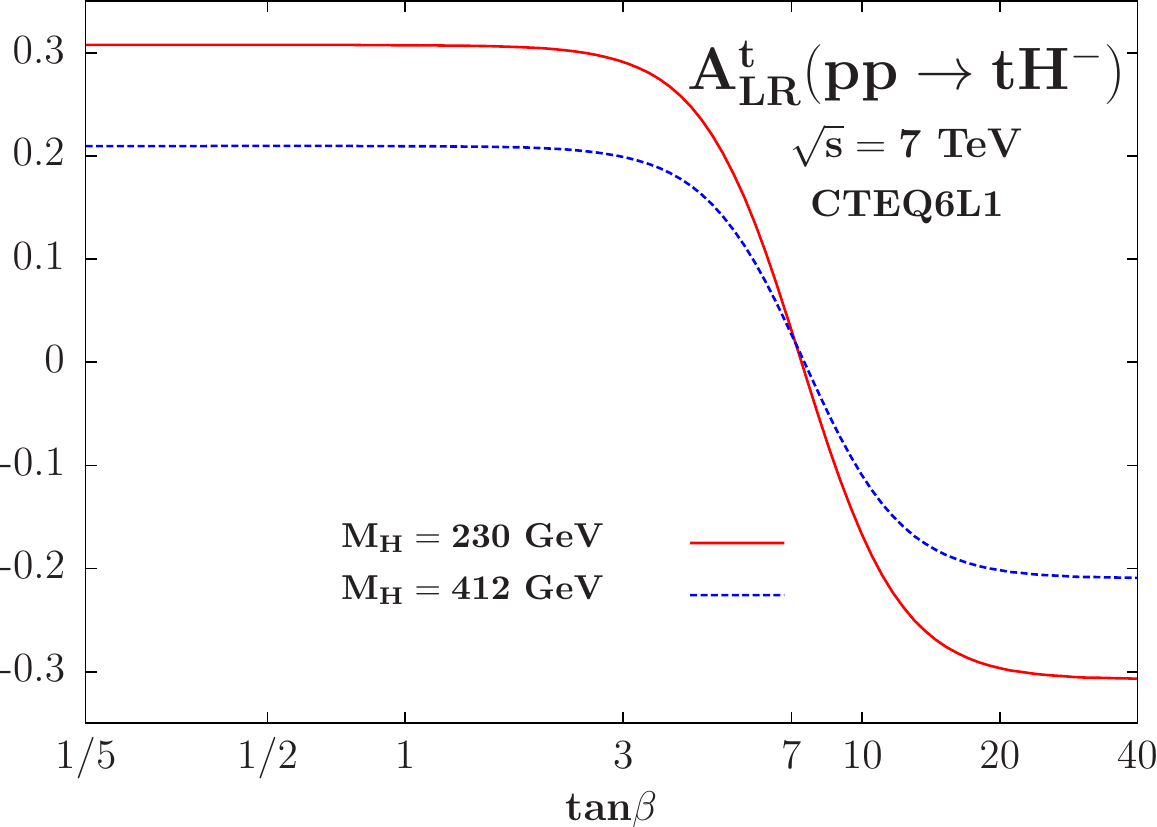}}
  \end{center}
\vspace{-2mm}
  \caption[LO $\sigma(g b \to t_{L,R} H^{-})$ cross section and
  polarization asymmetry at the lHC in the MSSM in two benchmark
  scenarios as a function of $\tb$]{The LO cross sections $\sigma_L$ and
    $\sigma_R$ (left) and the left--right top asymmetry $A_{LR}$ at
    leading order (right) at the LHC with $\sqrt s = 7$ TeV in two
    benchmark scenarios with $M_{H^\pm}=230$ and $412$ GeV in type II
    2HDMs.}
  \vspace*{-7mm}
  \label{fig:typeII}
\end{figure}

In Fig.~\ref{fig:typeII} we display the left-- and right-- handed cross
sections $\sigma_L$ and  $\sigma_R$ as well as the asymmetry
$A_{LR}^t$ at the LHC with $\sqrt s= 7$ TeV as a function of $\tb$. We
choose two values of $M_{H^\pm}$, $M_{H^\pm}=230$ and $412$ GeV
corresponding to the two MSSM scenarios proposed  in
Refs.~\cite{Beccaria:2009my} (LS2) and~\cite{Allanach:2002nj} (SPS1a)
respectively. We have adopted the CTEQ6L1 leading order
PDFs~\cite{Nadolsky:2008zw} with $\alpha_s(M_Z^2)=0.130$. The
factorisation scale $\mu_F$ has been set to the value $\mu_0=
(M_{H^\pm} +m_t)/6$ which according to Ref.~\cite{Plehn:2002vy} is
known to minimize the higher order  QCD corrections. For the $H^-tb$
coupling, we use the on--shell top mass value $m_t=173.1$ GeV and the
$\overline{\rm MS}$ mass of the bottom quark evaluated at a scale $\mu
= \mu_F$. In the analysis presented in this section, $m_b(\mu)$ is
approximately equal to 3 GeV depending on the values of $\mu_F$
considered~\cite{Baglio:2011ap}.

As we can see in Fig.~\ref{fig:typeII}, $\sigma_L$ and $\sigma_R$ have
the same order of magnitude: they are large at small $\tb$ values,
when the component $m_t\cot\beta$  of the $H^-tb$ coupling is
significant, as well as at large $\tb$ value when the $m_b\tb$
component of the coupling is enhanced. The cross sections are equal
and minimal at the value $\tan \beta =\sqrt{m_t /m_b} \simeq 7$ for
which the $H^-tb$ coupling is the smallest. This implies that in type
II 2HDM (such as the MSSM) $A_{LR}^t$ is maximal at low $\tb$ values
when the associated top quark is mostly left--handed  and minimal at
large $\tb$ values when the top quarks are right handed.  For a given
value of the charged Higgs mass, the absolute value of $A_{LR}^{t}$ is
the same in the $\tb \gg 1$ and in the $\tb \le 1$ region. In the
scenarios under consideration  $|A_{LR}^t|=0.31\; (0.21)$ for
$M_{H^\pm}= 230 \;(412)$~GeV. The two $\tb$ regions differ for the
sign of the asymmetry. Therefore the sign of  $A_{LR}^{t}$
helps to distinguish between the low and large $\tb$ scenarios. In the
intermediate $\tb$ region, $\tan \beta  \simeq 7$ for which $\sigma_L
\simeq \sigma_R$, the asymmetry goes through zero. 

The behaviour in a type I 2HDM is completely different: the left-- and
right-- components of the Yukawa coupling  $g_{H^\pm tb}$  are both
proportional to $\cot\beta$, and there is no $\tb$ dependence in
$A_{LR}^t$, which implies that the asymmetry is constant and simply
given by the $A_{LR}^t$ value in the corresponding type II model
evaluated at $\tb=1$.  For type I 2HDM characterized by
$M_{H^\pm}=230 \; (412)$~GeV the value of $A_{LR}^{t}$ can be read off
Fig.~\ref{fig:typeII},  $A_{LR}^{t}=0.31 \; (0.21)$. Combining this
value  with the value of $\sigma_{\rm tot} \propto \cot^2\beta$,  the
predictions of 2HDMs of type I and II can  eventually be discriminated.

It is worth mentioning that while $\sigma_L, \sigma_R$ and thus
$\sigma_{\rm tot}$ strongly depend on the hadronic center-of-mass
energy, the asymmetry dependence of $A_{LR}^{t}$  is mild. The
asymmetry  is comparable for $\sqrt s=7$ and $14$ TeV.  For instance
at $\sqrt s = 14$ TeV in the type I model one obtains
$A_{LR}^t=0.27\;(0.18)$ for $M_{H^\pm}=230\;(412)$ GeV.

\paragraph{Scale and PDF dependence; impact of  the SUSY NLO
  corrections\newline}

\begin{figure}[!h]
  \begin{center}
    \mbox{
      \includegraphics[scale=0.65]{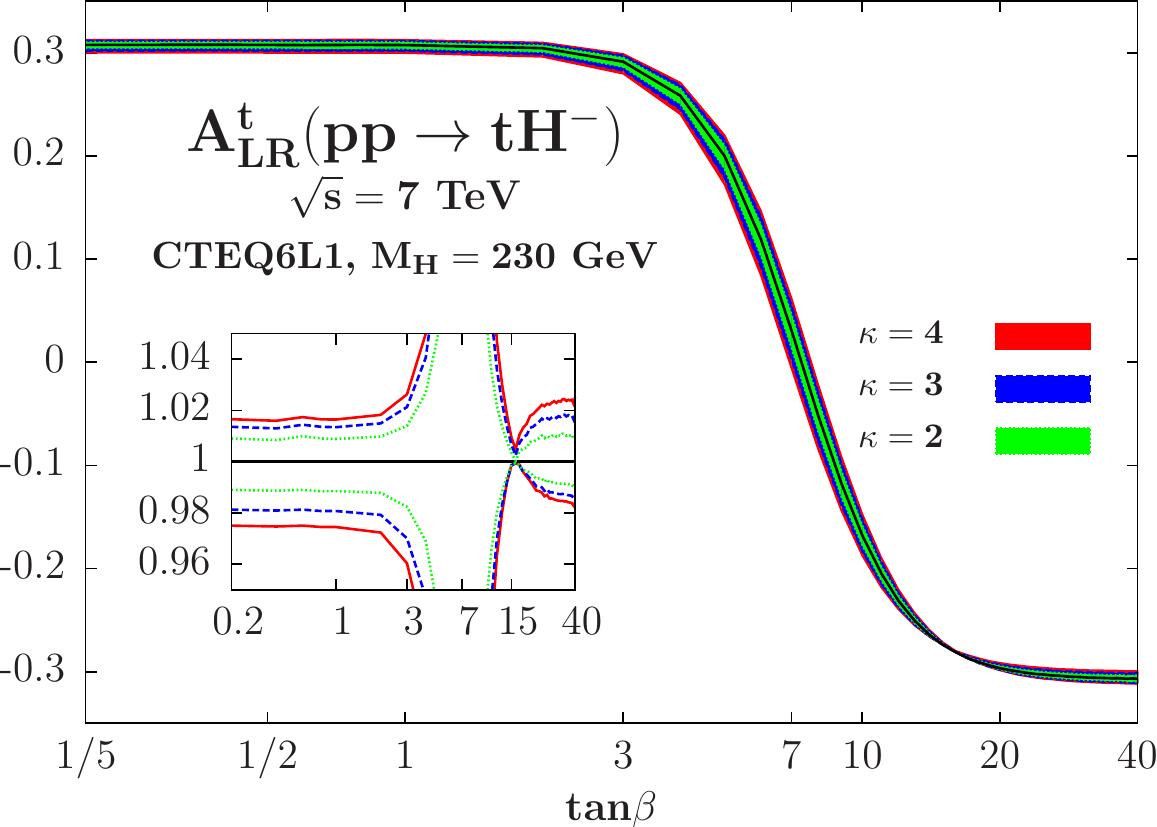}
      \includegraphics[scale=0.65]{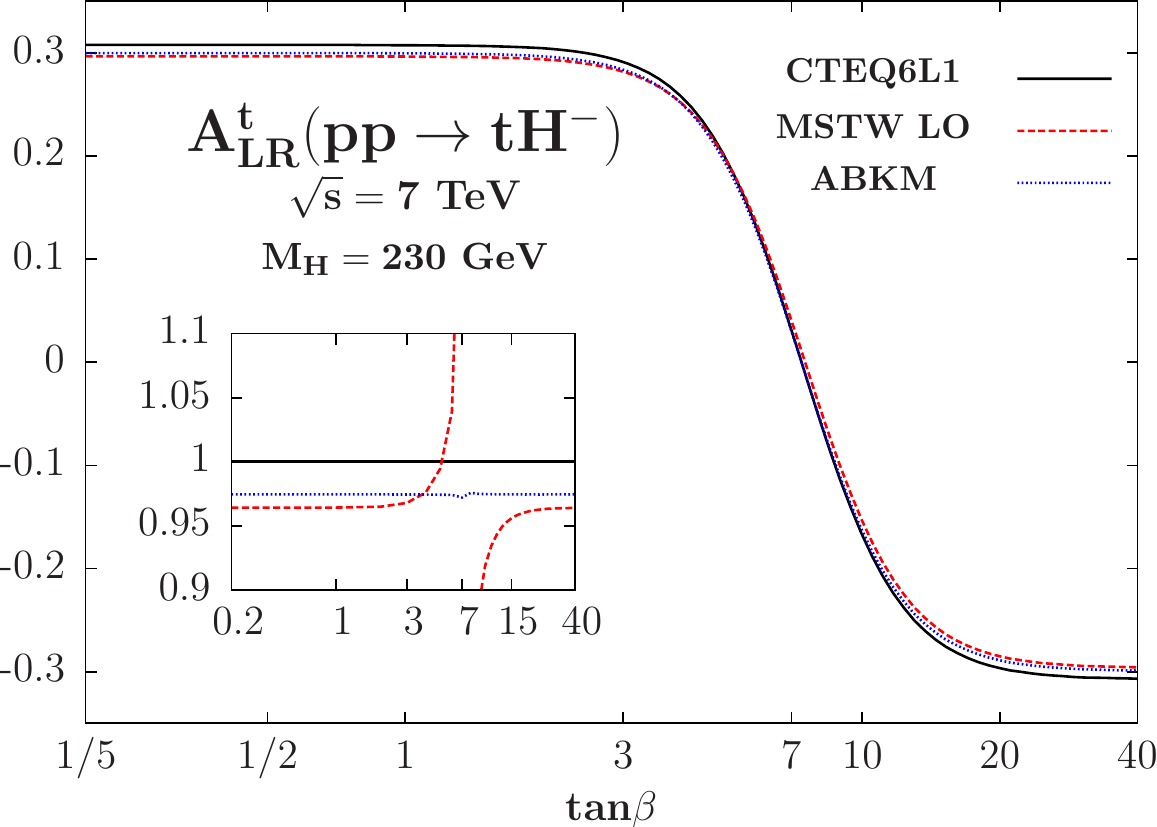}}
  \end{center}
  \caption[Scale and PDF dependence on top--charged Higgs asymmetry at
  the lHC]{The scale variation (left) and the PDF dependence (right) of
    the asymmetry $A_{LR}^t$ at leading order at the LHC with $\sqrt
    s=7$ TeV as a function of $\tb$. We consider the type II 2HDM
    characterized by  $M_{H^\pm}=230$ GeV. In the inserts, shown are the
    variations with respect to the central CTEQ value.}
  \vspace*{-2mm}
  \label{fig:scale}
\end{figure}

Up until now we have calculated the asymmetry only at leading order,
and made central predictions. The yet uncalculated higher order QCD
contributions on this observable can be estimated from its dependence
on the factorisation scale $\mu_F$ at which the process is
evaluated. Starting from our reference scale $\mu_0$ we vary $\mu_F$
within the range $\mu_0/\kappa \!\le \!  \mu_F \!\le \! \kappa \mu_0$
with the constant factor chosen to be $\kappa\!=\!2,3$ or $4$.
The left panel  of Fig.~\ref{fig:scale} shows the variation of the
polarisation asymmetry for the choices $\kappa=2,3$ and $4$. The
insert shows the scale variation relative to the asymmetry value when
the central scale is adopted.

The main output of this calculation is that the scale dependence is
very low. Indeed, in the low and in the high $\tb$ region, it is at
most at the level of $2\%$, even for the high $\kappa=4$ choice for
the scale interval.  At moderate values of $\tb$,  $\tb \simeq 7$,
the relative variation is much larger since the asymmetry
vanishes\footnote{This is nothing more than a numerical effect: as the
asymmetry vanishes, the precision of the numerical calculation is less
than the true scale uncertainty.}. However the absolute impact of the
scale variation is comparable to the one obtained for low and high
$\tb$ values, and thus small in absolute terms. It is worth mentioning
that the NLO QCD total cross section $\sigma_{\rm tot}$ exhibits a
bigger residual scale uncertainty estimated to be of the order of
10--20\% at the LHC with $\sqrt s=7$ TeV, see Ref.~\cite{Dittmaier:2011ti}.

The second potential source of uncertainty comes from from the
presently not satisfactory determination of the gluon and bottom quark
PDFs. We have chosen to estimate the impact of the different choice of
PDFs collaborations together with some differences in the choice of
the value of $\alpha_s(M_Z^2)$. As already discussed in the case of SM
Higgs production this is a way to estimate the PDF uncertainties. In
the right panel of Fig.~\ref{fig:scale} we show the dependence of the
asymmetry on $\tb$ when the CTEQ, the MSTW~\cite{Martin:2009iq}, and the
ABKM~\cite{Alekhin:2009ni} PDF sets are used. We consider the type II 2HDM
characterized by $M_{H^\pm}=230$ GeV. As usual the asymmetry has been
computed at the LHC with $\sqrt s=7$ TeV. In the insert we show the
relative deviation from the CTEQ central prediction. The difference
between the various predictions is found to be rather small as
displayed in the right panel of Fig.~\ref{fig:scale}, less than 
few percents at low and high $\tb$ values.  Again the peaks in the
insert for $\tb \simeq 7$ correspond to the vanishing of $A_{LR}^t$
and are nothing more than numerical effects. On the contrary, the
impact of the PDF variation on the total cross section $\sigma_{\rm
  tot}$ is expected to be much larger. For instance,
Ref.~\cite{Dittmaier:2011ti} has found that at NLO the PDF uncertainty
is expected to be of the order of $10\%$.

The last piece that remains to be discussed is the impact of radiative
corrections on the polarization asymmetry, in particular in the case
of supersymmetric scenarios. In the MSSM, the process $g b\to t H^{-}$
is affected by radiative corrections involving the supersymmetric
particle spectrum. The NLO QCD and electroweak corrections have been
discussed in Ref.~\cite{Plehn:2002vy} and in Ref.~\cite{Beccaria:2009my}
respectively. Some of these corrections are known to be large  for
high values of $\tb$ and some other parameters such as the higgsino
mass parameter $\mu$. It turns out that the bulk of these radiative
corrections can be accounted for by modifying the Yukawa coupling of
Eq~.\ref{eq:mycoupling} as described in Ref.~\cite{Carena:1999py}, in
much the same way as we have done in the case of MSSM neutral Higgs
production in previous sections using the $\Delta_b$
approximation. The approximation is rather good  for the SUSY--QCD
corrections (in particular when the SUSY spectrum is rather heavy),
and slightly worse in the case of the electroweak ones.

\begin{figure}[!h]
  \begin{center}
    \mbox{
      \includegraphics[scale=0.35]{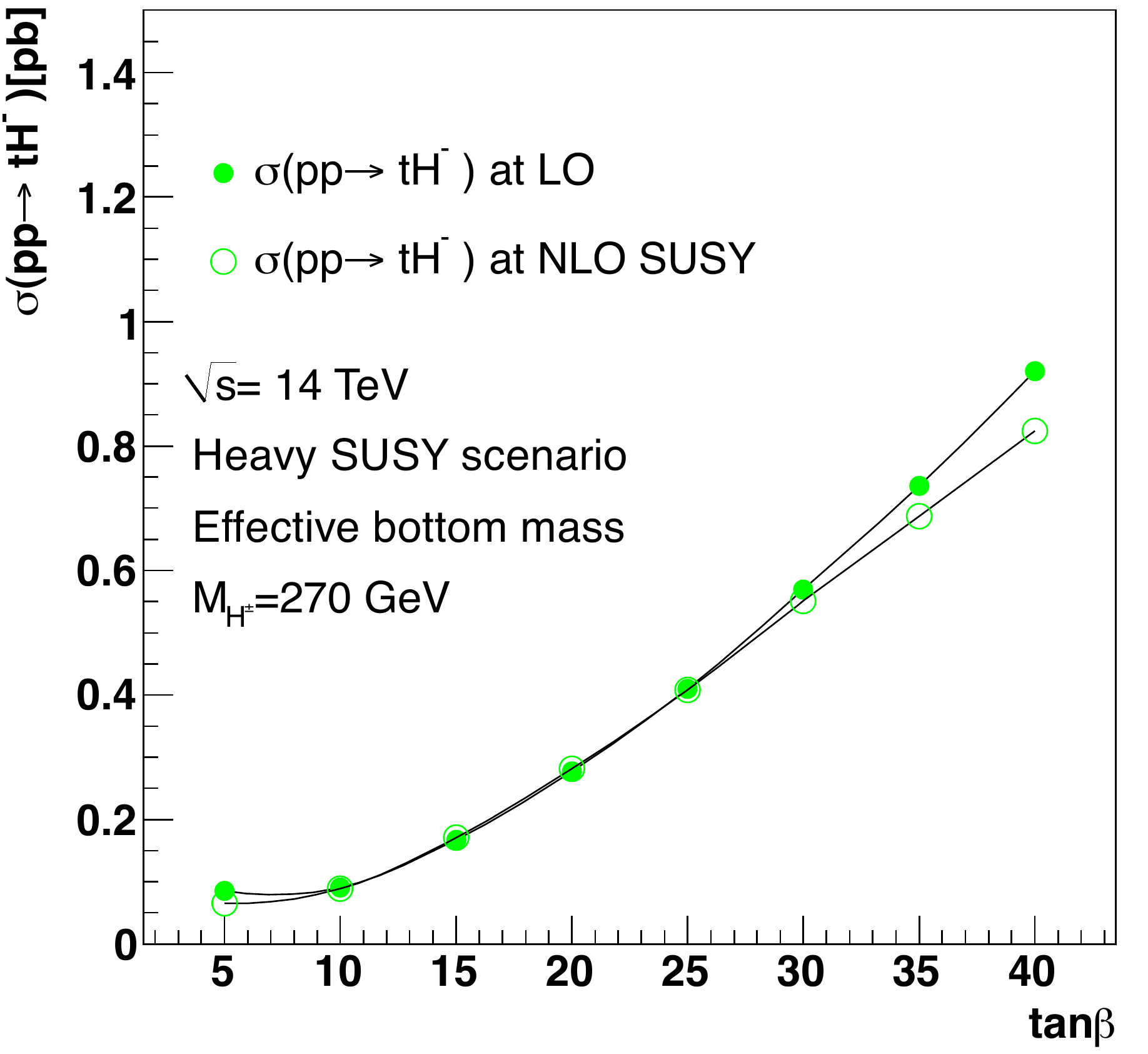}
      \includegraphics[scale=0.35]{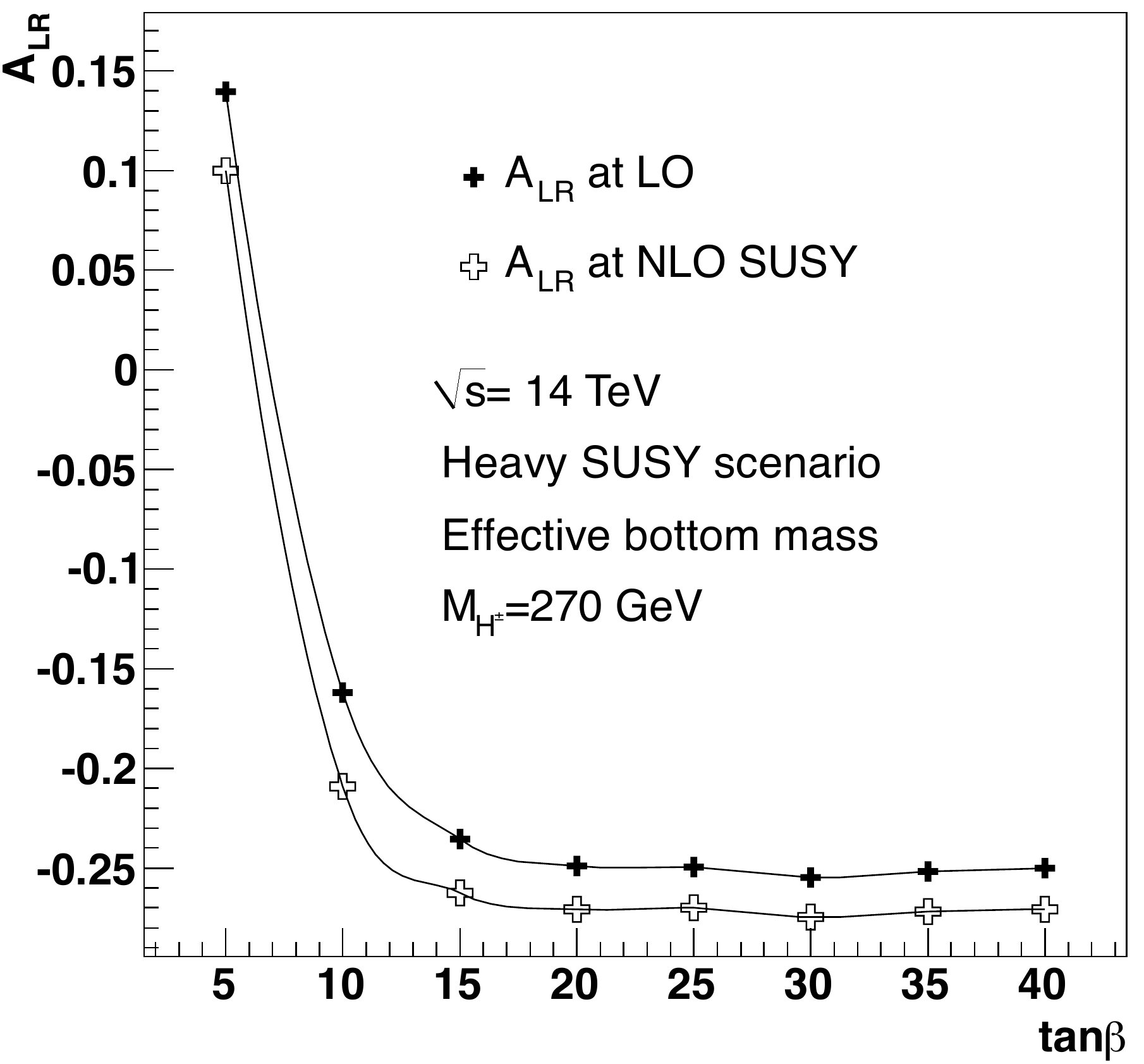}}
  \end{center}
  \caption[The impact of the NLO SUSY corrections on the top--charged
  Higgs asymmetry at the LHC with $\sqrt s = 14$ TeV]{The total
    production cross section  (left) and the asymmetry $A_{LR}^t$
    (right) at leading order and including the NLO  SUSY corrections at
    the LHC with $\sqrt s=14$ TeV.  We consider the MSSM scenario of
    Ref.~\cite{Beccaria:2009my} characterized by  a heavy sparticle
    spectrum and $M_{H^\pm}=270$~GeV.  $\tb$ is varied from $5$ to
    $40$. Figure taken from Ref.~\cite{Baglio:2011ap}.}
  \vspace*{-2mm}
  \label{fig:chargeddeltab}
\end{figure}

In Fig.~\ref{fig:chargeddeltab}, we display the impact of these NLO SUSY
radiative corrections within the MSSM on both the total cross section
and the left--right asymmetry as a function of $\tb$, obtained in
Ref.~\cite{Baglio:2011ap}. The other SUSY parameters are fixed according to the
scenario presented in Ref.~\cite{Beccaria:2009my}, characterized by a heavy
superparticle spectrum and  $M_{H^\pm}=270$ GeV. The SUSY QCD
corrections are included in the $\Delta_b$ approximation, while the
electroweak and the (very small) QED corrections are computed
exactly. In the $\tb$ range considered the approximation for the SUSY
QCD contributions  is expected to be valid.

Fig.~\ref{fig:chargeddeltab} shows that the NLO corrections can indeed
be large in both the cross section and the asymmetry.  In the case of
the latter observable the effect is of the order of $10\%$ in the
$\tb \ge 15$ region, where the asymmetry dependence  on $\tb$ is
almost flat. Therefore the asymmetry is sensitive to the quantum
contributions of the superparticle spectrum:  a precise measurement of
the asymmetry could then allow to probe these additional
supersymmetric corrections and, hence, could help to discriminate
between supersymmetric and non--supersymmetric 2HDM of type
II.\bigskip

We have then seen that the associated top quark and charged Higgs
production $b g\to t H^{-}$ at the LHC is a very interesting channel
to measure the parameter $\tb$ with some accuracy through the
left--right top asymmetry, obtained by identifying the polarisation of
the top quarks. This asymmetry has been proved to be essentially free
from scale and PDF uncertainties, and still sensitive to radiative
effects from new physics scenarios such as supersymmetry. The combined
measurement of the production cross section and the polarisation
asymmetry could discriminate between various new physics scenarios:
two--Higgs doublet models of type I versus type  II and the MSSM
versus non-supersymmetric models, at least for intermediate values of
$\tb$. For $\tb \gg 1$ or $\tb \le 1$ the method allows for the
determination of the region of $\tan \beta$ but not for the  exact
value of $\tb$, since in this two regions $A_{LR}^{t}$ has a
plateau. The only region where the distinction between type I and II
2HDM becomes challenging is for $\tb \le 1$ where the asymmetry is the
same in both models. This polarisation asymmetry is thus worth
investigating theoretically and experimentally in more detail.

This concludes the production of the MSSM Higgs bosons at the LHC. We
will now end this part~\ref{part:four} by giving the consequences of
these theoretical results on the MSSM parameter space and also on the
SM Higgs boson search at the LHC.

\vfill
\pagebreak

\section{Higgs$\to \tau \tau$ channel and limits on the MSSM parameter
  space}

After having conducted a detailed study of the two main production
channels for the MSSM neutral Higgs bosons at the Tevatron and the
LHC, taking into account all QCD related uncertainties, we are nearly
ready to study the impact of our predictions on the MSSM parameter
space. The last step that remains to be done is the study of the main
decay branching fraction that is of interest in the context of the
thesis, that is the decay $\Phi\to \tau^+\tau^-$.

Indeed this decay channel together with its parent $\Phi\to b\bar b$
is the most sensitive channel for MSSM neutral Higgs bosons searches
at hadron colliders~\cite{Benjamin:2010xb, Collaboration:2011rv,
  Chatrchyan:2011nx}. We will reproduce the set--up that has been used
in the case of the SM Higgs boson branching fraction in
section~\ref{section:SMHiggsDecay} and take into account the
uncertainties related to the $b$ quark mass and the value of the
strong coupling constant $\alpha_s$. In this case the role of the $c$
quark is negligible and will not be taken into account. We will see
that even if the MSSM adds lots of new particles in the spectrum the
major decay channels are up to a very good approximation the leptonic
$\Phi\to \tau^+\tau^-$ channel and the hadronic $\Phi\to b\bar b$
channel.

We will then combine the decay branching fraction with the production
channels at the Tevatron then at the lHC, giving the detailed tables
of the central predictions for the two channels together with the
different sources of uncertainties, the final uncertainty and the
total uncertainty taking into account the $\Phi\to\tau^+\tau^-$
branching fraction, with all the relevant numbers for the two
colliders. We will then give the impact of the theoretical
uncertainties on the MSSM $[M_A,\tan\beta]$ plane and in particular
revisit the limits obtained by the CDF and D0
collaborations~\cite{Benjamin:2010xb}. We will also study this impact
at the lHC and give some prospects for the higher luminosity in the 7
TeV run. We want to note that during the final stage of the writing of
the thesis new results have been presented in HEP--EPS 2011 conference
in July; they will not be commented in this section and postponed for
the conclusion of this work. We will end the section by giving a very
interesting consequence of the MSSM CMS search on the SM Higgs search
in the channel $gg\to\Phi\to\tau^+\tau^-$ that has been presented in
Ref.~\cite{Baglio:2011xz}. Again we will comment this important output
in the light of the new HEP--EPS 2011 results in the conclusion of the
thesis.

\label{section:MSSMHiggsExp}

\subsection{The main MSSM Higgs branching
  ratios \label{section:MSSMHiggsExpDecay}}

In the most general case, the decay pattern of the MSSM Higgs particles can be
rather complicated, in particular for the heavy states. Indeed, besides the
standard decays into pairs of fermions and gauge bosons, the latter can have
mixed  decays into gauge and Higgs bosons (and the $H$ bosons can decay into
$hh$ states)  and, if some superparticles are light, SUSY decays would also
occur. However,  for the large values of $\tb$ that we are interested in here,
$\tb \gsim 10$, the couplings of the non--SM like Higgs particles to bottom
quarks and $\tau$ leptons are so strongly enhanced and those to top quarks and
gauge bosons    suppressed, that the pattern becomes very simple. To a very
good approximation, the $\Phi=A$ or $H(h)$  bosons  will  decay  almost
exclusively into $b\bar b$ and $\tau^+\tau^-$ pairs with branching ratios of,
respectively, $\approx 90\%$ and $ \approx 10\%$, with the $t\bar t$ decay
channel and the decay involving gauge or Higgs bosons suppressed to a level
where the branching ratios are less than 1\%. The $CP$--even  $h$ or $H$ boson,
depending on whether we are in the decoupling or anti--decoupling regime, will
have the same decays as the SM Higgs boson in the mass range below
$M_h^{\rm max} \lsim 135$ GeV. The results presented in
section~\ref{section:SMHiggsDecay} can therefore be used for the study
of the decay pattern of the neutral SM--like Higgs boson\footnote{It
  is interesting to note that for $M_h^{\rm max}  \sim 135$ GeV, we
  are in the regime where the decays into $b \bar b$ and $WW$  are
  comparable and, thus, the uncertainties in the Higgs branching
  ratios are the largest.}

\begin{table}[!h]{\small%
\let\lbr\{\def\{{\char'173}%
\let\rbr\}\def\}{\char'175}%
\renewcommand{\arraystretch}{1.27}
\vspace*{1mm}
\begin{center}
\begin{tabular}{|c||cccc|cccc|}\hline
$\!M_\Phi\!$ &$\!b\bar b\!$   & $\!\Delta m_b\!$ &
$\!\Delta\alpha_s\!$ & ~tot~ & $\!\tau\tau\!$ & $\!\Delta m_b\!$ &
$\!\Delta\alpha_s\!$ & ~tot~ \\ \hline
$90$ & 90.40 & $^{+0.9\%}_{-0.3\%}$ &
$^{+0.2\%}_{-0.2\%}$ & $^{+0.9\%}_{-0.4\%}$ &
9.60 & $^{+3.2\%}_{-8.6\%}$ &
$^{+1.9\%}_{-1.8\%}$ & $^{+3.8\%}_{-8.8\%}$ \\ \hline
$100$ & 90.21 & $^{+0.9\%}_{-0.4\%}$ &
$^{+0.2\%}_{-0.2\%}$ & $^{+1.0\%}_{-0.4\%}$ &
9.79 & $^{+3.3\%}_{-8.6\%}$ &
$^{+1.9\%}_{-1.8\%}$ & $^{+3.8\%}_{-8.8\%}$ \\ \hline
$110$ & 90.04 & $^{+0.9\%}_{-0.4\%}$ &
$^{+0.2\%}_{-0.2\%}$ & $^{+1.0\%}_{-0.4\%}$ &
9.96 & $^{+3.2\%}_{-8.6\%}$ &
$^{+2.0\%}_{-1.9\%}$ & $^{+3.8\%}_{-8.8\%}$ \\ \hline
$120$ & 89.88 & $^{+1.0\%}_{-0.4\%}$ &
$^{+0.2\%}_{-0.2\%}$ & $^{+1.0\%}_{-0.4\%}$ &
10.12 & $^{+3.3\%}_{-8.6\%}$ &
$^{+2.0\%}_{-1.9\%}$ & $^{+3.8\%}_{-8.8\%}$ \\ \hline
$130$ & 89.74 & $^{+1.0\%}_{-0.4\%}$ &
$^{+0.2\%}_{-0.2\%}$ & $^{+1.0\%}_{-0.4\%}$ &
10.26 & $^{+3.2\%}_{-8.5\%}$ &
$^{+2.0\%}_{-1.9\%}$ & $^{+3.8\%}_{-8.7\%}$ \\ \hline
$140$ & 89.61 & $^{+1.0\%}_{-0.4\%}$ &
$^{+0.2\%}_{-0.2\%}$ & $^{+1.0\%}_{-0.4\%}$ &
10.39 & $^{+3.3\%}_{-8.5\%}$ &
$^{+2.0\%}_{-1.9\%}$ & $^{+3.8\%}_{-8.7\%}$ \\ \hline
$150$ & 89.48 & $^{+1.0\%}_{-0.4\%}$ &
$^{+0.2\%}_{-0.2\%}$ & $^{+1.0\%}_{-0.4\%}$ &
10.52 & $^{+3.1\%}_{-8.6\%}$ &
$^{+2.0\%}_{-2.0\%}$ & $^{+3.7\%}_{-8.8\%}$ \\ \hline
$160$ & 89.37 & $^{+1.0\%}_{-0.4\%}$ &
$^{+0.2\%}_{-0.2\%}$ & $^{+1.0\%}_{-0.5\%}$ &
10.63 & $^{+3.2\%}_{-8.5\%}$ &
$^{+2.1\%}_{-2.0\%}$ & $^{+3.8\%}_{-8.7\%}$ \\ \hline
$170$ & 89.26 & $^{+1.0\%}_{-0.4\%}$ &
$^{+0.2\%}_{-0.2\%}$ & $^{+1.1\%}_{-0.5\%}$ &
10.74 & $^{+3.2\%}_{-8.5\%}$ &
$^{+2.0\%}_{-2.0\%}$ & $^{+3.8\%}_{-8.8\%}$ \\ \hline
$180$ & 89.16 & $^{+1.0\%}_{-0.4\%}$ &
$^{+0.2\%}_{-0.3\%}$ & $^{+1.1\%}_{-0.5\%}$ &
10.84 & $^{+3.2\%}_{-8.5\%}$ &
$^{+2.1\%}_{-1.9\%}$ & $^{+3.9\%}_{-8.7\%}$ \\ \hline
$190$ & 89.05 & $^{+1.0\%}_{-0.4\%}$ &
$^{+0.2\%}_{-0.2\%}$ & $^{+1.1\%}_{-0.5\%}$ &
10.95 & $^{+3.2\%}_{-8.5\%}$ &
$^{+2.0\%}_{-2.0\%}$ & $^{+3.8\%}_{-8.7\%}$ \\ \hline
$200$ & 88.96 & $^{+1.0\%}_{-0.4\%}$ &
$^{+0.2\%}_{-0.3\%}$ & $^{+1.1\%}_{-0.5\%}$ &
11.04 & $^{+3.3\%}_{-8.4\%}$ &
$^{+2.2\%}_{-2.0\%}$ & $^{+3.9\%}_{-8.7\%}$ \\ \hline
$250$ & 88.53 & $^{+1.1\%}_{-0.4\%}$ &
$^{+0.3\%}_{-0.3\%}$ & $^{+1.1\%}_{-0.5\%}$ &
11.47 & $^{+3.1\%}_{-8.5\%}$ &
$^{+2.1\%}_{-2.1\%}$ & $^{+3.8\%}_{-8.7\%}$ \\ \hline
$300$ & 88.19 & $^{+1.1\%}_{-0.4\%}$ &
$^{+0.3\%}_{-0.3\%}$ & $^{+1.2\%}_{-0.5\%}$ &
11.81 & $^{+3.1\%}_{-8.5\%}$ &
$^{+2.1\%}_{-2.1\%}$ & $^{+3.8\%}_{-8.7\%}$ \\ \hline
$350$ & 87.90 & $^{+1.2\%}_{-0.4\%}$ &
$^{+0.3\%}_{-0.3\%}$ & $^{+1.2\%}_{-0.5\%}$ &
12.10 & $^{+3.1\%}_{-8.4\%}$ &
$^{+2.1\%}_{-2.1\%}$ & $^{+3.8\%}_{-8.7\%}$ \\ \hline
$400$ & 87.66 & $^{+1.2\%}_{-0.4\%}$ &
$^{+0.3\%}_{-0.3\%}$ & $^{+1.2\%}_{-0.5\%}$ &
12.34 & $^{+3.2\%}_{-8.3\%}$ &
$^{+2.2\%}_{-2.1\%}$ & $^{+3.8\%}_{-8.6\%}$ \\ \hline
$450$ & 87.44 & $^{+1.2\%}_{-0.4\%}$ &
$^{+0.3\%}_{-0.3\%}$ & $^{+1.2\%}_{-0.5\%}$ &
12.56 & $^{+3.1\%}_{-8.4\%}$ &
$^{+2.2\%}_{-2.1\%}$ & $^{+3.8\%}_{-8.6\%}$ \\ \hline
$500$ & 87.25 & $^{+1.2\%}_{-0.5\%}$ &
$^{+0.3\%}_{-0.3\%}$ & $^{+1.3\%}_{-0.6\%}$ &
12.75 & $^{+3.1\%}_{-8.3\%}$ &
$^{+2.3\%}_{-2.1\%}$ & $^{+3.9\%}_{-8.6\%}$ \\ \hline
\end{tabular} 
\end{center} 
\vspace*{-3mm}
\caption[The main MSSM $CP$--odd like Higgs bosons decay branching
fractions together with their uncertainties]{The MSSM $CP$--odd like
  Higgs bosons decay branching ratios into $b\bar b$ and $\tau^+\tau^-$
  final states (in \%) for given Higgs mass values (in GeV) with the
  corresponding individual uncertainties as well as the global
  uncertainties assuming $1\sigma$ uncertainties on the inputs
  $\overline{m}_b(\overline{m}_b)$ and $\alpha_s(M_Z^2)$.}
\label{table:BR-mssm}
\vspace*{-4mm}
}
\end{table}

For the evaluation of the theoretical uncertainties in the $b\bar b$ and $\tau^+
\tau^-$ decay branching ratios of the $\Phi$ states, the analysis of
section~\ref{section:SMHiggsDecay} for the SM Higgs boson can be
straightforwardly extended to the MSSM case. In this case we will
ignore the experimental error on the imput charm quark mass: indeed
the decay $\Phi\to c\bar c$ being strongly suppressed the impact of
the $c$ quark mass on the $\Phi\to\tau^+\tau^-$ and $\Phi\to b\bar b$
branching fractions is totally negligible as its impact was mainly on
the $\Phi\to c\bar c$ decay width which would translate on the
branching fractions. As the QCD corrections to the dominant $\Phi  \to
b\bar b$ decays are large, they are resummed by switching from the
$b$--quark pole mass $M_b$ which appears at tree--level to the running
quark mass in the $\overline{\rm MS}$   scheme evaluated at the scale
of the Higgs mass, $\overline{m}_b(M_\Phi )$. We will then consider
only the uncertainties coming from the two other sources: the inputs
$\overline{m}_b(\overline{m}_b)=4.19^{+0.18} _{-0.06}$
GeV\cite{Nakamura:2010zzi} where the central value corresponds to a
pole mass of $M_b=4.71$ GeV and the experimental errors on the QCD
coupling constant $\alpha_s(M_Z^2)\!=\! 0.1171 \pm 0.0014$ at NNLO
(the value adopted in the cross sections)~\cite{Martin:2009iq,
  Martin:2009bu, Martin:2010db} which is used to run the $b$--quark
mass from $\overline{m}_b$ up to $M_\Phi$. Again, the
impact of a scale variation in the range $\frac12 M_\Phi \leq \mu \leq
2M_\Phi$ is negligibly small. The uncertainties on the two branching
ratios are displayed in Table~\ref{table:BR-mssm} for some values of the
Higgs boson mass, together with the total uncertainties when the
individual uncertainties resulting from the ``1$\sigma$" errors on the
inputs $\overline{m}_b(\overline{m}_b)$ and $\alpha_s(M_Z^2)$ are
added in quadrature.

The $b\bar b$ branching ratios of the $\Phi$ states, BR$(\Phi \to
b\bar b)\approx 3 \overline{m}_b^2(M_\Phi)/[3\overline{m}_b^2(M_\Phi)+m_\tau^2]$,
slightly decrease with increasing $M_\Phi$ as a result of the higher
scale which reduces the $b$--quark mass $\overline{m}_b (M_\Phi)$, but
the total uncertainty is practically constant on the entire mass range
$90\leq M_\Phi\leq 200$ GeV and amounts to approximately $+1\%,-0.5\%$
as a consequence of the almost complete cancellation of the
uncertainties in the numerator and denominator. In contrast, there is
no such a cancellation in the branching fraction for Higgs decays into
$\tau^+\tau^-$ pairs, BR$(\Phi \to \tau^+ \tau^-) \approx m_\tau^2/[3
\overline{m}_b^2(M_\Phi)+m_\tau^2]$, and the total uncertainty, that
is dominated by  the error on the $b$--quark mass, reaches the level
of $+4\%,-9\%$ for all Higgs masses. If the error on the input
$b$--quark mass is ignored, the total uncertainty  in the $\tau^+
\tau^-$ branching ratio will reduce to $\approx 3\%$.

Before closing this subsection it should be noted that in the MSSM at
high $\tb$, the total decay widths of the $\Phi$ particles should be
taken into account. Indeed, they rise as $\Gamma(\Phi) \propto M_\Phi
\tan^2\beta$ and thus, reach the level of ${\cal O}(10~{\rm GeV})$ for
$M_\Phi \approx 200$ GeV and $\tb \approx 50$. The total Higgs width
can thus possibly be larger than the experimental resolution on the
$\tau^+ \tau^-$ and $b\bar b$ invariant masses when decays into these
final states are analyzed. If the total width has to be taken into
account in the experimental analyses, the uncertainties that affect it
should also be considered. These uncertainties are in fact simply
those affecting the $\Phi \to b\bar b$ partial widths (being the
dominant channel) and thus, to a good approximation, the $\Phi \to
\tau^+ \tau^-$ branching ratio. The numbers given in Table
\ref{table:BR-mssm} for the uncertainties of BR$(\Phi \to \tau^+
\tau^-)$ thus correspond (when multiplied by a factor $\approx 1.1$)
to the uncertainties on the total width $\Gamma(\Phi)$ with a good
accuracy.

\subsection{Combination of production cross section and Higgs$\to
  \tau \tau$ decay \label{section:MSSMHiggsExpTotal}}

This subsection will combine all the results obtained in the previous
section in order to obtain the final predictions at the Tevatron and
the lHC for the process $gg+b\bar b \to \Phi\to \tau^+\tau^-$ which is
used by experimental collaborations to set up limit on the
$[M_A,\tan\beta]$ MSSM parameter space. We will start by the Tevatron
study and then give the lHC results.

\subsubsection{The combination at the Tevatron}

To obtain the total uncertainty on the cross section times branching
ratio $\Delta ({\rm \sigma \times BR})$, we use the simple scheme
which consists in the addition of the total uncertainties on the
production cross sections and the uncertainties on the branching
fraction in Higgs decays into $\tau^+ \tau^-$ pairs. In this addition,
at least in the $b\bar b\to \Phi$ process where one defines the $\Phi
b\bar b$ coupling at the scale $\mu_R$, the uncertainty on the input
$b$--mass, which is common  to $\sigma$  and to BR, almost cancels
out; only $\approx 10\%$ of the error is left 
out as this uncertainty is anti--correlated between the production
cross section and the decay branching fraction\footnote{We remind that
  this is also the case of the $\Delta_b$ SUSY correction which enters
  both $\sigma(gg, b\bar b\to  \Phi)$ and BR$(\Phi \to \tau^+ \tau^-)$
  and which thus cancels in the product as shown in
  section~\ref{section:MSSMHiggsIntroModel}.}. In the case of $gg \to
\Phi$ where the Yukawa coupling is defined at the scale $m_b$ in
contrast to the Higgs decay widths, the errors on  $\alpha_s$ used for
the running from $m_b$ to $M_\Phi$ will induce a remaining
uncertainty. This $\alpha_s$ uncertainty is correlated in $\sigma(gg
\to \Phi)$ and BR$(\Phi \to \tau^+ \tau^-)$:  smaller $\alpha_s$
values lower $\sigma(gg \to \Phi) \propto \alpha_s^2$ at LO, and also
BR$(\Phi \to \tau^+ \tau^-)$ as the resulting $\bar m_b(M_\Phi)$ is
higher, thus enhancing/reducing the $\Phi \to b \bar b/\tau^+ \tau^-$
rates. This uncertainty should thus be added linearly to the overall
scale+scheme+PDF uncertainty of the cross section. 

We display the numerical results for the two processes $gg\to\Phi$ and
$b\bar b\to\Phi$ in the Tables.~\ref{table:ggPhiTev}
and~\ref{table:bbPhiTev} which include the individual uncertainties as
well as the total uncertainty, including or not the impact of the
branching fraction. This impact is modest but sizable in the upper
direction for both processes, and negligible in the lower direction in
the case of the bottom quarks fusion.

\begin{table}[!h]{\small%
\let\lbr\{\def\{{\char'173}%
\let\rbr\}\def\}{\char'175}%
\renewcommand{\arraystretch}{1.45}
\begin{center}
\begin{tabular}{|c||c|ccccc||cc|}\hline
$M_\Phi$ & $\sigma^{\rm NLO}_{\rm gg \to \Phi}$ & scale & scheme &
PDFs & param & total & $\sigma \times
$BR$ \hspace*{-9mm}$ &
 \\ \hline
$90$ & $49.46$ & $^{+22.0\%}_{-18.5\%}$ & $^{+5.6\%}_{-5.6\%}$ &
$^{+10.7\%}_{-10.3\%}$ & $^{+13.0\%}_{-4.0\%}$ & $^{+55.2\%}_{-35.9\%}$ &
$4.75$ & $^{+49.7\%}_{-35.7\%}$ \\ \hline
$95$ & $36.07$ & $^{+22.0\%}_{-18.6\%}$ & $^{+5.6\%}_{-5.6\%}$ &
$^{+10.8\%}_{-10.5\%}$ & $^{+13.0\%}_{-4.0\%}$ & $^{+55.3\%}_{-36.1\%}$ &
$3.50$ & $^{+49.9\%}_{-35.9\%}$ \\ \hline
$100$ & $26.62$ & $^{+21.9\%}_{-18.6\%}$ & $^{+5.5\%}_{-5.5\%}$ &
$^{+10.9\%}_{-10.7\%}$ & $^{+13.1\%}_{-4.0\%}$ & $^{+55.4\%}_{-36.3\%}$ &
$2.61$ & $^{+50.0\%}_{-36.0\%}$ \\ \hline
$105$ & $19.90$ & $^{+21.9\%}_{-18.7\%}$ & $^{+5.5\%}_{-5.5\%}$ &
$^{+11.1\%}_{-10.8\%}$ & $^{+13.0\%}_{-4.0\%}$ & $^{+55.5\%}_{-36.5\%}$ &
$1.97$ & $^{+50.1\%}_{-36.3\%}$ \\ \hline
$110$ & $15.03$ & $^{+21.8\%}_{-18.7\%}$ & $^{+5.5\%}_{-5.5\%}$ &
$^{+11.2\%}_{-11.0\%}$ & $^{+13.1\%}_{-4.1\%}$ & $^{+55.7\%}_{-36.7\%}$ &
$1.50$ & $^{+50.3\%}_{-36.6\%}$ \\ \hline
$115$ & $11.46$ & $^{+21.7\%}_{-18.8\%}$ & $^{+5.5\%}_{-5.5\%}$ &
$^{+11.4\%}_{-11.2\%}$ & $^{+13.1\%}_{-4.0\%}$ & $^{+55.7\%}_{-36.9\%}$ &
$1.15$ & $^{+50.4\%}_{-36.7\%}$ \\ \hline
$120$ & $8.82$ & $^{+21.6\%}_{-18.8\%}$ & $^{+5.5\%}_{-5.5\%}$ &
$^{+11.5\%}_{-11.4\%}$ & $^{+13.2\%}_{-4.1\%}$ & $^{+55.8\%}_{-37.1\%}$ &
$0.89$ & $^{+50.5\%}_{-37.0\%}$ \\ \hline
$125$ & $6.84$ & $^{+21.6\%}_{-18.8\%}$ & $^{+5.4\%}_{-5.4\%}$ &
$^{+11.7\%}_{-11.7\%}$ & $^{+13.2\%}_{-4.1\%}$ & $^{+56.0\%}_{-37.3\%}$ &
$0.70$ & $^{+50.6\%}_{-37.1\%}$ \\ \hline
$130$ & $5.35$ & $^{+21.5\%}_{-18.9\%}$ & $^{+5.4\%}_{-5.4\%}$ &
$^{+11.8\%}_{-11.9\%}$ & $^{+13.2\%}_{-4.1\%}$ & $^{+56.0\%}_{-37.6\%}$ &
$0.55$ & $^{+50.9\%}_{-37.4\%}$ \\ \hline
$135$ & $4.21$ & $^{+21.5\%}_{-18.9\%}$ & $^{+5.4\%}_{-5.4\%}$ &
$^{+12.0\%}_{-12.1\%}$ & $^{+13.3\%}_{-4.1\%}$ & $^{+56.3\%}_{-37.8\%}$ &
$0.43$ & $^{+51.0\%}_{-37.8\%}$ \\ \hline
$140$ & $3.34$ & $^{+21.4\%}_{-19.0\%}$ & $^{+5.4\%}_{-5.4\%}$ &
$^{+12.2\%}_{-12.3\%}$ & $^{+13.3\%}_{-4.1\%}$ & $^{+56.4\%}_{-38.0\%}$ &
$0.35$ & $^{+51.2\%}_{-37.9\%}$ \\ \hline
$145$ & $2.66$ & $^{+21.4\%}_{-19.0\%}$ & $^{+5.4\%}_{-5.4\%}$ &
$^{+12.3\%}_{-12.5\%}$ & $^{+13.3\%}_{-4.1\%}$ & $^{+56.6\%}_{-38.2\%}$ &
$0.28$ & $^{+51.3\%}_{-38.4\%}$ \\ \hline
$150$ & $2.13$ & $^{+21.3\%}_{-19.1\%}$ & $^{+5.4\%}_{-5.4\%}$ &
$^{+12.5\%}_{-12.8\%}$ & $^{+13.3\%}_{-4.1\%}$ & $^{+56.8\%}_{-38.4\%}$ &
$0.22$ & $^{+51.5\%}_{-38.6\%}$ \\ \hline
$155$ & $1.72$ & $^{+21.3\%}_{-19.1\%}$ & $^{+5.4\%}_{-5.4\%}$ &
$^{+12.7\%}_{-13.0\%}$ & $^{+13.3\%}_{-4.1\%}$ & $^{+56.9\%}_{-38.7\%}$ &
$0.18$ & $^{+51.8\%}_{-38.5\%}$ \\ \hline
$160$ & $1.40$ & $^{+21.2\%}_{-19.2\%}$ & $^{+5.4\%}_{-5.4\%}$ &
$^{+12.9\%}_{-13.2\%}$ & $^{+13.4\%}_{-4.1\%}$ & $^{+57.1\%}_{-38.9\%}$ &
$0.15$ & $^{+52.0\%}_{-38.9\%}$ \\ \hline
$165$ & $1.14$ & $^{+21.2\%}_{-19.2\%}$ & $^{+5.4\%}_{-5.4\%}$ &
$^{+13.1\%}_{-13.4\%}$ & $^{+13.4\%}_{-4.1\%}$ & $^{+57.4\%}_{-39.1\%}$ &
$0.12$ & $^{+52.3\%}_{-39.1\%}$ \\ \hline
$170$ & $0.93$ & $^{+21.2\%}_{-19.3\%}$ & $^{+5.4\%}_{-5.4\%}$ &
$^{+13.3\%}_{-13.6\%}$ & $^{+13.4\%}_{-4.1\%}$ & $^{+57.6\%}_{-39.3\%}$ &
$0.10$ & $^{+52.4\%}_{-39.4\%}$ \\ \hline
$175$ & $0.76$ & $^{+21.2\%}_{-19.3\%}$ & $^{+5.4\%}_{-5.4\%}$ &
$^{+13.5\%}_{-13.9\%}$ & $^{+13.4\%}_{-4.2\%}$ & $^{+57.8\%}_{-39.6\%}$ &
$0.08$ & $^{+52.7\%}_{-39.5\%}$ \\ \hline
$180$ & $0.63$ & $^{+21.1\%}_{-19.4\%}$ & $^{+5.4\%}_{-5.4\%}$ &
$^{+13.7\%}_{-14.1\%}$ & $^{+13.4\%}_{-4.1\%}$ & $^{+58.1\%}_{-39.8\%}$ &
$0.07$ & $^{+53.1\%}_{-39.7\%}$ \\ \hline
$185$ & $0.52$ & $^{+21.1\%}_{-19.4\%}$ & $^{+5.4\%}_{-5.4\%}$ &
$^{+14.0\%}_{-14.3\%}$ & $^{+13.4\%}_{-4.2\%}$ & $^{+58.3\%}_{-40.0\%}$ &
$0.06$ & $^{+53.1\%}_{-40.2\%}$ \\ \hline
$190$ & $0.43$ & $^{+21.1\%}_{-19.5\%}$ & $^{+5.4\%}_{-5.4\%}$ &
$^{+14.3\%}_{-14.5\%}$ & $^{+13.5\%}_{-4.2\%}$ & $^{+58.5\%}_{-40.3\%}$ &
$0.05$ & $^{+53.3\%}_{-40.4\%}$ \\ \hline
$195$ & $0.36$ & $^{+21.1\%}_{-19.5\%}$ & $^{+5.4\%}_{-5.4\%}$ &
$^{+14.6\%}_{-14.7\%}$ & $^{+13.4\%}_{-4.2\%}$ & $^{+58.8\%}_{-40.5\%}$ &
$0.04$ & $^{+53.7\%}_{-40.6\%}$ \\ \hline
$200$ & $0.30$ & $^{+21.1\%}_{-19.6\%}$ & $^{+5.4\%}_{-5.4\%}$ &
$^{+14.9\%}_{-15.0\%}$ & $^{+13.4\%}_{-4.2\%}$ & $^{+59.0\%}_{-40.7\%}$ &
$0.03$ & $^{+54.2\%}_{-40.7\%}$ \\ \hline
\end{tabular}
\end{center}
\caption[The central predictions in the MSSM $g g\to\Phi$ channel at
the Tevatron together with the detailed uncertainties and the impact
of the $\Phi\to\tau^+\tau^-$ branching fraction]{The production cross
  sections in the $\protect{gg\to \Phi}$ process at the Tevatron (in
  fb) for given $A$ masses (in GeV) at a scale $\mu_F=\mu_R =\frac12
  M_\Phi$ with MSTW PDFs. Shown also are the corresponding
  uncertainties from the various sources discussed as well as the total
  uncertainty. In the last column is displayed the product $\sigma(gg\to
  \Phi)\times$BR$(\Phi\to \tau^+\tau^-)$ together with its
  total uncertainty.}
\label{table:ggPhiTev}
\vspace*{-1mm}
}
\end{table}

\begin{table}[!h]{\small%
\let\lbr\{\def\{{\char'173}%
\let\rbr\}\def\}{\char'175}%
\renewcommand{\arraystretch}{1.45}
\begin{center}
\begin{tabular}{|c||c|cccc||cc|}\hline
$~~M_\Phi~~$ & $\sigma^{\rm NNLO}_{\rm b\bar b \to \Phi}$ & scale &
PDFs & param & total & $\sigma \times
$BR$ \hspace*{-5mm}$ &
 \\ \hline
$90$ & $26.31$ & $^{+35.1\%}_{-29.9\%}$ &
$^{+18.2\%}_{-17.2\%}$ & $^{+10.5\%}_{-3.3\%}$ & $^{+70.4\%}_{-44.9\%}$ &
$2.53$ & $^{+63.4\%}_{-44.0\%}$ \\ \hline
$95$ & $21.04$ & $^{+30.6\%}_{-29.0\%}$ &
$^{+19.0\%}_{-16.6\%}$ & $^{+9.9\%}_{-3.7\%}$ & $^{+64.1\%}_{-44.9\%}$ &
$2.04$ & $^{+57.1\%}_{-43.9\%}$ \\ \hline
$100$ & $16.96$ & $^{+26.6\%}_{-27.0\%}$ &
$^{+18.7\%}_{-17.5\%}$ & $^{+10.2\%}_{-3.4\%}$ & $^{+61.5\%}_{-42.9\%}$ &
$1.66$ & $^{+54.5\%}_{-42.0\%}$ \\ \hline
$105$ & $13.78$ & $^{+22.6\%}_{-27.2\%}$ &
$^{+19.2\%}_{-17.6\%}$ & $^{+10.7\%}_{-3.4\%}$ & $^{+57.8\%}_{-43.4\%}$ &
$1.36$ & $^{+50.8\%}_{-42.4\%}$ \\ \hline
$110$ & $11.22$ & $^{+20.3\%}_{-25.5\%}$ &
$^{+19.4\%}_{-17.9\%}$ & $^{+10.1\%}_{-3.6\%}$ & $^{+54.7\%}_{-42.6\%}$ &
$1.12$ & $^{+47.9\%}_{-41.7\%}$ \\ \hline
$115$ & $9.18$ & $^{+17.6\%}_{-24.8\%}$ &
$^{+20.1\%}_{-17.9\%}$ & $^{+10.6\%}_{-3.6\%}$ & $^{+52.7\%}_{-41.6\%}$ &
$0.92$ & $^{+45.7\%}_{-40.6\%}$ \\ \hline
$120$ & $7.57$ & $^{+15.1\%}_{-24.1\%}$ &
$^{+21.0\%}_{-17.8\%}$ & $^{+10.9\%}_{-3.0\%}$ & $^{+50.2\%}_{-40.3\%}$ &
$0.77$ & $^{+43.1\%}_{-40.0\%}$ \\ \hline
$125$ & $6.29$ & $^{+13.6\%}_{-23.5\%}$ &
$^{+21.5\%}_{-18.1\%}$ & $^{+10.2\%}_{-3.2\%}$ & $^{+47.9\%}_{-40.4\%}$ &
$0.64$ & $^{+40.7\%}_{-39.5\%}$ \\ \hline
$130$ & $5.24$ & $^{+12.8\%}_{-22.9\%}$ &
$^{+21.7\%}_{-18.6\%}$ & $^{+10.5\%}_{-3.2\%}$ & $^{+48.0\%}_{-40.4\%}$ &
$0.54$ & $^{+41.0\%}_{-39.5\%}$ \\ \hline
$135$ & $4.36$ & $^{+12.4\%}_{-21.9\%}$ &
$^{+21.8\%}_{-19.0\%}$ & $^{+10.5\%}_{-3.3\%}$ & $^{+48.5\%}_{-39.9\%}$ &
$0.45$ & $^{+41.6\%}_{-38.9\%}$ \\ \hline
$140$ & $3.66$ & $^{+11.6\%}_{-21.5\%}$ &
$^{+22.5\%}_{-19.1\%}$ & $^{+10.2\%}_{-3.6\%}$ & $^{+46.9\%}_{-40.2\%}$ &
$0.38$ & $^{+39.8\%}_{-39.3\%}$ \\ \hline
$145$ & $3.08$ & $^{+11.0\%}_{-21.2\%}$ &
$^{+23.0\%}_{-19.5\%}$ & $^{+10.4\%}_{-3.3\%}$ & $^{+46.7\%}_{-40.3\%}$ &
$0.32$ & $^{+39.5\%}_{-39.4\%}$ \\ \hline
$150$ & $2.60$ & $^{+10.8\%}_{-20.3\%}$ &
$^{+24.1\%}_{-19.3\%}$ & $^{+10.4\%}_{-3.4\%}$ & $^{+47.3\%}_{-39.7\%}$ &
$0.27$ & $^{+39.9\%}_{-38.9\%}$ \\ \hline
$155$ & $2.20$ & $^{+10.8\%}_{-20.0\%}$ &
$^{+24.3\%}_{-20.1\%}$ & $^{+10.3\%}_{-3.5\%}$ & $^{+47.8\%}_{-39.3\%}$ &
$0.23$ & $^{+40.8\%}_{-38.2\%}$ \\ \hline
$160$ & $1.88$ & $^{+10.5\%}_{-20.0\%}$ &
$^{+24.9\%}_{-20.4\%}$ & $^{+10.4\%}_{-3.4\%}$ & $^{+47.8\%}_{-40.0\%}$ &
$0.20$ & $^{+40.5\%}_{-39.1\%}$ \\ \hline
$165$ & $1.60$ & $^{+10.6\%}_{-18.7\%}$ &
$^{+25.1\%}_{-20.6\%}$ & $^{+10.4\%}_{-3.3\%}$ & $^{+49.2\%}_{-38.5\%}$ &
$0.17$ & $^{+41.8\%}_{-37.4\%}$ \\ \hline
$170$ & $1.37$ & $^{+10.6\%}_{-18.4\%}$ &
$^{+25.3\%}_{-21.1\%}$ & $^{+10.5\%}_{-3.1\%}$ & $^{+49.8\%}_{-38.7\%}$ &
$0.15$ & $^{+42.4\%}_{-37.8\%}$ \\ \hline
$175$ & $1.17$ & $^{+10.5\%}_{-18.2\%}$ &
$^{+26.0\%}_{-21.5\%}$ & $^{+10.4\%}_{-3.4\%}$ & $^{+50.5\%}_{-38.8\%}$ &
$0.13$ & $^{+43.0\%}_{-37.6\%}$ \\ \hline
$180$ & $1.01$ & $^{+10.3\%}_{-18.1\%}$ &
$^{+26.7\%}_{-21.6\%}$ & $^{+10.4\%}_{-3.4\%}$ & $^{+50.7\%}_{-38.6\%}$ &
$0.11$ & $^{+43.2\%}_{-37.4\%}$ \\ \hline
$185$ & $0.87$ & $^{+10.4\%}_{-17.6\%}$ &
$^{+27.8\%}_{-21.5\%}$ & $^{+10.3\%}_{-3.2\%}$ & $^{+51.4\%}_{-38.6\%}$ &
$0.09$ & $^{+43.8\%}_{-37.6\%}$ \\ \hline
$190$ & $0.75$ & $^{+10.3\%}_{-17.3\%}$ &
$^{+27.2\%}_{-22.7\%}$ & $^{+10.4\%}_{-3.3\%}$ & $^{+52.1\%}_{-39.2\%}$ &
$0.08$ & $^{+44.4\%}_{-38.2\%}$ \\ \hline
$195$ & $0.65$ & $^{+10.5\%}_{-16.2\%}$ &
$^{+28.5\%}_{-22.6\%}$ & $^{+10.1\%}_{-3.4\%}$ & $^{+52.3\%}_{-38.9\%}$ &
$0.07$ & $^{+44.7\%}_{-37.7\%}$ \\ \hline
$200$ & $0.56$ & $^{+10.4\%}_{-16.3\%}$ &
$^{+28.5\%}_{-23.6\%}$ & $^{+10.6\%}_{-3.2\%}$ & $^{+53.2\%}_{-38.9\%}$ &
$0.06$ & $^{+45.8\%}_{-37.9\%}$ \\ \hline
\end{tabular}
\end{center}
\caption[The central predictions in the MSSM $b\bar b\to\Phi$ channel
at the Tevatron together with the detailed uncertainties and the
impact of the $\Phi\to\tau^+\tau^-$ branching fraction]{Same as in
  Table~\ref{table:ggPhiTev} for the $b\bar b\to\Phi$ channel.}
\label{table:bbPhiTev}
\vspace*{-1mm}
}
\end{table}

\begin{figure}[!h]
\begin{center}
\vspace*{2mm}
\includegraphics[scale=0.75]{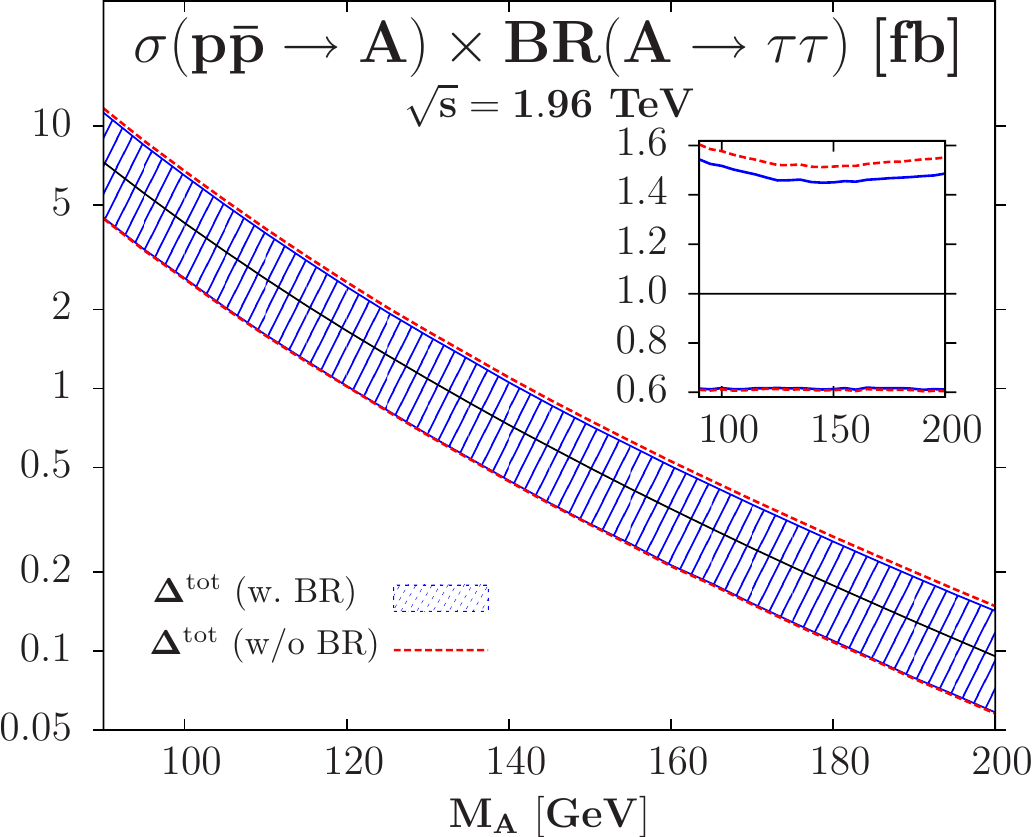}
\end{center}
\vspace*{-4mm}
\caption[$CP$--odd $A$ boson production in the $p\bar
p\to A\to\tau^+\tau^-$ channel at the Tevatron together with the
total uncertainty]{$\sigma(p\bar p\! \to\! A)\times {\rm BR}(A \!\to\!
  \tau^+ \tau^-)$ as a function of $M_A$ at the Tevatron, together
  with the associated overall theoretical uncertainty;  the
  uncertainty when excluding that on the branching ratio is also
  displayed. In the inserts, shown are the  relative deviations from
  the central values.}
\vspace*{-1mm}
\label{fig:TotalMSSMTev}
\end{figure}

We are now left with the final combination which is of utmost
interest: the combination of the two production times decay
processes. Indeed, to consider the final state topology that has been
searched for by the D0 and CDF collaborations~\cite{Benjamin:2010xb},
that is $p\bar p\! \to\! {\rm Higgs}\! \to\! \tau^+\tau^-$,  we first
have to add the cross sections for the two channels $gg \to A$ and
$b\bar b \to A$, and then to multiply the resulting production cross
section by the Higgs branching ratio BR($A \to \tau^+\tau^-) \approx
10\%$\footnote{We simply add all uncertainties
linearly, in contrast to Ref.~\cite{Dittmaier:2011ti} in which the PDF
uncertainties in $gg$ and $b\bar b \to A$ are added in quadrature,
with the total PDF+$\alpha_s$ uncertainty added linearly to the scale
uncertainty.}. The resulting $\sigma(p\bar p\!\to \! A)\!\times\! {\rm
BR}(A\!\to\! \tau^+ \tau^-)$ at the Tevatron is shown in
Fig.~\ref{fig:TotalMSSMTev} as a function of $M_A$. We stress again
that to obtain the true rate for $\Phi\!=\!A\!+\!H(h)$, one  has to
multiply the values given in the figure by a factor $2\tan^2\beta$.

In Fig.~\ref{fig:TotalMSSMTev} we also display the associated overall
theoretical uncertainties. The uncertainty from the cross section
alone is dominated by that of $gg\!\to\!A$ at low Higgs masses and
$b\bar b\! \to\! A$ at high masses as the corresponding cross sections
are largest. In the product $\sigma(p\bar p \to A) \times {\rm
  BR}(A\to \tau^+ \tau^-)$, the parametric uncertainty that  its
common to the production and decay rates almost cancels out as shown
by the solid curves in Fig.~\ref{fig:TotalMSSMTev} and only a few
percent are left. This leads to a smaller uncertainty in $\sigma
(p\bar p \!\to \! A) \times {\rm BR}(A\!\to\! \tau^+\tau^-) $ than in
$\sigma (p\bar p \!\to \!A)$ alone. The final theoretical uncertainty
for $p\bar p \! \to\! A \! \to\! \tau^+ \tau^-$ at the Tevatron is of
order  $+50\%, -40\%$.

\subsubsection{The combination at the lHC}

\begin{figure}[!h]
\begin{center}
\vspace*{-.2mm}
\mbox{
\includegraphics[scale=0.65]{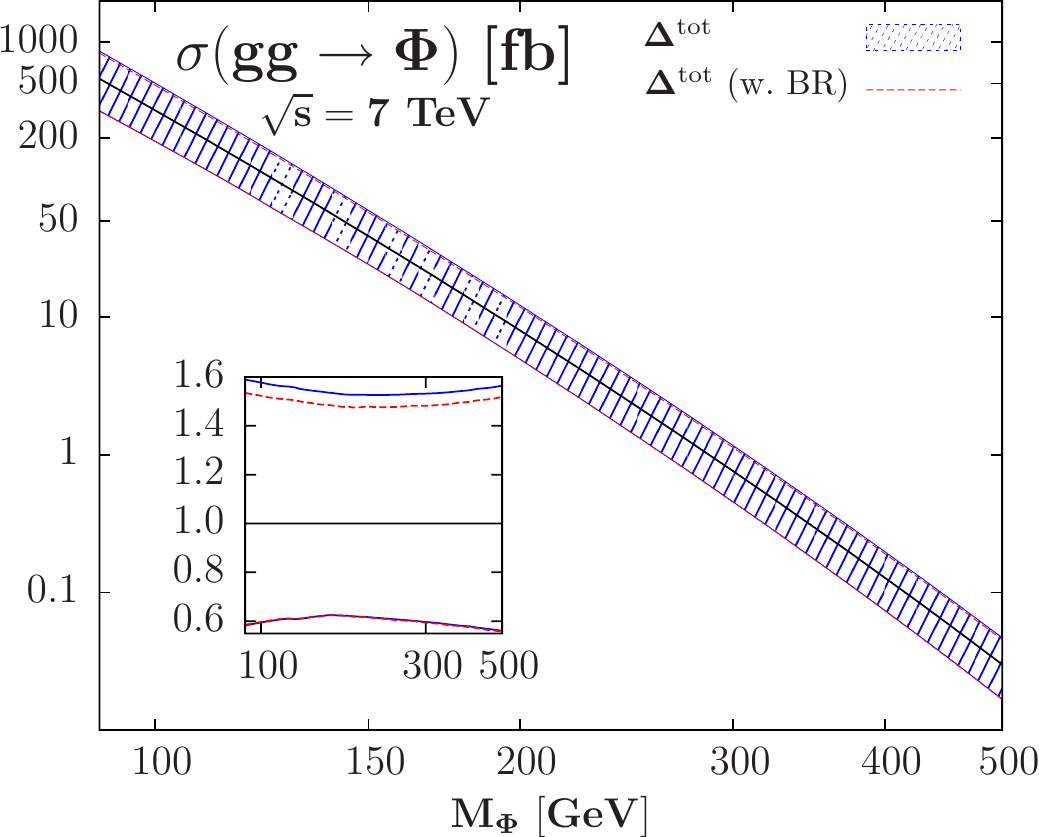}
\includegraphics[scale=0.65]{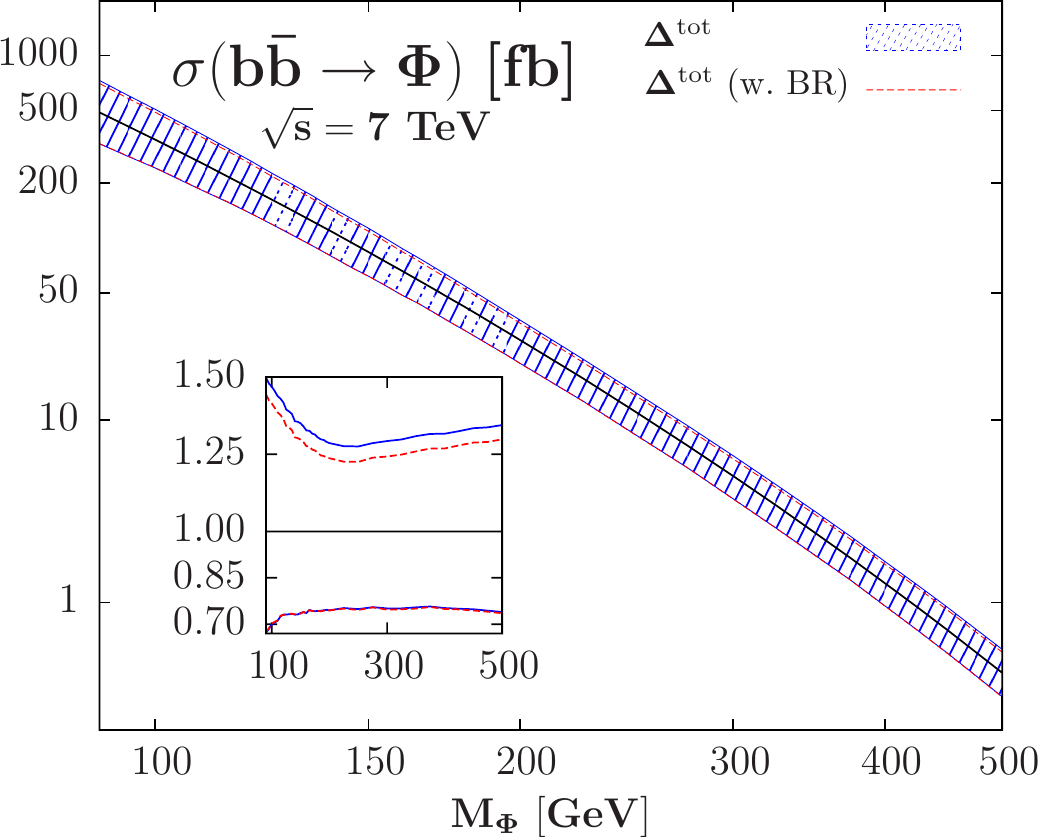}
}
\end{center}
\vspace*{-4mm}
\caption[The total uncertainties on the MSSM Higgs production in the
$gg\to\Phi$ and $b\bar b\to\Phi$ channels at the lHC including the
impact of the $\Phi\to\tau^+\tau^-$ branching fraction]{The total
  uncertainties due to the scale, PDF and $b$-quark mass in the $gg
  \to \Phi$ (left) and $b\bar b  \to \Phi$ (right) cross sections at
  the $\lhc$ with 7 TeV as a function of $M_\Phi$. The dotted lines
  show the uncertainties when those on the branching rations for Higgs
  decays into $\tau^+ \tau^-$ final states are added. In the inserts,
  the relative deviations are shown.}
\label{fig:MSSM-finalall1LHC}
\vspace*{-6mm}
\end{figure}

We do the same exercice in the case of the lHC, reproducing the lines
of argument presented in the Tevatron study above: the total
uncertainty on the cross section times branching ratio, $\Delta ({\rm
  \sigma \times BR})$ is obtained by adding the total uncertainties on
the production cross sections and the uncertainties on the branching
fraction in Higgs decays into $\tau^+ \tau^-$ pairs. Again the
uncertainty due to the parametric $b$ quark mass will almost cancel
out in the cross section times branching fraction because this
uncertainty is anti-correlated between the production and the
decay. Nevertheless, especially in the case of the bottom quarks
fusion channel, there will remain a residual uncertainty related to
the $\alpha_s$ running of the $b$ quark mass: it will be added
linearly to the scale+PDF+scheme uncertainty. The impact of the
production cross section times branching fraction uncertainty is shown
by the dotted lines of Fig.~\ref{fig:MSSM-finalall1LHC} where the
uncertainty $\Delta ({\rm   \sigma \times BR})$ for tau decays is
displayed for the two production processes. 

\vspace{5mm}
We display the numerical results for the two processes $gg\to\Phi$ and
$b\bar b\to\Phi$ in the Tables.~\ref{table:ggPhiLHC}
and~\ref{table:bbPhiLHC} which include the individual uncertainties as
well as the total uncertainty, including or not the impact of the
branching fraction. Again the impact of the branching fraction
uncertainty is modest, sizable in the upper direction and nearly
negligible in the lower direction. It should not be a surprise that
the total uncertainty may sometimes be higher (of some 0.2\%) in the
cross section section times branching fraction than in the production
cross section alone: indeed the additional $\alpha_s$ uncertainty on
the branching fraction remains and is added to the total uncertainty
on the cross section, thus sometimes overcompensating the cancellation
of the parametric $b$ quark mass uncertainty.

\begin{table}
\begin{bigcenter}
\small
\begin{tabular}{|c|ccccccc|ccc|}\hline
$M_\Phi$ & $\sigma^{\rm NLO}_{\rm gg \to \Phi}$ & Scale [\%] &
PDF+$\Delta^{\rm exp+th}_{\alpha_s}$ [\%] & $\Delta \overline{m}_b$
[\%] & Total [\%] & $\sigma \times$ BR & Total [\%] \\ \hline
$90$ & $542.28$ & ${+21.7}\;{-18.3}$ & ${+7.9}\;{-8.1}$ &
${+26.7}\;{-17.6}$ & ${+58.9}\;{-41.7}$ & $52.05$ &
${+53.5}\;{-41.5}$\\
$95$ & $414.48$ & ${+21.1}\;{-17.6}$ & ${+7.8}\;{-8.0}$ &
${+26.8}\;{-17.7}$ & ${+58.3}\;{-41.1}$ & $40.19$ &
${+52.8}\;{-40.9}$\\
$100$ & $320.31$ & ${+20.4}\;{-16.9}$ & ${+7.8}\;{-7.9}$ &
${+27.0}\;{-17.8}$ & ${+57.7}\;{-40.5}$ & $31.36$ &
${+52.3}\;{-40.3}$\\
$105$ & $250.43$ & ${+19.9}\;{-16.3}$ & ${+7.8}\;{-7.9}$ &
${+27.0}\;{-17.9}$ & ${+57.1}\;{-40.0}$ & $24.74$ &
${+51.7}\;{-39.8}$\\
$110$ & $197.61$ & ${+19.4}\;{-15.9}$ & ${+7.7}\;{-7.8}$ &
${+27.1}\;{-17.9}$ & ${+56.6}\;{-39.6}$ & $19.69$ &
${+51.3}\;{-39.4}$\\
$115$ & $157.34$ & ${+18.9}\;{-15.4}$ & ${+7.7}\;{-7.8}$
&${+27.2}\;{-18.0}$ & ${+56.2}\;{-39.2}$ & $15.80$ &
${+51.0}\;{-39.0}$\\
$120$ & $126.32$ & ${+18.5}\;{-15.1}$ & ${+7.7}\;{-7.8}$ &
${+27.3}\;{-18.0}$ & ${+56.1}\;{-39.0}$ & $12.78$ &
${+50.7}\;{-38.9}$\\
$125$ & $102.14$ & ${+18.1}\;{-15.3}$ & ${+7.7}\;{-7.7}$ &
${+27.4}\;{-18.1}$ & ${+55.8}\;{-39.2}$ & $10.41$ &
${+50.4}\;{-39.1}$\\
$130$ & $83.24$ & ${+17.7}\;{-15.1}$ & ${+7.7}\;{-7.7}$ &
${+27.4}\;{-18.1}$ & ${+55.1}\;{-39.0}$ & $8.54$ &
${+50.0}\;{-38.9}$\\
$135$ & $68.25$ & ${+17.4}\;{-14.7}$ & ${+7.7}\;{-7.7}$ &
${+27.5}\;{-18.2}$ & ${+54.8}\;{-38.7}$ & $7.05$ &
${+49.6}\;{-38.7}$\\
$140$ & $56.31$ & ${+17.1}\;{-14.2}$ & ${+7.7}\;{-7.7}$ &
${+27.6}\;{-18.2}$ & ${+54.5}\;{-38.4}$ & $5.85$ &
${+49.3}\;{-38.3}$\\
$145$ & $46.73$ & ${+16.7}\;{-13.8}$ & ${+7.6}\;{ -7.7}$ &
${+27.6}\;{-18.3}$ & ${+54.1}\;{-38.1}$ & $4.89$ &
${+48.9}\;{-38.2}$\\
$150$ & $38.98$ & ${+16.4}\;{-13.5}$ & ${+7.6}\;{-7.7}$ &
${+27.7}\;{-18.4}$ & ${+54.0}\;{-37.8}$ & $4.10$ &
${+48.7}\;{-37.9}$\\
$155$ & $32.68$ & ${+16.1}\;{-13.1}$ & ${+7.6}\;{-7.7}$ &
${+27.8}\;{-18.4}$ & ${+53.7}\;{-37.6}$ & $3.45$ &
${+48.6}\;{-37.5}$\\
$160$ & $27.52$ & ${+15.9}\;{-13.0}$ & ${+7.6}\;{-7.7}$ &
${+27.9}\;{-18.5}$ & ${+53.5}\;{-37.5}$ & $2.93$ &
${+48.3}\;{-37.5}$\\
$165$ & $23.29$ & ${+15.7}\;{-13.0}$ & ${+7.6}\;{ -7.7}$ &
${+27.9}\;{-18.5}$ & ${+53.3}\;{-37.6}$ & $2.49$ &
${+48.2}\;{-37.5}$\\
$170$ & $19.79$ & ${+15.4}\;{-13.0}$ & ${+7.6}\;{-7.7}$ &
${+28.0}\;{-18.6}$ & ${+53.0}\;{-37.7}$ & $2.13$ &
${+47.9}\;{-37.7}$\\
$175$ & $16.88$ & ${+15.2}\;{-13.1}$ & ${+7.6}\;{-7.7}$ &
${+28.0}\;{-18.6}$ & ${+52.8}\;{-37.8}$ & $1.82$ &
${+47.7}\;{-37.7}$\\
$180$ & $14.45$ & ${+15.0}\;{-13.1}$ & ${+7.7}\;{-7.7}$ &
${+28.1}\;{-18.6}$ & ${+52.7}\;{-37.8}$ & $1.57$ &
${+47.7}\;{-37.8}$\\
$185$ & $12.42$ & ${+15.0}\;{-13.1}$ & ${+7.7}\;{-7.7}$ &
${+28.1}\;{-18.7}$ & ${+52.7}\;{-38.0}$ & $1.35$ &
${+47.5}\;{-38.2}$\\
$190$ & $10.71$ & ${+14.9}\;{-13.2}$ & ${+7.6}\;{-7.8}$ &
${+28.2}\;{-18.8}$ & ${+52.8}\;{-38.1}$ & $1.17$ &
${+47.6}\;{-38.2}$\\
$195$ & $9.27$ & ${+14.9}\;{-13.2}$ & ${+7.7}\;{-7.8}$ &
${+28.2}\;{-18.8}$ & ${+52.7}\;{-38.2}$ & $1.02$ &
${+47.6}\;{-38.3}$\\
$200$ & $8.04$ & ${+14.8}\;{-13.2}$ & ${+7.7}\;{-7.8}$ &
${+28.0}\;{-18.7}$ & ${+52.7}\;{-38.3}$ & $0.89$ &
${+47.8}\;{-38.3}$\\
$225$ & $4.12$ & ${+14.7}\;{-13.4}$ & ${+7.8}\;{-8.0}$ &
${+28.3}\;{-19.0}$ & ${+52.6}\;{-38.8}$ & $0.46$ &
${+47.6}\;{-39.0}$\\
$250$ & $2.24$ & ${+14.5}\;{-13.6}$ & ${+7.9}\;{ -8.2}$ &
${+28.4}\;{-19.2}$ & ${+52.8}\;{-39.3}$ & $0.26$ &
${+47.8}\;{-39.6}$\\
$275$ & $1.28$ & ${+14.4}\;{-13.7}$ & ${+8.0}\;{-8.4}$ &
${+28.7}\;{-19.4}$ & ${+53.0}\;{-39.8}$ & $0.15$ &
${+48.3}\;{-39.9}$\\
$300$ & $0.76$ & ${+14.2}\;{-13.8}$ & ${+8.2}\;{-8.6}$ &
${+28.9}\;{-19.5}$ & ${+53.2}\;{-40.3}$ & $0.09$ &
${+48.2}\;{-40.6}$\\
$325$ & $0.47$ & ${+14.2}\;{-13.9}$ & ${+8.4}\;{-8.9}$ &
${+29.0}\;{-19.6}$ & ${+53.4}\;{-40.7}$ & $0.06$ &
${+48.5}\;{-41.0}$\\
$350$ & $0.30$ & ${+14.2}\;{-14.0}$ & ${+8.7}\;{-9.2}$ &
${+29.1}\;{-19.8}$ & ${+53.8}\;{-41.3}$ & $0.04$ &
${+48.9}\;{-41.7}$\\
$375$ & $0.19$ & ${+14.1}\;{-14.1}$ & ${+9.0}\;{-9.5}$ &
${+29.4}\;{-19.9}$ & ${+54.2}\;{-41.8}$ & $0.02$ &
${+49.5}\;{-41.9}$\\
$400$ & $0.13$ & ${+14.1}\;{-14.2}$ & ${+9.3}\;{-9.8}$ &
${+29.4}\;{-20.0}$ & ${+54.5}\;{-42.1}$ & $0.02$ &
${+49.8}\;{-42.4}$\\
$425$ & $0.09$ & ${+14.1}\;{-14.3}$ & ${+9.6}\;{-10.1}$ &
${+29.7}\;{-20.2}$ & ${+55.1}\;{-42.6}$ & $0.01$ &
${+50.5}\;{-42.9}$\\
$450$ & $0.06$ & ${+14.1}\;{-14.4}$ & ${+9.9}\;{-10.4}$ &
${+29.8}\;{-20.3}$ & ${+55.4}\;{-43.1}$ & $0.01$ &
${+50.8}\;{-43.5}$\\
$500$ & $0.03$ & ${+14.2}\;{-14.5}$ & ${+10.6}\;{-11.0}$ &
${+30.0}\;{-20.5}$ & ${+56.5}\;{-44.0}$ & $0.00$ &
${+51.9}\;{-44.3}$\\ \hline
\end{tabular}
\caption[The central predictions in the MSSM $g g\to\Phi$ channel at
the lHC together with the detailed uncertainties and the impact of the
$\Phi\to\tau^+\tau^-$ branching fraction]{The MSSM Higgs production
  cross sections in the $gg\to \Phi$
  channel (for $\tan\beta=1$) at the lHC with $\sqrt s = 7$ TeV as
  well as as the individual uncertainties (first from scale, then from
  PDF+$\Delta^{\rm \exp+th}\alpha_s$ at 90\%CL and from the input mass
  $\overline{m}_b$ at 1$\sigma$) and the total uncertainties for
  selected values of the Higgs mass. The last column displays the
  cross section times $\Phi\to \tau^+\tau^-$ branching fraction
  together with the total uncertainty when including the combination
  with the total uncertainty on $\Phi \to \tau^+\tau^-$ branching
  ratio.}
\label{table:ggPhiLHC}
\end{bigcenter}
\end{table}

\clearpage

\begin{table}
\begin{bigcenter}
\small
\begin{tabular}{|c|ccccccc|ccc|}\hline
$M_\Phi$ & $\sigma^{\rm NLO}_{\rm gg \to \Phi}$ & Scale [\%] &
PDF+$\Delta^{\rm exp+th}_{\alpha_s}$ [\%] & $\Delta \overline{m}_b$
[\%] & Total [\%] & $\sigma \times$ BR & Total [\%] \\ \hline
$90$ & $487.83$ & ${+26.8}\;{-21.1}$ & ${+6.2}\;{-8.0}$ &
${+17.8}\;{-10.7}$ & ${+49.8}\;{-32.8}$ & $46.83$ &${+44.3}\;{-32.5}$\\
$95$ & $409.59$ & ${+25.4}\;{-19.5}$ & ${+7.1}\;{-7.3}$ &
${+17.7}\;{-10.3}$ & ${+47.8}\;{-31.5}$ & $39.72$& ${+42.4}\;{-31.3}$\\
$100$ & $346.50$ & ${+23.9}\;{-18.4}$ & ${+6.4}\;{-7.8}$ &
${+17.8}\;{-9.9}$ & ${+46.9}\;{-29.9}$ & $33.92$& ${+41.4}\;{-29.8}$\\
$105$ & $294.57$ & ${+22.2}\;{-17.8}$ & ${+6.3}\;{-8.3}$ &
${+18.1}\;{-9.6}$ & ${+45.4}\;{-29.5}$ & $29.10$& ${+40.0}\;{-29.3}$\\
$110$ & $252.03$ & ${+21.3}\;{-16.7}$ & ${+6.9}\;{-7.7}$ &
${+17.2}\;{-10.1}$ & ${+43.8}\;{-28.9}$ & $25.11$& ${+38.5}\;{-28.8}$\\
$115$ & $215.89$ & ${+20.3}\;{-15.9}$ & ${+6.8}\;{-8.0}$ &
${+17.5}\;{-9.5}$ & ${+43.0}\;{-27.5}$ & $21.68$& ${+37.7}\;{-27.3}$\\
$120$ & $186.56$ & ${+18.8}\;{-15.4}$ & ${+7.2}\;{-7.7}$ &
${+17.6}\;{-9.3}$ & ${+41.7}\;{-26.9}$ & $18.88$& ${+36.4}\;{-27.0}$\\
$125$ & $161.84$ & ${+17.4}\;{-15.2}$ & ${+7.7}\;{-7.6}$ &
${+17.1}\;{-9.4}$ & ${+39.4}\;{-26.9}$ & $16.49$& ${+34.1}\;{-26.8}$\\
$130$ & $140.97$ & ${+16.4}\;{-14.9}$ & ${+7.6}\;{-7.8}$ &
${+17.5}\;{-9.5}$ & ${+38.8}\;{-26.8}$ & $14.46$& ${+33.7}\;{-26.7}$\\
$135$ & $122.98$ & ${+15.5}\;{-14.6}$ & ${+7.3}\;{-8.0}$ &
${+17.2}\;{-9.2}$ & ${+38.0}\;{-26.6}$ & $12.70$& ${+32.8}\;{-26.6}$\\
$140$ & $108.08$ & ${+14.1}\;{-14.7}$ & ${+7.5}\;{-8.0}$ &
${+16.6}\;{-9.5}$ & ${+35.6}\;{-26.9}$ & $11.23$& ${+30.5}\;{-26.8}$\\
$145$ & $94.85$ & ${+13.7}\;{-14.7}$ & ${+7.7}\;{-8.2}$ &
${+17.1}\;{-9.2}$ & ${+35.5}\;{-26.9}$ & $9.92$& ${+30.2}\;{-27.0}$\\
$150$ & $83.83$ & ${+13.0}\;{-14.3}$ & ${+8.2}\;{-8.0}$ &
${+17.0}\;{-9.1}$ & ${+35.0}\;{-26.4}$ & $8.82$& ${+29.7}\;{-26.5}$\\
$155$ & $74.13$ & ${+12.0}\;{-14.2}$ & ${+8.2}\;{-8.2}$ &
${+17.1}\;{-9.0}$ & ${+34.0}\;{-26.2}$ & $7.84$& ${+28.9}\;{-26.1}$\\
$160$ & $65.84$ & ${+11.0}\;{-14.1}$ & ${+8.3}\;{-8.2}$ &
${+16.9}\;{-9.1}$ & ${+32.7}\;{-26.4}$ & $7.00$& ${+27.6}\;{-26.5}$\\
$165$ & $58.60$ & ${+10.4}\;{-13.6}$ & ${+8.2}\;{-8.5}$ &
${+16.9}\;{-8.8}$ & ${+32.6}\;{-25.4}$ & $6.26$& ${+27.5}\;{-25.5}$\\
$170$ & $52.28$ & ${+9.6}\;{-13.6}$ & ${+8.5}\;{-8.3}$ &
${+16.6}\;{-9.2}$ & ${+31.7}\;{-25.7}$ & $5.61$& ${+26.5}\;{-25.7}$\\
$175$ & $46.75$ & ${+9.1}\;{-13.4}$ & ${+8.4}\;{-8.6}$ &
${+16.7}\;{-8.9}$ & ${+31.3}\;{-25.8}$ & $5.04$& ${+26.2}\;{-25.8}$\\
$180$ & $41.97$ & ${+8.3}\;{-13.8}$ & ${+8.5}\;{-8.7}$ &
${+16.9}\;{-8.7}$ & ${+30.4}\;{-25.7}$ & $4.55$& ${+25.4}\;{-25.7}$\\
$185$ & $37.69$ & ${+7.5}\;{-13.3}$ & ${+9.2}\;{-8.2}$ &
${+16.8}\;{-8.8}$ & ${+29.8}\;{-25.6}$ & $4.11$& ${+24.6}\;{-25.8}$\\
$190$ & $33.87$ & ${+7.5}\;{-12.8}$ & ${+8.6}\;{-9.2}$ &
${+16.5}\;{-8.9}$ & ${+29.6}\;{-25.5}$ & $3.71$& ${+24.4}\;{-25.6}$\\
$195$ & $30.55$ & ${+7.0}\;{-12.3}$ & ${+8.9}\;{-9.1}$ &
${+16.4}\;{-9.1}$ & ${+29.0}\;{-25.4}$ & $3.36$& ${+23.9}\;{-25.5}$\\
$200$ & $27.61$ & ${+6.3}\;{-12.5}$ & ${+8.7}\;{-9.6}$ &
${+16.4}\;{-8.8}$ & ${+28.6}\;{-25.5}$ & $3.05$& ${+23.7}\;{-25.6}$\\
$225$ & $17.10$ & ${+5.1}\;{-11.9}$ & ${+9.6}\;{-9.7}$ &
${+16.4}\;{-8.9}$ & ${+27.6}\;{-24.8}$ & $1.93$& ${+22.6}\;{-25.0}$\\
$250$ & $10.97$ & ${+5.0}\;{-11.5}$ & ${+10.6}\;{-10.0}$ &
${+16.1}\;{-8.8}$ & ${+27.5}\;{-25.1}$ & $1.26$& ${+22.5}\;{-25.4}$\\
$275$ & $7.27$ & ${+5.0}\;{-10.6}$ & ${+10.9}\;{-10.8}$ &
${+16.3}\;{-8.6}$ & ${+28.6}\;{-24.5}$ & $0.85$& ${+23.9}\;{-24.6}$\\
$300$ & $4.94$ & ${+4.8}\;{-10.5}$ & ${+11.6}\;{-11.4}$ &
${+16.4}\;{-8.4}$ & ${+29.3}\;{-24.9}$ & $0.58$& ${+24.3}\;{-25.3}$\\
$325$ & $3.43$ & ${+4.8}\;{-9.8}$ & ${+12.4}\;{-11.8}$ &
${+15.9}\;{-8.6}$ & ${+29.7}\;{-24.9}$ & $0.41$& ${+24.9}\;{-25.2}$\\
$350$ & $2.43$ & ${+5.0}\;{-9.4}$ & ${+13.1}\;{-12.3}$ &
${+16.2}\;{-8.4}$ & ${+30.8}\;{-24.6}$ & $0.29$& ${+25.9}\;{-25.0}$\\
$375$ & $1.75$ & ${+4.9}\;{-9.1}$ & ${+14.1}\;{-12.5}$ &
${+16.0}\;{-8.2}$ & ${+31.6}\;{-24.3}$ & $0.21$& ${+26.9}\;{-24.5}$\\
$400$ & $1.28$ & ${+4.7}\;{-8.9}$ & ${+14.9}\;{-13.1}$ &
${+15.9}\;{-8.2}$ & ${+31.6}\;{-24.8}$ & $0.16$& ${+26.9}\;{-25.1}$\\
$425$ & $0.95$ & ${+4.7}\;{-8.6}$ & ${+15.5}\;{-13.6}$ &
${+16.0}\;{-8.3}$ & ${+32.5}\;{-25.0}$ & $0.09$& ${+27.8}\;{-25.3}$\\
$450$ & $0.71$ & ${+4.7}\;{-8.2}$ & ${+16.1}\;{-14.2}$ &
${+16.1}\;{-8.2}$ & ${+33.4}\;{-25.2}$ & $0.07$& ${+28.8}\;{-25.6}$\\
$500$ & $0.41$ & ${+4.7}\;{-7.9}$ & ${+17.8}\;{-15.1}$ &
${+15.8}\;{-8.2}$ & ${+34.4}\;{-26.1}$ & $0.05$& ${+29.8}\;{-26.5}$\\ \hline
\end{tabular}
\caption[The central predictions in the MSSM $b\bar b\to\Phi$ channel
at the lHC together with the detailed uncertainties and the impact of
the $\Phi\to\tau^+\tau^-$ branching fraction]{The same as in
  Table.\ref{table:ggPhiLHC} but for  the $b\bar b\to \Phi$ channel.}
\label{table:bbPhiLHC}
\end{bigcenter}
\end{table}

\clearpage

\begin{figure}[!h]
\begin{center}
\vspace*{2mm}
\includegraphics[scale=0.75]{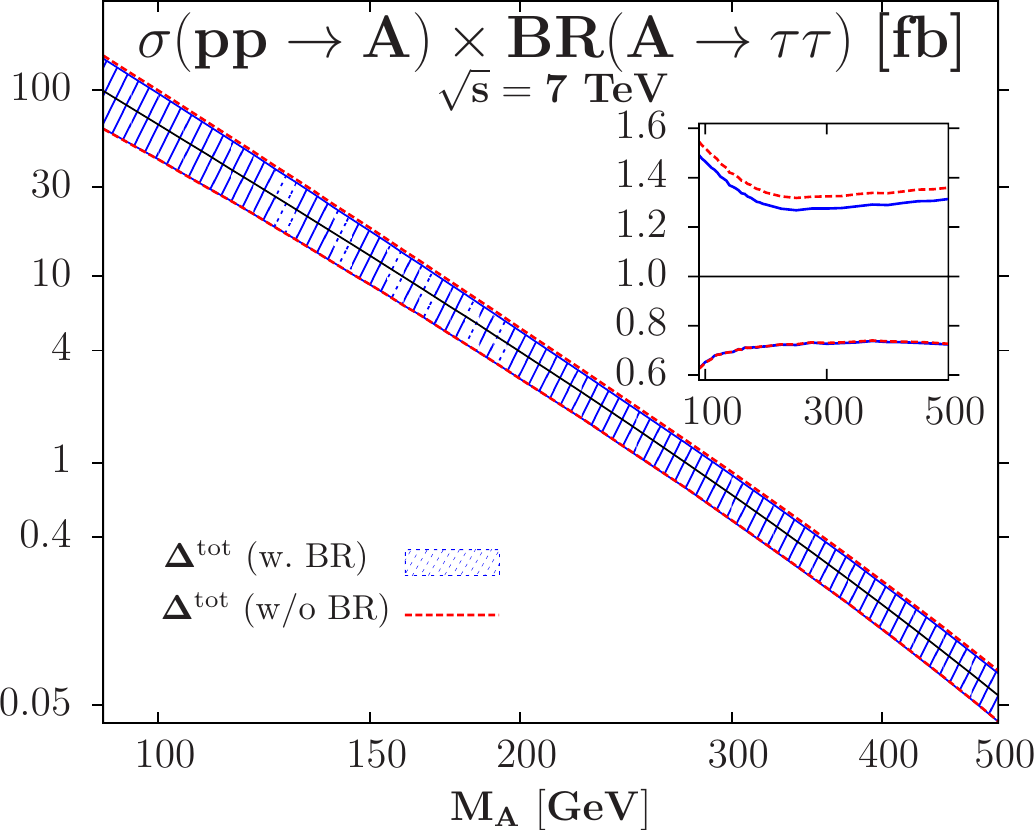}
\end{center}
\vspace*{-4mm}
\caption[$CP$--odd $A$ boson production in the $p
p\to A\to\tau^+\tau^-$ channel at the lHC together with the total
uncertainty]{$\sigma(p p\! \to\! A)\times{\rm BR}(A \!\to\! \tau^+
  \tau^-)$ as a function of $M_A$ at the lHC with $\sqrt s = 7$ TeV,
  together with the associated overall theoretical uncertainty;  the
  uncertainty when excluding that on the branching ratio is also
  displayed. In the inserts, shown are the relative deviations from
  the central values.}
\vspace*{-1mm}
\label{fig:TotalMSSMLHC}
\end{figure}

We finally combine the two production channel together to obtain the
final $p p\to\Phi\to\tau^+\tau^-$ central prediction and total
uncertainty. This is the prediction to be compared with ATLAS and CMS
experimental results as the search topology in the Higgs$\to
\tau^+\tau^-$ channel use the two production processes. The resulting
$\sigma(p p\!\to \! A)\!\times\! {\rm BR}(A\!\to\! \tau^+ \tau^-)$
at the lHC with $\sqrt s = 7$ TeV is shown in
Fig.~\ref{fig:TotalMSSMLHC} as a function of $M_A$. As stated in the
Tevatron study the true rates for the combined production of
$\Phi\!=\!A\!+\!H(h)$ are obtained when our results are multiplied by
a factor of $2\tan^2\beta$.

The impact of the theoretical uncertainties in both channels $b\bar b
, gg\to \Phi \to \tau^+ \tau^-$ for the MSSM $CP$--odd and one of the
$CP$--even Higgs particles are then large. We will see in the next
subsection that this has important consequences on the MSSM
$[\tan\beta, M_A]$ parameter space that can be probed at the lHC.

\subsection{Impact of the theoretical uncertainties on the limits on
  the MSSM parameter space \label{section:MSSMHiggsExpLimit}}

We are arrived nearly at the end of our journey. Thanks to the
subsection~\ref{section:MSSMHiggsExpTotal} we will apply our study on
the limits on the MSSM $[M_A,\tb]$ parameter space, first using the
Tevatron experimental results and then the early lHC experimental results.

\subsubsection{The Tevatron results}

The results that we have obtained in
section~\ref{section:MSSMHiggsExpTotal} have to be compared with the
experimental limits that have been obtained by the CDF/D0
collaborations in the $p\bar p \! \to \! \Phi \! \to \! \tau^+ \tau^-$
search topology~\cite{Benjamin:2010xb}. The major experimental output
that is of use for such a theoretical comparison is the production
cross section times branching fraction 95\%CL limit for various $A$
boson masses $M_A$. As the experiment has seen nothing, this
translates into limits on the MSSM [$M_A, \tan\beta$] parameter space
that is probed. We thus generate for each $(M_A,\tan\beta)$ value a
cross section by multiplying our results by $2\tan^2\beta$ as already
mentioned and then compare with the experimental results presented in
Table~X of ref~\cite{Benjamin:2010xb}; if our theoretical expectation
is above that means that the associated value $[M_A,\tan\beta]$ has
been excluded at 95\%CL.

\begin{figure}[!h]
\begin{center}
\includegraphics[scale=0.85]{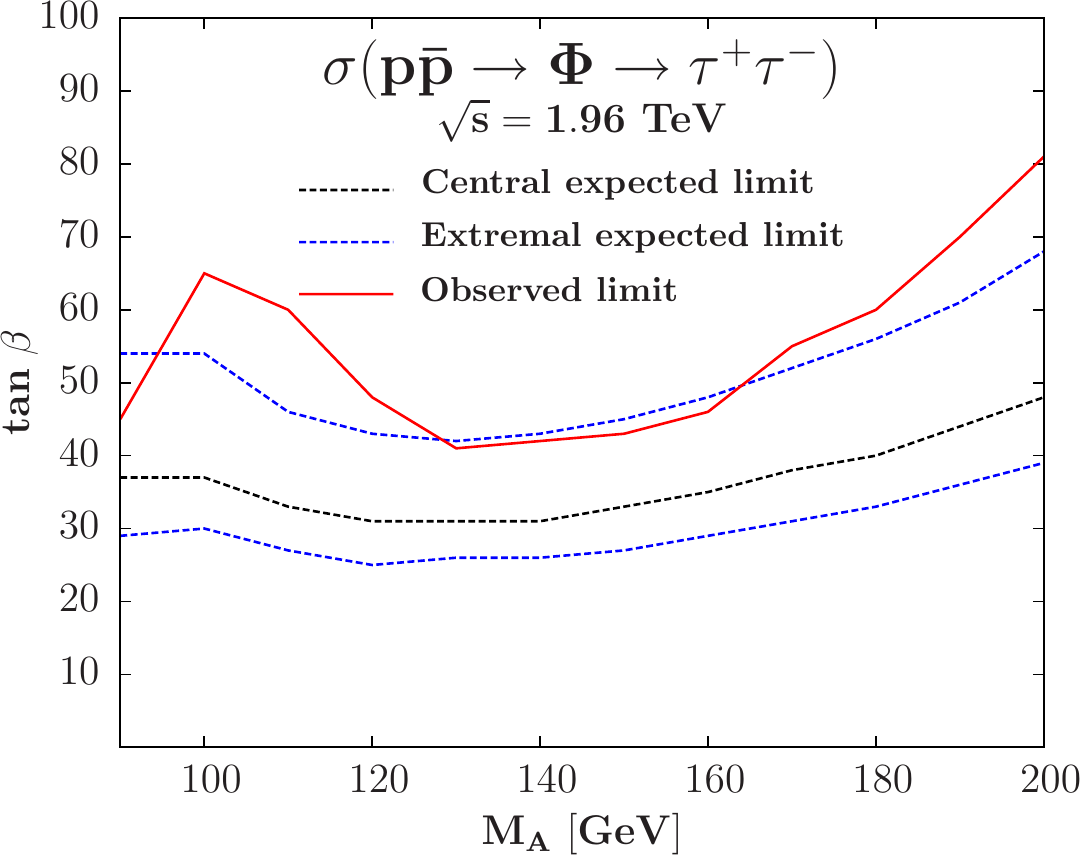} 
\end{center}
\caption[The 95\%CL limits on the MSSM parameter
space using our theoretical uncertainties confronted to the Tevatron
results]{Contours for the expected $\sigma(p\bar p\! \to\! \Phi  \!\to
  \!\tau^+ \tau^-)$ rate at the Tevatron in the [$M_A, \tb$] plane
  with the associated theory  uncertainties, confronted to the 95\%CL
  exclusion limit.}
\label{fig:ScanTev}
\end{figure}

To visualize the impact of these theoretical uncertainties on this
MSSM [$M_A, \tan\beta$] parameter space, we show in
Fig.~\ref{fig:ScanTev} the contour of the cross section times
branching ratio in this plane, together with the contours when the
uncertainties are included. We apply the model independent 95\%CL
expected and observed limits from the CDF/D0 analysis (Table X of
Ref.~\cite{Benjamin:2010xb}). We have displayed the expected limit
with our central prediction (black dotted line) together with the
expected limits using the maximal and minimal predictions (in blue
dotted lines) in order to show the impact of the theoretical
uncertainties on the $[M_A,\tb]$ plane. However, rather than applying
the observed limits on the central $\sigma \times$BR rate, we apply
them on the minimal one when the theory uncertainty is
included. Indeed, since the latter has a flat prior, the minimal
$\sigma \times$BR value is as respectable and likely as the central
value and the viable exclusion limit should take into account these
uncertainties: the final exclusion limit should be viewed as the one
obtained with the minimal cross section rather than the central cross
section with respect to theoretical
uncertainties.

Fig.~\ref{fig:ScanTev} displays the obtained observed limit in this
spirit in red solid line. Only values $\tb\gsim 50$ are excluded in
the mass ranges, $M_A \approx 95$--125 GeV and $M_A \gsim 165$ GeV. In
the intermediate range $M_A \approx 125$--165 GeV, the exclusion
limit is $\tb \gsim 40$--45, to be contrasted with the values $\tan
\beta \gsim 30$ excluded in the CDF/D0 analysis. Hence, the inclusion
of the theory uncertainties has a drastic impact on the allowed
$[M_A, \tan\beta]$ parameter space.

We conclude by mentioning that  there is a subleading channel which
has also been considered at the Tevatron, $bg\! \to\! \Phi b\! \to\!
3b$~\cite{CDF:10105}. The evaluation of the theory uncertainties in
$bg\! \to\! \Phi b$~\cite{Campbell:2002zm, Maltoni:2003pn} follows
that of the parent process $b\bar b\! \to\! \Phi$, being part of the
NLO contributions of the latter process. Similar results, that is an
uncertainty of approximately $\pm 40\%$, are expected. However, in
this case, it is the $\Phi\to b\bar b$ decay which is considered
experimentally and, since BR$(\Phi \! \to \! b\bar b)$ has a small
error, the uncertainties in $\sigma$ and $\sigma \times {\rm BR}$ are
almost the same. These uncertainties will thus also have an impact on
the excluded  [$M_A, \tb$] parameter space\footnote{Note that the
  $2\sigma$ excess observed by CDF in this channel cannot be a Higgs
  signal as it would correspond to a much larger excess in $p\bar p
  \to \tau^+\tau^-$ which has not been observed. In addition, for the
  huge $\tb$ values probed in this $3b$ channel, $\tb\! \gsim\! 100$,
  we are outside the theoretical favored range $\tb\!\lsim\!50$ and
  the $\Phi b\bar b$ Yukawa coupling become non--perturbative. For such
values, the total Higgs widths are to be included.}.

\subsubsection{The lHC results}

We now discuss the ATLAS and CMS limits~\cite{Collaboration:2011rv,
  Chatrchyan:2011nx} in the light of the theoretical uncertainties
that affect the Higgs production cross section and the decay branching
ratios. We use the same method presented in the Tevatron case a few
lines above, with a comparison to the date provided by the CMS
collaboration in
Ref.~\cite{Chatrchyan:2011nx}\footnote{Unfortunately, the ATLAS
  collaboration had not given this important information in its note
  ATLAS--CONF--2011--024 which is the preliminary version of
  Ref.~\cite{Collaboration:2011rv} that was available at the time of
  this analysis. But as the ATLAS exclusion limits are similar to
  those obtained by the CMS collaboration, the final result once
  the theory uncertainties have been included is the same as
  visualised in Ref.~\cite{Collaboration:2011rv}.}: we simply
rescale the $\sigma(gg+b\bar b \to A \to \tau\tau)$ numbers by a 
factor $2\times   \tan^2 \beta$ and then turn them into exclusion
limits on the MSSM $[M_A, \tan\beta]$ parameter space that has been
probed at the lHC thanks to the CMS numbers.

\begin{figure}[!h]
\begin{center}
\vspace*{-2mm}
\includegraphics[scale=0.85]{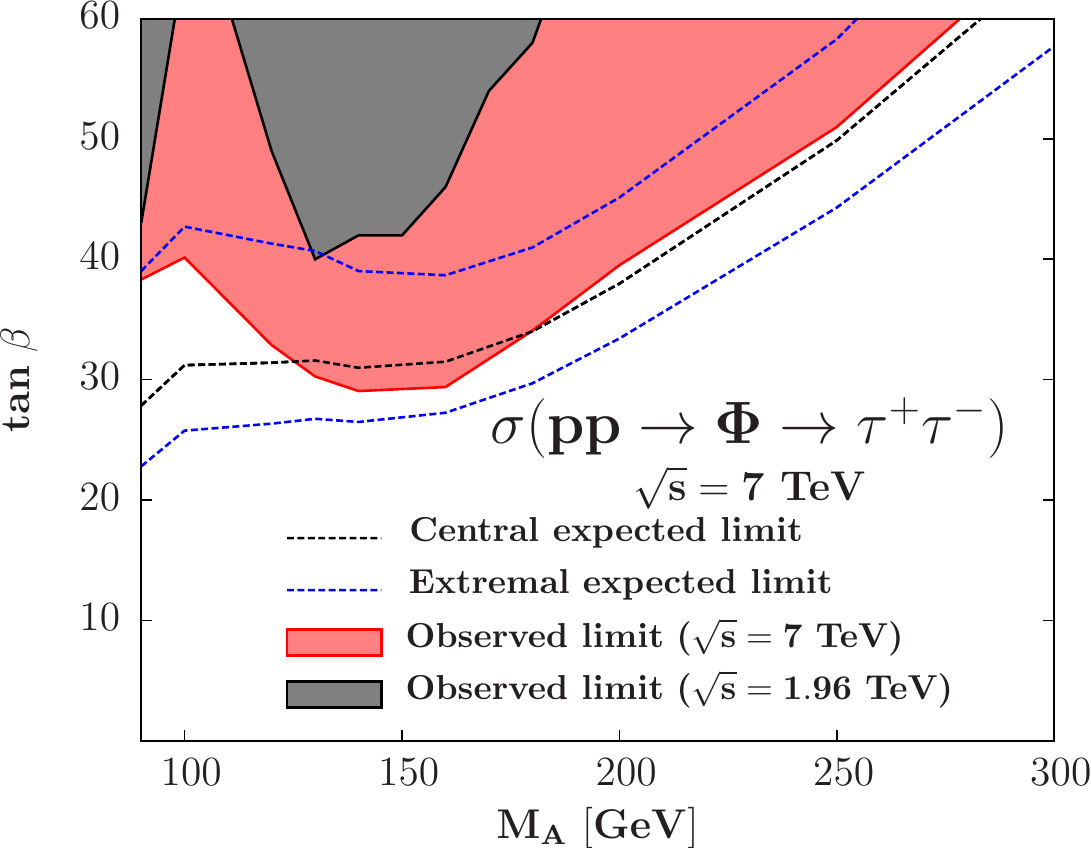}
\end{center}
\caption[The 95\%CL limits on the MSSM parameter
space using our theoretical uncertainties confronted to the lHC
results]{Contours for the expected $\sigma(p p\! \to\! \Phi \!\to
  \!\tau^+ \tau^-)$ exclusion limits at the lHC in the [$M_A, \tb$]
  plane with the associated theory  uncertainties, confronted to the
  95\%CL exclusion limits given by the CMS~\cite{Chatrchyan:2011nx}
  and also CDF/D0~\cite{Benjamin:2010xb} collaborations when our
  procedure is applied.} 
\label{fig:ScanLHC}
\end{figure}

Our results are shown in Fig.~\ref{fig:ScanLHC} where the contour of
the cross section times branching ratio in this plane is displayed,
together with the contours when the uncertainties are included. As in
Fig.~\ref{fig:ScanTev} we display the central expected limit (black
dotted line) obtained using our central predictions as well as the
maximal and minimal expected limits (blue dotted lines) when using
correspondingly the minimal and maximal theoretical rates.  We also
display in gray band the observed limit at the Tevatron according to
Fig.~\ref{fig:ScanTev}.

However, rather than applying the observed limits on the central $\sigma
\times$BR rate (as the CMS and also ATLAS collaborations do), we
apply them on the minimal one when the theory uncertainty is
included. In this case, only values $\tb\gsim 28$ are  excluded for a
Higgs mass $M_\Phi \!\approx \!130$ GeV, compared to $\tb \gsim 23$ if
the central prediction is considered as in the CMS analysis. The
inclusion of the theory uncertainties should lead to a slight
reduction of the excluded  $[M_A,\tan\beta]$ parameter space, which
can also be viewed in Fig.~3 of Ref.~\cite{Chatrchyan:2011nx}, though
our uncertainties are larger especially at low $A$ Higgs boson
masses.

\subsubsection{The expectations at the LHC}

The exclusion limits on the $[\tb, M_A$] MSSM parameter space obtained
by the ATLAS and CMS collaborations with only 36 pb$^{-1}$ data are
very strong: indeed when taking into account our theoretical
uncertainties we have just demonstrated that values $\tb \gsim 29$ are
excluded in the low mass range for the pseudoscalar Higgs boson,
$M_A=90$--200 GeV. We can draw some expectations in the view of higher
luminosity and some consequences. We note that the newest results from
HEP--EPS 2011 conference confirm and even surpass our expectations,
see below.

\begin{figure}[!h]
\begin{center}
\vspace*{-2mm}
\includegraphics[scale=0.85]{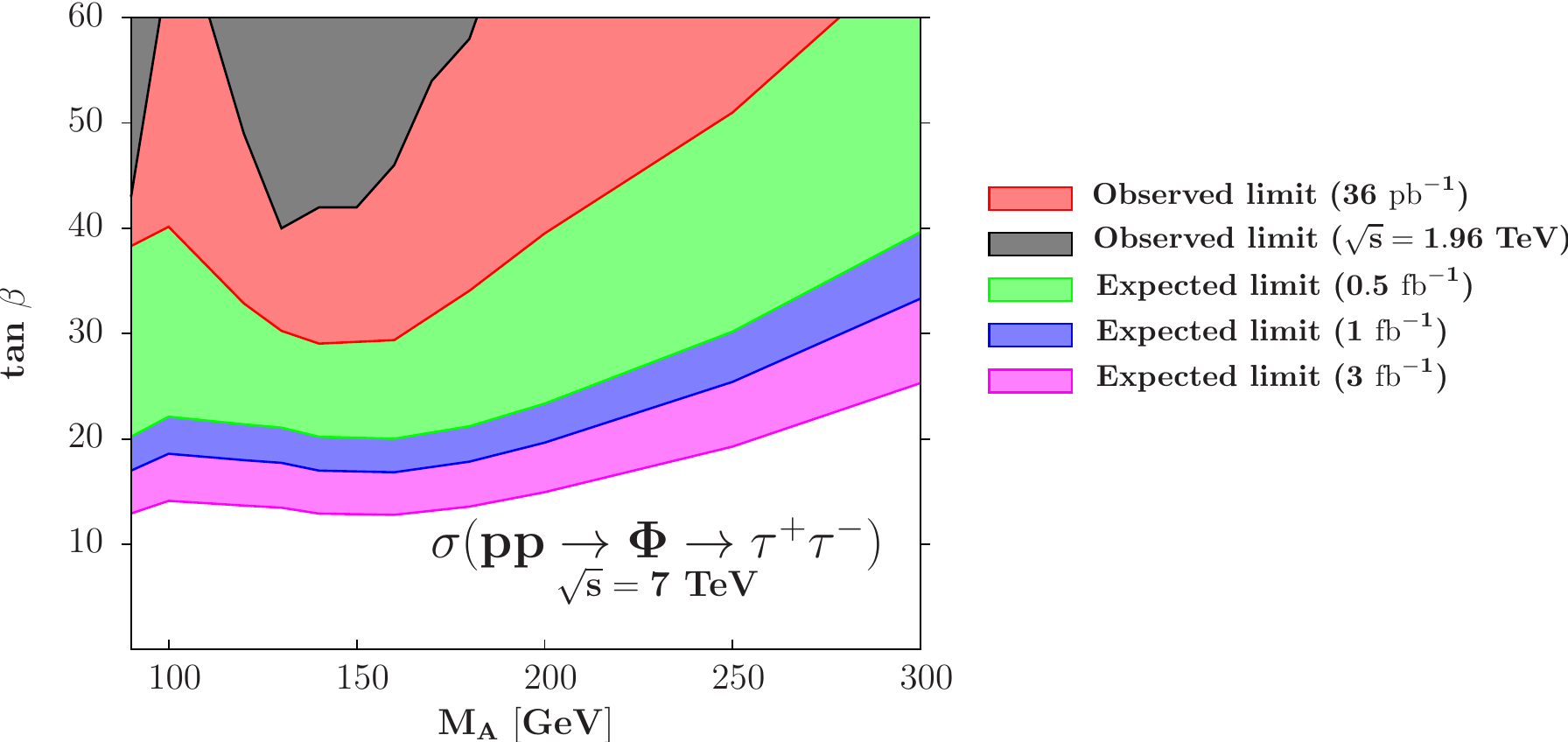}
\end{center}
\vspace*{-7mm}
\caption[Expectations at higher luminosity at the lHC for the 95\%CL
limits on the MSSM parameter space using our theoretical
calculation]{Contours for the ``expected" $\sigma(p p\! \to\! \Phi
  \!\to \!\tau^+ \tau^-)$ 95\%CL exclusion limits at the lHC with
  $\sqrt s=7$ TeV in the [$M_A, \tb$] plane for various integrated
  luminosities. The present limits from 36 pb$^{-1}$ CMS and the
  Tevatron are also displayed.}
\vspace*{-5mm}
\label{fig:MSSMprojection}
\end{figure}

First of all, if the luminosity is increased to the fb$^{-1}$ level as was
expected to be the case at the end of this year and is actually
already the case\footnote{The LHC technical team should be acclaimed
  for such an impressive effort.} at the level of 2
fb$^{-1}$~\cite{CERNbulletin:2011}, the values of $\tb$ 
which can be probed will be significantly lower. We have drawn
quantitative expectations assuming no improvement in the analysis,
which may be proved to be pessimistic\footnote{For
  example the SM $H\to\tau^+\tau^-$ search has been greatly improved
  from winter 2011 to July 2011, see Ref.~\cite{Bluj:2011hh}, page
  15.} and that the CMS sensitivity will simply scale as the square
root of the integrated luminosity, the region of the $[\tb,M_A]$
parameter space which can be excluded in the case where no signal is
observed is displayed in Fig.~\ref{fig:MSSMprojection} for several
values of the accumulated luminosity. With 3 fb$^{-1}$ data per
experiment (or with 1.5 fb$^{-1}$ when the ATLAS and CMS results are
combined), we have found that values $\tb \gsim 12$ could be excluded
in the entire mass range $M_A \lsim 200$ GeV; the exclusion reduces to
$\tb \gsim 20$ for the mass range $M_A \lsim 300$ GeV. The comparison
with Ref.~\cite{Tonelli:2011EPS} (page 24) from HEP--EPS 2011
conference with 1.1 fb$^{-1}$ data shows that we are not far from the
actual observations, thus nicely confirming our expectations.

\vspace{5mm}
We list below three ways to improve these limits by considering
additionnal production channels.

\begin{enumerate}[$a)$]
\item{{\it The process $gb \to \Phi b \to b b\bar b$ where the all final
    bottom quarks are detected}: the production cross section
  $\sigma(b g \to b\Phi+g\bar b\to \Phi \bar b)$ is one order of
  magnitude lower than that of the inclusive $gg+b\bar b\to \Phi$
  process (the $bg\to b \Phi$ process is part of the NLO corrections
  to $b\bar b \to \Phi$~\cite{Dicus:1988cx, Maltoni:2003pn}, see also
  Ref.~\cite{Assamagan:2004mu} page 5) but this is compensated by the
  larger fraction BR$(\Phi \to b\bar b) \approx 90\%$ compared to
  BR$(\Phi \to \tau^+ \tau^-) \approx 10\%$; the QCD background are
  much larger though\footnote{We estimate the theoretical
    uncertainties on the rate $\sigma(bg \to Ab \to b b\bar b)$  to be
    similar to that of the combined $gg+b \bar b \to \Phi$ process,
    that is $\pm 30$--40\%. However, the parametric and $\Delta_b$
    supersymmetric corrections do not cancel this time in the cross
    section times branching ratio and have to be taken into account.};}
\item{{\it the process  $pp \to \Phi \to \mu^+\mu^-$}:  its rate is
    simply given by $\sigma(pp\!\to\! \Phi \to \tau\tau )$ rescaled by
    BR$(\Phi \to \mu \mu)/$BR$(\Phi \to \tau \tau)= m_\mu^2/m_\tau^2
    \approx 4\times 10^{-3}$; the smallness of the rate is partly
    compensated by the much cleaner $\mu \mu$ final state and the
    better resolution on the $\mu\mu$ invariant mass. The efficiency
    of the $pp \to \Phi\to \tau \tau$ signal is estimated by the CMS
    collaboration (see Table 1 of Ref.~\cite{Chatrchyan:2011nx}) to be
    4.5\% when all $\tau$ decay channels are considered. This is only
    a factor 10 larger than the ratio BR$(\Phi \to \mu \mu)/$BR$(\Phi
    \to \tau \tau)$ (which  is approximately equal to the efficiency
    in the $\tau \to e\mu$  channel). Thus the $\Phi \to \mu \mu$
    decay channel might be useful. In particular, the small resolution
    that can achieved could allow to separate the three peaks of the
    almost degenerate $h,H$ and $A$ states in the intense coupling
    regime; see Ref.~\cite{Boos:2002ze,Boos:2003jt};}
\item{{\it the process $pp \to tbH^- \to tb \tau \nu$}: this leads to
    a cross section that is also proportional to $\tan^2\beta$ (and
    which might also be useful for very low $\tb$ values) but that is
    two orders of magnitude smaller than $\sigma(pp \to \Phi)$ for
    $M_A \approx 100$--300  GeV;}
\item{{\it charged Higgs production from top quark decays, $pp\! \to\!
      t\bar t$ with $t\to H^+b \to \tau^+ \nu b$}: this has also been
    recently analyzed by the CMS
    collaboration~\cite{MS-PAS-HIG-11-002}. With 36 pb$^{-1}$ data,
    values of the branching ratio BR$(t\to H^+b) \gsim 25\%$ are
    excluded which means that  only $\tb$ values larger than 60 are
    probed for the time being\footnote{We note that in this case the
      theoretical  uncertainties have not been estimated in
      Ref.~\cite{Dittmaier:2011ti} (contrary to the channel  $pp\!
      \to\! tbH^-\! \to\! tb \tau \nu$ where an uncertainty of $\pm
      30\%$  has been found). We have evaluated them with {\tt HATHOR}
      \cite{Aliev:2010zk} and, in the production channel  $\sigma(p
      p\! \to\! t\bar t)$ and for $m_t\!=\! 173.3\! \pm  \! 1.1$GeV,
      we  find  $\sigma(p  p \to t\bar t)\!=
      \!163~^{+2.5\%}_{-5.6\%}~{(\rm
        factor~2~from~central~m_t~scale)}~^{+10.4\%}_{-10.1\%}~ ({\rm
        PDF}\!+\!\Delta^{\rm exp+th}\alpha_s@90\%{\rm CL})~{\pm
        3.3\%}~ ( \Delta m_t)$ pb, which leads  using the procedure of
      Ref.~\cite{Baglio:2010um, Baglio:2011wn} presented in
      part~\ref{part:two} to a total uncertainty of  $\Delta
      \sigma/\sigma= ^{+16\%}_{-19\%}$, that is three times larger
      than the one assumed in the CMS analysis. To that, one should
      add the uncertainty on the branching ratio BR$(t\to H^+b)$  for
      which the parametric one (from the input $\overline{m}_b$ and
      $\alpha_s$ values) is about $+10\%, -4\%$ \cite{Baglio:2010ae}.}.}
\end{enumerate}

However the rates are small and would allow only for modest
improvements over the limit obtained with the $\Phi\to\tau^+\tau^-$
search. Indeed according to the projections by the ATLAS and CMS
collaborations in their technical reports~\cite{Aad:2008aa,
  Ball:2007zza}, in the context of the full LHC with $\sqrt s = 14$
TeV and 30 fb$^{-1}$ data, the observation of these processes is only
possible for not too large values of $M_A$ and values of $\tb
\gsim 20$ that have mostly been excluded by the newest 1.1 fb$^{-1}$
searches~\cite{Tonelli:2011EPS} (page 24). However, as is the case for
the $pp\to \tau \tau$ channel, some (hopefully significant)
improvement over these projections might be achieved.

\subsection{Consequences on the SM $H\to \tau \tau$
  search at the LHC \label{section:SMHiggsTauTau}}

We now conclude our study of the consequences of our theoretical
estimations by giving the very interesting consequence of the MSSM
analysis on the SM Higgs boson search that has been presented in
Ref.~\cite{Baglio:2011xz}.

Indeed the ATLAS and CMS $pp\to \tau \tau$ inclusive analyses open the
possibility of using this channel in the case of the Standard Model
Higgs boson $H$. Indeed, in this case, the main production process is
by far the $gg\to H$ channel which dominantly proceeds via a top quark
loop with a small contribution of the bottom quark loop as studied in
section~\ref{section:SMHiggsLHC} where the cross section for this
process has been discussed in detail (see also
Refs.~\cite{Baglio:2010ae,Dittmaier:2011ti}). In the mass range
$M_H\!=\!115$--140 GeV, the SM rate is at the level of 10 to 20 pb. The
branching ratio for the decay  $H\! \to\! \tau^+\tau^-$ ranges from 8\% at
$M_H\!=\!115$ GeV to 4\% at $M_H\!=\!140$ GeV, see
section~\ref{section:SMHiggsDecay}. The cross section times branching
ratio $\sigma(gg\!\to\! H\! \to \! \tau^+\tau^-)$ is thus rather
substantial at low Higgs masses\footnote{The cross section in the SM
  is comparable to that of $A+H(h)$ production in the MSSM with values
  $\tan\beta \approx 4$ (and not $\tb=1$), a consequence of the
  dominance of the top--quark loop (in the SM)  compared to
  bottom--quark loop (as is in general the case in the MSSM) in the
  $gg$ fusion process.}.

\begin{figure}[!h]
\begin{center}
\vspace*{-2mm}
\includegraphics[scale=0.85]{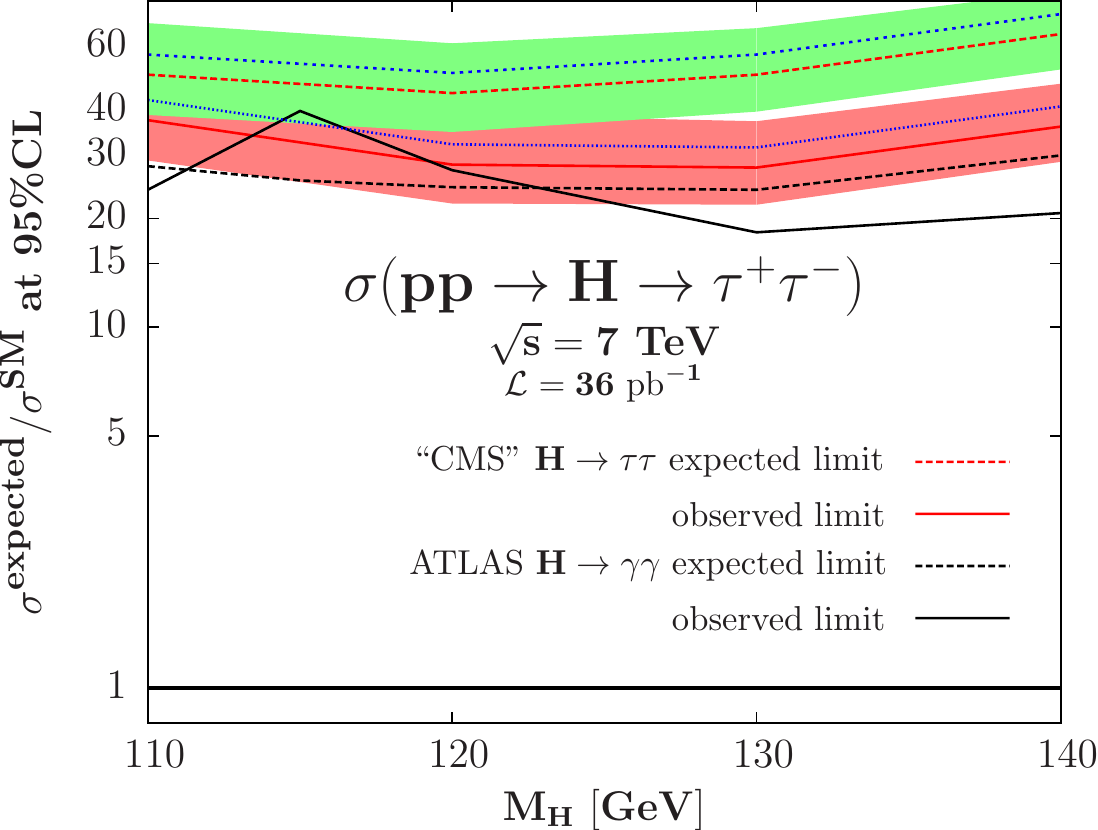}
\end{center}
\vspace*{-6mm}
\caption[The MSSM Higgs analysis applied to the SM $H\to\tau^+\tau^-$
search channel compared to the ATLAS $H\to\gamma\gamma$ limits]{The
  median expected and observed cross sections at 95\% CL for the
  production of the SM Higgs boson in the channels $gg\! \to\! H\to
  \!\tau \tau$ (blue lines) and $pp\! \to\! H \to 
  \!\tau^+\tau^-+X$ (red lines captioned) from an extrapolation of the
  CMS analysis with 36 pb$^{-1}$ data~\cite{Chatrchyan:2011nx}
  normalised to the SM cross section. The green and red bands display
  the impact of our theoretical uncertainties on the ``expected'' and
  ``observed'' limits, respectively, using all the production
  channels.  It is compared to the case of $H\!\to \!\gamma \gamma$ as
  recently analyzed by the ATLAS
  collaboration~\cite{ATLAS-CONF-2011-025} with 37.6 pb$^{-1}$ data.}
\vspace*{-2mm}
\label{fig:SMHiggsMSSM}
\end{figure}

We have then applied in Ref.~\cite{Baglio:2011xz} our numerical
results of the SM Higgs cross section $\sigma(gg\to H)$ times
$H\to\tau^+\tau^-$ branching fraction~\cite{Baglio:2010ae} to the
median expected' and observed  95\%CL limits obtained in the
CMS analysis at $\sqrt s=7$ TeV with 36 pb$^{-1}$ data. This will be
interpeted as ``expected'' and ``observed'' 95\%CL limits on the SM
process. This is displayed in blue (large dotted line for the
``expected'' limit, small dotted line for the ``observed'' limit) in
Fig.~\ref{fig:SMHiggsMSSM} where the 95\%CL observed and expected
cross sections are normalized to our SM prediction. With the small
amount of data that is available in a 36 pb$^{-1}$ analysis it is
shown that we are approximately 60 times above the expected SM rate in
the mass range $M_H = 110$--140 GeV. It is nevertheless not that bad
if we compare with the expected and observed rates in the
$H\to\gamma\gamma$ analysis of ATLAS
experiment~\cite{ATLAS-CONF-2011-025} which is the most important and
promising channel for the low SM Higgs mass searches. Indeed, the
$gg\to H\to \tau\tau$ inclusive channel, which has not been considered
neither by the ATLAS nor the CMS collaborations before HEP-EPS 2011
results, is in fact rather powerful and competes rather well with the
long celebrated  $H\to \gamma \gamma$ detection channel, as it has a
sensitivity that is only a factor of two smaller than the latter
channel.

Thus, the $gg\!\to\! H\!\to\! \tau^+\tau^-$ channel could be also used
to search for the SM Higgs  in the very difficult mass range $M_H\!
\approx\! 115$--130 GeV where only the $H\to \gamma \gamma$ channel
was considered to be effective. The two channels could be combined to
reach a better sensitivity. In addition, while little improvement
should be expected in $H\to \gamma \gamma$ (for which the analyses
have been tuned and optimised for more than twenty years now), a
better sensitivity could be achieved in the  $H\to \tau^+\tau^-$
channel. Indeed, on the one hand, a possible improvement might come
from the experimental side: novel and better mass reconstruction
techniques of the $\tau\tau$ resonance can be used\footnote{In
  Ref.~\cite{Elagin:2010aw}, a new mass technique for reconstructing
  resonances decaying into $\tau$ lepton pairs has been proposed and
  it is claimed that it allows for a major improvement in the search
  for the Higgs$\to\! \tau\tau$ signal.}, splitting of the analysis
into jet multiplicities as done for instance for the channel $H\!
\to\! WW \to \ell \ell \nu \nu$, inclusion of additional topologies
such as same sign $\ell= e,\mu$ final states,  etc\footnote{See for
  example the incredible improvement displayed in
  Ref.~\cite{Bluj:2011hh}, page 15, where some experimental
  uncertainties have been reduced from 23\% in winter 2011 down to the
  level of 6\% in July 2011.}. On the other
hand, one can render the $H\to \tau \tau$ channel more effective by
simply including the contribution of the other Higgs production
mechanisms such as: vector boson fusion, which up until now is the only
production channel that has been analyzed by experimental
collaborations in the $H\to\tau^+\tau^-$ search
topology~\cite{Aad:2008aa, Ball:2007zza} and first proposed in
Ref.~\cite{Plehn:1999xi}\footnote{Because of the two additional
  forward jets, the sensitivity of this channel (the only one
  involving $\tau$ leptons that has been considered in the SM so far)
  is much larger than its contribution ($\approx 10\%$) to the total
  $pp\to H\to \tau \tau+X$ inclusive rate indicates.}; associated
production with a $W$ and $Z$ bosons, which will increase the cross
section for the inclusive $pp\to \tau \tau+X$ production mechanism by
15 to 20\%. We have displayed in Fig.~\ref{fig:SMHiggsMSSM} the effect
of including all these production channels (naively, that is without
making use of the specific cut for vector boson fusion which
significantly increases the sensitivity) in the inclusive $pp\to H\to
\tau^+ \tau^-+X$ signal, in dashed red line for the ``expected'' limit
and in solid red line for the ``observed'' limit. We have also
displayed correspondingly in green and red bands the effect of the
theoretical uncertainties. Hence, the $pp\to {\rm Higgs} \to
\tau^+\tau^-$ inclusive channel turns out to be a very interesting and
potentially very competitive Higgs detection channel also in the
Standard Model.

\subsection{Summary and outlook}

We have presented two kinds of results in this section: the numerical
theoretical predictions for the MSSM neutral Higgs bosons branching
fractions and their combination with the production cross sections on
the one hand, the consequence of the latter on the MSSM parameter
space as well as a consequence on the SM Higgs search on the other
hand.

We recall that we have updated the cross sections for the
production of the MSSM $CP$--odd like Higgs bosons $\Phi$ at the
Tevatron in the processes $gg\to \Phi$ and $b\bar b \to \Phi$ and
found smaller rates in the high Higgs mass range compared to those
assumed by the CDF and D0 experiments. We have then evaluated the
associated theoretical uncertainties together with the impact of the
$\Phi\!\to \!\tau^+ \tau^-$ decay branching fractions and find that
they are very large. These uncertainties, together with the correct
normalization, affect significantly the exclusion limits set on the
MSSM parameter space from the negative Higgs searches in the channel
$p\bar p \to \Phi \to \tau^+ \tau^-$ at the Tevatron.

We have done the same exercice at the lHC with $\sqrt s = 7$ TeV. We
have investigated the production of the $CP$--odd like particles,
$\Phi=A$ as well as one of the $CP$--even $H$ or $h$ particles
(depending on whether we are in the decoupling or anti--decoupling
regimes) that are degenerate in mass with  $A$ and have the same
couplings,  in particular, enhanced couplings to bottom quarks for the
high values of $\tb$ than can be  probed at the lHC. The two main
production processes $gg \to \Phi$ and $b \bar b \to \Phi$, which
could have cross sections that are order of magnitude larger than in
the SM Higgs case, have been considered. Numerical results at
energies $\sqrt s=7$--14 TeV have been given and the associated
theoretical uncertainties have been evaluated. The latter are due not
only to the scale and PDF+$\alpha_s$ uncertainties which appear in the
SM case, but there are also uncertainties associated to the $b$--quark
which plays a major role in the MSSM. The error on the input $b$-quark
mass, the scheme dependence in the renormalization of the mass of the
$b$--quark in its contribution to the $gg \to \Phi$ amplitude and the
effect of $m_b$ on the bottom quark densities in the process $b\bar b
\to \Phi$ will induce additional uncertainties. To these results we
have to add the parametric uncertainties in the branching ratio of the
Higgs decay in tau lepton pairs, which is the cleanest detection
channel at hadron colliders, that is found to be at the level of 10\%.

The overall theoretical uncertainty in these two processes turn out to be 
extremely large at lHC energies, of the order of 50\%. This large
uncertainty have some impact on the MSSM parameter space that can be
probed and this has been also discussed when comparing with the 36
pb$^{-1}$ data from CMS collaboration. Still the results lead to very
strong constraints on the $[\tan\beta, M_A]$ MSSM parameter space. We
also recall that these constraints are essentially model independent
(for the values of $M_A$ and $\tb$ that are being probed  so far) and
slightly less effective when the theoretical uncertainties in the
predictions for the Higgs cross sections and branching ratios are
properly included, $\tb \gsim 29$ excluded. We have also estimated
that these limits can be significantly improved in the absence of a
Higgs signal with a few inverse femtobarn data: $tb \gsim 20$ can be
excluded with 1 fb$^{-1}$ data, a result that has been confirmed by
the newest CMS analysis presented in the HEP--EPS 2011
conference~\cite{Tonelli:2011EPS} (page 24), $\tb \lsim 10$ can be
excluded for not too heavy neutral Higgs bosons and a luminosity of
order 3 fb$^{-1}$. At this stage, many channels such as  $pp\! \to\!
\Phi\! \to \mu^+ \mu^-, pp \!\to \!btH^- \to\! bt \tau\nu$, $gb \!\to
\!\Phi b\!\to \!3b$ and $t \to H^+b$ will not be viable anymore even
at the design energy and luminosity of the LHC.

The final most important consequence that we have drawn in that the
inclusive $pp\to {\rm Higgs} \to \tau \tau$ process is also a very
promising channel in the search for the Standard Model Higgs
boson if the production rate $gg\to H$ is included. Indeed, this
channel has a sensitivity that is only a factor of two smaller than
the expected sensitivity of the main  $pp \to H \to \gamma \gamma$
channel when compared to the ATLAS analysis with a luminosity of 36
pb$^{-1}$ in the $H\to\gamma\gamma$ search channel. While little
improvement is expected in the later channel, there are ways to
significantly enhance the sensitivity of the $pp\to H\to \tau\tau$
signal and render it a very powerful discovery channel for the SM
Higgs boson in the difficult $M_H=115$--130 GeV mass range. Indeed,
this $gg\to H\to\tau^+\tau^-$ channel is now considered by the
experimental collaborations in their latest results presented at the
HEP--EPS 2011 conference~\cite{Gennai:2011EPS} (see also
Ref.~\cite{Murray:2011EPS} page 13).

\vfill
\pagebreak

\part{Perspectives}
\label{part:five}

\section{Exclusive study of the gluon--gluon fusion
  channel \label{section:Exclusive}}

This last section is devoted to the perspectives of the work that has
been done for the last three years. We have completed a detailed
analysis of the inclusive production cross section times decay
branching fractions at the Tevatron and LHC colliders both in the
Standard Model and its minimal supersymmetric extension. The next step
that is of utmost interest for experimentalists is the exclusive study
of all the production and decay channels. Indeed, any experimental
search is conducted in a specific channel with specific search
topologies and cuts in order to maximize the signal over the
background. We, as a theorist, are to give to experimental
collaborations the most precise and detailed analysis of such search
topologies. In addition the same work has to be done on the background
analysis.

There are other perspectives to think of, for example:
\begin{itemize}
\item{the study of the impact of the Higgs decay width on the
    predictions, as for heavy Higgs $M_H\gsim 500$ GeV this really has
  to be taken into account. Related to that effect is the study of the
interference between the signal and the background, e.g. in the
channel $pp\to H\to WW$ that means to make a detailed analysis of the
$pp\to WW$ cross section in which the process $pp\to H\to WW$ is
considered as part of the amplitude and not an isolated process;}
\item{the impact of the cuts on the theoretical uncertainties in an
    exclusive study. Indeed it is well known that cuts and jet bin
    seperation often enhance the scale uncertainties, see
    Ref.~\cite{Anastasiou:2009bt} in the case of the Tevatron; it may
    be interesting to make a study of the cuts that both improve the
    significance of the signal over thebackground and minimize the
    scale uncertainty.}
\end{itemize}

We will present in this section some preliminary results on the
Standard Model exclusive study of the channel $pp\to H\to W^+W^-$ at
the lHC with $\sqrt s = 7$ TeV, in which well educated cuts on the
phase space are implemented in order to maximize the signal
significance over the backgrounds, which is the spirit of an exclusive
study in comparison with an inclusive calculation where all the phase
space is taken into account, see
Refs.~\cite{Anastasiou:2009bt,Grazzini:2008tf} for studies of this
channel at the Tevatron and the LHC. We will also give some results on
the background study with the same set of cuts. This work is an
on--going collaboration~\cite{Jeremie} in the context of the LHC Higgs
Cross Section Working Group.

\subsection{Exclusive SM Higgs
  production \label{section:ExclusiveProduction}}

We study the exclusive channel at the lHC $gg\to H\to W^+W^- \to \ell
\nu \ell \nu$ in which two leptons have to be found together with
missing transverse energy. This channel is a silver channel for Higgs
discovery with a mass $M_H \ge 135$ GeV as stated in
subsection~\ref{section:SMHiggsDecayIntro}. We will make use of the
CMS cuts of Ref.~\cite{Chatrchyan:2011tz} and summarized in
Table~\ref{table:ggHcuts} below:

\begin{table}[!h]{
\let\lbr\{\def\{{\char'173}%
\let\rbr\}\def\}{\char'175}%
\begin{center}
\begin{tabular}{|c||c|c|l|l|}
\hline
$M_H$ [GeV] & $p_T^{\ell, \rm max}$ [GeV] & $p_T^{\ell, \rm min}$
[GeV] & $m_{\ell \ell}$ [GeV] & $\Delta \Phi_{\ell \ell}$ [deg] \\
\hline 
$\geq 130$ & $> 25$ & $> 20$ & $<45$ & $<60$ \\
$\geq 160$ & $> 30$ & $> 25$ & $<50$ & $<60$ \\
$\geq 200$ & $> 40$ & $> 25$ & $<90$ & $<100$ \\
$\geq 210$ & $> 44$ & $> 25$ & $<110$ & $<110$ \\
$\geq 400$ & $> 90$ & $> 25$ & $<300$ & $<175$ \\ \hline
\end{tabular}
\caption[CMS cuts used in the SM exclusive study $gg\to H\to WW\to
\ell \nu \ell \nu$ at the lHC]{CMS cuts on the charged leptons for the
  $gg\to H\to WW \to \ell \ell \nu \nu$ final state.} 
\label{table:ggHcuts}
\end{center}
}
\end{table}

Other cuts can further reduce the background: we can define a $Z$ veto
in the $l^{+}l^{-}$ final states by which events with an invariant
dilepton mass within 15 GeV around the $Z$ mass are discarded, which
reduces the Drell--Yan background. In addition we also impose the
following cuts:
\begin{itemize}
\item{$E_T^{\rm miss} > 20$ GeV;}
\item{$E_T^{\rm j} > 25$ GeV for the jets, with $|\eta^{\rm j}|<5$ and
    a distance $R=0.5$ between jets.}
\end{itemize}

To calculate the signal cross section with the cuts defined in
Table~\ref{table:ggHcuts} we use the program {\tt
  HNNLO}\footnote{The program can be found in
  Ref.~\cite{Grazzinipage}.}~\cite{Catani:2008me}. We will use for the
central prediction the MSTW2008 NNLO PDFs set if nothing else is
indicated and the central scales are set to $\mu_R=\mu_F=\mu_0=\frac12
M_H$ as in the inclusive study. This is known to absorb part of the
higher order corrections already at the LO thus improving the
perturbative expansion~\cite{Anastasiou:2009bt,Grazzini:2008tf}. The
results presented in this perspective section are preliminary in the
sense that the calculation is heavily time consuming: for the total
Higgs cross section $\sigma_{\rm   tot}^{\rm NNLO}$ and  the
Higgs+zero--jet (H+0j) $\sigma_{0j}^{\rm   NNLO}$ the program runs for
more than 45 hours. The results that we display need to be checked
against numerical instabilities and are not to be taken as serious but
rather an indication at this stage.

For the H+1j and H+2j cross sections, the NLO and LO MSTW sets are
adopted. We display in Table~\ref{table:ggHcutsResults} the results
for the most sensitive Higgs mass $M_H=160$ GeV together with the
associated numerical errors (note that for NNLO there are still at the
3\% level which is too much!). We have also calculated the Higgs+jet
cross sections at NLO consistently (i.e. all PDFs and $\alpha_s$ at
NLO) using the program {\tt MCFM}~\cite{Campbell:2010ff} (for the H+2j
cross section, there is an improvement here as the result is at NLO
while with HHNLO is only at LO); the results are shown in the last
column of Table~\ref{table:ggHcutsResults}.

\begin{table}[!h]
\begin{center}
\begin{tabular}{|l|l||c|c|c||c|}
\hline
$M_H$ & $\sigma_{\rm jets}$ [fb] & LO & NLO & NNLO & MCFM@NLO \\
\hline 
& H+0jets & $9.890 \pm 0.004$ & $11.598 \pm 0.043$ & $ 10.952 \pm
0.186$ & $11.173 \pm 0.010$ \\
160& H+1jet & $8.981 \pm 0.007$ & $7.048 \pm 0.090$  & -- & $  6.834
\pm 0.032$  \\
& H+2jets & $5.778 \pm 0.007$ & -- & -- &  $ 2.525  \pm 0.086 $ \\
\hline 
\end{tabular}
\caption[Results for the $gg\to H$+jet cross sections with $M_H=160$
GeV at the lHC with HNNLO and MCFM programs]{The $gg\to H$+jet cross
  sections at the 7 TeV lHC calculated with HNNLO at the various
  perturbative orders with MSTW PDFs and central scale
  $\mu_F=\mu_R=\frac12 M_H$. In the last colum, shown are the results
  when one uses the program MCFM at NLO.}
\label{table:ggHcutsResults}
\end{center}
\end{table}

Bearing in mind that these results as already mentioned are really
preliminary, we can raise some remarks:
\begin{enumerate}[$a)$]
\item{the K--factors are $K^{\rm NLO}=1.173 \pm 0.005$ and $K^{\rm
      NNLO}=1.107\pm 0.019$ which seems at first sight to be weird:
    $K^{\rm NLO} > K^{\rm NNLO}$ and both are really close to unity in
    constrast to what is known for the total exclusive cross section,
    even with $\mu_0=\frac12 M_H$ which improves the perturbative
    calculation;}
\item{the results between {\tt HNNLO} and {\tt MCFM} are consistent
    once the numerical errors are included. We note that for this
    comparison {\tt MCFM} is fully at NLO which means that the H+0j
    calculation misses one order compared to {\tt HNNLO}, the H+1j is
    at the same perturbative order and the H+2j is one perturbative
    order more than {\tt HNNLO} calculation;}
\item{the total Higgs cross section including the cuts is also given
    by HNNLO and is
    \beq
    \sigma_{cuts}^{\rm NNLO} ({\rm MSTWNNLO})=  20.778 \pm 0.174~{\rm
      fb}
    \eeq
    which is lower than what is obtained when summing the jet cross
    sections in Table~\ref{table:ggHcutsResults}, $\sigma_{0j}^{\rm
      NNLO}+ \sigma_{1j}^{\rm NLO}+\sigma_{2j}^{\rm LO}= 23.778  \pm
    0.21$ fb (with a quadratic sum of the errors). The difference is
    due of course to the fact that in the total cross section
    $\sigma_{cuts}^{\rm NNLO}$, the PDF for the 1j and 2j cross
    sections are both taken at NNLO consistently throughout the
    calculation. We need to think how to interpret this difference in
    term of PDF uncertainties on the cross section.}
\end{enumerate}

The next step is to apply the procedure that we have developed in
part~\ref{part:two} to deal with the various sources of theoretical
uncertainties that affect our predictions. In the case of this
exclusive study we deal with the scale uncertaintiy, the
PDF+$\alpha_s$ uncertainty and the specific jet cut uncertainty. All
of these are calculated as stated below:
\begin{enumerate}[$i)$]
\item{for the scale variation, rather than performing a scan in the
    usual domain  $\frac12 \mu _0 \leq \mu_R, \mu_F \leq 2\mu_0$ with
    $\mu_0=\frac 12 M_H$ which is time consuming, we simply take the
    cross section values for $\mu_F=\mu_R= \frac12\mu_0$ and
    $\mu_R=\mu_F= 2\mu_0$ which, in the case of the total inclusive
    cross section, provide the maximal and minimal values. This is
    nothing more than a mere assumption and it might indeed
    underestimate the scale uncertainty as in the case of the
    exclusive production two characteristic scales at least are
    present: the mass of the Higgs boson on the one hand and the $p_T$
    of the distribution on the other hand. Nevertheless our assumption
    will give an indication of the scale uncertainty that is
    sufficient for this preliminary study;}
\item{The PDF4LHC recommendation~\cite{PDF4LHC} is adopted for the
    evaluation of the PDF+$\alpha_s$ uncertainty: we use the
    5$\times$40 MSTW grids to obtain the 68\%CL combined
    PDF+$\alpha_s$ uncertainty of the various jet cross sections and
    multiply the result by a factor of two, see also
    Ref.~\cite{Dittmaier:2011ti};}
\item{the uncertainty on the jet cut is calculated assuming simply
    that there is a 10\% uncertainty on the jet energy. We thus
    calculate the cross section when the jet cut is either $E_T^{\rm
      jet}=22.5$ GeV or 27.5 GeV. We keep the same values for the
    rapidity $\eta^{\rm jet}$ and the distance R. We do not consider
    for the moment the problem that the errors for the various jet
    cross sections are correlated and that there should be no error in
    the sum;}
\item{the total uncertainty is eventually calculated by a linear sum
    of the three individual uncertainties: for scale and
    PDF+$\alpha_s$, this follows the LHC Higgs working group
    recipe~\cite{Dittmaier:2011ti}, while for the uncertainty on the
    jet energy which is of experimental nature, the situation might be
    slightly more complicated and we may have to think about
    correlations more carefully.}
\end{enumerate}

The results for the individual and total uncertainties are given (in
\%) in Table~\ref{table:ggHcutsUncertainty1} below.

\begin{table}[!h]
\begin{center}
\begin{tabular}{|c|c||c|c|c||c|}
\hline
$M_H$ & $\sigma_{\rm jets} $ & $\Delta \mu$ [\%] & PDF+$\Delta
\alpha_s$ [\%] & $\Delta E_T^j$ [\%] & total [\%]  \\ \hline
   & H+0jets & ${-0.1}\;{+0.2}$ & ${-55.7}\;{+50.0}$ &
   ${-7.6}\;{+6.7}$ & ${-63.4}\;{+56.9}$ \\
160 & H+1jet  & ${-7.3}\;{-9.9}$ & ${-16.2}\;{+12.7}$ &
${-4.1}\;{+4.2}$ & to be calculated \\
   & H+2jets & ${-43.3}\;{+89.5}$ & ${-16.4}\;{+18.1}$ &
   ${-13.9}\;{+16.5}$ & ${-73.6}\;{+124.1}$ \\
\hline 
\end{tabular}
\caption[Uncertainties on the exclusive production $gg\to H\to WW\to
\ell \nu \ell \nu$ with $M_H=160$ GeV at the lHC with {\tt HNNLO}
program]{The scale. PDF+$\alpha_s$ and jet energy uncertainty on the
  various $gg\to H\to \ell \nu \ell \nu$ jet cross sections at the lHC
  with $\sqrt s = 7$ TeV for $M_H=160$ GeV as well as the total
  uncertainty as obtained with {\tt HNNLO}.}
\label{table:ggHcutsUncertainty1}
\end{center}
\end{table}

It has been checked that for the H+1j, H+2j and also H+0j cross
sections, all evaluated at NLO, the same results are obtained with
{\tt MCFM} as shown in Table~\ref{table:ggHcutsUncertainty2} below:

\begin{table}[!h]
\begin{center}
\begin{tabular}{|c|c||c|c|c||c|}
\hline
$M_H$ & $\sigma_{\rm jets}$ & $\Delta \mu$ [\%] & PDF+$\Delta\alpha_s$
[\%] & $\Delta E_T^j$ [\%] & total [\%]  \\ \hline
   & H+0jets &  ${-9.7}\;{+8.5}$ & ${-7.5}\;{+8.1}$ &
   ${-5.0}\;{+4.0}$ & ${-22.2}\;{+20.6}$ \\
160 & H+1jet & ${-7.7}\;{-7.9}$ & ${-11.8}\;{+11.6}$ &
${-5.3}\;{+3.9}$ & to be calculated \\
   & H+2jets & ${-56.5}\;{+30.1}$ & to be calculated &
   ${-4.0}\;{+3.0}$ & to be calculated \\
\hline 
\end{tabular}
\caption[Uncertainties on the exclusive production $gg\to H\to WW\to
\ell \nu \ell \nu$ with $M_H=160$ GeV at the lHC with {\tt MCFM}
program]{The scale. PDF+$\alpha_s$ and jet energy uncertainty on the
  various $gg\to H\to \ell \nu \ell \nu$ jet cross sections at the lHC
  with $\sqrt s = 7$ TeV for $M_H=160$ GeV as well as the total
  uncertainty as obtained with {\tt MCFM}.}
\label{table:ggHcutsUncertainty2}
\end{center}
\end{table}

Some calculations remain to be done, and everything need to be checked
against numerical instabilities in particular the PDFs uncertainties
that are the most time consuming calculations.

\subsection{SM Backgrounds \label{section:ExclusiveBack}}

In addition to the study of the signal cross section it is of utmost
importance to study as well the backgrounds processes together with
the selection cuts. We will consider four background processes leading
to $\ell \ell \nu \nu$+X  final states with the cuts of
Table~\ref{table:ggHcuts}:

\begin{enumerate}[$1.$]
\item{$pp \to WW \to \ell \ell \nu \nu$: we evaluate this process at
    NLO using {\tt MCFM}  and we do not consider the interference
    between the signal and the background. As stated at the beginning
    of this perspective section, this is a separate study that
    remains to be done. The central scale is taken to be  $\mu_0=
    M_W$;}
\item{$pp \to ZZ\to \ell \ell \nu \nu$: this channel is also evaluted
    with {\tt MCFM} at a central scale $\mu_0=M_Z$. We recall that we
    use the $Z$ jet veto that removes the contribution of a resonant
    $Z$ boson;}
\item{$pp \to t\bar t$: this production channel will mainly affect the
    H+2jet cross section. The process is evaluated at NLO using {\tt
      MCFM} and renormalised to NNLO by using the approximate
    inclusive NNLO to NLO rates obtained with the program {\tt
      HATHOR}~\cite{Aliev:2010zk}. The scales are set to
    $\mu_F=\mu_R=m_t$;}
\item{single top production: we take into account the two processes
    $pp \to W t$ and $pp \to W t+1j$ that we evaluate again using {\tt
      MCFM} with central scales $\mu_0=\frac{M_W+m_t}{2}$.}
\end{enumerate}

There are many other backgrounds processes, involving the $Z$ or the
$\gamma$ bosons, not to mention the huge QCD background that is
inevitable at hadron colliders. However they will not have as much
impact on the Higgs search in the diboson channel that we study in
this section as we require the observation of two leptons.

The cross sections, using the cuts of Table~\ref{table:ggHcuts},
together with the cut on the transverse missing energy $E_T^{\rm
  miss}$ and the $E_T,\eta$ and $R$ of the jet, are displayed in
Table~\ref{table:backgroundcut} using MSTW2008 NLO PDFs\footnote{The
  numerical errors are very small and have been omitted.}. Together
with the central cross sections, we also dispaly the uncertainties on
the cross sections due to scale variation (within a factor of two from
the central scale) and due to the PDF+$\alpha_s$, applying again the
PDF4LHC recipe~\cite{PDF4LHC} by taking 68\%CL PDF+$\alpha_s$
uncertainty multiplied by a factor of two. Whenever relevant, we also
estimate the unceratinty on the jet energy.  The total uncertainty is
obtained by added linearly the three uncertainties as in the case of
the signal cross section.

\begin{table}[!h]
\begin{center}
\begin{tabular}{|c|c||c||c|c|c||c|}
\hline
$M_H$ & process & $\sigma_{\rm bkg}$ [fb]  & $\Delta\mu$ [\%] &
PDF+$\Delta\alpha_s$ [\%] & $\Delta E_T^j$ [\%] & total [\%] \\
\hline
   & $WW$ & $8.85$ & ${-1.1}\;{+2.1} $ & ${-4.0}\;{+5.6}$ &
   ${-3.4}\;{+3.3}$& ${-8.5}\;{+11.0}$ \\
160 & $ZZ$  & $0.26$ & ${-1.5}\;{+2.3}$ & ${-7.2}\;{+4.5}$ &
${-4.9}\;{+4.2}$& ${-13.6}\;{+11.0}$ \\
     & $t\bar t $  & $41.79$ & ${-23.2}\;{+10.2}$ & ${-6.6}\;{+6.9}$ &
     ${-4.5}\;{+3.3}$& ${-34.3}\;{+20.4}$ \\
       & $tW$  & $4.71$ & ${-9.9}\;{+8.6}$ & ${-24.7}\;{+12.8}$ &
       ${-9.7}\;{+8.4}$& ${-44.3}\;{+29.8}$ \\
       \hline 
\end{tabular}
\caption[Central values and uncertainties for the $H\to WW$ SM
backgrounds exclusive cross sections at the lHC]{The central values
  and the associated individual and total uncertainties for the
  various background cross sections at the lHC with 7 TeV using the
  cuts of Table~\ref{table:ggHcuts}. {\tt MCFM} has been used with the
  default scales discussed in the text and the MSTW2008 NLO PDF sets
  adopted.}
\label{table:backgroundcut}
\end{center}
\end{table}

\section*{Conclusion\markboth{Conclusion}{Conclusion}}
\addcontentsline{toc}{section}{\protect\numberline{}Conclusion}

The red line we have been guided by throughout this thesis has been
the wish to unravel the mechanism that is behind the electroweak
symmetry breaking, focusing of its realization through the Higgs
mechanism of spontaneous symmetry breaking either in the Standard
Model (SM) or in its minimal supersymmetric extension called the
MSSM. The Standard Model itself has been the subject of a short review
in part~\ref{part:one} while the MSSM and supersymmetry in general has
been the subject of part~\ref{part:three}. We noted that if one Higgs
boson is present in the SM, we have five Higgs bosons in the MSSM
coming from the use of two Higgs doublet: two $CP$--odd neutral Higgs
boson $h$ and $H$, one $CP$--odd neutral Higgs boson $A$ and two
charged Higgs bosons $H^{\pm}$.

The two current best places we have to study this fascinating subject
are the high energy hadron colliders, the Fermilab Tevatron collider
on the one hand and the CERN Large Hadron Collider (LHC) on the other
hand which has started its activities with a 7 TeV center--of--mass
energy. We thus have focused on giving predictions for the Higgs
bosons production and decay branching fractions at both colliders
together with a detailed analysis of the theoretical uncertainties
affecting those predictions.

The study of the SM Higgs boson production was the subject of
part~\ref{part:two}. We started by a review of the current theoretical
and experimental bounds on the Higgs boson mass in
section~\ref{section:HiggsBounds} and then we moved on to the study of
the SM Higgs boson production at the Tevatron in
section~\ref{section:SMHiggsTev}. We focused on the two main
production channels that are the gluon--gluon fusion with triangular
top and bottom quarks loops and the Higgs--strahlung production with
either a $W^{\pm}$ or a $Z$ boson in association with the Higgs
boson. The first production channel is the major search channel for
high Higgs mass searches $M_H\gsim 135$ GeV where the Tevatron is the
most sensitive to a Higgs signal while the latter is used in low Higgs
mass searches $M_H\lsim 135$ GeV.

We have studied three kinds of uncertainties.
\begin{enumerate}[$\bullet$]
\item{The first is the scale
    uncertainty that estimates the effect of unknown higher--order terms
    in the perturbative expansion by varying the renormalization and
    factorization scales $\mu_R,\mu_F$ within a definite interval
    $1/\kappa \mu_0 \leq \mu_R,\mu_F \leq \kappa \mu_0$ with $\mu_0$ as
    the central scale and we exposed a way to define its (subjective)
    choice in subsection~\ref{section:SMHiggsTevScale}: choose the
    constant factor $\kappa$ as the one that enables
    the lower order scale uncertainties bands to catch the central
    prediction of the highest order of the calculation. We then found that
    if $\kappa=2$ is enough for the Higgs--strahlung process at
    next--to--next--to--leading order (NNLO) with a central scale
    $\mu_0=M_{HV}$, $M_{HV}$ being the invariant mass of the $W$ and $H$
    boson system, $\kappa=3$ would be recommanded for the gluon--gluon
    process at NNLO even with the central scale choice $\mu_0=\frac12 M_H$
    which is known to minimize the impact of higher order terms.}
\item{The second source of uncertainties is related to the PDF
    puzzle. Indeed it is well known that there are many PDF
    collaborations on the market and they do not give the same
    predictions with sometimes dramatic variations up to $30\%$. A way
    to reconcile these different predictions is to take into account
    both the experimental and theoretical uncertainties on the value
    of $\alpha_s$ and we gave in
    subsection~\ref{section:SMHiggsTevPDF} numerical results within
    this scheme.}
\item{We then introduced in subsection~\ref{section:SMHiggsTevEFT} the
    specific uncertainties of the gluon--gluon fusion channel related
    to the use of an effective theory approach to simplify the hard
    calculation: the infinite top mass approximation at NNLO which is
    excellent for the top loop but implies that we have to estimate
    the absence of the bottom loop; the use of an effective approach
    for the calculation of the mixed QCD--electroweak corrections at
    NNLO. We gave sizeble numerical results}
\end{enumerate}
We closed off the section~\ref{section:SMHiggsTev} by the
subsection~\ref{section:SMHiggsTevTotal} which was devoted to the
combination of all the uncertainties into a single total
uncertainty. We defined our method as to calculate the
PDF+$\Delta\alpha_s$ uncertainty on the maximal and minimal cross
sections regarding to the scale uncertainty and then add linearly the
small effective field theory uncertainty. This has to be contrasted
with the quadratic addition used by the experimental collaborations,
and has to be compared with the recommanded linear addition of the LHC
Higgs Cross Section Working Group
recommendation~\cite{Dittmaier:2011ti}. We found an uncertainy that
amounts to nearly $\pm 20\%$ in the gluon--gluon fusion channel in the
Higgs mass range $115\leq M_H\leq 200$ GeV that is of Tevatron
interest and approximately $\pm 8\%$ in the case of the
Higgs--strahlung channel. 

We then moved on to section~\ref{section:SMHiggsLHC} when the same
outlines were repeated for the study of the gluon--gluon fusion main
production channel. We gave numerical results both for the LHC at 7
TeV that we call lHC for littler Hadron Collider, and for the LHC at
14 TeV. We found much more controled uncertainties as they amount up
to approximately $\pm 25\%$, either a bit more or a bit less depending
on the Higgs mass range. This was followed by
section~\ref{section:SMHiggsDecay} where we started by a short review
of the most interesting decay channels for experimental searches in
subsection~\ref{section:SMHiggsDecayIntro} before turning on to the
study of the SM Higgs decay branching fractions in
subsection~\ref{section:SMHiggsDecayResult} where we found that their
uncertainties should be taken into account in experimental searches
even if they are quite small. We finished this section by the most
important consequence of the combination of our results on the cross
section and decay branching fractions that deals with the current
Tevatron exclusion limit at 95\%CL for Higgs masses $M_H$ between 156
GeV and 177 GeV~\cite{Tevatron:2011cb}. We have estimated that if our
uncertainties are properly taken into account the exclusion band
obtained in Ref.~\cite{Aaltonen:2011gs} should be revisited and in
particular in view of the PDF puzzle that is still pending in the
community. We estimated that doubling the current luminosity would be
required to obtain the current claimed sensitivity.

Section~\ref{section:MSSMHiggsIntro} opened our study of the MSSM
neutral Higgs bosons production and decay at the hadron colliders. We
showed that the predictions we presented in the following
section~\ref{section:MSSMHiggsTev} and~\ref{section:MSSMHiggsLHC} are
somewhat model independant, we then move on to the study of the Higgs
bosons production at the Tevatron in
section~\ref{section:MSSMHiggsTev} focusing on the two main production
channels in the context of the MSSM at relatively high $\tb$ values,
$\tb$ being the ratio of the two Higgs doublets: the gluon--gluon
fusion channel through a triangular bottom quark loop, the top loop
being suppressed, and the bottom quarks fusion channel. Our study of
the uncertainties showed that they are quite large in the MSSM,
including a new source of uncertainty compared to the SM study and
coming from the $b$ quark. We reproduced the same study in the case of
the lHC in section~\ref{section:MSSMHiggsLHC} and then closed off this
part~\ref{part:four} by section~\ref{section:MSSMHiggsExp} where we
combined our results with the major Higgs to ditau decay branching
fraction. We compared our predictions to the experimental results
obtained both at the Tevatron and the lHC colliders and found that our
uncertainties have a significant impact on the MSSM $[M_A,\tb]$
parameter space that is probed by CDF/D0 and ATLAS/CMS experiments. We
also gave some predictions for the expected exclusion by CMS with
increased LHC luminosity that have been nicely confirmed by the latest
HEP--EPS 2011 conference results~\cite{Tonelli:2011EPS} (page 24). We
finished this section by giving one of the most important output of
the thesis: the possibility to use the $gg\to H\to\tau^+\tau^-$ search
channel for SM Higgs boson by comparing the limits obtained in the CMS
MSSM search with the SM predictions. We found that this channel is not
as bad compared to the long--celebrated $H\to\gamma\gamma$ for low
Higgs mass searches $110\leq M_H \leq 140$ GeV and proposed that it
should be used by the experimental collaborations. This is now the
case according to the latest HEP--EPS 2011 conference
results~\cite{Gennai:2011EPS}.

We finally arrived at the end by the perspective
section~\ref{section:Exclusive} where we gave some ideas for future
work and in particular the exclusive study of the Higgs bosons
production that is of utmost importance for experimental search as
specific topology with dedicated cuts on the kinematic variables are
actually used in a Higgs search. We gave some preliminary results from
an ongoing study that remains to be finished.

What is the roadmap for the very close future? The latest HEP--EPS
2011 conference results~\cite{Tonelli:2011EPS,Murray:2011EPS} have
dramatically changed the hideouts for the SM Higgs boson: indeed
thanks to the 1 fb$^{-1}$ analysis of the ATLAS and CMS experiments
the SM Higgs boson mass is excluded at 95\%CL in the mass ranges $149
\leq M_H \leq 206$ GeV and $270\leq M_H \leq 450$ GeV which is a
dramatic improvement over both the old LEP results $M_H\leq 114.4$ GeV
and the (disputed) Tevatron results $157\leq M_H\leq 177$ GeV. The
latest HEP--EPS 2011 conference results are also very
intriguing as both ATLAS and CMS experiments show some hints of a
Higgs signal near $M_H=140$ GeV, at the level of $2.9\sigma$. This of
course has to be carefully checked against statistical fluctuations
but the fact that this is observed by the two experiments, not only in
the $H\to WW$ channel but apparently also in the $H\to ZZ\to 4\ell$
channel, produces excitations within the community as we may be on the
edge of finding the most elusive and searched particles of the past
thirty years, which actually is the only thing that matters! 

In any case, much work remains to be done: if these signals are indeed
the first hints of a Higgs boson discovery we have to work hard to
unravel its true nature: is it a SM Higgs boson, a MSSM Higgs boson
and in this case which one of the three neutrals? Is it something more
exotic as composite Higgs, or even not a Higgs at all but rather a
scalar signal of something more exotic? And even if it is really the
SM Higgs boson, lots remain to be done for the precise measurements of
its mass, couplings and $CP$ nature. On a theoretical side,
theoretical uncertainties on these precision observables will be a critical
issue, while this does not matter so much for a discovery as a peak in
kinematic distribution would be clearly established if we have $5\sigma$
significance in the amount of data.  On a more general view lots of
mysteries remain to be solved, just to mention a few: the exact nature
of dark matter, the matter--antimatter asymetry, the origin of the
three families, the unification of the electroweak and strong
interactions and even with gravity, etc. We are on the very edges of
exciting times!

\vfill

\appendix

\selectlanguage{french}
\section{Appendix: Synopsis}

\subsection{Introduction}
 
\subsubsection{Le Modèle Standard}

Cette thèse s'intéresse à l'étude théorique de la production et de la
désintégration du boson de Higgs, la particule témoin du mécanisme de
brisure électrofaible et qui reste à découvrir, dans les grands
collisionneurs hadroniques actuels que sont le Tevatron au Fermilab et
le grand collisionneur de hadrons (LHC) au CERN. Nous commençons
l'étude par le Modèle Standard (SM), que nous décrivons dans les
quelques lignes qui suivent, avant de nous intéresser à son extension
supersymétrique minimale (MSSM).

Le Modèle Standard est un modèle des particules élémentaires et de
leurs interactions dans le cadre de la théorie quantique des champs,
et basé sur l'algèbre de Lie $SU(3)_c \times (SU(2)_L \times
U(1)_Y)$ décrivant les symétries de jauge s'appliquant sur les champs
fondamentaux du modèle.

Les champs de matière, fermioniques, se répartissent en deux catégories :
\begin{enumerate}[--]
\item{Les quarks sensibles à toutes les interactions et en particulier
    à l'interaction forte décrite par l'algèbre de Lie $SU(3)_c$ et
    dont le nombre quantique est la couleur et la théorie quantique
    associée est la ChromoDynamique Quantique (QCD);}
\item{Les leptons qui ne sont sensibles qu'à l'interaction
    électrofaible décrite par l'algèbre de Lie $SU(2)_L \times U(1)_Y$
 dont les nombres quantiques sont respectivement l'isospin faible $I$
 et l'hypercharge $Y$.}
\end{enumerate}

Les générateurs des interactions sont des bosons de jauge, chacun
associé aux algèbres de Lie correspondantes :
\begin{itemize}
\item{huit gluons $g$ pour l'interaction forte;}
\item{un boson $B_\mu$ pour l'algèbre $U(1)_Y$;}
\item{trois bosons $W^a_\mu$ pour l'algèbre $SU(2)_L$.}
\end{itemize}

L'interaction électrofaible viole la parité de façon maximale : les
fermions sont divisés en doublets d'isospin faible et de chiralité
gauche, et en singulets d'isospin faible et de chiralité droite. Seul
les neutrinos n'ont qu'une seule chiralité, gauche, dans le Modèle
Standard et sont considérés de masse nulle\footnote{Comme précisé dans
  le texte principal, nous savons aujourd'hui expérimentalement depuis
  1998 que les neutrinos ont une masse très faible mais cependant non
  nulle car oscillants entre eux. Ce point est laissé de côté au sein
  du Modèle Standard.}. 

Le tableau ci-dessous résume le contenu en
champ fermionique du Modèle Standard.
\begin{table}[!h]
{\small%
\let\lbr\{\def\{{\char'173}%
\let\rbr\}\def\}{\char'175}%
\renewcommand{\arraystretch}{1.35}
\vspace*{-2mm}
\begin{center}
\begin{tabular}{|c|c|c|ccc|c|}\hline
Type & Nom & Masse~[GeV] & Spin & Charge &
$\left(I^W,I^W_3\right)_{L}$ & rep. $SU(3)_c$ \\ \hline
 & $\nu_e$ & $< 2\times 10^{-6}$ & $1/2$ & $0$ & $(1/2,1/2)$
 & $\mathbf{1}$ \\
 & $e$ & $5.11\times 10^{-4}$ & $1/2$ & $-1$ & $(1/2,-1/2)$
 & $\mathbf{1}$ \\ \cline{2-7}
 & $\nu_\mu$ & $< 1.9\times 10^{-4}$ & $1/2$ & $0$ & $(1/2,1/2)$
 & $\mathbf{1}$ \\
 LEPTONS & $\mu$ & $1.06\times 10^{-1}$ & $1/2$ & $-1$ & $(1/2,-1/2)$
 & $\mathbf{1}$ \\ \cline{2-7}
 & $\nu_\tau$ & $< 1.82\times 10^{-2}$ & $1/2$ & $0$ & $(1/2,1/2)$
 & $\mathbf{1}$ \\
 & $\tau$ & $1.777$ & $1/2$ & $-1$ & $(1/2,-1/2)$
 & $\mathbf{1}$ \\ \hline
 & $u$ & $(1.5\leq m\leq 3.0)\times 10^{-3}~(\overline{\rm MS})$ &
 $1/2$ & $2/3$ & $(1/2,1/2)$
 & $\mathbf{3}$ \\
 & $d$ & $(3.0\leq m\leq 7.0)\times 10^{-3}~(\overline{\rm MS})$ &
 $1/2$ & $-1/3$ & $(1/2,-1/2)$
 & $\mathbf{3}$ \\ \cline{2-7}
 & $c$ & $1.28~(\overline{\rm MS})$ & $1/2$ & $2/3$ & $(1/2,1/2)$
 & $\mathbf{3}$ \\
 QUARKS & $s$ & $9.5\times 10^{-2}~(\overline{\rm MS})$ & $1/2$ &
 $-1/3$ & $(1/2,-1/2)$ & $\mathbf{3}$ \\ \cline{2-7}
 & $t$ & $173.1$ & $1/2$ & $2/3$ & $(1/2,1/2)$
 & $\mathbf{3}$ \\
 & $b$ & $4.16~(\overline{\rm MS})$ & $1/2$ & $-1/3$ & $(1/2,-1/2)$
 & $\mathbf{3}$ \\ \hline
\end{tabular} 
\end{center} 
\caption[Contenu fermionique du Modèle Standard]{Description des
  champs fermioniques du Modèle Standard organisé en trois
  générations. Tous les champs ont une chiralité gauche et une
  chiralité droite, cette dernière étant singulet de $SU(2)_L$, sauf
  les neutrinos qui ne sont que des champs gauches.}
\label{table:SMcontenu}
}
\end{table}

Afin de respecter l'invariance de jauge et les symétries du Modèle
Standard, il est nécessaire que les masses et des fermions et des
bosons de jauge soient nulles. Ceci est incompatible avec
l'expérience : il est donc nécessaire d'avoir un mécanisme générant la
masse aux particules tout en préservant l'invariance de jauge, ce qui
est précisément le mécanisme de Brout--Englert--Higgs qui permet de
briser spontanément la symétrie électrofaible : la symétrie est
présente dans le lagrangien de la théorie mais brisée par la
configuration du vide de la théorie. De ce mécanisme
qui sera décrit ci-dessous émerge une particule supplémentaire au sein
du Modèle Standard : le boson de Higgs, seule particule restante d'un
isodoublet gauche scalaire appelé champs de Higgs.

\subsubsection{Le mécanisme de Brout--Englert--Higgs}

Le lagrangien du Modèle Standard complet est disponible dans le texte
principal de la thèse, Eq.~\ref{eq:smlagrangian}. Nous ne reprenons
dans ce synopsis que la partie intéressante pour décrire le mécanisme
de Brout--Englert--Higgs, donnée dans l'équation ci-dessous et qui ne
concerne que la partie scalaire du lagrangien du Modèle Standard :
\beq
{\cal L}_{\rm scalaire} =
\left(D_{\mu}\Phi\right)^{\dagger}\left(D_\mu\Phi\right) -
V(\Phi),~V(\Phi) = \mu^2 \Phi^{\dagger}\Phi +
\lambda\left(\Phi^{\dagger}\Phi\right)^2
\label{eq:scalaireSM}
\eeq
avec $D_\mu$ la dérivée covariante, construite de telle sorte que
$\left(D_{\mu}\Phi\right)^{\dagger}\left(D_\mu\Phi\right)$ soit
invariant de jauge :
\beq
D_{\mu} \psi = \left( \partial_\mu -\imath g_s U_a G^a_\mu -\imath g
  T_a W^a_\mu   -\imath
g_Y \frac{Y_q}{2} B_\mu \right) \psi
\label{eq:covariante}
\eeq
Dans les équations~\ref{eq:scalaireSM} et~\ref{eq:covariante} le champ
$\Phi$ décrit l'isodoublet faible scalaire de Higgs suivant, avec
hypercharge $Y=+1$ :
\beq
\Phi = \left( \begin{matrix}\phi^+\\
    \phi^0\end{matrix}\right)\nonumber
\eeq

Le mécanisme de Brout--Englert--Higgs repose sur la forme donnée au
potentiel scalaire $V(\Phi)=\mu^2 \Phi^{\dagger}\Phi + \lambda
\left(\Phi^{\dagger}\Phi\right)^2$. Son rôle est d'une part de donner
la masse aux bosons de jauge faibles et aux fermions, d'autre part de
permettre au Modèle Standard de respecter l'unitarité, c'est-à-dire la
conservation des probabilités nécessaire dans toute théorie quantique
cohérente, dans les processus de diffusion mettant en jeu justement
les bosons de jauge faibles. Ces deux raisons sont développées en
détail dans le texte principal, dans les
sous-chapitres~\ref{section:UnitarityWW}
et~\ref{section:UnitarityMasses}.

\begin{figure}[!t]
\begin{center}
\includegraphics[scale=0.6]{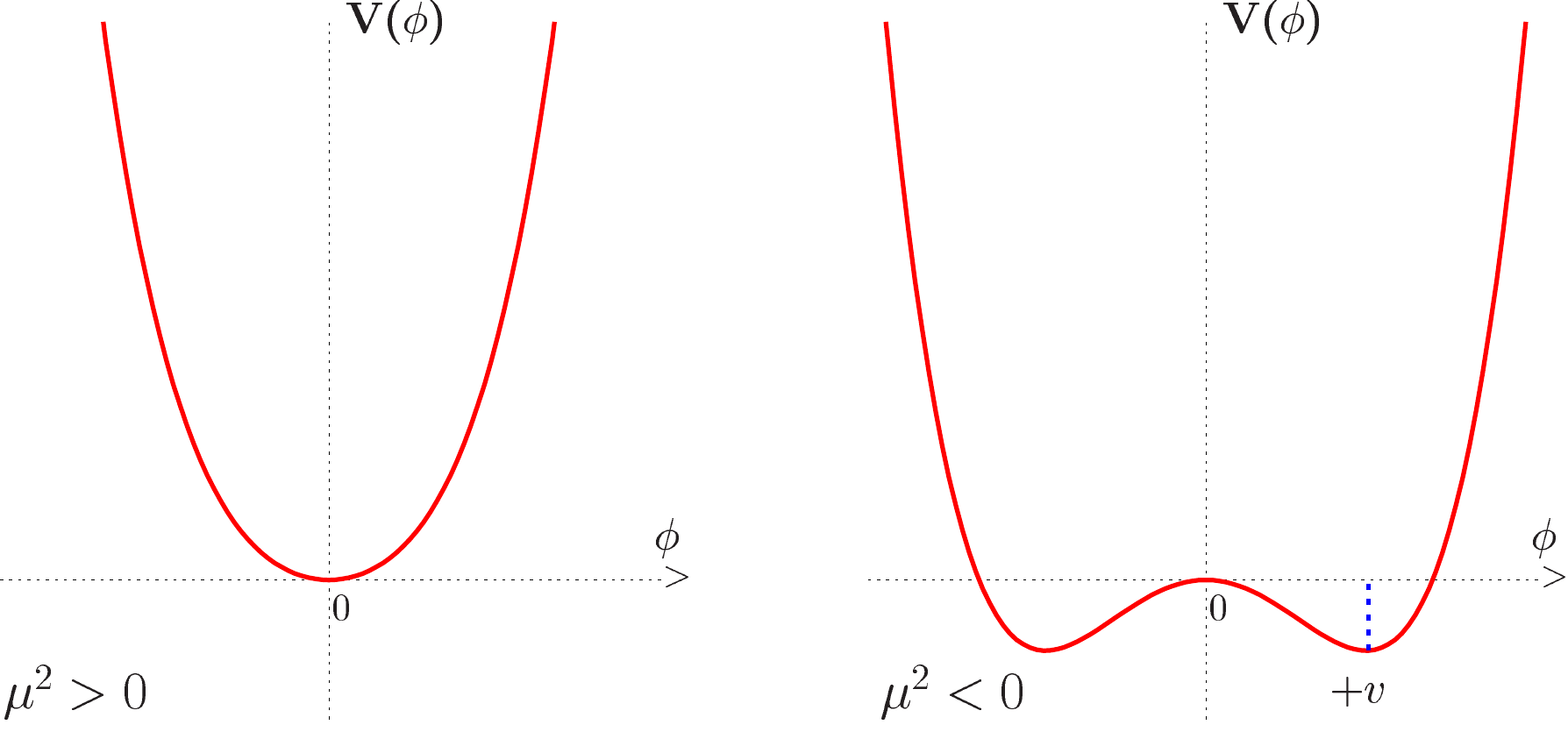} 
\end{center}
\vspace*{-.5cm}
\caption[Potentiel de Higgs dans le cas d'un champ scalaire réel selon
le signe du terme de masse]{Potentiel de Higgs dans le cas simplifié
d'un seul champ scalaire réel selon le signe du terme de mass
$\mu^2$. Cette figure est tirée de la référence~\cite{Djouadi:2005gi}.}
\label{fig:higgspotentiel} 
\end{figure}

La valeur moyenne dans le vide (vev) du doublet de Higgs est donnée
par le minimum du potentiel scalaire $V(\Phi)$. Afin d'avoir un
minimum il faut que le potentiel soit borné inférieurement, donc
$\lambda >0$. Selon la valeur du terme $\mu^2$, deux situations se
présentent, décrites dans la figure~\ref{fig:higgspotentiel}.

Pour un terme de masse positif $\mu^2>0$ le minimum est pour $\langle
\Phi \rangle = 0$ ce qui paraît attendu. L'hypothèse du mécanisme de
Brout--Englert--Higgs est de prendre $\mu^2 < 0$ : dans ce cas il y a
deux minima possibles, en réalité pour le doublet du Modèle Standard
une infinité invariante sous $U(1)$. La vev (et donc le vide) est alors une
configuration parmi toutes celles possibles et équivalentes : la
symétrie de départ est alors brisée par le vide de la théorie, ce
qu'on appelle une brisure spontanée de symétrie, cette dernière étant
encore présente dans les équations de la théorie mais ``cachée'' par le
vide qui lui brise la symétrie de jauge. Dans le cas du Modèle
Standard on dit que la symétrie électrofaible $SU(2)_L\times U(1)_Y$
se brise en la symétrie $U(1)_Q$ restante qui est l'électromagnétisme,
non brisée. La relation entre charge électrique et isospin/hypercharge
est donnée par $Q=I_3 + \frac12 Y$.

On fait un développement linéaire de l'isodoublet autour du minimum,
qui fait apparaître le boson de Higgs comme seule composante réelle du
doubet scalaire non absorbée par les bosons de jauge faibles. Dans la
jauge dite unitaire, ceci donne :
\beq
\Phi = \left( \begin{matrix}0\\
    \displaystyle \frac{v+H}{\sqrt 2}\end{matrix}\right)
\eeq
où la vev $v$ est donnée par
\beq
v^2 = - \frac{\mu^2}{\lambda} 
\eeq

En introduisant les mélanges suivants entre les divers champs de
jauge électrofaibles :
\beq
W^\pm = \frac{1}{\sqrt{2}} (W^1_\mu \mp \imath W^2_\mu) \  , \ 
Z_\mu = \frac{g W^3_\mu- g_Y B_\mu}{\sqrt{g^2+g_Y^2}} \ , \ 
A_\mu = \frac{g W^3_\mu+ g_Y B_\mu}{\sqrt{g^2+g_Y^2}} \  
\label{eq:WZA-champs}
\eeq
ces derniers sont les champs physiques et ont acquis une masse grâce à
la vev $v$ du champ de Higgs dans le développement du lagrangien
scalaire~\ref{eq:scalaireSM} :
\beq
M_W =\frac{gv}{2} \  , \ M_Z= \frac{v  \sqrt{g^2+g_Y^2}}{2}\ , 
 \ M_A=0 
\label{eq:massesfaibles}
\eeq
Le champ électromagnétique (champ du photon) est bien de masse nulle,
comme attendu pour une symétrie non brisée. Les termes de masses pour
les fermions sont donnés grâce à l'interaction de Yukawa suivante :
\beq
{\cal L}_{\rm Yukawa} &=& -\left(\frac{\lambda_e v}{\sqrt{2}}\right)
\bar{e} e - \left(\frac{\lambda_u v}{\sqrt{2}}\right)
\bar{u} u - \left(\frac{\lambda_d v}{\sqrt{2}}\right)
\bar{d} d \nonumber\\
 & & -\left(\frac{\lambda_e}{\sqrt{2}}\right)
\bar{e} H e - \left(\frac{\lambda_u}{\sqrt{2}}\right)
\bar{u} H u - \left(\frac{\lambda_d}{\sqrt{2}}\right)
\bar{d} H d
\label{eq:yukawabrisé}
\eeq
ce qui donne les masses suivantes :
\beq
m_e= \frac{\lambda_e\, v}{\sqrt{2}} \ \ , \ 
m_u= \frac{\lambda_u\, v}{\sqrt{2} }\ \ , \ 
m_d= \frac{\lambda_d\, v}{\sqrt{2}} 
\eeq

Les couplages entre le boson de Higgs et les autres particules du
Modèle Standard sont donnés dans le texte principal, notamment dans
l'équation~\ref{eq:higgscouplingsSMbosons}. Rappelons notamment
l'existence d'un couplage très important pour les collisionneurs
hadroniques, qui apparaît à l'ordre des boucles et qui n'est pas
présent directement dans le lagrangien du Modèle Standard : le
couplage Higgs--gluon--gluon via une boucle triangulaire de quarks
notamment les plus lourds, le top et le bottom.

Avant de conclure cette introduction nous indiquons au lecteur qu'il
peut trouver dans le corps principal du texte,
chapitre~\ref{section:HiggsBounds}, une étude des bornes théoriques et
expérimentales sur la masse du boson de Higgs du Modèle
Standard. Mentionnons simplement que cette masse n'est pas prédite par
la théorie ni même protégée des corrections radiatives\footnote{Ce que
  l'on appelle le problème de la hiérarchie
  des masses du Modèle Standard et qui est une (parmi beaucoup
  d'autres) des raisons motivant l'étude de théories allant au-delà du
  Modèle Standard.} et que les bornes théoriques donnent $50\lsim
M_H\lsim 750$ GeV si le Modèle Standard est une théorie valide
jusqu'à l'échelle du TeV, cet intervalle se restreignant à $130\lsim
M_H\lsim 180$ GeV si le Modèle Standard est valide jusqu'à l'échelle
de Planck $\Lambda_P \simeq 10^{18}$ GeV\footnote{Comme indiqué dans
  le texte principal, le sentiment prévalant dans la communauté est
  que le Modèle Standard n'est pas valide jusqu'à de telles échelles ;
  on peut penser au problème de la matière noire en cosmologie qui
  requiert l'existence d'une nouvelle particule non décrite par le
  Modèle Standard, si d'aventure la relativité générale est bien
  valide aux échelles cosmologiques. Il s'agit aussi d'avoir du
  travail pour le futur en tant que jeune physicien, espérons donc que
  la Nature ne nous a pas joué un vilain tour !}. Les bornes
expérimentales qui étaient disponibles lors de l'établissement de ce
travail de thèse sont $M_H > 114.4$ GeV (borne du LEP) ainsi que
l'exclusion de la bande $158\leq M_H\leq 177$ GeV au Tevatron, à 95\%
de niveau de confiance (CL).

\subsection{Production et désintégration du boson de Higgs du Modèle
  Standard}

\subsubsection{Le cas du Tevatron}

Le c\oe{}ur de la thèse est l'étude de la production inclusive du
boson de Higgs dans ses canaux principaux au Tevatron et au LHC et
notamment l'étude exhaustive des incertitudes théoriques affectant le
calcul. Nous allons voir qu'elles ont un impact significatif sur les
prédictions théoriques et qu'elles vont jouer un rôle important lors
de la comparaison de ces résultats avec les données expérimentales.

Deux canaux de découverte ont un rôle majeur au Tevatron :
\begin{itemize}
\item{pour des masses de Higgs petites, $M_H\lsim 135$ GeV, le canal
    de recherche dominant est la production du boson de Higgs par
    Higgs--strahlung, $p\bar p\to V H$ qui est une production associée
    avec un boson de jauge faible $V=W^{\pm}, Z$, suivie de la
    désintégration du boson de Higgs $H$ en paire de quarks bottom
    $H\to b\bar b$; même si le canal de production dominant est la
    fusion de gluons, ce dernier est noyé par un bruit de fond
    hadronique beaucoup trop important pour une recherche
    expérimentale efficace. Le canal de production $p\bar p\to VH$ est
    connue jusqu'à l'ordre $\alpha_s^2$ c'est-à-dire au
    next-to-next-to-leading order (NNLO) en QCD où $\alpha_s$ est
    la constante de couplage fort, et aussi au NNLO en ce qui concerne
    les corrections électrofaibles;}
\item{pour des masses de Higgs ``grandes'', $M_H\gsim 135$ GeV, le
    canal de recherche dominant est la production par fusion de gluons
  $gg\to H$ suivie de la désintégration du boson de Higgs en paires de
  bosons $W$, $H\to W W^*$. Ces derniers vont ensuite se désintégrer
  et le produit de désintégration le plus intéressant est une paire de
  leptons accompagnée d'énergie manquante emportée par les neutrinos,
  un état final particulièrement propre, ce qui d'ailleurs est le
  canal le plus sensible au boson de Higgs au Tevatron. Le canal
  $gg\to H$ est connu jusqu'au NLO en QCD de manière exacte pour les 
  contributions des boucles du quark top et du quark bottom, et
  jusqu'au NNLO en QCD dans une approche effective pour la boucle du
  quark top où ce dernier est pris de masse très grande par rapport à
  la masse du boson de Higgs. Les corrections électrofaibles sont
  connues jusqu'au NLO de façon exacte et jusqu'au NNLO de façon
  effective, dans un mélange entre corrections QCD et
  électrofaibles. Il existe aussi des corrections dite molles qui sont
  prises en compte dans une technique de resommation (NNLL) mais dont
  on ne tient pas directement compte dans cette thèse pour des raisons
  évoquées dans le texte principal,
  sous-chapitre~\ref{section:SMHiggsTevIntro}.}
\end{itemize}

Nous ne discuterons pas ici des deux autres canaux sous-dominants que 
sont la fusion de bosons faibles ainsi que la production en
association avec une paire de quarks lourds (principalement de quarks
top), quelques détails supplémentaires sont disponibles dans le corps
principal de la thèse, sous-chapitre~\ref{section:SMHiggsTevIntro}.

Les incertitudes théoriques se répartissent en trois grands groupes :
\begin{enumerate}[$1.$]
\item{La première incertitude est reliée au calcul perturbatif que
    l'on effectue. En effet, d'une part l'on effectue un calcul
    jusqu'à un ordre donné dans une expansion en la constante de
    couplage fort $\alpha_s$, d'autre part la section efficace
    hadronique est une convolution de ce calcul perturbatif partonique
  avec les distributions de partons (PDFs) dans le proton/l'antiproton, ce
  qui est un processus non perturbatif. Le calcul partonique fait
  intervenir l'échelle de renormalisation $\mu_R$, la convolution avec
les PDFs fait intervenir l'échelle de factorisation $\mu_F$. Ces deux
échelles ne sont pas physiques et ne sont que le reflet de notre
approximation dans le calcul : l'étude de la dépendance du résultat en
fonction de ces deux échelles est ce que l'on nomme l'incertitude
d'échelle et permet de donner une estimation de l'impact des
corrections quantiques d'ordres supérieurs.

L'usage est de faire varier ces deux échelles dans l'intervalle de
variation $\mu_0 / \kappa \leq \mu_R,\mu_F\leq \kappa \mu_0$ où
$\mu_0$ est l'échelle de la prédiction centrale (ou meilleure
prédiction) et $\kappa$ un coefficient constant à choisir. Dans le cas
de la fusion de gluons nous avons choisi $\mu_0=\frac12 M_H$ ce qui
permet une meilleure convergence des ordres supérieurs et permet de
reproduire l'effet des corrections NNLL; nous avons choisi
$\mu_0=M_{HV}$ pour la production en association avec les bosons de
jauge faibles, où $M_{HV}$ est la masse invariante de la paire
$H+V$. Il est important de bien comprendre que le choix de
l'intervalle est totalement subjectif, puisque en théorie cette
dépendance devrait être nulle si on pouvait prendre en compte tous les
ordres.

Nous avons défini un critère permettant de choisir $\kappa$ de
façon ``raisonnable'' : le coefficient que nous prendrons sera celui
permettant d'atteindre la prédiction centrale à l'ordre le plus élevé
dans le calcul avec les bandes d'incertitudes pour la prédiction à
l'ordre le plus bas (LO) ou l'ordre d'après (NLO). Le calcul au LO pour le
Higgs--strahlung étant un processus purement électrofaible, l'ordre le
plus bas qui est choisi est le NLO, et prendre $\kappa=2$ suffira. Par
contre, comme le montre la partie gauche de la
figure~\ref{fig:ggHTeV_echelle}, le processus $gg\to H$ demande
$\kappa=4$ si on utilise ce critère en prenant comme ordre le plus bas
l'ordre LO ; ceci dit nous voyons que la prédiction NNLO touche la
bande $\kappa=3$ à haute masse, et se trouve complètement dans cette
bande si on compare la prédiction NNLO aux bandes NLO. Nous
utiliserons donc $\kappa=3$ pour la fusion de gluons et nous obtenons
ainsi une incertitude d'ordre $+15\%,-20\%$ sur l'intervalle de masse
du boson de Higgs pertinent au Tevatron dans le cas du canal $gg\to
H$.
\begin{figure}[!h]
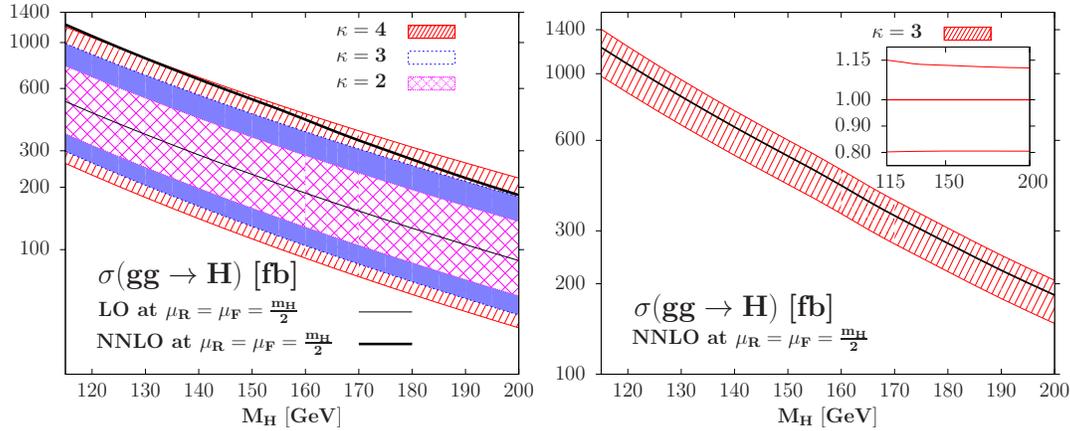

\begin{center}
\mbox{
\hspace*{-1.cm}
\includegraphics[scale=0.65]{./Figs/secondpart/tev/ggH_scale_errors1.pdf}
\hspace{-0.1cm}
\includegraphics[scale=0.65]{./Figs/secondpart/tev/ggH_scale_errors2.pdf}
}
\end{center}
\caption[Incertitude d'échelle dans le processus $gg\to H$ au
Tevatron]{Gauche : incertitude d'échelle au LO pour le canal $gg\to H$
  au Tevatron dans le domaine $ \mu_0/\kappa \leq \mu_R=\mu_F \leq
  \kappa \mu_0$ avec $\mu_0= \frac12 M_H$, pour des choix du
  coefficient $\kappa=2,3$ et $4$ comparée à la prédiction centrale au
  NNLO. Droite : l'incertitude d'échelle au NNLO du canal $gg\to H$
  avec $\kappa=3$ en fonction de $M_H$ au Tevatron. Dans les deux
  figures les encarts donnent les incertitudes relatives à la
  prédiction centrale.}
\vspace*{-5mm}
\label{fig:ggHTeV_echelle}
\end{figure}
\vspace{2mm}}
\item{Le second type d'incertitude est celui qui concentre le plus de
    questions brûlantes dans la communauté à l'heure actuelle :
    l'incertitude due aux fonctions de distribution des partons (PDFs)
  dans le proton/l'antiproton. En effet ces variables physiques sont
  non--perturbatives et ne sont donc pas prédites par la QCD à partir
  des premiers principes, mais résultent d'un ajustement à un ensemble
de données expérimentales. Il existe plusieurs collaborations qui
fournissent des prédictions variées pour ces PDFs, qui ne sont pas
toutes nécessairement en accord, et cela dans des proportions parfois
frappantes comme le montre la figure~\ref{fig:ggHTeV_pdf_french1}
ci-dessus. L'incertitude due aux erreurs aussi bien théoriques
qu'expérimentales sur la valeur de $\alpha_s(M_Z^2)$ est reliée à
cette incertitude sur les PDFs, tout autant qu'à l'ajustement des PDFs
: en effet toutes les collaborations ne donnent pas une même valeur
pour la constante de couplage forte.
\begin{figure}[!t]
\begin{center}
\includegraphics[scale=0.7]{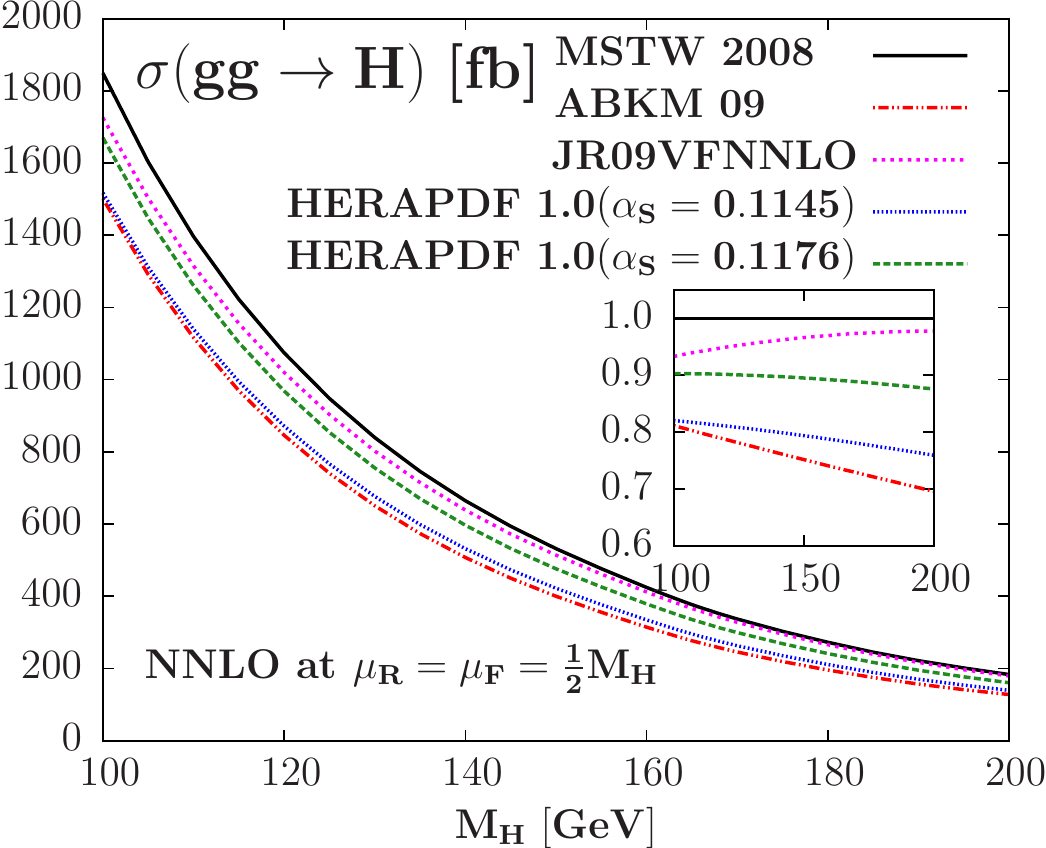}
\end{center}
\caption[Comparaison entre les prédictions des différentes
  collaborations de PDFs pour le canal $gg\to H$ au NNLO en
  QCD]{Section efficace du canal $gg\to H$ au Tevatron en fonction de
    $M_H$ en utilisant les quatre collaborations de PDFs au NNLO qui
    existent : MSTW, JR, HERA et ABKM. L'encart donne la déviation
    relative à la prédiction centrale obtenue avec MSTW.}
\vspace*{-2mm}
\label{fig:ggHTeV_pdf_french1}
\end{figure}
Comprendre pourquoi il existe tant de différences entre ces prédictions
est encore à l'heure actuelle une question ouverte parmi les
spécialistes, que nous ne sommes certainement pas à même de
trancher. Nous invitons le lecteur intéressé à se référer au texte
principal de la thèse où ces questions sont abordées en détail ainsi
qu'aux références associées, voir le
sous-chapitre~\ref{section:SMHiggsTevPDF}. Nous allons nous contenter
ici de résumer le résultat final que nous obtenons dans le cadre de
cette problématique. Nous utilisons la collaboration la plus utilisée
dans la communauté pour les prédictions centrales, c'est-à-dire la
collaboration MSTW, puis nous calculons dans son schéma l'incertitude
PDF+$\Delta^{\rm exp+th}\alpha_s$ en utilisant leurs données publiques
PDF et incertitudes expérimentales sur $\alpha_s$ corrélées d'une
part, et en utilisant $\Delta^{\rm th}\alpha_s=0.004$ dans le cas du
canal $gg\to H$, afin de réconcilier les prédictions d'ABKM et de MSTW
(au moins en tenant compte de leurs bandes d'incertitudes PDFs
respectives), et $\Delta^{\rm th}\alpha_s=0.002$ pour le canal
Higgs--strahlung, d'autre part. Le résultat final est décrit dans la
figure~\ref{fig:ggHTeV_pdf_french2} ci-dessous. Nous trouvons une
incertitude combinée de l'ordre de $\pm 15\%$ à $\pm 20\%$ dans le cas
de la fusion de gluons et de l'ordre de $\pm 7\%$ dans le cas du
Higgs--strahlung au Tevatron.
\begin{figure}[!t]
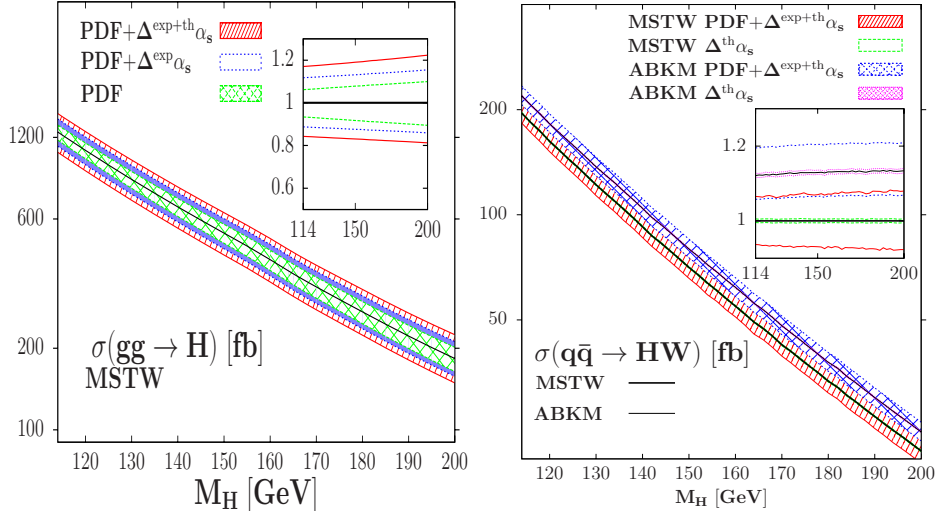

\begin{center}
\includegraphics[width=6.084cm, height=6.78cm]{./Figs/secondpart/tev/ggH_pdf_errors2.pdf}
\includegraphics[scale=0.6]{./Figs/secondpart/tev/ppHv_pdf_study_wboson2.pdf}
\end{center}
\caption[Incertitude PDF+$\Delta\alpha_s$ dans les canaux de
production $gg\to H$ et $p\bar p\to H W$ au Tevatron]{Gauche : bandes
  d'incertitude PDF, PDF+$\Delta^{\rm exp}\alpha_s$ et
  PDF+$\Delta^{\rm exp+th}\alpha_s$ au Tevatron en fonction de $M_H$
  dans le canal $gg\to H$. Droite : la même chose mais pour le canal
  $p\bar p \to H W$ au Tevatron. Les incertitudes relatives à la
  prédiction centrale sont données dans les encarts.}
\vspace*{-5mm}
\label{fig:ggHTeV_pdf_french2}
\end{figure}} 
\item{La dernière classe d'incertitude ne concerne que le canal de
    production $gg\to H$. En effet le calcul jusqu'à l'ordre le plus
    élevé, le NNLO, requiert l'utilisation d'une théorie effective
    dans laquelle on fait tendre la masse du quark top vers
    l'infini. Il a été prouvé par différents groupes que cette
    approximation est excellente pour la boucle du quark top, pour
    $M_H\lsim 300$ GeV, mais il est certain qu'elle ne l'est pas pour
    la boucle du quark bottom. Nous avons donc quantifié son absence
    au NNLO, et aussi étudié l'impact du schéma de renormalisation de
    la masse du quark b dans le calcul. En parallèle, et reliée à cette
    problématique de théorie effective, nous avons estimé
    l'incertitude des corrections électrofaibles au NNLO, qui sont
    aussi calculées de manière effective en supposant $M_H \ll
    M_W$. Les incertitudes obtenues sont de l'ordre de quelques
    pourcents, 5\% au maximum.}
\end{enumerate}

\begin{figure}[h]
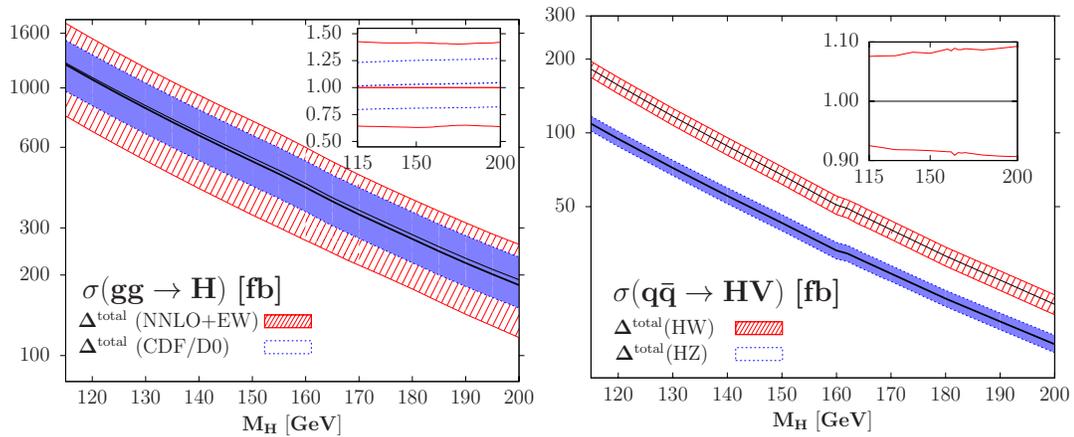

\vspace*{-1mm}
\begin{center}
\includegraphics[scale=0.65]{./Figs/secondpart/tev/ggH_allerrors.pdf}
\includegraphics[scale=0.65]{./Figs/secondpart/tev/ppHv_allerrors_mstw.pdf}
\end{center}
\vspace*{-6mm}
\caption[Sections efficaces de production inclusives des canaux $gg\to
H$ et $p\bar p\to HV$ au Tevatron ainsi que les incertitudes
théoriques totales associées]{Gauche : section efficace de production
  inclusive du canal $gg\to H$ (en fb) au Tevatron en fonction de
  $M_H$ (en GeV) ainsi que la bande d'incertitude totale obtenue avec
  notre combinaison des incertitudes individuelles. Droite : la même
  chose au Tevatron pour le canal $p\bar p\to VH$, $V=W,Z$. Les
  incertitudes relatives à la prédiction centrale sont données dans
  les encarts.}
\label{fig:TeVTotalfrench}
\end{figure}

Il s'agit enfin de combiner ces incertitudes pour obtenir
l'incertitude finale. C'est une problématique qui n'est pas
complètement résolue, car une combinaison quadratique suppose que
toutes les incertitudes sont totalement décorrélées, alors que
l'alternative linéaire est clairement trop conservative. Nous avons
choisi de combiner de la façon suivante : calculer l'incertitude
PDF+$\Delta\alpha_s$ sur les extrema obtenus avec l'incertitude
d'échelle, puis rajouter linéairement les incertitudes purement
théoriques dues à l'usage d'une théorie effective dans le cas de
$gg\to H$. Ceci est un bon compromis entre addition linéaire et
addition quadratique. Nous obtenons ainsi une incertitude de l'ordre
de $\pm 38\%$ dans le cas de la fusion de gluons et de l'ordre de 
$\pm 8\%$ dans le cas du Higgs--strahlung comme représenté dans la
figure~\ref{fig:TeVTotalfrench} ci-dessous. C'est quasiment deux fois
plus que l'incertitude utilisée par CDF et D0 dans leur interprétation
des résultats expérimentaux qu'ils obtiennent. Le lecteur trouvera les
résultats numériques détaillés dans les tables~\ref{table:ggHTeV}
et~\ref{table:ppHV}.

\subsubsection{Le cas du LHC}

Après avoir mené une étude détaillée des incertitudes théoriques
pesant sur les prédictions au Tevatron, nous avons réalisé le même
travail dans le cas du LHC. L'énergie au centre de masse actuelle
étant de 7 TeV nous avons concentré notre étude sur cette phase
actuelle des recherches et nous avons distingué le LHC dans son
fonctionnement nominal à 14 TeV de ce fonctionnement à 7 TeV en le
nommant lHC. Le chapitre~\ref{section:SMHiggsLHC} aborde en détail
cette étude pour le lecteur intéressé.

\begin{figure}[h]
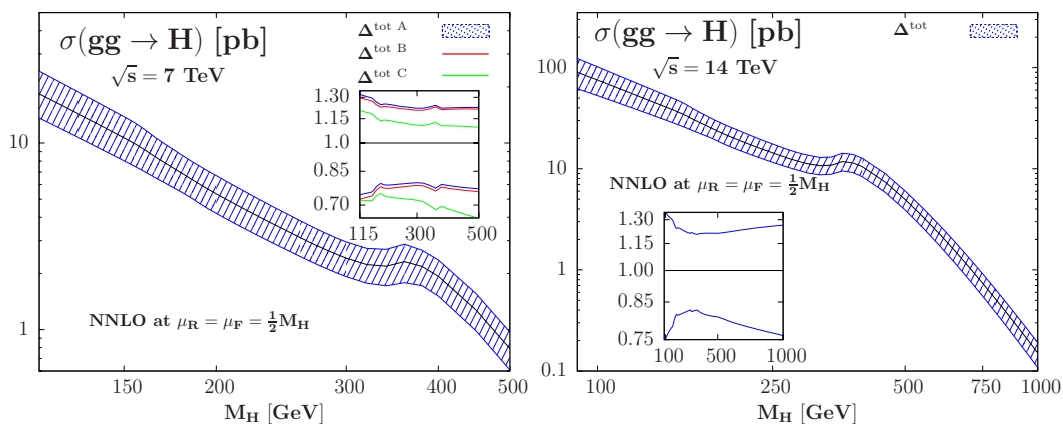

\begin{center}
\includegraphics[scale=0.65]{./Figs/secondpart/lhc/all_error_lhc7.pdf} 
\includegraphics[scale=0.65]{./Figs/secondpart/lhc/all_error_lhc14.pdf}
\end{center}
\vspace*{-5mm}
\caption[Sections efficaces de production inclusives du canal $gg\to
H$ au LHC à 7 et 14 TeV  ainsi que les incertitudes
théoriques totales associées]{Gauche : section efficace de production
  inclusive du canal $gg\to H$ (en pb) au lHC à 7 TeV en fonction de
  $M_H$ (en GeV) ainsi que la bande d'incertitude totale obtenue avec
  notre combinaison des incertitudes individuelles. Droite : la même
  chose au LHC à 14 TeV ainsi que la bande d'incertitude totale selon
  notre procédure. Les incertitudes relatives à la prédiction centrale
  sont données dans les encarts ainsi que les trois manières de
  combiner les incertitudes à 7 TeV présentées dans le corps principal
  de la thèse au sous-chapitre~\ref{section:SMHiggsLHCTotal}.}
\label{fig:LHCTotalfrench}
\end{figure}

Nous avons effectué la même démarche que pour le cas du Tevatron, en
se concentrant sur le canal dominant de production du boson de Higgs
du Modèle Standard qu'est la fusion de gluons. Nous n'allons donner
ici que le résultat final, pour le cas du lHC à 7 TeV ainsi que le cas
du LHC à 14 TeV. Nous avons aussi donné dans le
chapitre~\ref{section:SMHiggsLHC} des résultats pour des énergies au
centre de masse intermédiaires. Les incertitudes ne changent que
marginalement entre 7 et 14 TeV, mais sont par contre
significativement réduites en comparaison des résultats obtenus au
Tevatron, pour deux raisons essentiellement : une meilleure
convergence des ordres supérieurs, conduisant à une réduction de
l'incertitude d'échelle (et expliquant le choix du coefficient constant
$\kappa=2$ pour l'intervalle de variation choisi), d'une part ; une
meilleure connaissance de la fonction de distribution de gluons à
cette fraction du moment du proton dans cette gamme d'énergie,
réduisant l'incertitude due aux PDFs, d'autre part. Nous avons étudié
trois façons distinctes de définir l'incertitude totale, détaillées
dans le texte principal au
sous-chapitre~\ref{section:SMHiggsLHCTotal} et notées A, B et C ;
l'incertitude que nous choisirons sera l'incertitude A qui suit la
procédure présentée dans le cas du Tevatron. Nous obtenons au final
une incertitude totale de l'ordre de $\pm 25\%$ pour le lHC,
légèrement plus à basses masses $M_H\simeq 120$ GeV et légèrement
moins à haute masse $M_H\simeq 500$ GeV. La
figure~\ref{fig:LHCTotalfrench} ci-desous résume les résultats, le
lecteur intéressé pouvant trouver les résultats numériques détaillés
dans les tables~\ref{table:ggH_lhc7_sm} et~\ref{table:lhc14}.

\subsubsection{Conséquences sur les résultats expérimentaux}

Après avoir étudié les sections efficaces de production, nous avons
tourné notre attention vers les rapports d'embranchement des
désintégrations du boson de Higgs du Modèle Standard dans le
chapitre~\ref{section:SMHiggsDecay}. Nous avons, ici aussi, étudié
l'impact de certaines incertitudes pesant sur les prédictions
théoriques, et notamment l'impact des erreurs expérimentales sur la
masse des quarks bottom et charmé, ainsi que l'impact des erreurs
expérimentales sur la détermination de la constante de couplage
$\alpha_s(M_Z^2)$. Nous avons ainsi trouvé que cet impact peut être
significatif dans les deux canaux $H\to b\bar b$ et $H\to WW$ autour
de $M_H\approx 135$ GeV, où ces deux canaux de désintégration ont des
rapports d'embranchement similaires. {\it A contrario}, à haute masse, où
les canaux de désintégration en deux bosons faibles $W, Z$ dominent,
ces incertitudes sont quasiment nulles. Les résultats numériques sont
donnés dans les tables~\ref{table:Hdecay1} et~\ref{table:Hdecay2} du
corps principal de la thèse.

Il s'agit ensuite de combiner les sections efficaces de production et
les rapports d'embranchement des canaux de désintégration du boson de
Higgs. Nous avons choisi de travailler dans l'approximation dite de
largeur étroite de désintégration, où cette combinaison se fait en
multipliant simplement la section efficace par le rapport
d'embranchement. C'est une approximation excellente pour les basses
masses du boson de Higgs, mais il faut garder en tête que lorsque
ladite masse approche les 500 GeV, cela devient une mauvaise
approximation ; une étude plus fine serait nécessaire. Nous avons
trouvé que l'impact sur le résultat final des erreurs sur les rapports
d'embranchement sont négligeables à haute masse, et augmentent
légèrement l'incertitude à basse masse.

\begin{figure}[!h]
\begin{center}
\includegraphics[scale=0.6]{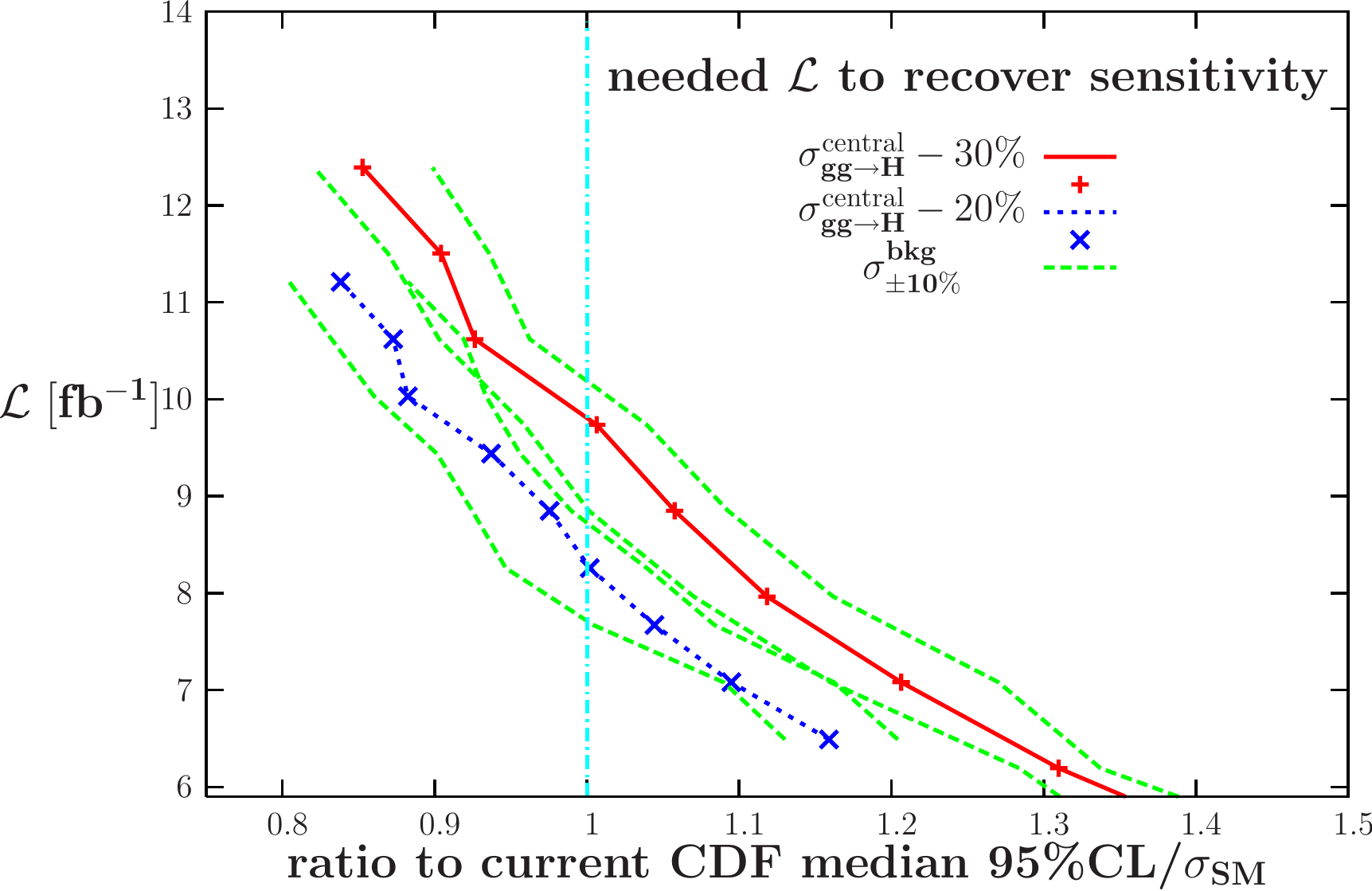}
\end{center}
\caption[Luminosité nécessaire à l'expérience CDF afin qu'elle
obtienne la sensibilité qu'elle prétend avoir actuellement, en tenant
compte de nos incertitudes théoriques]{Luminosité nécessaire à
  l'expérience CDF afin qu'elle obtienne la sensibilité qu'elle
  prétend avoir actuellement au boson de Higgs d'une masse de 160 GeV,
  lorsque le calcul théorique de la section efficace du canal
  $gg\!\to\! H\! \to \!\ell \ell \nu \nu$ est diminuée de 20\% et de
  30\% ; l'impact d'un changement de 10\% dans le bruit de fond
  dominant $p\bar p\to WW$ est aussi indiqué. Cette figure est
  extraite de la référence~\cite{Baglio:2011wn}.}
\vspace*{-5mm}
\label{fig:tev_lumi_resultsfrench}
\end{figure}

L'étape finale du chapitre~\ref{section:SMHiggsDecay} a été la
comparaison de nos calculs avec les résultats expérimentaux, en
particulier les limites d'exclusion obtenues par les collaborations
CDF et D0 au Tevatron. Une discussion beaucoup plus complète est
disponible dans le corps principal du texte dans le
sous-chapitre~\ref{section:SMHiggsTevExclusion} ; nous n'indiquons
ici que le résultat final de cette discussion. Une comparaison naïve
indique que la bande d'exclusion $158\leq M_H\leq 173$ GeV obtenue
avec 7.1 fb$^{-1}$ de luminosité est à questionner, mais afin d'avoir
un résultat plus quantitatif nous avons cherché à obtenir la
luminosité nécessaire pour avoir la sensibilité actuelle que
prétendent les deux expériences en tenant compte de nos incertitudes
qui sont le double de celles utilisées par les deux
collaborations. Pour ce faire nous avons recalculé cette sensibilité
en baissant de 20\% à 30\% la valeur centrale du calcul théorique
comparé aux données en présupposant que les 20\% restants sont déjà pris en
compte par les expériences. Ceci est résumé dans la
figure~\ref{fig:tev_lumi_resultsfrench} ci-dessous, où nous avons
effectué cet exercice pour une des masses du boson de Higgs à
lesquelles le Tevatron est le plus sensible, avec la procédure
détaillée dans le sous-chapitre~\ref{section:SMHiggsTevExclusion}.
Nous voyons clairement que la luminosité nécessaire afin d'exclure
effectivement le boson de Higgs à cette masse est de l'ordre de 10 à
11 fb$^{-1}$ particulièrement si on tient compte des incertitudes qui
pèsent aussi sur le calcul du bruit de fond du Modèle Standard. C'est
pourquoi nous questionnons cette exclusion à la lumière de nos
résultats théoriques, d'autant plus que cette exclusion est fortement
dépendante de la PDF choisie. Nous rappelons aussi que cette exclusion
est débattue dans la communauté tout particulièrement à cause de
l'impact des PDFs, dont la question n'est toujours pas résolue à
l'heure actuelle parmi les spécialistes et dont on espère que les
données futures du LHC permettrons de mieux comprendre le problème
d'une telle disparité de prédictions entre les différentes
collaborations de PDFs.

\subsection{Le Modèle Standard Supersymétrique Minimal (MSSM)}

\subsubsection{Bref aspect de la supersymétrie et du MSSM}

Après avoir étudié la production du boson de Higgs du Modèle Standard
au Tevatron et au LHC nous nous tournons vers l'étude de la brisure de
symétrie électrofaible dans le cadre de l'extension supersymétrique
minimale du Modèle Standard. Comme indiqué en détail dans le corps de
la thèse, chapitre~\ref{section:SUSYIntro}, il existe plusieurs
raisons qui poussent les théoriciens à envisager des modèles allant
au-delà du Modèle Standard, et en particulier à s'intéresser à la
supersymétrie (SUSY). Nous allons brièvement présenter ces raisons
ainsi que les aspects principaux de la supersymétrie, puis introduire
le contenu en champs de l'extension supersymétrique minimale du Modèle
Standard (MSSM) ainsi que ses caractéristiques.

Le chapitre~\ref{section:SUSYIntro} donne trois raisons (parmi
beaucoup d'autres !) qui poussent les physiciens des hautes énergies à
envisager des théories étendant le Modèle Standard et en particulier
les théories supersymétriques :
\begin{itemize}
\item{le problème de la hiérarchie : rien ne protège la masse du boson
    de Higgs dans le Modèle Standard, ce qui signifie que sa masse est
    naturellement de l'ordre de grandeur de l'échelle d'énergie la plus
    élevée de la théorie, sauf si l'on ajuste très finement les
    contre--termes dans les corrections radiatives. Techniquement cela
    signifie que les divergences ultraviolettes sont
    quadratiques. Or la supersymétrie, qui est une symétrie entre
    bosons et fermions, implique que les boucles bosoniques et les
    boucles fermioniques se compensent, stabilisant ainsi la masse du
    boson de Higgs;}
\item{dans le Modèle Standard, si on fait évoluer à l'aide des
    équations du groupe de renormalisation les constantes de couplage
    des trois algèbres de Lie $SU(2)_L$, $U(1)_Y$ et $SU(3)_c$, on
    constate qu'elles ne convergent pas tout à fait à haute énergie ;
    cependant si l'on rajoute de nouvelles particules dans le spectre,
    comme c'est le cas avec une extension supersymétrique, leurs
    contributions au groupe de renormalisation peuvent permettre une
    telle unification ; c'est bien le cas dans l'extension
    supersymétrique minimale du Modèle Standard, à une échelle de
    l'ordre de 10$^{16}$ GeV, comme le montre la
    figure~\ref{fig:gauge_unification} du
    sous-chapitre~\ref{section:IntroGUT};}
\item{il existe en cosmologie le problème dit de la matière noire : en
    effet les observations cosmologiques\footnote{Si l'on présuppose que
      la relativité générale est valide aux échelles cosmologiques ;
      cette hypothèse n'a pas été testée expérimentalement, et
      certains modèles alternatifs tentent d'expliquer les
      observations cosmologiques sans matière noire en modifiant la
      relativité générale. Néanmoins nous ne nous lançons pas ici dans
    une telle discussion.} pointent vers l'existence d'une composante
  de la matière de l'ordre de 23\% du contenu énergétique de
  l'Univers, que l'on ne connaît pas. La particule supersymétrique la
  plus légère pourrait parfaitement jouer le rôle de cette matière
  noire.}
\end{itemize}

Nous n'allons pas rentrer dans les détails techniques de la
supersymétrie dans ce synopsis, le lecteur intéressé pourra lire avec
profit le chapitre~\ref{section:FormalSUSY}. Nous allons mentionner
brièvement ses caractéristiques principales et notamment comment
l'appliquer au Modèle Standard.

Dans une théorie supersymétrique les champs sont groupés en
(super)multiplets associants bosons et fermions. Un lagrangien est
supersymétrique lorsque l'échange entre bosons et fermions d'un même
(super)multiplet laisse le lagrangien invariant à une dérivée totale
près. Afin que la supersymétrie soit une symétrie présente dans le
lagrangien sans tenir compte des équations du mouvement, il est alors
nécessaire au sein des supermultiplets d'introduire des champs
auxiliaires, permettant de respecter l'égalité entre les nombres de
degrés de liberté fermioniques et bosoniques : ces champs auxiliaires
ne sont pas dynamiques et peuvent être éliminés à l'aide des équations
de Lagrange. Voici un exemple de lagrangien supersymétrique, le modèle
de Wess--Zumino libre :
\beq
{\cal L} = \imath \bar{\psi} \bar{\sigma}^{\mu} \partial_\mu \psi
- \partial^\mu \phi^* \partial_{\mu} \phi + F^* F
\label{eq:WessZuminoFR}
\eeq
Il y a un seul supermultiplet composé du champ $F$ qui est un champ
scalaire complexe jouant le rôle de champ auxiliaire, du champ
scalaire complexe $\phi$ et du spineur à quatre composantes $\psi$. Il
y a bien quatre degrés de liberté bosoniques et quatre degrés de
liberté fermioniques. Si l'on applique les transformations
supersymétriques suivantes, paramétrisées par la variable
anti-commutante infinitésimale $\epsilon$ :
\beq
\delta_\epsilon \phi & = & \sqrt{2} \epsilon \psi\nonumber\\
\delta_\epsilon \psi & = & \imath \sqrt{2} \sigma^{\mu}
\bar{\epsilon} \partial_\mu \phi + \sqrt{2} \epsilon F\nonumber\\
\delta_\epsilon F & = & \imath \sqrt{2} \bar{\epsilon}
\bar{\sigma}^{\mu} \partial_\mu \psi
\label{eq:WessZuminoSUSYFrench}
\eeq
le lagrangien~\ref{eq:WessZuminoFR} est bien invariant à une dérivée
totale près : c'est un lagrangien supersymétrique.

Afin de traiter les interactions dans un cadre supersymétrique nous
avons introduit dans le sous-chapitre~\ref{section:FormalSuperspace}
le concept de superespace, qui étend l'espace-temps classique en
rajoutant des variables anticommutantes qui vont permettre de réécrire
les supermultiplets comme des superchamps vivant dans ce
superespace. Nous invitons le lecteur intéressé à parcourir ce
sous-chapitre qui décrit en détail la technique mathématique
sous-jacente. L'étape suivante est d'appliquer ces concepts au Modèle
Standard et notamment d'étendre ce dernier de manière minimale en
champ : c'est le Modèle Standard Supersymétrique Minimal ou MSSM.

Si la supersymétrie était une symétrie exacte de la Nature, l'électron
par exemple aurait un partenaire bosonique avec exactement la même
charge et la même masse, mais de spin zéro ; l'expérience prouve que
ce n'est pas le cas, ce qui signifie que la supersymétrie doit être
une symétrie brisée. Cette problématique est majeure dans le champ
d'étude de la supersymétrie et reste encore à l'heure actuelle un
problème ouvert, notamment si l'on souhaite le faire de manière
dynamique en réutilisant les concepts développés dans le cadre de la
brisure électrofaible du Modèle Standard. Dans le cadre du MSSM, la
brisure de la supersymétrie a été choisie de manière explicite, mais
tout en préservant les qualités fondamentales de la supersymétrie
comme par exemple de résoudre le problème de la hiérarchie du Modèle
Standard : c'est ce que l'on appelle la brisure ``douce''. On
introduit alors non seulement de nouvelles particules, partenaires des
particules standards, mais aussi des termes explicites de brisure SUSY
douce. Le tableau~\ref{table:MSSMcontenu} ci-dessous résume les
nouveaux champs du MSSM :
\begin{table}[!h]
{\small%
\let\lbr\{\def\{{\char'173}%
\let\rbr\}\def\}{\char'175}%
\renewcommand{\arraystretch}{1.35}
\begin{center}
\begin{tabular}{|c|c|c|ccc|}\hline
Type & Nom & Spin & rep. $SU(3)_c$ & rep. $SU(2)_L$ & 
 charge $U(1)_Y$\\ \hline
 & $\tilde{\nu}_e, \tilde{e}_L, \tilde{\nu}_\mu, \tilde{\mu}_L,
 \tilde{\nu}_{\tau}, \tilde{\tau}_L$ & $0$ & $\mathbf{1}$ &
 $\mathbf{2}$ & $-1$ \\
SLEPTONS  & $\tilde{e}_R, \tilde{\mu}_R, \tilde{\tau}_R$ & $0$ &
$\mathbf{1}$ & $\mathbf{1}$ & $2$\\ \hline
 & $\tilde{u}_L, \tilde{d}_L, \tilde{s}_L, \tilde{c}_L,
 \tilde{t}_L, \tilde{b}_L$ & $0$ & $\mathbf{3}$ &
 $\mathbf{2}$ & $1/3$ \\
SQUARKS  & $\tilde{u}_R, \tilde{c}_R, \tilde{t}_R$ & $0$ &
$\mathbf{3}$ & $\mathbf{1}$ & $-4/3$\\
& $\tilde{d}_R, \tilde{s}_R, \tilde{b}_R$ & $0$ &
$\mathbf{3}$ & $\mathbf{1}$ & $2/3$\\ \hline
 & $\tilde{B}$ & $1/2$ & $\mathbf{1}$ &
 $\mathbf{1}$ & $1$ \\
JAUGINOS  & $\tilde{W}_1, \tilde{W}_2, \tilde{W}_3$ & $1/2$ &
$\mathbf{1}$ & $\mathbf{3}$ & $0$\\
& $\tilde{g}$ & $1/2$ &
$\mathbf{8}$ & $\mathbf{1}$ & $0$\\ \hline
 & $\tilde{h}_u^+, \tilde{h}_u^0$ & $1/2$ & $\mathbf{1}$ &
 $\mathbf{2}$ & $1$ \\
HIGGSINOS  & $\tilde{h}_d^0, \tilde{h}_d^-$ & $1/2$ &
$\mathbf{1}$ & $\mathbf{2}$ & $-1$\\ \hline
 & $h_u^+, h_u^0$ & $0$ & $\mathbf{1}$ &
 $\mathbf{2}$ & $1$ \\
HIGGS  & $h_d^0, h_d^-$ & $0$ &
$\mathbf{1}$ & $\mathbf{2}$ & $-1$\\ \hline
\end{tabular} 
\end{center} 
\caption[Les superparticules et champs de Higgs du MSSM avant brisure
électrofaible]{Les superparticules et champs de Higgs du MSSM avant
  brisure électrofaible.}
\label{table:MSSMcontenu}
\vspace*{-2mm}
}
\end{table}

La brisure électrofaible nécessite maintenant deux doublets de Higgs
et non plus un seul, qui vont fournir cinq bosons de Higgs physiques :
deux bosons neutres $h$ et $H$, un boson neutre pseudoscalaire $A$ et
deux bosons chargés $H^{\pm}$. Elle induit aussi un mélange entre les
partenaires supersymétriques des bosons de jauge et des bosons de
Higgs, donnant naissance à quatre fermions neutres appelés neutralinos
$\tilde{\chi}_i^0$ et quatre fermions chargés appelés charginos
$\tilde{\chi}_{1,2}^{\pm}$. Nous expliquons plus en détails
ci-dessous le secteur de Higgs du MSSM qui nous intéresse tout
particulièrement dans cette thèse.

\subsubsection{Secteur de Higgs du MSSM}

Pour des raisons techniques il est nécessaire d'avoir deux doublets de
Higgs au sein du MSSM. Ces raisons sont données en détails dans le
sous-chapitre~\ref{section:MSSMAnomaly}, pour n'en citer qu'une seule
nous rappelons que l'un des doublets donne la masse aux fermions de
type up, le second doublet donne la masse aux fermions de type down,
ceci afin d'à la fois respecter l'expérience qui nous indique qu'il
est nécessaire de donner une masse aux fermions, et respecter la
structure mathématique de la supersymétrie et notamment la nécessité
d'annuler les anomalies de jauge. Nous avons donc les deux doublets
suivants :
\beq
h_u = \left(\begin{matrix} h_u^+\\h_u^0\end{matrix}\right),~ h_d =
\left(\begin{matrix} h_d^0\\h_d^-\end{matrix}\right)
\eeq
qui ont tous les deux une vev non--nulle :
\beq
\langle \, h_u \, \rangle = \frac{1}{\sqrt 2}
\left(\begin{matrix}0\\v_u\end{matrix}\right),~ \langle \, h_d \,
\rangle = \frac{1}{\sqrt 2} \left(\begin{matrix}
    v_d\\0\end{matrix}\right)
\eeq

Les détails du potentiel scalaire du MSSM sont donnés dans le
sous-chapitre~\ref{section:MSSMAnomaly} ; nous résumons les résultats
en rappelant que de la brisure électrofaible, intimement reliée à la
brisure de supersymétrie, émergent cinq bosons de Higgs physiques qui
sont :
\begin{enumerate}[--]
\item{deux bosons de Higgs neutres scalaires $h$ et $H$ ;}
\item{un boson de Higgs pseudoscalaire $A$ ;}
\item{deux bosons de Higgs chargés $H^{\pm}$.}
\end{enumerate}
dont les masses et couplages à l'ordre le plus bas (des arbres) sont
paramétrés par la masse $M_A$ du boson de Higgs pseudoscalaire et le
rapport des vev $\displaystyle \tan\beta \equiv \frac{v_u}{v_d}$.

Nous nous concentrons dans la suite sur l'étude de la production et de
la désintégration des bosons de Higgs neutres ; leurs couplages aux
bosons de jauge standards et aux fermions du Modèle Standard sont une
donnée essentielle dans l'optique de leur production au sein des
collisionneurs hadroniques. Nous avons donné dans le
sous-chapitre~\ref{section:MSSMAnomaly} les couplages des bosons de
Higgs aux fermions en particulier. Les couplages aux fermions de type
up sont supprimés d'un facteur $\tan\beta$ alors qu'au contraire les
couplages aux fermions de type down sont renforcés de ce même facteur :
cela signifie que la production des bosons de Higgs neutres à grand
$\tan\beta$ va faire principalement intervenir le quark bottom et non
plus le quark top. Ceci va être d'une importance cruciale dans la suite.

\subsection{Production et désintégration des bosons de Higgs
  supersymétriques}

La dernière partie de la thèse s'intéresse à l'étude de la production
et de la désintégration des bosons de Higgs neutres du MSSM au
Tevatron et au lHC. Du fait de la modification des couplages d'un
facteur $\tan\beta$ comparés à ceux du Modèle Standard, les chances
d'observer de tels bosons au sein des collisionneurs hadroniques sont
renforcées comparativement au boson de Higgs standard. Nous
reproduisons le même schéma d'étude développé dans le cas du Modèle
Standard : l'étude exhaustive des incertitudes théoriques pesant sur
la prédiction, dont les sources sont identiques à celles jouant un
rôle dans le cas du Modèle Standard et en y rajoutant l'impact des
incertitudes sur le quark bottom qui joue cette fois-ci un rôle
prépondérant.

Nous avons démontré dans le chapitre~\ref{section:MSSMHiggsIntro} que
si les corrections QCD jouent évidemment un grand rôle dans les
prédictions des sections efficace de production et les rapports
d'embranchement des canaux de désintégration, les corrections
supersymétriques dans une certaine classe de modèles assez large ont
un impact relativement faible, quand on s'intéresse à la chaîne
complète production--désintégration des bosons de Higgs neutres. Cette
classe de modèles possède la particularité de présenter un spectre
organisé ainsi : un des deux bosons scalaires se comporte comme le
boson de Higgs du Modèle Standard tandis que le second se comporte
comme le boson pseudoscalaire. L'étude en est ainsi simplifiée et nous
présenterons nos résultats pour le boson de Higgs générique $\Phi$
représentant alternativement les deux particles semblables $A$ et
$h/H$. Nous négligerons les corrections SUSY\footnote{Il est bon de
  rappeler que si ces corrections sont prises en compte, non seulement
  elles sont faibles mais en plus elles génèrent aussi une incertitude
  théorique liée aux paramètres supersymétriques sous-jacents ; ces
  corrections sont ainsi noyées dans les incertitudes QCD standards et
  les incertitudes SUSY à estimer, ces dernières étant loin d'être
  négligeables.} et nous nous placerons dans l'hypothèse d'étude des
collisionneurs hadroniques à savoir $\tan\beta$ relativement élevé ce
qui permet de négliger toute contribution due au quark top. Nous
présenterons nos résultats avec $\tan\beta=1$ : il est entendu que le
lecteur doit multiplier ces résultats par un facteur $2\tan^2\beta$
afin d'obtenir la prédiction réelle pour la production de $A+h/H$.

\subsubsection{Les Higgs neutres du MSSM au Tevatron}

Dans le schéma d'étude résumé ci-dessus les deux canaux principaux de
production du boson $\Phi$ au Tevatron (et au LHC) sont la fusion de
gluons ainsi que la fusion de quarks bottom. Ce dernier canal devient
même le canal dominant pour des masses $M_\Phi \gsim 135$ GeV. L'étude
détaillée est disponible dans le chapitre~\ref{section:MSSMHiggsTev}.

Le calcul perturbatif de la section efficace inclusive de production
du canal $b\bar b\to \Phi$ est connue jusqu'à l'ordre NNLO dans une
modélisation dite ``à 5 saveurs'' pour les PDFs, en supposant que
l'ordre le plus bas est effectivement une fusion des deux quarks
bottom que l'on arrive à extraire de la mer des protons/antiprotons,
au contraire d'une modélisation dite ``à 4 saveurs'' où les quarks
bottom n'apparaissent qu'au NLO après séparation des gluons de la mer
en paires bottom/antibottom. En ce qui concerne la fusion de gluons,
dans la mesure où seule la boucle triangulaire de quarks bottom est
prise en compte, le calcul s'effectue jusqu'à l'ordre NLO seulement
contrairement au cas du Modèle Standard. Nous prenons en compte toutes
les incertitudes théoriques déjà discutées dans le cas du Modèle
Standard, hormis bien évidemment l'incertitude générée par l'usage
d'une théorie effective puisque ici le calcul est exact à un ordre
perturbatif donné. Par contre, une nouvelle incertitude joue un rôle
majeur : celle reliée aux erreurs expérimentales de la masse du quark
bottom ainsi que les incertitudes liées au choix du schéma de
renormalisation de ladite masse.

En ce qui concerne les incertitudes d'échelle, nous utilisons un
facteur d'intervalle $\kappa=2$ dans le cas de la fusion de gluons,
les contributions dues aux quarks bottom entraînant une plus grande
stabilité de la prédiction comparativement au cas du Modèle Standard à
boucle top dominante et où $\kappa=3$ au Tevatron était
nécessaire. Nous utilisons l'échelle centrale $\mu_R=\mu_F=\frac12
M_H$ afin de rester cohérent avec l'étude du Modèle Standard qui est
valide pour le boson de Higgs neutre du MSSM se comportant comme le
boson de Higgs du Modèle Standard. Par contre, comme indiqué dans le
sous-chapitre~\ref{section:MSSMHiggsTevScale} et illustré dans la
figure~\ref{fig:MSSM-scaleTev}, le choix $\kappa=3$ assorti de la restriction
$1/3 \leq \mu_R/\mu_F\leq 3$ est nécessaire dans le cas de la fusion
de quarks bottom où l'incertitude d'échelle est instable ; ceci permet
aussi de prendre en compte l'incertitude reliée au schéma de
renormalisation du quark bottom puisque dans le schéma $\overline{\rm
  MS}$ qui est choisi pour la prédiction centrale la masse du quark
bottom dans le processus $b\bar b \to \Phi$ est prise à l'échelle de
renormalisation. L'échelle centrale de ce processus est
$\mu_R=\mu_F=\frac14 M_H$.

\begin{figure}[!h] 
  \begin{center} 
    \mbox{
      \includegraphics[scale=0.65]{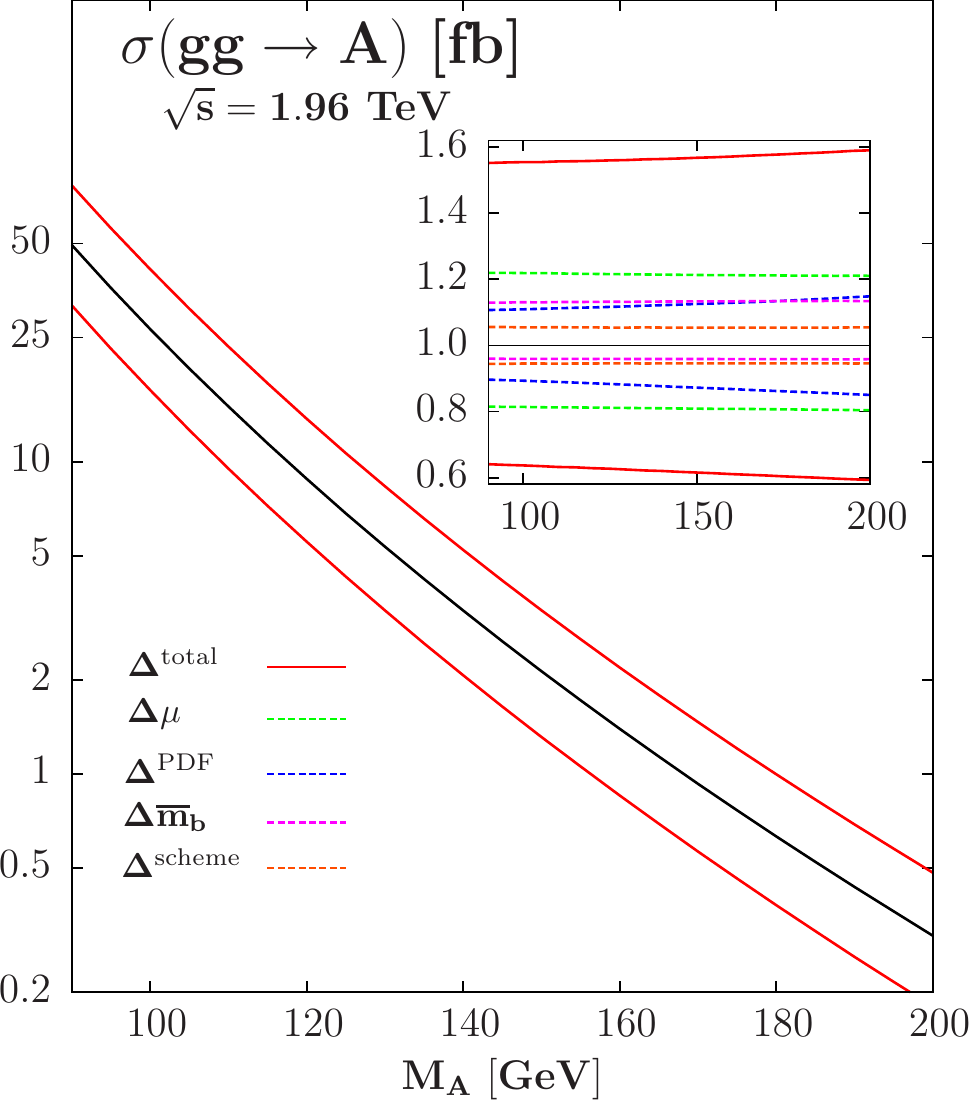}
      \includegraphics[scale=0.65]{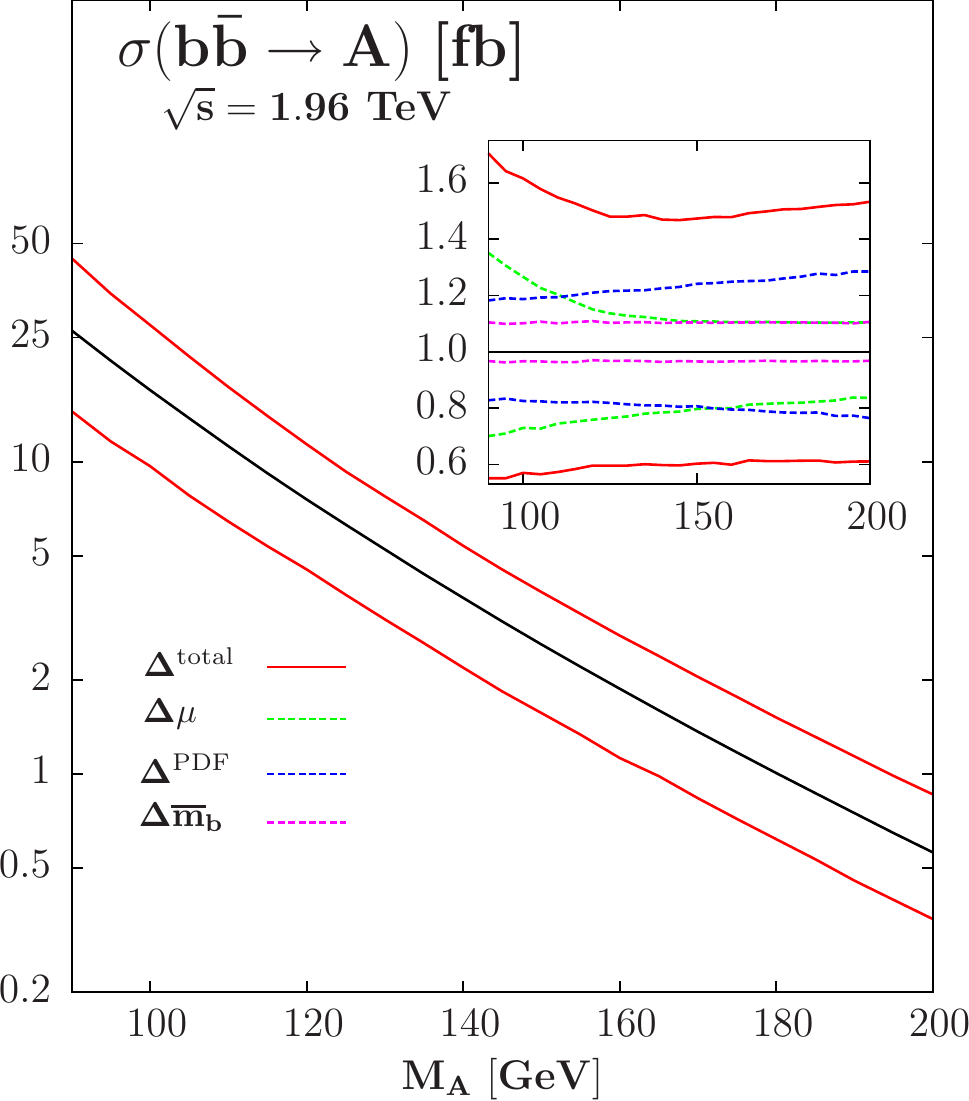}}
  \end{center} 
  \vspace*{-5mm}
  \caption[Les sections efficaces de production inclusives du boson de
  Higgs $A$ du MSSM au Tevatron dans les canaux $gg\to A$ et
  $b\bar b\to A$ accompagnées des incertitudes théoriques]{Les
    prédictions centrales des sections efficaces inclusives de
    production $\sigma^{\rm NLO}_{gg\!\to \!A}$ (gauche) et
    $\sigma^{\rm NNLO}_{b\bar b \!\to\!A}$ (droite) au Tevatron en
    fonction de $M_A$ avec les PDFs de MSTW et des couplages $Ab\bar
    b$ unitaires, accompagnées des bandes d'incertitude théorique
    totale. Les différentes sources d'incertitude théorique ainsi que
    l'incertitude totale, relativement à la prédiction centrale, sont
    données dans les encarts pour chacun des canaux.}
\label{fig:MSSMTotalTevfrench}
\end{figure}

Le résultat final est décrit dans la
figure~\ref{fig:MSSMTotalTevfrench} ci-dessous, les résultats
numériques étant disponibles dans le corps principal de la thèse dans
les tables~\ref{table:ggPhiTev} et~\ref{table:bbPhiTev}. L'impact de
ces incertitudes est significatif, puisque de l'ordre de $\pm 50\%$
dans les deux canaux, un peu plus pour la fusion de gluons et un peu
moins pour la fusion de quarks bottom.

\subsubsection{Les Higgs neutres du MSSM au lHC}

\begin{figure}[!h] 
  \begin{center} 
    \mbox{
      \includegraphics[scale=0.65]{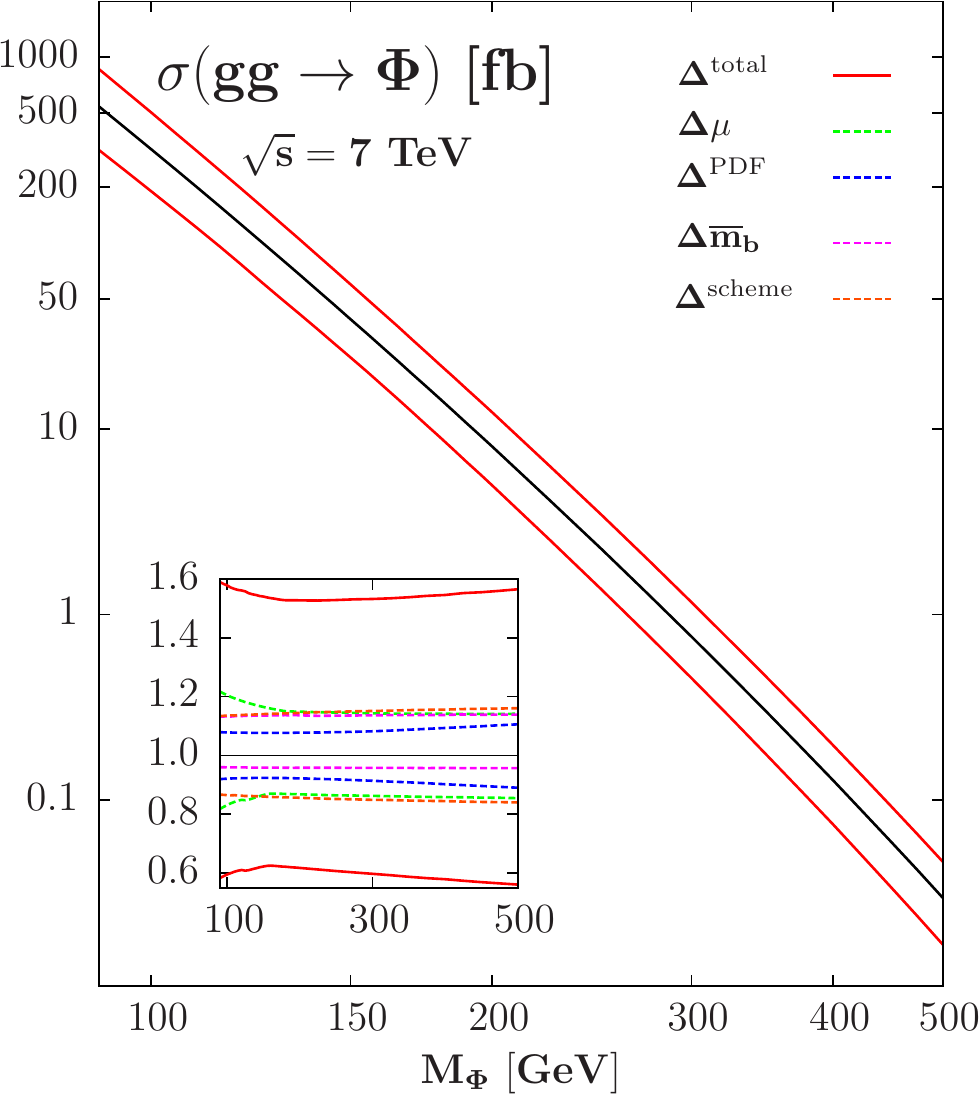}
      \includegraphics[scale=0.65]{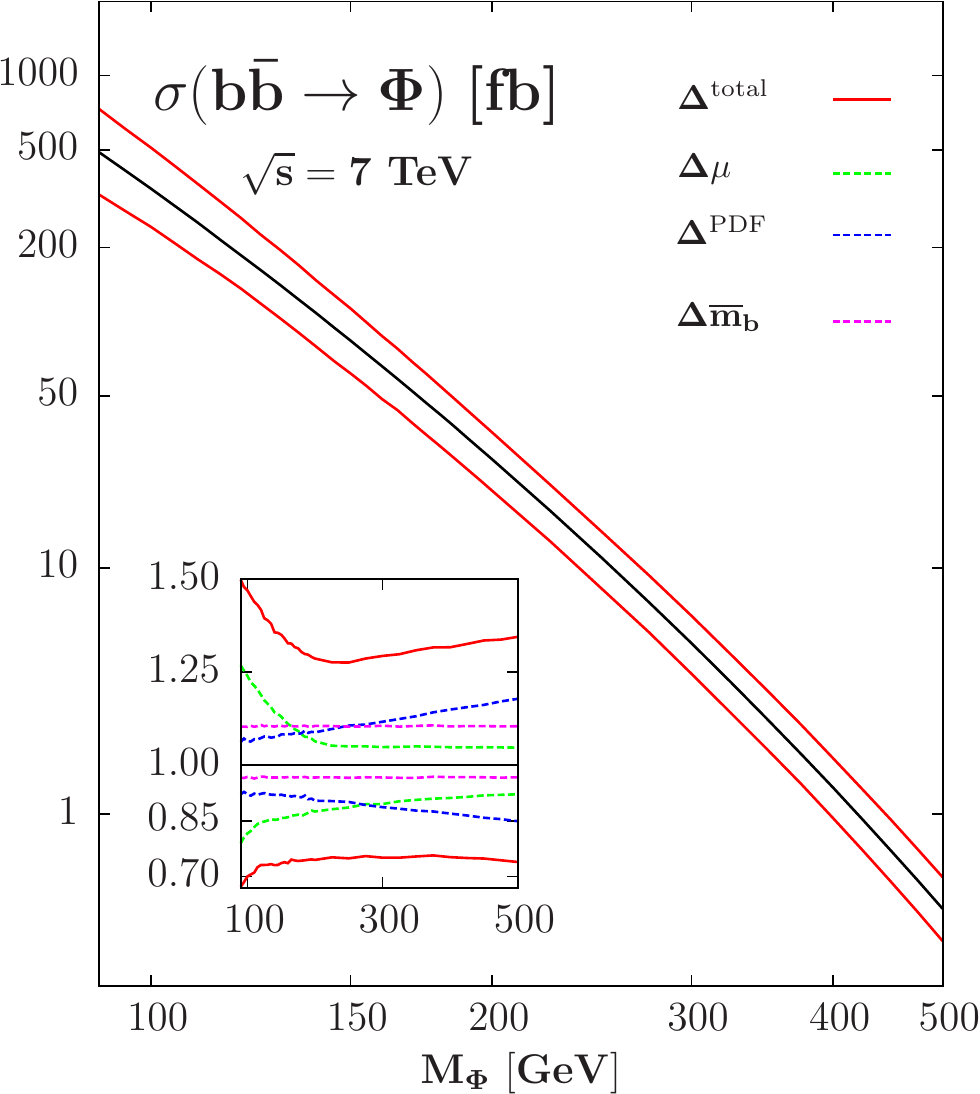}}
  \end{center} 
  \vspace*{-5mm}
  \caption[Les sections efficaces de production inclusives du boson de
  Higgs $\Phi$ du MSSM au lHC dans les canaux $gg\to \Phi$ et
  $b\bar b\to \Phi$ accompagnées des incertitudes théoriques]{Les
    prédictions centrales des sections efficaces inclusives de
    production $\sigma^{\rm NLO}_{gg\!\to \!\Phi}$ (gauche) et
    $\sigma^{\rm NNLO}_{b\bar b \!\to\!\Phi}$ (droite) au lHC avec $\sqrt
    s = 7$ TeV en fonction de $M_\Phi$ avec les PDFs de MSTW et des
    couplages $\Phi b\bar b$ unitaires, accompagnées des bandes
    d'incertitude théorique totale. Les différentes sources
    d'incertitude théorique ainsi que l'incertitude totale,
    relativement à la prédiction centrale, sont données dans les
    encarts pour chacun des canaux.}
\label{fig:MSSMTotalLHCfrench}
\end{figure}

L'étude présentée ci-dessus a été répétée dans le cas du lHC dans le
même cadre théorique qui a été présenté en introduction de l'étude du
MSSM. L'étude détaillée est disponible dans le
chapitre~\ref{section:MSSMHiggsLHC}. Plutôt qu'effectuer une
répétition fastidieuse et inutilement longue dans le cadre d'un
synopsis, nous allons donner les figures finales ainsi que l'incertitude
totale dans les deux canaux $gg\to\Phi$ et $b\bar b\to \Phi$. La
figure~\ref{fig:MSSMTotalLHCfrench} résume l'étude, les données
numériques étant disponibles dans le corps de la thèse,
tables~\ref{table:ggPhiLHC} et~\ref{table:bbPhiLHC}.
La comparaison avec le Tevatron montre que les incertitudes sont
fortement réduites dans le cas du canal (dominant) $b\bar b\to \Phi$,
ce qui n'est pas une surprise car ne serait-ce que les incertitudes
dues aux PDFs sont réduites du fait que l'on sonde des domaines de
fraction d'impulsion mieux maîtrisés. Par contre assez paradoxalement
l'incertitude totale dans le canal $gg\to\Phi$ ne diminue que
marginalement, car en fait dominée par l'incertitude due au quark
bottom qui ne dépend que peu de l'énergie au centre de masse. En
résumé, l'incertitude totale est significative et peut avoir un
certain impact sur la comparaison avec les données expérimentales
comme le montre le paragraphe qui suit.

\subsubsection{L'impact des incertitudes théoriques sur l'espace des
  paramètres du MSSM}

Il nous reste maintenant une dernière étape avant la comparaison
finale avec les données expérimentales aussi bien du Tevatron que du
lHC : la combinaison des sections efficaces de production avec les
rapports d'embranchement des canaux de désintégration. Dans le cas du
MSSM, et en particulier dans les classes de modèles que nous étudions
où il y a une dégénérescence entre le boson de Higgs pseudoscalaire $A$ et
l'un des bosons de Higgs $h/H$, seuls deux canaux sont significatifs :
le canal $\Phi\to b\bar b$ avec un rapport d'embranchement de l'ordre
de 90\% et le canal $\Phi\to \tau^+\tau^-$ avec un rapport
d'embranchement de l'ordre de 10\%. La table~\ref{table:BR-mssm} du
sous-chapitre~\ref{section:MSSMHiggsExpDecay} fournit les données
numériques de notre étude et en particulier les incertitudes
théoriques qui pèsent sur ces rapports d'embranchement, dues aux
incertitudes sur la masse du quark bottom et sur la valeur de
$\alpha_s(M_Z^2)$. Le canal de désintégration qui nous intéressera
tout particulièrement dans l'optique d'une comparaison à l'expérience
est le canal $\Phi\to\tau^+\tau^-$.

Il est très important de remarquer que l'incertitude sur la masse du
quark bottom est anticorrélée entre section efficace de production et
rapport d'embranchement : en définitive cette incertitude va quasiment
s'annuler dans le processus final de production du boson $\Phi$ suivi
de sa désintégration en paire de leptons taus. Ceci est illustré par
les figures~\ref{fig:TotalMSSMTev} et~\ref{fig:MSSM-finalall1LHC} dans
le corps principal du texte.

Nous pouvons maintenant comparer nos prédictions combinant les deux
canaux de production ainsi qu'avec le rapport d'embranchement
$\Phi\to\tau^+\tau^-$ aux résultats expérimentaux du Tevatron et du
lHC. Nous multiplions par un facteur $2\tan^2\beta$ les résultats que
nous avons présentés afin d'avoir une prédiction donnée ainsi que son
incertitude pour chaque couple de valeur $(M_A,\tan\beta)$. Comparées
aux données expérimentales, ces prédictions théoriques fournissent une
limite sur l'espace des paramètres $[M_A,\tan\beta]$ du MSSM. Les
détails sont donnés dans le
sous-chapitre~\ref{section:MSSMHiggsExpLimit} ; mentionons que nous
soutenons qu'une exclusion à 95\% de niveau de confiance doit se faire
entre les données expérimentales et la prédiction théorique minimale
et non la prédiction centrale, c'est-à-dire qu'il faut tenir compte
des incertitudes théoriques séparées des erreurs expérimentales sur
les données.
\begin{figure}[!h]
\begin{center}
\vspace*{-2mm}
\includegraphics[scale=0.75]{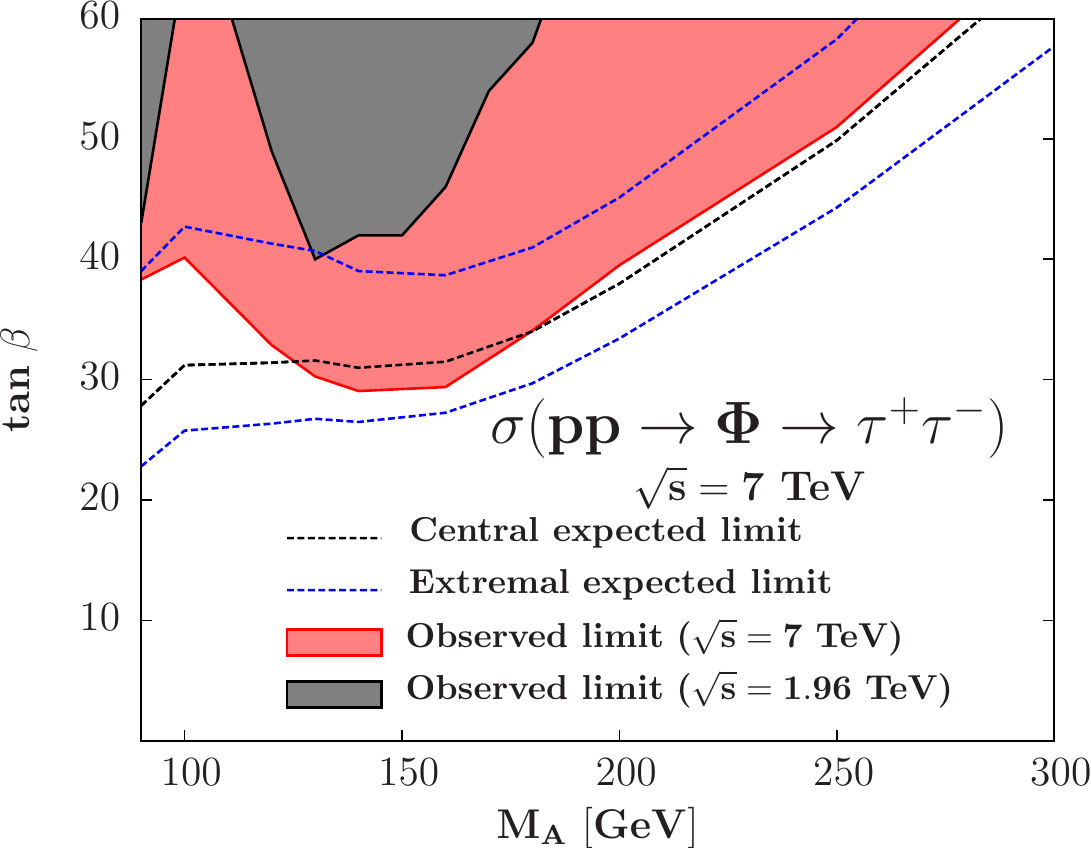}
\end{center}
\caption[Les limites à 95\% de niveau de confiance sur l'espace des
  paramètres du MSSM en tenant compte de nos incertitudes théoriques
  confrontées aux données du Tevatron et du lHC]{Limites d'exclusion
    attendues et obtenues dans le canal $\sigma(p p\! \to\! \Phi \!\to
  \!\tau^+ \tau^-)$ au Tevatron et au lHC sur l'espace des
    paramètres du MSSM [$M_A, \tb$] en tenant compte de nos
    incertitudes théoriques, comparées aux résultats de CMS et de
    CDF/D0 lorsque notre procédure est appliquée.}
\label{fig:Scanfrench}
\end{figure}

Nous voyons donc dans la figure~\ref{fig:Scanfrench} que l'impact des
incertitudes théoriques est important dans le cas du Tevatron, bien
moindre dans le cas du lHC mais cependant significatif. Avec nos
incertitudes, le lHC exclut donc des valeurs $\tan\beta \geq
29$. Nous avons aussi donné quelques prédictions sur ce que l'on
pourrait attendre comme exclusion avec une plus grande luminosité au
lHC, nous invitons le lecteur intéressé à regarder la
figure~\ref{fig:MSSMprojection}.

\subsubsection{Le canal $H\to \tau\tau$ dans le Modèle Standard revisité}

Avant de terminer ce synopsis par une liste de perspectives à donner à
cette thèse, nous présentons une des conséquences les plus
intéressantes de ce travail pour la recherche du boson de Higgs du
Modèle Standard. En effet, comme le présente en détail le
sous-chapitre~\ref{section:SMHiggsTauTau}, nous avons eu l'idée de
comparer les résultats expérimentaux obtenus par CMS dans la recherche
des bosons de Higgs neutres du MSSM et notamment leurs limites à
95\% de niveau de confiance sur la section efficace $pp\to \Phi\to
\tau^+\tau^-$ via fusion de gluons et fusion de quarks bottom, à nos
prédictions au sein du Modèle Standard : en effet les états finaux à
étudier expérimentalement sont les mêmes dans les deux cas, le seul
changement significatif étant la comparaison des taux de production
entre Modèle Standard et MSSM lorsque l'on cherche à comparer les
données expérimentales aux prédictions théoriques.

\begin{figure}[!h]
\begin{center}
\vspace*{-2mm}
\includegraphics[scale=0.75]{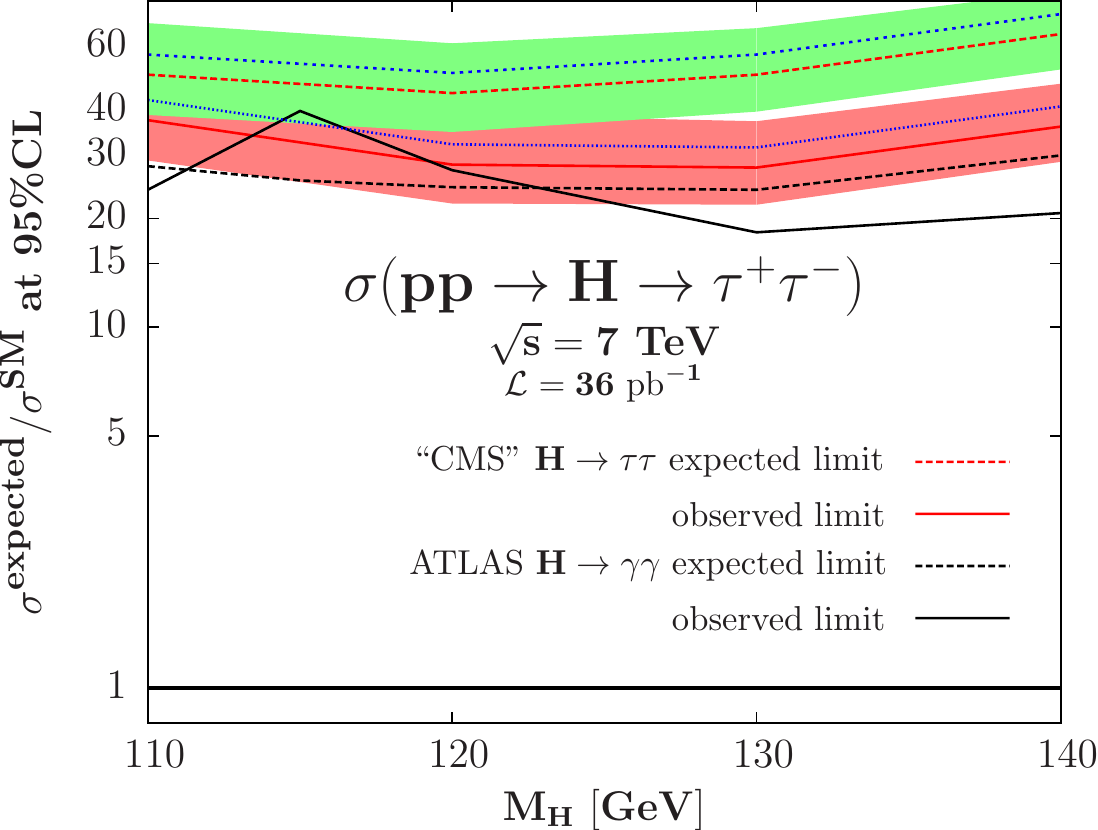}
\end{center}
\vspace*{-6mm}
\caption[L'analyse MSSM des bosons de Higgs neutres appliquée au canal
  de recherche $H\to\tau^+\tau^-$ du Modèle Standard, comparée aux
  résultats obtenus par ATLAS dans le canal $H\to\gamma\gamma$]{Les
    sensibilités attendues et observées à 95\% de niveau de confiance
    pour la production du boson de Higgs du Modèle Standard dans le
    canal $gg\! \to\! H\to \!\tau \tau$ (lignes bleues) et $pp\! \to\!
    H \to \!\tau^+\tau^-+X$ (lignes rouges et légende de la figure) à
    partir d'une extrapolation de l'analyse CMS du MSSM avec une
    luminosité de 36 pb$^{-1}$. Les bandes vertes et rouges décrivent
    l'impact de nos incertitudes théoriques sur les limites
    ``attendues'' et ``observées''. Ces résultats sont comparés à
    l'analyse du canal $H\!\to \!\gamma \gamma$ faite par la
    collaboration ATLAS avec une luminosité de 37 pb$^{-1}$.}
\label{fig:SMHiggsMSSMfrench}
\end{figure}

La figure~\ref{fig:SMHiggsMSSMfrench} ci-dessous montre le résultat de
cette étude, à comparer avec la sensibilité qu'obtient ATLAS dans le
canal $H\to\gamma\gamma$ qui est considéré comme le meilleur canal de 
recherche pour des bosons de Higgs du Modèle Standard dans la gamme
115--140 GeV. Le canal $gg\to H\to\tau^+\tau^-$, qui n'est que peu ou
pas considéré par les collaborations expérimentales, se révèle plutôt
intéressant et compétitif, d'autant plus lorsque il est combiné avec
le canal de la fusion de bosons faibles suivie de la
désintégration $H\to\tau^+\tau^-$, le seul réellement considéré par
les collaborations expérimentales dans le canal de recherche en
di--taus. Nous invitons donc les expériences à utiliser ce nouveau
moyen de détection du boson de Higgs du Modèle Standard, et cela est
d'ailleurs ce qui se fait dans les derniers résultats présentés à
Grenoble en juillet 2011 à la conférence HEP--EPS.

\subsection{Perspectives}

Nous avons vu qu'aussi bien dans le Modèle Standard que dans le MSSM,
l'impact des incertitudes théoriques est certain. Les expériences
ATLAS et CMS en cours au LHC n'ont toujours pas observé le boson de
Higgs, même si quelques signaux intéressants ont fait leur apparition
durant la fin de la rédaction de cette thèse, présentés à la
conférence HEP--EPS à Grenoble en juillet 2011. Peut-être que le
champagne pourra être débouché avant la fin de l'année ! En attendant,
nous pouvons donner quelques perspectives pour un travail futur :
\begin{itemize}
\item{La suite importante à donner à ce travail est une étude
    exclusive où l'on calcule en particulier les incertitudes
    théoriques canal par canal en tenant compte des coupures
    cinématiques imposées pour maximiser le rapport signal sur
    bruit. Quelques résultats préliminaires sont donnés dans le texte
    principal de la thèse, section~\ref{section:Exclusive}. Il faut
    aussi faire de même pour l'étude des bruits de fond standards
    eux-mêmes.}
\item{Étudier l'impact de la largeur finie du boson de Higgs : tout le
    travail présenté dans cette thèse se place dans l'approximation
    d'une largeur étroite, ce qui a permis de faire le produit entre
    section efficace et rapport de branchement pour étudier le signal
    complet, or pour des masses élevées du(des) boson(s) de Higgs la
    largeur totale de désintégration n'est plus négligeable.}
\item{Il existe une étude complémentaire reliée au problème précédent
    : quel peut-être l'impact des interférences entre signal et bruit
    de fond ? En effet la production du boson de Higgs suivi de sa
    désintégration peut être vue comme n'étant qu'un processus
    intermédiaire parmi tant d'autres pour produire les produits de
    désintégration du boson de Higgs à partir des partons initiaux. Il
    s'agit donc d'étudier l'effet de l'interférence entre l'amplitude
    du signal et l'amplitude du bruit de fond sur l'observation du
    premier.}
\item{Il pourrait être intéressant de faire une étude précise des
    coupures cinématiques vis-à-vis des incertitudes d'échelles, ces
    dernières étant sensibles aux coupures. Ainsi, quelles sont les
    coupures optimales pour à la fois maximiser le rapport signal sur
    bruit et minimiser les incertitudes d'échelle ?}
\item{Dans l'hypothèse d'une découverte du boson de Higgs, il reste
    énormément de travail aux physiciens des hautes énergies : quelles
  sont ses propriétés exactes, est-il exactement celui du Modèle
  Standard, quelles informations nous apporte-t-il sur la brisure
  électrofaible ? Nous avons encore beaucoup de choses fascinantes à
  découvrir !}
\end{itemize}

\selectlanguage{english}



\addcontentsline{toc}{part}{References}

\bibliography{these}

\bibliographystyle{these}

\end{document}